\pdfoutput=1
\documentclass[11pt]{article}
\usepackage[margin=1in]{geometry}
\usepackage{amsmath,amssymb,amsthm}
\usepackage{booktabs}
\usepackage{longtable}
\usepackage{graphicx}
\usepackage{url}
\usepackage{tikz}
\usetikzlibrary{arrows.meta,positioning,circuits.logic.US,shapes.geometric,calc}
\usepackage{algorithm}
\usepackage{algpseudocode}
\usepackage[section]{placeins}

\usepackage[hidelinks]{hyperref}
\newif\ifarxiv
\arxivtrue
\newif\iffull
\fulltrue

\newtheorem{theorem}{Theorem}
\newtheorem{proposition}{Proposition}
\newtheorem{lemma}{Lemma}
\newtheorem{corollary}{Corollary}
\theoremstyle{remark}
\newtheorem{remark}{Remark}
\newtheorem{example}{Example}
\newcommand{\F}{\mathbb{F}}
\newcommand{\GL}{\mathrm{GL}}
\newcommand{\AGL}{\mathrm{AGL}}
\newcommand{\EA}{\mathrm{EA}}
\newcommand{\Var}{\operatorname{Var}}
\newcommand{\rank}{\operatorname{rank}}
\newcommand{\spn}{\operatorname{span}}
\newcommand{\supp}{\operatorname{supp}}
\newcommand{\RG}{R_G}
\newcommand{\Sh}{\widehat{S}}
\newcommand{\one}{\mathbf{1}}
\newcommand{\deff}{d_{\mathrm{eff}}}
\newcommand{\Stab}{\operatorname{Stab}}
\newcommand{\Wr}{\mathcal W}

\newcommand{\AbRatioEight}{30}
\newcommand{\AbSharedClasses}{2}
\newcommand{\AbSharedMags}{0, 4, 8, 12, 16, 32, 96}
\newcommand{\AbSharedText}{one group of seven inputs}
\newcommand{\AbUsBrute}{306}
\newcommand{\AbUsElevenFour}{$<$0.35}
\newcommand{\AbUsElevenSix}{0.27}
\newcommand{\AbUsPerFour}{0.29}
\newcommand{\AbUsPerSix}{2.24}
\newcommand{\AnChernoffOver}{29}
\newcommand{\AnChernoffWorst}{0.42}
\newcommand{\AnLTableC}{23}
\newcommand{\AnLTableCfour}{6}
\newcommand{\AnLTableCsixteen}{2}
\newcommand{\AnMajBitErrB}{0.0035}
\newcommand{\AnMajLevels}{$0$, $16$ and $144$}
\newcommand{\AnMajOrbits}{6}
\newcommand{\AnMajRawTableB}{0.81}
\newcommand{\AnMajTableB}{0.10}
\newcommand{\CcChowFour}{68}
\newcommand{\CcClassesFive}{616{,}126}
\newcommand{\CcClassesFour}{222}
\newcommand{\CcMagAllFour}{8}
\newcommand{\CcMagFive}{40}
\newcommand{\CcOrdFive}{80{,}527}
\newcommand{\CcOrdFour}{186}
\newcommand{\CcOrdPctFive}{13.1}
\newcommand{\CcPerFive}{535{,}371}
\newcommand{\CcPerFour}{219}
\newcommand{\ClAGLClasses}{18}
\newcommand{\ClAGLOwn}{9}
\newcommand{\ClBBisp}{222}
\newcommand{\ClBClasses}{222}
\newcommand{\ClBCollisionSize}{4}
\newcommand{\ClBOwn}{145}
\newcommand{\ClBTper}{219}
\newcommand{\ClEAClasses}{8}
\newcommand{\CoFailClusA}{0.34}
\newcommand{\CoFailHullA}{0.65}
\newcommand{\CoFailRandA}{1.69}
\newcommand{\CoFallbackText}{for none of the estimates with uniform solutions and for 45 of the 249{,}300 estimates with clustered solutions}
\newcommand{\CoOutputs}{277}
\newcommand{\CoOverClusA}{0.0}
\newcommand{\CoOverHullA}{0.0}
\newcommand{\CoOverRandA}{1.1}
\newcommand{\CoRuns}{100}
\newcommand{\CoTightClusA}{3.11}
\newcommand{\CoTightClusB}{0.37}
\newcommand{\CoTightHullA}{4.14}
\newcommand{\CoTightHullB}{0.41}
\newcommand{\CoTightOverClusA}{20.6}
\newcommand{\CoTightOverHullA}{34.3}
\newcommand{\CoTightOverRandA}{39.7}
\newcommand{\CoTightRandA}{5.16}
\newcommand{\CoTightRandB}{0.53}
\newcommand{\CuBentFour}{1}
\newcommand{\CuCircuits}{11}
\newcommand{\CuClassFive}{138}
\newcommand{\CuClassFour}{58}
\newcommand{\CuClassSix}{160}
\newcommand{\CuFunctions}{8{,}778}
\newcommand{\CuPerNode}{24}
\newcommand{\CuTopFour}{18}
\newcommand{\GjClauses}{400}
\newcommand{\GjDefault}{0.49}
\newcommand{\GjLimit}{120}
\newcommand{\GjVertices}{100}
\newcommand{\HcBlindRatio}{1.01}
\newcommand{\HcBlindSd}{0.141}
\newcommand{\HcCircuits}{22}
\newcommand{\HcHullCat}{3.1}
\newcommand{\HcHullRatio}{0.50}
\newcommand{\HcHullSevere}{0.01}
\newcommand{\HcNmax}{24}
\newcommand{\HcNmin}{9}
\newcommand{\HcOracleSd}{0.003}
\newcommand{\HcOutputs}{277}
\newcommand{\HcPilotCatB}{2.6}
\newcommand{\HcPilotCatD}{1.3}
\newcommand{\HcPilotRatioB}{0.40}
\newcommand{\HcPilotRatioD}{0.22}
\newcommand{\HcPool}{256}
\newcommand{\HcRandCatPct}{4.2}
\newcommand{\HcRandMedSd}{0.013}
\newcommand{\HcRandSevere}{0.78}
\newcommand{\HcRpilot}{40}
\newcommand{\HcRrand}{500}
\newcommand{\HcTailOnePct}{93}
\newcommand{\LuBitSixB}{0.26}
\newcommand{\LuHalvedSixB}{75}
\newcommand{\LuNontrivSix}{99.1}
\newcommand{\LuRawSixB}{0.47}
\newcommand{\LuTableFiveB}{0.49}
\newcommand{\LuTableFourB}{0.60}
\newcommand{\LuTableSixB}{0.29}
\newcommand{\MhCluHalf}{58.9}
\newcommand{\MhCluMed}{0.396}
\newcommand{\MhCluZero}{221}
\newcommand{\MhCoreAffine}{34}
\newcommand{\MhCoreBeyondText}{0.01 percent of the clustered calls and none of the uniform calls}
\newcommand{\MhCoreCluMax}{2.5}
\newcommand{\MhCoreCluMiss}{0.24}
\newcommand{\MhCoreExactMiss}{0.62}
\newcommand{\MhCoreOutputs}{203}
\newcommand{\MhCoreProper}{117}
\newcommand{\MhCoreRandMax}{5.0}
\newcommand{\MhCoreRandMiss}{1.04}
\newcommand{\MhCoreRuns}{200}
\newcommand{\MhCoreSmax}{2^{17}}
\newcommand{\MhCoreUniMax}{4.0}
\newcommand{\MhCoreUniMiss}{0.55}
\newcommand{\MhEOneCases}{143}
\newcommand{\MhEOneSets}{60}
\newcommand{\MhETwoDiff}{$2\times10^{-16}$}
\newcommand{\MhETwoDraws}{5}
\newcommand{\MhETwoOutputs}{277}
\newcommand{\MhEps}{0.8}
\newcommand{\MhMaxRatio}{1.0007}
\newcommand{\MhThresh}{73}
\newcommand{\MhUniHalf}{35.0}
\newcommand{\MhUniMed}{0.746}
\newcommand{\MhUniZero}{200}
\newcommand{\NpClasses}{616{,}126}
\newcommand{\NpGroupOrder}{7{,}680}
\newcommand{\NpResidClasses}{23}
\newcommand{\NpResidGroups}{5}
\newcommand{\NpSepBisp}{616{,}108}
\newcommand{\NpSepPer}{535{,}371}
\newcommand{\NpSepPerPct}{86.9}
\newcommand{\NyBelowC}{70}
\newcommand{\NyBigCount}{66}
\newcommand{\NyBigRatioC}{0.007}
\newcommand{\NyBigTestC}{0.017}
\newcommand{\NyCircuits}{34}
\newcommand{\NyFunctions}{306}
\newcommand{\NyLevel}{0.01}
\newcommand{\NyNmax}{10}
\newcommand{\NyNmin}{5}
\newcommand{\NyRatioA}{0.32}
\newcommand{\NyRatioB}{0.39}
\newcommand{\NyRatioC}{0.42}
\newcommand{\NyRatioMidA}{0.87}
\newcommand{\NyRatioMidB}{0.57}
\newcommand{\NyRatioSmallC}{1.41}
\newcommand{\NyRestC}{144}
\newcommand{\NyRestOnErrC}{0.94}
\newcommand{\NyTestA}{0.22}
\newcommand{\NyTestB}{0.28}
\newcommand{\NyTestBetterC}{59}
\newcommand{\NyTestC}{0.34}
\newcommand{\NyTestLevel}{0.01}
\newcommand{\NyTestMidA}{0.35}
\newcommand{\NyTestMidB}{0.32}
\newcommand{\NyTestSmallC}{1.26}
\newcommand{\NyTestWorseC}{21}
\newcommand{\NyTrials}{50}
\newcommand{\PsCmsMax}{1.5}
\newcommand{\PsMiterDeadInputs}{60}
\newcommand{\PsMiterDeadVars}{178}
\newcommand{\PsMiterMaxVars}{958}
\newcommand{\PsSeeds}{3}
\newcommand{\PsSpecMax}{0.11}
\newcommand{\PsTimeout}{60}
\newcommand{\PsTseitinDeadVars}{180}
\newcommand{\PsTseitinMaxVars}{600}
\newcommand{\PsXorDeadVars}{800}
\newcommand{\PsXorMaxVars}{1000}
\newcommand{\ScAmcSentence}{ApproxMC is not run as a reference here, since its Python package on this machine (4.1.24) fails the self-test of the counter, a projected count of 8 with a free sampling variable.}
\newcommand{\ScCallLimit}{30}
\newcommand{\ScConstRate}{0.25}
\newcommand{\ScExactTimeout}{60}
\newcommand{\ScExamined}{316}
\newcommand{\ScFamCap}{8}
\newcommand{\ScFamilies}{12}
\newcommand{\ScFormulas}{1{,}896}
\newcommand{\ScHullCalls}{49}
\newcommand{\ScHullLimit}{120}
\newcommand{\ScHullSec}{0.19}
\newcommand{\ScInstances}{74}
\newcommand{\ScMatchMiss}{0.9}
\newcommand{\ScMatchMissThree}{0.0}
\newcommand{\ScMatchRow}{30.9}
\newcommand{\ScMedLog}{38}
\newcommand{\ScMedN}{72}
\newcommand{\ScMedRatio}{0.97}
\newcommand{\ScNmax}{1{,}000}
\newcommand{\ScOnlyMatched}{2}
\newcommand{\ScOnlyMatchedReduced}{1}
\newcommand{\ScOnlyRandom}{0}
\newcommand{\ScPaired}{63}
\newcommand{\ScPairedReduced}{17}
\newcommand{\ScProper}{47}
\newcommand{\ScProperDense}{11}
\newcommand{\ScRandMiss}{0.5}
\newcommand{\ScRandMissThree}{0.0}
\newcommand{\ScRandRow}{45.2}
\newcommand{\ScReduced}{19}
\newcommand{\ScRuns}{9}
\newcommand{\ScSkipCounter}{97}
\newcommand{\ScSkipExact}{105}
\newcommand{\ScSkipSmall}{40}
\newcommand{\ScTimeRatio}{0.98}
\newcommand{\ScTimeRatioReduced}{0.90}
\newcommand{\SvAffinePct}{25}
\newcommand{\SvBerkeley}{19}
\newcommand{\SvBerkeleyCoverNames}{\texttt{addm4}, \texttt{mlp4} and \texttt{risc}}
\newcommand{\SvBerkeleyCoverWord}{three}
\newcommand{\SvCircuits}{51}
\newcommand{\SvCollapsedNames}{\texttt{alu2c}, \texttt{cm85a}, \texttt{majority} and \texttt{x2}}
\newcommand{\SvCollapsedWord}{four}
\newcommand{\SvCordicK}{14}
\newcommand{\SvCordicUsed}{23}
\newcommand{\SvCubeOfAffinePct}{87}
\newcommand{\SvDtwoLowPct}{92}
\newcommand{\SvHidden}{50}
\newcommand{\SvHiddenPct}{7.3}
\newcommand{\SvHullMedian}{4}
\newcommand{\SvHullPct}{64}
\newcommand{\SvLarge}{12}
\newcommand{\SvLargeNmax}{24}
\newcommand{\SvLargeNmin}{14}
\newcommand{\SvLgsPla}{16}
\newcommand{\SvOutputs}{688}
\newcommand{\SvPairPct}{78}
\newcommand{\SvSmall}{39}
\newcommand{\SvTransPct}{59}
\makeatletter
\newcommand{\Rg}[1]{\@ifundefined{rg@#1}{\textbf{??#1??}}{\csname rg@#1\endcsname}}
\expandafter\def\csname rg@8div.equalities\endcsname{2,065}
\expandafter\def\csname rg@8div.rank.16\endcsname{2,062}
\expandafter\def\csname rg@8div.rank.2\endcsname{1,049}
\expandafter\def\csname rg@8div.rank.4\endcsname{1,094}
\expandafter\def\csname rg@8div.rank.8\endcsname{1,215}
\expandafter\def\csname rg@8div.rank.Q\endcsname{2,063}
\expandafter\def\csname rg@bipalo.native.exact:pb.n\endcsname{160}
\expandafter\def\csname rg@bipalo.native.exact:pb.s\endcsname{0.01}
\expandafter\def\csname rg@bipalo.native.highs:pb.n\endcsname{160}
\expandafter\def\csname rg@bipalo.native.highs:pb.s\endcsname{0.02}
\expandafter\def\csname rg@bipalo.native.n\endcsname{160}
\expandafter\def\csname rg@bipalo.native.roundingsat:pb.n\endcsname{160}
\expandafter\def\csname rg@bipalo.native.roundingsat:pb.s\endcsname{0.01}
\expandafter\def\csname rg@bipalo.native.sat4j-cp:pb.n\endcsname{160}
\expandafter\def\csname rg@bipalo.native.sat4j-cp:pb.s\endcsname{0.38}
\expandafter\def\csname rg@bipalo.native.scip:pb.n\endcsname{160}
\expandafter\def\csname rg@bipalo.native.scip:pb.s\endcsname{0.03}
\expandafter\def\csname rg@bipalo.native.smax\endcsname{0.38}
\expandafter\def\csname rg@bipalo.native.smin\endcsname{0.01}
\expandafter\def\csname rg@bipalo.native.tools\endcsname{all five}
\expandafter\def\csname rg@bipalo.native.toolsp\endcsname{all five native-form solvers}
\expandafter\def\csname rg@bipalo.pairwise.best.n\endcsname{160}
\expandafter\def\csname rg@bipalo.pairwise.bve+hull.n\endcsname{160}
\expandafter\def\csname rg@bipalo.pairwise.bve+hull.s\endcsname{3.2}
\expandafter\def\csname rg@bipalo.pairwise.bve+int.n\endcsname{--}
\expandafter\def\csname rg@bipalo.pairwise.bve+int.s\endcsname{--}
\expandafter\def\csname rg@bipalo.pairwise.bve+win.n\endcsname{--}
\expandafter\def\csname rg@bipalo.pairwise.bve+win.s\endcsname{--}
\expandafter\def\csname rg@bipalo.pairwise.cadical.n\endcsname{20}
\expandafter\def\csname rg@bipalo.pairwise.cadical.s\endcsname{5.4}
\expandafter\def\csname rg@bipalo.pairwise.cms.n\endcsname{20}
\expandafter\def\csname rg@bipalo.pairwise.cms.s\endcsname{9.2}
\expandafter\def\csname rg@bipalo.pairwise.cnf.n\endcsname{160}
\expandafter\def\csname rg@bipalo.pairwise.cnf.s\endcsname{0.01}
\expandafter\def\csname rg@bipalo.pairwise.cnf.tools\endcsname{Lingeling}
\expandafter\def\csname rg@bipalo.pairwise.cnfx.n\endcsname{160}
\expandafter\def\csname rg@bipalo.pairwise.cnfx.tools\endcsname{Lingeling}
\expandafter\def\csname rg@bipalo.pairwise.exact:cnf.n\endcsname{--}
\expandafter\def\csname rg@bipalo.pairwise.exact:cnf.s\endcsname{--}
\expandafter\def\csname rg@bipalo.pairwise.hull.n\endcsname{160}
\expandafter\def\csname rg@bipalo.pairwise.hull.s\endcsname{0.24}
\expandafter\def\csname rg@bipalo.pairwise.int.n\endcsname{--}
\expandafter\def\csname rg@bipalo.pairwise.int.s\endcsname{--}
\expandafter\def\csname rg@bipalo.pairwise.kissat.n\endcsname{20}
\expandafter\def\csname rg@bipalo.pairwise.kissat.s\endcsname{2.7}
\expandafter\def\csname rg@bipalo.pairwise.lingeling.n\endcsname{160}
\expandafter\def\csname rg@bipalo.pairwise.lingeling.s\endcsname{0.01}
\expandafter\def\csname rg@bipalo.pairwise.pgpbs-2.n\endcsname{80}
\expandafter\def\csname rg@bipalo.pairwise.pgpbs-2.s\endcsname{14.9}
\expandafter\def\csname rg@bipalo.pairwise.pgpbs-3.n\endcsname{80}
\expandafter\def\csname rg@bipalo.pairwise.pgpbs-3.s\endcsname{15.1}
\expandafter\def\csname rg@bipalo.pairwise.pgpbs-a.n\endcsname{80}
\expandafter\def\csname rg@bipalo.pairwise.pgpbs-a.s\endcsname{14.7}
\expandafter\def\csname rg@bipalo.pairwise.pgpbs-i.n\endcsname{80}
\expandafter\def\csname rg@bipalo.pairwise.pgpbs-i.s\endcsname{15.3}
\expandafter\def\csname rg@bipalo.pairwise.read.fastest\endcsname{--}
\expandafter\def\csname rg@bipalo.pairwise.read.n\endcsname{--}
\expandafter\def\csname rg@bipalo.pairwise.read.s\endcsname{--}
\expandafter\def\csname rg@bipalo.pairwise.read.smax\endcsname{--}
\expandafter\def\csname rg@bipalo.pairwise.read.tool\endcsname{--}
\expandafter\def\csname rg@bipalo.pairwise.roundingsat:cnf.n\endcsname{--}
\expandafter\def\csname rg@bipalo.pairwise.roundingsat:cnf.s\endcsname{--}
\expandafter\def\csname rg@bipalo.pairwise.sat4j-cp:cnf.n\endcsname{--}
\expandafter\def\csname rg@bipalo.pairwise.sat4j-cp:cnf.s\endcsname{--}
\expandafter\def\csname rg@bipalo.pairwise.scip:cnf.n\endcsname{20}
\expandafter\def\csname rg@bipalo.pairwise.scip:cnf.s\endcsname{0.07}
\expandafter\def\csname rg@bipalo.pairwise.top\endcsname{160}
\expandafter\def\csname rg@bipalo.pairwise.win.n\endcsname{--}
\expandafter\def\csname rg@bipalo.pairwise.win.s\endcsname{--}
\expandafter\def\csname rg@bipalo.pairwise.xor.n\endcsname{--}
\expandafter\def\csname rg@bipalo.pairwise.xor.s\endcsname{--}
\expandafter\def\csname rg@bipalo.pairwise.xorcle-dev.n\endcsname{--}
\expandafter\def\csname rg@bipalo.pairwise.xorcle-dev.s\endcsname{--}
\expandafter\def\csname rg@bipalo.pairwise.xorcle.n\endcsname{--}
\expandafter\def\csname rg@bipalo.pairwise.xorcle.s\endcsname{--}
\expandafter\def\csname rg@bipalo.seq.best.n\endcsname{160}
\expandafter\def\csname rg@bipalo.seq.bve+hull.n\endcsname{160}
\expandafter\def\csname rg@bipalo.seq.bve+hull.s\endcsname{6.8}
\expandafter\def\csname rg@bipalo.seq.bve+int.n\endcsname{--}
\expandafter\def\csname rg@bipalo.seq.bve+int.s\endcsname{--}
\expandafter\def\csname rg@bipalo.seq.bve+win.n\endcsname{--}
\expandafter\def\csname rg@bipalo.seq.bve+win.s\endcsname{--}
\expandafter\def\csname rg@bipalo.seq.cadical.n\endcsname{20}
\expandafter\def\csname rg@bipalo.seq.cadical.s\endcsname{5.2}
\expandafter\def\csname rg@bipalo.seq.cms.n\endcsname{20}
\expandafter\def\csname rg@bipalo.seq.cms.s\endcsname{11.9}
\expandafter\def\csname rg@bipalo.seq.cnf.n\endcsname{80}
\expandafter\def\csname rg@bipalo.seq.cnf.s\endcsname{15.2}
\expandafter\def\csname rg@bipalo.seq.cnf.tools\endcsname{pgpbs, pgpbs mod 2, pgpbs mod 3, pgpbs integer}
\expandafter\def\csname rg@bipalo.seq.cnfx.n\endcsname{80}
\expandafter\def\csname rg@bipalo.seq.cnfx.tools\endcsname{pgpbs, pgpbs mod 2, pgpbs mod 3, pgpbs integer}
\expandafter\def\csname rg@bipalo.seq.exact:cnf.n\endcsname{--}
\expandafter\def\csname rg@bipalo.seq.exact:cnf.s\endcsname{--}
\expandafter\def\csname rg@bipalo.seq.hull.n\endcsname{160}
\expandafter\def\csname rg@bipalo.seq.hull.s\endcsname{2.7}
\expandafter\def\csname rg@bipalo.seq.int.n\endcsname{--}
\expandafter\def\csname rg@bipalo.seq.int.s\endcsname{--}
\expandafter\def\csname rg@bipalo.seq.kissat.n\endcsname{20}
\expandafter\def\csname rg@bipalo.seq.kissat.s\endcsname{2.1}
\expandafter\def\csname rg@bipalo.seq.lingeling.n\endcsname{20}
\expandafter\def\csname rg@bipalo.seq.lingeling.s\endcsname{4.0}
\expandafter\def\csname rg@bipalo.seq.pgpbs-2.n\endcsname{80}
\expandafter\def\csname rg@bipalo.seq.pgpbs-2.s\endcsname{15.2}
\expandafter\def\csname rg@bipalo.seq.pgpbs-3.n\endcsname{80}
\expandafter\def\csname rg@bipalo.seq.pgpbs-3.s\endcsname{15.0}
\expandafter\def\csname rg@bipalo.seq.pgpbs-a.n\endcsname{80}
\expandafter\def\csname rg@bipalo.seq.pgpbs-a.s\endcsname{14.9}
\expandafter\def\csname rg@bipalo.seq.pgpbs-i.n\endcsname{80}
\expandafter\def\csname rg@bipalo.seq.pgpbs-i.s\endcsname{15.0}
\expandafter\def\csname rg@bipalo.seq.read.fastest\endcsname{--}
\expandafter\def\csname rg@bipalo.seq.read.n\endcsname{--}
\expandafter\def\csname rg@bipalo.seq.read.s\endcsname{--}
\expandafter\def\csname rg@bipalo.seq.read.smax\endcsname{--}
\expandafter\def\csname rg@bipalo.seq.read.tool\endcsname{--}
\expandafter\def\csname rg@bipalo.seq.roundingsat:cnf.n\endcsname{--}
\expandafter\def\csname rg@bipalo.seq.roundingsat:cnf.s\endcsname{--}
\expandafter\def\csname rg@bipalo.seq.sat4j-cp:cnf.n\endcsname{--}
\expandafter\def\csname rg@bipalo.seq.sat4j-cp:cnf.s\endcsname{--}
\expandafter\def\csname rg@bipalo.seq.scip:cnf.n\endcsname{20}
\expandafter\def\csname rg@bipalo.seq.scip:cnf.s\endcsname{24.2}
\expandafter\def\csname rg@bipalo.seq.top\endcsname{160}
\expandafter\def\csname rg@bipalo.seq.win.n\endcsname{--}
\expandafter\def\csname rg@bipalo.seq.win.s\endcsname{--}
\expandafter\def\csname rg@bipalo.seq.xor.n\endcsname{--}
\expandafter\def\csname rg@bipalo.seq.xor.s\endcsname{--}
\expandafter\def\csname rg@bipalo.seq.xorcle-dev.n\endcsname{--}
\expandafter\def\csname rg@bipalo.seq.xorcle-dev.s\endcsname{--}
\expandafter\def\csname rg@bipalo.seq.xorcle.n\endcsname{--}
\expandafter\def\csname rg@bipalo.seq.xorcle.s\endcsname{--}
\expandafter\def\csname rg@bipmatch.binary.best.n\endcsname{160}
\expandafter\def\csname rg@bipmatch.binary.bve+hull.n\endcsname{160}
\expandafter\def\csname rg@bipmatch.binary.bve+hull.s\endcsname{2.6}
\expandafter\def\csname rg@bipmatch.binary.bve+int.n\endcsname{160}
\expandafter\def\csname rg@bipmatch.binary.bve+int.s\endcsname{0.47}
\expandafter\def\csname rg@bipmatch.binary.bve+win.n\endcsname{160}
\expandafter\def\csname rg@bipmatch.binary.bve+win.s\endcsname{0.47}
\expandafter\def\csname rg@bipmatch.binary.cadical.n\endcsname{20}
\expandafter\def\csname rg@bipmatch.binary.cadical.s\endcsname{0.08}
\expandafter\def\csname rg@bipmatch.binary.cms.n\endcsname{20}
\expandafter\def\csname rg@bipmatch.binary.cms.s\endcsname{0.23}
\expandafter\def\csname rg@bipmatch.binary.cnf.n\endcsname{160}
\expandafter\def\csname rg@bipmatch.binary.cnf.s\endcsname{49.2}
\expandafter\def\csname rg@bipmatch.binary.cnf.tools\endcsname{Xorcle}
\expandafter\def\csname rg@bipmatch.binary.cnfx.n\endcsname{20}
\expandafter\def\csname rg@bipmatch.binary.cnfx.tools\endcsname{CDCL, Sat4j, RoundingSat, Exact, SCIP}
\expandafter\def\csname rg@bipmatch.binary.exact:cnf.n\endcsname{20}
\expandafter\def\csname rg@bipmatch.binary.exact:cnf.s\endcsname{1.7}
\expandafter\def\csname rg@bipmatch.binary.hull.n\endcsname{160}
\expandafter\def\csname rg@bipmatch.binary.hull.s\endcsname{2.6}
\expandafter\def\csname rg@bipmatch.binary.int.n\endcsname{160}
\expandafter\def\csname rg@bipmatch.binary.int.s\endcsname{0.84}
\expandafter\def\csname rg@bipmatch.binary.kissat.n\endcsname{20}
\expandafter\def\csname rg@bipmatch.binary.kissat.s\endcsname{0.18}
\expandafter\def\csname rg@bipmatch.binary.lingeling.n\endcsname{20}
\expandafter\def\csname rg@bipmatch.binary.lingeling.s\endcsname{0.15}
\expandafter\def\csname rg@bipmatch.binary.pgpbs-2.n\endcsname{--}
\expandafter\def\csname rg@bipmatch.binary.pgpbs-2.s\endcsname{--}
\expandafter\def\csname rg@bipmatch.binary.pgpbs-3.n\endcsname{--}
\expandafter\def\csname rg@bipmatch.binary.pgpbs-3.s\endcsname{--}
\expandafter\def\csname rg@bipmatch.binary.pgpbs-a.n\endcsname{--}
\expandafter\def\csname rg@bipmatch.binary.pgpbs-a.s\endcsname{--}
\expandafter\def\csname rg@bipmatch.binary.pgpbs-i.n\endcsname{--}
\expandafter\def\csname rg@bipmatch.binary.pgpbs-i.s\endcsname{--}
\expandafter\def\csname rg@bipmatch.binary.read.fastest\endcsname{field after elim.}
\expandafter\def\csname rg@bipmatch.binary.read.n\endcsname{160}
\expandafter\def\csname rg@bipmatch.binary.read.s\endcsname{0.47}
\expandafter\def\csname rg@bipmatch.binary.read.smax\endcsname{0.84}
\expandafter\def\csname rg@bipmatch.binary.read.tool\endcsname{field, ring, field after elim., ring after elim.}
\expandafter\def\csname rg@bipmatch.binary.roundingsat:cnf.n\endcsname{20}
\expandafter\def\csname rg@bipmatch.binary.roundingsat:cnf.s\endcsname{1.0}
\expandafter\def\csname rg@bipmatch.binary.sat4j-cp:cnf.n\endcsname{20}
\expandafter\def\csname rg@bipmatch.binary.sat4j-cp:cnf.s\endcsname{1.9}
\expandafter\def\csname rg@bipmatch.binary.scip:cnf.n\endcsname{20}
\expandafter\def\csname rg@bipmatch.binary.scip:cnf.s\endcsname{12.1}
\expandafter\def\csname rg@bipmatch.binary.top\endcsname{160}
\expandafter\def\csname rg@bipmatch.binary.win.n\endcsname{160}
\expandafter\def\csname rg@bipmatch.binary.win.s\endcsname{0.68}
\expandafter\def\csname rg@bipmatch.binary.xor.n\endcsname{160}
\expandafter\def\csname rg@bipmatch.binary.xor.s\endcsname{46.9}
\expandafter\def\csname rg@bipmatch.binary.xorcle-dev.n\endcsname{160}
\expandafter\def\csname rg@bipmatch.binary.xorcle-dev.s\endcsname{49.2}
\expandafter\def\csname rg@bipmatch.binary.xorcle.n\endcsname{160}
\expandafter\def\csname rg@bipmatch.binary.xorcle.s\endcsname{46.9}
\expandafter\def\csname rg@bipmatch.commander.best.n\endcsname{160}
\expandafter\def\csname rg@bipmatch.commander.bve+hull.n\endcsname{160}
\expandafter\def\csname rg@bipmatch.commander.bve+hull.s\endcsname{2.8}
\expandafter\def\csname rg@bipmatch.commander.bve+int.n\endcsname{160}
\expandafter\def\csname rg@bipmatch.commander.bve+int.s\endcsname{0.75}
\expandafter\def\csname rg@bipmatch.commander.bve+win.n\endcsname{160}
\expandafter\def\csname rg@bipmatch.commander.bve+win.s\endcsname{0.75}
\expandafter\def\csname rg@bipmatch.commander.cadical.n\endcsname{20}
\expandafter\def\csname rg@bipmatch.commander.cadical.s\endcsname{0.10}
\expandafter\def\csname rg@bipmatch.commander.cms.n\endcsname{20}
\expandafter\def\csname rg@bipmatch.commander.cms.s\endcsname{0.14}
\expandafter\def\csname rg@bipmatch.commander.cnf.n\endcsname{160}
\expandafter\def\csname rg@bipmatch.commander.cnf.s\endcsname{48.5}
\expandafter\def\csname rg@bipmatch.commander.cnf.tools\endcsname{Lingeling, pgpbs, pgpbs mod 2, pgpbs mod 3, pgpbs integer, SCIP, Xorcle}
\expandafter\def\csname rg@bipmatch.commander.cnfx.n\endcsname{160}
\expandafter\def\csname rg@bipmatch.commander.cnfx.tools\endcsname{Lingeling, pgpbs, pgpbs mod 2, pgpbs mod 3, pgpbs integer, SCIP}
\expandafter\def\csname rg@bipmatch.commander.exact:cnf.n\endcsname{20}
\expandafter\def\csname rg@bipmatch.commander.exact:cnf.s\endcsname{1.4}
\expandafter\def\csname rg@bipmatch.commander.hull.n\endcsname{160}
\expandafter\def\csname rg@bipmatch.commander.hull.s\endcsname{2.0}
\expandafter\def\csname rg@bipmatch.commander.int.n\endcsname{160}
\expandafter\def\csname rg@bipmatch.commander.int.s\endcsname{1.4}
\expandafter\def\csname rg@bipmatch.commander.kissat.n\endcsname{20}
\expandafter\def\csname rg@bipmatch.commander.kissat.s\endcsname{0.12}
\expandafter\def\csname rg@bipmatch.commander.lingeling.n\endcsname{160}
\expandafter\def\csname rg@bipmatch.commander.lingeling.s\endcsname{0.01}
\expandafter\def\csname rg@bipmatch.commander.pgpbs-2.n\endcsname{160}
\expandafter\def\csname rg@bipmatch.commander.pgpbs-2.s\endcsname{2.2}
\expandafter\def\csname rg@bipmatch.commander.pgpbs-3.n\endcsname{160}
\expandafter\def\csname rg@bipmatch.commander.pgpbs-3.s\endcsname{3.4}
\expandafter\def\csname rg@bipmatch.commander.pgpbs-a.n\endcsname{160}
\expandafter\def\csname rg@bipmatch.commander.pgpbs-a.s\endcsname{2.4}
\expandafter\def\csname rg@bipmatch.commander.pgpbs-i.n\endcsname{160}
\expandafter\def\csname rg@bipmatch.commander.pgpbs-i.s\endcsname{48.5}
\expandafter\def\csname rg@bipmatch.commander.read.fastest\endcsname{ring after elim.}
\expandafter\def\csname rg@bipmatch.commander.read.n\endcsname{160}
\expandafter\def\csname rg@bipmatch.commander.read.s\endcsname{0.75}
\expandafter\def\csname rg@bipmatch.commander.read.smax\endcsname{1.4}
\expandafter\def\csname rg@bipmatch.commander.read.tool\endcsname{field, ring, field after elim., ring after elim.}
\expandafter\def\csname rg@bipmatch.commander.roundingsat:cnf.n\endcsname{20}
\expandafter\def\csname rg@bipmatch.commander.roundingsat:cnf.s\endcsname{0.32}
\expandafter\def\csname rg@bipmatch.commander.sat4j-cp:cnf.n\endcsname{20}
\expandafter\def\csname rg@bipmatch.commander.sat4j-cp:cnf.s\endcsname{1.5}
\expandafter\def\csname rg@bipmatch.commander.scip:cnf.n\endcsname{160}
\expandafter\def\csname rg@bipmatch.commander.scip:cnf.s\endcsname{0.16}
\expandafter\def\csname rg@bipmatch.commander.top\endcsname{160}
\expandafter\def\csname rg@bipmatch.commander.win.n\endcsname{160}
\expandafter\def\csname rg@bipmatch.commander.win.s\endcsname{1.2}
\expandafter\def\csname rg@bipmatch.commander.xor.n\endcsname{160}
\expandafter\def\csname rg@bipmatch.commander.xor.s\endcsname{31.2}
\expandafter\def\csname rg@bipmatch.commander.xorcle-dev.n\endcsname{160}
\expandafter\def\csname rg@bipmatch.commander.xorcle-dev.s\endcsname{33.9}
\expandafter\def\csname rg@bipmatch.commander.xorcle.n\endcsname{160}
\expandafter\def\csname rg@bipmatch.commander.xorcle.s\endcsname{31.2}
\expandafter\def\csname rg@bipmatch.native.exact:pb.n\endcsname{160}
\expandafter\def\csname rg@bipmatch.native.exact:pb.s\endcsname{0.01}
\expandafter\def\csname rg@bipmatch.native.highs:pb.n\endcsname{160}
\expandafter\def\csname rg@bipmatch.native.highs:pb.s\endcsname{0.01}
\expandafter\def\csname rg@bipmatch.native.n\endcsname{160}
\expandafter\def\csname rg@bipmatch.native.roundingsat:pb.n\endcsname{160}
\expandafter\def\csname rg@bipmatch.native.roundingsat:pb.s\endcsname{0.01}
\expandafter\def\csname rg@bipmatch.native.sat4j-cp:pb.n\endcsname{160}
\expandafter\def\csname rg@bipmatch.native.sat4j-cp:pb.s\endcsname{0.37}
\expandafter\def\csname rg@bipmatch.native.scip:pb.n\endcsname{160}
\expandafter\def\csname rg@bipmatch.native.scip:pb.s\endcsname{0.02}
\expandafter\def\csname rg@bipmatch.native.smax\endcsname{0.37}
\expandafter\def\csname rg@bipmatch.native.smin\endcsname{0.01}
\expandafter\def\csname rg@bipmatch.native.tools\endcsname{all five}
\expandafter\def\csname rg@bipmatch.native.toolsp\endcsname{all five native-form solvers}
\expandafter\def\csname rg@bipmatch.pairwise.best.n\endcsname{160}
\expandafter\def\csname rg@bipmatch.pairwise.bve+hull.n\endcsname{160}
\expandafter\def\csname rg@bipmatch.pairwise.bve+hull.s\endcsname{2.4}
\expandafter\def\csname rg@bipmatch.pairwise.bve+int.n\endcsname{160}
\expandafter\def\csname rg@bipmatch.pairwise.bve+int.s\endcsname{0.77}
\expandafter\def\csname rg@bipmatch.pairwise.bve+win.n\endcsname{160}
\expandafter\def\csname rg@bipmatch.pairwise.bve+win.s\endcsname{0.76}
\expandafter\def\csname rg@bipmatch.pairwise.cadical.n\endcsname{20}
\expandafter\def\csname rg@bipmatch.pairwise.cadical.s\endcsname{0.09}
\expandafter\def\csname rg@bipmatch.pairwise.cms.n\endcsname{20}
\expandafter\def\csname rg@bipmatch.pairwise.cms.s\endcsname{0.15}
\expandafter\def\csname rg@bipmatch.pairwise.cnf.n\endcsname{160}
\expandafter\def\csname rg@bipmatch.pairwise.cnf.s\endcsname{45.7}
\expandafter\def\csname rg@bipmatch.pairwise.cnf.tools\endcsname{pgpbs, pgpbs mod 2, pgpbs mod 3, pgpbs integer, SCIP, Xorcle}
\expandafter\def\csname rg@bipmatch.pairwise.cnfx.n\endcsname{160}
\expandafter\def\csname rg@bipmatch.pairwise.cnfx.tools\endcsname{pgpbs, pgpbs mod 2, pgpbs mod 3, pgpbs integer, SCIP}
\expandafter\def\csname rg@bipmatch.pairwise.exact:cnf.n\endcsname{20}
\expandafter\def\csname rg@bipmatch.pairwise.exact:cnf.s\endcsname{1.1}
\expandafter\def\csname rg@bipmatch.pairwise.hull.n\endcsname{160}
\expandafter\def\csname rg@bipmatch.pairwise.hull.s\endcsname{0.29}
\expandafter\def\csname rg@bipmatch.pairwise.int.n\endcsname{160}
\expandafter\def\csname rg@bipmatch.pairwise.int.s\endcsname{1.6}
\expandafter\def\csname rg@bipmatch.pairwise.kissat.n\endcsname{20}
\expandafter\def\csname rg@bipmatch.pairwise.kissat.s\endcsname{0.09}
\expandafter\def\csname rg@bipmatch.pairwise.lingeling.n\endcsname{20}
\expandafter\def\csname rg@bipmatch.pairwise.lingeling.s\endcsname{0.17}
\expandafter\def\csname rg@bipmatch.pairwise.pgpbs-2.n\endcsname{160}
\expandafter\def\csname rg@bipmatch.pairwise.pgpbs-2.s\endcsname{2.0}
\expandafter\def\csname rg@bipmatch.pairwise.pgpbs-3.n\endcsname{160}
\expandafter\def\csname rg@bipmatch.pairwise.pgpbs-3.s\endcsname{3.1}
\expandafter\def\csname rg@bipmatch.pairwise.pgpbs-a.n\endcsname{160}
\expandafter\def\csname rg@bipmatch.pairwise.pgpbs-a.s\endcsname{2.1}
\expandafter\def\csname rg@bipmatch.pairwise.pgpbs-i.n\endcsname{160}
\expandafter\def\csname rg@bipmatch.pairwise.pgpbs-i.s\endcsname{45.7}
\expandafter\def\csname rg@bipmatch.pairwise.read.fastest\endcsname{field after elim.}
\expandafter\def\csname rg@bipmatch.pairwise.read.n\endcsname{160}
\expandafter\def\csname rg@bipmatch.pairwise.read.s\endcsname{0.76}
\expandafter\def\csname rg@bipmatch.pairwise.read.smax\endcsname{1.6}
\expandafter\def\csname rg@bipmatch.pairwise.read.tool\endcsname{field, ring, field after elim., ring after elim.}
\expandafter\def\csname rg@bipmatch.pairwise.roundingsat:cnf.n\endcsname{20}
\expandafter\def\csname rg@bipmatch.pairwise.roundingsat:cnf.s\endcsname{1.3}
\expandafter\def\csname rg@bipmatch.pairwise.sat4j-cp:cnf.n\endcsname{20}
\expandafter\def\csname rg@bipmatch.pairwise.sat4j-cp:cnf.s\endcsname{1.7}
\expandafter\def\csname rg@bipmatch.pairwise.scip:cnf.n\endcsname{160}
\expandafter\def\csname rg@bipmatch.pairwise.scip:cnf.s\endcsname{0.06}
\expandafter\def\csname rg@bipmatch.pairwise.top\endcsname{160}
\expandafter\def\csname rg@bipmatch.pairwise.win.n\endcsname{160}
\expandafter\def\csname rg@bipmatch.pairwise.win.s\endcsname{1.5}
\expandafter\def\csname rg@bipmatch.pairwise.xor.n\endcsname{160}
\expandafter\def\csname rg@bipmatch.pairwise.xor.s\endcsname{12.4}
\expandafter\def\csname rg@bipmatch.pairwise.xorcle-dev.n\endcsname{160}
\expandafter\def\csname rg@bipmatch.pairwise.xorcle-dev.s\endcsname{12.4}
\expandafter\def\csname rg@bipmatch.pairwise.xorcle.n\endcsname{160}
\expandafter\def\csname rg@bipmatch.pairwise.xorcle.s\endcsname{12.8}
\expandafter\def\csname rg@bipmatch.seq.best.n\endcsname{160}
\expandafter\def\csname rg@bipmatch.seq.bve+hull.n\endcsname{160}
\expandafter\def\csname rg@bipmatch.seq.bve+hull.s\endcsname{2.6}
\expandafter\def\csname rg@bipmatch.seq.bve+int.n\endcsname{160}
\expandafter\def\csname rg@bipmatch.seq.bve+int.s\endcsname{0.80}
\expandafter\def\csname rg@bipmatch.seq.bve+win.n\endcsname{160}
\expandafter\def\csname rg@bipmatch.seq.bve+win.s\endcsname{0.80}
\expandafter\def\csname rg@bipmatch.seq.cadical.n\endcsname{20}
\expandafter\def\csname rg@bipmatch.seq.cadical.s\endcsname{0.08}
\expandafter\def\csname rg@bipmatch.seq.cms.n\endcsname{20}
\expandafter\def\csname rg@bipmatch.seq.cms.s\endcsname{0.19}
\expandafter\def\csname rg@bipmatch.seq.cnf.n\endcsname{160}
\expandafter\def\csname rg@bipmatch.seq.cnf.s\endcsname{47.3}
\expandafter\def\csname rg@bipmatch.seq.cnf.tools\endcsname{pgpbs, pgpbs mod 2, pgpbs mod 3, pgpbs integer, Xorcle}
\expandafter\def\csname rg@bipmatch.seq.cnfx.n\endcsname{160}
\expandafter\def\csname rg@bipmatch.seq.cnfx.tools\endcsname{pgpbs, pgpbs mod 2, pgpbs mod 3, pgpbs integer}
\expandafter\def\csname rg@bipmatch.seq.exact:cnf.n\endcsname{20}
\expandafter\def\csname rg@bipmatch.seq.exact:cnf.s\endcsname{1.7}
\expandafter\def\csname rg@bipmatch.seq.hull.n\endcsname{160}
\expandafter\def\csname rg@bipmatch.seq.hull.s\endcsname{2.1}
\expandafter\def\csname rg@bipmatch.seq.int.n\endcsname{160}
\expandafter\def\csname rg@bipmatch.seq.int.s\endcsname{1.1}
\expandafter\def\csname rg@bipmatch.seq.kissat.n\endcsname{20}
\expandafter\def\csname rg@bipmatch.seq.kissat.s\endcsname{0.10}
\expandafter\def\csname rg@bipmatch.seq.lingeling.n\endcsname{20}
\expandafter\def\csname rg@bipmatch.seq.lingeling.s\endcsname{0.17}
\expandafter\def\csname rg@bipmatch.seq.pgpbs-2.n\endcsname{160}
\expandafter\def\csname rg@bipmatch.seq.pgpbs-2.s\endcsname{2.4}
\expandafter\def\csname rg@bipmatch.seq.pgpbs-3.n\endcsname{160}
\expandafter\def\csname rg@bipmatch.seq.pgpbs-3.s\endcsname{3.6}
\expandafter\def\csname rg@bipmatch.seq.pgpbs-a.n\endcsname{160}
\expandafter\def\csname rg@bipmatch.seq.pgpbs-a.s\endcsname{2.6}
\expandafter\def\csname rg@bipmatch.seq.pgpbs-i.n\endcsname{160}
\expandafter\def\csname rg@bipmatch.seq.pgpbs-i.s\endcsname{47.3}
\expandafter\def\csname rg@bipmatch.seq.read.fastest\endcsname{field after elim.}
\expandafter\def\csname rg@bipmatch.seq.read.n\endcsname{160}
\expandafter\def\csname rg@bipmatch.seq.read.s\endcsname{0.80}
\expandafter\def\csname rg@bipmatch.seq.read.smax\endcsname{1.1}
\expandafter\def\csname rg@bipmatch.seq.read.tool\endcsname{field, ring, field after elim., ring after elim.}
\expandafter\def\csname rg@bipmatch.seq.roundingsat:cnf.n\endcsname{20}
\expandafter\def\csname rg@bipmatch.seq.roundingsat:cnf.s\endcsname{0.25}
\expandafter\def\csname rg@bipmatch.seq.sat4j-cp:cnf.n\endcsname{20}
\expandafter\def\csname rg@bipmatch.seq.sat4j-cp:cnf.s\endcsname{1.2}
\expandafter\def\csname rg@bipmatch.seq.scip:cnf.n\endcsname{20}
\expandafter\def\csname rg@bipmatch.seq.scip:cnf.s\endcsname{0.11}
\expandafter\def\csname rg@bipmatch.seq.top\endcsname{160}
\expandafter\def\csname rg@bipmatch.seq.win.n\endcsname{160}
\expandafter\def\csname rg@bipmatch.seq.win.s\endcsname{1.0}
\expandafter\def\csname rg@bipmatch.seq.xor.n\endcsname{160}
\expandafter\def\csname rg@bipmatch.seq.xor.s\endcsname{40.7}
\expandafter\def\csname rg@bipmatch.seq.xorcle-dev.n\endcsname{160}
\expandafter\def\csname rg@bipmatch.seq.xorcle-dev.s\endcsname{43.3}
\expandafter\def\csname rg@bipmatch.seq.xorcle.n\endcsname{160}
\expandafter\def\csname rg@bipmatch.seq.xorcle.s\endcsname{40.7}
\expandafter\def\csname rg@cert.check.max\endcsname{4.8}
\expandafter\def\csname rg@cert.families\endcsname{10}
\expandafter\def\csname rg@cert.farkas.max\endcsname{192}
\expandafter\def\csname rg@cert.farkas.min\endcsname{17}
\expandafter\def\csname rg@cert.files\endcsname{10}
\expandafter\def\csname rg@cert.generated\endcsname{22}
\expandafter\def\csname rg@cert.ring.max\endcsname{1,023}
\expandafter\def\csname rg@cert.ring.min\endcsname{18}
\expandafter\def\csname rg@cert.total\endcsname{32}
\expandafter\def\csname rg@cert.verified\endcsname{32}
\expandafter\def\csname rg@count3.best3.text\endcsname{80 points on each of the pairwise, commander and binary encodings}
\expandafter\def\csname rg@count3.binary.best.n\endcsname{80}
\expandafter\def\csname rg@count3.binary.bve+hull.n\endcsname{--}
\expandafter\def\csname rg@count3.binary.bve+hull.s\endcsname{--}
\expandafter\def\csname rg@count3.binary.bve+int.n\endcsname{320}
\expandafter\def\csname rg@count3.binary.bve+int.s\endcsname{2.0}
\expandafter\def\csname rg@count3.binary.bve+win.n\endcsname{320}
\expandafter\def\csname rg@count3.binary.bve+win.s\endcsname{1.5}
\expandafter\def\csname rg@count3.binary.cadical.n\endcsname{80}
\expandafter\def\csname rg@count3.binary.cadical.s\endcsname{0.50}
\expandafter\def\csname rg@count3.binary.cms.n\endcsname{80}
\expandafter\def\csname rg@count3.binary.cms.s\endcsname{0.73}
\expandafter\def\csname rg@count3.binary.cnf.n\endcsname{80}
\expandafter\def\csname rg@count3.binary.cnf.s\endcsname{56.9}
\expandafter\def\csname rg@count3.binary.cnf.tools\endcsname{CDCL, Sat4j, RoundingSat, Exact, SCIP}
\expandafter\def\csname rg@count3.binary.cnfx.n\endcsname{80}
\expandafter\def\csname rg@count3.binary.cnfx.tools\endcsname{CDCL, Sat4j, RoundingSat, Exact, SCIP}
\expandafter\def\csname rg@count3.binary.exact:cnf.n\endcsname{80}
\expandafter\def\csname rg@count3.binary.exact:cnf.s\endcsname{6.7}
\expandafter\def\csname rg@count3.binary.hull.n\endcsname{--}
\expandafter\def\csname rg@count3.binary.hull.s\endcsname{--}
\expandafter\def\csname rg@count3.binary.int.n\endcsname{320}
\expandafter\def\csname rg@count3.binary.int.s\endcsname{2.0}
\expandafter\def\csname rg@count3.binary.kissat.n\endcsname{80}
\expandafter\def\csname rg@count3.binary.kissat.s\endcsname{0.73}
\expandafter\def\csname rg@count3.binary.lingeling.n\endcsname{80}
\expandafter\def\csname rg@count3.binary.lingeling.s\endcsname{1.2}
\expandafter\def\csname rg@count3.binary.pgpbs-2.n\endcsname{--}
\expandafter\def\csname rg@count3.binary.pgpbs-2.s\endcsname{--}
\expandafter\def\csname rg@count3.binary.pgpbs-3.n\endcsname{--}
\expandafter\def\csname rg@count3.binary.pgpbs-3.s\endcsname{--}
\expandafter\def\csname rg@count3.binary.pgpbs-a.n\endcsname{--}
\expandafter\def\csname rg@count3.binary.pgpbs-a.s\endcsname{--}
\expandafter\def\csname rg@count3.binary.pgpbs-i.n\endcsname{--}
\expandafter\def\csname rg@count3.binary.pgpbs-i.s\endcsname{--}
\expandafter\def\csname rg@count3.binary.read.fastest\endcsname{field}
\expandafter\def\csname rg@count3.binary.read.n\endcsname{320}
\expandafter\def\csname rg@count3.binary.read.s\endcsname{1.5}
\expandafter\def\csname rg@count3.binary.read.smax\endcsname{2.0}
\expandafter\def\csname rg@count3.binary.read.tool\endcsname{field, ring, field after elim., ring after elim.}
\expandafter\def\csname rg@count3.binary.roundingsat:cnf.n\endcsname{80}
\expandafter\def\csname rg@count3.binary.roundingsat:cnf.s\endcsname{5.1}
\expandafter\def\csname rg@count3.binary.sat4j-cp:cnf.n\endcsname{80}
\expandafter\def\csname rg@count3.binary.sat4j-cp:cnf.s\endcsname{2.7}
\expandafter\def\csname rg@count3.binary.scip:cnf.n\endcsname{80}
\expandafter\def\csname rg@count3.binary.scip:cnf.s\endcsname{56.9}
\expandafter\def\csname rg@count3.binary.top\endcsname{320}
\expandafter\def\csname rg@count3.binary.win.n\endcsname{320}
\expandafter\def\csname rg@count3.binary.win.s\endcsname{1.5}
\expandafter\def\csname rg@count3.binary.xor.n\endcsname{40}
\expandafter\def\csname rg@count3.binary.xor.s\endcsname{0.38}
\expandafter\def\csname rg@count3.binary.xorcle-dev.n\endcsname{40}
\expandafter\def\csname rg@count3.binary.xorcle-dev.s\endcsname{0.38}
\expandafter\def\csname rg@count3.binary.xorcle.n\endcsname{40}
\expandafter\def\csname rg@count3.binary.xorcle.s\endcsname{0.45}
\expandafter\def\csname rg@count3.commander.best.n\endcsname{80}
\expandafter\def\csname rg@count3.commander.bve+hull.n\endcsname{--}
\expandafter\def\csname rg@count3.commander.bve+hull.s\endcsname{--}
\expandafter\def\csname rg@count3.commander.bve+int.n\endcsname{320}
\expandafter\def\csname rg@count3.commander.bve+int.s\endcsname{1.4}
\expandafter\def\csname rg@count3.commander.bve+win.n\endcsname{320}
\expandafter\def\csname rg@count3.commander.bve+win.s\endcsname{1.3}
\expandafter\def\csname rg@count3.commander.cadical.n\endcsname{80}
\expandafter\def\csname rg@count3.commander.cadical.s\endcsname{0.33}
\expandafter\def\csname rg@count3.commander.cms.n\endcsname{80}
\expandafter\def\csname rg@count3.commander.cms.s\endcsname{0.79}
\expandafter\def\csname rg@count3.commander.cnf.n\endcsname{80}
\expandafter\def\csname rg@count3.commander.cnf.s\endcsname{44.1}
\expandafter\def\csname rg@count3.commander.cnf.tools\endcsname{CDCL, Sat4j, RoundingSat, Exact, SCIP}
\expandafter\def\csname rg@count3.commander.cnfx.n\endcsname{80}
\expandafter\def\csname rg@count3.commander.cnfx.tools\endcsname{CDCL, Sat4j, RoundingSat, Exact, SCIP}
\expandafter\def\csname rg@count3.commander.exact:cnf.n\endcsname{80}
\expandafter\def\csname rg@count3.commander.exact:cnf.s\endcsname{9.3}
\expandafter\def\csname rg@count3.commander.hull.n\endcsname{--}
\expandafter\def\csname rg@count3.commander.hull.s\endcsname{--}
\expandafter\def\csname rg@count3.commander.int.n\endcsname{320}
\expandafter\def\csname rg@count3.commander.int.s\endcsname{2.4}
\expandafter\def\csname rg@count3.commander.kissat.n\endcsname{80}
\expandafter\def\csname rg@count3.commander.kissat.s\endcsname{0.49}
\expandafter\def\csname rg@count3.commander.lingeling.n\endcsname{80}
\expandafter\def\csname rg@count3.commander.lingeling.s\endcsname{0.52}
\expandafter\def\csname rg@count3.commander.pgpbs-2.n\endcsname{--}
\expandafter\def\csname rg@count3.commander.pgpbs-2.s\endcsname{--}
\expandafter\def\csname rg@count3.commander.pgpbs-3.n\endcsname{--}
\expandafter\def\csname rg@count3.commander.pgpbs-3.s\endcsname{--}
\expandafter\def\csname rg@count3.commander.pgpbs-a.n\endcsname{--}
\expandafter\def\csname rg@count3.commander.pgpbs-a.s\endcsname{--}
\expandafter\def\csname rg@count3.commander.pgpbs-i.n\endcsname{--}
\expandafter\def\csname rg@count3.commander.pgpbs-i.s\endcsname{--}
\expandafter\def\csname rg@count3.commander.read.fastest\endcsname{field after elim.}
\expandafter\def\csname rg@count3.commander.read.n\endcsname{320}
\expandafter\def\csname rg@count3.commander.read.s\endcsname{1.3}
\expandafter\def\csname rg@count3.commander.read.smax\endcsname{2.4}
\expandafter\def\csname rg@count3.commander.read.tool\endcsname{field, ring, field after elim., ring after elim.}
\expandafter\def\csname rg@count3.commander.roundingsat:cnf.n\endcsname{80}
\expandafter\def\csname rg@count3.commander.roundingsat:cnf.s\endcsname{4.7}
\expandafter\def\csname rg@count3.commander.sat4j-cp:cnf.n\endcsname{80}
\expandafter\def\csname rg@count3.commander.sat4j-cp:cnf.s\endcsname{3.1}
\expandafter\def\csname rg@count3.commander.scip:cnf.n\endcsname{80}
\expandafter\def\csname rg@count3.commander.scip:cnf.s\endcsname{44.1}
\expandafter\def\csname rg@count3.commander.top\endcsname{320}
\expandafter\def\csname rg@count3.commander.win.n\endcsname{320}
\expandafter\def\csname rg@count3.commander.win.s\endcsname{1.9}
\expandafter\def\csname rg@count3.commander.xor.n\endcsname{40}
\expandafter\def\csname rg@count3.commander.xor.s\endcsname{0.46}
\expandafter\def\csname rg@count3.commander.xorcle-dev.n\endcsname{40}
\expandafter\def\csname rg@count3.commander.xorcle-dev.s\endcsname{0.46}
\expandafter\def\csname rg@count3.commander.xorcle.n\endcsname{40}
\expandafter\def\csname rg@count3.commander.xorcle.s\endcsname{0.57}
\expandafter\def\csname rg@count3.native.exact:pb.n\endcsname{80}
\expandafter\def\csname rg@count3.native.exact:pb.s\endcsname{9.4}
\expandafter\def\csname rg@count3.native.highs:pb.n\endcsname{80}
\expandafter\def\csname rg@count3.native.highs:pb.s\endcsname{3.7}
\expandafter\def\csname rg@count3.native.n\endcsname{80}
\expandafter\def\csname rg@count3.native.roundingsat:pb.n\endcsname{80}
\expandafter\def\csname rg@count3.native.roundingsat:pb.s\endcsname{5.9}
\expandafter\def\csname rg@count3.native.sat4j-cp:pb.n\endcsname{40}
\expandafter\def\csname rg@count3.native.sat4j-cp:pb.s\endcsname{0.62}
\expandafter\def\csname rg@count3.native.scip:pb.n\endcsname{80}
\expandafter\def\csname rg@count3.native.scip:pb.s\endcsname{6.4}
\expandafter\def\csname rg@count3.native.smax\endcsname{9.4}
\expandafter\def\csname rg@count3.native.smin\endcsname{3.7}
\expandafter\def\csname rg@count3.native.tools\endcsname{RoundingSat, Exact, SCIP, HiGHS}
\expandafter\def\csname rg@count3.native.toolsp\endcsname{RoundingSat, Exact, SCIP and HiGHS}
\expandafter\def\csname rg@count3.pairwise.best.n\endcsname{80}
\expandafter\def\csname rg@count3.pairwise.bve+hull.n\endcsname{--}
\expandafter\def\csname rg@count3.pairwise.bve+hull.s\endcsname{--}
\expandafter\def\csname rg@count3.pairwise.bve+int.n\endcsname{320}
\expandafter\def\csname rg@count3.pairwise.bve+int.s\endcsname{2.2}
\expandafter\def\csname rg@count3.pairwise.bve+win.n\endcsname{320}
\expandafter\def\csname rg@count3.pairwise.bve+win.s\endcsname{1.8}
\expandafter\def\csname rg@count3.pairwise.cadical.n\endcsname{80}
\expandafter\def\csname rg@count3.pairwise.cadical.s\endcsname{0.31}
\expandafter\def\csname rg@count3.pairwise.cms.n\endcsname{80}
\expandafter\def\csname rg@count3.pairwise.cms.s\endcsname{0.41}
\expandafter\def\csname rg@count3.pairwise.cnf.n\endcsname{80}
\expandafter\def\csname rg@count3.pairwise.cnf.s\endcsname{8.3}
\expandafter\def\csname rg@count3.pairwise.cnf.tools\endcsname{CDCL, Sat4j, RoundingSat, Exact, SCIP}
\expandafter\def\csname rg@count3.pairwise.cnfx.n\endcsname{80}
\expandafter\def\csname rg@count3.pairwise.cnfx.tools\endcsname{CDCL, Sat4j, RoundingSat, Exact, SCIP}
\expandafter\def\csname rg@count3.pairwise.exact:cnf.n\endcsname{80}
\expandafter\def\csname rg@count3.pairwise.exact:cnf.s\endcsname{4.8}
\expandafter\def\csname rg@count3.pairwise.hull.n\endcsname{--}
\expandafter\def\csname rg@count3.pairwise.hull.s\endcsname{--}
\expandafter\def\csname rg@count3.pairwise.int.n\endcsname{320}
\expandafter\def\csname rg@count3.pairwise.int.s\endcsname{2.1}
\expandafter\def\csname rg@count3.pairwise.kissat.n\endcsname{80}
\expandafter\def\csname rg@count3.pairwise.kissat.s\endcsname{0.45}
\expandafter\def\csname rg@count3.pairwise.lingeling.n\endcsname{80}
\expandafter\def\csname rg@count3.pairwise.lingeling.s\endcsname{0.41}
\expandafter\def\csname rg@count3.pairwise.pgpbs-2.n\endcsname{--}
\expandafter\def\csname rg@count3.pairwise.pgpbs-2.s\endcsname{--}
\expandafter\def\csname rg@count3.pairwise.pgpbs-3.n\endcsname{--}
\expandafter\def\csname rg@count3.pairwise.pgpbs-3.s\endcsname{--}
\expandafter\def\csname rg@count3.pairwise.pgpbs-a.n\endcsname{--}
\expandafter\def\csname rg@count3.pairwise.pgpbs-a.s\endcsname{--}
\expandafter\def\csname rg@count3.pairwise.pgpbs-i.n\endcsname{--}
\expandafter\def\csname rg@count3.pairwise.pgpbs-i.s\endcsname{--}
\expandafter\def\csname rg@count3.pairwise.read.fastest\endcsname{field}
\expandafter\def\csname rg@count3.pairwise.read.n\endcsname{320}
\expandafter\def\csname rg@count3.pairwise.read.s\endcsname{1.7}
\expandafter\def\csname rg@count3.pairwise.read.smax\endcsname{2.2}
\expandafter\def\csname rg@count3.pairwise.read.tool\endcsname{field, ring, field after elim., ring after elim.}
\expandafter\def\csname rg@count3.pairwise.roundingsat:cnf.n\endcsname{80}
\expandafter\def\csname rg@count3.pairwise.roundingsat:cnf.s\endcsname{3.9}
\expandafter\def\csname rg@count3.pairwise.sat4j-cp:cnf.n\endcsname{80}
\expandafter\def\csname rg@count3.pairwise.sat4j-cp:cnf.s\endcsname{3.2}
\expandafter\def\csname rg@count3.pairwise.scip:cnf.n\endcsname{80}
\expandafter\def\csname rg@count3.pairwise.scip:cnf.s\endcsname{8.3}
\expandafter\def\csname rg@count3.pairwise.top\endcsname{320}
\expandafter\def\csname rg@count3.pairwise.win.n\endcsname{320}
\expandafter\def\csname rg@count3.pairwise.win.s\endcsname{1.7}
\expandafter\def\csname rg@count3.pairwise.xor.n\endcsname{40}
\expandafter\def\csname rg@count3.pairwise.xor.s\endcsname{0.74}
\expandafter\def\csname rg@count3.pairwise.xorcle-dev.n\endcsname{40}
\expandafter\def\csname rg@count3.pairwise.xorcle-dev.s\endcsname{0.74}
\expandafter\def\csname rg@count3.pairwise.xorcle.n\endcsname{40}
\expandafter\def\csname rg@count3.pairwise.xorcle.s\endcsname{0.78}
\expandafter\def\csname rg@count3.seq.best.n\endcsname{320}
\expandafter\def\csname rg@count3.seq.bve+hull.n\endcsname{--}
\expandafter\def\csname rg@count3.seq.bve+hull.s\endcsname{--}
\expandafter\def\csname rg@count3.seq.bve+int.n\endcsname{320}
\expandafter\def\csname rg@count3.seq.bve+int.s\endcsname{1.4}
\expandafter\def\csname rg@count3.seq.bve+win.n\endcsname{320}
\expandafter\def\csname rg@count3.seq.bve+win.s\endcsname{1.2}
\expandafter\def\csname rg@count3.seq.cadical.n\endcsname{80}
\expandafter\def\csname rg@count3.seq.cadical.s\endcsname{0.33}
\expandafter\def\csname rg@count3.seq.cms.n\endcsname{80}
\expandafter\def\csname rg@count3.seq.cms.s\endcsname{0.57}
\expandafter\def\csname rg@count3.seq.cnf.n\endcsname{320}
\expandafter\def\csname rg@count3.seq.cnf.s\endcsname{17.4}
\expandafter\def\csname rg@count3.seq.cnf.tools\endcsname{pgpbs mod 3}
\expandafter\def\csname rg@count3.seq.cnfx.n\endcsname{320}
\expandafter\def\csname rg@count3.seq.cnfx.tools\endcsname{pgpbs mod 3}
\expandafter\def\csname rg@count3.seq.exact:cnf.n\endcsname{80}
\expandafter\def\csname rg@count3.seq.exact:cnf.s\endcsname{7.4}
\expandafter\def\csname rg@count3.seq.hull.n\endcsname{--}
\expandafter\def\csname rg@count3.seq.hull.s\endcsname{--}
\expandafter\def\csname rg@count3.seq.int.n\endcsname{--}
\expandafter\def\csname rg@count3.seq.int.s\endcsname{--}
\expandafter\def\csname rg@count3.seq.kissat.n\endcsname{80}
\expandafter\def\csname rg@count3.seq.kissat.s\endcsname{0.62}
\expandafter\def\csname rg@count3.seq.lingeling.n\endcsname{80}
\expandafter\def\csname rg@count3.seq.lingeling.s\endcsname{0.64}
\expandafter\def\csname rg@count3.seq.pgpbs-2.n\endcsname{--}
\expandafter\def\csname rg@count3.seq.pgpbs-2.s\endcsname{--}
\expandafter\def\csname rg@count3.seq.pgpbs-3.n\endcsname{320}
\expandafter\def\csname rg@count3.seq.pgpbs-3.s\endcsname{17.4}
\expandafter\def\csname rg@count3.seq.pgpbs-a.n\endcsname{--}
\expandafter\def\csname rg@count3.seq.pgpbs-a.s\endcsname{--}
\expandafter\def\csname rg@count3.seq.pgpbs-i.n\endcsname{--}
\expandafter\def\csname rg@count3.seq.pgpbs-i.s\endcsname{--}
\expandafter\def\csname rg@count3.seq.read.fastest\endcsname{field after elim.}
\expandafter\def\csname rg@count3.seq.read.n\endcsname{320}
\expandafter\def\csname rg@count3.seq.read.s\endcsname{1.2}
\expandafter\def\csname rg@count3.seq.read.smax\endcsname{1.4}
\expandafter\def\csname rg@count3.seq.read.tool\endcsname{field after elim., ring after elim.}
\expandafter\def\csname rg@count3.seq.roundingsat:cnf.n\endcsname{80}
\expandafter\def\csname rg@count3.seq.roundingsat:cnf.s\endcsname{4.7}
\expandafter\def\csname rg@count3.seq.sat4j-cp:cnf.n\endcsname{80}
\expandafter\def\csname rg@count3.seq.sat4j-cp:cnf.s\endcsname{3.6}
\expandafter\def\csname rg@count3.seq.scip:cnf.n\endcsname{80}
\expandafter\def\csname rg@count3.seq.scip:cnf.s\endcsname{43.7}
\expandafter\def\csname rg@count3.seq.top\endcsname{320}
\expandafter\def\csname rg@count3.seq.win.n\endcsname{--}
\expandafter\def\csname rg@count3.seq.win.s\endcsname{--}
\expandafter\def\csname rg@count3.seq.xor.n\endcsname{40}
\expandafter\def\csname rg@count3.seq.xor.s\endcsname{0.78}
\expandafter\def\csname rg@count3.seq.xorcle-dev.n\endcsname{40}
\expandafter\def\csname rg@count3.seq.xorcle-dev.s\endcsname{0.78}
\expandafter\def\csname rg@count3.seq.xorcle.n\endcsname{40}
\expandafter\def\csname rg@count3.seq.xorcle.s\endcsname{0.93}
\expandafter\def\csname rg@count4.binary.best.n\endcsname{162}
\expandafter\def\csname rg@count4.binary.bve+hull.n\endcsname{--}
\expandafter\def\csname rg@count4.binary.bve+hull.s\endcsname{--}
\expandafter\def\csname rg@count4.binary.bve+int.n\endcsname{642}
\expandafter\def\csname rg@count4.binary.bve+int.s\endcsname{16.8}
\expandafter\def\csname rg@count4.binary.bve+win.n\endcsname{--}
\expandafter\def\csname rg@count4.binary.bve+win.s\endcsname{--}
\expandafter\def\csname rg@count4.binary.cadical.n\endcsname{162}
\expandafter\def\csname rg@count4.binary.cadical.s\endcsname{0.08}
\expandafter\def\csname rg@count4.binary.cms.n\endcsname{162}
\expandafter\def\csname rg@count4.binary.cms.s\endcsname{0.15}
\expandafter\def\csname rg@count4.binary.cnf.n\endcsname{162}
\expandafter\def\csname rg@count4.binary.cnf.s\endcsname{52.1}
\expandafter\def\csname rg@count4.binary.cnf.tools\endcsname{CDCL, Sat4j, RoundingSat, Exact, SCIP}
\expandafter\def\csname rg@count4.binary.cnfx.n\endcsname{162}
\expandafter\def\csname rg@count4.binary.cnfx.tools\endcsname{CDCL, Sat4j, RoundingSat, Exact, SCIP}
\expandafter\def\csname rg@count4.binary.exact:cnf.n\endcsname{162}
\expandafter\def\csname rg@count4.binary.exact:cnf.s\endcsname{1.6}
\expandafter\def\csname rg@count4.binary.hull.n\endcsname{--}
\expandafter\def\csname rg@count4.binary.hull.s\endcsname{--}
\expandafter\def\csname rg@count4.binary.int.n\endcsname{642}
\expandafter\def\csname rg@count4.binary.int.s\endcsname{16.7}
\expandafter\def\csname rg@count4.binary.kissat.n\endcsname{162}
\expandafter\def\csname rg@count4.binary.kissat.s\endcsname{0.15}
\expandafter\def\csname rg@count4.binary.lingeling.n\endcsname{162}
\expandafter\def\csname rg@count4.binary.lingeling.s\endcsname{0.20}
\expandafter\def\csname rg@count4.binary.pgpbs-2.n\endcsname{--}
\expandafter\def\csname rg@count4.binary.pgpbs-2.s\endcsname{--}
\expandafter\def\csname rg@count4.binary.pgpbs-3.n\endcsname{--}
\expandafter\def\csname rg@count4.binary.pgpbs-3.s\endcsname{--}
\expandafter\def\csname rg@count4.binary.pgpbs-4.n\endcsname{--}
\expandafter\def\csname rg@count4.binary.pgpbs-4.s\endcsname{--}
\expandafter\def\csname rg@count4.binary.pgpbs-a.n\endcsname{--}
\expandafter\def\csname rg@count4.binary.pgpbs-a.s\endcsname{--}
\expandafter\def\csname rg@count4.binary.pgpbs-i.n\endcsname{--}
\expandafter\def\csname rg@count4.binary.pgpbs-i.s\endcsname{--}
\expandafter\def\csname rg@count4.binary.read.fastest\endcsname{ring}
\expandafter\def\csname rg@count4.binary.read.n\endcsname{642}
\expandafter\def\csname rg@count4.binary.read.s\endcsname{16.7}
\expandafter\def\csname rg@count4.binary.read.smax\endcsname{16.8}
\expandafter\def\csname rg@count4.binary.read.tool\endcsname{ring, ring after elim.}
\expandafter\def\csname rg@count4.binary.roundingsat:cnf.n\endcsname{162}
\expandafter\def\csname rg@count4.binary.roundingsat:cnf.s\endcsname{0.20}
\expandafter\def\csname rg@count4.binary.sat4j-cp:cnf.n\endcsname{162}
\expandafter\def\csname rg@count4.binary.sat4j-cp:cnf.s\endcsname{0.89}
\expandafter\def\csname rg@count4.binary.scip:cnf.n\endcsname{162}
\expandafter\def\csname rg@count4.binary.scip:cnf.s\endcsname{52.1}
\expandafter\def\csname rg@count4.binary.top\endcsname{642}
\expandafter\def\csname rg@count4.binary.win.n\endcsname{--}
\expandafter\def\csname rg@count4.binary.win.s\endcsname{--}
\expandafter\def\csname rg@count4.binary.xor.n\endcsname{82}
\expandafter\def\csname rg@count4.binary.xor.s\endcsname{0.35}
\expandafter\def\csname rg@count4.binary.xorcle-dev.n\endcsname{82}
\expandafter\def\csname rg@count4.binary.xorcle-dev.s\endcsname{0.35}
\expandafter\def\csname rg@count4.binary.xorcle.n\endcsname{82}
\expandafter\def\csname rg@count4.binary.xorcle.s\endcsname{0.44}
\expandafter\def\csname rg@count4.cnf3.text\endcsname{162 points on each of the pairwise, commander and sequential encodings}
\expandafter\def\csname rg@count4.cnf4.text\endcsname{162 points on each of the pairwise, commander, binary and sequential encodings}
\expandafter\def\csname rg@count4.commander.best.n\endcsname{162}
\expandafter\def\csname rg@count4.commander.bve+hull.n\endcsname{--}
\expandafter\def\csname rg@count4.commander.bve+hull.s\endcsname{--}
\expandafter\def\csname rg@count4.commander.bve+int.n\endcsname{642}
\expandafter\def\csname rg@count4.commander.bve+int.s\endcsname{9.7}
\expandafter\def\csname rg@count4.commander.bve+win.n\endcsname{--}
\expandafter\def\csname rg@count4.commander.bve+win.s\endcsname{--}
\expandafter\def\csname rg@count4.commander.cadical.n\endcsname{162}
\expandafter\def\csname rg@count4.commander.cadical.s\endcsname{0.11}
\expandafter\def\csname rg@count4.commander.cms.n\endcsname{162}
\expandafter\def\csname rg@count4.commander.cms.s\endcsname{0.14}
\expandafter\def\csname rg@count4.commander.cnf.n\endcsname{162}
\expandafter\def\csname rg@count4.commander.cnf.s\endcsname{20.9}
\expandafter\def\csname rg@count4.commander.cnf.tools\endcsname{CDCL, Sat4j, RoundingSat, Exact, SCIP}
\expandafter\def\csname rg@count4.commander.cnfx.n\endcsname{162}
\expandafter\def\csname rg@count4.commander.cnfx.tools\endcsname{CDCL, Sat4j, RoundingSat, Exact, SCIP}
\expandafter\def\csname rg@count4.commander.exact:cnf.n\endcsname{162}
\expandafter\def\csname rg@count4.commander.exact:cnf.s\endcsname{1.7}
\expandafter\def\csname rg@count4.commander.hull.n\endcsname{--}
\expandafter\def\csname rg@count4.commander.hull.s\endcsname{--}
\expandafter\def\csname rg@count4.commander.int.n\endcsname{642}
\expandafter\def\csname rg@count4.commander.int.s\endcsname{20.7}
\expandafter\def\csname rg@count4.commander.kissat.n\endcsname{162}
\expandafter\def\csname rg@count4.commander.kissat.s\endcsname{0.14}
\expandafter\def\csname rg@count4.commander.lingeling.n\endcsname{162}
\expandafter\def\csname rg@count4.commander.lingeling.s\endcsname{0.11}
\expandafter\def\csname rg@count4.commander.pgpbs-2.n\endcsname{--}
\expandafter\def\csname rg@count4.commander.pgpbs-2.s\endcsname{--}
\expandafter\def\csname rg@count4.commander.pgpbs-3.n\endcsname{--}
\expandafter\def\csname rg@count4.commander.pgpbs-3.s\endcsname{--}
\expandafter\def\csname rg@count4.commander.pgpbs-4.n\endcsname{--}
\expandafter\def\csname rg@count4.commander.pgpbs-4.s\endcsname{--}
\expandafter\def\csname rg@count4.commander.pgpbs-a.n\endcsname{--}
\expandafter\def\csname rg@count4.commander.pgpbs-a.s\endcsname{--}
\expandafter\def\csname rg@count4.commander.pgpbs-i.n\endcsname{--}
\expandafter\def\csname rg@count4.commander.pgpbs-i.s\endcsname{--}
\expandafter\def\csname rg@count4.commander.read.fastest\endcsname{ring after elim.}
\expandafter\def\csname rg@count4.commander.read.n\endcsname{642}
\expandafter\def\csname rg@count4.commander.read.s\endcsname{9.7}
\expandafter\def\csname rg@count4.commander.read.smax\endcsname{20.7}
\expandafter\def\csname rg@count4.commander.read.tool\endcsname{ring, ring after elim.}
\expandafter\def\csname rg@count4.commander.roundingsat:cnf.n\endcsname{162}
\expandafter\def\csname rg@count4.commander.roundingsat:cnf.s\endcsname{0.20}
\expandafter\def\csname rg@count4.commander.sat4j-cp:cnf.n\endcsname{162}
\expandafter\def\csname rg@count4.commander.sat4j-cp:cnf.s\endcsname{0.85}
\expandafter\def\csname rg@count4.commander.scip:cnf.n\endcsname{162}
\expandafter\def\csname rg@count4.commander.scip:cnf.s\endcsname{20.9}
\expandafter\def\csname rg@count4.commander.top\endcsname{642}
\expandafter\def\csname rg@count4.commander.win.n\endcsname{--}
\expandafter\def\csname rg@count4.commander.win.s\endcsname{--}
\expandafter\def\csname rg@count4.commander.xor.n\endcsname{82}
\expandafter\def\csname rg@count4.commander.xor.s\endcsname{0.33}
\expandafter\def\csname rg@count4.commander.xorcle-dev.n\endcsname{82}
\expandafter\def\csname rg@count4.commander.xorcle-dev.s\endcsname{0.33}
\expandafter\def\csname rg@count4.commander.xorcle.n\endcsname{82}
\expandafter\def\csname rg@count4.commander.xorcle.s\endcsname{0.43}
\expandafter\def\csname rg@count4.native.exact:pb.n\endcsname{162}
\expandafter\def\csname rg@count4.native.exact:pb.s\endcsname{2.0}
\expandafter\def\csname rg@count4.native.highs:pb.n\endcsname{162}
\expandafter\def\csname rg@count4.native.highs:pb.s\endcsname{1.2}
\expandafter\def\csname rg@count4.native.n\endcsname{162}
\expandafter\def\csname rg@count4.native.roundingsat:pb.n\endcsname{162}
\expandafter\def\csname rg@count4.native.roundingsat:pb.s\endcsname{0.16}
\expandafter\def\csname rg@count4.native.sat4j-cp:pb.n\endcsname{162}
\expandafter\def\csname rg@count4.native.sat4j-cp:pb.s\endcsname{4.7}
\expandafter\def\csname rg@count4.native.scip:pb.n\endcsname{162}
\expandafter\def\csname rg@count4.native.scip:pb.s\endcsname{1.8}
\expandafter\def\csname rg@count4.native.smax\endcsname{4.7}
\expandafter\def\csname rg@count4.native.smin\endcsname{0.16}
\expandafter\def\csname rg@count4.native.tools\endcsname{all five}
\expandafter\def\csname rg@count4.native.toolsp\endcsname{all five native-form solvers}
\expandafter\def\csname rg@count4.pairwise.best.n\endcsname{162}
\expandafter\def\csname rg@count4.pairwise.bve+hull.n\endcsname{--}
\expandafter\def\csname rg@count4.pairwise.bve+hull.s\endcsname{--}
\expandafter\def\csname rg@count4.pairwise.bve+int.n\endcsname{642}
\expandafter\def\csname rg@count4.pairwise.bve+int.s\endcsname{26.4}
\expandafter\def\csname rg@count4.pairwise.bve+win.n\endcsname{--}
\expandafter\def\csname rg@count4.pairwise.bve+win.s\endcsname{--}
\expandafter\def\csname rg@count4.pairwise.cadical.n\endcsname{162}
\expandafter\def\csname rg@count4.pairwise.cadical.s\endcsname{0.05}
\expandafter\def\csname rg@count4.pairwise.cms.n\endcsname{162}
\expandafter\def\csname rg@count4.pairwise.cms.s\endcsname{0.09}
\expandafter\def\csname rg@count4.pairwise.cnf.n\endcsname{162}
\expandafter\def\csname rg@count4.pairwise.cnf.s\endcsname{2.4}
\expandafter\def\csname rg@count4.pairwise.cnf.tools\endcsname{CDCL, Sat4j, RoundingSat, Exact, SCIP}
\expandafter\def\csname rg@count4.pairwise.cnfx.n\endcsname{162}
\expandafter\def\csname rg@count4.pairwise.cnfx.tools\endcsname{CDCL, Sat4j, RoundingSat, Exact, SCIP}
\expandafter\def\csname rg@count4.pairwise.exact:cnf.n\endcsname{162}
\expandafter\def\csname rg@count4.pairwise.exact:cnf.s\endcsname{1.1}
\expandafter\def\csname rg@count4.pairwise.hull.n\endcsname{--}
\expandafter\def\csname rg@count4.pairwise.hull.s\endcsname{--}
\expandafter\def\csname rg@count4.pairwise.int.n\endcsname{642}
\expandafter\def\csname rg@count4.pairwise.int.s\endcsname{25.9}
\expandafter\def\csname rg@count4.pairwise.kissat.n\endcsname{162}
\expandafter\def\csname rg@count4.pairwise.kissat.s\endcsname{0.11}
\expandafter\def\csname rg@count4.pairwise.lingeling.n\endcsname{162}
\expandafter\def\csname rg@count4.pairwise.lingeling.s\endcsname{0.12}
\expandafter\def\csname rg@count4.pairwise.pgpbs-2.n\endcsname{--}
\expandafter\def\csname rg@count4.pairwise.pgpbs-2.s\endcsname{--}
\expandafter\def\csname rg@count4.pairwise.pgpbs-3.n\endcsname{--}
\expandafter\def\csname rg@count4.pairwise.pgpbs-3.s\endcsname{--}
\expandafter\def\csname rg@count4.pairwise.pgpbs-4.n\endcsname{--}
\expandafter\def\csname rg@count4.pairwise.pgpbs-4.s\endcsname{--}
\expandafter\def\csname rg@count4.pairwise.pgpbs-a.n\endcsname{--}
\expandafter\def\csname rg@count4.pairwise.pgpbs-a.s\endcsname{--}
\expandafter\def\csname rg@count4.pairwise.pgpbs-i.n\endcsname{--}
\expandafter\def\csname rg@count4.pairwise.pgpbs-i.s\endcsname{--}
\expandafter\def\csname rg@count4.pairwise.read.fastest\endcsname{ring}
\expandafter\def\csname rg@count4.pairwise.read.n\endcsname{642}
\expandafter\def\csname rg@count4.pairwise.read.s\endcsname{25.9}
\expandafter\def\csname rg@count4.pairwise.read.smax\endcsname{26.4}
\expandafter\def\csname rg@count4.pairwise.read.tool\endcsname{ring, ring after elim.}
\expandafter\def\csname rg@count4.pairwise.roundingsat:cnf.n\endcsname{162}
\expandafter\def\csname rg@count4.pairwise.roundingsat:cnf.s\endcsname{0.11}
\expandafter\def\csname rg@count4.pairwise.sat4j-cp:cnf.n\endcsname{162}
\expandafter\def\csname rg@count4.pairwise.sat4j-cp:cnf.s\endcsname{0.69}
\expandafter\def\csname rg@count4.pairwise.scip:cnf.n\endcsname{162}
\expandafter\def\csname rg@count4.pairwise.scip:cnf.s\endcsname{2.4}
\expandafter\def\csname rg@count4.pairwise.top\endcsname{642}
\expandafter\def\csname rg@count4.pairwise.win.n\endcsname{--}
\expandafter\def\csname rg@count4.pairwise.win.s\endcsname{--}
\expandafter\def\csname rg@count4.pairwise.xor.n\endcsname{82}
\expandafter\def\csname rg@count4.pairwise.xor.s\endcsname{0.16}
\expandafter\def\csname rg@count4.pairwise.xorcle-dev.n\endcsname{82}
\expandafter\def\csname rg@count4.pairwise.xorcle-dev.s\endcsname{0.16}
\expandafter\def\csname rg@count4.pairwise.xorcle.n\endcsname{82}
\expandafter\def\csname rg@count4.pairwise.xorcle.s\endcsname{0.18}
\expandafter\def\csname rg@count4.seq.best.n\endcsname{162}
\expandafter\def\csname rg@count4.seq.bve+hull.n\endcsname{--}
\expandafter\def\csname rg@count4.seq.bve+hull.s\endcsname{--}
\expandafter\def\csname rg@count4.seq.bve+int.n\endcsname{642}
\expandafter\def\csname rg@count4.seq.bve+int.s\endcsname{12.0}
\expandafter\def\csname rg@count4.seq.bve+win.n\endcsname{--}
\expandafter\def\csname rg@count4.seq.bve+win.s\endcsname{--}
\expandafter\def\csname rg@count4.seq.cadical.n\endcsname{162}
\expandafter\def\csname rg@count4.seq.cadical.s\endcsname{0.09}
\expandafter\def\csname rg@count4.seq.cms.n\endcsname{162}
\expandafter\def\csname rg@count4.seq.cms.s\endcsname{0.17}
\expandafter\def\csname rg@count4.seq.cnf.n\endcsname{162}
\expandafter\def\csname rg@count4.seq.cnf.s\endcsname{39.5}
\expandafter\def\csname rg@count4.seq.cnf.tools\endcsname{CDCL, Sat4j, RoundingSat, Exact, SCIP}
\expandafter\def\csname rg@count4.seq.cnfx.n\endcsname{162}
\expandafter\def\csname rg@count4.seq.cnfx.tools\endcsname{CDCL, Sat4j, RoundingSat, Exact, SCIP}
\expandafter\def\csname rg@count4.seq.exact:cnf.n\endcsname{162}
\expandafter\def\csname rg@count4.seq.exact:cnf.s\endcsname{1.7}
\expandafter\def\csname rg@count4.seq.hull.n\endcsname{--}
\expandafter\def\csname rg@count4.seq.hull.s\endcsname{--}
\expandafter\def\csname rg@count4.seq.int.n\endcsname{--}
\expandafter\def\csname rg@count4.seq.int.s\endcsname{--}
\expandafter\def\csname rg@count4.seq.kissat.n\endcsname{162}
\expandafter\def\csname rg@count4.seq.kissat.s\endcsname{0.12}
\expandafter\def\csname rg@count4.seq.lingeling.n\endcsname{162}
\expandafter\def\csname rg@count4.seq.lingeling.s\endcsname{0.11}
\expandafter\def\csname rg@count4.seq.pgpbs-2.n\endcsname{--}
\expandafter\def\csname rg@count4.seq.pgpbs-2.s\endcsname{--}
\expandafter\def\csname rg@count4.seq.pgpbs-3.n\endcsname{--}
\expandafter\def\csname rg@count4.seq.pgpbs-3.s\endcsname{--}
\expandafter\def\csname rg@count4.seq.pgpbs-4.n\endcsname{--}
\expandafter\def\csname rg@count4.seq.pgpbs-4.s\endcsname{--}
\expandafter\def\csname rg@count4.seq.pgpbs-a.n\endcsname{--}
\expandafter\def\csname rg@count4.seq.pgpbs-a.s\endcsname{--}
\expandafter\def\csname rg@count4.seq.pgpbs-i.n\endcsname{--}
\expandafter\def\csname rg@count4.seq.pgpbs-i.s\endcsname{--}
\expandafter\def\csname rg@count4.seq.read.fastest\endcsname{ring after elim.}
\expandafter\def\csname rg@count4.seq.read.n\endcsname{642}
\expandafter\def\csname rg@count4.seq.read.s\endcsname{12.0}
\expandafter\def\csname rg@count4.seq.read.smax\endcsname{12.0}
\expandafter\def\csname rg@count4.seq.read.tool\endcsname{ring after elim.}
\expandafter\def\csname rg@count4.seq.roundingsat:cnf.n\endcsname{162}
\expandafter\def\csname rg@count4.seq.roundingsat:cnf.s\endcsname{0.16}
\expandafter\def\csname rg@count4.seq.sat4j-cp:cnf.n\endcsname{162}
\expandafter\def\csname rg@count4.seq.sat4j-cp:cnf.s\endcsname{0.89}
\expandafter\def\csname rg@count4.seq.scip:cnf.n\endcsname{162}
\expandafter\def\csname rg@count4.seq.scip:cnf.s\endcsname{39.5}
\expandafter\def\csname rg@count4.seq.top\endcsname{642}
\expandafter\def\csname rg@count4.seq.win.n\endcsname{--}
\expandafter\def\csname rg@count4.seq.win.s\endcsname{--}
\expandafter\def\csname rg@count4.seq.xor.n\endcsname{82}
\expandafter\def\csname rg@count4.seq.xor.s\endcsname{0.23}
\expandafter\def\csname rg@count4.seq.xorcle-dev.n\endcsname{82}
\expandafter\def\csname rg@count4.seq.xorcle-dev.s\endcsname{0.23}
\expandafter\def\csname rg@count4.seq.xorcle.n\endcsname{82}
\expandafter\def\csname rg@count4.seq.xorcle.s\endcsname{0.28}
\expandafter\def\csname rg@count9.binary.best.n\endcsname{327}
\expandafter\def\csname rg@count9.binary.bve+hull.n\endcsname{--}
\expandafter\def\csname rg@count9.binary.bve+hull.s\endcsname{--}
\expandafter\def\csname rg@count9.binary.bve+int.n\endcsname{--}
\expandafter\def\csname rg@count9.binary.bve+int.s\endcsname{--}
\expandafter\def\csname rg@count9.binary.bve+win.n\endcsname{--}
\expandafter\def\csname rg@count9.binary.bve+win.s\endcsname{--}
\expandafter\def\csname rg@count9.binary.cadical.n\endcsname{327}
\expandafter\def\csname rg@count9.binary.cadical.s\endcsname{0.14}
\expandafter\def\csname rg@count9.binary.cms.n\endcsname{327}
\expandafter\def\csname rg@count9.binary.cms.s\endcsname{0.13}
\expandafter\def\csname rg@count9.binary.cnf.n\endcsname{327}
\expandafter\def\csname rg@count9.binary.cnf.s\endcsname{1.4}
\expandafter\def\csname rg@count9.binary.cnf.tools\endcsname{CDCL, Sat4j, RoundingSat, Exact}
\expandafter\def\csname rg@count9.binary.cnfx.n\endcsname{327}
\expandafter\def\csname rg@count9.binary.cnfx.tools\endcsname{CDCL, Sat4j, RoundingSat, Exact}
\expandafter\def\csname rg@count9.binary.exact:cnf.n\endcsname{327}
\expandafter\def\csname rg@count9.binary.exact:cnf.s\endcsname{1.3}
\expandafter\def\csname rg@count9.binary.hull.n\endcsname{--}
\expandafter\def\csname rg@count9.binary.hull.s\endcsname{--}
\expandafter\def\csname rg@count9.binary.int.n\endcsname{--}
\expandafter\def\csname rg@count9.binary.int.s\endcsname{--}
\expandafter\def\csname rg@count9.binary.kissat.n\endcsname{327}
\expandafter\def\csname rg@count9.binary.kissat.s\endcsname{0.16}
\expandafter\def\csname rg@count9.binary.lingeling.n\endcsname{327}
\expandafter\def\csname rg@count9.binary.lingeling.s\endcsname{1.1}
\expandafter\def\csname rg@count9.binary.pgpbs-2.n\endcsname{--}
\expandafter\def\csname rg@count9.binary.pgpbs-2.s\endcsname{--}
\expandafter\def\csname rg@count9.binary.pgpbs-3.n\endcsname{--}
\expandafter\def\csname rg@count9.binary.pgpbs-3.s\endcsname{--}
\expandafter\def\csname rg@count9.binary.pgpbs-9.n\endcsname{--}
\expandafter\def\csname rg@count9.binary.pgpbs-9.s\endcsname{--}
\expandafter\def\csname rg@count9.binary.pgpbs-a.n\endcsname{--}
\expandafter\def\csname rg@count9.binary.pgpbs-a.s\endcsname{--}
\expandafter\def\csname rg@count9.binary.pgpbs-i.n\endcsname{--}
\expandafter\def\csname rg@count9.binary.pgpbs-i.s\endcsname{--}
\expandafter\def\csname rg@count9.binary.read.fastest\endcsname{--}
\expandafter\def\csname rg@count9.binary.read.n\endcsname{--}
\expandafter\def\csname rg@count9.binary.read.s\endcsname{--}
\expandafter\def\csname rg@count9.binary.read.smax\endcsname{--}
\expandafter\def\csname rg@count9.binary.read.tool\endcsname{--}
\expandafter\def\csname rg@count9.binary.roundingsat:cnf.n\endcsname{327}
\expandafter\def\csname rg@count9.binary.roundingsat:cnf.s\endcsname{0.37}
\expandafter\def\csname rg@count9.binary.sat4j-cp:cnf.n\endcsname{327}
\expandafter\def\csname rg@count9.binary.sat4j-cp:cnf.s\endcsname{1.4}
\expandafter\def\csname rg@count9.binary.scip:cnf.n\endcsname{165}
\expandafter\def\csname rg@count9.binary.scip:cnf.s\endcsname{2.0}
\expandafter\def\csname rg@count9.binary.top\endcsname{327}
\expandafter\def\csname rg@count9.binary.win.n\endcsname{--}
\expandafter\def\csname rg@count9.binary.win.s\endcsname{--}
\expandafter\def\csname rg@count9.binary.xor.n\endcsname{165}
\expandafter\def\csname rg@count9.binary.xor.s\endcsname{3.7}
\expandafter\def\csname rg@count9.binary.xorcle-dev.n\endcsname{165}
\expandafter\def\csname rg@count9.binary.xorcle-dev.s\endcsname{3.7}
\expandafter\def\csname rg@count9.binary.xorcle.n\endcsname{165}
\expandafter\def\csname rg@count9.binary.xorcle.s\endcsname{3.9}
\expandafter\def\csname rg@count9.commander.best.n\endcsname{327}
\expandafter\def\csname rg@count9.commander.bve+hull.n\endcsname{--}
\expandafter\def\csname rg@count9.commander.bve+hull.s\endcsname{--}
\expandafter\def\csname rg@count9.commander.bve+int.n\endcsname{--}
\expandafter\def\csname rg@count9.commander.bve+int.s\endcsname{--}
\expandafter\def\csname rg@count9.commander.bve+win.n\endcsname{--}
\expandafter\def\csname rg@count9.commander.bve+win.s\endcsname{--}
\expandafter\def\csname rg@count9.commander.cadical.n\endcsname{327}
\expandafter\def\csname rg@count9.commander.cadical.s\endcsname{0.22}
\expandafter\def\csname rg@count9.commander.cms.n\endcsname{327}
\expandafter\def\csname rg@count9.commander.cms.s\endcsname{0.23}
\expandafter\def\csname rg@count9.commander.cnf.n\endcsname{327}
\expandafter\def\csname rg@count9.commander.cnf.s\endcsname{2.0}
\expandafter\def\csname rg@count9.commander.cnf.tools\endcsname{CDCL, Sat4j, RoundingSat, Exact}
\expandafter\def\csname rg@count9.commander.cnfx.n\endcsname{327}
\expandafter\def\csname rg@count9.commander.cnfx.tools\endcsname{CDCL, Sat4j, RoundingSat, Exact}
\expandafter\def\csname rg@count9.commander.exact:cnf.n\endcsname{327}
\expandafter\def\csname rg@count9.commander.exact:cnf.s\endcsname{1.5}
\expandafter\def\csname rg@count9.commander.hull.n\endcsname{--}
\expandafter\def\csname rg@count9.commander.hull.s\endcsname{--}
\expandafter\def\csname rg@count9.commander.int.n\endcsname{--}
\expandafter\def\csname rg@count9.commander.int.s\endcsname{--}
\expandafter\def\csname rg@count9.commander.kissat.n\endcsname{327}
\expandafter\def\csname rg@count9.commander.kissat.s\endcsname{0.18}
\expandafter\def\csname rg@count9.commander.lingeling.n\endcsname{327}
\expandafter\def\csname rg@count9.commander.lingeling.s\endcsname{0.47}
\expandafter\def\csname rg@count9.commander.pgpbs-2.n\endcsname{--}
\expandafter\def\csname rg@count9.commander.pgpbs-2.s\endcsname{--}
\expandafter\def\csname rg@count9.commander.pgpbs-3.n\endcsname{--}
\expandafter\def\csname rg@count9.commander.pgpbs-3.s\endcsname{--}
\expandafter\def\csname rg@count9.commander.pgpbs-9.n\endcsname{--}
\expandafter\def\csname rg@count9.commander.pgpbs-9.s\endcsname{--}
\expandafter\def\csname rg@count9.commander.pgpbs-a.n\endcsname{--}
\expandafter\def\csname rg@count9.commander.pgpbs-a.s\endcsname{--}
\expandafter\def\csname rg@count9.commander.pgpbs-i.n\endcsname{--}
\expandafter\def\csname rg@count9.commander.pgpbs-i.s\endcsname{--}
\expandafter\def\csname rg@count9.commander.read.fastest\endcsname{--}
\expandafter\def\csname rg@count9.commander.read.n\endcsname{--}
\expandafter\def\csname rg@count9.commander.read.s\endcsname{--}
\expandafter\def\csname rg@count9.commander.read.smax\endcsname{--}
\expandafter\def\csname rg@count9.commander.read.tool\endcsname{--}
\expandafter\def\csname rg@count9.commander.roundingsat:cnf.n\endcsname{327}
\expandafter\def\csname rg@count9.commander.roundingsat:cnf.s\endcsname{0.55}
\expandafter\def\csname rg@count9.commander.sat4j-cp:cnf.n\endcsname{327}
\expandafter\def\csname rg@count9.commander.sat4j-cp:cnf.s\endcsname{2.0}
\expandafter\def\csname rg@count9.commander.scip:cnf.n\endcsname{165}
\expandafter\def\csname rg@count9.commander.scip:cnf.s\endcsname{6.1}
\expandafter\def\csname rg@count9.commander.top\endcsname{327}
\expandafter\def\csname rg@count9.commander.win.n\endcsname{--}
\expandafter\def\csname rg@count9.commander.win.s\endcsname{--}
\expandafter\def\csname rg@count9.commander.xor.n\endcsname{165}
\expandafter\def\csname rg@count9.commander.xor.s\endcsname{3.0}
\expandafter\def\csname rg@count9.commander.xorcle-dev.n\endcsname{165}
\expandafter\def\csname rg@count9.commander.xorcle-dev.s\endcsname{3.0}
\expandafter\def\csname rg@count9.commander.xorcle.n\endcsname{165}
\expandafter\def\csname rg@count9.commander.xorcle.s\endcsname{4.3}
\expandafter\def\csname rg@count9.native.exact:pb.n\endcsname{327}
\expandafter\def\csname rg@count9.native.exact:pb.s\endcsname{1.6}
\expandafter\def\csname rg@count9.native.highs:pb.n\endcsname{327}
\expandafter\def\csname rg@count9.native.highs:pb.s\endcsname{9.5}
\expandafter\def\csname rg@count9.native.n\endcsname{327}
\expandafter\def\csname rg@count9.native.roundingsat:pb.n\endcsname{327}
\expandafter\def\csname rg@count9.native.roundingsat:pb.s\endcsname{0.36}
\expandafter\def\csname rg@count9.native.sat4j-cp:pb.n\endcsname{165}
\expandafter\def\csname rg@count9.native.sat4j-cp:pb.s\endcsname{0.70}
\expandafter\def\csname rg@count9.native.scip:pb.n\endcsname{327}
\expandafter\def\csname rg@count9.native.scip:pb.s\endcsname{4.4}
\expandafter\def\csname rg@count9.native.smax\endcsname{9.5}
\expandafter\def\csname rg@count9.native.smin\endcsname{0.36}
\expandafter\def\csname rg@count9.native.tools\endcsname{RoundingSat, Exact, SCIP, HiGHS}
\expandafter\def\csname rg@count9.native.toolsp\endcsname{RoundingSat, Exact, SCIP and HiGHS}
\expandafter\def\csname rg@count9.pairwise.best.n\endcsname{327}
\expandafter\def\csname rg@count9.pairwise.bve+hull.n\endcsname{84}
\expandafter\def\csname rg@count9.pairwise.bve+hull.s\endcsname{2.0}
\expandafter\def\csname rg@count9.pairwise.bve+int.n\endcsname{327}
\expandafter\def\csname rg@count9.pairwise.bve+int.s\endcsname{19.2}
\expandafter\def\csname rg@count9.pairwise.bve+win.n\endcsname{--}
\expandafter\def\csname rg@count9.pairwise.bve+win.s\endcsname{--}
\expandafter\def\csname rg@count9.pairwise.cadical.n\endcsname{327}
\expandafter\def\csname rg@count9.pairwise.cadical.s\endcsname{0.12}
\expandafter\def\csname rg@count9.pairwise.cms.n\endcsname{327}
\expandafter\def\csname rg@count9.pairwise.cms.s\endcsname{0.12}
\expandafter\def\csname rg@count9.pairwise.cnf.n\endcsname{327}
\expandafter\def\csname rg@count9.pairwise.cnf.s\endcsname{7.4}
\expandafter\def\csname rg@count9.pairwise.cnf.tools\endcsname{CDCL, Sat4j, RoundingSat, Exact, SCIP}
\expandafter\def\csname rg@count9.pairwise.cnfx.n\endcsname{327}
\expandafter\def\csname rg@count9.pairwise.cnfx.tools\endcsname{CDCL, Sat4j, RoundingSat, Exact, SCIP}
\expandafter\def\csname rg@count9.pairwise.exact:cnf.n\endcsname{327}
\expandafter\def\csname rg@count9.pairwise.exact:cnf.s\endcsname{1.1}
\expandafter\def\csname rg@count9.pairwise.hull.n\endcsname{84}
\expandafter\def\csname rg@count9.pairwise.hull.s\endcsname{2.0}
\expandafter\def\csname rg@count9.pairwise.int.n\endcsname{327}
\expandafter\def\csname rg@count9.pairwise.int.s\endcsname{19.3}
\expandafter\def\csname rg@count9.pairwise.kissat.n\endcsname{327}
\expandafter\def\csname rg@count9.pairwise.kissat.s\endcsname{0.15}
\expandafter\def\csname rg@count9.pairwise.lingeling.n\endcsname{327}
\expandafter\def\csname rg@count9.pairwise.lingeling.s\endcsname{0.23}
\expandafter\def\csname rg@count9.pairwise.pgpbs-2.n\endcsname{--}
\expandafter\def\csname rg@count9.pairwise.pgpbs-2.s\endcsname{--}
\expandafter\def\csname rg@count9.pairwise.pgpbs-3.n\endcsname{--}
\expandafter\def\csname rg@count9.pairwise.pgpbs-3.s\endcsname{--}
\expandafter\def\csname rg@count9.pairwise.pgpbs-9.n\endcsname{--}
\expandafter\def\csname rg@count9.pairwise.pgpbs-9.s\endcsname{--}
\expandafter\def\csname rg@count9.pairwise.pgpbs-a.n\endcsname{--}
\expandafter\def\csname rg@count9.pairwise.pgpbs-a.s\endcsname{--}
\expandafter\def\csname rg@count9.pairwise.pgpbs-i.n\endcsname{--}
\expandafter\def\csname rg@count9.pairwise.pgpbs-i.s\endcsname{--}
\expandafter\def\csname rg@count9.pairwise.read.fastest\endcsname{ring after elim.}
\expandafter\def\csname rg@count9.pairwise.read.n\endcsname{327}
\expandafter\def\csname rg@count9.pairwise.read.s\endcsname{19.2}
\expandafter\def\csname rg@count9.pairwise.read.smax\endcsname{19.3}
\expandafter\def\csname rg@count9.pairwise.read.tool\endcsname{ring, ring after elim.}
\expandafter\def\csname rg@count9.pairwise.roundingsat:cnf.n\endcsname{327}
\expandafter\def\csname rg@count9.pairwise.roundingsat:cnf.s\endcsname{0.25}
\expandafter\def\csname rg@count9.pairwise.sat4j-cp:cnf.n\endcsname{327}
\expandafter\def\csname rg@count9.pairwise.sat4j-cp:cnf.s\endcsname{1.3}
\expandafter\def\csname rg@count9.pairwise.scip:cnf.n\endcsname{327}
\expandafter\def\csname rg@count9.pairwise.scip:cnf.s\endcsname{7.4}
\expandafter\def\csname rg@count9.pairwise.top\endcsname{327}
\expandafter\def\csname rg@count9.pairwise.win.n\endcsname{--}
\expandafter\def\csname rg@count9.pairwise.win.s\endcsname{--}
\expandafter\def\csname rg@count9.pairwise.xor.n\endcsname{165}
\expandafter\def\csname rg@count9.pairwise.xor.s\endcsname{3.1}
\expandafter\def\csname rg@count9.pairwise.xorcle-dev.n\endcsname{165}
\expandafter\def\csname rg@count9.pairwise.xorcle-dev.s\endcsname{3.1}
\expandafter\def\csname rg@count9.pairwise.xorcle.n\endcsname{165}
\expandafter\def\csname rg@count9.pairwise.xorcle.s\endcsname{3.4}
\expandafter\def\csname rg@ctl.circuit\endcsname{30}
\expandafter\def\csname rg@ctl.cnf\endcsname{144}
\expandafter\def\csname rg@ctl.n\endcsname{174}
\expandafter\def\csname rg@ctl.wrong\endcsname{0}
\expandafter\def\csname rg@deg.binary.10.w12\endcsname{-}
\expandafter\def\csname rg@deg.binary.10.w12e\endcsname{-}
\expandafter\def\csname rg@deg.binary.10.w16\endcsname{2}
\expandafter\def\csname rg@deg.binary.10.w16e\endcsname{2}
\expandafter\def\csname rg@deg.binary.10.w8\endcsname{-}
\expandafter\def\csname rg@deg.binary.10.w8e\endcsname{-}
\expandafter\def\csname rg@deg.binary.12.w12\endcsname{-}
\expandafter\def\csname rg@deg.binary.12.w12e\endcsname{-}
\expandafter\def\csname rg@deg.binary.12.w16\endcsname{2}
\expandafter\def\csname rg@deg.binary.12.w16e\endcsname{2}
\expandafter\def\csname rg@deg.binary.12.w8\endcsname{-}
\expandafter\def\csname rg@deg.binary.12.w8e\endcsname{-}
\expandafter\def\csname rg@deg.binary.14.w12\endcsname{-}
\expandafter\def\csname rg@deg.binary.14.w12e\endcsname{-}
\expandafter\def\csname rg@deg.binary.14.w16\endcsname{-}
\expandafter\def\csname rg@deg.binary.14.w16e\endcsname{-}
\expandafter\def\csname rg@deg.binary.14.w8\endcsname{-}
\expandafter\def\csname rg@deg.binary.14.w8e\endcsname{-}
\expandafter\def\csname rg@deg.binary.16.w12\endcsname{-}
\expandafter\def\csname rg@deg.binary.16.w12e\endcsname{-}
\expandafter\def\csname rg@deg.binary.16.w16\endcsname{-}
\expandafter\def\csname rg@deg.binary.16.w16e\endcsname{-}
\expandafter\def\csname rg@deg.binary.16.w8\endcsname{-}
\expandafter\def\csname rg@deg.binary.16.w8e\endcsname{-}
\expandafter\def\csname rg@deg.binary.4.w12\endcsname{2}
\expandafter\def\csname rg@deg.binary.4.w12e\endcsname{2}
\expandafter\def\csname rg@deg.binary.4.w16\endcsname{2}
\expandafter\def\csname rg@deg.binary.4.w16e\endcsname{2}
\expandafter\def\csname rg@deg.binary.4.w8\endcsname{2}
\expandafter\def\csname rg@deg.binary.4.w8e\endcsname{-}
\expandafter\def\csname rg@deg.binary.6.w12\endcsname{2}
\expandafter\def\csname rg@deg.binary.6.w12e\endcsname{2}
\expandafter\def\csname rg@deg.binary.6.w16\endcsname{2}
\expandafter\def\csname rg@deg.binary.6.w16e\endcsname{2}
\expandafter\def\csname rg@deg.binary.6.w8\endcsname{-}
\expandafter\def\csname rg@deg.binary.6.w8e\endcsname{-}
\expandafter\def\csname rg@deg.binary.8.w12\endcsname{2}
\expandafter\def\csname rg@deg.binary.8.w12e\endcsname{2}
\expandafter\def\csname rg@deg.binary.8.w16\endcsname{2}
\expandafter\def\csname rg@deg.binary.8.w16e\endcsname{2}
\expandafter\def\csname rg@deg.binary.8.w8\endcsname{-}
\expandafter\def\csname rg@deg.binary.8.w8e\endcsname{-}
\expandafter\def\csname rg@deg.binary.w12.max\endcsname{8}
\expandafter\def\csname rg@deg.binary.w16.max\endcsname{12}
\expandafter\def\csname rg@deg.binary.w8.max\endcsname{4}
\expandafter\def\csname rg@deg.pairwise.10.w12\endcsname{2}
\expandafter\def\csname rg@deg.pairwise.10.w12e\endcsname{2}
\expandafter\def\csname rg@deg.pairwise.10.w16\endcsname{2}
\expandafter\def\csname rg@deg.pairwise.10.w16e\endcsname{2}
\expandafter\def\csname rg@deg.pairwise.10.w8\endcsname{-}
\expandafter\def\csname rg@deg.pairwise.10.w8e\endcsname{-}
\expandafter\def\csname rg@deg.pairwise.12.w12\endcsname{2}
\expandafter\def\csname rg@deg.pairwise.12.w12e\endcsname{2}
\expandafter\def\csname rg@deg.pairwise.12.w16\endcsname{2}
\expandafter\def\csname rg@deg.pairwise.12.w16e\endcsname{2}
\expandafter\def\csname rg@deg.pairwise.12.w8\endcsname{-}
\expandafter\def\csname rg@deg.pairwise.12.w8e\endcsname{-}
\expandafter\def\csname rg@deg.pairwise.14.w12\endcsname{-}
\expandafter\def\csname rg@deg.pairwise.14.w12e\endcsname{-}
\expandafter\def\csname rg@deg.pairwise.14.w16\endcsname{2}
\expandafter\def\csname rg@deg.pairwise.14.w16e\endcsname{2}
\expandafter\def\csname rg@deg.pairwise.14.w8\endcsname{-}
\expandafter\def\csname rg@deg.pairwise.14.w8e\endcsname{-}
\expandafter\def\csname rg@deg.pairwise.16.w12\endcsname{-}
\expandafter\def\csname rg@deg.pairwise.16.w12e\endcsname{-}
\expandafter\def\csname rg@deg.pairwise.16.w16\endcsname{2}
\expandafter\def\csname rg@deg.pairwise.16.w16e\endcsname{2}
\expandafter\def\csname rg@deg.pairwise.16.w8\endcsname{-}
\expandafter\def\csname rg@deg.pairwise.16.w8e\endcsname{-}
\expandafter\def\csname rg@deg.pairwise.4.w12\endcsname{2}
\expandafter\def\csname rg@deg.pairwise.4.w12e\endcsname{2}
\expandafter\def\csname rg@deg.pairwise.4.w16\endcsname{2}
\expandafter\def\csname rg@deg.pairwise.4.w16e\endcsname{2}
\expandafter\def\csname rg@deg.pairwise.4.w8\endcsname{2}
\expandafter\def\csname rg@deg.pairwise.4.w8e\endcsname{2}
\expandafter\def\csname rg@deg.pairwise.6.w12\endcsname{2}
\expandafter\def\csname rg@deg.pairwise.6.w12e\endcsname{2}
\expandafter\def\csname rg@deg.pairwise.6.w16\endcsname{2}
\expandafter\def\csname rg@deg.pairwise.6.w16e\endcsname{2}
\expandafter\def\csname rg@deg.pairwise.6.w8\endcsname{2}
\expandafter\def\csname rg@deg.pairwise.6.w8e\endcsname{-}
\expandafter\def\csname rg@deg.pairwise.8.w12\endcsname{2}
\expandafter\def\csname rg@deg.pairwise.8.w12e\endcsname{2}
\expandafter\def\csname rg@deg.pairwise.8.w16\endcsname{2}
\expandafter\def\csname rg@deg.pairwise.8.w16e\endcsname{2}
\expandafter\def\csname rg@deg.pairwise.8.w8\endcsname{2}
\expandafter\def\csname rg@deg.pairwise.8.w8e\endcsname{2}
\expandafter\def\csname rg@deg.pairwise.w12.max\endcsname{12}
\expandafter\def\csname rg@deg.pairwise.w16.max\endcsname{16}
\expandafter\def\csname rg@deg.pairwise.w8.max\endcsname{8}
\expandafter\def\csname rg@enc.bve+int.refuted\endcsname{29}
\expandafter\def\csname rg@enc.bve+win.refuted\endcsname{21}
\expandafter\def\csname rg@enc.combos\endcsname{29}
\expandafter\def\csname rg@enc.count3.count\endcsname{8}
\expandafter\def\csname rg@enc.count3.n\endcsname{80}
\expandafter\def\csname rg@enc.count4.count\endcsname{8}
\expandafter\def\csname rg@enc.count4.n\endcsname{82}
\expandafter\def\csname rg@enc.int.refuted\endcsname{18}
\expandafter\def\csname rg@enc.lingeling.refuted\endcsname{16}
\expandafter\def\csname rg@enc.oddfactor.count\endcsname{5}
\expandafter\def\csname rg@enc.oddfactor.n\endcsname{41}
\expandafter\def\csname rg@enc.pgpbsmod.refuted\endcsname{5}
\expandafter\def\csname rg@enc.pmodd.count\endcsname{8}
\expandafter\def\csname rg@enc.pmodd.n\endcsname{81}
\expandafter\def\csname rg@enc.scip.refuted\endcsname{19}
\expandafter\def\csname rg@enc.scip.some\endcsname{21}
\expandafter\def\csname rg@enc.win.refuted\endcsname{14}
\expandafter\def\csname rg@fam.count\endcsname{10}
\expandafter\def\csname rg@fam.count.word\endcsname{10}
\expandafter\def\csname rg@guarded.binary.base.n\endcsname{41}
\expandafter\def\csname rg@guarded.binary.basex.n\endcsname{41}
\expandafter\def\csname rg@guarded.binary.probe.n\endcsname{161}
\expandafter\def\csname rg@guarded.binary.probe.s\endcsname{21.1}
\expandafter\def\csname rg@guarded.binary.scip.n\endcsname{21}
\expandafter\def\csname rg@guarded.binary.xor.n\endcsname{81}
\expandafter\def\csname rg@guarded.pairwise.base.n\endcsname{41}
\expandafter\def\csname rg@guarded.pairwise.basex.n\endcsname{41}
\expandafter\def\csname rg@guarded.pairwise.probe.n\endcsname{161}
\expandafter\def\csname rg@guarded.pairwise.probe.s\endcsname{11.0}
\expandafter\def\csname rg@guarded.pairwise.scip.n\endcsname{21}
\expandafter\def\csname rg@guarded.pairwise.xor.n\endcsname{161}
\expandafter\def\csname rg@guarded3.binary.base.n\endcsname{80}
\expandafter\def\csname rg@guarded3.binary.basex.n\endcsname{80}
\expandafter\def\csname rg@guarded3.binary.probe.n\endcsname{160}
\expandafter\def\csname rg@guarded3.binary.probe.s\endcsname{41.3}
\expandafter\def\csname rg@guarded3.binary.scip.n\endcsname{40}
\expandafter\def\csname rg@guarded3.binary.xor.n\endcsname{40}
\expandafter\def\csname rg@guarded3.pairwise.base.n\endcsname{80}
\expandafter\def\csname rg@guarded3.pairwise.basex.n\endcsname{80}
\expandafter\def\csname rg@guarded3.pairwise.probe.n\endcsname{160}
\expandafter\def\csname rg@guarded3.pairwise.probe.s\endcsname{20.2}
\expandafter\def\csname rg@guarded3.pairwise.scip.n\endcsname{40}
\expandafter\def\csname rg@guarded3.pairwise.xor.n\endcsname{40}
\expandafter\def\csname rg@limit\endcsname{60}
\expandafter\def\csname rg@mchess.binary.best.n\endcsname{24}
\expandafter\def\csname rg@mchess.binary.bve+hull.n\endcsname{24}
\expandafter\def\csname rg@mchess.binary.bve+hull.s\endcsname{2.5}
\expandafter\def\csname rg@mchess.binary.bve+int.n\endcsname{24}
\expandafter\def\csname rg@mchess.binary.bve+int.s\endcsname{0.49}
\expandafter\def\csname rg@mchess.binary.bve+win.n\endcsname{24}
\expandafter\def\csname rg@mchess.binary.bve+win.s\endcsname{0.50}
\expandafter\def\csname rg@mchess.binary.cadical.n\endcsname{16}
\expandafter\def\csname rg@mchess.binary.cadical.s\endcsname{44.3}
\expandafter\def\csname rg@mchess.binary.cms.n\endcsname{12}
\expandafter\def\csname rg@mchess.binary.cms.s\endcsname{0.66}
\expandafter\def\csname rg@mchess.binary.cnf.n\endcsname{16}
\expandafter\def\csname rg@mchess.binary.cnf.s\endcsname{44.3}
\expandafter\def\csname rg@mchess.binary.cnf.tools\endcsname{CaDiCaL}
\expandafter\def\csname rg@mchess.binary.cnfx.n\endcsname{16}
\expandafter\def\csname rg@mchess.binary.cnfx.tools\endcsname{CaDiCaL}
\expandafter\def\csname rg@mchess.binary.exact:cnf.n\endcsname{12}
\expandafter\def\csname rg@mchess.binary.exact:cnf.s\endcsname{13.7}
\expandafter\def\csname rg@mchess.binary.hull.n\endcsname{24}
\expandafter\def\csname rg@mchess.binary.hull.s\endcsname{2.5}
\expandafter\def\csname rg@mchess.binary.int.n\endcsname{24}
\expandafter\def\csname rg@mchess.binary.int.s\endcsname{1.1}
\expandafter\def\csname rg@mchess.binary.kissat.n\endcsname{12}
\expandafter\def\csname rg@mchess.binary.kissat.s\endcsname{0.46}
\expandafter\def\csname rg@mchess.binary.lingeling.n\endcsname{12}
\expandafter\def\csname rg@mchess.binary.lingeling.s\endcsname{0.69}
\expandafter\def\csname rg@mchess.binary.pgpbs-2.n\endcsname{--}
\expandafter\def\csname rg@mchess.binary.pgpbs-2.s\endcsname{--}
\expandafter\def\csname rg@mchess.binary.pgpbs-3.n\endcsname{--}
\expandafter\def\csname rg@mchess.binary.pgpbs-3.s\endcsname{--}
\expandafter\def\csname rg@mchess.binary.pgpbs-a.n\endcsname{--}
\expandafter\def\csname rg@mchess.binary.pgpbs-a.s\endcsname{--}
\expandafter\def\csname rg@mchess.binary.pgpbs-i.n\endcsname{--}
\expandafter\def\csname rg@mchess.binary.pgpbs-i.s\endcsname{--}
\expandafter\def\csname rg@mchess.binary.read.fastest\endcsname{ring after elim.}
\expandafter\def\csname rg@mchess.binary.read.n\endcsname{24}
\expandafter\def\csname rg@mchess.binary.read.s\endcsname{0.49}
\expandafter\def\csname rg@mchess.binary.read.smax\endcsname{1.1}
\expandafter\def\csname rg@mchess.binary.read.tool\endcsname{field, ring, field after elim., ring after elim.}
\expandafter\def\csname rg@mchess.binary.roundingsat:cnf.n\endcsname{8}
\expandafter\def\csname rg@mchess.binary.roundingsat:cnf.s\endcsname{0.02}
\expandafter\def\csname rg@mchess.binary.sat4j-cp:cnf.n\endcsname{12}
\expandafter\def\csname rg@mchess.binary.sat4j-cp:cnf.s\endcsname{14.2}
\expandafter\def\csname rg@mchess.binary.scip:cnf.n\endcsname{8}
\expandafter\def\csname rg@mchess.binary.scip:cnf.s\endcsname{0.14}
\expandafter\def\csname rg@mchess.binary.top\endcsname{24}
\expandafter\def\csname rg@mchess.binary.win.n\endcsname{24}
\expandafter\def\csname rg@mchess.binary.win.s\endcsname{0.89}
\expandafter\def\csname rg@mchess.binary.xor.n\endcsname{8}
\expandafter\def\csname rg@mchess.binary.xor.s\endcsname{1.5}
\expandafter\def\csname rg@mchess.binary.xorcle-dev.n\endcsname{8}
\expandafter\def\csname rg@mchess.binary.xorcle-dev.s\endcsname{1.5}
\expandafter\def\csname rg@mchess.binary.xorcle.n\endcsname{8}
\expandafter\def\csname rg@mchess.binary.xorcle.s\endcsname{1.6}
\expandafter\def\csname rg@mchess.commander.best.n\endcsname{24}
\expandafter\def\csname rg@mchess.commander.bve+hull.n\endcsname{24}
\expandafter\def\csname rg@mchess.commander.bve+hull.s\endcsname{4.8}
\expandafter\def\csname rg@mchess.commander.bve+int.n\endcsname{24}
\expandafter\def\csname rg@mchess.commander.bve+int.s\endcsname{0.63}
\expandafter\def\csname rg@mchess.commander.bve+win.n\endcsname{24}
\expandafter\def\csname rg@mchess.commander.bve+win.s\endcsname{0.63}
\expandafter\def\csname rg@mchess.commander.cadical.n\endcsname{12}
\expandafter\def\csname rg@mchess.commander.cadical.s\endcsname{0.36}
\expandafter\def\csname rg@mchess.commander.cms.n\endcsname{16}
\expandafter\def\csname rg@mchess.commander.cms.s\endcsname{45.8}
\expandafter\def\csname rg@mchess.commander.cnf.n\endcsname{24}
\expandafter\def\csname rg@mchess.commander.cnf.s\endcsname{0.14}
\expandafter\def\csname rg@mchess.commander.cnf.tools\endcsname{SCIP}
\expandafter\def\csname rg@mchess.commander.cnfx.n\endcsname{24}
\expandafter\def\csname rg@mchess.commander.cnfx.tools\endcsname{SCIP}
\expandafter\def\csname rg@mchess.commander.exact:cnf.n\endcsname{12}
\expandafter\def\csname rg@mchess.commander.exact:cnf.s\endcsname{6.9}
\expandafter\def\csname rg@mchess.commander.hull.n\endcsname{24}
\expandafter\def\csname rg@mchess.commander.hull.s\endcsname{3.6}
\expandafter\def\csname rg@mchess.commander.int.n\endcsname{24}
\expandafter\def\csname rg@mchess.commander.int.s\endcsname{2.1}
\expandafter\def\csname rg@mchess.commander.kissat.n\endcsname{12}
\expandafter\def\csname rg@mchess.commander.kissat.s\endcsname{0.42}
\expandafter\def\csname rg@mchess.commander.lingeling.n\endcsname{16}
\expandafter\def\csname rg@mchess.commander.lingeling.s\endcsname{0.01}
\expandafter\def\csname rg@mchess.commander.pgpbs-2.n\endcsname{--}
\expandafter\def\csname rg@mchess.commander.pgpbs-2.s\endcsname{--}
\expandafter\def\csname rg@mchess.commander.pgpbs-3.n\endcsname{--}
\expandafter\def\csname rg@mchess.commander.pgpbs-3.s\endcsname{--}
\expandafter\def\csname rg@mchess.commander.pgpbs-a.n\endcsname{--}
\expandafter\def\csname rg@mchess.commander.pgpbs-a.s\endcsname{--}
\expandafter\def\csname rg@mchess.commander.pgpbs-i.n\endcsname{--}
\expandafter\def\csname rg@mchess.commander.pgpbs-i.s\endcsname{--}
\expandafter\def\csname rg@mchess.commander.read.fastest\endcsname{ring after elim.}
\expandafter\def\csname rg@mchess.commander.read.n\endcsname{24}
\expandafter\def\csname rg@mchess.commander.read.s\endcsname{0.63}
\expandafter\def\csname rg@mchess.commander.read.smax\endcsname{2.1}
\expandafter\def\csname rg@mchess.commander.read.tool\endcsname{field, ring, field after elim., ring after elim.}
\expandafter\def\csname rg@mchess.commander.roundingsat:cnf.n\endcsname{8}
\expandafter\def\csname rg@mchess.commander.roundingsat:cnf.s\endcsname{0.02}
\expandafter\def\csname rg@mchess.commander.sat4j-cp:cnf.n\endcsname{12}
\expandafter\def\csname rg@mchess.commander.sat4j-cp:cnf.s\endcsname{5.8}
\expandafter\def\csname rg@mchess.commander.scip:cnf.n\endcsname{24}
\expandafter\def\csname rg@mchess.commander.scip:cnf.s\endcsname{0.14}
\expandafter\def\csname rg@mchess.commander.top\endcsname{24}
\expandafter\def\csname rg@mchess.commander.win.n\endcsname{24}
\expandafter\def\csname rg@mchess.commander.win.s\endcsname{1.6}
\expandafter\def\csname rg@mchess.commander.xor.n\endcsname{8}
\expandafter\def\csname rg@mchess.commander.xor.s\endcsname{0.64}
\expandafter\def\csname rg@mchess.commander.xorcle-dev.n\endcsname{8}
\expandafter\def\csname rg@mchess.commander.xorcle-dev.s\endcsname{0.64}
\expandafter\def\csname rg@mchess.commander.xorcle.n\endcsname{8}
\expandafter\def\csname rg@mchess.commander.xorcle.s\endcsname{0.69}
\expandafter\def\csname rg@mchess.native.exact:pb.n\endcsname{24}
\expandafter\def\csname rg@mchess.native.exact:pb.s\endcsname{0.02}
\expandafter\def\csname rg@mchess.native.highs:pb.n\endcsname{24}
\expandafter\def\csname rg@mchess.native.highs:pb.s\endcsname{0.02}
\expandafter\def\csname rg@mchess.native.n\endcsname{24}
\expandafter\def\csname rg@mchess.native.roundingsat:pb.n\endcsname{24}
\expandafter\def\csname rg@mchess.native.roundingsat:pb.s\endcsname{0.01}
\expandafter\def\csname rg@mchess.native.sat4j-cp:pb.n\endcsname{24}
\expandafter\def\csname rg@mchess.native.sat4j-cp:pb.s\endcsname{0.41}
\expandafter\def\csname rg@mchess.native.scip:pb.n\endcsname{24}
\expandafter\def\csname rg@mchess.native.scip:pb.s\endcsname{0.05}
\expandafter\def\csname rg@mchess.native.smax\endcsname{0.41}
\expandafter\def\csname rg@mchess.native.smin\endcsname{0.01}
\expandafter\def\csname rg@mchess.native.tools\endcsname{all five}
\expandafter\def\csname rg@mchess.native.toolsp\endcsname{all five native-form solvers}
\expandafter\def\csname rg@mchess.pairwise.best.n\endcsname{24}
\expandafter\def\csname rg@mchess.pairwise.bve+hull.n\endcsname{24}
\expandafter\def\csname rg@mchess.pairwise.bve+hull.s\endcsname{5.1}
\expandafter\def\csname rg@mchess.pairwise.bve+int.n\endcsname{24}
\expandafter\def\csname rg@mchess.pairwise.bve+int.s\endcsname{0.62}
\expandafter\def\csname rg@mchess.pairwise.bve+win.n\endcsname{24}
\expandafter\def\csname rg@mchess.pairwise.bve+win.s\endcsname{0.64}
\expandafter\def\csname rg@mchess.pairwise.cadical.n\endcsname{12}
\expandafter\def\csname rg@mchess.pairwise.cadical.s\endcsname{0.37}
\expandafter\def\csname rg@mchess.pairwise.cms.n\endcsname{16}
\expandafter\def\csname rg@mchess.pairwise.cms.s\endcsname{53.0}
\expandafter\def\csname rg@mchess.pairwise.cnf.n\endcsname{24}
\expandafter\def\csname rg@mchess.pairwise.cnf.s\endcsname{3.4}
\expandafter\def\csname rg@mchess.pairwise.cnf.tools\endcsname{Lingeling, pgpbs, pgpbs mod 3, pgpbs integer, SCIP}
\expandafter\def\csname rg@mchess.pairwise.cnfx.n\endcsname{24}
\expandafter\def\csname rg@mchess.pairwise.cnfx.tools\endcsname{Lingeling, pgpbs, pgpbs mod 3, pgpbs integer, SCIP}
\expandafter\def\csname rg@mchess.pairwise.exact:cnf.n\endcsname{12}
\expandafter\def\csname rg@mchess.pairwise.exact:cnf.s\endcsname{12.1}
\expandafter\def\csname rg@mchess.pairwise.hull.n\endcsname{24}
\expandafter\def\csname rg@mchess.pairwise.hull.s\endcsname{0.68}
\expandafter\def\csname rg@mchess.pairwise.int.n\endcsname{24}
\expandafter\def\csname rg@mchess.pairwise.int.s\endcsname{2.0}
\expandafter\def\csname rg@mchess.pairwise.kissat.n\endcsname{12}
\expandafter\def\csname rg@mchess.pairwise.kissat.s\endcsname{0.60}
\expandafter\def\csname rg@mchess.pairwise.lingeling.n\endcsname{24}
\expandafter\def\csname rg@mchess.pairwise.lingeling.s\endcsname{0.01}
\expandafter\def\csname rg@mchess.pairwise.pgpbs-2.n\endcsname{--}
\expandafter\def\csname rg@mchess.pairwise.pgpbs-2.s\endcsname{--}
\expandafter\def\csname rg@mchess.pairwise.pgpbs-3.n\endcsname{24}
\expandafter\def\csname rg@mchess.pairwise.pgpbs-3.s\endcsname{1.3}
\expandafter\def\csname rg@mchess.pairwise.pgpbs-a.n\endcsname{24}
\expandafter\def\csname rg@mchess.pairwise.pgpbs-a.s\endcsname{1.4}
\expandafter\def\csname rg@mchess.pairwise.pgpbs-i.n\endcsname{24}
\expandafter\def\csname rg@mchess.pairwise.pgpbs-i.s\endcsname{3.4}
\expandafter\def\csname rg@mchess.pairwise.read.fastest\endcsname{ring after elim.}
\expandafter\def\csname rg@mchess.pairwise.read.n\endcsname{24}
\expandafter\def\csname rg@mchess.pairwise.read.s\endcsname{0.62}
\expandafter\def\csname rg@mchess.pairwise.read.smax\endcsname{2.0}
\expandafter\def\csname rg@mchess.pairwise.read.tool\endcsname{field, ring, field after elim., ring after elim.}
\expandafter\def\csname rg@mchess.pairwise.roundingsat:cnf.n\endcsname{8}
\expandafter\def\csname rg@mchess.pairwise.roundingsat:cnf.s\endcsname{0.02}
\expandafter\def\csname rg@mchess.pairwise.sat4j-cp:cnf.n\endcsname{12}
\expandafter\def\csname rg@mchess.pairwise.sat4j-cp:cnf.s\endcsname{21.4}
\expandafter\def\csname rg@mchess.pairwise.scip:cnf.n\endcsname{24}
\expandafter\def\csname rg@mchess.pairwise.scip:cnf.s\endcsname{0.12}
\expandafter\def\csname rg@mchess.pairwise.top\endcsname{24}
\expandafter\def\csname rg@mchess.pairwise.win.n\endcsname{24}
\expandafter\def\csname rg@mchess.pairwise.win.s\endcsname{1.7}
\expandafter\def\csname rg@mchess.pairwise.xor.n\endcsname{8}
\expandafter\def\csname rg@mchess.pairwise.xor.s\endcsname{0.66}
\expandafter\def\csname rg@mchess.pairwise.xorcle-dev.n\endcsname{8}
\expandafter\def\csname rg@mchess.pairwise.xorcle-dev.s\endcsname{0.66}
\expandafter\def\csname rg@mchess.pairwise.xorcle.n\endcsname{8}
\expandafter\def\csname rg@mchess.pairwise.xorcle.s\endcsname{0.67}
\expandafter\def\csname rg@mchess.seq.best.n\endcsname{24}
\expandafter\def\csname rg@mchess.seq.bve+hull.n\endcsname{24}
\expandafter\def\csname rg@mchess.seq.bve+hull.s\endcsname{4.9}
\expandafter\def\csname rg@mchess.seq.bve+int.n\endcsname{24}
\expandafter\def\csname rg@mchess.seq.bve+int.s\endcsname{0.64}
\expandafter\def\csname rg@mchess.seq.bve+win.n\endcsname{24}
\expandafter\def\csname rg@mchess.seq.bve+win.s\endcsname{0.66}
\expandafter\def\csname rg@mchess.seq.cadical.n\endcsname{12}
\expandafter\def\csname rg@mchess.seq.cadical.s\endcsname{0.47}
\expandafter\def\csname rg@mchess.seq.cms.n\endcsname{12}
\expandafter\def\csname rg@mchess.seq.cms.s\endcsname{1.0}
\expandafter\def\csname rg@mchess.seq.cnf.n\endcsname{24}
\expandafter\def\csname rg@mchess.seq.cnf.s\endcsname{4.5}
\expandafter\def\csname rg@mchess.seq.cnf.tools\endcsname{pgpbs, pgpbs mod 3, pgpbs integer}
\expandafter\def\csname rg@mchess.seq.cnfx.n\endcsname{24}
\expandafter\def\csname rg@mchess.seq.cnfx.tools\endcsname{pgpbs, pgpbs mod 3, pgpbs integer}
\expandafter\def\csname rg@mchess.seq.exact:cnf.n\endcsname{12}
\expandafter\def\csname rg@mchess.seq.exact:cnf.s\endcsname{20.4}
\expandafter\def\csname rg@mchess.seq.hull.n\endcsname{24}
\expandafter\def\csname rg@mchess.seq.hull.s\endcsname{4.9}
\expandafter\def\csname rg@mchess.seq.int.n\endcsname{24}
\expandafter\def\csname rg@mchess.seq.int.s\endcsname{2.4}
\expandafter\def\csname rg@mchess.seq.kissat.n\endcsname{12}
\expandafter\def\csname rg@mchess.seq.kissat.s\endcsname{0.62}
\expandafter\def\csname rg@mchess.seq.lingeling.n\endcsname{12}
\expandafter\def\csname rg@mchess.seq.lingeling.s\endcsname{1.0}
\expandafter\def\csname rg@mchess.seq.pgpbs-2.n\endcsname{--}
\expandafter\def\csname rg@mchess.seq.pgpbs-2.s\endcsname{--}
\expandafter\def\csname rg@mchess.seq.pgpbs-3.n\endcsname{24}
\expandafter\def\csname rg@mchess.seq.pgpbs-3.s\endcsname{2.5}
\expandafter\def\csname rg@mchess.seq.pgpbs-a.n\endcsname{24}
\expandafter\def\csname rg@mchess.seq.pgpbs-a.s\endcsname{2.5}
\expandafter\def\csname rg@mchess.seq.pgpbs-i.n\endcsname{24}
\expandafter\def\csname rg@mchess.seq.pgpbs-i.s\endcsname{4.5}
\expandafter\def\csname rg@mchess.seq.read.fastest\endcsname{ring after elim.}
\expandafter\def\csname rg@mchess.seq.read.n\endcsname{24}
\expandafter\def\csname rg@mchess.seq.read.s\endcsname{0.64}
\expandafter\def\csname rg@mchess.seq.read.smax\endcsname{2.4}
\expandafter\def\csname rg@mchess.seq.read.tool\endcsname{field, ring, field after elim., ring after elim.}
\expandafter\def\csname rg@mchess.seq.roundingsat:cnf.n\endcsname{8}
\expandafter\def\csname rg@mchess.seq.roundingsat:cnf.s\endcsname{0.02}
\expandafter\def\csname rg@mchess.seq.sat4j-cp:cnf.n\endcsname{12}
\expandafter\def\csname rg@mchess.seq.sat4j-cp:cnf.s\endcsname{7.1}
\expandafter\def\csname rg@mchess.seq.scip:cnf.n\endcsname{8}
\expandafter\def\csname rg@mchess.seq.scip:cnf.s\endcsname{0.09}
\expandafter\def\csname rg@mchess.seq.top\endcsname{24}
\expandafter\def\csname rg@mchess.seq.win.n\endcsname{24}
\expandafter\def\csname rg@mchess.seq.win.s\endcsname{2.2}
\expandafter\def\csname rg@mchess.seq.xor.n\endcsname{8}
\expandafter\def\csname rg@mchess.seq.xor.s\endcsname{0.53}
\expandafter\def\csname rg@mchess.seq.xorcle-dev.n\endcsname{8}
\expandafter\def\csname rg@mchess.seq.xorcle-dev.s\endcsname{0.53}
\expandafter\def\csname rg@mchess.seq.xorcle.n\endcsname{8}
\expandafter\def\csname rg@mchess.seq.xorcle.s\endcsname{0.59}
\expandafter\def\csname rg@neos.baseline.text\endcsname{neither SCIP nor HiGHS decided it within 300~s}
\expandafter\def\csname rg@neos.consistent\endcsname{2, 3 and 5}
\expandafter\def\csname rg@neos.consistent.all\endcsname{2, 4, 8, 16, 3, 9, 27 and 5}
\expandafter\def\csname rg@neos.equalities\endcsname{168}
\expandafter\def\csname rg@neos.highs\endcsname{timeout}
\expandafter\def\csname rg@neos.modulus\endcsname{25}
\expandafter\def\csname rg@neos.relations\endcsname{157}
\expandafter\def\csname rg@neos.rhs\endcsname{15}
\expandafter\def\csname rg@neos.scip\endcsname{timeout}
\expandafter\def\csname rg@neos.seconds\endcsname{0.76}
\expandafter\def\csname rg@neos.timeout\endcsname{300}
\expandafter\def\csname rg@neos.vars\endcsname{31,329}
\expandafter\def\csname rg@oddfactor.direct.best.n\endcsname{161}
\expandafter\def\csname rg@oddfactor.direct.bve+hull.n\endcsname{--}
\expandafter\def\csname rg@oddfactor.direct.bve+hull.s\endcsname{--}
\expandafter\def\csname rg@oddfactor.direct.bve+int.n\endcsname{161}
\expandafter\def\csname rg@oddfactor.direct.bve+int.s\endcsname{1.0}
\expandafter\def\csname rg@oddfactor.direct.bve+win.n\endcsname{161}
\expandafter\def\csname rg@oddfactor.direct.bve+win.s\endcsname{1.0}
\expandafter\def\csname rg@oddfactor.direct.cadical.n\endcsname{15}
\expandafter\def\csname rg@oddfactor.direct.cadical.s\endcsname{0.83}
\expandafter\def\csname rg@oddfactor.direct.cms.n\endcsname{15}
\expandafter\def\csname rg@oddfactor.direct.cms.s\endcsname{1.7}
\expandafter\def\csname rg@oddfactor.direct.cnf.n\endcsname{41}
\expandafter\def\csname rg@oddfactor.direct.cnf.s\endcsname{56.1}
\expandafter\def\csname rg@oddfactor.direct.cnf.tools\endcsname{Xorcle}
\expandafter\def\csname rg@oddfactor.direct.cnfx.n\endcsname{19}
\expandafter\def\csname rg@oddfactor.direct.cnfx.tools\endcsname{Kissat}
\expandafter\def\csname rg@oddfactor.direct.exact:cnf.n\endcsname{11}
\expandafter\def\csname rg@oddfactor.direct.exact:cnf.s\endcsname{1.2}
\expandafter\def\csname rg@oddfactor.direct.hull.n\endcsname{--}
\expandafter\def\csname rg@oddfactor.direct.hull.s\endcsname{--}
\expandafter\def\csname rg@oddfactor.direct.int.n\endcsname{161}
\expandafter\def\csname rg@oddfactor.direct.int.s\endcsname{1.0}
\expandafter\def\csname rg@oddfactor.direct.kissat.n\endcsname{19}
\expandafter\def\csname rg@oddfactor.direct.kissat.s\endcsname{52.7}
\expandafter\def\csname rg@oddfactor.direct.lingeling.n\endcsname{15}
\expandafter\def\csname rg@oddfactor.direct.lingeling.s\endcsname{1.6}
\expandafter\def\csname rg@oddfactor.direct.pgpbs-2.n\endcsname{--}
\expandafter\def\csname rg@oddfactor.direct.pgpbs-2.s\endcsname{--}
\expandafter\def\csname rg@oddfactor.direct.pgpbs-3.n\endcsname{--}
\expandafter\def\csname rg@oddfactor.direct.pgpbs-3.s\endcsname{--}
\expandafter\def\csname rg@oddfactor.direct.pgpbs-a.n\endcsname{--}
\expandafter\def\csname rg@oddfactor.direct.pgpbs-a.s\endcsname{--}
\expandafter\def\csname rg@oddfactor.direct.pgpbs-i.n\endcsname{--}
\expandafter\def\csname rg@oddfactor.direct.pgpbs-i.s\endcsname{--}
\expandafter\def\csname rg@oddfactor.direct.read.fastest\endcsname{ring after elim.}
\expandafter\def\csname rg@oddfactor.direct.read.n\endcsname{161}
\expandafter\def\csname rg@oddfactor.direct.read.s\endcsname{1.0}
\expandafter\def\csname rg@oddfactor.direct.read.smax\endcsname{1.0}
\expandafter\def\csname rg@oddfactor.direct.read.tool\endcsname{field, ring, field after elim., ring after elim.}
\expandafter\def\csname rg@oddfactor.direct.roundingsat:cnf.n\endcsname{11}
\expandafter\def\csname rg@oddfactor.direct.roundingsat:cnf.s\endcsname{1.1}
\expandafter\def\csname rg@oddfactor.direct.sat4j-cp:cnf.n\endcsname{15}
\expandafter\def\csname rg@oddfactor.direct.sat4j-cp:cnf.s\endcsname{5.9}
\expandafter\def\csname rg@oddfactor.direct.scip:cnf.n\endcsname{11}
\expandafter\def\csname rg@oddfactor.direct.scip:cnf.s\endcsname{19.3}
\expandafter\def\csname rg@oddfactor.direct.top\endcsname{161}
\expandafter\def\csname rg@oddfactor.direct.win.n\endcsname{161}
\expandafter\def\csname rg@oddfactor.direct.win.s\endcsname{1.0}
\expandafter\def\csname rg@oddfactor.direct.xor.n\endcsname{41}
\expandafter\def\csname rg@oddfactor.direct.xor.s\endcsname{55.4}
\expandafter\def\csname rg@oddfactor.direct.xorcle-dev.n\endcsname{41}
\expandafter\def\csname rg@oddfactor.direct.xorcle-dev.s\endcsname{55.4}
\expandafter\def\csname rg@oddfactor.direct.xorcle.n\endcsname{41}
\expandafter\def\csname rg@oddfactor.direct.xorcle.s\endcsname{56.1}
\expandafter\def\csname rg@oddfactor.native.exact:pb.n\endcsname{161}
\expandafter\def\csname rg@oddfactor.native.exact:pb.s\endcsname{0.28}
\expandafter\def\csname rg@oddfactor.native.highs:pb.n\endcsname{161}
\expandafter\def\csname rg@oddfactor.native.highs:pb.s\endcsname{2.6}
\expandafter\def\csname rg@oddfactor.native.n\endcsname{161}
\expandafter\def\csname rg@oddfactor.native.roundingsat:pb.n\endcsname{161}
\expandafter\def\csname rg@oddfactor.native.roundingsat:pb.s\endcsname{0.20}
\expandafter\def\csname rg@oddfactor.native.sat4j-cp:pb.n\endcsname{11}
\expandafter\def\csname rg@oddfactor.native.sat4j-cp:pb.s\endcsname{0.62}
\expandafter\def\csname rg@oddfactor.native.scip:pb.n\endcsname{161}
\expandafter\def\csname rg@oddfactor.native.scip:pb.s\endcsname{0.86}
\expandafter\def\csname rg@oddfactor.native.smax\endcsname{2.6}
\expandafter\def\csname rg@oddfactor.native.smin\endcsname{0.20}
\expandafter\def\csname rg@oddfactor.native.tools\endcsname{RoundingSat, Exact, SCIP, HiGHS}
\expandafter\def\csname rg@oddfactor.native.toolsp\endcsname{RoundingSat, Exact, SCIP and HiGHS}
\expandafter\def\csname rg@oddfactor.seq.best.n\endcsname{161}
\expandafter\def\csname rg@oddfactor.seq.bve+hull.n\endcsname{--}
\expandafter\def\csname rg@oddfactor.seq.bve+hull.s\endcsname{--}
\expandafter\def\csname rg@oddfactor.seq.bve+int.n\endcsname{161}
\expandafter\def\csname rg@oddfactor.seq.bve+int.s\endcsname{0.86}
\expandafter\def\csname rg@oddfactor.seq.bve+win.n\endcsname{161}
\expandafter\def\csname rg@oddfactor.seq.bve+win.s\endcsname{0.88}
\expandafter\def\csname rg@oddfactor.seq.cadical.n\endcsname{15}
\expandafter\def\csname rg@oddfactor.seq.cadical.s\endcsname{1.2}
\expandafter\def\csname rg@oddfactor.seq.cms.n\endcsname{15}
\expandafter\def\csname rg@oddfactor.seq.cms.s\endcsname{3.8}
\expandafter\def\csname rg@oddfactor.seq.cnf.n\endcsname{19}
\expandafter\def\csname rg@oddfactor.seq.cnf.s\endcsname{57.4}
\expandafter\def\csname rg@oddfactor.seq.cnf.tools\endcsname{Kissat}
\expandafter\def\csname rg@oddfactor.seq.cnfx.n\endcsname{19}
\expandafter\def\csname rg@oddfactor.seq.cnfx.tools\endcsname{Kissat}
\expandafter\def\csname rg@oddfactor.seq.exact:cnf.n\endcsname{11}
\expandafter\def\csname rg@oddfactor.seq.exact:cnf.s\endcsname{1.5}
\expandafter\def\csname rg@oddfactor.seq.hull.n\endcsname{--}
\expandafter\def\csname rg@oddfactor.seq.hull.s\endcsname{--}
\expandafter\def\csname rg@oddfactor.seq.int.n\endcsname{--}
\expandafter\def\csname rg@oddfactor.seq.int.s\endcsname{--}
\expandafter\def\csname rg@oddfactor.seq.kissat.n\endcsname{19}
\expandafter\def\csname rg@oddfactor.seq.kissat.s\endcsname{57.4}
\expandafter\def\csname rg@oddfactor.seq.lingeling.n\endcsname{15}
\expandafter\def\csname rg@oddfactor.seq.lingeling.s\endcsname{2.0}
\expandafter\def\csname rg@oddfactor.seq.pgpbs-2.n\endcsname{--}
\expandafter\def\csname rg@oddfactor.seq.pgpbs-2.s\endcsname{--}
\expandafter\def\csname rg@oddfactor.seq.pgpbs-3.n\endcsname{--}
\expandafter\def\csname rg@oddfactor.seq.pgpbs-3.s\endcsname{--}
\expandafter\def\csname rg@oddfactor.seq.pgpbs-a.n\endcsname{--}
\expandafter\def\csname rg@oddfactor.seq.pgpbs-a.s\endcsname{--}
\expandafter\def\csname rg@oddfactor.seq.pgpbs-i.n\endcsname{--}
\expandafter\def\csname rg@oddfactor.seq.pgpbs-i.s\endcsname{--}
\expandafter\def\csname rg@oddfactor.seq.read.fastest\endcsname{ring after elim.}
\expandafter\def\csname rg@oddfactor.seq.read.n\endcsname{161}
\expandafter\def\csname rg@oddfactor.seq.read.s\endcsname{0.86}
\expandafter\def\csname rg@oddfactor.seq.read.smax\endcsname{0.88}
\expandafter\def\csname rg@oddfactor.seq.read.tool\endcsname{field after elim., ring after elim.}
\expandafter\def\csname rg@oddfactor.seq.roundingsat:cnf.n\endcsname{11}
\expandafter\def\csname rg@oddfactor.seq.roundingsat:cnf.s\endcsname{2.7}
\expandafter\def\csname rg@oddfactor.seq.sat4j-cp:cnf.n\endcsname{15}
\expandafter\def\csname rg@oddfactor.seq.sat4j-cp:cnf.s\endcsname{54.5}
\expandafter\def\csname rg@oddfactor.seq.scip:cnf.n\endcsname{11}
\expandafter\def\csname rg@oddfactor.seq.scip:cnf.s\endcsname{35.7}
\expandafter\def\csname rg@oddfactor.seq.top\endcsname{161}
\expandafter\def\csname rg@oddfactor.seq.win.n\endcsname{--}
\expandafter\def\csname rg@oddfactor.seq.win.s\endcsname{--}
\expandafter\def\csname rg@oddfactor.seq.xor.n\endcsname{--}
\expandafter\def\csname rg@oddfactor.seq.xor.s\endcsname{--}
\expandafter\def\csname rg@oddfactor.seq.xorcle-dev.n\endcsname{--}
\expandafter\def\csname rg@oddfactor.seq.xorcle-dev.s\endcsname{--}
\expandafter\def\csname rg@oddfactor.seq.xorcle.n\endcsname{--}
\expandafter\def\csname rg@oddfactor.seq.xorcle.s\endcsname{--}
\expandafter\def\csname rg@php.native.exact:pb.n\endcsname{32}
\expandafter\def\csname rg@php.native.exact:pb.s\endcsname{0.02}
\expandafter\def\csname rg@php.native.highs:pb.n\endcsname{32}
\expandafter\def\csname rg@php.native.highs:pb.s\endcsname{0.01}
\expandafter\def\csname rg@php.native.n\endcsname{32}
\expandafter\def\csname rg@php.native.roundingsat:pb.n\endcsname{32}
\expandafter\def\csname rg@php.native.roundingsat:pb.s\endcsname{0.01}
\expandafter\def\csname rg@php.native.sat4j-cp:pb.n\endcsname{32}
\expandafter\def\csname rg@php.native.sat4j-cp:pb.s\endcsname{0.36}
\expandafter\def\csname rg@php.native.scip:pb.n\endcsname{32}
\expandafter\def\csname rg@php.native.scip:pb.s\endcsname{0.03}
\expandafter\def\csname rg@php.native.smax\endcsname{0.36}
\expandafter\def\csname rg@php.native.smin\endcsname{0.01}
\expandafter\def\csname rg@php.native.tools\endcsname{all five}
\expandafter\def\csname rg@php.native.toolsp\endcsname{all five native-form solvers}
\expandafter\def\csname rg@php.pairwise.best.n\endcsname{32}
\expandafter\def\csname rg@php.pairwise.bve+hull.n\endcsname{32}
\expandafter\def\csname rg@php.pairwise.bve+hull.s\endcsname{13.0}
\expandafter\def\csname rg@php.pairwise.bve+int.n\endcsname{--}
\expandafter\def\csname rg@php.pairwise.bve+int.s\endcsname{--}
\expandafter\def\csname rg@php.pairwise.bve+win.n\endcsname{--}
\expandafter\def\csname rg@php.pairwise.bve+win.s\endcsname{--}
\expandafter\def\csname rg@php.pairwise.cadical.n\endcsname{8}
\expandafter\def\csname rg@php.pairwise.cadical.s\endcsname{0.20}
\expandafter\def\csname rg@php.pairwise.cms.n\endcsname{8}
\expandafter\def\csname rg@php.pairwise.cms.s\endcsname{0.29}
\expandafter\def\csname rg@php.pairwise.cnf.n\endcsname{32}
\expandafter\def\csname rg@php.pairwise.cnf.s\endcsname{5.4}
\expandafter\def\csname rg@php.pairwise.cnf.tools\endcsname{Lingeling, SCIP}
\expandafter\def\csname rg@php.pairwise.cnfx.n\endcsname{32}
\expandafter\def\csname rg@php.pairwise.cnfx.tools\endcsname{Lingeling, SCIP}
\expandafter\def\csname rg@php.pairwise.exact:cnf.n\endcsname{8}
\expandafter\def\csname rg@php.pairwise.exact:cnf.s\endcsname{2.2}
\expandafter\def\csname rg@php.pairwise.hull.n\endcsname{32}
\expandafter\def\csname rg@php.pairwise.hull.s\endcsname{13.0}
\expandafter\def\csname rg@php.pairwise.int.n\endcsname{--}
\expandafter\def\csname rg@php.pairwise.int.s\endcsname{--}
\expandafter\def\csname rg@php.pairwise.kissat.n\endcsname{8}
\expandafter\def\csname rg@php.pairwise.kissat.s\endcsname{0.05}
\expandafter\def\csname rg@php.pairwise.lingeling.n\endcsname{32}
\expandafter\def\csname rg@php.pairwise.lingeling.s\endcsname{0.03}
\expandafter\def\csname rg@php.pairwise.pgpbs-2.n\endcsname{16}
\expandafter\def\csname rg@php.pairwise.pgpbs-2.s\endcsname{12.9}
\expandafter\def\csname rg@php.pairwise.pgpbs-3.n\endcsname{16}
\expandafter\def\csname rg@php.pairwise.pgpbs-3.s\endcsname{13.0}
\expandafter\def\csname rg@php.pairwise.pgpbs-a.n\endcsname{16}
\expandafter\def\csname rg@php.pairwise.pgpbs-a.s\endcsname{13.1}
\expandafter\def\csname rg@php.pairwise.pgpbs-i.n\endcsname{16}
\expandafter\def\csname rg@php.pairwise.pgpbs-i.s\endcsname{13.2}
\expandafter\def\csname rg@php.pairwise.read.fastest\endcsname{--}
\expandafter\def\csname rg@php.pairwise.read.n\endcsname{--}
\expandafter\def\csname rg@php.pairwise.read.s\endcsname{--}
\expandafter\def\csname rg@php.pairwise.read.smax\endcsname{--}
\expandafter\def\csname rg@php.pairwise.read.tool\endcsname{--}
\expandafter\def\csname rg@php.pairwise.roundingsat:cnf.n\endcsname{8}
\expandafter\def\csname rg@php.pairwise.roundingsat:cnf.s\endcsname{4.9}
\expandafter\def\csname rg@php.pairwise.sat4j-cp:cnf.n\endcsname{8}
\expandafter\def\csname rg@php.pairwise.sat4j-cp:cnf.s\endcsname{0.88}
\expandafter\def\csname rg@php.pairwise.scip:cnf.n\endcsname{32}
\expandafter\def\csname rg@php.pairwise.scip:cnf.s\endcsname{5.4}
\expandafter\def\csname rg@php.pairwise.top\endcsname{32}
\expandafter\def\csname rg@php.pairwise.win.n\endcsname{--}
\expandafter\def\csname rg@php.pairwise.win.s\endcsname{--}
\expandafter\def\csname rg@php.pairwise.xor.n\endcsname{--}
\expandafter\def\csname rg@php.pairwise.xor.s\endcsname{--}
\expandafter\def\csname rg@php.pairwise.xorcle-dev.n\endcsname{--}
\expandafter\def\csname rg@php.pairwise.xorcle-dev.s\endcsname{--}
\expandafter\def\csname rg@php.pairwise.xorcle.n\endcsname{--}
\expandafter\def\csname rg@php.pairwise.xorcle.s\endcsname{--}
\expandafter\def\csname rg@php.product.best.n\endcsname{32}
\expandafter\def\csname rg@php.product.bve+hull.n\endcsname{8}
\expandafter\def\csname rg@php.product.bve+hull.s\endcsname{1.3}
\expandafter\def\csname rg@php.product.bve+int.n\endcsname{--}
\expandafter\def\csname rg@php.product.bve+int.s\endcsname{--}
\expandafter\def\csname rg@php.product.bve+win.n\endcsname{--}
\expandafter\def\csname rg@php.product.bve+win.s\endcsname{--}
\expandafter\def\csname rg@php.product.cadical.n\endcsname{8}
\expandafter\def\csname rg@php.product.cadical.s\endcsname{0.01}
\expandafter\def\csname rg@php.product.cms.n\endcsname{8}
\expandafter\def\csname rg@php.product.cms.s\endcsname{0.03}
\expandafter\def\csname rg@php.product.cnf.n\endcsname{16}
\expandafter\def\csname rg@php.product.cnf.s\endcsname{3.5}
\expandafter\def\csname rg@php.product.cnf.tools\endcsname{SCIP}
\expandafter\def\csname rg@php.product.cnfx.n\endcsname{16}
\expandafter\def\csname rg@php.product.cnfx.tools\endcsname{SCIP}
\expandafter\def\csname rg@php.product.exact:cnf.n\endcsname{12}
\expandafter\def\csname rg@php.product.exact:cnf.s\endcsname{16.4}
\expandafter\def\csname rg@php.product.hull.n\endcsname{8}
\expandafter\def\csname rg@php.product.hull.s\endcsname{1.3}
\expandafter\def\csname rg@php.product.int.n\endcsname{--}
\expandafter\def\csname rg@php.product.int.s\endcsname{--}
\expandafter\def\csname rg@php.product.kissat.n\endcsname{12}
\expandafter\def\csname rg@php.product.kissat.s\endcsname{0.78}
\expandafter\def\csname rg@php.product.lingeling.n\endcsname{12}
\expandafter\def\csname rg@php.product.lingeling.s\endcsname{1.8}
\expandafter\def\csname rg@php.product.pgpbs-2.n\endcsname{--}
\expandafter\def\csname rg@php.product.pgpbs-2.s\endcsname{--}
\expandafter\def\csname rg@php.product.pgpbs-3.n\endcsname{--}
\expandafter\def\csname rg@php.product.pgpbs-3.s\endcsname{--}
\expandafter\def\csname rg@php.product.pgpbs-a.n\endcsname{--}
\expandafter\def\csname rg@php.product.pgpbs-a.s\endcsname{--}
\expandafter\def\csname rg@php.product.pgpbs-i.n\endcsname{--}
\expandafter\def\csname rg@php.product.pgpbs-i.s\endcsname{--}
\expandafter\def\csname rg@php.product.read.fastest\endcsname{--}
\expandafter\def\csname rg@php.product.read.n\endcsname{--}
\expandafter\def\csname rg@php.product.read.s\endcsname{--}
\expandafter\def\csname rg@php.product.read.smax\endcsname{--}
\expandafter\def\csname rg@php.product.read.tool\endcsname{--}
\expandafter\def\csname rg@php.product.roundingsat:cnf.n\endcsname{8}
\expandafter\def\csname rg@php.product.roundingsat:cnf.s\endcsname{0.06}
\expandafter\def\csname rg@php.product.sat4j-cp:cnf.n\endcsname{8}
\expandafter\def\csname rg@php.product.sat4j-cp:cnf.s\endcsname{0.83}
\expandafter\def\csname rg@php.product.scip:cnf.n\endcsname{16}
\expandafter\def\csname rg@php.product.scip:cnf.s\endcsname{3.5}
\expandafter\def\csname rg@php.product.top\endcsname{32}
\expandafter\def\csname rg@php.product.win.n\endcsname{--}
\expandafter\def\csname rg@php.product.win.s\endcsname{--}
\expandafter\def\csname rg@php.product.xor.n\endcsname{8}
\expandafter\def\csname rg@php.product.xor.s\endcsname{4.5}
\expandafter\def\csname rg@php.product.xorcle-dev.n\endcsname{8}
\expandafter\def\csname rg@php.product.xorcle-dev.s\endcsname{4.6}
\expandafter\def\csname rg@php.product.xorcle.n\endcsname{8}
\expandafter\def\csname rg@php.product.xorcle.s\endcsname{4.5}
\expandafter\def\csname rg@php.seq.best.n\endcsname{32}
\expandafter\def\csname rg@php.seq.bve+hull.n\endcsname{32}
\expandafter\def\csname rg@php.seq.bve+hull.s\endcsname{19.2}
\expandafter\def\csname rg@php.seq.bve+int.n\endcsname{--}
\expandafter\def\csname rg@php.seq.bve+int.s\endcsname{--}
\expandafter\def\csname rg@php.seq.bve+win.n\endcsname{--}
\expandafter\def\csname rg@php.seq.bve+win.s\endcsname{--}
\expandafter\def\csname rg@php.seq.cadical.n\endcsname{12}
\expandafter\def\csname rg@php.seq.cadical.s\endcsname{0.57}
\expandafter\def\csname rg@php.seq.cms.n\endcsname{12}
\expandafter\def\csname rg@php.seq.cms.s\endcsname{0.91}
\expandafter\def\csname rg@php.seq.cnf.n\endcsname{16}
\expandafter\def\csname rg@php.seq.cnf.s\endcsname{13.3}
\expandafter\def\csname rg@php.seq.cnf.tools\endcsname{pgpbs, pgpbs mod 2, pgpbs mod 3, pgpbs integer}
\expandafter\def\csname rg@php.seq.cnfx.n\endcsname{16}
\expandafter\def\csname rg@php.seq.cnfx.tools\endcsname{pgpbs, pgpbs mod 2, pgpbs mod 3, pgpbs integer}
\expandafter\def\csname rg@php.seq.exact:cnf.n\endcsname{12}
\expandafter\def\csname rg@php.seq.exact:cnf.s\endcsname{11.1}
\expandafter\def\csname rg@php.seq.hull.n\endcsname{32}
\expandafter\def\csname rg@php.seq.hull.s\endcsname{19.2}
\expandafter\def\csname rg@php.seq.int.n\endcsname{--}
\expandafter\def\csname rg@php.seq.int.s\endcsname{--}
\expandafter\def\csname rg@php.seq.kissat.n\endcsname{12}
\expandafter\def\csname rg@php.seq.kissat.s\endcsname{1.4}
\expandafter\def\csname rg@php.seq.lingeling.n\endcsname{12}
\expandafter\def\csname rg@php.seq.lingeling.s\endcsname{2.0}
\expandafter\def\csname rg@php.seq.pgpbs-2.n\endcsname{16}
\expandafter\def\csname rg@php.seq.pgpbs-2.s\endcsname{13.3}
\expandafter\def\csname rg@php.seq.pgpbs-3.n\endcsname{16}
\expandafter\def\csname rg@php.seq.pgpbs-3.s\endcsname{13.2}
\expandafter\def\csname rg@php.seq.pgpbs-a.n\endcsname{16}
\expandafter\def\csname rg@php.seq.pgpbs-a.s\endcsname{13.1}
\expandafter\def\csname rg@php.seq.pgpbs-i.n\endcsname{16}
\expandafter\def\csname rg@php.seq.pgpbs-i.s\endcsname{13.1}
\expandafter\def\csname rg@php.seq.read.fastest\endcsname{--}
\expandafter\def\csname rg@php.seq.read.n\endcsname{--}
\expandafter\def\csname rg@php.seq.read.s\endcsname{--}
\expandafter\def\csname rg@php.seq.read.smax\endcsname{--}
\expandafter\def\csname rg@php.seq.read.tool\endcsname{--}
\expandafter\def\csname rg@php.seq.roundingsat:cnf.n\endcsname{8}
\expandafter\def\csname rg@php.seq.roundingsat:cnf.s\endcsname{0.09}
\expandafter\def\csname rg@php.seq.sat4j-cp:cnf.n\endcsname{8}
\expandafter\def\csname rg@php.seq.sat4j-cp:cnf.s\endcsname{0.64}
\expandafter\def\csname rg@php.seq.scip:cnf.n\endcsname{8}
\expandafter\def\csname rg@php.seq.scip:cnf.s\endcsname{0.12}
\expandafter\def\csname rg@php.seq.top\endcsname{32}
\expandafter\def\csname rg@php.seq.win.n\endcsname{--}
\expandafter\def\csname rg@php.seq.win.s\endcsname{--}
\expandafter\def\csname rg@php.seq.xor.n\endcsname{8}
\expandafter\def\csname rg@php.seq.xor.s\endcsname{16.2}
\expandafter\def\csname rg@php.seq.xorcle-dev.n\endcsname{8}
\expandafter\def\csname rg@php.seq.xorcle-dev.s\endcsname{16.7}
\expandafter\def\csname rg@php.seq.xorcle.n\endcsname{8}
\expandafter\def\csname rg@php.seq.xorcle.s\endcsname{16.2}
\expandafter\def\csname rg@plug.cert.refuted\endcsname{24}
\expandafter\def\csname rg@plug.cert.verified\endcsname{24}
\expandafter\def\csname rg@plug.count3.clause.best\endcsname{80}
\expandafter\def\csname rg@plug.count3.clause.besttools\endcsname{RoundingSat, Sat4j, HiGHS, Exact, CryptoMiniSat, Lingeling and Kissat}
\expandafter\def\csname rg@plug.count3.clause.read\endcsname{640}
\expandafter\def\csname rg@plug.count3.clause.scip\endcsname{80}
\expandafter\def\csname rg@plug.count3.native.best\endcsname{80}
\expandafter\def\csname rg@plug.count3.native.besttools\endcsname{RoundingSat, HiGHS, Exact, CryptoMiniSat, Lingeling and Kissat}
\expandafter\def\csname rg@plug.count3.native.read\endcsname{640}
\expandafter\def\csname rg@plug.count3.native.scip\endcsname{80}
\expandafter\def\csname rg@plug.count3.top\endcsname{640}
\expandafter\def\csname rg@plug.count3_sat.top\endcsname{120}
\expandafter\def\csname rg@plug.count4.clause.best\endcsname{322}
\expandafter\def\csname rg@plug.count4.clause.besttools\endcsname{Lingeling}
\expandafter\def\csname rg@plug.count4.clause.read\endcsname{642}
\expandafter\def\csname rg@plug.count4.clause.scip\endcsname{162}
\expandafter\def\csname rg@plug.count4.native.best\endcsname{322}
\expandafter\def\csname rg@plug.count4.native.besttools\endcsname{Kissat}
\expandafter\def\csname rg@plug.count4.native.read\endcsname{642}
\expandafter\def\csname rg@plug.count4.native.scip\endcsname{162}
\expandafter\def\csname rg@plug.count4.top\endcsname{642}
\expandafter\def\csname rg@plug.limit\endcsname{60}
\expandafter\def\csname rg@plug.others.cnf\endcsname{CryptoMiniSat, Lingeling and Kissat}
\expandafter\def\csname rg@plug.others.pb\endcsname{RoundingSat, Sat4j, HiGHS and Exact}
\expandafter\def\csname rg@plug.pm_sat.top\endcsname{80}
\expandafter\def\csname rg@plug.pmodd.clause.best\endcsname{41}
\expandafter\def\csname rg@plug.pmodd.clause.besttools\endcsname{RoundingSat, Sat4j, HiGHS, Exact, CryptoMiniSat, Lingeling and Kissat}
\expandafter\def\csname rg@plug.pmodd.clause.read\endcsname{321}
\expandafter\def\csname rg@plug.pmodd.clause.scip\endcsname{321}
\expandafter\def\csname rg@plug.pmodd.native.best\endcsname{321}
\expandafter\def\csname rg@plug.pmodd.native.besttools\endcsname{HiGHS and Exact}
\expandafter\def\csname rg@plug.pmodd.native.read\endcsname{321}
\expandafter\def\csname rg@plug.pmodd.native.scip\endcsname{321}
\expandafter\def\csname rg@plug.pmodd.top\endcsname{321}
\expandafter\def\csname rg@plug.seeds\endcsname{1}
\expandafter\def\csname rg@plug.seeds.word\endcsname{one}
\expandafter\def\csname rg@plug.veripb.n\endcsname{0}
\expandafter\def\csname rg@plug.veripb.ok\endcsname{0}
\expandafter\def\csname rg@pmodd.binary.best.n\endcsname{321}
\expandafter\def\csname rg@pmodd.binary.bve+hull.n\endcsname{--}
\expandafter\def\csname rg@pmodd.binary.bve+hull.s\endcsname{--}
\expandafter\def\csname rg@pmodd.binary.bve+int.n\endcsname{321}
\expandafter\def\csname rg@pmodd.binary.bve+int.s\endcsname{0.46}
\expandafter\def\csname rg@pmodd.binary.bve+win.n\endcsname{321}
\expandafter\def\csname rg@pmodd.binary.bve+win.s\endcsname{0.46}
\expandafter\def\csname rg@pmodd.binary.cadical.n\endcsname{41}
\expandafter\def\csname rg@pmodd.binary.cadical.s\endcsname{0.18}
\expandafter\def\csname rg@pmodd.binary.cms.n\endcsname{41}
\expandafter\def\csname rg@pmodd.binary.cms.s\endcsname{0.43}
\expandafter\def\csname rg@pmodd.binary.cnf.n\endcsname{321}
\expandafter\def\csname rg@pmodd.binary.cnf.s\endcsname{46.8}
\expandafter\def\csname rg@pmodd.binary.cnf.tools\endcsname{Xorcle}
\expandafter\def\csname rg@pmodd.binary.cnfx.n\endcsname{41}
\expandafter\def\csname rg@pmodd.binary.cnfx.tools\endcsname{CDCL, Sat4j, RoundingSat, Exact, SCIP}
\expandafter\def\csname rg@pmodd.binary.exact:cnf.n\endcsname{41}
\expandafter\def\csname rg@pmodd.binary.exact:cnf.s\endcsname{2.7}
\expandafter\def\csname rg@pmodd.binary.hull.n\endcsname{--}
\expandafter\def\csname rg@pmodd.binary.hull.s\endcsname{--}
\expandafter\def\csname rg@pmodd.binary.int.n\endcsname{321}
\expandafter\def\csname rg@pmodd.binary.int.s\endcsname{0.83}
\expandafter\def\csname rg@pmodd.binary.kissat.n\endcsname{41}
\expandafter\def\csname rg@pmodd.binary.kissat.s\endcsname{0.23}
\expandafter\def\csname rg@pmodd.binary.lingeling.n\endcsname{41}
\expandafter\def\csname rg@pmodd.binary.lingeling.s\endcsname{0.23}
\expandafter\def\csname rg@pmodd.binary.pgpbs-2.n\endcsname{--}
\expandafter\def\csname rg@pmodd.binary.pgpbs-2.s\endcsname{--}
\expandafter\def\csname rg@pmodd.binary.pgpbs-3.n\endcsname{--}
\expandafter\def\csname rg@pmodd.binary.pgpbs-3.s\endcsname{--}
\expandafter\def\csname rg@pmodd.binary.pgpbs-a.n\endcsname{--}
\expandafter\def\csname rg@pmodd.binary.pgpbs-a.s\endcsname{--}
\expandafter\def\csname rg@pmodd.binary.pgpbs-i.n\endcsname{--}
\expandafter\def\csname rg@pmodd.binary.pgpbs-i.s\endcsname{--}
\expandafter\def\csname rg@pmodd.binary.read.fastest\endcsname{ring after elim.}
\expandafter\def\csname rg@pmodd.binary.read.n\endcsname{321}
\expandafter\def\csname rg@pmodd.binary.read.s\endcsname{0.46}
\expandafter\def\csname rg@pmodd.binary.read.smax\endcsname{0.83}
\expandafter\def\csname rg@pmodd.binary.read.tool\endcsname{field, ring, field after elim., ring after elim.}
\expandafter\def\csname rg@pmodd.binary.roundingsat:cnf.n\endcsname{41}
\expandafter\def\csname rg@pmodd.binary.roundingsat:cnf.s\endcsname{2.5}
\expandafter\def\csname rg@pmodd.binary.sat4j-cp:cnf.n\endcsname{41}
\expandafter\def\csname rg@pmodd.binary.sat4j-cp:cnf.s\endcsname{3.2}
\expandafter\def\csname rg@pmodd.binary.scip:cnf.n\endcsname{41}
\expandafter\def\csname rg@pmodd.binary.scip:cnf.s\endcsname{25.2}
\expandafter\def\csname rg@pmodd.binary.top\endcsname{321}
\expandafter\def\csname rg@pmodd.binary.win.n\endcsname{321}
\expandafter\def\csname rg@pmodd.binary.win.s\endcsname{0.61}
\expandafter\def\csname rg@pmodd.binary.xor.n\endcsname{321}
\expandafter\def\csname rg@pmodd.binary.xor.s\endcsname{42.3}
\expandafter\def\csname rg@pmodd.binary.xorcle-dev.n\endcsname{321}
\expandafter\def\csname rg@pmodd.binary.xorcle-dev.s\endcsname{46.8}
\expandafter\def\csname rg@pmodd.binary.xorcle.n\endcsname{321}
\expandafter\def\csname rg@pmodd.binary.xorcle.s\endcsname{42.3}
\expandafter\def\csname rg@pmodd.commander.best.n\endcsname{321}
\expandafter\def\csname rg@pmodd.commander.bve+hull.n\endcsname{--}
\expandafter\def\csname rg@pmodd.commander.bve+hull.s\endcsname{--}
\expandafter\def\csname rg@pmodd.commander.bve+int.n\endcsname{321}
\expandafter\def\csname rg@pmodd.commander.bve+int.s\endcsname{0.75}
\expandafter\def\csname rg@pmodd.commander.bve+win.n\endcsname{321}
\expandafter\def\csname rg@pmodd.commander.bve+win.s\endcsname{0.75}
\expandafter\def\csname rg@pmodd.commander.cadical.n\endcsname{41}
\expandafter\def\csname rg@pmodd.commander.cadical.s\endcsname{0.22}
\expandafter\def\csname rg@pmodd.commander.cms.n\endcsname{41}
\expandafter\def\csname rg@pmodd.commander.cms.s\endcsname{0.51}
\expandafter\def\csname rg@pmodd.commander.cnf.n\endcsname{321}
\expandafter\def\csname rg@pmodd.commander.cnf.s\endcsname{25.2}
\expandafter\def\csname rg@pmodd.commander.cnf.tools\endcsname{pgpbs mod 2, SCIP, Xorcle}
\expandafter\def\csname rg@pmodd.commander.cnfx.n\endcsname{321}
\expandafter\def\csname rg@pmodd.commander.cnfx.tools\endcsname{pgpbs mod 2, SCIP}
\expandafter\def\csname rg@pmodd.commander.exact:cnf.n\endcsname{41}
\expandafter\def\csname rg@pmodd.commander.exact:cnf.s\endcsname{2.5}
\expandafter\def\csname rg@pmodd.commander.hull.n\endcsname{--}
\expandafter\def\csname rg@pmodd.commander.hull.s\endcsname{--}
\expandafter\def\csname rg@pmodd.commander.int.n\endcsname{321}
\expandafter\def\csname rg@pmodd.commander.int.s\endcsname{1.4}
\expandafter\def\csname rg@pmodd.commander.kissat.n\endcsname{41}
\expandafter\def\csname rg@pmodd.commander.kissat.s\endcsname{0.30}
\expandafter\def\csname rg@pmodd.commander.lingeling.n\endcsname{41}
\expandafter\def\csname rg@pmodd.commander.lingeling.s\endcsname{0.20}
\expandafter\def\csname rg@pmodd.commander.pgpbs-2.n\endcsname{321}
\expandafter\def\csname rg@pmodd.commander.pgpbs-2.s\endcsname{1.4}
\expandafter\def\csname rg@pmodd.commander.pgpbs-3.n\endcsname{--}
\expandafter\def\csname rg@pmodd.commander.pgpbs-3.s\endcsname{--}
\expandafter\def\csname rg@pmodd.commander.pgpbs-a.n\endcsname{--}
\expandafter\def\csname rg@pmodd.commander.pgpbs-a.s\endcsname{--}
\expandafter\def\csname rg@pmodd.commander.pgpbs-i.n\endcsname{--}
\expandafter\def\csname rg@pmodd.commander.pgpbs-i.s\endcsname{--}
\expandafter\def\csname rg@pmodd.commander.read.fastest\endcsname{ring after elim.}
\expandafter\def\csname rg@pmodd.commander.read.n\endcsname{321}
\expandafter\def\csname rg@pmodd.commander.read.s\endcsname{0.75}
\expandafter\def\csname rg@pmodd.commander.read.smax\endcsname{1.4}
\expandafter\def\csname rg@pmodd.commander.read.tool\endcsname{field, ring, field after elim., ring after elim.}
\expandafter\def\csname rg@pmodd.commander.roundingsat:cnf.n\endcsname{41}
\expandafter\def\csname rg@pmodd.commander.roundingsat:cnf.s\endcsname{2.6}
\expandafter\def\csname rg@pmodd.commander.sat4j-cp:cnf.n\endcsname{41}
\expandafter\def\csname rg@pmodd.commander.sat4j-cp:cnf.s\endcsname{2.8}
\expandafter\def\csname rg@pmodd.commander.scip:cnf.n\endcsname{321}
\expandafter\def\csname rg@pmodd.commander.scip:cnf.s\endcsname{2.9}
\expandafter\def\csname rg@pmodd.commander.top\endcsname{321}
\expandafter\def\csname rg@pmodd.commander.win.n\endcsname{321}
\expandafter\def\csname rg@pmodd.commander.win.s\endcsname{1.2}
\expandafter\def\csname rg@pmodd.commander.xor.n\endcsname{321}
\expandafter\def\csname rg@pmodd.commander.xor.s\endcsname{24.3}
\expandafter\def\csname rg@pmodd.commander.xorcle-dev.n\endcsname{321}
\expandafter\def\csname rg@pmodd.commander.xorcle-dev.s\endcsname{25.2}
\expandafter\def\csname rg@pmodd.commander.xorcle.n\endcsname{321}
\expandafter\def\csname rg@pmodd.commander.xorcle.s\endcsname{24.3}
\expandafter\def\csname rg@pmodd.native.exact:pb.n\endcsname{321}
\expandafter\def\csname rg@pmodd.native.exact:pb.s\endcsname{24.4}
\expandafter\def\csname rg@pmodd.native.highs:pb.n\endcsname{321}
\expandafter\def\csname rg@pmodd.native.highs:pb.s\endcsname{0.85}
\expandafter\def\csname rg@pmodd.native.n\endcsname{321}
\expandafter\def\csname rg@pmodd.native.roundingsat:pb.n\endcsname{161}
\expandafter\def\csname rg@pmodd.native.roundingsat:pb.s\endcsname{0.21}
\expandafter\def\csname rg@pmodd.native.sat4j-cp:pb.n\endcsname{41}
\expandafter\def\csname rg@pmodd.native.sat4j-cp:pb.s\endcsname{2.1}
\expandafter\def\csname rg@pmodd.native.scip:pb.n\endcsname{321}
\expandafter\def\csname rg@pmodd.native.scip:pb.s\endcsname{7.0}
\expandafter\def\csname rg@pmodd.native.smax\endcsname{24.4}
\expandafter\def\csname rg@pmodd.native.smin\endcsname{0.85}
\expandafter\def\csname rg@pmodd.native.tools\endcsname{Exact, SCIP, HiGHS}
\expandafter\def\csname rg@pmodd.native.toolsp\endcsname{Exact, SCIP and HiGHS}
\expandafter\def\csname rg@pmodd.pairwise.best.n\endcsname{321}
\expandafter\def\csname rg@pmodd.pairwise.bve+hull.n\endcsname{--}
\expandafter\def\csname rg@pmodd.pairwise.bve+hull.s\endcsname{--}
\expandafter\def\csname rg@pmodd.pairwise.bve+int.n\endcsname{321}
\expandafter\def\csname rg@pmodd.pairwise.bve+int.s\endcsname{0.75}
\expandafter\def\csname rg@pmodd.pairwise.bve+win.n\endcsname{321}
\expandafter\def\csname rg@pmodd.pairwise.bve+win.s\endcsname{0.73}
\expandafter\def\csname rg@pmodd.pairwise.cadical.n\endcsname{41}
\expandafter\def\csname rg@pmodd.pairwise.cadical.s\endcsname{0.21}
\expandafter\def\csname rg@pmodd.pairwise.cms.n\endcsname{41}
\expandafter\def\csname rg@pmodd.pairwise.cms.s\endcsname{0.26}
\expandafter\def\csname rg@pmodd.pairwise.cnf.n\endcsname{321}
\expandafter\def\csname rg@pmodd.pairwise.cnf.s\endcsname{35.6}
\expandafter\def\csname rg@pmodd.pairwise.cnf.tools\endcsname{pgpbs mod 2, SCIP, Xorcle}
\expandafter\def\csname rg@pmodd.pairwise.cnfx.n\endcsname{321}
\expandafter\def\csname rg@pmodd.pairwise.cnfx.tools\endcsname{pgpbs mod 2, SCIP}
\expandafter\def\csname rg@pmodd.pairwise.exact:cnf.n\endcsname{41}
\expandafter\def\csname rg@pmodd.pairwise.exact:cnf.s\endcsname{1.9}
\expandafter\def\csname rg@pmodd.pairwise.hull.n\endcsname{--}
\expandafter\def\csname rg@pmodd.pairwise.hull.s\endcsname{--}
\expandafter\def\csname rg@pmodd.pairwise.int.n\endcsname{321}
\expandafter\def\csname rg@pmodd.pairwise.int.s\endcsname{1.5}
\expandafter\def\csname rg@pmodd.pairwise.kissat.n\endcsname{41}
\expandafter\def\csname rg@pmodd.pairwise.kissat.s\endcsname{0.19}
\expandafter\def\csname rg@pmodd.pairwise.lingeling.n\endcsname{41}
\expandafter\def\csname rg@pmodd.pairwise.lingeling.s\endcsname{0.20}
\expandafter\def\csname rg@pmodd.pairwise.pgpbs-2.n\endcsname{321}
\expandafter\def\csname rg@pmodd.pairwise.pgpbs-2.s\endcsname{1.1}
\expandafter\def\csname rg@pmodd.pairwise.pgpbs-3.n\endcsname{--}
\expandafter\def\csname rg@pmodd.pairwise.pgpbs-3.s\endcsname{--}
\expandafter\def\csname rg@pmodd.pairwise.pgpbs-a.n\endcsname{--}
\expandafter\def\csname rg@pmodd.pairwise.pgpbs-a.s\endcsname{--}
\expandafter\def\csname rg@pmodd.pairwise.pgpbs-i.n\endcsname{--}
\expandafter\def\csname rg@pmodd.pairwise.pgpbs-i.s\endcsname{--}
\expandafter\def\csname rg@pmodd.pairwise.read.fastest\endcsname{field after elim.}
\expandafter\def\csname rg@pmodd.pairwise.read.n\endcsname{321}
\expandafter\def\csname rg@pmodd.pairwise.read.s\endcsname{0.73}
\expandafter\def\csname rg@pmodd.pairwise.read.smax\endcsname{1.5}
\expandafter\def\csname rg@pmodd.pairwise.read.tool\endcsname{field, ring, field after elim., ring after elim.}
\expandafter\def\csname rg@pmodd.pairwise.roundingsat:cnf.n\endcsname{41}
\expandafter\def\csname rg@pmodd.pairwise.roundingsat:cnf.s\endcsname{1.5}
\expandafter\def\csname rg@pmodd.pairwise.sat4j-cp:cnf.n\endcsname{41}
\expandafter\def\csname rg@pmodd.pairwise.sat4j-cp:cnf.s\endcsname{2.3}
\expandafter\def\csname rg@pmodd.pairwise.scip:cnf.n\endcsname{321}
\expandafter\def\csname rg@pmodd.pairwise.scip:cnf.s\endcsname{35.6}
\expandafter\def\csname rg@pmodd.pairwise.top\endcsname{321}
\expandafter\def\csname rg@pmodd.pairwise.win.n\endcsname{321}
\expandafter\def\csname rg@pmodd.pairwise.win.s\endcsname{1.4}
\expandafter\def\csname rg@pmodd.pairwise.xor.n\endcsname{321}
\expandafter\def\csname rg@pmodd.pairwise.xor.s\endcsname{5.9}
\expandafter\def\csname rg@pmodd.pairwise.xorcle-dev.n\endcsname{321}
\expandafter\def\csname rg@pmodd.pairwise.xorcle-dev.s\endcsname{5.9}
\expandafter\def\csname rg@pmodd.pairwise.xorcle.n\endcsname{321}
\expandafter\def\csname rg@pmodd.pairwise.xorcle.s\endcsname{6.0}
\expandafter\def\csname rg@pmodd.seq.best.n\endcsname{321}
\expandafter\def\csname rg@pmodd.seq.bve+hull.n\endcsname{--}
\expandafter\def\csname rg@pmodd.seq.bve+hull.s\endcsname{--}
\expandafter\def\csname rg@pmodd.seq.bve+int.n\endcsname{321}
\expandafter\def\csname rg@pmodd.seq.bve+int.s\endcsname{0.77}
\expandafter\def\csname rg@pmodd.seq.bve+win.n\endcsname{321}
\expandafter\def\csname rg@pmodd.seq.bve+win.s\endcsname{0.75}
\expandafter\def\csname rg@pmodd.seq.cadical.n\endcsname{41}
\expandafter\def\csname rg@pmodd.seq.cadical.s\endcsname{0.16}
\expandafter\def\csname rg@pmodd.seq.cms.n\endcsname{41}
\expandafter\def\csname rg@pmodd.seq.cms.s\endcsname{0.44}
\expandafter\def\csname rg@pmodd.seq.cnf.n\endcsname{321}
\expandafter\def\csname rg@pmodd.seq.cnf.s\endcsname{36.5}
\expandafter\def\csname rg@pmodd.seq.cnf.tools\endcsname{pgpbs mod 2, Xorcle}
\expandafter\def\csname rg@pmodd.seq.cnfx.n\endcsname{321}
\expandafter\def\csname rg@pmodd.seq.cnfx.tools\endcsname{pgpbs mod 2}
\expandafter\def\csname rg@pmodd.seq.exact:cnf.n\endcsname{41}
\expandafter\def\csname rg@pmodd.seq.exact:cnf.s\endcsname{2.6}
\expandafter\def\csname rg@pmodd.seq.hull.n\endcsname{--}
\expandafter\def\csname rg@pmodd.seq.hull.s\endcsname{--}
\expandafter\def\csname rg@pmodd.seq.int.n\endcsname{321}
\expandafter\def\csname rg@pmodd.seq.int.s\endcsname{1.0}
\expandafter\def\csname rg@pmodd.seq.kissat.n\endcsname{41}
\expandafter\def\csname rg@pmodd.seq.kissat.s\endcsname{0.32}
\expandafter\def\csname rg@pmodd.seq.lingeling.n\endcsname{41}
\expandafter\def\csname rg@pmodd.seq.lingeling.s\endcsname{0.20}
\expandafter\def\csname rg@pmodd.seq.pgpbs-2.n\endcsname{321}
\expandafter\def\csname rg@pmodd.seq.pgpbs-2.s\endcsname{1.4}
\expandafter\def\csname rg@pmodd.seq.pgpbs-3.n\endcsname{--}
\expandafter\def\csname rg@pmodd.seq.pgpbs-3.s\endcsname{--}
\expandafter\def\csname rg@pmodd.seq.pgpbs-a.n\endcsname{--}
\expandafter\def\csname rg@pmodd.seq.pgpbs-a.s\endcsname{--}
\expandafter\def\csname rg@pmodd.seq.pgpbs-i.n\endcsname{--}
\expandafter\def\csname rg@pmodd.seq.pgpbs-i.s\endcsname{--}
\expandafter\def\csname rg@pmodd.seq.read.fastest\endcsname{field after elim.}
\expandafter\def\csname rg@pmodd.seq.read.n\endcsname{321}
\expandafter\def\csname rg@pmodd.seq.read.s\endcsname{0.75}
\expandafter\def\csname rg@pmodd.seq.read.smax\endcsname{1.0}
\expandafter\def\csname rg@pmodd.seq.read.tool\endcsname{field, ring, field after elim., ring after elim.}
\expandafter\def\csname rg@pmodd.seq.roundingsat:cnf.n\endcsname{41}
\expandafter\def\csname rg@pmodd.seq.roundingsat:cnf.s\endcsname{2.5}
\expandafter\def\csname rg@pmodd.seq.sat4j-cp:cnf.n\endcsname{41}
\expandafter\def\csname rg@pmodd.seq.sat4j-cp:cnf.s\endcsname{2.8}
\expandafter\def\csname rg@pmodd.seq.scip:cnf.n\endcsname{41}
\expandafter\def\csname rg@pmodd.seq.scip:cnf.s\endcsname{23.3}
\expandafter\def\csname rg@pmodd.seq.top\endcsname{321}
\expandafter\def\csname rg@pmodd.seq.win.n\endcsname{321}
\expandafter\def\csname rg@pmodd.seq.win.s\endcsname{1.0}
\expandafter\def\csname rg@pmodd.seq.xor.n\endcsname{321}
\expandafter\def\csname rg@pmodd.seq.xor.s\endcsname{33.7}
\expandafter\def\csname rg@pmodd.seq.xorcle-dev.n\endcsname{321}
\expandafter\def\csname rg@pmodd.seq.xorcle-dev.s\endcsname{36.5}
\expandafter\def\csname rg@pmodd.seq.xorcle.n\endcsname{321}
\expandafter\def\csname rg@pmodd.seq.xorcle.s\endcsname{33.7}
\expandafter\def\csname rg@pub.n\endcsname{53}
\expandafter\def\csname rg@pub.par\endcsname{30}
\expandafter\def\csname rg@pub.par.wrong\endcsname{0}
\expandafter\def\csname rg@pub.pgpbs.n\endcsname{10}
\expandafter\def\csname rg@pub.pgpbs.read\endcsname{10}
\expandafter\def\csname rg@real.allowance\endcsname{60}
\expandafter\def\csname rg@real.certified\endcsname{78}
\expandafter\def\csname rg@real.gate\endcsname{FAIL}
\expandafter\def\csname rg@real.instances\endcsname{555}
\expandafter\def\csname rg@real.issues\endcsname{0}
\expandafter\def\csname rg@real.lattice\endcsname{11}
\expandafter\def\csname rg@real.miplib.control.anydiv\endcsname{1}
\expandafter\def\csname rg@real.miplib.control.certified\endcsname{0}
\expandafter\def\csname rg@real.miplib.control.div\endcsname{1}
\expandafter\def\csname rg@real.miplib.control.eq\endcsname{17}
\expandafter\def\csname rg@real.miplib.control.mod2\endcsname{2}
\expandafter\def\csname rg@real.miplib.control.mod3\endcsname{1}
\expandafter\def\csname rg@real.miplib.control.mod4\endcsname{2}
\expandafter\def\csname rg@real.miplib.control.modother\endcsname{0}
\expandafter\def\csname rg@real.miplib.control.n\endcsname{29}
\expandafter\def\csname rg@real.miplib.control.rankdiff\endcsname{2}
\expandafter\def\csname rg@real.miplib.control.read\endcsname{29}
\expandafter\def\csname rg@real.miplib.control.refE\endcsname{0}
\expandafter\def\csname rg@real.miplib.control.refW\endcsname{0}
\expandafter\def\csname rg@real.miplib.control.two\endcsname{1}
\expandafter\def\csname rg@real.miplib.test-application.anydiv\endcsname{2}
\expandafter\def\csname rg@real.miplib.test-application.certified\endcsname{2}
\expandafter\def\csname rg@real.miplib.test-application.div\endcsname{2}
\expandafter\def\csname rg@real.miplib.test-application.eq\endcsname{20}
\expandafter\def\csname rg@real.miplib.test-application.mod2\endcsname{1}
\expandafter\def\csname rg@real.miplib.test-application.mod3\endcsname{0}
\expandafter\def\csname rg@real.miplib.test-application.mod4\endcsname{1}
\expandafter\def\csname rg@real.miplib.test-application.modother\endcsname{0}
\expandafter\def\csname rg@real.miplib.test-application.n\endcsname{34}
\expandafter\def\csname rg@real.miplib.test-application.rankdiff\endcsname{1}
\expandafter\def\csname rg@real.miplib.test-application.read\endcsname{32}
\expandafter\def\csname rg@real.miplib.test-application.refE\endcsname{2}
\expandafter\def\csname rg@real.miplib.test-application.refW\endcsname{0}
\expandafter\def\csname rg@real.miplib.test-application.two\endcsname{2}
\expandafter\def\csname rg@real.miplib.test-crafted.anydiv\endcsname{0}
\expandafter\def\csname rg@real.miplib.test-crafted.certified\endcsname{4}
\expandafter\def\csname rg@real.miplib.test-crafted.div\endcsname{0}
\expandafter\def\csname rg@real.miplib.test-crafted.eq\endcsname{5}
\expandafter\def\csname rg@real.miplib.test-crafted.mod2\endcsname{1}
\expandafter\def\csname rg@real.miplib.test-crafted.mod3\endcsname{0}
\expandafter\def\csname rg@real.miplib.test-crafted.mod4\endcsname{1}
\expandafter\def\csname rg@real.miplib.test-crafted.modother\endcsname{0}
\expandafter\def\csname rg@real.miplib.test-crafted.n\endcsname{11}
\expandafter\def\csname rg@real.miplib.test-crafted.rankdiff\endcsname{1}
\expandafter\def\csname rg@real.miplib.test-crafted.read\endcsname{9}
\expandafter\def\csname rg@real.miplib.test-crafted.refE\endcsname{3}
\expandafter\def\csname rg@real.miplib.test-crafted.refW\endcsname{1}
\expandafter\def\csname rg@real.miplib.test-crafted.two\endcsname{0}
\expandafter\def\csname rg@real.pb.test-application.anydiv\endcsname{4}
\expandafter\def\csname rg@real.pb.test-application.certified\endcsname{1}
\expandafter\def\csname rg@real.pb.test-application.div\endcsname{4}
\expandafter\def\csname rg@real.pb.test-application.eq\endcsname{88}
\expandafter\def\csname rg@real.pb.test-application.mod2\endcsname{6}
\expandafter\def\csname rg@real.pb.test-application.mod3\endcsname{0}
\expandafter\def\csname rg@real.pb.test-application.mod4\endcsname{4}
\expandafter\def\csname rg@real.pb.test-application.modother\endcsname{0}
\expandafter\def\csname rg@real.pb.test-application.n\endcsname{121}
\expandafter\def\csname rg@real.pb.test-application.rankdiff\endcsname{6}
\expandafter\def\csname rg@real.pb.test-application.read\endcsname{121}
\expandafter\def\csname rg@real.pb.test-application.refE\endcsname{0}
\expandafter\def\csname rg@real.pb.test-application.refW\endcsname{1}
\expandafter\def\csname rg@real.pb.test-application.two\endcsname{4}
\expandafter\def\csname rg@real.pb.test-crafted.anydiv\endcsname{12}
\expandafter\def\csname rg@real.pb.test-crafted.certified\endcsname{65}
\expandafter\def\csname rg@real.pb.test-crafted.div\endcsname{12}
\expandafter\def\csname rg@real.pb.test-crafted.eq\endcsname{109}
\expandafter\def\csname rg@real.pb.test-crafted.mod2\endcsname{0}
\expandafter\def\csname rg@real.pb.test-crafted.mod3\endcsname{0}
\expandafter\def\csname rg@real.pb.test-crafted.mod4\endcsname{0}
\expandafter\def\csname rg@real.pb.test-crafted.modother\endcsname{0}
\expandafter\def\csname rg@real.pb.test-crafted.n\endcsname{179}
\expandafter\def\csname rg@real.pb.test-crafted.rankdiff\endcsname{0}
\expandafter\def\csname rg@real.pb.test-crafted.read\endcsname{179}
\expandafter\def\csname rg@real.pb.test-crafted.refE\endcsname{49}
\expandafter\def\csname rg@real.pb.test-crafted.refW\endcsname{16}
\expandafter\def\csname rg@real.pb.test-crafted.two\endcsname{12}
\expandafter\def\csname rg@real.ref.mod2\endcsname{74}
\expandafter\def\csname rg@real.ref.other\endcsname{1}
\expandafter\def\csname rg@real.ref.other.word\endcsname{one}
\expandafter\def\csname rg@real.ref.otherlist\endcsname{neos-952987 (mod 25)}
\expandafter\def\csname rg@real.ref.total\endcsname{80}
\expandafter\def\csname rg@real.refE\endcsname{54}
\expandafter\def\csname rg@real.refW\endcsname{26}
\expandafter\def\csname rg@real.refW.app\endcsname{3}
\expandafter\def\csname rg@real.sat.control.anydiv\endcsname{9}
\expandafter\def\csname rg@real.sat.control.certified\endcsname{0}
\expandafter\def\csname rg@real.sat.control.div\endcsname{9}
\expandafter\def\csname rg@real.sat.control.eq\endcsname{0}
\expandafter\def\csname rg@real.sat.control.mod2\endcsname{9}
\expandafter\def\csname rg@real.sat.control.mod3\endcsname{0}
\expandafter\def\csname rg@real.sat.control.mod4\endcsname{9}
\expandafter\def\csname rg@real.sat.control.modother\endcsname{0}
\expandafter\def\csname rg@real.sat.control.n\endcsname{21}
\expandafter\def\csname rg@real.sat.control.rankdiff\endcsname{9}
\expandafter\def\csname rg@real.sat.control.read\endcsname{21}
\expandafter\def\csname rg@real.sat.control.refE\endcsname{0}
\expandafter\def\csname rg@real.sat.control.refW\endcsname{0}
\expandafter\def\csname rg@real.sat.control.two\endcsname{9}
\expandafter\def\csname rg@real.sat.test-application.anydiv\endcsname{36}
\expandafter\def\csname rg@real.sat.test-application.certified\endcsname{0}
\expandafter\def\csname rg@real.sat.test-application.div\endcsname{36}
\expandafter\def\csname rg@real.sat.test-application.eq\endcsname{0}
\expandafter\def\csname rg@real.sat.test-application.mod2\endcsname{36}
\expandafter\def\csname rg@real.sat.test-application.mod3\endcsname{0}
\expandafter\def\csname rg@real.sat.test-application.mod4\endcsname{36}
\expandafter\def\csname rg@real.sat.test-application.modother\endcsname{0}
\expandafter\def\csname rg@real.sat.test-application.n\endcsname{76}
\expandafter\def\csname rg@real.sat.test-application.rankdiff\endcsname{36}
\expandafter\def\csname rg@real.sat.test-application.read\endcsname{76}
\expandafter\def\csname rg@real.sat.test-application.refE\endcsname{0}
\expandafter\def\csname rg@real.sat.test-application.refW\endcsname{2}
\expandafter\def\csname rg@real.sat.test-application.two\endcsname{36}
\expandafter\def\csname rg@real.sat.test-crafted.anydiv\endcsname{24}
\expandafter\def\csname rg@real.sat.test-crafted.certified\endcsname{6}
\expandafter\def\csname rg@real.sat.test-crafted.div\endcsname{24}
\expandafter\def\csname rg@real.sat.test-crafted.eq\endcsname{0}
\expandafter\def\csname rg@real.sat.test-crafted.mod2\endcsname{22}
\expandafter\def\csname rg@real.sat.test-crafted.mod3\endcsname{0}
\expandafter\def\csname rg@real.sat.test-crafted.mod4\endcsname{19}
\expandafter\def\csname rg@real.sat.test-crafted.modother\endcsname{0}
\expandafter\def\csname rg@real.sat.test-crafted.n\endcsname{84}
\expandafter\def\csname rg@real.sat.test-crafted.rankdiff\endcsname{22}
\expandafter\def\csname rg@real.sat.test-crafted.read\endcsname{84}
\expandafter\def\csname rg@real.sat.test-crafted.refE\endcsname{0}
\expandafter\def\csname rg@real.sat.test-crafted.refW\endcsname{6}
\expandafter\def\csname rg@real.sat.test-crafted.two\endcsname{24}
\expandafter\def\csname rg@real.satapp.div\endcsname{36}
\expandafter\def\csname rg@real.satapp.n\endcsname{76}
\expandafter\def\csname rg@real.tests\endcsname{505}
\expandafter\def\csname rg@real.timeout\endcsname{300}
\expandafter\def\csname rg@real.uncert\endcsname{2}
\expandafter\def\csname rg@real.uncert.app\endcsname{2}
\expandafter\def\csname rg@real.uncert.facts\endcsname{2}
\expandafter\def\csname rg@real.uncert.word\endcsname{two}
\expandafter\def\csname rg@real.unverified\endcsname{6}
\expandafter\def\csname rg@real.unverified.bv\endcsname{yes}
\expandafter\def\csname rg@real.unverified.contra\endcsname{1}
\expandafter\def\csname rg@real.unverified.contra.tools\endcsname{HiGHS}
\expandafter\def\csname rg@real.unverified.contra.word\endcsname{one}
\expandafter\def\csname rg@real.unverified.inst\endcsname{4}
\expandafter\def\csname rg@real.unverified.inst.word\endcsname{four}
\expandafter\def\csname rg@real.unverified.pb\endcsname{yes}
\expandafter\def\csname rg@real.unverified.readref\endcsname{0}
\expandafter\def\csname rg@real.unverified.runs\endcsname{6}
\expandafter\def\csname rg@real.unverified.scip\endcsname{yes}
\expandafter\def\csname rg@real.unverified.tools\endcsname{SCIP}
\expandafter\def\csname rg@real.window\endcsname{3}
\expandafter\def\csname rg@real.window.app\endcsname{0}
\expandafter\def\csname rg@real.window.crafted\endcsname{3}
\expandafter\def\csname rg@run.arch\endcsname{arm64}
\expandafter\def\csname rg@run.awake.h\endcsname{7.4}
\expandafter\def\csname rg@run.legacy\endcsname{no}
\expandafter\def\csname rg@run.macos\endcsname{15.3.2}
\expandafter\def\csname rg@run.slept.h\endcsname{3.7}
\expandafter\def\csname rg@scale.allslow.text\endcsname{the triple partition and perfect matching}
\expandafter\def\csname rg@scale.both.n\endcsname{18}
\expandafter\def\csname rg@scale.count3.1280.vars\endcsname{2,561}
\expandafter\def\csname rg@scale.count3.160.vars\endcsname{321}
\expandafter\def\csname rg@scale.count3.2560.vars\endcsname{5,121}
\expandafter\def\csname rg@scale.count3.320.vars\endcsname{641}
\expandafter\def\csname rg@scale.count3.640.vars\endcsname{1,281}
\expandafter\def\csname rg@scale.count3.80.vars\endcsname{161}
\expandafter\def\csname rg@scale.count3.int.1280.s\endcsname{32.7}
\expandafter\def\csname rg@scale.count3.int.1280.status\endcsname{unsat}
\expandafter\def\csname rg@scale.count3.int.160.s\endcsname{0.71}
\expandafter\def\csname rg@scale.count3.int.160.status\endcsname{unsat}
\expandafter\def\csname rg@scale.count3.int.2560.s\endcsname{255.7}
\expandafter\def\csname rg@scale.count3.int.2560.status\endcsname{unsat}
\expandafter\def\csname rg@scale.count3.int.320.s\endcsname{2.0}
\expandafter\def\csname rg@scale.count3.int.320.status\endcsname{unsat}
\expandafter\def\csname rg@scale.count3.int.640.s\endcsname{5.1}
\expandafter\def\csname rg@scale.count3.int.640.status\endcsname{unsat}
\expandafter\def\csname rg@scale.count3.int.80.s\endcsname{0.27}
\expandafter\def\csname rg@scale.count3.int.80.status\endcsname{unsat}
\expandafter\def\csname rg@scale.count3.int.exp\endcsname{1.93}
\expandafter\def\csname rg@scale.count3.int.fitpts\endcsname{6}
\expandafter\def\csname rg@scale.count3.int.max.n\endcsname{2,560}
\expandafter\def\csname rg@scale.count3.int.max.s\endcsname{255.7}
\expandafter\def\csname rg@scale.count3.int.max.vars\endcsname{5,121}
\expandafter\def\csname rg@scale.count3.int.within.n\endcsname{1,280}
\expandafter\def\csname rg@scale.count3.int.within.s\endcsname{32.7}
\expandafter\def\csname rg@scale.count3.win.1280.s\endcsname{21.5}
\expandafter\def\csname rg@scale.count3.win.1280.status\endcsname{unsat}
\expandafter\def\csname rg@scale.count3.win.160.s\endcsname{0.67}
\expandafter\def\csname rg@scale.count3.win.160.status\endcsname{unsat}
\expandafter\def\csname rg@scale.count3.win.2560.s\endcsname{112.8}
\expandafter\def\csname rg@scale.count3.win.2560.status\endcsname{unsat}
\expandafter\def\csname rg@scale.count3.win.320.s\endcsname{1.6}
\expandafter\def\csname rg@scale.count3.win.320.status\endcsname{unsat}
\expandafter\def\csname rg@scale.count3.win.640.s\endcsname{4.4}
\expandafter\def\csname rg@scale.count3.win.640.status\endcsname{unsat}
\expandafter\def\csname rg@scale.count3.win.80.s\endcsname{0.25}
\expandafter\def\csname rg@scale.count3.win.80.status\endcsname{unsat}
\expandafter\def\csname rg@scale.count3.win.exp\endcsname{1.74}
\expandafter\def\csname rg@scale.count3.win.fitpts\endcsname{6}
\expandafter\def\csname rg@scale.count3.win.max.n\endcsname{2,560}
\expandafter\def\csname rg@scale.count3.win.max.s\endcsname{112.8}
\expandafter\def\csname rg@scale.count3.win.max.vars\endcsname{5,121}
\expandafter\def\csname rg@scale.count3.win.within.n\endcsname{1,280}
\expandafter\def\csname rg@scale.count3.win.within.s\endcsname{21.5}
\expandafter\def\csname rg@scale.count4.1282.vars\endcsname{1,924}
\expandafter\def\csname rg@scale.count4.162.vars\endcsname{244}
\expandafter\def\csname rg@scale.count4.322.vars\endcsname{484}
\expandafter\def\csname rg@scale.count4.642.vars\endcsname{964}
\expandafter\def\csname rg@scale.count4.82.vars\endcsname{124}
\expandafter\def\csname rg@scale.count4.int.1282.s\endcsname{150.5}
\expandafter\def\csname rg@scale.count4.int.1282.status\endcsname{unsat}
\expandafter\def\csname rg@scale.count4.int.162.s\endcsname{1.3}
\expandafter\def\csname rg@scale.count4.int.162.status\endcsname{unsat}
\expandafter\def\csname rg@scale.count4.int.322.s\endcsname{4.0}
\expandafter\def\csname rg@scale.count4.int.322.status\endcsname{unsat}
\expandafter\def\csname rg@scale.count4.int.642.s\endcsname{23.3}
\expandafter\def\csname rg@scale.count4.int.642.status\endcsname{unsat}
\expandafter\def\csname rg@scale.count4.int.82.s\endcsname{0.42}
\expandafter\def\csname rg@scale.count4.int.82.status\endcsname{unsat}
\expandafter\def\csname rg@scale.count4.int.exp\endcsname{2.14}
\expandafter\def\csname rg@scale.count4.int.fitpts\endcsname{5}
\expandafter\def\csname rg@scale.count4.int.max.n\endcsname{1,282}
\expandafter\def\csname rg@scale.count4.int.max.s\endcsname{150.5}
\expandafter\def\csname rg@scale.count4.int.max.vars\endcsname{1,924}
\expandafter\def\csname rg@scale.count4.int.within.n\endcsname{642}
\expandafter\def\csname rg@scale.count4.int.within.s\endcsname{23.3}
\expandafter\def\csname rg@scale.count4.win.1282.s\endcsname{62.5}
\expandafter\def\csname rg@scale.count4.win.1282.status\endcsname{open}
\expandafter\def\csname rg@scale.count4.win.162.s\endcsname{0.90}
\expandafter\def\csname rg@scale.count4.win.162.status\endcsname{open}
\expandafter\def\csname rg@scale.count4.win.322.s\endcsname{2.8}
\expandafter\def\csname rg@scale.count4.win.322.status\endcsname{open}
\expandafter\def\csname rg@scale.count4.win.642.s\endcsname{10.6}
\expandafter\def\csname rg@scale.count4.win.642.status\endcsname{open}
\expandafter\def\csname rg@scale.count4.win.82.s\endcsname{0.37}
\expandafter\def\csname rg@scale.count4.win.82.status\endcsname{open}
\expandafter\def\csname rg@scale.mixed.count\endcsname{1}
\expandafter\def\csname rg@scale.mixed.pct\endcsname{4}
\expandafter\def\csname rg@scale.mixed.text\endcsname{the odd 3-factor}
\expandafter\def\csname rg@scale.oddfactor.1281.vars\endcsname{3,844}
\expandafter\def\csname rg@scale.oddfactor.161.vars\endcsname{484}
\expandafter\def\csname rg@scale.oddfactor.321.vars\endcsname{964}
\expandafter\def\csname rg@scale.oddfactor.41.vars\endcsname{124}
\expandafter\def\csname rg@scale.oddfactor.641.vars\endcsname{1,924}
\expandafter\def\csname rg@scale.oddfactor.81.vars\endcsname{244}
\expandafter\def\csname rg@scale.oddfactor.int.1281.s\endcsname{11.2}
\expandafter\def\csname rg@scale.oddfactor.int.1281.status\endcsname{unsat}
\expandafter\def\csname rg@scale.oddfactor.int.161.s\endcsname{1.0}
\expandafter\def\csname rg@scale.oddfactor.int.161.status\endcsname{unsat}
\expandafter\def\csname rg@scale.oddfactor.int.321.s\endcsname{2.2}
\expandafter\def\csname rg@scale.oddfactor.int.321.status\endcsname{unsat}
\expandafter\def\csname rg@scale.oddfactor.int.41.s\endcsname{0.17}
\expandafter\def\csname rg@scale.oddfactor.int.41.status\endcsname{unsat}
\expandafter\def\csname rg@scale.oddfactor.int.641.s\endcsname{5.0}
\expandafter\def\csname rg@scale.oddfactor.int.641.status\endcsname{unsat}
\expandafter\def\csname rg@scale.oddfactor.int.81.s\endcsname{0.35}
\expandafter\def\csname rg@scale.oddfactor.int.81.status\endcsname{unsat}
\expandafter\def\csname rg@scale.oddfactor.int.exp\endcsname{1.24}
\expandafter\def\csname rg@scale.oddfactor.int.fitpts\endcsname{6}
\expandafter\def\csname rg@scale.oddfactor.int.max.n\endcsname{1,281}
\expandafter\def\csname rg@scale.oddfactor.int.max.s\endcsname{11.2}
\expandafter\def\csname rg@scale.oddfactor.int.max.vars\endcsname{3,844}
\expandafter\def\csname rg@scale.oddfactor.int.within.n\endcsname{1,281}
\expandafter\def\csname rg@scale.oddfactor.int.within.s\endcsname{11.2}
\expandafter\def\csname rg@scale.oddfactor.win.1281.s\endcsname{11.1}
\expandafter\def\csname rg@scale.oddfactor.win.1281.status\endcsname{unsat}
\expandafter\def\csname rg@scale.oddfactor.win.161.s\endcsname{0.95}
\expandafter\def\csname rg@scale.oddfactor.win.161.status\endcsname{unsat}
\expandafter\def\csname rg@scale.oddfactor.win.321.s\endcsname{2.2}
\expandafter\def\csname rg@scale.oddfactor.win.321.status\endcsname{unsat}
\expandafter\def\csname rg@scale.oddfactor.win.41.s\endcsname{0.16}
\expandafter\def\csname rg@scale.oddfactor.win.41.status\endcsname{unsat}
\expandafter\def\csname rg@scale.oddfactor.win.641.s\endcsname{5.0}
\expandafter\def\csname rg@scale.oddfactor.win.641.status\endcsname{unsat}
\expandafter\def\csname rg@scale.oddfactor.win.81.s\endcsname{0.35}
\expandafter\def\csname rg@scale.oddfactor.win.81.status\endcsname{unsat}
\expandafter\def\csname rg@scale.oddfactor.win.exp\endcsname{1.25}
\expandafter\def\csname rg@scale.oddfactor.win.fitpts\endcsname{6}
\expandafter\def\csname rg@scale.oddfactor.win.max.n\endcsname{1,281}
\expandafter\def\csname rg@scale.oddfactor.win.max.s\endcsname{11.1}
\expandafter\def\csname rg@scale.oddfactor.win.max.vars\endcsname{3,844}
\expandafter\def\csname rg@scale.oddfactor.win.within.n\endcsname{1,281}
\expandafter\def\csname rg@scale.oddfactor.win.within.s\endcsname{11.1}
\expandafter\def\csname rg@scale.pmodd.1281.vars\endcsname{5,125}
\expandafter\def\csname rg@scale.pmodd.161.vars\endcsname{645}
\expandafter\def\csname rg@scale.pmodd.2561.vars\endcsname{10,245}
\expandafter\def\csname rg@scale.pmodd.321.vars\endcsname{1,285}
\expandafter\def\csname rg@scale.pmodd.641.vars\endcsname{2,565}
\expandafter\def\csname rg@scale.pmodd.81.vars\endcsname{325}
\expandafter\def\csname rg@scale.pmodd.int.1281.s\endcsname{3.2}
\expandafter\def\csname rg@scale.pmodd.int.1281.status\endcsname{unsat}
\expandafter\def\csname rg@scale.pmodd.int.161.s\endcsname{0.39}
\expandafter\def\csname rg@scale.pmodd.int.161.status\endcsname{unsat}
\expandafter\def\csname rg@scale.pmodd.int.2561.s\endcsname{7.1}
\expandafter\def\csname rg@scale.pmodd.int.2561.status\endcsname{unsat}
\expandafter\def\csname rg@scale.pmodd.int.321.s\endcsname{0.76}
\expandafter\def\csname rg@scale.pmodd.int.321.status\endcsname{unsat}
\expandafter\def\csname rg@scale.pmodd.int.641.s\endcsname{1.6}
\expandafter\def\csname rg@scale.pmodd.int.641.status\endcsname{unsat}
\expandafter\def\csname rg@scale.pmodd.int.81.s\endcsname{0.18}
\expandafter\def\csname rg@scale.pmodd.int.81.status\endcsname{unsat}
\expandafter\def\csname rg@scale.pmodd.int.exp\endcsname{1.06}
\expandafter\def\csname rg@scale.pmodd.int.fitpts\endcsname{6}
\expandafter\def\csname rg@scale.pmodd.int.max.n\endcsname{2,561}
\expandafter\def\csname rg@scale.pmodd.int.max.s\endcsname{7.1}
\expandafter\def\csname rg@scale.pmodd.int.max.vars\endcsname{10,245}
\expandafter\def\csname rg@scale.pmodd.int.within.n\endcsname{2,561}
\expandafter\def\csname rg@scale.pmodd.int.within.s\endcsname{7.1}
\expandafter\def\csname rg@scale.pmodd.win.1281.s\endcsname{2.3}
\expandafter\def\csname rg@scale.pmodd.win.1281.status\endcsname{unsat}
\expandafter\def\csname rg@scale.pmodd.win.161.s\endcsname{0.29}
\expandafter\def\csname rg@scale.pmodd.win.161.status\endcsname{unsat}
\expandafter\def\csname rg@scale.pmodd.win.2561.s\endcsname{4.8}
\expandafter\def\csname rg@scale.pmodd.win.2561.status\endcsname{unsat}
\expandafter\def\csname rg@scale.pmodd.win.321.s\endcsname{0.58}
\expandafter\def\csname rg@scale.pmodd.win.321.status\endcsname{unsat}
\expandafter\def\csname rg@scale.pmodd.win.641.s\endcsname{1.2}
\expandafter\def\csname rg@scale.pmodd.win.641.status\endcsname{unsat}
\expandafter\def\csname rg@scale.pmodd.win.81.s\endcsname{0.14}
\expandafter\def\csname rg@scale.pmodd.win.81.status\endcsname{unsat}
\expandafter\def\csname rg@scale.pmodd.win.exp\endcsname{1.02}
\expandafter\def\csname rg@scale.pmodd.win.fitpts\endcsname{6}
\expandafter\def\csname rg@scale.pmodd.win.max.n\endcsname{2,561}
\expandafter\def\csname rg@scale.pmodd.win.max.s\endcsname{4.8}
\expandafter\def\csname rg@scale.pmodd.win.max.vars\endcsname{10,245}
\expandafter\def\csname rg@scale.pmodd.win.within.n\endcsname{2,561}
\expandafter\def\csname rg@scale.pmodd.win.within.s\endcsname{4.8}
\expandafter\def\csname rg@scale.ringslower.n\endcsname{17}
\expandafter\def\csname rg@search.fam.count\endcsname{3}
\expandafter\def\csname rg@search.fam.word\endcsname{three}
\expandafter\def\csname rg@search.probe.smax\endcsname{41.3}
\expandafter\def\csname rg@search.probe.smin\endcsname{5.8}
\expandafter\def\csname rg@seeds\endcsname{2}
\expandafter\def\csname rg@select2.binary.base.n\endcsname{41}
\expandafter\def\csname rg@select2.binary.basex.n\endcsname{41}
\expandafter\def\csname rg@select2.binary.probe.n\endcsname{161}
\expandafter\def\csname rg@select2.binary.probe.s\endcsname{11.9}
\expandafter\def\csname rg@select2.binary.scip.n\endcsname{41}
\expandafter\def\csname rg@select2.binary.xor.n\endcsname{41}
\expandafter\def\csname rg@select2.pairwise.base.n\endcsname{41}
\expandafter\def\csname rg@select2.pairwise.basex.n\endcsname{41}
\expandafter\def\csname rg@select2.pairwise.probe.n\endcsname{161}
\expandafter\def\csname rg@select2.pairwise.probe.s\endcsname{5.8}
\expandafter\def\csname rg@select2.pairwise.scip.n\endcsname{41}
\expandafter\def\csname rg@select2.pairwise.xor.n\endcsname{41}
\expandafter\def\csname rg@tseitincnf.xor.best.n\endcsname{320}
\expandafter\def\csname rg@tseitincnf.xor.bve+hull.n\endcsname{--}
\expandafter\def\csname rg@tseitincnf.xor.bve+hull.s\endcsname{--}
\expandafter\def\csname rg@tseitincnf.xor.bve+int.n\endcsname{320}
\expandafter\def\csname rg@tseitincnf.xor.bve+int.s\endcsname{1.1}
\expandafter\def\csname rg@tseitincnf.xor.bve+win.n\endcsname{320}
\expandafter\def\csname rg@tseitincnf.xor.bve+win.s\endcsname{1.0}
\expandafter\def\csname rg@tseitincnf.xor.cadical.n\endcsname{--}
\expandafter\def\csname rg@tseitincnf.xor.cadical.s\endcsname{--}
\expandafter\def\csname rg@tseitincnf.xor.cms.n\endcsname{320}
\expandafter\def\csname rg@tseitincnf.xor.cms.s\endcsname{0.59}
\expandafter\def\csname rg@tseitincnf.xor.cnf.n\endcsname{320}
\expandafter\def\csname rg@tseitincnf.xor.cnf.s\endcsname{2.1}
\expandafter\def\csname rg@tseitincnf.xor.cnf.tools\endcsname{CryptoMiniSat, Lingeling, pgpbs, pgpbs mod 2, pgpbs mod 3, pgpbs integer, Xorcle}
\expandafter\def\csname rg@tseitincnf.xor.cnfx.n\endcsname{320}
\expandafter\def\csname rg@tseitincnf.xor.cnfx.tools\endcsname{CryptoMiniSat, Lingeling, pgpbs, pgpbs mod 2, pgpbs mod 3, pgpbs integer}
\expandafter\def\csname rg@tseitincnf.xor.exact:cnf.n\endcsname{--}
\expandafter\def\csname rg@tseitincnf.xor.exact:cnf.s\endcsname{--}
\expandafter\def\csname rg@tseitincnf.xor.hull.n\endcsname{--}
\expandafter\def\csname rg@tseitincnf.xor.hull.s\endcsname{--}
\expandafter\def\csname rg@tseitincnf.xor.int.n\endcsname{320}
\expandafter\def\csname rg@tseitincnf.xor.int.s\endcsname{1.0}
\expandafter\def\csname rg@tseitincnf.xor.kissat.n\endcsname{40}
\expandafter\def\csname rg@tseitincnf.xor.kissat.s\endcsname{2.6}
\expandafter\def\csname rg@tseitincnf.xor.lingeling.n\endcsname{320}
\expandafter\def\csname rg@tseitincnf.xor.lingeling.s\endcsname{0.27}
\expandafter\def\csname rg@tseitincnf.xor.pgpbs-2.n\endcsname{320}
\expandafter\def\csname rg@tseitincnf.xor.pgpbs-2.s\endcsname{2.0}
\expandafter\def\csname rg@tseitincnf.xor.pgpbs-3.n\endcsname{320}
\expandafter\def\csname rg@tseitincnf.xor.pgpbs-3.s\endcsname{2.0}
\expandafter\def\csname rg@tseitincnf.xor.pgpbs-a.n\endcsname{320}
\expandafter\def\csname rg@tseitincnf.xor.pgpbs-a.s\endcsname{2.1}
\expandafter\def\csname rg@tseitincnf.xor.pgpbs-i.n\endcsname{320}
\expandafter\def\csname rg@tseitincnf.xor.pgpbs-i.s\endcsname{2.1}
\expandafter\def\csname rg@tseitincnf.xor.read.fastest\endcsname{field}
\expandafter\def\csname rg@tseitincnf.xor.read.n\endcsname{320}
\expandafter\def\csname rg@tseitincnf.xor.read.s\endcsname{0.92}
\expandafter\def\csname rg@tseitincnf.xor.read.smax\endcsname{1.1}
\expandafter\def\csname rg@tseitincnf.xor.read.tool\endcsname{field, ring, field after elim., ring after elim.}
\expandafter\def\csname rg@tseitincnf.xor.roundingsat:cnf.n\endcsname{--}
\expandafter\def\csname rg@tseitincnf.xor.roundingsat:cnf.s\endcsname{--}
\expandafter\def\csname rg@tseitincnf.xor.sat4j-cp:cnf.n\endcsname{--}
\expandafter\def\csname rg@tseitincnf.xor.sat4j-cp:cnf.s\endcsname{--}
\expandafter\def\csname rg@tseitincnf.xor.scip:cnf.n\endcsname{--}
\expandafter\def\csname rg@tseitincnf.xor.scip:cnf.s\endcsname{--}
\expandafter\def\csname rg@tseitincnf.xor.top\endcsname{320}
\expandafter\def\csname rg@tseitincnf.xor.win.n\endcsname{320}
\expandafter\def\csname rg@tseitincnf.xor.win.s\endcsname{0.92}
\expandafter\def\csname rg@tseitincnf.xor.xor.n\endcsname{320}
\expandafter\def\csname rg@tseitincnf.xor.xor.s\endcsname{0.89}
\expandafter\def\csname rg@tseitincnf.xor.xorcle-dev.n\endcsname{320}
\expandafter\def\csname rg@tseitincnf.xor.xorcle-dev.s\endcsname{0.89}
\expandafter\def\csname rg@tseitincnf.xor.xorcle.n\endcsname{320}
\expandafter\def\csname rg@tseitincnf.xor.xorcle.s\endcsname{1.1}
\expandafter\def\csname rg@ver.cms\endcsname{5.17.0}
\expandafter\def\csname rg@ver.highs\endcsname{1.15.1}
\expandafter\def\csname rg@ver.lingeling\endcsname{1.0.0}
\expandafter\def\csname rg@ver.numpy\endcsname{2.5.1}
\expandafter\def\csname rg@ver.pysat\endcsname{1.9.dev17}
\expandafter\def\csname rg@ver.pyscipopt\endcsname{6.3.0}
\expandafter\def\csname rg@ver.sat4j\endcsname{2.3.5}
\expandafter\def\csname rg@ver.scip\endcsname{10.0}
\expandafter\def\csname rg@ver.scipy\endcsname{1.17.1}
\expandafter\def\csname rg@xor.awake.min\endcsname{33}
\expandafter\def\csname rg@xor.calls\endcsname{720}
\expandafter\def\csname rg@xor.ctl.failed\endcsname{0}
\expandafter\def\csname rg@xor.ctl.models\endcsname{244}
\expandafter\def\csname rg@xor.ctl.n\endcsname{144}
\expandafter\def\csname rg@xor.ctl.wrong\endcsname{0}
\expandafter\def\csname rg@xor.date\endcsname{2026-10-30}
\expandafter\def\csname rg@xor.grids\endcsname{match}
\expandafter\def\csname rg@xor.issues\endcsname{0}
\expandafter\def\csname rg@xor.limit\endcsname{60}
\expandafter\def\csname rg@xor.present\endcsname{yes}
\expandafter\def\csname rg@xor.seeds\endcsname{2}
\expandafter\def\csname rg@xor.versions.same\endcsname{yes}
\expandafter\def\csname rg@xor.xorcle-dev.commit\endcsname{af93ea6}
\expandafter\def\csname rg@xor.xorcle-dev.version\endcsname{0.2.0-dev}
\expandafter\def\csname rg@xor.xorcle.commit\endcsname{daa0125}
\expandafter\def\csname rg@xor.xorcle.version\endcsname{0.1}
\makeatother
\newcommand{\IxPoints}{16}
\newcommand{\IxDiffWord}{eight}
\newcommand{\IxLevels}{0, 16 and 64}
\newcommand{\IxTop}{64}
\newcommand{\IxAgreeMSBWord}{twelve}
\newcommand{\IxZerosWord}{six}
\newcommand{\IxPatterns}{16}
\newcommand{\OdAGLFourSets}{2}
\newcommand{\OdBentFns}{65{,}536}
\newcommand{\OdBentCount}{896}
\newcommand{\OdNzFns}{33{,}664}
\newcommand{\OdNzClasses}{2{,}104}
\newcommand{\OdMomCases}{36}
\newcommand{\OdAGLSixFour}{373}
\newcommand{\OdAGLSixSix}{374}
\newcommand{\FoOutputs}{277}
\newcommand{\FoLevels}{1{,}174}
\newcommand{\FoBetter}{65}
\newcommand{\FoBetterPct}{6}
\newcommand{\FoRepsSaved}{142}
\newcommand{\FoWorse}{765}
\newcommand{\FoTPair}{5}
\newcommand{\FoTVar}{3}
\newcommand{\FoWorstBetter}{17}
\newcommand{\FoEnergyMedian}{6.6}
\newcommand{\FoMuLo}{8}
\newcommand{\FoMuHi}{128}
\newcommand{\SaOutputs}{139}
\newcommand{\SaCoefs}{6{,}720}
\newcommand{\SaGain}{2.02}
\newcommand{\SaRatio}{1.000}
\newcommand{\SaRatioLo}{0.94}
\newcommand{\SaRatioHi}{1.07}
\newcommand{\SaBis}{939}
\newcommand{\SaBisGain}{1.99}
\newcommand{\SaBisRatio}{0.97}
\newcommand{\SaMaxGain}{110}
\newcommand{\SaMaxOrbit}{120}
\newcommand{\SaBelow}{0}
\newcommand{\SaL}{1{,}536}
\newcommand{\SaR}{400}
\newcommand{\SaLref}{2{,}097{,}152}
\newcommand{\SaCensus}{16{,}384}
\newcommand{\SaSources}{5}
\newcommand{\SyStabTotal}{24}
\newcommand{\SyStabPartial}{8}
\newcommand{\SyStabNone}{1}
\newcommand{\SyBentCount}{896}
\newcommand{\SyBentHist}{480 of order 1, 360 of order 2, 24 of order 4, 16 of order 6, 12 of order 8, 4 of order 24}
\newcommand{\SyBTrivFour}{22{,}656}
\newcommand{\SySubgroupsSix}{1{,}455}
\newcommand{\EdInstances}{17{,}347}
\newcommand{\EdSmallOne}{65.6}
\newcommand{\EdSmallTwo}{24.2}
\newcommand{\EdSmallLeTwo}{90}
\newcommand{\EdOrbitFour}{$1.6\times10^{-3}$}
\newcommand{\EdSecdedFour}{$2.5\times10^{-5}$}
\newcommand{\EdTmrFour}{$1.9\times10^{-6}$}
\newcommand{\EdBeatsLow}{10}
\newcommand{\EdCodeFour}{$5.9\times10^{-5}$}
\newcommand{\EdCodeTwo}{$5.1\times10^{-3}$}
\newcommand{\EdCodeSix}{$5.9\times10^{-7}$}
\newcommand{\EdSecdedSix}{$2.6\times10^{-9}$}
\newcommand{\EdBeatsHigh}{74}
\newcommand{\VoReduceAll}{2.13}
\newcommand{\VoReduceFour}{5.38}
\newcommand{\VoReduceFive}{15.07}
\newcommand{\VoSiftSix}{98}
\newcommand{\VoHaarSix}{77}
\newcommand{\VoSiftAll}{83}
\newcommand{\VoHaarAll}{19}
\newcommand{\VoRevAll}{36}
\newcommand{\LcCostFour}{0.87}
\newcommand{\LcCostFive}{0.71}
\newcommand{\LcSymSix}{94}
\newcommand{\LcBitsSix}{0.43}
\newcommand{\LcCostSix}{0.54}
\newcommand{\BgFour}{65{,}536}
\newcommand{\BgClasses}{222}
\newcommand{\BgAffineMaps}{322{,}560}
\newcommand{\BgLoose}{7.8}
\newcommand{\BgThirdExact}{100}
\newcommand{\BgSecondMean}{2.49}
\newcommand{\BgProjMean}{2.13}
\newcommand{\BgMsFour}{0.22}
\newcommand{\BgTrivSixB}{0.9}
\newcommand{\BgMedSixB}{36}
\newcommand{\BgMedSixAGL}{14{,}450{,}688}
\newcommand{\BgMaxMsSixAGL}{115}
\newcommand{\BgMedMsSixAGL}{3.32}
\newcommand{\BgTrials}{400}
\newcommand{\BgVecOne}{72}
\newcommand{\BgDiagOne}{93}
\newcommand{\BgExactOne}{57}
\newcommand{\BgVecTwo}{55}
\newcommand{\BgDiagTwo}{81}
\newcommand{\BgExactTwo}{40}
\newcommand{\BgVecFour}{32}
\newcommand{\BgDiagFour}{56}
\newcommand{\BgExactFour}{22}
\newcommand{\BgVecEight}{12}
\newcommand{\BgDiagEight}{24}
\newcommand{\BgExactEight}{12}
\newcommand{\BgMaxNodesSixAGL}{1{,}033}
\newcommand{\ClWreathFour}{32{,}768}
\newcommand{\EeCarryBound}{0.50}
\newcommand{\EePairsFour}{917{,}504}
\newcommand{\EeEqualFour}{33.5}
\newcommand{\EeSubFuncs}{2{,}000}
\newcommand{\EeFunctions}{2{,}168}
\newcommand{\EeInstances}{76{,}286}
\newcommand{\EeHitOne}{100.0}
\newcommand{\EeCovOne}{0.37}
\newcommand{\EeHitTwo}{99.4}
\newcommand{\EeCovTwo}{0.59}
\newcommand{\EeHitThree}{100.0}
\newcommand{\EeCovThree}{0.75}
\newcommand{\EeLossTwo}{0.0016}
\newcommand{\CrAllFour}{65{,}536}
\newcommand{\CrSac}{4{,}128}
\newcommand{\CrSacBal}{1{,}368}
\newcommand{\CrPcTwo}{896}
\newcommand{\CrBestRate}{0.236}
\newcommand{\CrRandomPair}{0.0039}
\newcommand{\CrAboveFour}{2.4}
\newcommand{\CrAesPr}{0.492}
\newcommand{\CrAesNl}{112}
\newcommand{\CrAesDu}{4}
\newcommand{\CrRandEightPr}{0.336}
\newcommand{\CrRandEightCount}{50}
\newcommand{\CrPresentNl}{4}
\newcommand{\CrPresentPrMin}{0.25}
\newcommand{\CrPresentPrMean}{0.37}
\newcommand{\CrPcgSerial}{242.4}
\newcommand{\CrBiasW}{0.2016, 0.0387, 0.0077}
\newcommand{\CrNoiseFloor}{0.0039}
\newcommand{\CrLfsrMask}{193}
\newcommand{\TsSizes}{8, 12, 16, 20, 24, 28, 32, 40, 48, 60, 100}
\newcommand{\TsInstances}{33}
\newcommand{\TsSeeds}{3}
\newcommand{\TsTimeout}{60}
\newcommand{\TsTopVars}{200}
\newcommand{\TsTopClauses}{1{,}074}
\newcommand{\TsReadMax}{0.14}
\newcommand{\TsCadLast}{48}
\newcommand{\TsCadFirst}{60}
\newcommand{\TsCadTimeouts}{5}
\newcommand{\TsCadForty}{0.38}
\newcommand{\TsCmsLast}{48}
\newcommand{\TsCmsTimeouts}{6}
\newcommand{\TsCmsForty}{0.66}
\newcommand{\TsReadForty}{0.04}
\newcommand{\ScrubInstances}{17{,}347}
\newcommand{\ScrubRawFour}{$5.0\times10^{-5}$}
\newcommand{\ScrubVoteFour}{$1.3\times10^{-5}$}
\newcommand{\ScrubTmrFour}{$1.0\times10^{-8}$}
\newcommand{\ScrubFrameFour}{$2.5\times10^{-5}$}
\newcommand{\ScrubRatio}{0.25}
\newcommand{\ScrubSmall}{0.25}
\newcommand{\ScrubTenX}{38.6}
\newcommand{\ScrubRawTwo}{$5.0\times10^{-3}$}
\newcommand{\ScrubVoteTwo}{$1.3\times10^{-3}$}
\newcommand{\ScrubFrameTwo}{0.16}
\newcommand{\EeXorOne}{5.1}
\newcommand{\EeXorGainOne}{0.04}
\newcommand{\EeXorSetsOne}{15}
\newcommand{\EeXorTwo}{17.1}
\newcommand{\EeXorGainTwo}{0.09}
\newcommand{\EeXorSetsTwo}{35}
\newcommand{\EeUnitSetsTwo}{6}
\newcommand{\AsgMmax}{5}
\newcommand{\AsgThreeBurn}{10}
\newcommand{\AsgThreeMU}{7}
\newcommand{\AsgFourBurn}{38}
\newcommand{\AsgFourMU}{35}
\newcommand{\AsgFourRows}{3}
\newcommand{\AsgFiveRows}{140}
\newcommand{\AsgCases}{56}
\newcommand{\CrOptClasses}{16}
\newcommand{\CrOptLow}{8}
\newcommand{\CrOptHigh}{8}
\newcommand{\AmFns}{2{,}379}
\newcommand{\AmClasses}{160}
\newcommand{\AmBrute}{1.04}
\newcommand{\AmBdOne}{0.094}
\newcommand{\AmEnumOne}{0.007}
\newcommand{\AmBdTwo}{0.175}
\newcommand{\AmEnumTwo}{0.038}
\newcommand{\AmBdFour}{0.481}
\newcommand{\AmEnumFour}{9.94}
\newcommand{\AmBdEight}{1.02}
\newcommand{\AmEnumEightCount}{5.1}
\newcommand{\AmNear}{0.137}
\newcommand{\AmNearEnum}{21.9}
\newcommand{\AmNearEnumN}{96}
\newcommand{\AmFar}{4.7}
\newcommand{\DmDcFour}{4.4}
\newcommand{\DmMatchFour}{9.9}
\newcommand{\DmDcFive}{11.3}
\newcommand{\DmMatchFive}{13.5}
\newcommand{\DmDcSix}{27.1}
\newcommand{\DmMatchSix}{14.4}
\newcommand{\DmGainSix}{73}
\newcommand{\DmIsfs}{1{,}500}
\newcommand{\DmSat}{100{,}519}
\newcommand{\DmBrute}{1.39}
\newcommand{\DmBdZero}{1.08}
\newcommand{\DmOverZero}{61}
\newcommand{\DmBdTwo}{1.43}
\newcommand{\DmBdFour}{1.14}
\newcommand{\DmNear}{2.45}
\newcommand{\DmTight}{4.8}
\newcommand{\DmCompMedian}{32}
\newcommand{\AxTightLSix}{9.5}
\newcommand{\AxTightBSix}{32.6}
\newcommand{\AxTightLSeven}{7.6}
\newcommand{\AxTightBSeven}{29.1}
\newcommand{\AxTightLEight}{5.7}
\newcommand{\AxTightBEight}{25.8}
\newcommand{\AxCSeven}{88}
\newcommand{\AxDcBaseSeven}{1.3}
\newcommand{\AxDcFirstSeven}{309}
\newcommand{\AxDcFirstXSeven}{4}
\newcommand{\AxQEight}{80}
\newcommand{\AxCEight}{57}
\newcommand{\AxBaseEight}{13.5}
\newcommand{\AxFirstEight}{40.6}
\newcommand{\AxFirstXEight}{334}
\newcommand{\AxNearEight}{273}
\newcommand{\AxNearXEight}{50}
\newcommand{\AxTopEEight}{32}
\newcommand{\AxTopXEight}{4.7}
\newcommand{\AxLevelOnlyEight}{653}
\newcommand{\AxDcBaseEight}{15.1}
\newcommand{\AxDcFirstEight}{3{,}557}
\newcommand{\AxDcFirstXEight}{4}
\newcommand{\AxOrbitEight}{5.8}
\newcommand{\AxPairsEight}{600}
\newcommand{\CmFunctions}{77}
\newcommand{\CmCuts}{600}
\newcommand{\CmExactFour}{27.3}
\newcommand{\CmOdcFour}{50.0}
\newcommand{\CmSmallFour}{18.5}
\newcommand{\CmSaveFour}{8.6}
\newcommand{\CmExactFive}{16.8}
\newcommand{\CmOdcFive}{32.5}
\newcommand{\CmSmallFive}{16.8}
\newcommand{\CmSaveFive}{7.7}
\newcommand{\CmExactSix}{1.5}
\newcommand{\CmOdcSix}{26.3}
\newcommand{\CmSmallSix}{12.5}
\newcommand{\CmSaveSix}{7.5}
\newcommand{\CdCircuits}{10}
\newcommand{\CdCells}{81}
\newcommand{\CdFlows}{48}
\newcommand{\CdAbcPct}{4.14}
\newcommand{\CdMfsPct}{1.88}
\newcommand{\CdOursPct}{1.37}
\newcommand{\CdAfterPct}{0.55}
\newcommand{\CdAfterMin}{0.00}
\newcommand{\CdAfterMax}{2.03}
\newcommand{\CdNetlists}{40}
\newcommand{\CpCircuits}{10}
\newcommand{\CpPatterns}{1{,}024}
\newcommand{\CpDetected}{58{,}100}
\newcommand{\CpRandomTwo}{5{,}903}
\newcommand{\CpGroupsTwo}{145}
\newcommand{\AfClasses}{48}
\newcommand{\AfMag}{40}
\newcommand{\PbClassesSix}{808}
\newcommand{\PbPairsSix}{242{,}400}
\newcommand{\PbFullSix}{0.0316}
\newcommand{\PbMsOneSix}{0.52}
\newcommand{\PbGainOneSix}{61}
\newcommand{\PbMsEightSix}{5.46}
\newcommand{\PbGainEightSix}{5.8}
\newcommand{\PbMsThirtyTwoSix}{71.87}
\newcommand{\PbGainThirtyTwoSix}{0.4}
\newcommand{\PbMsNearSix}{0.66}
\newcommand{\PbClassesSeven}{561}
\newcommand{\PbPairsSeven}{112{,}200}
\newcommand{\PbFullSeven}{0.303}
\newcommand{\PbMsOneSeven}{0.57}
\newcommand{\PbGainOneSeven}{530}
\newcommand{\PbMsEightSeven}{3.00}
\newcommand{\PbGainEightSeven}{101}
\newcommand{\PbMsThirtyTwoSeven}{22.75}
\newcommand{\PbGainThirtyTwoSeven}{13}
\newcommand{\PbMsNearSeven}{1.00}
\newcommand{\PbClassesEight}{184}
\newcommand{\PbPairsEight}{589}
\newcommand{\PbFullEight}{1.61}
\newcommand{\PbMsOneEight}{0.64}
\newcommand{\PbGainOneEight}{2{,}514}
\newcommand{\PbMsEightEight}{1.58}
\newcommand{\PbGainEightEight}{1{,}016}
\newcommand{\PbMsThirtyTwoEight}{4.10}
\newcommand{\PbGainThirtyTwoEight}{392}
\newcommand{\PbMsNearEight}{0.86}
\newcommand{\PbClassesNine}{18}
\newcommand{\PbPairsNine}{40}
\newcommand{\PbFullNine}{19}
\newcommand{\PbMsOneNine}{0.99}
\newcommand{\PbGainOneNine}{18{,}930}
\newcommand{\PbMsEightNine}{1.87}
\newcommand{\PbGainEightNine}{9{,}992}
\newcommand{\PbMsThirtyTwoNine}{1.87}
\newcommand{\PbGainThirtyTwoNine}{9{,}984}
\newcommand{\PbMsNearNine}{1.02}
\newcommand{\PbClassesTen}{40}
\newcommand{\PbPairsTen}{12}
\newcommand{\PbFullTen}{823}
\newcommand{\PbMsOneTen}{1.21}
\newcommand{\PbGainOneTen}{679{,}062}
\newcommand{\PbMsEightTen}{1.74}
\newcommand{\PbGainEightTen}{474{,}197}
\newcommand{\PbMsThirtyTwoTen}{1.77}
\newcommand{\PbGainThirtyTwoTen}{463{,}875}
\newcommand{\PbMsNearTen}{1.43}
\newcommand{\PbTightOnesSix}{70}
\newcommand{\PbTightEnergySix}{11}
\newcommand{\PbTightCoupledSix}{20}
\newcommand{\PbCoupledAboveEnergySix}{68}
\newcommand{\PbEqualCoupledSix}{97}
\newcommand{\PbCellNodesSix}{13}
\newcommand{\PbCellTimeSix}{3.4}
\newcommand{\PbTightOnesSeven}{74}
\newcommand{\PbTightEnergySeven}{11}
\newcommand{\PbTightCoupledSeven}{16}
\newcommand{\PbCoupledAboveEnergySeven}{70}
\newcommand{\PbEqualCoupledSeven}{100}
\newcommand{\PbCellNodesSeven}{13}
\newcommand{\PbCellTimeSeven}{3.7}
\newcommand{\PbTightOnesEight}{69}
\newcommand{\PbTightEnergyEight}{15}
\newcommand{\PbTightCoupledEight}{22}
\newcommand{\PbCoupledAboveEnergyEight}{73}
\newcommand{\PbEqualCoupledEight}{100}
\newcommand{\PbCellNodesEight}{8}
\newcommand{\PbCellTimeEight}{4.2}
\newcommand{\SdGapMixedFive}{0.34}
\newcommand{\SdGapSdpFive}{0.24}
\newcommand{\SdAboveFive}{6}
\newcommand{\SdGapMixedSix}{1.09}
\newcommand{\SdGapSdpSix}{0.72}
\newcommand{\SdAboveSix}{12}
\newcommand{\SdGapMixedSeven}{2.17}
\newcommand{\SdGapSdpSeven}{1.40}
\newcommand{\SdAboveSeven}{15}
\newcommand{\SdGapMixedEight}{3.00}
\newcommand{\SdGapSdpEight}{2.27}
\newcommand{\SdAboveEight}{12}
\newcommand{\SdParamsEight}{165}
\newcommand{\CrFunctionsFour}{2{,}168}
\newcommand{\CrPermClassesFour}{379}
\newcommand{\CrPermNodesFour}{1.9}
\newcommand{\CrNpnClassesFour}{58}
\newcommand{\CrNpnNodesFour}{3.4}
\newcommand{\CrFunctionsFive}{4{,}231}
\newcommand{\CrPermClassesFive}{1{,}033}
\newcommand{\CrPermNodesFive}{2.5}
\newcommand{\CrNpnClassesFive}{138}
\newcommand{\CrNpnNodesFive}{4.1}
\newcommand{\CrFunctionsSix}{2{,}379}
\newcommand{\CrPermClassesSix}{808}
\newcommand{\CrPermNodesSix}{3.3}
\newcommand{\CrNpnClassesSix}{160}
\newcommand{\CrNpnNodesSix}{6.3}
\newcommand{\CrFunctionsSeven}{1{,}614}
\newcommand{\CrPermClassesSeven}{561}
\newcommand{\CrPermNodesSeven}{4.1}
\newcommand{\CrNpnClassesSeven}{143}
\newcommand{\CrNpnNodesSeven}{8.8}
\newcommand{\CrFunctionsEight}{421}
\newcommand{\CrPermClassesEight}{184}
\newcommand{\CrPermNodesEight}{4.8}
\newcommand{\CrNpnClassesEight}{67}
\newcommand{\CrNpnNodesEight}{10.3}
\newcommand{\CrFunctionsNine}{36}
\newcommand{\CrPermClassesNine}{18}
\newcommand{\CrPermNodesNine}{8.0}
\newcommand{\CrNpnClassesNine}{5}
\newcommand{\CrNpnNodesNine}{11.3}
\newcommand{\CrFunctionsTen}{159}
\newcommand{\CrPermClassesTen}{40}
\newcommand{\CrPermNodesTen}{7.5}
\newcommand{\CrNpnClassesTen}{13}
\newcommand{\CrNpnNodesTen}{13.0}
\newcommand{\CrInvChecked}{2{,}390}
\newcommand{\CrcFunctionsFour}{14.1}
\newcommand{\CrcSubFour}{14.1}
\newcommand{\CrcNpnUsFour}{1.08}
\newcommand{\CrcAbcElevenUsFour}{$<$0.71}
\newcommand{\CrcPUsFour}{0.49}
\newcommand{\CrcAbcTwelveUsFour}{n/a}
\newcommand{\CrcPSubUsFour}{0.79}
\newcommand{\CrcKpermUsFour}{0.61}
\newcommand{\CrcFunctionsFive}{253}
\newcommand{\CrcSubFive}{100}
\newcommand{\CrcNpnUsFive}{2.45}
\newcommand{\CrcAbcElevenUsFive}{0.40}
\newcommand{\CrcPUsFive}{1.02}
\newcommand{\CrcAbcTwelveUsFive}{n/a}
\newcommand{\CrcPSubUsFive}{1.13}
\newcommand{\CrcKpermUsFive}{2.23}
\newcommand{\CrcFunctionsSix}{1{,}000}
\newcommand{\CrcSubSix}{100}
\newcommand{\CrcNpnUsSix}{5.99}
\newcommand{\CrcAbcElevenUsSix}{0.80}
\newcommand{\CrcPUsSix}{2.18}
\newcommand{\CrcAbcTwelveUsSix}{0.14}
\newcommand{\CrcPSubUsSix}{2.41}
\newcommand{\CrcKpermUsSix}{8.21}
\newcommand{\CrcFunctionsSeven}{1{,}000}
\newcommand{\CrcSubSeven}{20.0}
\newcommand{\CrcNpnUsSeven}{18.4}
\newcommand{\CrcAbcElevenUsSeven}{1.12}
\newcommand{\CrcPUsSeven}{4.79}
\newcommand{\CrcAbcTwelveUsSeven}{0.23}
\newcommand{\CrcPSubUsSeven}{5.46}
\newcommand{\CrcKpermUsSeven}{32.1}
\newcommand{\CrcFunctionsEight}{1{,}000}
\newcommand{\CrcSubEight}{20.0}
\newcommand{\CrcNpnUsEight}{39.3}
\newcommand{\CrcAbcElevenUsEight}{1.16}
\newcommand{\CrcPUsEight}{11.0}
\newcommand{\CrcAbcTwelveUsEight}{0.31}
\newcommand{\CrcPSubUsEight}{11.4}
\newcommand{\CrcKpermUsEight}{192}
\newcommand{\CrcKpermFactorEight}{17}
\newcommand{\CrcFunctionsNine}{200}
\newcommand{\CrcSubNine}{2.0}
\newcommand{\CrcNpnUsNine}{90.7}
\newcommand{\CrcAbcElevenUsNine}{0.85}
\newcommand{\CrcPUsNine}{28.9}
\newcommand{\CrcAbcTwelveUsNine}{0.40}
\newcommand{\CrcPSubUsNine}{37.9}
\newcommand{\CrcKpermUsNine}{170}
\newcommand{\CrcFunctionsTen}{100}
\newcommand{\CrcSubTen}{0.5}
\newcommand{\CrcNpnUsTen}{161}
\newcommand{\CrcAbcElevenUsTen}{1.20}
\newcommand{\CrcPUsTen}{67.1}
\newcommand{\CrcAbcTwelveUsTen}{0.60}
\newcommand{\CrcPSubUsTen}{68.5}
\newcommand{\CrcKpermUsTen}{8{,}208}
\newcommand{\CrcKpermFactorTen}{120}
\newcommand{\CrcAbcTwelveClassesSix}{8{,}270}
\newcommand{\CrcPClassesSix}{8{,}230}
\newcommand{\CrcNpnRatioMin}{6.2}
\newcommand{\CrcNpnRatioMax}{134}
\newcommand{\CrcPRatioMin}{16}
\newcommand{\CrcPRatioMax}{112}
\newcommand{\CrcKpermFrom}{five}
\newcommand{\PlFunctionsFour}{2{,}168}
\newcommand{\PlPbPctFour}{13.9}
\newcommand{\PlConcFour}{0.39}
\newcommand{\PlConcRatioFour}{6}
\newcommand{\PlOffwhiteFour}{0.50}
\newcommand{\PlBentPulseFour}{38}
\newcommand{\PlFunctionsFive}{4{,}231}
\newcommand{\PlPbPctFive}{3.2}
\newcommand{\PlConcFive}{0.39}
\newcommand{\PlBentPulseFive}{76}
\newcommand{\PlFunctionsSix}{2{,}379}
\newcommand{\PlPbPctSix}{1.6}
\newcommand{\PlConcSix}{0.52}
\newcommand{\PlBentPulseSix}{2}
\newcommand{\PlFunctionsSeven}{1{,}614}
\newcommand{\PlPbPctSeven}{1.5}
\newcommand{\PlConcSeven}{0.66}
\newcommand{\PlPbPctEight}{1.7}
\newcommand{\PlConcRatioEight}{153}
\newcommand{\PlOffwhiteEight}{0.97}
\newcommand{\PlBentFour}{38}
\newcommand{\PlSinglePctFour}{71.6}
\newcommand{\PlMixPctFour}{8.3}
\newcommand{\PlMixOnlyFour}{4}
\newcommand{\PlNpnSplitFour}{17}
\newcommand{\PlNpnClassesFour}{58}
\newcommand{\PlSinglePctFive}{83.8}
\newcommand{\PlMixPctFive}{27.7}
\newcommand{\PlMixOnlyFive}{10}
\newcommand{\PlNpnSplitFive}{37}
\newcommand{\PlNpnClassesFive}{138}
\newcommand{\PlSinglePctSix}{90.5}
\newcommand{\PlMixPctSix}{49.3}
\newcommand{\PlMixOnlySix}{28}
\newcommand{\PlNpnSplitSix}{49}
\newcommand{\PlNpnClassesSix}{160}
\newcommand{\PlSinglePctSeven}{91.3}
\newcommand{\PlMixPctSeven}{60.8}
\newcommand{\PlMixOnlySeven}{35}
\newcommand{\PlGroupsSeven}{367}
\newcommand{\PlPairsSeven}{13{,}846}
\newcommand{\PlMixSTSeven}{70}
\newcommand{\PlNpnSplitSeven}{41}
\newcommand{\PlNpnClassesSeven}{143}
\newcommand{\PlPClassesSeven}{561}
\newcommand{\PlPShConstSix}{52.6}
\newcommand{\PlPShAffSix}{18.4}
\newcommand{\PlPShSymSix}{0.6}
\newcommand{\PlPShPartSix}{25.5}
\newcommand{\PlPShOtherSix}{2.8}
\newcommand{\PlPShBentSix}{5}
\newcommand{\PlPDvConstSix}{45.7}
\newcommand{\PlPDvAffSix}{13.7}
\newcommand{\PlPDvSymSix}{4.4}
\newcommand{\PlPDvPartSix}{33.3}
\newcommand{\PlPDvOtherSix}{2.9}
\newcommand{\PlPDvBentSix}{2}
\newcommand{\PlPKfConstSix}{53.5}
\newcommand{\PlPKfAffSix}{18.5}
\newcommand{\PlPKfSymSix}{0.8}
\newcommand{\PlPKfPartSix}{25.0}
\newcommand{\PlPKfOtherSix}{2.3}
\newcommand{\PlPKfBentSix}{2}
\newcommand{\PlPShConstSeven}{64.5}
\newcommand{\PlPShAffSeven}{20.9}
\newcommand{\PlPShSymSeven}{0.4}
\newcommand{\PlPShPartSeven}{12.8}
\newcommand{\PlPShOtherSeven}{1.4}
\newcommand{\PlPShBentSeven}{0}
\newcommand{\PlPDvConstSeven}{63.9}
\newcommand{\PlPDvAffSeven}{7.3}
\newcommand{\PlPDvSymSeven}{9.1}
\newcommand{\PlPDvPartSeven}{18.3}
\newcommand{\PlPDvOtherSeven}{1.4}
\newcommand{\PlPDvBentSeven}{0}
\newcommand{\PlPKfConstSeven}{68.1}
\newcommand{\PlPKfAffSeven}{18.0}
\newcommand{\PlPKfSymSeven}{0.5}
\newcommand{\PlPKfPartSeven}{12.1}
\newcommand{\PlPKfOtherSeven}{1.3}
\newcommand{\PlPKfBentSeven}{0}
\newcommand{\PlPShConstEight}{69.3}
\newcommand{\PlPShAffEight}{24.0}
\newcommand{\PlPShSymEight}{0.4}
\newcommand{\PlPShPartEight}{5.7}
\newcommand{\PlPShOtherEight}{0.6}
\newcommand{\PlPShBentEight}{0}
\newcommand{\PlPDvConstEight}{77.1}
\newcommand{\PlPDvAffEight}{5.7}
\newcommand{\PlPDvSymEight}{9.6}
\newcommand{\PlPDvPartEight}{6.8}
\newcommand{\PlPDvOtherEight}{0.7}
\newcommand{\PlPDvBentEight}{0}
\newcommand{\PlPKfConstEight}{69.4}
\newcommand{\PlPKfAffEight}{24.1}
\newcommand{\PlPKfSymEight}{0.5}
\newcommand{\PlPKfPartEight}{5.4}
\newcommand{\PlPKfOtherEight}{0.5}
\newcommand{\PlPKfBentEight}{0}
\newcommand{\PlDlWholeSix}{32}
\newcommand{\PlDlCofSix}{12}
\newcommand{\PlDlCofSmallerSix}{2{,}363}
\newcommand{\PlDlWholeIdxSix}{44}
\newcommand{\PlDlCofIdxSix}{48}
\newcommand{\PlDlWholeSmallerIdxSix}{1{,}561}
\newcommand{\PlDlWholeSmallerIdxPctSix}{66}
\newcommand{\PlIndexBitsSix}{12}
\newcommand{\PlChoiceBitsSix}{8}
\newcommand{\PlDlWholeSmallerIdxPctSeven}{77}
\newcommand{\PlPulseIndexBits}{7}
\newcommand{\PlChartsFive}{42{,}310}
\newcommand{\PlChartsSix}{47{,}580}
\newcommand{\PlNontrivSix}{40{,}701}
\newcommand{\PlPeelLeastPctSix}{98.3}
\newcommand{\PlPeelAbovePctSix}{0.5}
\newcommand{\DcPairs}{21{,}155}
\newcommand{\DcDavio}{9{,}505}
\newcommand{\DcShannon}{1{,}433}
\newcommand{\DcBasePct}{86.9}
\newcommand{\DcAucDcv}{0.79}
\newcommand{\DcAucRho}{0.56}
\newcommand{\DcAucPsi}{0.38}
\newcommand{\DcAucPsiRev}{0.62}
\newcommand{\DcBestDcvPct}{86.9}
\newcommand{\DcBestRhoPct}{86.9}
\newcommand{\DcBestPsiPct}{86.9}
\newcommand{\DcZero}{2{,}692}
\newcommand{\DcFunctions}{4{,}231}
\newcommand{\DcWholeShorter}{2{,}205}
\newcommand{\DcRouteBasePct}{52}
\newcommand{\DcWholeIndexBits}{9}
\newcommand{\DcPulseIndexBits}{7}
\newcommand{\DcChoiceBits}{4}
\newcommand{\DcAucRoute}{0.64}
\newcommand{\DcRouteBestPct}{64}
\newcommand{\DcAucConc}{0.50}
\newcommand{\DcRouteConcBestPct}{53}
\newcommand{\DcRouteZero}{1{,}434}
\newcommand{\DcRouteZeroWhole}{1{,}062}

\newcommand{\AcBentDl}{47.5}
\newcommand{\AcBentRaw}{73.5}
\newcommand{\AcCutPeeledEight}{236}
\newcommand{\AcCutPeeledSeven}{244}
\newcommand{\AcCutPeeledSix}{210}
\newcommand{\AcCutResEight}{1.2}
\newcommand{\AcCutResSeven}{2.3}
\newcommand{\AcCutResSix}{4.7}
\newcommand{\AcCutRouteEight}{5}
\newcommand{\AcCutRouteNSix}{60}
\newcommand{\AcCutRouteSeven}{12}
\newcommand{\AcCutRouteSix}{0}
\newcommand{\AcCutSaveEight}{114.5}
\newcommand{\AcCutSaveSeven}{41.1}
\newcommand{\AcCutSaveSix}{10.8}
\newcommand{\AcCutSix}{300}
\newcommand{\AcCutTwoEight}{51}
\newcommand{\AcCutTwoSeven}{17}
\newcommand{\AcCutTwoSix}{14}
\newcommand{\AcIdentPairEight}{238,262}
\newcommand{\AcIdentPairSeven}{51,131}
\newcommand{\AcIdentPairSix}{3,909}
\newcommand{\AcMajDl}{32.1}
\newcommand{\AcMajRaw}{139.7}
\newcommand{\AcMuxCofDl}{55.9}
\newcommand{\AcMuxDl}{56.5}
\newcommand{\AcNoiseEight}{0, 1 and 8}
\newcommand{\AcNoiseSeven}{0, 1 and 4}
\newcommand{\AcNoiseSix}{0, 1 and 2}
\newcommand{\AcNoisyScenesSix}{100}
\newcommand{\AcNullSeven}{0}
\newcommand{\AcNullSix}{0}
\newcommand{\AcNullTablesSeven}{30}
\newcommand{\AcNullTablesSix}{50}
\newcommand{\AcOptEight}{80}
\newcommand{\AcOptPairEight}{35}
\newcommand{\AcOptSeven}{90}
\newcommand{\AcOptSingleEight}{45}
\newcommand{\AcOptSix}{149}
\newcommand{\AcPairScenesEight}{45}
\newcommand{\AcPairZeroStructEight}{15}
\newcommand{\AcRefAtoms}{770}
\newcommand{\AcRefMaxDiff}{$1.1\times10^{-16}$}
\newcommand{\AcRefRes}{608}
\newcommand{\AcScenesEight}{90}
\newcommand{\AcScenesSeven}{90}
\newcommand{\AcScenesSix}{150}
\newcommand{\AcSingleScenesEight}{45}
\newcommand{\AcStructNoisySix}{68}
\newcommand{\AcSumDl}{46.5}
\newcommand{\AcSumRaw}{74.2}
\newcommand{\AcTrialsSeven}{15}
\newcommand{\AcTrialsSix}{25}

\title{Algebraic Diversity of Switching Functions}
\author{Mitchell A. Thornton\\[2pt]
\small Darwin Deason Institute for Cyber Security and\\
\small Department of Electrical and Computer Engineering\\
\small Southern Methodist University, Dallas, Texas, USA\\
\small mitch@smu.edu, ORCID 0000-0003-3559-9511}
\date{November 2026}

\begin{document}
\maketitle

\begin{abstract}
Algebraic Diversity generalizes averaging over many observations to averaging one
observation over a group of transformations matched to its structure. This paper
develops the framework for the Boolean cube, the vector space over the binary field, and
its classical groups of input transformations, namely the
translations, the coordinate permutations, the hyperoctahedral group, and the general
linear and affine groups. It determines the commutant of each group action, gives
closed forms for the group-averaged second-order statistic of a single truth table,
and shows how that statistic collapses as the group grows, to the Walsh periodogram
for the translations, to per-weight averages for the hyperoctahedral group, to the
weight of the function up to complementation for the affine group and to a constant for the extended affine
group. A normalizer covariance makes the translation periodogram, canonicalized under
the quotient group, an invariant that refines every average over a group containing
the translations, bent functions are exactly the functions whose translation average
is white, and the group gain of the translations is half their order because every
nonzero translation is an involution. The classical theorems of Haar, Maschke, Schur, Peter and Weyl,
Burnside and Schur and Weyl take explicit forms on the cube, and the Haar transform along an order of
the inputs is the multiplexer tree of Shannon's expansion, matched to the iterated wreath product as
the Walsh transform is matched to the translations. Averages of higher order see only the weight below a
first informative order at which the group stops being homogeneous on sets of points, which is
two for the translations and four for the affine group, where the fourth Walsh moment, the
sum-of-squares indicator and the additive energy coincide and the bent functions are the minima,
and where the spectrum has no zero the bispectrum determines the function up to translation. The
moments of a hashed cell count show the same threshold, and a symmetry found by a sampled census
lowers the variance of spectra estimated from simulation by the orbit size. A matched stabilizer stands in for repeated observations, since
one evaluation per orbit identifies a function, a vote over orbits of $s$ points divides by
$s$ the number of noisy truth tables a target error requires, and the error of the vote
decays exponentially at a rate proportional to the orbit size. The same viewpoint gives the
linear support and the affine hull of a function as matched coordinates, an exact identity for
the variance of exclusive-OR (XOR) hashing whose random mean depends only on the number of
solutions, the exact variance of hashes matched to a sample of solutions, which never exceeds
the variance of a pairwise independent hash, an identity between majority voting over orbits and repeated
evaluation, and the squared-error-optimal shrinkage between an observed bit and its orbit
average, which leads to a vote tested orbit by orbit whose loss on a matched orbit is
bounded by the level of the test. The applications on public benchmark circuits are a census of the
group structure of their outputs, the choice of hashes for approximate model counting
from a few solutions, a hashing counter run end to end with them and one call of a prefix
counter in the form of ApproxMC2, a counter run with a satisfiability solver on the benchmark
formulas of approximate model counting with the affine hull of the solutions certified by the
solver, the decision of
parity-structured formulas by reading implied parities from the spectra of clause blocks,
the refutation of counting formulas modulo 3 and 4 by reading the same windows over the rings of
residues modulo prime powers, with checkable certificates, at about four times the sizes the
tested solvers decide, and with no advantage on the application families of three public
libraries, the separation of equivalence classes by canonical spectra, exhaustively for small functions,
on the cut functions of the ISCAS-85 circuits and on all classes of five-input functions,
where third order leaves a few near-bent classes together, compared with the classical spectral
signatures and, for speed, with the exact canonical forms of ABC, the correction of noisy truth tables by
voting over discovered symmetries, where the tested vote limits the damage of
near-symmetries, and the protection of lookup-table contents against bit upsets by voting
over recorded symmetries, compared with SECDED and triple modular redundancy, against which it
wins only at upset rates far above terrestrial ones. Further applications are the exact reduction of
the variable-order search for multiplexer trees by the stabilizer, where a greedy Haar order is worse
than sifting, a blind group matching algorithm that finds the stabilizer of one truth table in the
hyperoctahedral or the affine group from the periodogram and the spectral signs, with a closed-form
double-commutator eigenvalue problem for near symmetries, the counts of symmetric, partially symmetric
and asymmetric functions, and the compression of symmetric tables as transform coding over GF(2). The
paper is written for readers with one course in digital logic, and each group-theoretic term is
translated into switching terms, from Karnaugh maps as Cayley graphs of the translation group to
hash cells, annihilators and affine hulls, and a table lists which results restate or extend
classical procedures. The same averages grade cryptographic primitives, from the strict avalanche
criterion and S-boxes to collisions and bit-stream tests, and bound the coverage of early
evaluation in self-timed circuits from one Walsh spectrum. For approximate matching and matching
with satisfiability don't cares, spectral bounds on the distance to an NPN class and a search over
the permutations alone, one Walsh transform per coset of the translations, give exact answers up to
two orders of magnitude faster than search at eight inputs.
An analog of the CLEAN algorithm writes a truth table as the XOR of atoms in the fixed spaces of
groups, chosen by description length, and finds such atoms in most of the sampled cut functions of
six to eight inputs. It also peels a bent function, whose translation average is white, into two
atoms with no residue.
All proofs are given in full, and the appendices report the
per-output data.
\end{abstract}


\section{Introduction}
\label{sec:intro}

Many tasks in the design, verification and protection of digital
circuits are estimation or decision problems over the space of switching functions.
Model counting is the problem of counting the input assignments that satisfy a Boolean
formula, which for a circuit designer is the number of minterms of a function given as a product of
clauses. The word model is the logician's name for a satisfying assignment and has nothing to do
with a simulation model. Approximate model counting estimates that number by
intersecting the solution set with random affine subspaces defined by exclusive-OR (XOR)
constraints \cite{valiant1986unique,stockmeyer1983,gomes2006xor,chakraborty2013approxmc}, and
a counter that draws an unlucky hash returns a count far from the truth. The same count
measures what an attacker still does not know about a locked circuit, since the keys consistent
with a set of oracle observations are the solutions of a formula over the key inputs, and
they can be counted exactly when the gate-level factor graph of the observations has small
induced width \cite{thornton2026versionspace}.\iffull{} Satisfiability solvers struggle on formulas
whose constraints are parities, a class that includes equivalence checking of parity-based
circuits and the Tseitin formulas \cite{tseitin1983complexity,urquhart1987hard,bensasson2001narrow}.\fi{}
Technology mappers and rewriting engines compute the functions of thousands of small cuts of
a netlist and match each one against a library under negation and permutation of its inputs
and output \cite{mohnke1993,chai2006matcher,abdollahi2008,mishchenko2006dag}, so the signature
that decides a match must be both inexpensive and discriminating.\iffull{} The configuration memory of
a field-programmable gate array stores each lookup table as a truth table, and radiation
flips its bits \cite{wirthlin2015fpga}, so a design must either recompute a corrupted
table from a stored copy or correct it in place.\fi{} Symmetry detection finds the input
transformations that leave a function unchanged \cite{pomeranz1994symm,mishchenko2003symm}.
Each of these problems carries a group of transformations of the Boolean cube, and each
method for it uses, explicitly or not, statistics that are invariant or covariant under some
group.

A small example shows what choosing the group buys. The majority function of five inputs is
unchanged by every permutation of its inputs. Its $32$ input points fall into \AnMajOrbits\
orbits, one for each input weight, so \AnMajOrbits\ evaluations identify it instead of $32$.
\iffull If its stored truth table is read through independent bit flips of probability $0.05$, a
majority vote over each orbit lowers the expected bit error from $0.05$ to \AnMajBitErrB, and
the probability that the table holds any wrong bit falls from \AnMajRawTableB\ to
\AnMajTableB, without reading the table a second time.\fi{} Its Walsh spectrum is supported on the
odd-weight indices and its squared values take only the levels \AnMajLevels, which is the
signature a matcher compares. Each of these effects is the average of a statistic over a
group matched to the function, and each has a counterpart in the applications of this paper.
Table~\ref{tab:map} lists them.

The squared Walsh spectrum is the periodogram of the function, and since many of the
statistics in this paper are built from it, we recall what a periodogram measures. For a signal
sampled at $N$ instants, the periodogram is the squared magnitude of the discrete Fourier
transform, the energy that the signal carries at each frequency \cite{stoica2005spectral}.
Shifting the signal in time multiplies each Fourier coefficient by a phase of modulus one, so the
periodogram does not change, and the Wiener--Khinchin theorem makes it the Fourier transform of
the autocorrelation, the sum of the products of the signal with a copy of itself shifted by each
lag. A switching function $f$ of $n$ inputs supplies the same three ingredients once a shift in
time is replaced by the complementation of some inputs. Its signal is $F(x)=(-1)^{f(x)}$ over the
$N=2^n$ input points $x$. The sinusoids become the $N$ parity functions $(-1)^{a\cdot x}$, where
$a\cdot x$ is the XOR of the inputs selected by the ones of $a$. The Walsh coefficient
$W_f(a)=\sum_xF(x)(-1)^{a\cdot x}$ equals $N$ minus twice the number of points on which $f$
differs from that parity, so $W_f(a)^2$ is large when $f$ is close to the parity or to its
complement and zero when $f$ agrees with it on exactly half the points
\cite{karpovsky1976,hurst1985spectral}. Complementing the inputs selected by the ones of $c$
multiplies $W_f(a)$ by the sign $(-1)^{a\cdot c}$, so the periodogram does not change, and it is
the Walsh transform of the autocorrelation $r_f(d)=\sum_xF(x)F(x\oplus d)$.

\begin{figure}[t]
\centering
\includegraphics[width=0.62\textwidth]{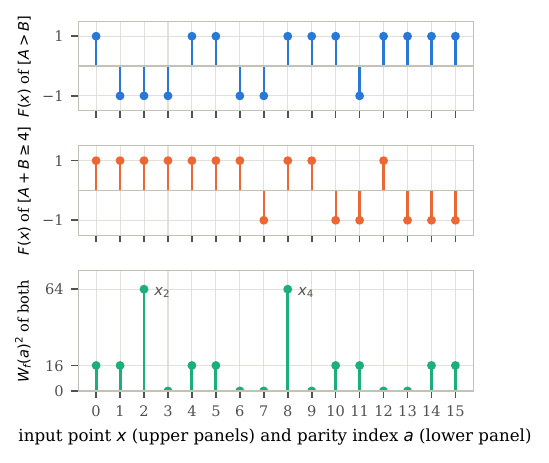}
\caption{The sign vectors $F(x)=(-1)^{f(x)}$ of the two-bit comparator $[A>B]$ (top) and of the
carry out of the two-bit adder $[A+B\ge4]$ (middle), with $A=x_1+2x_2$ and $B=x_3+2x_4$, and
their common periodogram $W_f(a)^2$ (bottom). The input point $x$ and the parity index $a$ are the
integers $\sum_ix_i2^{i-1}$ and $\sum_ia_i2^{i-1}$, and the labels mark the indices of the single
inputs $x_2$ and $x_4$.}
\label{fig:intro}
\end{figure}

Fig.~\ref{fig:intro} shows two circuits on the two-bit numbers $A=x_1+2x_2$ and $B=x_3+2x_4$, the
comparator that outputs $[A>B]$ and the carry out of the adder, $[A+B\ge4]$, where $[\cdot]$ is
one when its argument holds. Their truth tables differ on \IxDiffWord\ of the \IxPoints\ input
points, yet their periodograms coincide. The periodogram takes the values \IxLevels. Its largest
value, \IxTop, sits at the indices of the most significant bits $x_2$ and $x_4$, since the
comparator agrees with $x_2$ and with the complement of $x_4$ on \IxAgreeMSBWord\ of the
\IxPoints\ points, and the high bits decide most comparisons. The periodograms coincide because
$A>B$ exactly when $A+\bar B\ge4$, where $\bar B=3-B$ is the number whose bits are those of $B$
complemented, the identity behind subtraction by adding the complement. The carry circuit is
therefore the comparator with $x_3$ and $x_4$ complemented. Since the autocorrelation sums the
product $F(x)F(x\oplus d)$ over every translation $x\mapsto x\oplus c$ of the input points, the
periodogram is an average over the group of input complementations and cannot see which inputs
are complemented. A matcher that compares the \IxPoints\ periodogram values recognizes the carry
circuit as the comparator with two inputs complemented without trying the \IxPatterns\
complementation patterns, and Section~\ref{sec:normalizer} extends the comparison to permutations
of the inputs.

\begin{table*}[t]
\centering
\caption{The tasks of this paper, the group matched to each, the statistic averaged over it and
what the average provides.}
\label{tab:map}
\setlength{\tabcolsep}{4pt}
\begin{tabular}{@{}p{3.0cm}p{3.3cm}p{4.0cm}p{4.9cm}@{}}
\toprule
Task & Matched group & Averaged statistic & What the average provides\\
\midrule
Hashed model counting & affine group of the input space & cell occupancy of an XOR hash & exact hash variance; matched hashes that avoid implied parities (Sections~\ref{sec:hashvar}, \ref{sec:matched}, \ref{sec:hash}, \ref{sec:counter}, \ref{sec:core})\\
\iffull Parity-structured and counting satisfiability & translations of each block, over GF(2) and over the residues modulo prime powers & indicator and Chrestenson spectra of a block's local solutions & implied parities and congruences, elimination over fields and rings, certificates (Sections~\ref{sec:support}, \ref{sec:parity}, \ref{sec:rings})\\\fi
Boolean matching of cut functions & negations and permutations & Walsh periodogram, canonicalized over the quotient & signatures that separate equivalence classes (Sections~\ref{sec:normalizer}, \ref{sec:classify}, \ref{sec:cuts})\\
\iffull Identification and noisy tables & stabilizer of the function & bit values on each orbit & one evaluation per orbit; voting without repeated reads (Sections~\ref{sec:voting}, \ref{sec:observations}, \ref{sec:noisy}, \ref{sec:lut})\\\fi
\iffull Variable order and blind matching & stabilizer in $B$ or $\AGL(n,2)$; iterated wreath product & periodogram and spectral signs; Haar level energies & verified stabilizer generators; order classes for exact search (Sections~\ref{sec:haar}, \ref{sec:bgm}, \ref{sec:varorder}, \ref{sec:bgmexp})\\
Compression of symmetric tables & exchanges within blocks of inputs & coefficients in the elementary symmetric basis & table of block weights; transform coding over GF(2) (Section~\ref{sec:lutcomp})\\
Decomposition by peeling & symmetries, independence and joint complementation of sets of inputs; one input translation & tables constant on orbits, over GF(2) & mixtures of two groups; Shannon, Davio and Ashenhurst decompositions as peels; the white residue; cross-cofactor consistency as the pointer to the expansion; AD-CLEAN over the cube and its cascades (Sections~\ref{sec:peel}, \ref{sec:adcleancube})\\\fi
\bottomrule
\end{tabular}
\end{table*}

\iffull
Six practical results follow from choosing the group deliberately. First, the variance of a
hashed solution count is the energy of the indicator spectrum of the solution set on the row
space of the hash. Random hashes achieve a mean variance that depends only on the number of
solutions, while the variance over individual random hashes is heavy-tailed on the outputs of
real circuits. A few sampled solutions suffice to choose hashes with a lower mean variance and
fewer catastrophic draws. Drawn uniformly among the row spaces that contain no parity constant
on the sample, such hashes never exceed the variance of a pairwise independent hash, so a
counter that uses them keeps its guarantee, and the parities can also be found from the
formula itself with a few satisfiability calls. A hashing counter run end to end with such
hashes misses the true count less often when it uses a single hash, even when the sampled
solutions are clustered, and so does one call of a prefix counter in the form of ApproxMC2.
Run with a satisfiability solver on the benchmark set of ApproxMC, the counter certifies the
affine hull of the projected solutions and draws its rows on the free coordinates of the hull.
Second, the parities implied by each block of clauses are read from the indicator spectrum of
the block's local solution set, and Gaussian elimination over GF(2) then decides Tseitin
formulas and equivalence checks of parity trees in polynomial time when the blocks are small.
A conflict-driven solver without parity reasoning does not decide them within a
\PsTimeout-second limit beyond small sizes, while a solver with built-in parity reasoning
does. Read over the rings of residues modulo prime powers, the same windows refute partitions
into triples and into 4-sets at about four times the sizes that the tested conflict-driven,
pseudo-Boolean and integer-programming solvers decide, with checkable certificates, while on the
application families of three public libraries they show no advantage. Third, for Boolean matching, the second-order average over the full group separates far
fewer equivalence classes than the Walsh power spectrum canonicalized under the quotient by
the translations, which separates nearly all, and adding the bispectrum separates every class
of four-variable functions under negation and permutation. Fourth, on the cut functions of
the ISCAS-85 circuits the canonical periodogram alone separates every equivalence class of
four, five and six inputs, though the exact NPN algorithms of ABC compute a canonical form
faster. The same decomposition gives two lower bounds on the distance from a function, or from
its values on a care set, to every member of an NPN class, one from the spectrum sorted within each
weight and one from the cofactors, and an exact search that optimizes all negations of a
permutation by one transform. These make approximate matching and matching with satisfiability
don't cares exact and several times to two orders of magnitude faster than search, at sizes where
flipping bits or enumerating completions cannot finish.
Fifth, a matched group stands in for repeated observations. Setting each bit of a noisy
truth table to the majority over its orbit under an exact symmetry group corrects errors at
the rate that repeated evaluation would achieve with as many repetitions as the orbit has
points, so the number of noisy tables a target error requires falls by the orbit size and
the error exponent grows in proportion to it. The symmetries can be discovered from the noisy
table itself when their orbits are large and the minority value is not rarer than the
flipped bits, and testing each orbit against a constant value, with shrinkage toward the
orbit average only as far as the bias and variance of the average allow, reduces the damage
that near-symmetries do to the vote. Sixth, when the symmetries of a lookup-table function are
recorded at mapping time, voting over them corrects upset bits of the stored table in place,
and for the six-input cut functions of the ISCAS-85 circuits it lowers the probability that a
table is wrong to a fraction of its unprotected value. Measured against SECDED and triple
modular redundancy, though, the vote wins only at upset rates far above terrestrial ones, because
most tables have orbits of one or two points.

Five further results connect the framework with logic synthesis. The Haar transform along an order
of the inputs is the multiplexer tree of Shannon's expansion, its coefficients are the imbalances of
the cofactors, and the group matched to it is the iterated wreath product, exactly as the translations
are matched to the Walsh transform. The stabilizer of a function cuts the exact search over variable
orders by the order of its permutation part, while a greedy order built from the Haar energy is worse
than sifting. A blind matching algorithm finds the stabilizer of one truth table in the group of input
negations and permutations or in the affine group from the periodogram and the signs of the Walsh
spectrum, exactly and with verified generators, and its soft second-order form, the double-commutator
eigenvalue problem, ranks a planted symmetry first more often than the exact test still finds a
symmetry once bits are flipped. A table fixed by exchanges of its inputs can be stored by the weights of the blocks
of exchangeable inputs, which is transform coding over GF(2) in the basis of the elementary symmetric
functions. Finally, the iterative peel of group matching carries to the cube, where components combine
by XOR, the white residue of the translations is a bent function, and the Shannon, Davio and
Ashenhurst decompositions are peels of the smallest groups. On the cut functions of real circuits,
mixtures of two symmetry groups are common and grow with the input count, bent residues are rare, and
whether to peel the whole function or its cofactors is settled function by function by description
length. The cross-snapshot consistency of group matching, with the two cofactors of an input as the
two observations, ranks the choice between Shannon and Davio expansion and between the whole function
and its cofactors better than the single-observation measures do.

The same averages reach two further areas. In cryptography the strict avalanche and propagation
criteria are the vanishing of the translation average on Hamming spheres, the collisions of a vector
function at a given difference are an average of the autocorrelations of its components, S-boxes
are graded by the participation ratio of their components, and the serial test of a bit stream is
the energy of a periodogram whose largest coefficient names a linear defect. In self-timed design the
trigger of early evaluation fires on the orbits of the late-input translations on which the master
function is constant, and the Walsh energy on the early inputs bounds its coverage for every
candidate set from one spectrum, with equality whenever the cofactors are constant or balanced.
\fi

\begin{figure}[t]
\centering
\begin{tikzpicture}[
  box/.style={draw, rounded corners=2pt, align=center, font=\footnotesize, inner sep=3pt,
              minimum height=7mm},
  arr/.style={-{Stealth[length=2mm]}, thick}]
\node[box] (obs) {one truth table\\[-1pt] sign vector $F$};
\node[box, right=4mm of obs] (img) {images $P_gF$\\[-1pt] for $g\in G$};
\node[box, right=4mm of img] (avg) {average $R_G(f)$,\\[-1pt] projection onto $\mathcal C(G)$};
\node[box, below=6mm of img] (tr) {matched transform\\[-1pt] diagonalizes $R_G$};
\node[box, below=6mm of avg] (use) {\iffull counting, parities,\\[-1pt] matching, voting\fi};
\draw[arr] (obs) -- (img);
\draw[arr] (img) -- (avg);
\draw[arr] (avg) -- (use);
\draw[arr] (img) -- (tr);
\draw[arr] (tr) -- (use);
\end{tikzpicture}
\caption{The construction of the framework for switching functions. A single truth table and
its images under a group $G$ acting on the inputs are averaged into $R_G(f)$, the orthogonal
projection of $FF^{\mathsf T}$ onto the commutant of $G$. When $G$ is matched to the function,
its character transform diagonalizes the average, and the applications read their
statistics from it.}
\label{fig:pipeline}
\end{figure}
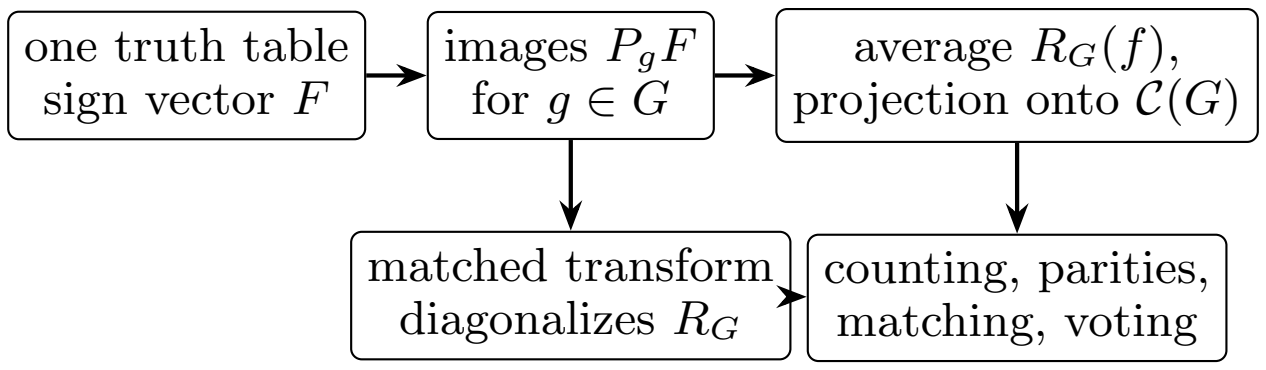

The common thread is Algebraic Diversity, a framework in which group averaging
generalizes temporal averaging \cite{thornton2026ad,thornton2026principles}. An
estimator built from a single observation and its images under a group
acting on the observation has the variance of an average over many independent
observations when the group is matched to the structure of the data. The matched
group fixes the transform that diagonalizes the averaged statistic, and the group gain
is governed by the effective dimension of the commutant of the group action
\cite{thornton2026ad,thornton2026principles,thornton2026unification}. The framework
has been developed for spectral estimation, transform theory, Lie group actions,
optimal group selection by a double-commutator eigenvalue problem, matched
generators of the Karhunen--Lo\`eve transform, covariance shrinkage and quantum state
estimation
\cite{thornton2026continuous,thornton2026gevp,thornton2026generators,thornton2026shrinkage,thornton2026quantum}.
The translation group of $\F_2^n$ is matched to the Walsh--Hadamard transform
\cite{thornton2026unification}. The present paper is the counterpart of the
foundations of the framework for the vector space $\F_2^n$ and the affine group over
GF(2), the setting of switching algebra. Fig.~\ref{fig:pipeline} shows the construction.
\iffull Two results of the framework for signals have direct counterparts here. The group gain,
the factor by which a matched average reduces the variance of an estimate, becomes the
number of independent repetitions that one orbit supplies to a vote, and it adds to the
error exponent in proportion to the orbit size. The reduction in the number of snapshots
that a covariance estimate needs becomes a reduction in the number of evaluations, or of
noisy truth tables, that identify a function or correct it to a target error.\fi

The contributions are the following. The paper determines the commutants of the
classical groups acting on $\F_2^n$, gives closed forms for their group-averaged
second-order statistics, and shows how those statistics collapse as the group grows,
down to the weight up to complementation for the affine group and to a constant for
the extended affine group. It proves a normalizer covariance that makes the
translation periodogram, canonicalized under the quotient group, an invariant that
refines every average over a group containing the translations, and it measures the
refinement exhaustively for small functions and on the cut functions of real circuits.

\iffull
The paper shows that the realized group gain of the translations is half their order,
because every nonzero translation is an involution. It proves that a stabilizer with $r$
orbits identifies a function with $r$ evaluations, that voting over orbits of $s$ points
divides by $s$ the number of noisy truth tables a target error requires, and that the
error of the vote decays exponentially at a rate proportional to the orbit size. It
restates classical facts, the linear structures and linear support of a function, the
affine hull of a set, the characterization of bent functions and the bounds on the
sum-of-squares indicator, as statements about matched group averages, and it uses the
Poisson summation form of the hash variance to separate the structure-blind random mean from
the variance of an individual hash and to give the exact variance of a hash matched to a
sample of solutions. For voting on noisy tables it derives the
squared-error-optimal shrinkage between each observed bit and its orbit average, with an
unbiased estimate of the weight, and a tested vote whose loss on a matched orbit is bounded
by the level of the test. The classical sources are cited where each fact appears.
\fi

The paper also orders the group averages by the power of the truth table they average. It
counts the parameters of the average of order $\varkappa$ with the lemma of Burnside and proves
that a permutation group sees only the weight until the order at which it stops being homogeneous
on sets of input points, which is two for the translations and the hyperoctahedral group and four
for the affine group. At that order the extended affine average is the fourth Walsh moment, which
is the sum-of-squares indicator, the additive energy and the Gowers $U^2$ norm under three names,
and the bent functions are its minima. Where the Walsh spectrum has no zero, which includes every
bent function, the bispectrum determines the function up to complementation of inputs, and averages
of order up to twice the group order separate every orbit. The same threshold governs the moments
of a hashed cell count, of which the first three depend only on the number of solutions and the
fourth also on their additive energy. Table~\ref{tab:ladder} places the classical statistics of
switching theory on this scale.

\iffull
The applications in Section~\ref{sec:apps} are evaluated on public benchmark circuits. They cover hashed counting, both the variance of
individual hashes and a counter run end to end, parity-structured satisfiability,
classification of all functions of three and four variables and of the cut functions of the
ISCAS-85 circuits, error correction of noisy truth tables, and the protection of lookup-table
contents against bit upsets. They also peel group structure from truth tables, up to an analog of the
CLEAN algorithm that writes a table as the XOR of atoms in fixed spaces of groups and draws the
result as a cascade (Section~\ref{sec:adcleancube}). The selection steps in them need no training data and no neural
network.
\fi

Section~\ref{sec:prelim} fixes notation. Section~\ref{sec:theory} develops the
theory. Section~\ref{sec:apps} applies it. Section~\ref{sec:related} discusses related work,
and Section~\ref{sec:limits} states the limitations and open problems.

\section{Switching Function Spectra and Symmetry}
\label{sec:prelim}

A switching function of $n$ inputs is a map $f:\F_2^n\to\F_2$, where $\F_2=\{0,1\}$
is the field with two elements. We write $N=2^n$, identify $x\in\F_2^n$ with the
integer whose binary digits are its coordinates, and write $a\cdot x=\sum_i a_ix_i$
over $\F_2$ for the inner product. The truth vector of $f$ is $t\in\{0,1\}^N$ with
$t(x)=f(x)$, the on-set is $S=\{x:f(x)=1\}$, the weight is $|S|$, and the sign vector is
$F(x)=(-1)^{f(x)}$, a vector in $\{\pm1\}^N$.

In the vocabulary of logic design a point $x\in\F_2^n$ is a fully specified input assignment,
one row of the truth table without its output value, and $S$ is the ON-set, with the OFF-set as its
complement. An incompletely specified function also has a don't-care set $D$ and is defined only on
the rows outside $D$. A group is a set of transformations of the rows that contains the identity
and the inverse of each of its members and is closed under composition. The translations
$x\mapsto x\oplus c$, one for each $c\in\F_2^n$, form the simplest example, the translation group
$T$. The translation by $c$ complements the inputs where $c$ has a one, two translations compose to
the translation by the XOR of their vectors, and each translation is its own inverse. Applied to a
function, the translation by $c$ changes the polarity of the literals of the inputs in $c$. A
product term keeps its variables and complements those literals, and the fixed-polarity
Reed--Muller expansion of $f(x\oplus c)$ is the expansion of $f$ with polarity vector $c$. The
orbit of a row under a group is the set of rows that the group can reach from it. Under $T$ every
row reaches every other one, and under the translations that complement only the inputs of a set
$L$ the orbit of $x$ is the subcube of rows that agree with $x$ outside $L$, the rows covered by
the product term whose literals are the inputs outside $L$ at their values in $x$. More generally a
coset of a subgroup of $T$ is an affine subspace, the set of rows that satisfy a system of XOR
constraints. The Hamming weight of a row is its number of ones, the Hamming distance of two rows
is the weight of their XOR, the orbits of $S_n$ are the classes of rows of equal weight, and the
transformations of $B$ preserve the distance between every pair of rows. For functions the
weight is $|S|=(N-W_f(0))/2$, and the distance of $f$ from $g$ is $(N-\sum_xF(x)G(x))/2$, where
$G$ is the sign vector of $g$. In particular, the distance from $f$ to the affine function
$a\cdot x\oplus\nu$ is $(N-(-1)^\nu W_f(a))/2$, so the largest $|W_f(a)|$ measures how close $f$
comes to a single XOR of inputs.

A Karnaugh map \cite{karnaugh1953map} is a picture of these definitions. It draws the truth table of four
inputs on a torus, with rows and columns in Gray-code order that wraps around at the edges, so two
cells are neighbors exactly when their rows differ by a single translation $e_i$, the unit vector
of input $i$, so the map is the Cayley graph of $T$ generated by the unit vectors.\iffull{}
Fig.~\ref{fig:kmap} shows the four neighbors of a cell and two of the groupings below.\fi{} Grouping
$2^k$ adjacent cells into a rectangle, wrapping around the edges if needed, is choosing a coset of a
subgroup of $T$ spanned by $k$ unit vectors, a subcube on which $f$ is one, and a prime implicant
is a maximal such coset. Allowing cosets of every subgroup of $T$, the affine subspaces, enlarges
the classical groupings to the solution sets of XOR constraints, which reappear below as the cells
of XOR hashes and as affine hulls.
\iffull
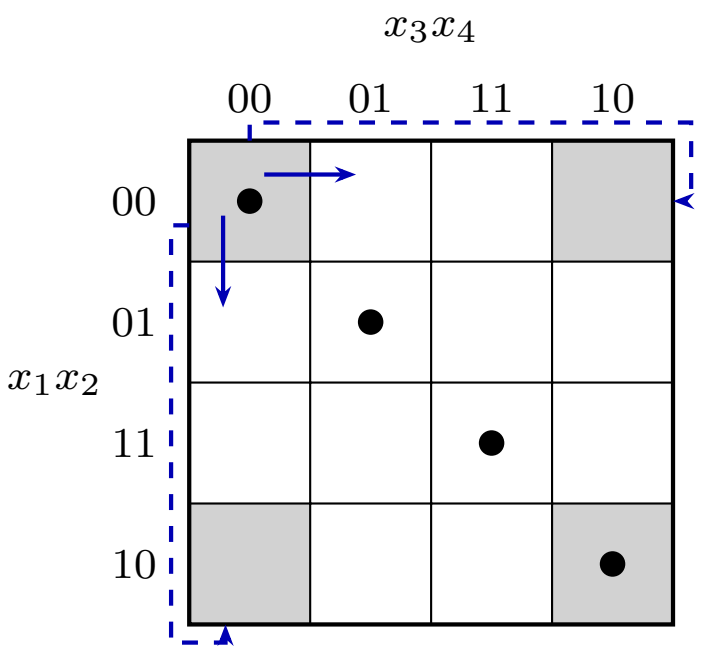
\begin{figure}[t]
\centering
\begin{tikzpicture}[font=\scriptsize, x=0.8cm, y=0.8cm]
\foreach \c [count=\i from 0] in {00,01,11,10} {
  \node at (\i+0.5, 4.35) {\c};
  \node[anchor=east] at (-0.1, 3.5-\i) {\c};
}
\node[anchor=south] at (2, 4.65) {$x_3x_4$};
\node[anchor=east] at (-0.55, 2) {$x_1x_2$};
\foreach \r/\cc in {0/0,0/3,3/0,3/3} \fill[gray!35] (\cc, 3-\r) rectangle ++(1,1);
\foreach \k in {0,1,2,3} \fill (\k+0.5, 3.5-\k) circle (2.4pt);
\draw[step=1] (0,0) grid (4,4);
\draw[thick] (0,0) rectangle (4,4);
\draw[-{Stealth[length=4pt]}, thick, blue!70!black] (0.62,3.72) -- (1.38,3.72);
\draw[-{Stealth[length=4pt]}, thick, blue!70!black] (0.28,3.38) -- (0.28,2.62);
\draw[-{Stealth[length=4pt]}, thick, dashed, blue!70!black] (0.5,4.0) -- (0.5,4.15) -| (4.15,3.5) -- (4.0,3.5);
\draw[-{Stealth[length=4pt]}, thick, dashed, blue!70!black] (0.0,3.3) -- (-0.15,3.3) |- (0.3,-0.15) -- (0.3,0.0);
\end{tikzpicture}
\caption{A Karnaugh map as the Cayley graph of the translation group. Neighboring cells differ by
one unit translation, including across the edges. The arrows show the four neighbors of the cell
$0000$, two of them (dashed) reached by wrapping around. The four gray corners form a subcube, a coset
of the subgroup spanned by $e_1$ and $e_3$ and the classical wraparound grouping. The four dots form
a coset of the subgroup spanned by $e_1\oplus e_3$ and $e_2\oplus e_4$, the rows that satisfy two
XOR constraints, which is not a rectangle of the map.}
\label{fig:kmap}
\end{figure}

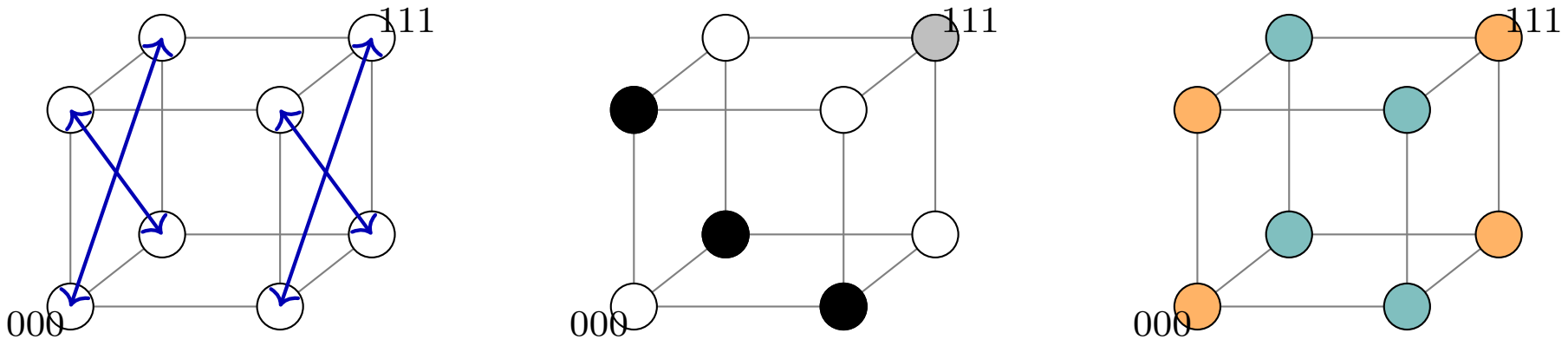
\begin{figure}[t]
\centering
\begin{tikzpicture}[font=\scriptsize, x=1cm, y=1cm,
  v/.style={circle, draw, inner sep=1pt, minimum size=10pt}]
\def\cubeat#1#2{%
  \foreach \a in {0,1} \foreach \b in {0,1} \foreach \c in {0,1} {
    \coordinate (#2\a\b\c) at ($(#1)+(1.6*\a+0.7*\b, 1.5*\c+0.55*\b)$);
  }
  \foreach \b in {0,1} \foreach \c in {0,1} \draw[gray] (#20\b\c) -- (#21\b\c);
  \foreach \a in {0,1} \foreach \c in {0,1} \draw[gray] (#2\a0\c) -- (#2\a1\c);
  \foreach \a in {0,1} \foreach \b in {0,1} \draw[gray] (#2\a\b0) -- (#2\a\b1);
}
\cubeat{0,0}{A}
\foreach \a in {0,1} \foreach \b in {0,1} \foreach \c in {0,1}
  \node[v, fill=white] at (A\a\b\c) {};
\foreach \a in {0,1} {
  \draw[<->, thick, blue!70!black] (A\a00) -- (A\a11);
  \draw[<->, thick, blue!70!black] (A\a01) -- (A\a10);
}
\node[anchor=north] at ($(A000)+(1.2,-0.25)$) {(a) translation by $c=011$};
\cubeat{4.3,0}{B}
\foreach \a in {0,1} \foreach \b in {0,1} \foreach \c in {0,1}
  \node[v, fill=white] at (B\a\b\c) {};
\foreach \p in {100,010,001} \node[v, fill=black] at (B\p) {};
\node[v, fill=gray!50] at (B111) {};
\node[anchor=north] at ($(B000)+(1.2,-0.25)$) {(b) hull of $\{100,010,001\}$};
\cubeat{8.6,0}{C}
\foreach \a/\b in {0/0,1/1} \foreach \c in {0,1} \node[v, fill=orange!60] at (C\a\b\c) {};
\foreach \a/\b in {0/1,1/0} \foreach \c in {0,1} \node[v, fill=teal!50] at (C\a\b\c) {};
\node[anchor=north] at ($(C000)+(1.2,-0.25)$) {(c) cells of $x_1\oplus x_2$};
\node[anchor=west] at ($(A000)+(-0.35,-0.05)$) {};
\node[anchor=north east, inner sep=1pt] at (A000) {$000$};
\node[anchor=south west, inner sep=1pt] at (A111) {$111$};
\node[anchor=north east, inner sep=1pt] at (B000) {$000$};
\node[anchor=south west, inner sep=1pt] at (B111) {$111$};
\node[anchor=north east, inner sep=1pt] at (C000) {$000$};
\node[anchor=south west, inner sep=1pt] at (C111) {$111$};
\end{tikzpicture}
\caption{The 3-cube with rows written $x_1x_2x_3$, $x_1$ to the right, $x_2$ into the page and $x_3$
up. (a) The translation by $c=011$ complements $x_2$ and $x_3$ and pairs the rows as the arrows show, and
each pair is an orbit of the subgroup $\{0,c\}$. (b) The affine hull of the three rows of weight one
(black) adds their XOR $111$ (gray), giving the set of rows that satisfy the only XOR constraint the
three share, $x_1\oplus x_2\oplus x_3=1$. (c) The XOR gate $x_1\oplus x_2$ is a hash with two cells,
the rows where it outputs 0 (orange) and 1 (teal), each a coset of the kernel spanned by $110$ and
$001$.}
\label{fig:cubes}
\end{figure}
\fi

The Walsh spectrum of $f$ is $W_f(a)=\sum_x(-1)^{f(x)+a\cdot x}$ and the indicator
spectrum of a set $S$ is $\Sh(a)=\sum_{x\in S}(-1)^{a\cdot x}$. The two are related by
$W_f(a)=N\,[a=0]-2\Sh(a)$, where $[\cdot]$ is one when its argument holds and zero
otherwise. With the Sylvester--Hadamard matrix $H$ of order $N$, $H_{a,x}=(-1)^{a\cdot
x}$, we have $W_f=HF$ and $H^2=NI$. Parseval's identity reads $\sum_aW_f(a)^2=N^2$ and
$\sum_a\Sh(a)^2=N|S|$. The autocorrelation is $r_f(d)=\sum_xF(x)F(x\oplus d)$, and the
dyadic Wiener--Khinchin relation is $W_f^2=Hr_f$ \cite{karpovsky1976,hurst1985spectral}.

\begin{figure}[t]
\centering
\begin{tikzpicture}[
  g/.style={draw, rounded corners=2pt, align=center, font=\footnotesize, inner sep=2.5pt,
            minimum width=17mm},
  l/.style={thick}]
\node[g] (T) at (0,0) {$T$\\[-1pt] $\dim\mathcal C=N$};
\node[g] (S) at (3.6,0) {$S_n$\\[-1pt] $\dim\mathcal C=\binom{n+3}{3}$};
\node[g] (B) at (0,1.45) {$B=T\rtimes S_n$\\[-1pt] $\dim\mathcal C=n+1$};
\node[g] (GL) at (3.6,1.45) {$\GL(n,2)$\\[-1pt] $\dim\mathcal C=5$};
\node[g] (AGL) at (1.8,2.9) {$\AGL(n,2)$\\[-1pt] $\dim\mathcal C=2$};
\node[g] (EA) at (1.8,4.25) {$\EA$\\[-1pt] $\dim\mathcal C=1$};
\draw[l] (T) -- (B); \draw[l] (S) -- (B); \draw[l] (S) -- (GL);
\draw[l] (B) -- (AGL); \draw[l] (GL) -- (AGL); \draw[l] (AGL) -- (EA);
\node[font=\scriptsize, align=left, anchor=west] at (-1.3,-0.75) {keeps $W_f^2$};
\node[font=\scriptsize, align=left, anchor=west] at (4.3,2.9) {keeps $|S|$ up to\\[-1pt] complementation};
\end{tikzpicture}
\caption{The groups acting on the inputs, ordered by inclusion from bottom to top, with the
dimension of the commutant of each for $n\ge2$. The second-order average over a larger group
keeps less of the function, from the full Walsh periodogram for $T$ to the weight for
$\AGL(n,2)$ and nothing for $\EA$.}
\label{fig:lattice}
\end{figure}
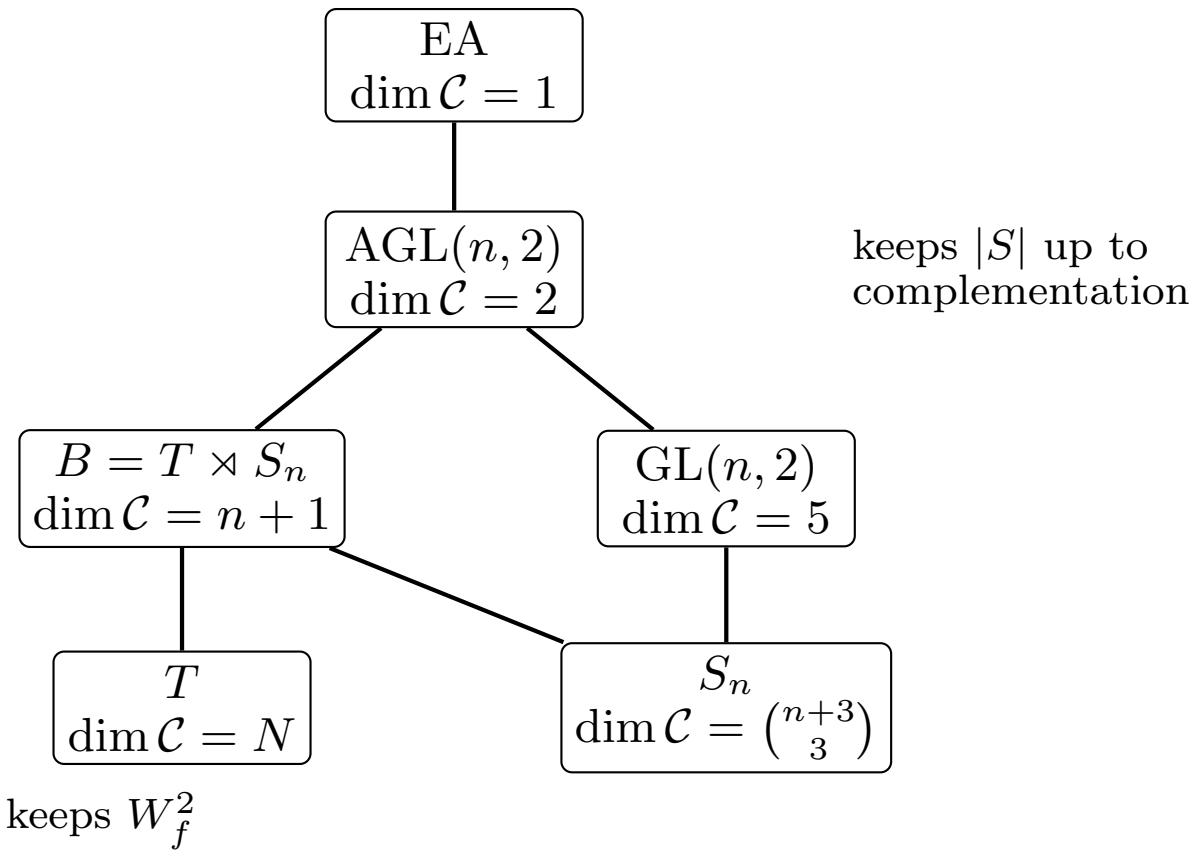

The groups considered act on the inputs. The translation group $T=\{x\mapsto x\oplus
c\}$ is isomorphic to $\mathbb{Z}_2^n$ and realizes input complementation. The
symmetric group $S_n$ permutes coordinates, the hyperoctahedral group $B=T\rtimes S_n$
realizes negation and permutation of inputs, the general linear group $\GL(n,2)$ acts
by $x\mapsto Ax$ with $A$ invertible over $\F_2$, and the affine group
$\AGL(n,2)=T\rtimes\GL(n,2)$ acts by $x\mapsto Ax\oplus c$. A group element $g$ acts on a
function by $(g\cdot f)(x)=f(g^{-1}x)$, so that $f\circ g=g^{-1}\cdot f$, and on truth and
sign vectors by a permutation matrix $P_g$. Output negation, $f\mapsto f\oplus1$, sends
$F$ to $-F$. The extended affine group $\EA$ adds affine functions to the output,
$f\mapsto f\circ g\oplus b\cdot x\oplus\nu$ with $b\in\F_2^n$ and $\nu\in\F_2$, and acts on
$F$ by signed permutation matrices. These groups and
their classification problems go back to
\cite{golomb1959,harrison1964,edwards1975,lechner1971,maiorana1991}, and their invariants are
standard tools of spectral logic design
\cite{karpovsky1976,hurst1985spectral,thornton2001spectral,thornton2014vsim,carlet2021book}.
Fig.~\ref{fig:lattice} shows how the groups contain one another, together with the dimension
of the commutant of each, which Section~\ref{sec:commutant} derives.
\iffull
Table~\ref{tab:groups} defines every group of the paper twice, once in plain words and once
formally. Three objects appear there that the rest of the paper also uses. The symmetric group $S_N$
rearranges the $N$ points of the cube without any regard for their coordinates, so the only
property of a function that it keeps is the weight. The iterated wreath product
$\Wr_n=\mathbb Z_2\wr\cdots\wr\mathbb Z_2$ of an order of the inputs is the group of symmetries of
the complete decision tree that tests the inputs in that order, generated by exchanging the two
subtrees below any node. Section~\ref{sec:haar} shows that it is matched to the Haar transform as
$T$ is matched to the Walsh transform \cite{foote2000wreath}. The stabilizer of $f$ in a group $G$ is
$\Stab_G(f)=\{g\in G:f\circ g=f\}$, the transformations in $G$ that leave $f$ unchanged, and the
orbit of a point $x$ under it is the set of points $gx$ on which every function with that
stabilizer must take the value $f(x)$.

\begin{table*}[t]
\centering
\caption{The groups of the paper. Each acts on the inputs $x\in\F_2^n$, except NPN and $\EA$,
which also change the output. $Q$ is a permutation matrix, $A$ an invertible matrix over $\F_2$,
$c,b\in\F_2^n$ and $\nu\in\F_2$. The order of $\GL(n,2)$ is $\prod_{i=0}^{n-1}(2^n-2^i)$.}
\label{tab:groups}
\small
\renewcommand{\arraystretch}{1.15}
\setlength{\tabcolsep}{4pt}
\begin{tabular}{@{}p{0.11\textwidth}p{0.33\textwidth}p{0.31\textwidth}p{0.16\textwidth}@{}}
\toprule
group & in words & formally & order\\
\midrule
$T$ & complement any subset of the inputs & $\{x\mapsto x\oplus c\}\cong\mathbb Z_2^n$ & $2^n$\\
$S_n$ & relabel the inputs & $\{x\mapsto Qx\}$ & $n!$\\
$B$ (hyperoctahedral) & relabel and complement inputs; the symmetries of the $n$-dimensional cube & $\{x\mapsto Qx\oplus c\}=T\rtimes S_n$ & $2^nn!$\\
$\GL(n,2)$ & map the inputs to $n$ independent parities of inputs; the all-zero input stays fixed & $\{x\mapsto Ax\}$ & $|\GL(n,2)|$\\
$\AGL(n,2)$ & the same, followed by complementing inputs & $\{x\mapsto Ax\oplus c\}=T\rtimes\GL(n,2)$ & $2^n|\GL(n,2)|$\\
NPN & relabel and complement inputs, and complement the output & $\{f\mapsto f\circ g\oplus\nu:g\in B\}$ & $2^{n+1}n!$\\
$\EA$ & an affine map of the inputs, then add an affine function to the output & $\{f\mapsto f\circ g\oplus b\cdot x\oplus\nu:$\newline\hspace*{1em}$g\in\AGL(n,2)\}$ & $2^{2n+1}|\GL(n,2)|$\\
$S_N$ & any rearrangement of the $N$ points & all permutations of $\F_2^n$ & $N!$\\
$\Wr_n$ & for a fixed order of the inputs, exchange the two halves of the table below any node of the decision tree & $\Wr_1=\mathbb Z_2$, $\Wr_n=(\Wr_{n-1}\times\Wr_{n-1})\rtimes\mathbb Z_2$ & $2^{N-1}$\\
$\Stab_G(f)$ & the elements of $G$ that leave $f$ unchanged & $\{g\in G:f\circ g=f\}$ & divides $|G|$\\
\bottomrule
\end{tabular}
\end{table*}
\fi

\begin{example}[Majority of five inputs]
\label{ex:maj}
Let $f(x)=1$ exactly when at least three of the five coordinates of $x$ are one. Then
$N=32$, $|S|=16$, and $f$ is fixed by every coordinate permutation, so its stabilizer contains
$S_5$, whose orbits are the \AnMajOrbits\ weight classes of sizes $1,5,10,10,5,1$. Since
$f(x\oplus\one)=f(x)\oplus1$, the Walsh spectrum vanishes on the even-weight indices, and its
squared values take only the levels \AnMajLevels. The translation periodogram is therefore
constant on each weight class of the index, the form that Proposition~\ref{prop:collapse}
gives for every function fixed by $S_n$.
\end{example}

\iffull
\paragraph*{How many functions are symmetric}
Fig.~\ref{fig:symcircuits} shows three functions of four inputs, each built from three gates of two
inputs. The XOR tree (a) is unchanged by every permutation of its inputs, so it is
totally symmetric, with stabilizer of order \SyStabTotal\ in $S_4$. The function (b) is unchanged by
exchanging $x_1$ with $x_2$, $x_3$ with $x_4$, or the two pairs, a stabilizer of order
\SyStabPartial, so it is partially symmetric. The function (c) is changed by every exchange of two
inputs and by every other permutation, so its stabilizer is trivial. Following
\cite{rosenkrantz2019ncf} we call such functions strongly asymmetric, and keep the classical term
totally asymmetric for functions that no exchange of two inputs fixes. The two notions differ, since
a function can be fixed by a cyclic shift of three inputs without being fixed by any exchange.
Table~\ref{tab:symcount} counts both. The classical counts follow in closed form from M\"obius
inversion over the set partitions of the inputs, because a set of exchanges generates the product
of the symmetric groups on the connected components of its graph, whose orbits on $\F_2^n$ number
$\prod(b+1)$ over the component sizes $b$. The group counts follow from M\"obius inversion over the
lattice of subgroups of $S_n$, the \SySubgroupsSix\ subgroups of $S_6$ for $n=6$, with
$2^{\text{number of orbits}}$ functions fixed by each subgroup. Symmetry becomes rare quickly. At
$n=4$ about two thirds of all functions are strongly asymmetric, and from $n=7$ on the share of
functions with any symmetric pair is below $10^{-8}$. In the group of input negations and
permutations the trivial stabilizer is rarer, \SyBTrivFour\ of the $65{,}536$ functions of four
inputs.

\begin{figure}[t]
\centering
\begin{minipage}[c]{0.56\columnwidth}
\centering
\begin{tikzpicture}[circuit logic US, every circuit symbol/.style={thick}, font=\scriptsize,
  inp/.style={inner sep=1pt, anchor=east}]
\foreach \gA/\gB/\lab/\yy in {xor/xor/a/0, and/xor/b/-2.2} {
  \begin{scope}[yshift=\yy cm]
  \node[\gA\space gate, inputs=nn] (g1\lab) at (1.2,0.55) {};
  \node[\gA\space gate, inputs=nn] (g2\lab) at (1.2,-0.55) {};
  \node[\gB\space gate, inputs=nn] (g3\lab) at (2.4,0) {};
  \node[inp] (i1\lab) at (0.2,0.85) {$x_1$};
  \node[inp] (i2\lab) at (0.2,0.35) {$x_2$};
  \node[inp] (i3\lab) at (0.2,-0.35) {$x_3$};
  \node[inp] (i4\lab) at (0.2,-0.85) {$x_4$};
  \draw (i1\lab.east) -- ++(0.35,0) |- (g1\lab.input 1);
  \draw (i2\lab.east) -- ++(0.35,0) |- (g1\lab.input 2);
  \draw (i3\lab.east) -- ++(0.35,0) |- (g2\lab.input 1);
  \draw (i4\lab.east) -- ++(0.35,0) |- (g2\lab.input 2);
  \draw (g1\lab.output) -- ++(0.25,0) |- (g3\lab.input 1);
  \draw (g2\lab.output) -- ++(0.25,0) |- (g3\lab.input 2);
  \draw (g3\lab.output) -- ++(0.3,0) node[right] {(\lab)};
  \end{scope}
}
\begin{scope}[yshift=-4.5cm]
\node[and gate, inputs=in] (c1) at (1.1,0.6) {};
\node[or gate, inputs=nn] (c2) at (2.0,0.2) {};
\node[and gate, inputs=nn] (c3) at (2.9,-0.2) {};
\node[inp] (j1) at (0.2,0.85) {$x_1$};
\node[inp] (j2) at (0.2,0.35) {$x_2$};
\node[inp] (j3) at (0.2,-0.15) {$x_3$};
\node[inp] (j4) at (0.2,-0.65) {$x_4$};
\draw (j1.east) -- ++(0.35,0) |- (c1.input 1);
\draw (j2.east) -- ++(0.35,0) |- (c1.input 2);
\draw (c1.output) -- ++(0.2,0) |- (c2.input 1);
\draw (j3.east) -- ++(1.25,0) |- (c2.input 2);
\draw (c2.output) -- ++(0.2,0) |- (c3.input 1);
\draw (j4.east) -- ++(2.15,0) |- (c3.input 2);
\draw (c3.output) -- ++(0.3,0) node[right] {(c)};
\end{scope}
\end{tikzpicture}
\end{minipage}\hfill
\begin{minipage}[c]{0.4\columnwidth}
\centering\scriptsize
\begin{tabular}{@{}c@{\hspace{6pt}}ccc@{}}
\toprule
$x_1x_2x_3x_4$ & (a) & (b) & (c)\\
\midrule
0000 & 0 & 0 & 0\\
1000 & 1 & 0 & 0\\
0100 & 1 & 0 & 0\\
1100 & 0 & 1 & 0\\
0010 & 1 & 0 & 0\\
1010 & 0 & 0 & 0\\
0110 & 0 & 0 & 0\\
1110 & 1 & 1 & 0\\
0001 & 1 & 0 & 0\\
1001 & 0 & 0 & 0\\
0101 & 0 & 0 & 1\\
1101 & 1 & 1 & 0\\
0011 & 0 & 1 & 1\\
1011 & 1 & 1 & 1\\
0111 & 1 & 1 & 1\\
1111 & 0 & 0 & 1\\
\bottomrule
\end{tabular}

\end{minipage}
\caption{Three functions of four inputs built from three gates of two inputs, with their truth
tables. (a) $x_1\oplus x_2\oplus x_3\oplus x_4$, totally symmetric, stabilizer of order
\SyStabTotal\ in $S_4$. (b) $(x_1\wedge x_2)\oplus(x_3\wedge x_4)$, partially symmetric, stabilizer
of order \SyStabPartial, and bent. (c) $((\neg x_1\wedge x_2)\vee x_3)\wedge x_4$, strongly
asymmetric, stabilizer of order \SyStabNone.}
\label{fig:symcircuits}
\end{figure}

\begin{table*}[t]
\centering
\caption{Switching functions of $n$ inputs by symmetry. The classical definition counts input exchanges $x_i\leftrightarrow x_j$ that fix $f$: totally symmetric (all of them), partially symmetric (some, not all) and totally asymmetric (none). The group definition uses the stabilizer of $f$ in $S_n$: a proper nontrivial stabilizer (partial) and the trivial one (strongly asymmetric). Exact counts for $n\le6$; for $n\ge7$ the shares of all $2^{2^n}$ functions, where the stabilizer counts were not computed. Every count for $n\le4$ agrees with enumeration of all functions.}
\label{tab:symcount}
\setlength{\tabcolsep}{3pt}
\scriptsize
\begin{tabular}{@{}rrrrrr@{}}
\toprule
& \multicolumn{3}{c}{exchanges of two inputs} & \multicolumn{2}{c}{stabilizer in $S_n$}\\
\cmidrule(lr){2-4}\cmidrule(l){5-6}
$n$ & totally symmetric & partially symmetric & totally asymmetric & partial & trivial\\
\midrule
2 & 8 & 0 & 8 & 0 & 8\\
3 & 16 & 144 & 96 & 144 & 96\\
4 & 32 & 21{,}152 & 44{,}352 & 22{,}496 & 43{,}008\\
5 & 64 & 162{,}640{,}320 & 4{,}132{,}326{,}912 & 174{,}278{,}592 & 4{,}120{,}688{,}640\\
6 & 128 & 4{,}218{,}864{,}590{,}273{,}408 & 18{,}442{,}525{,}209{,}119{,}278{,}080 & 4{,}266{,}181{,}297{,}816{,}448 & 18{,}442{,}477{,}892{,}411{,}735{,}040\\
\midrule
7 & $7.50\times10^{-37}$ & $4.80\times10^{-9}$ & $1-4.80\times10^{-9}$ & -- & --\\
8 & $4.42\times10^{-75}$ & $1.51\times10^{-18}$ & $1-1.51\times10^{-18}$ & -- & --\\
9 & $7.60\times10^{-152}$ & $1.05\times10^{-37}$ & $1-1.05\times10^{-37}$ & -- & --\\
10 & $1.13\times10^{-305}$ & $3.88\times10^{-76}$ & $1-3.88\times10^{-76}$ & -- & --\\
11 & $1.26\times10^{-613}$ & $4.10\times10^{-153}$ & $1-4.10\times10^{-153}$ & -- & --\\
12 & $7.84\times10^{-1230}$ & $3.67\times10^{-307}$ & $1-3.67\times10^{-307}$ & -- & --\\
\bottomrule
\end{tabular}
\end{table*}

The examples connect symmetry with the bent functions of Corollary~\ref{cor:bent} below. Function
(b) is bent, and so are many functions that look less regular. Among the \SyBentCount\ bent
functions of four inputs the orders of the stabilizers in $S_4$ are distributed as \SyBentHist, so
most bent functions are strongly asymmetric. The four totally symmetric ones are
$\mathsf e_2$, $\mathsf e_2\oplus\mathsf e_1$ and their complements, where $\mathsf e_k$ is the
XOR of all products of $k$ distinct inputs (the elementary symmetric function of degree $k$), and
Savick\'y proved that these four are the only symmetric bent functions for every even $n$
\cite{savicky1994bent}. Symmetry and bentness are therefore independent properties. A flat Walsh
power spectrum says nothing about how many input permutations fix the function, which is one more
reason why second-order statistics cannot classify bent functions.
\fi

\section{Group Averages Organize the Structure of Switching Functions}
\label{sec:theory}

\subsection{The Group Average of One Truth Table Is a Projection onto a Commutant}
\label{sec:commutant}

The group-averaged estimator built from a single observation $F$ is
\begin{equation}
\label{eq:RG}
\RG(f)=\frac{1}{|G|}\sum_{g\in G}P_gFF^{\mathsf T}P_g^{\mathsf T}.
\end{equation}
It is the Reynolds operator of $G$ applied to the rank-one matrix $FF^{\mathsf T}$,
and therefore the orthogonal projection, in the Frobenius inner product, of
$FF^{\mathsf T}$ onto the commutant $\mathcal{C}(G)=\{M:P_gM=MP_g\ \forall g\}$
\cite{thornton2026ad,sturmfels2008,diaconis1988}. The matrix $\RG(f)$ is a complete
second-order invariant, since $\RG(g\cdot f)=\RG(f)$ for every $g\in G$, and it is
the matrix the framework uses in place of a sample covariance.

\begin{proposition}[Commutant lattice]
\label{prop:commutant}
For a permutation group $G$ on $\F_2^n$, $\dim\mathcal{C}(G)$ equals the number of
orbits of $G$ on ordered pairs $(x,y)$. For $n\ge2$ the dimensions are $N$ for $T$,
$\binom{n+3}{3}$ for $S_n$, $n+1$ for $B$, $5$ for $\GL(n,2)$ and $2$ for $\AGL(n,2)$.
For the signed permutation action of $\EA$ the commutant consists of the multiples of
the identity.
\end{proposition}

\begin{proof}
The indicator matrices of the orbits on pairs span the commutant of a permutation
representation and are linearly independent. The translations act regularly, so the
orbit of $(x,y)$ is determined by $x\oplus y$, which gives $N$ orbits. Under $S_n$
the orbit of $(x,y)$ is determined by how many coordinates take each of the four
values $(x_i,y_i)$, a multiset of size $n$ from four types. Under $B$ only the
Hamming distance of $x$ and $y$ survives, which takes $n+1$ values. Under $\GL(n,2)$ the
orbits are $\{(0,0)\}$, $\{(0,y):y\ne0\}$, $\{(x,0):x\ne0\}$, $\{(x,x):x\ne0\}$ and the
linearly independent pairs, since $\GL(n,2)$ is transitive on ordered bases of
two-dimensional subspaces. $\AGL(n,2)$ is two-transitive, which leaves the diagonal
and its complement. The signed permutation matrices of $\EA$ include the translations
$P_c$ and the modulations $Q_b=\mathrm{diag}((-1)^{b\cdot x})$, and the $N^2$
products $P_cQ_b$ form a basis of all $N\times N$ matrices, so only multiples of the
identity commute with all of them.
\end{proof}

The effective dimension $\deff(G)=N^2/\dim\mathcal{C}(G)$ of the framework
\cite{thornton2026ad} counts how many entries of $FF^{\mathsf T}$ are averaged into each
parameter of $\RG$. Table~\ref{tab:commutant} lists the group orders and commutant
dimensions for small $n$, obtained by enumerating the groups and counting orbits on
pairs. The translation group is the only one in the lattice whose commutant has the
dimension $N$ of the space, which is the regular case in which $\deff=|G|$.
\ifarxiv The orders for $n=5$ of $\GL$ and $\AGL$ are $9{,}999{,}360$ and
$319{,}979{,}520$, too many elements to enumerate, and their commutant dimensions
follow from the two-transitivity argument of the proof.\fi

\begin{table}[t]
\centering
\caption{Commutant dimensions (orbits on ordered pairs) on $\F_2^n$ and group orders, by
enumeration. $T$ has order $N$ and $S_n$ order $n!$. The extended affine group has a
one-dimensional commutant for every $n$.}
\label{tab:commutant}
\setlength{\tabcolsep}{4pt}
\begin{tabular}{@{}r r r rr rr rr@{}}
\toprule
 & $T$ & $S_n$ & \multicolumn{2}{c}{$B$} & \multicolumn{2}{c}{$\GL(n,2)$} & \multicolumn{2}{c}{$\AGL(n,2)$}\\
\cmidrule(lr){2-2}\cmidrule(lr){3-3}\cmidrule(lr){4-5}\cmidrule(lr){6-7}\cmidrule(lr){8-9}
$n$ & dim & dim & order & dim & order & dim & order & dim\\
\midrule
2 & 4 & 10 & 8 & 3 & 6 & 5 & 24 & 2 \\
3 & 8 & 20 & 48 & 4 & 168 & 5 & 1{,}344 & 2 \\
4 & 16 & 35 & 384 & 5 & 20{,}160 & 5 & 322{,}560 & 2 \\
5 & 32 & 56 & 3{,}840 & 6 & -- & -- & -- & -- \\
\bottomrule
\end{tabular}
\end{table}

\iffull
\subsection{Classical Theorems Carried to the Boolean Cube}
\label{sec:classical}

The averages of this paper rest on a handful of classical results about finite groups. Their
statements on the cube are short, and Table~\ref{tab:classical} collects them with the place where
each is used. On a finite group the Haar measure is the uniform measure $1/|G|$, so the Haar
integral of the general theory is the average \eqref{eq:RG}, and the Reynolds operator that
projects onto invariants is that average. Maschke's theorem splits every real or complex
representation of a finite group into irreducible ones, and Schur's lemma says that a matrix
commuting with an irreducible representation is a multiple of the identity
\cite{serre1977linear,fulton1991rep}. Together they give $\dim\mathcal C(G)=\sum_\varrho m_\varrho^2$
over the irreducible representations $\varrho$ that occur with multiplicity $m_\varrho$ in the
action on $\mathbb R^N$, and Burnside's lemma \cite{burnside1911,neumann1979burnside} gives the same
number as the mean of $\mathrm{fix}(g)^2$, the count of orbits on pairs used in
Proposition~\ref{prop:commutant}. When every $m_\varrho$ is one the commutant is commutative,
$\RG(f)$ is diagonal in a basis adapted to the decomposition, and that basis is the transform
matched to $G$ \cite{thornton2026ad,thornton2026unification}. The translations, the hyperoctahedral
group, the affine group and the iterated wreath product are all of this kind, while $S_n$ and $\GL(n,2)$
are not.

Over $\F_2$ Maschke's theorem fails for 2-groups. The group algebra $\F_2[\mathbb Z_2^n]$, which is the
space of truth tables with XOR as addition and translation as multiplication, has a single
irreducible representation, the trivial one, and no complement to the invariant constants. Berman
showed that the powers of its radical, the ideal generated by the elements $1+x_i$, are exactly the
Reed--Muller codes \cite{berman1967group}. The structure that the real Walsh transform
diagonalizes is therefore a filtration over $\F_2$, by algebraic degree, and Section~\ref{sec:lutcomp}
uses that filtration for compression.

\begin{table*}[t]
\centering
\caption{Classical theorems on finite groups and their form on the Boolean cube. The last column
gives where the paper uses each.}
\label{tab:classical}
\small
\renewcommand{\arraystretch}{1.15}
\setlength{\tabcolsep}{4pt}
\begin{tabular}{@{}p{0.14\textwidth}p{0.27\textwidth}p{0.39\textwidth}p{0.11\textwidth}@{}}
\toprule
theorem & general statement & on the Boolean cube & used in\\
\midrule
Haar measure, Reynolds operator & the invariant mean of a compact group; averaging projects onto invariants & uniform weight $1/|G|$; $\RG(f)$ of \eqref{eq:RG} is the projection of $FF^{\mathsf T}$ onto $\mathcal C(G)$ & Sec.~\ref{sec:commutant}\\
Maschke & complete reducibility when the characteristic does not divide $|G|$ & real actions on truth tables split; over $\F_2$ the group algebra of $T$ is local and Berman's theorem gives the Reed--Muller codes as its radical powers & Sec.~\ref{sec:lutcomp}\\
Schur's lemma & the commutant of an irreducible representation is the scalars & $\dim\mathcal C(G)=\sum m_\varrho^2$; multiplicity-free actions have commutative commutants diagonal in the matched transform & Prop.~\ref{prop:commutant}\\
Peter--Weyl (finite form) & matrix entries of the irreducibles form an orthogonal basis; the regular representation holds each with multiplicity equal to its dimension & $T$ acts regularly on truth tables, its $N$ irreducibles are the Walsh functions, and the Walsh transform is the Fourier transform of $\mathbb Z_2^n$ & Prop.~\ref{prop:T}\\
Burnside's lemma & orbits number the mean count of fixed points & $\dim\mathcal C_\varkappa(G)=|G|^{-1}\sum_g\mathrm{fix}(g)^\varkappa$; symmetric-function counts by M\"obius inversion & Sec.~\ref{sec:order}, Table~\ref{tab:symcount}\\
Schur--Weyl duality & on $(\mathbb C^2)^{\otimes n}$, $S_n$ and $\GL(2)$ generate each other's commutant & $\mathbb R^N=(\mathbb R^2)^{\otimes n}$; $S_n$ holds the irreducible of shape $(n-j,j)$ with multiplicity $n-2j+1$, so $\dim\mathcal C(S_n)=\sum_j(n-2j+1)^2=\binom{n+3}3$ & Prop.~\ref{prop:commutant}\\
Gelfand pairs & a multiplicity-free permutation action has spherical functions & $(B,S_n)$: the spherical functions are the Krawtchouk polynomials; $(\Wr_n,\text{leaf stabilizer})$: the Haar levels & Props.~\ref{prop:collapse}, \ref{prop:wreath}\\
Algebraic Diversity & a group average stands in for repeated observations; the matched group is the one whose commutant the transform diagonalizes & $T\leftrightarrow$ Walsh, $\Wr_n\leftrightarrow$ Haar, $B\leftrightarrow$ Krawtchouk; group gain and first informative order & Secs.~\ref{sec:T}, \ref{sec:haar}, \ref{sec:order}\\
\bottomrule
\end{tabular}
\end{table*}

The decompositions behind Proposition~\ref{prop:commutant} are these, for $n\ge2$. The translations
split $\mathbb R^N$ into the $N$ lines spanned by the Walsh functions, each once, which gives $N$. The
group $B$ splits it into the $n+1$ spans of the Walsh functions of each weight, of dimensions
$\binom nj$, each once, which gives $n+1$ \cite{diaconis1988,delsarte1973}. The permutations hold the
irreducible of shape $(n-j,j)$ for $0\le j\le n/2$ with multiplicity $n-2j+1$, by Young's rule on the
weight classes, which gives $\binom{n+3}3$. The linear group holds the trivial representation twice,
once on the origin and once on the sum of the other points, and an irreducible of dimension $N-2$
once, because it is two-transitive on the nonzero points, which gives $2^2+1=5$. The affine group is
two-transitive on all points and holds the trivial representation and an irreducible of dimension
$N-1$, which gives $2$. The signed action of $\EA$ is irreducible, which gives $1$. The reference
software checks the multiplicities of $S_n$ for $n\le6$ by character inner products and the identity
$\sum_j(n-2j+1)^2=\binom{n+3}3$ for $n\le12$.

Many statements of this paper restate a procedure that logic designers already use, and some
extend one. Table~\ref{tab:restated} lists them, so that a reader can map each group-theoretic
result onto a familiar method and see where the group adds something.

\begin{table*}[t]
\centering
\caption{Group-theoretic statements of the paper and the classical procedures of logic design that
they restate (same) or extend (superset).}
\label{tab:restated}
\small
\renewcommand{\arraystretch}{1.15}
\setlength{\tabcolsep}{4pt}
\begin{tabular}{@{}p{0.33\textwidth}p{0.33\textwidth}p{0.08\textwidth}p{0.18\textwidth}@{}}
\toprule
statement in this paper & classical procedure & relation & where\\
\midrule
neighbors are rows one unit translation apart; groupings are cosets of coordinate subgroups of $T$ & Karnaugh map adjacency and grouping, prime implicants & same & Sec.~\ref{sec:prelim}, Fig.~\ref{fig:kmap}\\
cosets of every subgroup of $T$ (affine subspaces) & groupings of the map & superset & Secs.~\ref{sec:prelim}, \ref{sec:hashvar}\\
translation by $c$ & complementing inputs; choice of literal polarities; fixed-polarity Reed--Muller forms & same & Sec.~\ref{sec:prelim}\\
$T$-average of $FF^{\mathsf T}$ & autocorrelation and Walsh power spectrum of spectral logic design & same & Prop.~\ref{prop:T}\\
orbits of $B$ with output negation & NPN classification for Boolean matching & same & Sec.~\ref{sec:matching}\\
stabilizer in $S_n$ & detection of symmetric variables; symmetric functions when it is all of $S_n$ & same & Sec.~\ref{sec:prelim}\\
stabilizer in $B$ or $\AGL$ & symmetries with input negation, linear symmetries & superset & Sec.~\ref{sec:bgm}\\
Haar transform and $\Wr_n$ along an order & Shannon expansion, multiplexer trees, ordered BDDs & same & Sec.~\ref{sec:haar}\\
order classes under the stabilizer & symmetric sifting groups symmetric variables & superset & Prop.~\ref{prop:cosets}\\
average over $\AGL$ of hash cells & random XOR hashing in approximate counters & same & Thm.~\ref{thm:hash}\\
matched hashes on the affine hull & hashing restricted to the free coordinates & superset & Sec.~\ref{sec:matched}\\
Reed--Muller codes as radical powers of $\F_2[\mathbb Z_2^n]$ & Reed--Muller (ANF) expansion by degree & same & Sec.~\ref{sec:classical}\\
autocorrelation zero on the unit translations & strict avalanche criterion & same & Sec.~\ref{sec:crypto}\\
constancy on cosets of the late-input translations & trigger functions of early evaluation & same & Sec.~\ref{sec:early}\\
constancy on cosets of any subgroup of $T$ & early evaluation from XORs of inputs & superset & Sec.~\ref{sec:early}\\
\bottomrule
\end{tabular}
\end{table*}
\fi

\subsection{The Translation Group Is Matched to the Walsh Transform}
\label{sec:T}

\begin{proposition}[Translation average]
\label{prop:T}
$R_T(f)_{x,y}=r_f(x\oplus y)/N$. Its eigenvectors are the Walsh functions
$h_a(x)=(-1)^{a\cdot x}$ and its eigenvalues are the periodogram values
$W_f(a)^2/N$, so that $R_T(f)=N^{-2}H\,\mathrm{diag}(W_f^2)\,H$.
\end{proposition}

\begin{proof}
$R_T(f)_{x,y}=N^{-1}\sum_cF(x\oplus c)F(y\oplus c)=N^{-1}r_f(x\oplus y)$ after the
substitution $z=x\oplus c$. A matrix that depends only on $x\oplus y$ is diagonalized by
the characters of $\mathbb{Z}_2^n$, and the Wiener--Khinchin relation gives the
eigenvalue $N^{-1}\sum_dr_f(d)(-1)^{a\cdot d}=W_f(a)^2/N$.
\end{proof}

Proposition~\ref{prop:T} is the GF(2) instance of the statement that a regular Abelian group is
matched to its character transform \cite{thornton2026ad,thornton2026unification}. The
periodogram $W_f^2$ is the classical Walsh power spectrum
\cite{karpovsky1976,hurst1985spectral,edwards1975}, here obtained as the spectrum of
the matched group average of a single truth table.

\iffull
\subsection{The Haar Transform Is a Multiplexer Tree Matched to the Iterated Wreath Product}
\label{sec:haar}

Shannon's expansion $f=(\neg x_i\wedge f|_{x_i=0})\oplus(x_i\wedge f|_{x_i=1})$
\cite{shannon1949synthesis} is a 2:1 multiplexer whose select input is $x_i$ and whose two data
inputs are the cofactors. Expanding the cofactors in turn along an order $o=(o_1,\dots,o_n)$ of the
inputs gives a complete tree of $N-1$ multiplexers whose leaves are the entries of the truth table.
Merging equal subtrees and bypassing every multiplexer whose two data inputs are equal gives the
reduced ordered binary decision diagram \cite{bryant1986bdd}, which is mapped one multiplexer per node
onto pass transistors or onto timed Shannon circuits \cite{lavagno1995timed,thornton2002timedshannon}.
For $0\le j<n$ and $u\in\F_2^j$ let $F_u$ be the sign vector of the cofactor of order $j$ that fixes
$x_{o_1},\dots,x_{o_j}$ to the bits of $u$, a vector over the remaining $n-j$ inputs. The Haar
transform of $F$ along the same order has the coefficient $W_f(0)$ and one coefficient per
multiplexer of the complete tree,
\begin{equation}
\label{eq:haar}
\mathcal D_j(u)=\sum_yF_{u0}(y)-\sum_yF_{u1}(y),
\end{equation}
where $u0$ and $u1$ extend $u$ by the value of $x_{o_{j+1}}$ and the sums run over the last
$n-j-1$ inputs. Each coefficient is the difference of the weights of the two data inputs of its
multiplexer, so a multiplexer that the reduction bypasses has a zero coefficient. The Haar wavelets of
level $j$ are the differences of cofactors of order $j+1$, that is, higher-order Shannon cofactors,
and Haar spectral diagrams carry these coefficients on the decision diagram itself
\cite{thornton2002haar}. Fig.~\ref{fig:haarmux} draws the tree for two inputs.

\begin{figure}[t]
\centering
\begin{tikzpicture}[font=\scriptsize,
  mux/.style={trapezium, trapezium angle=70, draw, thick, shape border rotate=270, minimum width=7mm,
              minimum height=5mm, inner sep=1pt},
  lab/.style={align=left, anchor=west}]
\node[mux] (r) at (2.6,0) {};
\node[mux] (m0) at (1.2,0.8) {};
\node[mux] (m1) at (1.2,-0.8) {};
\foreach \y/\t in {1.15/{$f(00)$},0.45/{$f(01)$},-0.45/{$f(10)$},-1.15/{$f(11)$}} \node[anchor=east] at (0.55,\y) {\t};
\draw (0.55,1.15) -- ($(m0.north west)!0.3!(m0.south west)$); \draw (0.55,0.45) -- ($(m0.north west)!0.7!(m0.south west)$);
\draw (0.55,-0.45) -- ($(m1.north west)!0.3!(m1.south west)$); \draw (0.55,-1.15) -- ($(m1.north west)!0.7!(m1.south west)$);
\draw (m0.east) -- ++(0.3,0) |- ($(r.north west)!0.3!(r.south west)$);
\draw (m1.east) -- ++(0.3,0) |- ($(r.north west)!0.7!(r.south west)$);
\draw (r.east) -- ++(0.4,0) node[right] {$f$};
\draw (m0.south) -- ++(0,-0.15) node[below, inner sep=1pt] {$x_{o_2}$};
\draw (m1.south) -- ++(0,-0.15) node[below, inner sep=1pt] {$x_{o_2}$};
\draw (r.south) -- ++(0,-0.25) node[below, inner sep=1pt] {$x_{o_1}$};
\node[lab] at (3.5,0.95) {$\mathcal D_1(0)=F(00)-F(01)$};
\node[lab] at (3.5,0.4) {$\mathcal D_1(1)=F(10)-F(11)$};
\node[lab] at (3.5,-0.15) {$\mathcal D_0=F(00)+F(01)$};
\node[lab] at (3.5,-0.5) {$\hphantom{\mathcal D_0=}-F(10)-F(11)$};
\node[lab] at (3.5,-1.0) {$W_f(0)=\textstyle\sum_xF(x)$};
\end{tikzpicture}
\caption{The multiplexer tree of Shannon's expansion along the order $(x_{o_1},x_{o_2})$, with the
Haar coefficient of each multiplexer, the difference of the sign sums of its two data inputs. The
leaves are written with $x_{o_1}$ as the first bit. A multiplexer with equal data inputs is removed by
the reduction and has a zero coefficient.}
\label{fig:haarmux}
\end{figure}
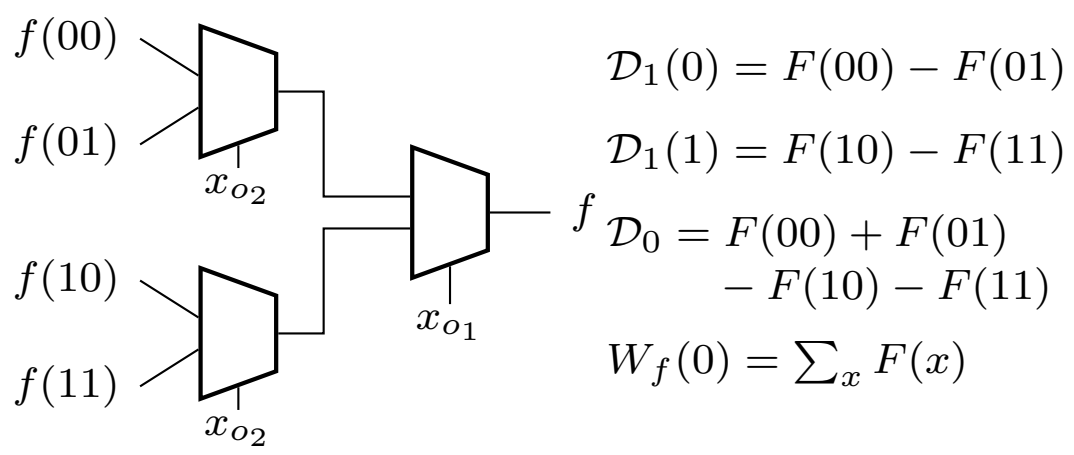

\begin{proposition}[Haar level energy from the Walsh spectrum]
\label{prop:haarenergy}
For an order $o$ and $0\le j<n$ let $A_j(o)$ be the set of $a\in\F_2^n$ supported on
$\{o_1,\dots,o_{j+1}\}$ with $a_{o_{j+1}}=1$. Then the energy of level $j$ is
\[
\mathcal L_j(o)=\sum_{u\in\F_2^j}\mathcal D_j(u)^2=2^{-j}\sum_{a\in A_j(o)}W_f(a)^2.
\]
\end{proposition}

\begin{proof}
Summing $F$ over the inputs outside $V=\{o_1,\dots,o_{j+1}\}$ gives a function $g$ on $\F_2^{j+1}$
whose Walsh coefficient at $b$ is $W_f(a)$ for the $a$ supported on $V$ that agrees with $b$ there,
because $a\cdot x$ depends only on the coordinates in $V$. Then $\mathcal D_j(u)=g(u0)-g(u1)$, and
$\sum_u(-1)^{b'\cdot u}\mathcal D_j(u)=W_f(a)$ for the $a$ that equals $b'$ on the first $j$
coordinates of $V$ and one on the last. Parseval's identity on $\F_2^j$ gives the claim.
\end{proof}

\begin{proposition}[The Haar transform is matched to the iterated wreath product]
\label{prop:wreath}
Let $\Wr_n$ act on the truth table through the tree of the order $o$. Its orbits on ordered pairs of
points are determined by the length of the common prefix of the two points along $o$, so
$\dim\mathcal C(\Wr_n)=n+1$. The constants and the $n$ Haar levels are its isotypic components, of
dimensions $1,1,2,\dots,2^{n-1}$, each irreducible and occurring once, and
\[
R_{\Wr_n}(f)=\frac{W_f(0)^2}{N}\,E_{-1}+\sum_{j=0}^{n-1}\frac{\mathcal L_j(o)}{N}\,E_j,
\]
where $E_{-1}$ projects onto the constants and $E_j$ onto the span of the wavelets of level $j$.
\end{proposition}

\begin{proof}
A tree automorphism can move a pair of leaves to any other pair whose deepest common ancestor lies
at the same depth, and to no other pair, which gives $n+1$ orbits on pairs and the dimension of the
commutant. Each Haar level is invariant, since an exchange of subtrees permutes the wavelets of each
level up to sign, and the $n+1$ levels are orthogonal and span $\mathbb R^N$. As
$\sum m_\varrho^2=n+1$ and there are $n+1$ invariant summands, each is irreducible and the
multiplicities are one. The projection of $FF^{\mathsf T}$ onto the span of the $E_j$ has the
coefficient $F^{\mathsf T}E_jF/\operatorname{tr}E_j$. A wavelet of level $j$ has squared norm
$2^{n-j}$, so $F^{\mathsf T}E_jF=2^{j-n}\mathcal L_j(o)$, and $\operatorname{tr}E_j=2^j$.
\end{proof}

Proposition~\ref{prop:wreath} is the binary case of the multiresolution analysis of Foote, Mirchandani,
Rockmore, Healy and Olson \cite{foote2000wreath}, and it places the Haar transform beside the Walsh
transform in the lattice of matched groups. The translations see the full periodogram, the wreath
product sees one energy per level of the decision tree, and the ordering of the inputs is part of the
group. The reference software enumerates $\Wr_n$ for $n\le4$, of order up to
\ClWreathFour, and checks the orbit count, the diagonal form and the eigenvalues, and checks on every
function of four inputs that the bypassed multiplexers carry zero coefficients.

The group view also bears on the choice of the order, which is NP-hard to optimize
\cite{bollig1996np} and is usually improved by sifting \cite{rudell1993sifting}, with symmetric inputs
moved together \cite{panda1994symmetry}.

\begin{proposition}[Orders on cosets of the stabilizer]
\label{prop:cosets}
Let $\mathcal S_f\subseteq S_n$ be the group of the permutation matrices $Q$ for which
$x\mapsto Qx\oplus c$ lies in $\Stab_B(f)$ for some $c$, and let $Q$ act on orders by relabeling the
inputs. The number of multiplexers of the reduced tree is the same for the orders $o$ and $Qo$ for
every $Q\in\mathcal S_f$, so an exact search over orders needs one order from each of the
$n!/|\mathcal S_f|$ classes.
\end{proposition}

\begin{proof}
Complementing an input exchanges the two data inputs of every multiplexer that it selects, which
changes no count. Relabeling the inputs of $f$ by $Q$ and reading along $o$ is the same as reading $f$
along $Qo$. For $g(x)=Qx\oplus c$ in the stabilizer, $f\circ g=f$, so the counts along $o$ and $Qo$
agree, and $\mathcal S_f$ acts freely on the $n!$ orders.
\end{proof}

Proposition~\ref{prop:haarenergy} also suggests a heuristic that builds the order greedily, at each
depth taking the input that puts the most Walsh energy into the coarse levels. Section~\ref{sec:varorder}
tests both the reduction of Proposition~\ref{prop:cosets} and this heuristic against exact optima and
sifting.
\fi

\subsection{Larger Groups Collapse the Second-Order Statistic}
\label{sec:collapse}

\begin{proposition}[Closed forms along the lattice]
\label{prop:collapse}
Let $w(a)$ be the Hamming weight of $a$.
\begin{enumerate}
\item $R_B(f)$ has the Walsh functions as eigenvectors, with the eigenvalue
\[
\lambda_j=\binom nj^{-1}\sum_{w(a)=j}\frac{W_f(a)^2}{N}
\]
on every $a$ of weight $j$. Its entries are
$R_B(f)_{x,y}=N^{-1}\sum_j\lambda_jK_j(w(x\oplus y))$, with $K_j$ the Krawtchouk
polynomials of the Hamming scheme.
\item $R_{\AGL}(f)=(1-\rho)I+\rho J$, where $J$ is the all-ones matrix and
$\rho=(W_f(0)^2-N)/(N(N-1))$.
\item $R_{\GL}(f)$ has unit diagonal, the entries $F(0)s/(N-1)$ in row and column zero
off the diagonal, and the entry $(s^2-N+1)/((N-1)(N-2))$ on every other off-diagonal
position, where $s=\sum_{x\ne0}F(x)$.
\item $R_{\EA}(f)=I$.
\end{enumerate}
\end{proposition}

\begin{proof}
Since $T\subset B$, $\mathcal{C}(B)\subset\mathcal{C}(T)$, so the projection onto
$\mathcal{C}(B)$ factors through the projection onto $\mathcal{C}(T)$, and $R_B(f)$ is the
average of $R_T(f)$ over $S_n$, which permutes the Walsh index by the same
coordinate permutation and averages the eigenvalues over each weight class. The entry
formula is the spectral decomposition of the Bose--Mesner algebra of the Hamming
scheme \cite{delsarte1973,delsarte1998,macwilliams1977}\ifarxiv. Explicitly,
$R_B(f)_{x,y}=N^{-1}\sum_a\lambda_{w(a)}(-1)^{a\cdot(x\oplus y)}$, and the inner sum over
the $a$ of weight $j$ is $K_j(i)=\sum_l(-1)^l\binom il\binom{n-i}{j-l}$ with
$i=w(x\oplus y)$, since $(-1)^{a\cdot d}$ depends only on how many of the $j$ ones of
$a$ fall in the support of $d$\fi. For the affine group,
two-transitivity makes every off-diagonal entry the mean of $F(x)F(y)$ over pairs
$x\ne y$, which is $((\sum_xF(x))^2-N)/(N(N-1))$, and $\sum_xF(x)=W_f(0)$. For
$\GL(n,2)$ the five orbits of Proposition~\ref{prop:commutant} give the stated means.
For $\EA$ the projection onto the multiples of the identity is $N^{-1}\operatorname{tr}(FF^{\mathsf
T})I=I$.
\end{proof}

The collapse is complete at the top of the lattice. The affine average sees only
$W_f(0)^2$, that is, the weight $|S|=(N-W_f(0))/2$ up to complementation, the linear
average adds the value at the origin up to the global sign, and the extended affine
average sees nothing. The signed permutation action of $\EA$ contains the $n$-qubit
Pauli group generated by the $P_c$ and $Q_b$, whose average alone collapses to the
identity, as in the twirl of quantum state estimation \cite{thornton2026quantum}. The
collapse has a practical reading. A second-order quantity averaged over a large group
can depend only on the invariants of that group's average. Random
XOR hashing with a uniformly chosen cell is an average over $\AGL(n,2)$ in
this sense, and Section~\ref{sec:hashvar} shows that its mean variance depends on
$|S|$ alone.

\subsection{The Normal Translation Subgroup Keeps What the Full Average Discards}
\label{sec:normalizer}

The groups $B$, $\AGL(n,2)$ and $\EA$ contain $T$ as a normal subgroup. Their own
averages project onto small commutants, but their action on the translation
periodogram is explicit.

\begin{theorem}[Normalizer covariance]
\label{thm:normalizer}
For $g(x)=Ax\oplus c$ in $\AGL(n,2)$,
\begin{equation}
\label{eq:normalizer}
W_{f\circ g}(a)=(-1)^{(A^{-\mathsf T}a)\cdot c}\,W_f(A^{-\mathsf T}a),
\end{equation}
so $W_{f\circ g}^2(a)=W_f^2(A^{-\mathsf T}a)$. Adding $b\cdot x\oplus\nu$ to $f$ sends
$W_f(a)$ to $(-1)^\nu W_f(a\oplus b)$. The bispectrum
$\beta_f(a,a')=W_f(a)W_f(a')W_f(a\oplus a')$ satisfies
$\beta_{f\circ g}(a,a')=\beta_f(A^{-\mathsf T}a,A^{-\mathsf T}a')$ and changes sign under
output negation.
\end{theorem}

\begin{proof}
Substitute $y=Ax\oplus c$ in the definition of $W_{f\circ g}$. Then $x=A^{-1}(y\oplus
c)$ and $a\cdot x=(A^{-\mathsf T}a)\cdot y\oplus(A^{-\mathsf T}a)\cdot c$. The sign
factors of the three spectral values in $\beta$ multiply to one because the three
indices sum to zero. The remaining statements follow from the definitions.
\end{proof}

Theorem~\ref{thm:normalizer} says that the quotient $G/T$ acts on the periodogram by
permutations of the index, by coordinate permutations for $B$, by the maps
$a\mapsto A^{-\mathsf T}a$ for $\AGL$, and by these together with index translations for
$\EA$. The canonical
periodogram, the lexicographically least image of $W_f^2$ under that index action, is
therefore a $G$-invariant. It determines $\RG(f)$ for each of the three groups, by
Proposition~\ref{prop:collapse}, and it is in general far finer. The rule is to average
over the normal Abelian subgroup, whose commutant is as large as the space, and to
remove the quotient by canonicalization instead of averaging.
Section~\ref{sec:classify} measures the difference exhaustively for $n=3$ and $n=4$.

Second-order statistics have two blind spots, both visible in these terms. Every
function of $FF^{\mathsf T}$ is unchanged by output negation. And when $n$ is even,
the functions whose periodogram is flat are the bent functions \cite{rothaus1976bent},
so that the following holds.

\begin{corollary}[Bent functions are white]
\label{cor:bent}
$R_T(f)=I$ if and only if $f$ is bent. All bent functions of $n$ variables have the
same translation periodogram, so no invariant that factors through $W_f^2$ separates
their classes.
\end{corollary}

\begin{proof}
By Proposition~\ref{prop:T}, $R_T(f)=I$ exactly when $W_f(a)^2=N$ for all $a$, which
is the definition of a bent function.
\end{proof}

Corollary~\ref{cor:bent} restates, in the language of group averages, the classical
characterization of bent functions by an autocorrelation that vanishes away from the
origin \cite{rothaus1976bent,carlet2021book}.

The third-order bispectrum is invariant under $T$, covariant under the linear index
action (though not under index translations) and sign-sensitive, so it resolves both blind spots when it is canonicalized jointly with
the periodogram. For Abelian groups the triple correlation determines a signal up to
translation when its spectrum has no zeros \cite{yellott1992,kakarala2012bispectrum}, and
Proposition~\ref{prop:bisp} gives the short proof over $\F_2^n$. Spectral zeros are common for
switching functions, the comparator of Fig.~\ref{fig:intro} has \IxZerosWord, so for the
classification we rely on the exhaustive result of Section~\ref{sec:classify}.

\iffull
\subsection{Blind Group Matching Recovers the Stabilizer from One Truth Table}
\label{sec:bgm}

Blind group matching asks for the group of a signal from that signal alone, with no labeled examples
and no list of candidate groups \cite{thornton2026ad,thornton2026gevp}. On the cube the question is
the stabilizer $\Stab_G(f)$ of one truth table in $G=B$ or $G=\AGL(n,2)$, possibly allowing the
output to be complemented, and Theorem~\ref{thm:normalizer} turns it into two problems of different
orders. Write $M=A^{-\mathsf T}$ for the map that $g(x)=Ax\oplus c$ induces on spectral indices,
$s(a)=1$ when $W_f(a)<0$ and $s(a)=0$ otherwise, and $\nu\in\F_2$ for the output complement.

\begin{proposition}[Second order finds the linear part, the signs find the negations]
\label{prop:bgmsigns}
The map $x\mapsto Ax\oplus c$, followed by adding $\nu$ to the output, fixes $f$ if and only if
$W_f(Ma)^2=W_f(a)^2$ for every $a$ and
\[
(Ma)\cdot c=s(Ma)\oplus s(a)\oplus\nu\quad\text{for every $a$ with $W_f(a)\ne0$.}
\]
For fixed $M$ and $\nu$ the solutions $c$ of this system over $\F_2$ form a coset of
$T_f=\{c:r_f(c)=N\}$, the translations that fix $f$.
\end{proposition}

\begin{proof}
By \eqref{eq:normalizer} the transformed function has the Walsh coefficient
$(-1)^{\nu+(Ma)\cdot c}W_f(Ma)$ at $a$, and two functions agree exactly when their spectra agree.
Equal magnitudes give the first condition and equal signs the second. For $M=I$ and $\nu=0$ the
system reads $a\cdot c=0$ on $\supp W_f$, whose solutions are the $c$ with $W_{f}(a)(-1)^{a\cdot
c}=W_f(a)$ for all $a$, that is, with $f(x\oplus c)=f(x)$, and $r_f(c)=N$ says the same.
\end{proof}

Every input negation is invisible at second order, and Proposition~\ref{prop:bgmsigns} recovers the
negations from the signs of the spectrum by Gaussian elimination over $\F_2$. The linear parts that
preserve the periodogram are the soft part of the problem, and the double-commutator generalized
eigenvalue problem (DC-GEVP) of the framework \cite{thornton2026gevp} finds them. For a dictionary of
candidate linear maps with spectral maps $M_1,\dots,M_K$ (for instance the $\binom n2$ input exchanges
and the $n(n-1)$ transvections $x_i\leftarrow x_i\oplus x_j$ that generate $\GL(n,2)$), it minimizes
$\|[\sum_kw_kP_k,R_T(f)]\|_F^2$ over $\mathbf w=(w_1,\dots,w_K)$ with $\mathbf w^{\mathsf T}\mathbf Y\mathbf w=1$,
where $P_k$ is the permutation matrix of the $k$-th candidate.

\begin{proposition}[DC-GEVP on the cube]
\label{prop:dcgevp}
The DC-GEVP is $\mathbf X\mathbf w=\iota\,\mathbf Y\mathbf w$ with
\[
\mathbf X_{kl}=\frac1{N^2}\!\!\sum_{a:\,M_ka=M_la}\!\!\bigl(W_f(a)^2-W_f(M_ka)^2\bigr)^2,\qquad
\mathbf Y_{kl}=\bigl|\{a:M_ka=M_la\}\bigr|,
\]
and its eigenvectors of eigenvalue zero span the combinations of candidates that commute with
$R_T(f)$, and a single candidate has $\mathbf X_{kk}=0$ exactly when it preserves the periodogram.
\end{proposition}

\begin{proof}
In the orthonormal Walsh basis $R_T(f)$ is the diagonal matrix $\mathbf D=\mathrm{diag}(W_f^2/N)$ by
Proposition~\ref{prop:T}, and $P_k$ becomes the permutation matrix that sends index $a$ to $M_ka$.
The commutator has the single nonzero entry $\mathbf D_{aa}-\mathbf D_{M_ka,M_ka}$ in column $a$, at row $M_ka$, so the
Frobenius inner product of two commutators collects the columns where both entries sit in the same
row. The same count gives $\mathbf Y_{kl}=\operatorname{tr}(P_k^{\mathsf T}P_l)$. The objective is
nonnegative, so its zero set is the kernel of $\mathbf X$.
\end{proof}

The closed form takes $O(K^2N)$ operations after one Walsh transform and needs no matrix of size
$N$. For exact symmetries Algorithm~\ref{alg:bgm} uses the whole group in place of the dictionary. It first
colors the spectral indices, starting from the power $W_f(a)^2$ and refining each color by the
multiset of pairs formed by the color of $a'$ and the bispectrum $\beta_f(a,a')$, until the colors
stop splitting. By Theorem~\ref{thm:normalizer} the linear part of every stabilizer element preserves
these colors, and with output complementation it preserves them when $\beta_f$ is replaced by
$|\beta_f|$. It then picks a base $b_1,\dots,b_n$ of $\F_2^n$ greedily from the smallest color classes
(from the unit vectors, for $B$), and builds $M$ by choosing the images $v_j=Mb_j$ in turn among the
indices of the color of $b_j$. Once $v_1,\dots,v_j$ are fixed, the image of every index in the span
of $b_1,\dots,b_j$ is known, and the $2^{j-1}$ new ones are checked for their color and contribute
their sign equations to an incremental system $\mathcal K$ over $\F_2$. The order of the stabilizer
follows from the chain of pointwise stabilizers of $b_1,\dots,b_j$ \cite{cameron1999permutation}. It
is $|T_f|$ times the product over $j$ of the size of the orbit $O_j$ of $b_j$ under the elements that
fix $b_1,\dots,b_{j-1}$, doubled when an element that complements the output exists. The levels are
searched from the deepest one up, so that the generators of the stabilizer of $b_1,\dots,b_j$ are
already known at level $j$. The images of $b_j$ that extend to an element, and those that do not,
each form unions of orbits of that stabilizer, so a single search settles a whole orbit. One element
per new orbit point is kept, which gives a strong generating set, and every generator is checked on
the truth table.

\begin{algorithm}[!htbp]
\caption{Blind group matching on the Boolean cube}
\label{alg:bgm}
\begin{algorithmic}[1]
\Require truth table of $f$ on $\F_2^n$, group $G\in\{B,\AGL(n,2)\}$, flag for output complement
\Ensure $|\Stab_G(f)|$ and verified strong generators
\State $W_f\gets$ fast Walsh transform of $F$; $s(a)\gets[W_f(a)<0]$ \Comment{$O(nN)$}
\State color the indices by $W_f(a)^2$, refined by $\beta_f(a,a')$ until stable \Comment{$O(N^2\log N)$ per round}
\State base $b_1,\dots,b_n$ from the smallest color classes; $\mathcal A_j\gets$ indices of the color of $b_j$
\State $T_f\gets$ solutions of $a\cdot c=0$, $a\in\supp W_f$ \Comment{$O(nN)$}
\Function{Extend}{$v_1,\dots,v_j$; $\mathcal K$; $\nu$}
  \If{$j=n$} \Return $(v_1,\dots,v_n;\mathcal K)$ \EndIf
  \For{$v\in\mathcal A_{j+1}\setminus\spn\{v_1,\dots,v_j\}$}
    \State $\mathcal K'\gets\mathcal K$; $v_{j+1}\gets v$
    \ForAll{$\alpha\in\F_2^{j+1}$ with $\alpha_{j+1}=1$} \Comment{$2^j$ new indices}
      \State $a\gets\sum_i\alpha_ib_i$; $b\gets\sum_i\alpha_iv_i$
      \State \textbf{if} $\mathrm{color}(b)\ne\mathrm{color}(a)$ \textbf{then} reject $v$
      \State \textbf{if} $W_f(a)\ne0$ \textbf{then} add $b\cdot c=s(b)\oplus s(a)\oplus\nu$ to $\mathcal K'$; reject $v$ if inconsistent
    \EndFor
    \State \textbf{if} \Call{Extend}{$v_1,\dots,v_{j+1}$; $\mathcal K'$; $\nu$} succeeds \textbf{then return} it
  \EndFor
  \State \Return fail
\EndFunction
\State order $\gets|T_f|$; $\mathcal G\gets\emptyset$
\For{$j=n,n-1,\dots,1$}
  \State $O_j\gets$ orbit of $b_j$ under $\mathcal G$; $X_j\gets\emptyset$
  \For{$v\in\mathcal A_j\setminus(O_j\cup X_j)$}
    \If{\Call{Extend}{$b_1,\dots,b_{j-1},v$; $\emptyset$; $0$} returns $g$}
      \State $\mathcal G\gets\mathcal G\cup\{g\}$; $O_j\gets$ orbit of $b_j$ under $\mathcal G$
    \Else\ $X_j\gets X_j\,\cup$ orbit of $v$ under $\mathcal G$
    \EndIf
  \EndFor
  \State order $\gets$ order $\cdot|O_j|$
\EndFor
\State \textbf{if} complement allowed and \Call{Extend}{$\,$; $\emptyset$; $1$} succeeds \textbf{then} order $\gets2\cdot$order
\State for each kept element: $A\gets M^{-\mathsf T}$, $c\gets$ a solution of $\mathcal K$; verify $f(Ax\oplus c)\oplus\nu=f(x)$
\end{algorithmic}
\end{algorithm}

\paragraph*{Complexity} The Walsh transform and $T_f$ take $O(nN)$ word operations, and each round
of the color refinement $O(N^2\log N)$, with at most $N$ rounds and in practice a few. A node of the
search at depth $j$ checks $2^{j-1}$ new indices, each with one comparison and one insertion into
$\mathcal K$ in $O(n)$ operations, so a path from the root to a leaf takes $O(nN)$. The order needs at
most $\sum_j|\mathcal A_j|$ calls of \textsc{Extend}, at most $n^2$ for $B$ and $n(N-1)$ for $\AGL$, and the
orbit pruning reduces this to about one call per orbit. When the colors single out the base vectors
every $\mathcal A_j$ has one element and the whole computation takes $O(n^2N)$ after the refinement.
Backtracking is exponential in the worst case, which functions whose colors do not split approach,
and Section~\ref{sec:bgmexp} reports the nodes visited on circuit functions.

Hoyos et al.\ enforce known symmetry groups exactly with group-structured tensor algebras and learn
symmetric models from labeled data in the regular representation \cite{hoyos2026exact}. The matching
here is different in kind. It is unsupervised, it uses one observation of one function, and it finds
the group from the commutant of the second-order average and the signs of the spectrum, with no
group given in advance.
\fi

\subsection{The Order of a Group Average}
\label{sec:order}

The average \eqref{eq:RG} is of second order, since it is built from $FF^{\mathsf T}$. For an
integer $\varkappa\ge1$ the average of order $\varkappa$ uses, in place of $FF^{\mathsf T}$, the
$\varkappa$-fold tensor power,
\begin{equation}
\label{eq:Rk}
R_G^{(\varkappa)}(f)=\frac1{|G|}\sum_{g\in G}(P_gF)^{\otimes\varkappa},
\end{equation}
an array with one entry for each $\varkappa$-tuple $\mathbf z=(z^{(1)},\dots,z^{(\varkappa)})$ of
input points, equal to the mean over $G$ of $\prod_i(P_gF)(z^{(i)})$. For a group of signed
permutations, $P_g$ carries the sign. We write $\mathcal C_\varkappa(G)$ for the space of arrays
of order $\varkappa$ that every $P_g^{\otimes\varkappa}$ leaves fixed, so that
$\mathcal C_2(G)$ is the commutant $\mathcal C(G)$ read as arrays. The second-order average is $R_G^{(2)}=\RG$, and the
bispectrum of Theorem~\ref{thm:normalizer} is built from entries of $R_T^{(3)}$. For a permutation
group, $\mathrm{fix}(g)$ denotes the number of points that $g$ fixes.

\begin{proposition}[Averages of order $\varkappa$]
\label{prop:orderk}
Let $G$ be a permutation group on $\F_2^n$. The entry of $R_G^{(\varkappa)}(f)$ at $\mathbf z$ is
the mean of $\prod_iF(y^{(i)})$ over the tuples $\mathbf y$ in the orbit of $\mathbf z$, and
$R_G^{(\varkappa)}$ is the orthogonal projection of $F^{\otimes\varkappa}$ onto the
$G$-invariant arrays. Their dimension is the number of orbits of $G$ on $\varkappa$-tuples,
\begin{equation}
\label{eq:burnside}
\dim\mathcal C_\varkappa(G)=\frac1{|G|}\sum_{g\in G}\mathrm{fix}(g)^{\varkappa}.
\end{equation}
For $\varkappa=2\varkappa'$, reading the first and the last $\varkappa'$ indices as a row and a
column turns $R_G^{(\varkappa)}(f)$ into the second-order average of $F^{\otimes\varkappa'}$ under
the representation $g\mapsto P_g^{\otimes\varkappa'}$, so an average of order four is a commutant
projection for the tensor square.
\end{proposition}

\begin{proof}
The map $g\mapsto g\mathbf z$ covers the orbit of $\mathbf z$ uniformly, each tuple being reached
by one coset of the stabilizer, which gives the orbit mean. The indicator arrays of the orbits
are orthogonal and span the invariant arrays, so the orbit means are the coefficients of the
orthogonal projection, and their number is given by the lemma of Burnside, since $g$ fixes
$\mathrm{fix}(g)^{\varkappa}$ tuples \cite{cameron1999permutation}. The last statement holds
because $(P_gF)^{\otimes2\varkappa'}$ reshaped is
$(P_g^{\otimes\varkappa'}F^{\otimes\varkappa'})(P_g^{\otimes\varkappa'}F^{\otimes\varkappa'})^{\mathsf
T}$, and \eqref{eq:RG} applied to it is the projection onto the commutant of
$P_g^{\otimes\varkappa'}$.
\end{proof}

Proposition~\ref{prop:collapse} shows that the second-order averages over $\AGL(n,2)$ and $\EA$
see little. The averages of higher order see more, but only from a certain order on. A group is
homogeneous on $j$-sets when it maps every set of $j$ input points onto every other
\cite{cameron1999permutation}.

\begin{theorem}[First informative order]
\label{thm:informative}
For a permutation group $G$ on $\F_2^n$, the average $R_G^{(\varkappa)}(f)$ depends on $f$ only
through its weight $|S|$, for every $f$, if and only if $G$ is homogeneous on $j$-sets for every
$j\le\varkappa$ with $j\equiv\varkappa\pmod2$. The first order at which this fails is $1$ for
$S_n$ and $\GL(n,2)$, $2$ for $T$ and $B$, and $4$ for $\AGL(n,2)$ with $n\ge3$, and it never
fails for the symmetric group $S_N$ of all permutations of the points.
\end{theorem}

\begin{proof}
Since $F(z)^2=1$, the product $\prod_iF(z^{(i)})$ equals $\prod_{v\in V}F(v)$, where $V$ is the
set of points that occur an odd number of times in $\mathbf z$, and $|V|\equiv\varkappa\pmod2$.
Every $j$-set with $j\le\varkappa$ and $j\equiv\varkappa$ occurs as such a $V$. By
Proposition~\ref{prop:orderk}, the entry is the mean of $\prod_{v\in V'}F(v)$ over the sets $V'$
of the orbit $O$ of $V$ under $G$, a multilinear polynomial in the $N$ variables $F(v)$ with
coefficient $1/|O|$ on each set of $O$ and zero elsewhere. If $G$ is homogeneous on $j$-sets, $O$
is every $j$-set, and the mean is a symmetric function of $F$, fixed by the number of entries
equal to $-1$, which is $|S|$. Conversely, if the mean depends only on $|S|$, it is invariant
under every permutation of the points, and since the multilinear representation of a function
on $\{\pm1\}^N$ is unique, its coefficients are equal on all $j$-sets, so $O$ is every $j$-set.
For the listed groups, $S_n$ and $\GL(n,2)$ are not transitive, since they fix the weight and the
point $0$. The translations and $B$ are transitive but keep $x\oplus y$, respectively the
distance between $x$ and $y$, so they have more than one orbit on $2$-sets. $\AGL(n,2)$ is
transitive on ordered triples of distinct points, since any three distinct points of $\F_2^n$ are
affinely independent, and a $4$-set that is an affine plane, $z^{(1)}\oplus z^{(2)}\oplus
z^{(3)}\oplus z^{(4)}=0$, cannot be mapped onto one that is not.
\end{proof}

\ifarxiv
The parity condition matters for the groups that are homogeneous on $j$-sets for odd $j$ but not
for even $j$, and for the affine group it means that $R_{\AGL}^{(3)}$ is still blind, although
$R_{\AGL}^{(2)}$ already sees the weight through $\rho$ in Proposition~\ref{prop:collapse}. A
uniformly random hash $x\mapsto\Phi x\oplus u$ with a uniformly random row space is an average of
this kind, and Theorem~\ref{thm:hash4} below shows the same threshold in its cell counts.
\fi

For the signed groups the sign removes entries. Output negation sends $F$ to $-F$, so the averages
of odd order over $\EA$ and over $B$ with output negation vanish. The modulations $Q_b$ of
$\EA$ multiply the entry at $\mathbf z$ by $(-1)^{b\cdot(z^{(1)}\oplus\dots\oplus z^{(\varkappa)})}$,
so only tuples with zero sum survive. For $\varkappa=2$ these are the pairs $z^{(1)}=z^{(2)}$,
which gives $R_{\EA}(f)=I$. For $\varkappa=4$ they are the tuples of two equal pairs and the
ordered affine planes, on which $\AGL(n,2)$ is transitive, so $R_{\EA}^{(4)}(f)$ has a single
informative value, the mean of $\prod_iF(z^{(i)})$ over the ordered affine planes. The next
proposition evaluates it.

\begin{proposition}[Fourth Walsh moment]
\label{prop:fourth}
For every switching function,
\begin{equation}
\label{eq:fourth}
\begin{split}
\sum_{z^{(1)}\oplus z^{(2)}\oplus z^{(3)}\oplus z^{(4)}=0}\ \prod_{i=1}^4F(z^{(i)})
&=\frac1N\sum_aW_f(a)^4\\
&=\sum_dr_f(d)^2 ,
\end{split}
\end{equation}
and for a set $S$ the additive energy
$\mathcal E(S)=|\{\mathbf z\in S^4:z^{(1)}\oplus\dots\oplus z^{(4)}=0\}|$ equals
$N^{-1}\sum_a\Sh(a)^4$. The order-four average $R_{\EA}^{(4)}(f)$ is determined by
$\sum_aW_f(a)^4$ and determines it, and for even $n$ its informative value is smallest exactly when
$f$ is bent.
\end{proposition}

\begin{proof}
Insert $[z^{(1)}\oplus\dots\oplus z^{(4)}=0]=N^{-1}\sum_a(-1)^{a\cdot(z^{(1)}\oplus\dots\oplus
z^{(4)})}$ and sum over the points. The second equality is Parseval's identity for
$W_f^2=Hr_f$, and the same computation with the indicator of $S$ gives $\mathcal E(S)$. The
zero-sum tuples with a repeated point number $3N^2-2N$ and contribute one each, so the mean over
the $N(N-1)(N-2)$ ordered affine planes is $(N^{-1}\sum_aW_f(a)^4-3N^2+2N)/(N(N-1)(N-2))$. By
Proposition~\ref{prop:pr}, $\sum_aW_f(a)^4\ge N^3$ with equality exactly for bent $f$.
\end{proof}

The quantity \eqref{eq:fourth} has several names, the sum-of-squares indicator
\cite{zhang1995gac}, the additive energy \cite{tao2006additive} and the fourth power of the
Gowers $U^2$ norm, $N^3\|F\|_{U^2}^4$ \cite{gowers2001szemeredi}, and
Proposition~\ref{prop:fourth} places all three at the first informative order of the affine
and extended affine groups. The bent functions are white at order two
(Corollary~\ref{cor:bent}) and minimal at order four, so every bent function has the same
averages over $T$ of order two and over $\EA$ of order four. Exhaustively for $n=4$, the minimum
$N^3$ of $\sum_aW_f(a)^4$ is attained by exactly the \OdBentCount\ bent functions among all
\OdBentFns. The bispectrum separates what these averages leave together.

\begin{proposition}[Bispectrum without spectral zeros]
\label{prop:bisp}
If $W_f(a)\ne0$ for every $a$, then $\beta_{f'}=\beta_f$ holds exactly when $f'(x)=f(x\oplus c)$
for some $c\in\F_2^n$. Every bent function has this property, so the bispectrum determines a
bent function up to translation of its inputs.
\end{proposition}

\begin{proof}
From $\beta(0,0)=W(0)^3$ and the reality of the cube root, $W_{f'}(0)=W_f(0)\ne0$. From
$\beta(a,0)=W(a)^2W(0)$, $W_{f'}(a)=\upsilon(a)W_f(a)$ with a sign $\upsilon(a)=\pm1$. From
$\beta(a,a')$, $\upsilon(a)\upsilon(a')\upsilon(a\oplus a')=1$, so $\upsilon$ is a character,
$\upsilon(a)=(-1)^{a\cdot c}$, and $f'(x)=f(x\oplus c)$ by Theorem~\ref{thm:normalizer}. A
bent function has $W_f(a)^2=N$ for all $a$.
\end{proof}

For $n=4$, the \OdNzFns\ functions without a spectral zero fall into \OdNzClasses\ classes of
equal bispectrum, and each class is one translation orbit. For the canonical forms of
Section~\ref{sec:classify}, the bispectrum resolves the bent classes that the periodogram leaves
together. Without restriction on the spectrum, averages of high enough order separate every
orbit.

\begin{proposition}[Completeness]
\label{prop:complete}
Let $G$ act on sign vectors by signed permutations. If $R_G^{(\varkappa)}(f)=R_G^{(\varkappa)}(f')$
for every $\varkappa\le2|G|$, then $f'=g\cdot f$ for some $g\in G$.
\end{proposition}

\begin{proof}
The polynomial $\mathcal Q(Y)=\prod_{g\in G}\|Y-P_gF'\|^2$ in $Y\in\mathbb R^N$ has degree $2|G|$, is invariant under $G$,
vanishes at $F'$ and is positive at every $Y$ outside the orbit of $F'$. The value of an invariant
polynomial at $F$ is its own mean over $G$, a linear function of the averages
$R_G^{(\varkappa)}(f)$ of the orders $\varkappa$ that occur in it, so equal averages give
$\mathcal Q(F)=\mathcal Q(F')=0$.
\end{proof}

Table~\ref{tab:order} lists the dimensions \eqref{eq:burnside}, computed from the distribution of
the fixed-space dimensions in $\GL(n,2)$ and checked by enumeration for $n=3$. The affine group
follows the Bell numbers of $S_N$ up to order three and leaves them at order four, where its
\OdAGLFourSets\ orbits on $4$-sets separate the affine planes from the other $4$-sets\iffull, and
its dimensions do not change with $n$ once $n\ge\varkappa-1$ in the computed range $n\le6$, \OdAGLSixFour\ for $n=4$ against
\OdAGLSixSix\ for $n=6$ at $\varkappa=6$\fi. Table~\ref{tab:ladder} places the statistics used in
switching theory and its neighbors by the order of the average they come from.

\iffull
\begin{table*}[t]
\centering
\caption{Dimension of the space $\mathcal C_\varkappa(G)$ of invariant arrays of order $\varkappa$, the number of orbits on $\varkappa$-tuples of input points (for $\EA$, of the signed invariant arrays), for $n=4$ and $n=6$. A dimension equal to that of $S_N$ means the average of that order sees the weight only.}
\label{tab:order}
\setlength{\tabcolsep}{3.2pt}
\begin{tabular}{@{}lrrrrrr@{\hspace{1.4em}}rrrrrr@{}}
\toprule
& \multicolumn{6}{c}{$n=4$, $\varkappa=1,\dots,6$} & \multicolumn{6}{c}{$n=6$, $\varkappa=1,\dots,6$}\\
\midrule
$T$ & 1 & 16 & 256 & 4{,}096 & 65{,}536 & 1{,}048{,}576 & 1 & 64 & 4{,}096 & 262{,}144 & 16{,}777{,}216 & 1{,}073{,}741{,}824\\
$S_n$ & 5 & 35 & 330 & 3{,}876 & 52{,}360 & 766{,}480 & 7 & 84 & 1{,}716 & 54{,}264 & 2{,}324{,}784 & 119{,}877{,}472\\
$B$ & 1 & 5 & 35 & 330 & 3{,}876 & 52{,}360 & 1 & 7 & 84 & 1{,}716 & 54{,}264 & 2{,}324{,}784\\
$\GL(n,2)$ & 2 & 5 & 16 & 67 & 373 & 2{,}761 & 2 & 5 & 16 & 67 & 374 & 2{,}825\\
$\AGL(n,2)$ & 1 & 2 & 5 & 16 & 67 & 373 & 1 & 2 & 5 & 16 & 67 & 374\\
$\EA$ & 0 & 1 & 0 & 5 & 0 & 67 & 0 & 1 & 0 & 5 & 0 & 67\\
$S_N$ & 1 & 2 & 5 & 15 & 52 & 203 & 1 & 2 & 5 & 15 & 52 & 203\\
\bottomrule
\end{tabular}
\end{table*}
\fi

\begin{table*}[t]
\centering
\caption{Statistics of switching functions ordered by the order $\varkappa$ of the group average
that produces them, with the group and the reason for the order. The rows above order four are
listed for completeness and are not used in this paper.}
\label{tab:ladder}
\setlength{\tabcolsep}{4pt}
\begin{tabular}{@{}cp{0.27\textwidth}p{0.15\textwidth}p{0.46\textwidth}@{}}
\toprule
$\varkappa$ & Statistic & Group & Why this order\\
\midrule
1 & weight $|S|$ and $W_f(0)$, weight profile, value $F(0)$ & $\AGL$, $S_n$, $\GL$ in turn & mean of $F$ over the orbits of points\iffull; orbit voting (Section~\ref{sec:voting})\fi\\
2 & periodogram $W_f^2$ and autocorrelation $r_f$ & $T$ & $R_T(f)_{x,y}=r_f(x\oplus y)/N$ (Proposition~\ref{prop:T})\\
2 & linear structures, strict avalanche and propagation criteria \cite{webster1985sbox} & $T$ & conditions $r_f(d)=\pm N$, and $r_f(d)=0$ for every $d$ of weight one, on the autocorrelation\\
2 & distance distribution, canonical periodogram & $B$, and $B$, $\AGL$, $\EA$ & Krawtchouk eigenvalues $\lambda_j$, index action of $G/T$ (Theorem~\ref{thm:normalizer})\\
2 & hash variance, affine hull and implied parities & $\AGL$, and $T$ & $\sum_{0\ne a\in\mathrm{row}(\Phi)}\Sh(a)^2$ (Theorem~\ref{thm:hash}), support of $\Sh$\iffull, Chrestenson periodogram over $\mathbb Z/m$ (Section~\ref{sec:rings})\fi\\
3 & bispectrum $\beta_f$ & $T$ & triple products with zero index sum (Theorem~\ref{thm:normalizer}, Proposition~\ref{prop:bisp})\\
3 & linearity test \cite{blum1990selftesting,bellare1995linearity} & $T$ & acceptance probability $\tfrac12+\tfrac12N^{-3}\sum_aW_f(a)^3$, a zero-sum triple average\\
3 & third moment of a hashed cell count & $\AGL$ & fixed by $|S|$, since $\AGL$ is homogeneous on $3$-sets (Theorem~\ref{thm:hash4})\\
4 & $\sum_aW_f^4$, the sum-of-squares indicator \cite{zhang1995gac}, additive energy \cite{tao2006additive}, $U^2$ norm \cite{gowers2001szemeredi}, participation ratio & $\EA$, $\AGL$ & first informative order (Theorem~\ref{thm:informative}, Proposition~\ref{prop:fourth}), with the bent functions as minima\\
4 & fourth moment of a hashed cell count & $\AGL$ & depends on $|S|$ and $\mathcal E(S)$ (Theorem~\ref{thm:hash4})\\
6 & $\sum_aW_f^6$ and the zero-sum $6$-tuples & $\EA$ & next order with new $\EA$ invariants (Table~\ref{tab:order})\\
8 & Gowers $U^3$ norm, testing of degree two \cite{alon2003lowdegree} & $T$ & averages over affine cubes of dimension three\\
16 & $U^4$ norm, testing of degree three \cite{alon2003lowdegree} & $T$ & averages over affine cubes of dimension four, and so on by powers of two\\
$\le2|G|$ & complete invariants & any $G$ & Proposition~\ref{prop:complete}\\
\bottomrule
\end{tabular}
\end{table*}

\subsection{Characteristic Two Halves the Group Gain of the Translations}
\label{sec:pairing}

The group gain of an average is the factor by which it reduces the variance of an
entry relative to the single unaveraged product. For the translations of $\F_2^n$ the
gain is half the group order.

\begin{proposition}[Pairing]
\label{prop:pairing}
Let the entries of $F$ be independent with $\mathbb{E}F(x)=\mu$. For $x\ne y$,
$\mathbb{E}R_T(f)_{x,y}=\mu^2$ and
\begin{equation}
\label{eq:pairing}
\Var R_T(f)_{x,y}=\frac{2(1-\mu^4)}{N},
\end{equation}
while the single product $F(x)F(y)$ has variance $1-\mu^4$. The group gain is $N/2$.
\end{proposition}

\begin{proof}
With $d=x\oplus y\ne0$, $NR_T(f)_{x,y}=\sum_zF(z)F(z\oplus d)$. The terms for $z$ and
$z\oplus d$ coincide, so the sum is twice a sum of $N/2$ products over disjoint pairs
$\{z,z\oplus d\}$. These products are independent with variance $1-\mu^4$.
\end{proof}

Over a cyclic group of the same order, a lag other than $N/2$ is not an involution,
and the corresponding average of a zero-mean white sequence has variance $1/N$. The
halving is specific to characteristic two, where every orbit of $T$ on pairs contains
each pair together with its transpose, so that the real product $F(x)F(y)$ is
averaged twice. The realized group gain of the translation average is therefore $N/2$,
half its effective dimension $N$, for second-order estimation from noisy or randomly
sampled truth tables.

\subsection{The Walsh Support Fixes the Matched Linear Coordinates}
\label{sec:support}

A function may depend on fewer linear forms than it has inputs. The forms, and the
affine subspace that contains a set, are read from spectra.

\begin{proposition}[Linear support]
\label{thm:support}
Let $U=\spn\supp W_f$ and $\ell=\dim U$. There is $A\in\GL(n,2)$ such that $f\circ A^{-1}$
depends on its first $\ell$ inputs only, and no invertible change of inputs gives fewer.
The translations that leave $f$ unchanged form the subspace $U^\perp$, so
$\ell=n-\log_2|\{d:r_f(d)=N\}|$.
\end{proposition}

\begin{proof}
$F(x\oplus d)=F(x)$ for all $x$ holds if and only if $W_f(a)(-1)^{a\cdot d}=W_f(a)$ for
all $a$, that is, if and only if $d\perp\supp W_f$. By inversion,
$F(x)=N^{-1}\sum_{a\in U}W_f(a)(-1)^{a\cdot x}$ depends on $x$ only through $a_i\cdot
x$ for a basis $a_1,\dots,a_\ell$ of $U$, and completing this basis to the rows of $A$
proves the first claim. If $f(x)=\varphi(Mx)$ with $M$ of $\ell'$ rows, then $\supp W_f$
lies in the row space of $M$, so $\ell\le\ell'$. Finally $r_f(d)=N$ exactly when $F(x\oplus
d)=F(x)$ for all $x$.
\end{proof}

Proposition~\ref{thm:support} collects classical facts on linear structures, the
translations with $|r_f(d)|=N$, of which those with $r_f(d)=-N$ complement the output
\cite{evertse1987,lai1994linear,dubuc2001linear,carlet2021book}. In the language of the framework, $U^\perp$ is the stabilizer
of $f$ in $T$, and $A$ is the matched change of coordinates that moves the
stabilizer onto coordinate translations, which removes the redundant inputs.

The affine subspace that contains a set is equally standard, and we state it in the
form used later. Two definitions come first. The annihilator of a set $\mathcal Z$ of vectors is
$\mathcal Z^\perp=\{a:a\cdot z=0\text{ for every }z\in\mathcal Z\}$, the parity checks that every vector of
$\mathcal Z$ passes, which in circuit terms are the XOR gates whose output is zero on every member of $\mathcal Z$. The
affine hull of a set $S$ of rows is the smallest affine subspace that contains it. Equivalently, it is
the set of rows that satisfy every XOR constraint $a\cdot x=\theta$ that all rows of $S$ satisfy, and
also the smallest set containing $S$ that contains $x\oplus y\oplus z$ whenever it contains $x$, $y$
and $z$. The parities that are constant on $S$ are the annihilator of the differences $x\oplus y$ of
its rows. For example, the rows covered by a product term form a subcube, which is its own hull. The
hull of two rows is the pair itself, and the three rows of weight one of a 3-cube share the single
constraint $x_1\oplus x_2\oplus x_3=1$, so their hull adds the fourth row of odd weight, $111$\iffull{}
(Fig.~\ref{fig:cubes}b)\fi. The ON-set of the majority of three shares no constraint, and its hull is
the whole cube.

\begin{proposition}[Affine hull]
\label{prop:hull}
For a nonempty set $S$, the set $U_S=\{a:|\Sh(a)|=|S|\}$ is a subspace. The affine
hull of $S$ is the solution set of the equations $a\cdot x=\theta_a$ for $a\in U_S$, with
$\theta_a=0$ when $\Sh(a)>0$ and $\theta_a=1$ otherwise. The set $S$ is an affine subspace if and
only if $|S|\cdot2^{\dim U_S}=N$.
\end{proposition}

\begin{proof}
$|\Sh(a)|=|S|$ holds if and only if $a\cdot x$ is constant on $S$, and the set of such
$a$ is closed under addition. An affine subspace contains $S$ if and only if it is cut
out by parities that are constant on $S$, so the hull is the solution set of all of
them, of size $N/2^{\dim U_S}$. The set $S$ is affine exactly when it equals its hull.
\end{proof}

\subsection{Hash Variance Is Spectral Energy on the Row Space}
\label{sec:hashvar}

A hash here is a map that sorts the rows of a truth table into a few bins, its cells. The hashes
used for counting are linear. A $k\times n$ matrix $\Phi$ over $\F_2$ is a bank of $k$ XOR gates,
each taking the parity of the inputs selected by one row of $\Phi$. The $k$-bit word they produce
names the cell of a row, so a cell is the set of rows on which the $k$ XOR gates produce one given
word, that is, the rows that satisfy $k$ XOR constraints with prescribed right-hand sides. A cell is a
subcube when every gate has a single input and an affine subspace in general\iffull{}
(Fig.~\ref{fig:cubes}c)\fi. These hashes are drawn at random from a family, and what matters is
statistical, the probability that two given rows share a cell. They are not cryptographic hashes,
which are single fixed functions such as SHA-256 designed so that producing two inputs with the
same output, a collision, or an input with a given output is computationally infeasible\iffull{}
(Section~\ref{sec:crypto} returns to them)\fi.

Let $\Phi$ be a $k\times n$ matrix of rank $k$ over $\F_2$. The cells of the hash
$x\mapsto\Phi x$ are the cosets $C_u=\{x:\Phi x=u\}$, $u\in\F_2^k$, each of size $N/2^k$. The
hashed count of a set $S$ is the estimator $Z=2^k|S\cap C_u|$ with $u$ uniform.

\begin{theorem}[Hash variance]
\label{thm:hash}
$\mathbb{E}Z=|S|$ and
\begin{equation}
\label{eq:hashvar}
\Var Z=\sum_{a\in\mathrm{row}(\Phi)\setminus\{0\}}\Sh(a)^2 .
\end{equation}
Over a uniformly random $k$-dimensional row space,
\begin{equation}
\label{eq:blind}
\mathbb{E}_\Phi\Var Z=\frac{2^k-1}{N-1}\bigl(N|S|-|S|^2\bigr),
\end{equation}
which depends on $S$ only through $|S|$.
\end{theorem}

\begin{proof}
Expanding the indicator of $\Phi x=u$ in characters of $\F_2^k$ gives
\[
|S\cap C_u|=2^{-k}\sum_{v\in\F_2^k}(-1)^{v\cdot u}\,\Sh(\Phi^{\mathsf T}v),
\]
so $Z$ is a Walsh series in $u$ with coefficients $\Sh(\Phi^{\mathsf T}v)$. Its mean over
uniform $u$ is the $v=0$ term $|S|$, and by Parseval its second moment is
$\sum_v\Sh(\Phi^{\mathsf T}v)^2$. Since $\Phi$ has full rank, $v\mapsto\Phi^{\mathsf T}v$
is a bijection onto the row space. For the random row space, each nonzero $a$ lies in it with probability
$(2^k-1)/(N-1)$, and $\sum_{a\ne0}\Sh(a)^2=N|S|-|S|^2$ by Parseval.
\end{proof}

The identity \eqref{eq:hashvar} is Poisson summation over the kernel of $\Phi$
\cite{odonnell2014,macwilliams1977}, and its value here is the separation of two regimes
that pairwise independence arguments treat together
\cite{carter1979universal,gomes2006xor,chakraborty2013approxmc}. The mean over random
row spaces, \eqref{eq:blind}, is at most the variance $(2^k-1)|S|$ that a pairwise
independent hash family guarantees, with near equality when $|S|$ is small compared
with $N$. A particular hash can do far better or far worse, according to how much
spectral energy of the solution set falls on its row space. Every parity implied
by $S$, an element of $U_S$ in Proposition~\ref{prop:hull}, contributes $|S|^2$, a
relative variance of at least one. Matching the row space to the solution set means
avoiding the large spectral values, and the relative variance of a chosen hash,
$\sum_{v\ne0}p_v^2$ with $p_v=\Sh(\Phi^{\mathsf T}v)/|S|$, can be estimated from $m$
independent uniform solutions $y_1,\dots,y_m$. The probability that two of them share a
cell is $2^{-k}(1+\sum_{v\ne0}p_v^2)$, so with $\Gamma$ the number of ordered pairs
$i\ne j$ with $\Phi y_i=\Phi y_j$,
\begin{equation}
\label{eq:pilot}
\widehat V=\frac{2^k\Gamma}{m(m-1)}-1
\end{equation}
is an unbiased estimate. Its standard deviation near a good hash is about
$\sqrt2\,2^{k/2}/m$\ifarxiv. Indeed, when the relative variance is near zero the cells
are nearly equally occupied, $\Gamma$ is twice the number of unordered colliding pairs, and
that number is close to a Poisson variable with mean $\binom m2 2^{-k}$, so that
$\Var\widehat V\approx 2^{2k}\cdot4\binom m22^{-k}/(m(m-1))^2=2^{k+1}/(m(m-1))$\fi, so the estimate resolves the relative variance only when $m$ is
of the order of $2^{k/2}$, a birthday bound. Implied parities are detected with far
fewer solutions. A parity constant on all of $S$ is constant on every sample, so the
rule that rejects any row space containing a parity constant on the sample rejects every
row space with an implied parity, whatever $m$ is, and needs only that the sample be
large enough to break the parities that are not implied.

\subsection{Matched Hashes Keep the Variance Bound of Pairwise Independence}
\label{sec:matched}

A rule that draws candidates and rejects some of them returns a hash that is not uniform over
the row spaces, so the mean \eqref{eq:blind} does not describe it. A matched hash can instead
be drawn uniformly from the row spaces that the hull rule accepts, and its variance then has a
closed form. Let $P$ be a set of solutions with first element $y_0$, fixed before the hash is
drawn. Let $U$ be the span of the differences $y\oplus y_0$ for $y\in P$, with $r=\dim U$, and
let $W=U^{\perp}$, the set of parities that are constant on $P$. A $k$-dimensional row space
$R$ is matched to $P$ when $R\cap W=\{0\}$, which is possible when $k\le r$.

\begin{proposition}[Matched hash variance]
\label{prop:matched}
Let $R$ be drawn uniformly among the $k$-dimensional row spaces that meet $W$ only in $0$, and
let the cell $u$ be uniform. Then $\mathbb{E}Z=|S|$ and
\begin{equation}
\label{eq:matched}
\begin{split}
\Var Z&=\frac{2^k-1}{N-2^{n-r}}\sum_{a\notin W}\Sh(a)^2\\
&=\frac{2^k-1}{2^r-1}\Bigl(2^r|S|-\sum_{c}|S\cap c|^2\Bigr),
\end{split}
\end{equation}
where the last sum runs over the $2^{n-r}$ cosets $c$ of $U$.
\end{proposition}

\begin{proof}
For every $R$, Theorem~\ref{thm:hash} gives $\mathbb{E}_uZ=|S|$, so the variance over the
joint draw is the mean over $R$ of $\sum_{0\ne a\in R}\Sh(a)^2$, that is,
$\sum_a\Sh(a)^2\Pr[a\in R]$. A nonzero vector of $W$ never lies in $R$. The invertible linear
maps that fix $W$ as a set preserve the distribution of $R$, and they act transitively on the
vectors outside $W$, since a basis of $W$ extended by $a$ can be mapped onto the same basis
extended by any other $a'\notin W$. Hence $\Pr[a\in R]$ is one value $p$ for every $a\notin W$.
The $2^k-1$ nonzero vectors of $R$ all lie outside $W$, so $(N-2^{n-r})p=2^k-1$, which gives
the first form. For the second, Poisson summation over $U$ gives
$\sum_{a\in W}\Sh(a)^2=2^{n-r}\sum_c|S\cap c|^2$, and Parseval gives $\sum_a\Sh(a)^2=N|S|$.
\end{proof}

When $P$ spans the whole space, $r=n$, $W=\{0\}$, and \eqref{eq:matched} is the blind mean
\eqref{eq:blind}. Two consequences make the draw usable in a counter.

\begin{corollary}[Pairwise bound]
\label{cor:pairwise}
Under the draw of Proposition~\ref{prop:matched}, $\Var Z\le(2^k-1)|S|$ for every set $S$ and
every pilot $P$. The count of the cell, $Z/2^k$, has mean $\mu=2^{-k}|S|$ and variance at most
$\mu(1-2^{-k})$.
\end{corollary}

\begin{proof}
Each solution lies in exactly one coset of $U$, so $\sum_c|S\cap c|^2\ge\sum_c|S\cap c|=|S|$,
and \eqref{eq:matched} is at most $(2^k-1)|S|$.
\end{proof}

\begin{corollary}[Zero variance and the certified hull]
\label{cor:zero}
Under the same draw, $\Var Z=0$ exactly when $S$ is a union of cosets of $U$. When $U$ is the
direction of the affine hull of $S$, of dimension $d$,
\begin{equation}
\label{eq:certhull}
\Var Z=(2^k-1)|S|\,\frac{2^d-|S|}{2^d-1},
\end{equation}
which is the blind mean with $N$ replaced by $2^d$.
\end{corollary}

\begin{proof}
Since $|S\cap c|\le2^r$, $\sum_c|S\cap c|^2\le2^r|S|$, with equality exactly when every
$|S\cap c|$ is $0$ or $2^r$. When $U$ is the direction of the hull, $r=d$ and $S$ lies in a
single coset, so the sum is $|S|^2$.
\end{proof}

The variance is the second moment of the cell count. The higher moments follow from the same
draw and show the threshold of Theorem~\ref{thm:informative}. Write $s=|S|$, let $n'$ be $n$ for
the uniformly random row space of Theorem~\ref{thm:hash} and $d$ for the matched draw of a
certified hull, and let
\[
\Upsilon_r=2^{-k}\binom{n'-r}{k}_{\!2}\Big/\binom{n'}{k}_{\!2},
\]
where the Gaussian binomial $\binom{n'}{k}_2$ counts the $k$-dimensional subspaces of
$\F_2^{n'}$, and likewise for $n'-r$, with the value zero when $n'-r<k$.

\begin{theorem}[Moments of a cell count]
\label{thm:hash4}
Under either draw, the cell count $c=|S\cap C_u|$ has
\begin{align*}
\mathbb Ec&=s\Upsilon_0,\qquad \mathbb Ec^2=s\Upsilon_0+s(s-1)\Upsilon_1,\\
\mathbb Ec^3&=s\Upsilon_0+3s(s-1)\Upsilon_1+s(s-1)(s-2)\Upsilon_2,\\
\mathbb Ec^4&=s\Upsilon_0+7s(s-1)\Upsilon_1+\bigl(6s(s-1)(s-2)+D_S\bigr)\Upsilon_2\\
&\quad+\bigl(s(s-1)(s-2)(s-3)-D_S\bigr)\Upsilon_3,
\end{align*}
where $D_S=\mathcal E(S)-3s^2+2s$ counts the ordered $4$-tuples of distinct points of $S$ with
zero sum. The moments up to the third depend on $S$ only through $s$, and the fourth through $s$
and the additive energy $\mathcal E(S)$.
\end{theorem}

\begin{proof}
$c^j$ counts the ordered $j$-tuples of $S$ that fall in the cell $u$. A tuple lies in one cell
exactly when the span of the differences $z^{(i)}\oplus z^{(1)}$, of some dimension $r$, lies in
$\ker\Phi$, that is, when every row of $\Phi$ gives parity zero on every difference, so that
$\mathrm{row}(\Phi)$ lies in the annihilator of the span of the differences. For a uniformly random
$k$-dimensional row space of an $n'$-dimensional space this has probability
$\binom{n'-r}{k}_2/\binom{n'}{k}_2$, and the common cell is $u$ with probability $2^{-k}$. Under
the matched draw with a certified hull, the row spaces that meet $W$ only in $0$ restrict to the
hull as the uniformly random $k$-dimensional row spaces of its $d$-dimensional dual, by the
transitivity argument of Proposition~\ref{prop:matched}, so the same count holds with $n'=d$. It
remains to sort the tuples by $r$. A tuple whose points are all equal has $r=0$, two distinct
points give $r=1$, and three distinct points give $r=2$, since their two differences are distinct
and nonzero. The patterns of equal points in a $j$-tuple are counted by the Stirling numbers of
the second kind, $1,3,1$ for $j=3$ and $1,7,6,1$ for $j=4$. Four distinct points give $r=2$ when
their sum is zero and $r=3$ otherwise. The ordered $4$-tuples of $S$ with zero sum number
$\mathcal E(S)$, and those with a repeated point number $3s^2-2s$.
\end{proof}

The moments up to the third are fixed by $s$ for the reason that $\AGL(n,2)$ averages of order
up to three are (Theorem~\ref{thm:informative}), and Proposition~\ref{prop:fourth} turns the
fourth into a spectral quantity, $\mathcal E(S)=N^{-1}\sum_a\Sh(a)^4$. The formulas agree with
enumeration over every row space and cell in \OdMomCases\ cases with $n=4$.
The equality of the first three moments with those of an ideal hash restates the 3-wise
independence of XOR hashes with a uniformly random offset, which XOR-based sampling and counting
already use \cite{gomes2006nips,chakraborty2013approxmc}, and the largest cell of a random linear
hash over GF(2) has been bounded separately \cite{alon1997linear,jaber2025linear}. The exact fourth
moment under both draws, with its excess over an ideal hash counted by the zero-sum quadruples
$D_S$, is the part we did not find in the papers we checked.
\ifarxiv
\paragraph*{Space compaction of test responses}
The same counting applies to a linear space compactor, a binary matrix applied to the outputs of
a circuit on every test pattern. A fault escapes exactly when the row space of the matrix is
orthogonal to the span of the error vectors of the fault over the test. A uniformly random
compactor with two parity outputs therefore misses a fault whose error vectors span one dimension
with probability one quarter, and the mean number of escapes is the sum of such terms over the
faults. On \CpCircuits\ ISCAS-85 circuits with \CpPatterns\ random patterns and \CpDetected\
detected single stuck-at faults, random compactors with two outputs let \CpRandomTwo\ faults
escape on average, two disjoint parity trees let \CpGroupsTwo\ escape, and compactors whose rows
are chosen against the error spaces let none escape. That two parity outputs suffice for these
circuits is known \cite{chakrabarty1998optimal,chakrabarty1998parity}. The identity places that
result beside the hash moments and adds no new compactor.
\fi
\iffull
\paragraph*{Generating functions and cumulants}
The moments of Theorem~\ref{thm:hash4} are the first terms of the moment generating function
$\mathcal M_c(\mathfrak t)=\mathbb Ee^{\mathfrak tc}=\sum_{j\ge0}\mathbb Ec^j\mathfrak t^j/j!$, and the cumulants $\mathrm{cum}_j(c)$ are
the coefficients of its logarithm, $\log\mathcal M_c(\mathfrak t)=\sum_{j\ge1}\mathrm{cum}_j(c)\mathfrak t^j/j!$. The first four
are the mean $\mathrm{cum}_1=\mathbb Ec$, the variance $\mathrm{cum}_2$, and $\mathrm{cum}_3=\mathbb E(c-\mathbb Ec)^3$ and
$\mathrm{cum}_4=\mathbb E(c-\mathbb Ec)^4-3\,\mathrm{cum}_2^2$, which measure asymmetry and tail weight. By the
theorem, the first three cumulants depend on $S$ only through $s$, and the additive energy enters
first in $\mathrm{cum}_4$, which is the threshold of Theorem~\ref{thm:informative} seen as a statement about
tails. The same objects describe a switching function read as a probability mass function. For a
random row $X$ with distribution $\mathsf p$ on $\F_2^n$, the characteristic function on the group is
$\phi_X(a)=\mathbb E(-1)^{a\cdot X}=\sum_x\mathsf p(x)(-1)^{a\cdot x}$, the Walsh transform of $\mathsf p$, and for $X$
uniform on the ON-set, $\phi_X(a)=\Sh(a)/|S|$. If $X$ and $Y$ are independent, the XOR $X\oplus Y$ has
$\phi_{X\oplus Y}=\phi_X\phi_Y$, the convolution theorem of $\mathbb Z_2^n$, so $\log|\phi|$ adds under
XOR as a cumulant generating function adds under sums. For single bits with $\mathbb P(X_i=1)=\mathsf p_i$
this is the piling-up lemma of linear cryptanalysis, $\mathbb E(-1)^{X_1\oplus\cdots\oplus
X_m}=\prod_i(1-2\mathsf p_i)$ \cite{matsui1993linear}, which Section~\ref{sec:crypto} uses for bit streams.
\fi
\ifarxiv
Theorem~\ref{thm:hash4} gives a fourth-moment bound on the probability that one estimate leaves
$[|S|/c,\,c|S|]$, and we tested whether it lowers the number of estimates whose median a counter
needs. On the \FoOutputs\ outputs of Section~\ref{sec:hash}, at \FoLevels\ levels with $k$ chosen
so that the mean cell count lies between \FoMuLo\ and \FoMuHi, with $c=1.8$ and a target miss
probability of $0.05$ for the median, the pairwise bound of Corollary~\ref{cor:pairwise} needs a
median of \FoTPair\ estimates and the exact variance \FoTVar. The fourth-moment bound alone needs
more than the exact variance at \FoWorse\ levels. Taking the smaller of the two bounds lowers the
count at \FoBetter\ levels (\FoBetterPct\ percent), by \FoRepsSaved\ estimates in all, and with
the largest additive energy that $s$ points allow, $s(s-1)(s-2)$ for $D_S$, the fourth-moment bound
improves on the pairwise bound at \FoWorstBetter\ levels. The outputs carry far more additive
structure than random sets, with a median additive energy \FoEnergyMedian\ times that of a random
set of the same size, and that structure inflates the fourth moment. The fourth moment therefore
explains the tails of a cell count but does not shorten a counter, and we keep it as this remark.
\fi

The analyses of hashing counters use the hash family to bound the variance of a cell count by
its mean \cite{chakraborty2016approxmc2,agrawal2020sparsity}, and Corollary~\ref{cor:pairwise}
gives that bound for matched hashes whatever the pilot, including the correlated solutions
that a solver returns, as long as the pilot is fixed before the hash is drawn. Counters in the
form of ApproxMC2 search over the prefixes of one hash and apply the bound level by level
\cite{chakraborty2016approxmc2}. Drawing row $i$ uniformly outside the span of $W$ and rows
$1$ to $i-1$ gives, at every prefix length $m\le r$, a uniform $m$-dimensional row space that
meets $W$ only in $0$, because the number of choices at each step does not depend on the
earlier choices. Proposition~\ref{prop:matched} therefore holds at every level $m\le r$, and
the cells stay nested as before. With a certified hull the search needs at most $d$ levels,
since a matched row space of dimension $d$ is injective on the hull and every cell then holds
at most one solution. The draw does not always lower the variance. By the Cauchy--Schwarz
inequality the ratio of \eqref{eq:matched} to \eqref{eq:blind} is at most
$2^r(N-1)/(N(2^r-1))$, about $1+2^{-r}$, which is approached when $S$ fills the cosets of $U$
evenly, so a smaller $U$ can remove more spectral energy than the exact hull does.

The implied parities can also be drawn from a formula rather than from sampled solutions,
with a satisfiability oracle.

\begin{proposition}[Affine hull from a satisfiability oracle]
\label{prop:oraclehull}
Let $\phi$ be a satisfiable formula whose solution set $S$ has an affine hull of dimension $d$.
Find a solution $y_0$ and set $V=\{0\}$. Repeatedly ask for a solution $y$ of $\phi$ with
$w\cdot y\ne w\cdot y_0$ for some $w$ in a basis of $V^{\perp}$, and add $y\oplus y_0$ to $V$
when one is found. The procedure stops after at most $d+2$ calls, and $y_0\oplus V$ is then the
affine hull of $S$.
\end{proposition}

\begin{proof}
A returned $y$ violates a parity of $V^{\perp}$ relative to $y_0$, so $y\oplus y_0\notin V$
and $\dim V$ grows by one. $V$ is spanned by differences of solutions, so it lies in the
direction of the hull and grows at most $d$ times. An unsatisfiable call shows that every
solution satisfies every parity of $V^{\perp}$ relative to $y_0$, so $S\subseteq y_0\oplus V$,
and $y_0\oplus V$, which is spanned by solutions, lies in the hull. Counting the call that
finds $y_0$, at most $d$ calls that succeed and the final call gives $d+2$.
\end{proof}

The query asks for a violation of at least one of $s$ parities. With a fresh variable $z_i$
equal to $w_i\cdot x\oplus w_i\cdot y_0$ for each basis vector $w_i$ and the clause
$z_1\vee\dots\vee z_s$, it is a formula with parity constraints, which a solver with Gaussian
elimination handles directly \cite{soos2009cms}. A pilot can seed $V$ and save calls.

\iffull
\subsection{Orbit Voting Realizes the Continuum between Group and Repetition}
\label{sec:voting}

Suppose the truth table is observed through independent bit flips, $t(x)\oplus\xi(x)$
with $\Pr[\xi(x)=1]=\epsilon<1/2$, and let $G$ stabilize $f$. Orbit voting sets each
observed bit to the majority of the observed bits on its orbit $Gx$, keeping the
observed bit when the vote is tied.

\begin{proposition}[Orbit voting]
\label{prop:voting}
The orbit-voted bit at $x$ is wrong with probability $e(|Gx|,\epsilon)$, where
$e(L,\epsilon)$ is the error probability of a majority over $L$ independent repeated
evaluations of $t(x)$ with the same tie rule,
\begin{equation}
\label{eq:vote}
\begin{split}
e(L,\epsilon)={}&\sum_{2j>L}\binom Lj\epsilon^j(1-\epsilon)^{L-j}\\
&+\tfrac12\binom{L}{L/2}\epsilon^{L/2}(1-\epsilon)^{L/2}\,[L\text{ even}].
\end{split}
\end{equation}
\end{proposition}

\begin{proof}
$t$ is constant on $Gx$, so the observed bits on the orbit are $|Gx|$ independent noisy
copies of $t(x)$. On a tie the kept bit is wrong exactly when it was flipped, which
happens for half of the tied flip patterns by symmetry.
\end{proof}

Proposition~\ref{prop:voting} is the switching-algebra form of the $(G,L)$ continuum of
the framework \cite{thornton2026principles}. The group supplies repetitions without
re-evaluating the function, and $L$ independent repeated tables combine with an orbit of
$|Gx|$ points as a majority over $|Gx|L$ copies. Since $e(2i,\epsilon)=e(2i-1,\epsilon)$
for every $i\ge1$, an orbit of two points gives no correction, so a single involution
acting alone does not help. A group that does not stabilize $f$
merges orbits with different values and biases the vote, which is why the symmetries
must be matched to the function. Section~\ref{sec:noisy} discovers them from the noisy
table itself.

A group discovered from noisy data can contain near-symmetries, and the vote then mixes
points of different true values. The bias of such a merge and the variance it removes can
be weighed orbit by orbit. Let $O$ be an orbit of $k\ge2$ points of any group $H$ acting
on the inputs, let $y(x)=t(x)\oplus\xi(x)$ be the observed bits, and let
$m=\frac1k\sum_{x\in O}y(x)$ be their mean on the orbit. The expected observed bit is
$q(x)=\epsilon+(1-2\epsilon)t(x)$, and $q(x)>1/2$ exactly when $t(x)=1$, so estimating
$q$ and thresholding at one half recovers $t$. The shrinkage family
\begin{equation}
\label{eq:shrinkvote}
\hat s_\beta(x)=(1-\beta)\,y(x)+\beta\,m,\qquad 0\le\beta\le1,
\end{equation}
runs from the observed bit at $\beta=0$ to the orbit average at $\beta=1$, whose threshold
is the orbit vote. Write $\nu=\epsilon(1-\epsilon)$ for the variance of one observed bit
and $p$ for the fraction of points of $O$ with $t(x)=1$.

\begin{proposition}[Shrinkage weight on an orbit]
\label{prop:shrinkvote}
The squared error of \eqref{eq:shrinkvote} summed over the orbit is
\begin{equation}
\label{eq:shrinkrisk}
\sum_{x\in O}\mathbb{E}\bigl[(\hat s_\beta(x)-q(x))^2\bigr]
=k\nu-(k-1)\nu\,(2\beta-\beta^2)+\beta^2B,
\end{equation}
where $B=(1-2\epsilon)^2k\,p(1-p)$ is the bias energy of the orbit average. It is
minimized at
\begin{equation}
\label{eq:betastar}
\beta^\star=\frac{(k-1)\nu}{(k-1)\nu+B},
\end{equation}
and $\mathbb{E}[k\,m(1-m)]=(k-1)\nu+B$, so
$\hat\beta=\min\{1,(k-1)\nu/(k\,m(1-m))\}$ is the plug-in estimate of $\beta^\star$.
\end{proposition}

\begin{proof}
Write $y(x)=q(x)+e(x)$, where the $e(x)$ are independent with mean zero and variance
$\nu$, and let $\bar q$ and $\bar e$ be the means of $q$ and $e$ over $O$, so that
$m=\bar q+\bar e$. Then
$\hat s_\beta(x)-q(x)=(1-\beta)e(x)+\beta\bar e-\beta\,(q(x)-\bar q)$. The last term is
deterministic and the first two have mean zero, so the expected cross term vanishes and
the summed squared error is the sum of a noise part and a bias part. For the noise part,
$\mathbb{E}[e(x)\bar e]=\mathbb{E}[\bar e^{\,2}]=\nu/k$, hence
\[
\begin{split}
\sum_{x\in O}\mathbb{E}\bigl[((1-\beta)e(x)+\beta\bar e)^2\bigr]
&=k(1-\beta)^2\nu+(2\beta-\beta^2)\nu\\
&=k\nu-(k-1)\nu(2\beta-\beta^2).
\end{split}
\]
For the bias part, $q(x)-\bar q=(1-2\epsilon)(t(x)-p)$ and
$\sum_{x\in O}(t(x)-p)^2=k\,p(1-p)$ because $t$ takes values in $\{0,1\}$, which gives
$\beta^2B$ and proves \eqref{eq:shrinkrisk}. Its derivative in $\beta$ is
$-2(k-1)\nu(1-\beta)+2\beta B$, which vanishes at \eqref{eq:betastar}, and its second
derivative $2(k-1)\nu+2B$ is positive. Finally, $y(x)^2=y(x)$ gives
$\sum_{x\in O}(y(x)-m)^2=k\,m(1-m)$, and the same decomposition gives
\[
\begin{split}
\mathbb{E}\sum_{x\in O}(y(x)-m)^2&=\sum_{x\in O}\mathbb{E}\bigl(e(x)-\bar e\bigr)^2\\
&\quad+\sum_{x\in O}\bigl(q(x)-\bar q\bigr)^2=(k-1)\nu+B.
\end{split}
\]
\end{proof}

Proposition~\ref{prop:shrinkvote} is the orbit form of the shrinkage between a statistic
and its group average in the framework \cite{thornton2026shrinkage}. The weight compares
the noise that the average removes, $(k-1)\nu$, with the bias it introduces, $B$. On an
orbit of a true symmetry $B=0$ and $\beta^\star=1$. The plug-in $\hat\beta$ falls below one
whenever the observed spread exceeds $(k-1)\nu$. On a matched orbit with
$(k-1)\nu<1/2$, a single flipped bit gives $k\,m(1-m)=(k-1)/k$, hence $\hat\beta=k\nu$
and $\hat s_{\hat\beta}=1-(k-1)\nu>1/2$ at the flipped point when its true value is zero,
so the plug-in alone would leave that flip uncorrected. The plug-in is therefore used only
on orbits whose observed bits are not consistent with a constant value. Under a constant value the number of flipped bits on
$O$ is $J\sim\mathrm{Bin}(k,\epsilon)$ and the observed minority count is
$\min(J,k-J)$. The tested vote compares the observed minority count with this exact
distribution. When the one-sided $p$-value exceeds a level $\alpha$ the orbit votes by
majority, and otherwise each bit becomes the threshold at one half of
$\hat s_{\hat\beta}(x)$, with ties keeping the observed bit.

\begin{proposition}[Tested vote on a matched orbit]
\label{prop:testedvote}
If $t$ is constant on an orbit of $k$ points, the tested vote at level $\alpha$ returns the
orbit majority with probability at least $1-\alpha$, and each of its bits on the orbit is
wrong with probability at most $e(k,\epsilon)+\alpha$.
\end{proposition}

\begin{proof}
Under a constant value the observed minority count has exactly the distribution of
$\min(J,k-J)$ used by the test, so the event that its $p$-value is at most $\alpha$ has
probability at most $\alpha$, as for any test with an exact null distribution. Outside
that event $\beta=1$ and the bit is the majority with ties kept, which is wrong with
probability $e(k,\epsilon)$ by Proposition~\ref{prop:voting}. The union bound over the two
events gives the second statement.
\end{proof}

On an orbit that merges points of different values the expected spread exceeds its noise
level by the bias energy $B$, so the test tends to reject such an orbit more often as $k$
and $(1-2\epsilon)^2p(1-p)$ grow. When it rejects, the plug-in weight is small and most
observed bits are kept, so the error on the minority points of the orbit is near
$\epsilon$ rather than near the error of a vote that overrules them. An orbit of two or three points cannot separate one flipped bit from
one point of the other value, because both produce the same observation.

\subsection{A Matched Group Stands in for Repeated Observations}
\label{sec:observations}

In signal processing a covariance estimate needs many snapshots, and a matched group average
of a single snapshot can stand in for most of them \cite{thornton2026ad,thornton2026principles}.
For switching functions the observations are evaluations of the function, or complete truth
tables read through noise, and the stabilizer of the function plays the role of the matched
group. Let $G$ be a group of input transformations with $f\circ g=f$ for every $g\in G$, let
$O_1,\dots,O_r$ be its orbits on $\F_2^n$, and for a flip rate $\epsilon<1/2$ and a target
error $\delta>0$ let $L_\delta$ be the least $L$ with $e(L,\epsilon)\le\delta$, where
$e(L,\epsilon)$ is the vote error of \eqref{eq:vote}.

\begin{lemma}[Monotone vote error]
\label{lem:monotone}
For $0<\epsilon<1/2$ and every $i\ge1$, $e(2i,\epsilon)=e(2i-1,\epsilon)$ and
$e(2i+1,\epsilon)<e(2i-1,\epsilon)$. Hence $e(L,\epsilon)$ does not increase with $L$.
\end{lemma}

\begin{proof}
Let $J$ be the number of flipped copies among the first $2i-1$ and let $X,Y$ indicate whether
two further copies are flipped. With $2i-1$ copies the vote is wrong exactly when $J\ge i$, and
with $2i+1$ copies exactly when $J+X+Y\ge i+1$. The second event differs from the first only
when $J=i-1$ and $X=Y=1$, where it becomes wrong, or when $J=i$ and $X=Y=0$, where it becomes
right. With $c=\binom{2i-1}{i-1}=\binom{2i-1}{i}$,
\[
\begin{split}
e(2i+1,\epsilon)-e(2i-1,\epsilon)&=c\,\epsilon^{i-1}(1-\epsilon)^{i}\epsilon^2\\
&\quad-c\,\epsilon^{i}(1-\epsilon)^{i-1}(1-\epsilon)^2\\
&=c\,\epsilon^{i}(1-\epsilon)^{i}\,(2\epsilon-1)<0 .
\end{split}
\]
For $2i$ copies, condition on the first $2i-1$. If they hold at least $i+1$ flips or at most
$i-2$ flips the last copy cannot change the outcome. If they hold $i$ flips the vote over $2i-1$
is wrong, and with the last copy it stays wrong when that copy is flipped and becomes a tie
when it is not, which keeps a wrong observed bit with probability one half by the tie rule.
If they hold $i-1$ flips the vote over $2i-1$ is right, and with the last copy it stays right
when that copy is not flipped and becomes a tie when it is. Summing,
$e(2i,\epsilon)-e(2i-1,\epsilon)=\tfrac12 c\,\epsilon^{i-1}(1-\epsilon)^{i}\epsilon
-\tfrac12 c\,\epsilon^{i}(1-\epsilon)^{i-1}(1-\epsilon)=0$.
\end{proof}

\begin{proposition}[Observations a matched group saves]
\label{prop:observations}
Let $G$ fix $f$, with orbits $O_1,\dots,O_r$.
\begin{enumerate}
\item The values of $f$ at one point of each orbit determine $f$, and
$r=|G|^{-1}\sum_{g\in G}|\{x:gx=x\}|$.
\item If $K$ independent truth tables of $f$ are read through flips of probability $\epsilon$
and every bit at $x$ is replaced by the majority, ties keeping the observed bit, over the $K|Gx|$
observed bits on the orbit of $x$ in all tables, the bit at $x$ is wrong with probability
$e(K|Gx|,\epsilon)$. It is at most $\delta$ for every $x$ when
$K\ge\lceil L_\delta/s\rceil$, where $s=\min_i|O_i|$, while without the group $K=L_\delta$
tables are required.
\item $e(L,\epsilon)\le(4\epsilon(1-\epsilon))^{L/2}$, so that
$L_\delta\le\bigl\lceil 2\ln(1/\delta)/\ln\bigl(1/(4\epsilon(1-\epsilon))\bigr)\bigr\rceil$.
\end{enumerate}
\end{proposition}

\begin{proof}
For the first statement, $f$ is constant on each orbit, so its value at one point of an orbit
is its value on the whole orbit. The orbit count is Burnside's lemma
\cite{burnside1911,neumann1979burnside}. For the second, $f$ is constant on $Gx$, so the
$K|Gx|$ observed bits on the orbit in the $K$ tables are independent noisy copies of $f(x)$,
and Proposition~\ref{prop:voting} applies with $L=K|Gx|$. If $K\ge\lceil L_\delta/s\rceil$ then
$K|Gx|\ge Ks\ge L_\delta$, and Lemma~\ref{lem:monotone} gives
$e(K|Gx|,\epsilon)\le e(L_\delta,\epsilon)\le\delta$. For the third, the vote is wrong only if the
number $J$ of flipped copies satisfies $J\ge L/2$, ties included. For every $\lambda\ge1$,
Markov's inequality applied to $\lambda^J$ gives
$\Pr[J\ge L/2]\le\lambda^{-L/2}\,\mathbb{E}\lambda^J=\lambda^{-L/2}(1-\epsilon+\epsilon\lambda)^L$.
The choice $\lambda=(1-\epsilon)/\epsilon$, which is at least one because $\epsilon<1/2$, gives
$\lambda^{-L/2}(2(1-\epsilon))^L=(4\epsilon(1-\epsilon))^{L/2}$. Solving
$(4\epsilon(1-\epsilon))^{L/2}\le\delta$ for $L$ gives the bound on $L_\delta$.
\end{proof}

The first statement is the reduction in the number of snapshots. A function fixed by a
stabilizer with $r$ orbits is a vector of $r$ free values rather than $N$, and the ratio $N/r$,
the mean orbit size, plays the part of the effective dimension. A totally symmetric function
of $n$ inputs has $r=n+1$, a function symmetric in two halves of its inputs has
$(\lfloor n/2\rfloor+1)(\lceil n/2\rceil+1)$ orbits, and a function with a $d$-dimensional
stabilizer among the translations has $2^{n-d}$. The second and third statements are the group
gain. Each point of an orbit of $s$ points contributes an independent repetition, the
per-observation exponent $\ln(1/(4\epsilon(1-\epsilon)))/2$ is multiplied by $s$, and the
number of noisy tables falls by the factor $s$. Fig.~\ref{fig:observations}(a) shows the
number of tables needed for a per-bit error of $10^{-6}$. At $\epsilon=0.1$ a vote without
symmetry needs \AnLTableC\ tables, orbits of $4$ points need \AnLTableCfour, and orbits of
$16$ points need \AnLTableCsixteen. Fig.~\ref{fig:observations}(b) shows the number of
evaluations that identify a function for several stabilizers. The exponential bound of the
third statement is conservative. For flip rates from $0.001$ to $0.4$ and votes of up to $512$
copies the exact error stays below \AnChernoffWorst\ times the bound, and for a target of
$10^{-6}$ the bound overstates $L_\delta$ by at most \AnChernoffOver\ percent.

\begin{figure}[t]
\centering
\includegraphics[width=\textwidth]{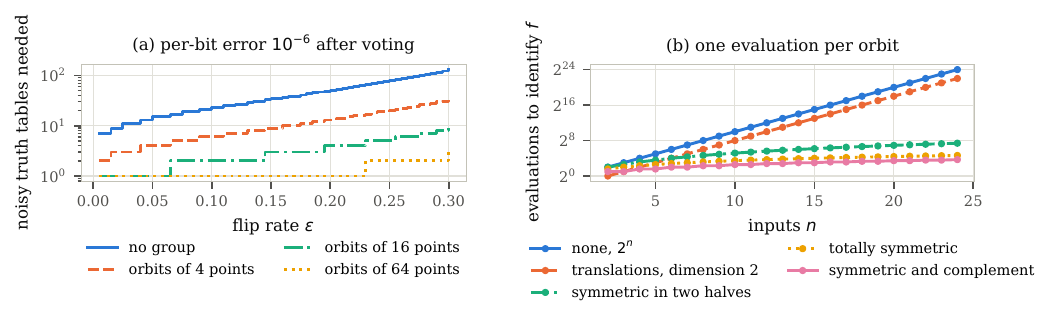}
\caption{Observations saved by a matched group. (a) Number of noisy truth tables that a vote
needs for a per-bit error of $10^{-6}$, against the flip rate, without symmetry and for orbits
of uniform size. (b) Number of evaluations that identify a function, the number of orbits of
its stabilizer, against the number of inputs.}
\label{fig:observations}
\end{figure}

\begin{example}[Majority of five inputs, continued]
\label{ex:majvote}
The stabilizer $S_5$ of Example~\ref{ex:maj} has \AnMajOrbits\ orbits, so \AnMajOrbits\
evaluations identify the function. Its smallest orbits, $\{0\}$ and $\{\one\}$, have one point
each, so the second statement of Proposition~\ref{prop:observations} gives no reduction for a
uniform target error. The mean error over the table falls nevertheless, because most points lie
in orbits of $5$ or $10$ points. With one noisy table at $\epsilon=0.05$ the expected bit error
after voting is \AnMajBitErrB, and the probability that the table holds a wrong bit falls from
\AnMajRawTableB\ to \AnMajTableB.
\end{example}

\fi

\subsection{The Participation Ratio Measures Spectral Spread}
\label{sec:pr}

For nonnegative weights $\mathfrak w_1,\dots,\mathfrak w_{\mathfrak q}$, not all zero, the participation ratio is
\[
\mathrm{PR}(\mathfrak w)=\frac{\bigl(\sum_i\mathfrak w_i\bigr)^2}{\sum_i\mathfrak w_i^2},
\]
which lies between $1$, when one weight carries everything, and the number of nonzero weights,
when they are equal, and it counts how many weights effectively participate. The same ratio
has a name in many fields. In the physics of disordered solids it counts the atoms that take part in
a vibration mode \cite{bell1970vitreous}, and its inverse measures localization \cite{thouless1974}.
In ecology it is the inverse Simpson index, the number of equally common species with the same
diversity, Hill's number of order two \cite{hill1973}. In survey statistics and Monte Carlo it is
Kish's effective sample size $(\sum_i\mathfrak w_i)^2/\sum_i\mathfrak w_i^2$ of a set of sampling weights \cite{kish1965}, which
sequential importance sampling and particle filters monitor to decide when to resample
\cite{kong1994sequential}. In statistical and array signal processing, applied to the eigenvalues of
a covariance matrix, it is an effective rank, the number of dimensions that the signals occupy, and
Roy and Vetterli's effective rank is the Shannon member of the same family \cite{roy2007effective}.
The eigenvalues $W_f(a)^2/N$ of $R_T(f)$ sum to $N$, and their participation ratio
\cite{hill1973} is
\begin{equation}
\label{eq:D2}
D_2(f)=\frac{N^4}{\sum_aW_f(a)^4}=\frac{N^3}{\sum_dr_f(d)^2}.
\end{equation}

\begin{proposition}[Range of the participation ratio]
\label{prop:pr}
$1\le D_2(f)\le N$, with $D_2(f)=1$ if and only if $f$ is affine and $D_2(f)=N$ if and
only if $f$ is bent.
\end{proposition}

\begin{proof}
$\sum_aW_f^4\le\max_aW_f^2\sum_aW_f^2\le N^4$, with equality if and only if a single
$|W_f(a)|$ equals $N$, which characterizes affine functions. By Cauchy--Schwarz,
$N^4=(\sum_aW_f^2)^2\le N\sum_aW_f^4$, with equality if and only if $W_f^2$ is
constant. The second form of \eqref{eq:D2} is Parseval's identity for $W_f^2=Hr_f$.
\end{proof}

The participation ratio $D_2(f)$ counts the Walsh components the function effectively
occupies. The sum $\sum_dr_f(d)^2$ is the sum-of-squares indicator of the global
avalanche characteristics \cite{zhang1995gac}, and Proposition~\ref{prop:pr} restates its
classical bounds. When $R_T(f)$ is taken as the covariance, the structural capacity
$(\operatorname{tr}R_T(f))^2/\|R_T(f)\|_F^2$ of \cite{thornton2026principles} is $D_2(f)$ itself. Affine functions sit at
its floor of one, where one component carries all the energy, and bent functions at the ceiling
$N$.
For a designer the ratio answers how many Walsh coefficients a function effectively uses. A
function with $D_2(f)$ near one is close to a single XOR of inputs, small when realized with XOR gates,
easy to approximate linearly and therefore a weak component for a cipher, while $D_2(f)$ near $N$
marks a bent-like function that no affine function approximates well. In hashed counting a small
$D_2$ of the on-set indicator warns that a random XOR hash can land on a heavy coefficient and give a
large variance (Theorem~\ref{thm:hash}).
\iffull

\paragraph*{The R\'enyi family}
The normalized power spectrum $q(a)=W_f(a)^2/N^2$ is a probability mass function on $\F_2^n$ by
Parseval's identity. R\'enyi's entropies of order $\alpha\ge0$ \cite{renyi1961measures},
\[
H_\alpha(q)=\frac1{1-\alpha}\log_2\sum_aq(a)^\alpha,
\]
with the limits $\alpha\to1$ and $\alpha\to\infty$, give one family of spectral spreads. Order $0$ is
the Hartley entropy, the logarithm of the number of nonzero Walsh coefficients, which is the size
of the support whose span is the linear support of Proposition~\ref{thm:support}. Order $1$ is
Shannon's entropy of the spectrum. Order $2$ is the collision entropy, and $D_2(f)=2^{H_2(q)}$ is the
participation ratio. Order $\infty$ is the min-entropy $-\log_2\max_aq(a)$, which is
$2\log_2(N/\max_a|W_f(a)|)$ and so a function of the nonlinearity
$\mathrm{NL}(f)=N/2-\max_a|W_f(a)|/2$, the distance from $f$ to the nearest affine function. The
numbers $2^{H_\alpha}$ are Hill's numbers of order $\alpha$ \cite{hill1973}. They decrease in
$\alpha$, all equal $1$ exactly for affine functions and all equal $N$ exactly for bent functions, so
every member of the family is a grade of distance from the affine functions, and the members differ
in how much weight they give to the largest coefficients.
\fi

\begin{remark}[Prime fields]
\label{rem:prime}
The constructions carry over to $\F_p^n$ for a prime $p$, with the characters of
$\mathbb{Z}_p^n$ (the Chrestenson transform) in place of the Walsh functions. The
affine group $\AGL(n,p)$ is still two-transitive, so its average depends only on how
many inputs take each output value, and $\GL(n,p)$ has $p+3$ orbits on pairs for $n\ge2$,
one for each nonzero scalar ratio $y=\alpha x$ in place of the single diagonal. For odd $p$ no
nonzero translation is an involution, and Proposition~\ref{prop:pairing} no longer
halves the gain.
\end{remark}

\section{Applications}
\label{sec:apps}

This section evaluates the application of the theoretical results to the practical switching theory
problems known as approximate model counting\iffull, satisfiability of parity-structured and counting
formulas, Boolean matching for technology mapping and the correction of bit errors in stored truth
tables\fi. Each application opens
with the problem, the methods in current use, what the matched group adds and how the experiments
test it. A census of benchmark circuits comes first, since every application assumes that the
functions met in practice carry structure that a group can match.

\subsection{Benchmark Outputs Carry Substantial Group Structure}
\label{sec:survey}

Each application relies on a structure of the function, a translation that fixes it, a parity
that is constant on its on-set or a pair of inputs that can be exchanged, and the gain of a matched
group vanishes when that structure is absent. Whether the functions of real circuits carry it is
therefore an empirical question, and the census answers it for a public benchmark set. The benchmark set consists of \SvSmall\ two-level circuits with 5 to 12 inputs and \SvLarge\
larger LGSynth91 circuits with \SvLargeNmin\ to \SvLargeNmax\ inputs. All the small circuits
are public. Of them, \SvLgsPla\ are distributed as PLAs with the LGSynth91 collection
\cite{yang1991lgsynth}, and \SvBerkeley\ belong to the Berkeley PLA test set distributed with
the Espresso minimizer \cite{brayton1984espresso}. For \SvBerkeleyCoverWord\ of these,
\SvBerkeleyCoverNames, the file used here is a different cover of the same functions. The other
\SvCollapsedWord, \SvCollapsedNames, are distributed with LGSynth91 as multilevel circuits,
\texttt{alu2c} is the six-output LGSynth91 circuit \texttt{alu2}, which differs from the
eight-output test-set PLA of that name. Each file that differs from its public source computes
the same functions output by output, as a comparison of binary decision diagrams confirms.
\ifarxiv Appendix~\ref{app:sources} lists the collection and the public copy of every
circuit.\fi{}
Every nonconstant output, \SvOutputs\ outputs
of \SvCircuits\ circuits, was analyzed exactly from its Walsh, indicator and
autocorrelation spectra. Table~\ref{tab:survey} summarizes the census.

\begin{table}[t]
\centering
\caption{Group structure of the 688 nonconstant outputs of 51 benchmark circuits.}
\label{tab:survey}
\begin{tabular}{@{}lrr@{}}
\toprule
Property of the output & outputs & percent\\
\midrule
fixed by a nonzero translation & 408 & 59\\
not using every input & 389 & 57\\
fewer linear forms than used inputs & 50 & 7\\
complemented by a translation & 40 & 6\\
on-set in a proper affine subspace & 437 & 64\\
on-set an affine subspace & 175 & 25\\
on-set a single cube & 152 & 22\\
symmetric pair of inputs ($n\le16$) & 421 & 78\\
participation ratio below $\sqrt N$ & 635 & 92\\
\bottomrule
\end{tabular}
\end{table}

Group structure is common in these outputs. \SvTransPct\ percent of the
outputs are fixed by a nonzero translation, mostly because they do not use every input of
their circuit, and \SvHiddenPct\ percent (\SvHidden\ outputs) depend on fewer linear forms
than the inputs they use, which Proposition~\ref{thm:support} detects and removes by a
change of coordinates. The sum outputs of adders, the symmetric function \texttt{9sym},
the parity outputs of the \texttt{rd} circuits and the \texttt{cordic} outputs are
examples, the last depending on \SvCordicK\ linear forms of
their \SvCordicUsed\ inputs. \SvHullPct\ percent of the on-sets lie in a proper affine
subspace, with \SvHullMedian\ independent implied parities at the median, and
\SvAffinePct\ percent are affine subspaces, \SvCubeOfAffinePct\ percent of these being single cubes. Among the outputs
with at most 16 inputs, \SvPairPct\ percent are unchanged by the exchange or the
complemented exchange of some pair of inputs. The spectra are concentrated. The
participation ratio of Section~\ref{sec:pr} is below $\sqrt N$ for \SvDtwoLowPct\
percent of the outputs. These are the structures the following applications exploit,
the implied parities in hashing and satisfiability, the translation stabilizer in
classification and the symmetric pairs in error correction.

\subsection{Approximate Model Counting}
\label{sec:counting}

A switching-theory problem that recurs in verification and security is counting the satisfying
assignments of a function given as a circuit or a formula. The count measures, for example, the
input patterns that expose a fault, the probability that a random input violates a property and the
keys of a locked circuit that agree with a set of oracle observations \cite{thornton2026versionspace},
and exact counting is \#P-complete \cite{valiant1979enumeration}. Current approximate counters
intersect the solution set with a random affine subspace defined by XOR constraints, count
the solutions in one cell with a satisfiability solver and scale the count up
\cite{gomes2006xor,chakraborty2013approxmc,chakraborty2016approxmc2,soos2019bird}. Their guarantee
rests on the variance bound of a pairwise independent hash family, and they repeat the estimate and
take the median to make an unlucky hash rare. Sparser hashes shorten the constraints and accept
a weaker bound \cite{ermon2014ldpc,achlioptas2015codes,agrawal2020sparsity}. Here, the variance of
one hash is the spectral energy of the solution set on the row space of the hash
(Theorem~\ref{thm:hash}), so the parities that are constant on the solutions, which the affine hull
of the solution set determines, are what make a draw catastrophic. Hashes drawn outside the span of
the parities constant on a few solutions avoid them and keep the bound of a pairwise independent
family (Corollary~\ref{cor:pairwise}), and a satisfiability solver certifies those parities
(Proposition~\ref{prop:oraclehull}). Four experiments test this with increasing realism, namely the exact
variance of single hashes on circuit outputs (Section~\ref{sec:hash}), a hashing counter
(Section~\ref{sec:counter}), the core of a prefix counter in the form of ApproxMC2
(Section~\ref{sec:core}) and the same core run with a satisfiability solver on the benchmark formulas
of ApproxMC (Section~\ref{sec:satcount}). On circuit outputs the selected hashes lower the share of
catastrophic single hashes from \HcRandCatPct\ to \HcHullCat\ percent and the miss rate of one call of
the prefix counter from \MhCoreRandMiss\ to \MhCoreUniMiss\ percent. On the formulas a random hash
rarely holds a constant parity, the single-call miss rates are close, \ScRandMiss\ percent for random
rows and \ScMatchMiss\ percent for matched ones, and the certified hull mainly shortens the rows, to a
median time per call of \ScTimeRatioReduced\ of the random rule on the formulas whose hull removes a
quarter of the coordinates.

\subsubsection{Selected Hashes Make Catastrophic Counts Rarer}
\label{sec:hash}

The model-counting application asks how well an XOR hash estimates the
number of solutions of a switching function. Its outputs are those with
$8\le n\le24$ inputs and $256\le|S|\le N-256$, \HcOutputs\ outputs of \HcCircuits\
circuits with \HcNmin\ to \HcNmax\ inputs. For each, $k=\lfloor\log_2(|S|/32)\rfloor$
rows give cells that hold between 32 and 64 solutions on average, and every variance
is computed exactly from Theorem~\ref{thm:hash} with the indicator spectrum of the
on-set. We report the relative standard deviation $\sqrt{\Var Z}/|S|$ and call a hash
catastrophic when it exceeds one quarter. The random baseline samples \HcRrand\
uniformly random row spaces per output. The selection rules choose from a pool of
\HcPool\ further random row spaces. The plug-in rule takes the candidate with the
smallest estimate \eqref{eq:pilot} from $m$ solutions drawn uniformly with replacement,
and the hull rule takes the first candidate, in a random order, that contains no parity
constant on $m=32$ solutions. Each rule is applied to \HcRpilot\ independent draws of the
solutions, and every chosen hash is evaluated exactly. The best candidate in the pool,
found from the exact variances, serves as a reference.

The closed form \eqref{eq:blind} agrees with the mean over the sampled random row
spaces, with a median ratio of \HcBlindRatio\ across outputs, and the mean is large,
a median relative standard deviation of \HcBlindSd. The typical random hash is much
better than this mean. Its relative standard deviation has a median of \HcRandMedSd,
and the mean is carried by a heavy tail. Over all outputs, \HcRandCatPct\ percent of
the random hashes are catastrophic, and on \HcTailOnePct\ percent of the outputs at least
one random hash in a hundred is. A counter that draws a fresh random hash for each query
relies on repetition to remove those draws.

\begin{table}[t]
\centering
\caption{Hashed counts on the 277 outputs. Median over outputs of the median
relative standard deviation over the hashes each choice produces, percent of
catastrophic hashes (relative standard deviation above 0.25) and of hashes at or above
1, and median ratio of the mean variance to the random mean.}
\label{tab:hash}
\setlength{\tabcolsep}{4pt}
\begin{tabular}{@{}lrrrr@{}}
\toprule
Hash & median & $>0.25$ (\%) & $\ge1$ (\%) & mean ratio\\
\midrule
random hash & 0.013 & 4.2 & 0.78 & 1.00\\
plug-in rule, $m=16$ & 0.013 & 3.0 & 0.42 & 0.58\\
plug-in rule, $m=32$ & 0.012 & 2.6 & 0.42 & 0.40\\
plug-in rule, $m=64$ & 0.013 & 1.5 & 0.30 & 0.31\\
plug-in rule, $m=128$ & 0.011 & 1.3 & 0.14 & 0.22\\
hull rule, $m=32$ & 0.013 & 3.1 & 0.01 & 0.50\\
best hash of the pool & 0.003 & 0.0 & 0.00 & 0.00\\
\bottomrule
\end{tabular}
\end{table}

\begin{figure}[t]
\centering
\includegraphics[width=0.62\textwidth]{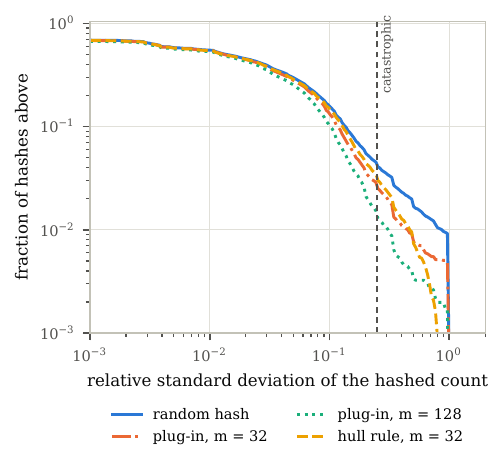}
\caption{Fraction of hashes whose relative standard deviation exceeds each value,
pooled over the \HcOutputs\ benchmark outputs with equal weight per output, for random
hashes, the plug-in rule with 32 and 128 solutions and the hull rule with 32 solutions.
The dashed line marks the catastrophic threshold of one quarter.}
\label{fig:hash}
\end{figure}

The selection rules act on the tail, as Table~\ref{tab:hash} and Fig.~\ref{fig:hash}
show. With 32 solutions the plug-in rule lowers the share of catastrophic hashes to
\HcPilotCatB\ percent and the mean variance to a median of \HcPilotRatioB\ of the random
mean, and with 128 solutions to \HcPilotCatD\ percent and \HcPilotRatioD. The hull rule
with 32 solutions gives \HcHullCat\ percent and \HcHullRatio, and it nearly removes the
most extreme hashes. A relative standard deviation of one or more, the level that a
single implied parity in the row space produces, occurs for \HcRandSevere\ percent of
the random hashes and for \HcHullSevere\ percent of those the hull rule chooses. The typical hash changes
little, since the median relative standard deviation of the chosen hashes stays close to
that of a random hash. The plug-in rule improves steadily with $m$, as the birthday bound
of Section~\ref{sec:hashvar} predicts. The hull rule needs no second-order estimate, but
it removes only the parities that the sample shows as constant, so its effect on the
moderate part of the tail is smaller. The best candidate in
the pool has a median relative standard deviation of \HcOracleSd, which shows how much
room the pool leaves for better selection.

The experiment evaluates the variance of the chosen hash exactly, and the plug-in rule assumes
solutions drawn uniformly, which a counter does not have for free. The hull rule is less
demanding, since it needs only solutions that span the affine hull of the solution set.
Section~\ref{sec:counter} runs a counter end to end with it.

\subsubsection{Selected Hashes Make a Hashing Counter Fail Less Often}
\label{sec:counter}

The counter runs on the same \CoOutputs\ outputs with the same number $k$ of rows. One estimate
is $2^k$ times the number of solutions in a uniformly chosen cell of a hash, computed exactly
from the indicator spectrum on the row space, and the counter reports the median of $t$
estimates, each with a fresh hash and cell. A run fails when its count lies outside
$[|S|/c,\,c|S|]$, with $c=1.8$, the tolerance usual for hashing counters
\cite{chakraborty2013approxmc}, and with the tighter $c=1.25$. Three rules choose each hash.
The random rule draws a uniformly random row space. The hull rule draws $32$ solutions
uniformly once per run and, for every estimate, draws a fresh matched row space as in
Section~\ref{sec:matched}, row by row, each row uniform outside the span of the parities
constant on the solutions and the earlier rows. When the solutions span fewer than $k$
dimensions no matched row space exists, and the rule draws a uniformly random one instead, which
happened \CoFallbackText. By Corollary~\ref{cor:pairwise} and
Theorem~\ref{thm:hash}, every estimate of either rule has the mean $|S|$ and at most the variance
of a pairwise independent hash, so the counter keeps the guarantee that rests on that bound. Its clustered variant
takes the $32$ solutions nearest in Hamming distance to a random solution, with ties broken
by the smaller solution index, a model of the correlated solutions that a solver returns. Each output is counted in \CoRuns\ independent runs
for every rule.

\begin{table}[t]
\centering
\caption{Percent of counter runs whose reported count lies outside $[|S|/c,\,c|S|]$, mean over
the 277 outputs, for the median of $t$ estimates.}
\label{tab:counter}
\setlength{\tabcolsep}{3.5pt}
\begin{tabular}{@{}lrrrrrr@{}}
\toprule
 & \multicolumn{3}{c}{$c=1.8$} & \multicolumn{3}{c}{$c=1.25$}\\
\cmidrule(lr){2-4}\cmidrule(l){5-7}
Hash rule & $t=1$ & $3$ & $9$ & $t=1$ & $3$ & $9$\\
\midrule
random & 1.69 & 0.03 & 0.00 & 5.16 & 0.53 & 0.00\\
hull rule, uniform solutions & 0.65 & 0.03 & 0.00 & 4.14 & 0.41 & 0.00\\
hull rule, clustered solutions & 0.34 & 0.01 & 0.00 & 3.11 & 0.37 & 0.00\\
\bottomrule
\end{tabular}
\end{table}

Table~\ref{tab:counter} shows the failure rates. With a single estimate at the usual tolerance,
the random rule fails in \CoFailRandA\ percent of the runs, and on \CoOverRandA\ percent of the
outputs it fails in more than one run in twenty. The hull rule lowers the failures to
\CoFailHullA\ percent with uniform solutions and to \CoFailClusA\ percent with clustered ones,
and the share of outputs above one failure in twenty to \CoOverHullA\ and \CoOverClusA\ percent.
The clustered sample shows more parities as constant than a uniform one, so its matched row
spaces avoid more of them, and on these outputs the stricter choice does not lose accuracy. The median of
three estimates removes nearly all failures at the usual tolerance for every rule, so at that
tolerance the selection saves repetitions. At the tighter tolerance a single estimate fails in
\CoTightRandA, \CoTightHullA\ and \CoTightClusA\ percent of the runs for the random rule, the
hull rule and its clustered variant, and the share of outputs with more than one failure in
twenty falls from \CoTightOverRandA\ to \CoTightOverHullA\ and \CoTightOverClusA\ percent.
The median of three estimates fails in \CoTightRandB, \CoTightHullB\ and \CoTightClusB\
percent of the runs.

\subsubsection{Matched Hashes Lower the Variance inside a Prefix Counter}
\label{sec:core}

The closed form of Proposition~\ref{prop:matched} was checked first. For $n=6$, enumerating every
ordered basis of every matched row space for $k=1$ to $3$, on \MhEOneSets\ random sets with
random pilots, gives the mean of \eqref{eq:hashvar} equal to \eqref{eq:matched} in all
\MhEOneCases\ cases. On the \MhETwoOutputs\ outputs of the counter, with \MhETwoDraws\ uniform
and \MhETwoDraws\ clustered pilots of $32$ solutions each, the closed form agrees with the
spectral sum over the parities outside $W$ to a relative difference of \MhETwoDiff, and no case
exceeds the bound of Corollary~\ref{cor:pairwise}. Against the blind mean \eqref{eq:blind}, the
matched variance has a median ratio of \MhUniMed\ with uniform pilots and \MhCluMed\ with
clustered ones, and \MhUniHalf\ and \MhCluHalf\ percent of the pilots lower it below one half.
\MhUniZero\ uniform and \MhCluZero\ clustered pilots give zero variance, on outputs whose on-set
is a union of cosets of the pilot span, often a single cube, as Corollary~\ref{cor:zero}
predicts. The largest ratio is \MhMaxRatio, within the bound of Section~\ref{sec:matched}.

The counter of Section~\ref{sec:counter} draws a fresh hash for every estimate. Counters in the
form of ApproxMC2 instead search the prefixes of one hash for the smallest number $m$ of rows at
which a cell holds fewer solutions than a threshold, and report that count times $2^m$
\cite{chakraborty2016approxmc2}. The core of such a counter, with $\varepsilon=\MhEps$ and its
threshold of \MhThresh\ solutions, was run on the \MhCoreOutputs\ outputs with at most
$\MhCoreSmax$ solutions, with \MhCoreRuns\ calls per output and rule. The rows were random,
matched to $32$ uniform solutions, matched to $32$ clustered solutions, or matched to the exact
affine hull read from the indicator spectrum by Proposition~\ref{prop:hull}. Of these outputs,
\MhCoreProper\ have a proper affine hull and \MhCoreAffine\ have an affine on-set.

\begin{table}[t]
\centering
\caption{One call of the core of a prefix-hash counter (ApproxMC2 form, $\varepsilon=0.8$,
threshold 73), 203 outputs with at most $2^{17}$ solutions: percent of
calls whose count lies outside $[|S|/1.8,\,1.8|S|]$ (miss), mean over outputs and largest for one
output, median of $|\log_2(\text{count}/|S|)|$ (error), and percent of calls that needed more than
$r$ levels. The pilots hold 32 solutions.}
\label{tab:core}
\setlength{\tabcolsep}{3pt}
\begin{tabular}{@{}lrrrr@{}}
\toprule
Rows & miss & largest & error & beyond $r$\\
\midrule
random & 1.04 & 5.0 & 0.018 & 0.00\\
matched, uniform pilot & 0.55 & 4.0 & 0.020 & 0.00\\
matched, clustered pilot & 0.24 & 2.5 & 0.017 & 0.01\\
matched, exact hull & 0.62 & 3.5 & 0.020 & 0.00\\
\bottomrule
\end{tabular}
\end{table}

Table~\ref{tab:core} shows that a single call with matched rows misses the tolerance less often
than one with random rows, \MhCoreUniMiss\ and \MhCoreCluMiss\ percent with uniform and
clustered solutions against \MhCoreRandMiss\ percent, and the largest miss rate on one output
falls from \MhCoreRandMax\ to \MhCoreUniMax\ and \MhCoreCluMax\ percent. The exact hull gives
\MhCoreExactMiss\ percent. A clustered pilot can do better than the exact hull because its
smaller span removes more spectral energy, as the bound of Section~\ref{sec:matched} allows.
Calls that needed more than $r$ levels, where the hash continues with rows that are not matched
and Proposition~\ref{prop:matched} does not apply, made up \MhCoreBeyondText. A certified hull removes them,
since the search then stops by level $d$.

\subsubsection{A Counter for Formulas with a Certified Hull}
\label{sec:satcount}

The counters above read every cell count from a truth table. For formulas, the core of
Section~\ref{sec:core} was run with a satisfiability solver on the benchmark set used in the
evaluations of ApproxMC \cite{soos2024benchmarks}, which holds \ScFormulas\ formulas from
probabilistic inference, planning, program synthesis, bit-blasted arithmetic and the ISCAS-89
circuits. The counter enumerates the projected solutions of a cell with CryptoMiniSat
\cite{soos2009cms} up to the threshold of \MhThresh\ by blocking clauses, adds each row of the hash
as an XOR constraint when a prefix first needs it, and finds the level by galloping from
the level of its previous call and then by bisection, as ApproxMC does
\cite{chakraborty2016approxmc2,soos2019bird}. Random rows include each sampling variable with
probability one half. Matched rows use the affine hull $y_0\oplus V$ of the projected solutions,
certified by Proposition~\ref{prop:oraclehull} with the solutions of the first enumeration as the
seed. A row drawn uniformly outside $\operatorname{span}(V^{\perp},\text{previous rows})$ acts on the
solutions through its residue modulo $V^{\perp}$, which is supported on the $d$ pivot coordinates of
$V$, so the matched rows are drawn on those coordinates, linearly independent, and the search ends
by level $d$. Formulas were taken in a seeded random order, at most \ScFamCap\ from each family,
with a sampling set of $12$ to \ScNmax\ variables, an exact projected count from Ganak
\cite{sharma2019ganak} within \ScExactTimeout\ seconds and at least $2^{10}$ solutions, until
\ScInstances\ formulas from \ScFamilies\ families had been counted. Of the \ScExamined\ formulas
read, \ScSkipExact\ had no exact count within the limit, \ScSkipSmall\ had fewer than $2^{10}$
solutions and \ScSkipCounter\ did not finish the hull within \ScHullLimit\ seconds or failed in the
counter. Each rule made \ScRuns\ calls per formula, alternating with the other rule, and a call
that took longer than \ScCallLimit\ seconds counts as a miss.

\begin{table}[t]
\centering
\caption{The core of a prefix-hash counter run with CryptoMiniSat on 63 formulas of the
ApproxMC benchmark set on which both rules finished every call (9 calls per rule and formula):
percent of single calls and of medians of three calls outside $[|S|/1.8,\,1.8|S|]$, median of
$|\log_2(\text{count}/|S|)|$, mean number of variables in a hash row, mean solver calls per call of
the core, median seconds per call, and calls over the time limit on all 74 formulas.}
\label{tab:satcount}
\setlength{\tabcolsep}{2.5pt}
\begin{tabular}{@{}lrrrrrrr@{}}
\toprule
Rows & miss & of 3 & error & row & calls & s & over\\
\midrule
random & 0.5 & 0.0 & 0.066 & 45.2 & 188 & 0.10 & 21\\
matched & 0.9 & 0.0 & 0.062 & 30.9 & 185 & 0.10 & 18\\
\bottomrule
\end{tabular}
\end{table}

The sampling sets have a median of \ScMedN\ variables and the counts a median of $2^{\ScMedLog}$.
The certified hull is proper on \ScProper\ of the \ScInstances\ formulas, with a median ratio $d/n$
of \ScMedRatio, and it took a median of \ScHullCalls\ solver calls and \ScHullSec\ seconds, computed once
per formula. Table~\ref{tab:satcount} compares the rules on the \ScPaired\ formulas on which both
finished every call. A single call misses the tolerance in \ScRandMiss\ percent of the random calls
and \ScMatchMiss\ percent of the matched ones, and the median of three calls in \ScRandMissThree\
and \ScMatchMissThree\ percent. A matched row holds \ScMatchRow\ variables on average against
\ScRandRow\ for a random row, and the median ratio of the time per call, matched over random, is
\ScTimeRatio. The hull removes at least a quarter of the sampling coordinates on \ScReduced\
formulas. On the \ScPairedReduced\ of them that both rules finished the median ratio is
\ScTimeRatioReduced, and on \ScOnlyMatchedReduced\ of them only the matched rule finished every
call within the limit. Over all formulas the matched rule alone finished every call on
\ScOnlyMatched\ and the random rule alone on \ScOnlyRandom. \ScAmcSentence

Two readings follow. Where the sampling set of a formula carries many affine dependences, as when
it is the whole set of variables, the hull shortens the rows and with them the solver time while
the variance bound of Corollary~\ref{cor:pairwise} still holds. Where the sampling set is already an
independent support, the hull is nearly the whole space and the matched rows behave like random
ones, as their distribution on the solutions then nearly coincides. How much variance the matched draw can remove depends
on the codimension $n-d$ of the hull, without which no parity is constant on the solutions, and
on the level at which the search stops. A prefix of $m$ random rows holds
$(2^m-1)(2^{n-d}-1)/(2^n-1)$ such parities on average. Evaluated at the levels the random calls
reached on the formulas with a proper hull, this is \ScConstRate\ percent per call, so on this set
a random hash rarely holds a parity that the matched draw would exclude, although \ScProperDense\
formulas have a proper hull and fewer than four bits between $d$ and $\log_2|S|$, the regime in
which the matched draw removes the most. A variable that is an affine
function of the others is definable from them, so independent-support reduction
\cite{soos2022arjun} removes it too. The hull is the linear and certified part of that reduction,
the part that gives the matched draw its variance bound.

\iffull
\subsection{Spectral Elimination Decides Parity-Structured Formulas}
\label{sec:parity}

Equivalence checking of circuits rich in XOR gates, for example restructured parity trees,
error-correcting encoders and the linear layers of ciphers, gives formulas whose constraints are
parities, and the Tseitin formulas show that resolution, the proof system behind conflict-driven
solvers, needs proofs of exponential size for some of them
\cite{tseitin1983complexity,urquhart1987hard,bensasson2001narrow}. Current solvers recover
XOR constraints from the clauses and reason about them by Gauss--Jordan elimination
\cite{han2012gauss,soos2009cms,soos2020tinted}. For arithmetic datapaths over GF($2^m$) and over the
integers, the leading methods are verifiers that rewrite the circuit into polynomials
\cite{lv2012gf,kaufmann2019sat}. Here, the matched group of a block of clauses is the translation
group of the affine hull of its local solution set, and the indicator spectrum of the block gives the
parities the block implies whether or not its clauses encode an XOR
(Proposition~\ref{prop:hull}). Gaussian elimination over the union of these parities then decides the
formula when every block is affine. The experiment runs this spectral elimination against CaDiCaL and
CryptoMiniSat on Tseitin formulas, miters of parity trees and planted 3-XORSAT. It shows that the
reading recovers the parities that XOR recovery finds. On the Tseitin family it decides every
instance up to \PsTseitinMaxVars\ variables, where the conflict-driven solver exceeds the limit from
\PsTseitinDeadVars\ variables, and the solver with XOR recovery decides them as well. The
experiment does not include the algebraic verifiers, whose domain is arithmetic.

Parity reasoning is the application of Proposition~\ref{prop:hull} to blocks of
clauses. The spectral elimination groups the clauses of a formula by their variable
sets, evaluates the local solution set of each block exhaustively, reads the parities
implied by the block from its indicator spectrum, and solves the union of these
parities by Gaussian elimination over GF(2). An inconsistent system proves the formula
unsatisfiable. A consistent system yields a candidate assignment, which is checked
against every clause. When every block has an affine solution set, the system is
equivalent to the formula and the procedure decides it in time polynomial in the
formula size and exponential only in the largest block, which our implementation
limits to 16 variables. Larger blocks are skipped, which leaves a sound relaxation.

Three families exercise it. Tseitin formulas on random three-regular graphs with odd
total charge are unsatisfiable and require resolution proofs of exponential size
\cite{tseitin1983complexity,urquhart1987hard,bensasson2001narrow}. The miter family
checks the equivalence of two parity trees over the same $n$ inputs, each built by
pairing its current nodes at random, with every two-input XOR gate encoded by
its four clauses and the two outputs required to differ. These are unsatisfiable
equivalence checks of the kind that arise when restructured parity logic is
verified. Planted random 3-XORSAT with as many equations as variables is satisfiable.
The comparison solvers are CaDiCaL, version 1.9.5 \cite{biere2024cadical}, called
through PySAT \cite{ignatiev2018pysat}, a conflict-driven solver without parity
reasoning, and CryptoMiniSat, version 5.17 \cite{soos2009cms,soos2020tinted}, a solver
designed to recover XOR constraints from clauses and to reason about them by
elimination, used in its default configuration. That configuration performs
Gauss-Jordan elimination on the recovered constraints. On an unsatisfiable Tseitin formula
with \GjVertices\ vertices and \GjClauses\ clauses it reports unsatisfiability in
\GjDefault~s, and with its elimination matrices switched off it does not finish within
\GjLimit~s. Every call runs in its own process with
a limit of \PsTimeout~s, on \PsSeeds\ seeds per size, and a solver is not called at
larger sizes once it has exceeded the limit on every seed of a size.

\begin{figure}[t]
\centering
\includegraphics[width=\textwidth]{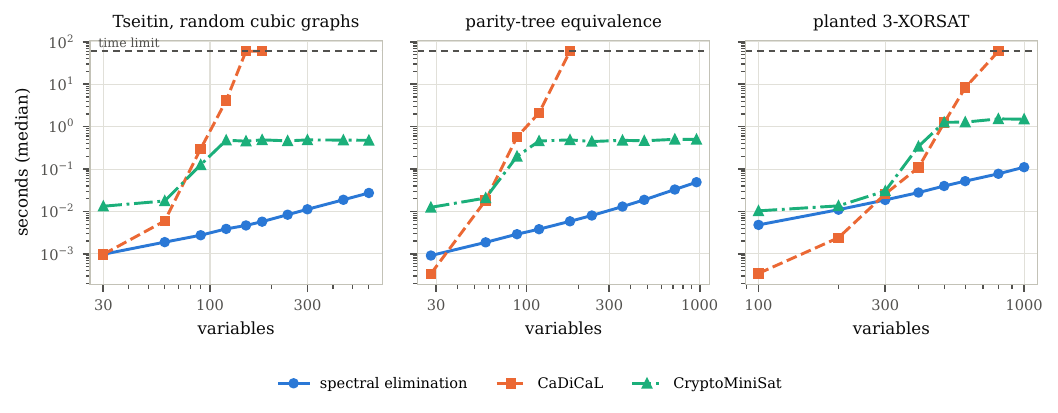}
\caption{Running time against the number of variables for the Tseitin, miter and
planted 3-XORSAT families, median over \PsSeeds\ seeds. Points at the time limit mark
calls that did not finish, and the CaDiCaL curve stops where the solver was no longer
called.}
\label{fig:parity}
\end{figure}

Fig.~\ref{fig:parity} shows the result. The conflict-driven solver exceeds the limit on
every seed at \PsTseitinDeadVars\ variables for the Tseitin family, at
\PsMiterDeadVars\ variables (\PsMiterDeadInputs\ inputs) for the miter family and at
\PsXorDeadVars\ variables for 3-XORSAT. The spectral elimination finds every block
affine and decides every instance, up to \PsTseitinMaxVars, \PsMiterMaxVars\ and
\PsXorMaxVars\ variables, in at most \PsSpecMax~s, with a running time that grows
slowly with size. CryptoMiniSat also decides every instance, in at most \PsCmsMax~s.
Both methods owe their behavior to linear algebra over GF(2). The instances are pure
XOR encodings, so the experiment shows that the group-theoretic reading of a
block, the affine hull of its local solution set, recovers the parities that the
elimination needs, as XOR recovery does. The reading also applies to blocks
whose solution set is not affine, where it yields the implied parities as a sound
relaxation of the block, a case these families do not exercise.

\subsection{Reading Windows over Rings Refutes Counting Formulas}
\label{sec:rings}

Formulas that encode a counting argument modulo an integer other than two, such as a partition of
points into triples or into 4-sets when the number of points forbids it, are standard hard cases
for satisfiability and proof systems, and the counting principles behind them separate proof
systems that reason modulo different primes \cite{buss1999,beame1998}. Current solvers decide
them only at small sizes. Conflict-driven solvers, pseudo-Boolean solvers with cutting planes
\cite{leberre2010,elffers2018,devriendt2021} and integer-programming solvers
\cite{hojny2025scip10,huangfu2018} search or relax, Gaussian elimination inside satisfiability
solvers covers parity only \cite{han2012gauss,soos2009cms}, and the one tool we found that reasons
modulo other integers, pgpbs, must be told the modulus \cite{bryant2022,bryant2022pgpbs}. Here,
the affine hull of a block's local solution set, which Section~\ref{sec:parity} reads over GF(2),
is read over the integers. One diagonal form of an integer matrix built from the local solutions
gives every affine relation they satisfy modulo every prime power at once
(Theorem~\ref{thm:diag}). These relations are the characters of the group of residue vectors modulo an
integer that are constant on the local solutions, and an unsolvable union of relations over a ring
of residues modulo a prime power
refutes the formula with a certificate that an independent program checks. The experiments
compare this ring reading with conflict-driven, pseudo-Boolean, integer-programming and
parity-aware solvers on counting families in four encodings, place it inside the presolve of an
integer-programming solver, and read \Rg{real.instances} instances of three public libraries.
The reading refutes the triple partition at \Rg{count3.pairwise.read.n} points and the 4-set
partition at \Rg{count4.pairwise.read.n} points in their pairwise encodings, the largest sizes run,
where the best tested solvers stop at \Rg{count3.pairwise.best.n} and \Rg{count4.pairwise.best.n}.
On the library instances it shows no advantage on any application family.

\paragraph*{Windows and their relations.} Let $\Xi$ be a formula in conjunctive normal form over
variables $x$. A \emph{window} is a set $\Omega$ of its variables, the window formula $\Xi_\Omega$ is
the conjunction of the clauses whose variables all lie in $\Omega$, and the \emph{local solution set}
$S_\Omega \subseteq \{0,1\}^\Omega$ holds the assignments to $\Omega$ that satisfy $\Xi_\Omega$. For
an integer $\kappa \ge 2$, a \emph{relation modulo $\kappa$} on $\Omega$ is a pair $(\chi,\gamma)$ of
an integer vector $\chi$ on $\Omega$ and an integer $\gamma$ with $\chi^{\mathsf T}x \equiv \gamma
\pmod\kappa$ for every $x \in S_\Omega$, and a rational relation is a pair with $\chi^{\mathsf T}x =
\gamma$ on $S_\Omega$. Every solution of $\Xi$ restricts to a point of $S_\Omega$, so every relation
on $\Omega$ holds on every solution. Let $\Lambda_\Omega$ be the integer matrix whose rows are the
points of $S_\Omega$, each followed by a one, so that $(\chi,\gamma)$ is a relation modulo $\kappa$
exactly when $\Lambda_\Omega(\chi;-\gamma) \equiv 0 \pmod\kappa$. Unimodular row and column operations
bring $\Lambda_\Omega$ to a diagonal form $\Psi_1\Lambda_\Omega\Psi_2 = \Delta$ whose nonzero entries
$\sigma_1,\dots,\sigma_\varrho$ fill the first $\varrho$ diagonal positions, where $\varrho$ is the
rank of $\Lambda_\Omega$, and the Smith normal form is one such form. Write $\psi_j$ for column $j$ of
$\Psi_2$.

\begin{theorem}[one diagonal form serves every modulus]
\label{thm:diag}
For every $\kappa \ge 2$, the integer vectors $(\chi;-\gamma)$ with $\Lambda_\Omega(\chi;-\gamma)
\equiv 0 \pmod\kappa$ are exactly the integer combinations of $\frac{\kappa}{\gcd(\kappa,\sigma_j)}
\psi_j$ for $j \le \varrho$ and of $\psi_j$ for $j > \varrho$. The rational relations correspond to the
span of $\psi_j$ for $j > \varrho$.
\end{theorem}

\begin{proof}
$\Psi_1$ has determinant $\pm1$ and is invertible modulo $\kappa$, so $\Lambda_\Omega(\chi;-\gamma)
\equiv 0$ holds exactly when $\Delta\zeta \equiv 0$ for $\zeta = \Psi_2^{-1}(\chi;-\gamma)$, that is,
when $\sigma_j\zeta_j \equiv 0 \pmod\kappa$ for $j \le \varrho$, so that $\zeta_j$ is a multiple of
$\kappa/\gcd(\kappa,\sigma_j)$, with $\zeta_j$ free for $j > \varrho$. Multiplying by $\Psi_2$ maps
these $\zeta$ onto the stated combinations. Over the rationals the same argument gives $\zeta_j = 0$
for $j \le \varrho$.
\end{proof}

\begin{corollary}[the diagonal entries name the moduli]
\label{cor:primes}
The window has a relation modulo $\kappa$ that is not the reduction of a rational relation exactly when
$\gcd(\kappa,\sigma_j) > 1$ for some $j \le \varrho$. The primes $p$ whose rings
$\mathbb{Z}/p^\tau\mathbb{Z}$ see more than the rationals on $\Omega$ are those dividing
$\sigma_1\cdots\sigma_\varrho$.
\end{corollary}

\begin{proof}
The columns of the unimodular $\Psi_2$ form a basis of the integer lattice and their residues a basis
modulo $\kappa$, and the reductions of rational relations span the residues of $\psi_j$ for $j >
\varrho$. By Theorem~\ref{thm:diag} a further relation exists exactly when
$\frac{\kappa}{\gcd(\kappa,\sigma_j)}\psi_j \not\equiv 0 \pmod\kappa$ for some $j \le \varrho$, which
holds exactly when $\gcd(\kappa,\sigma_j) > 1$, since the coordinates of that vector in the basis are
$\kappa/\gcd(\kappa,\sigma_j)$ at position $j$ and zero elsewhere.
Finally $\gcd(p^\tau,\sigma_j) > 1$ exactly when $p \mid \sigma_j$.
\end{proof}

For $\omega_\kappa = e^{2\pi\mathrm{i}/\kappa}$ and an integer vector $\chi$, the triangle
inequality gives $|\sum_{x\in S_\Omega}\omega_\kappa^{\chi^{\mathsf T}x}| = |S_\Omega|$ exactly when
$\chi^{\mathsf T}x$ is constant modulo $\kappa$ on $S_\Omega$. The relations modulo $\kappa$ are
therefore the characters of $(\mathbb{Z}/\kappa\mathbb{Z})^\Omega$ that are constant on $S_\Omega$,
the frequencies at which the Fourier transform of the indicator of $S_\Omega$ over that group has
full magnitude. For $\kappa=2$ this is the indicator spectrum that Proposition~\ref{prop:hull} reads,
and for larger $\kappa$ it is the Chrestenson spectrum of Remark~\ref{rem:prime}
\cite{thornton2001spectral}. Corollary~\ref{cor:primes} makes the choice of the matched group
arithmetic. The diagonal entries name every prime power whose group carries a constant character on
the window that no rational relation explains.

\paragraph*{Global systems and certificates.} For a set of windows and a modulus $\kappa$, the global
system $\Theta x \equiv \vartheta \pmod\kappa$ collects a generating set of the relations of every
window, and every solution of $\Xi$ solves it. Over $\mathbb{Z}/p^\tau\mathbb{Z}$ it is decided by
elimination on unit pivots. When no coefficient of the remaining rows is a unit modulo $p$, a row whose
right side is not divisible by $p$ is a contradiction, and otherwise every row is divided by $p$ and
the elimination continues modulo $p^{\tau-1}$. By the Chinese remainder theorem prime powers suffice.

\begin{proposition}[completeness and certificates]
\label{prop:ringcert}
The elimination reports a contradiction exactly when $\Theta x \equiv \vartheta \pmod{p^\tau}$ has no
solution, and it performs at most $\tau$ Gaussian eliminations. If a global system has no solution,
$\Xi$ is unsatisfiable, and there is an integer vector $\varpi$ with $\varpi^{\mathsf T}\Theta \equiv
0$ and $\varpi^{\mathsf T}\vartheta \not\equiv 0 \pmod\kappa$, which a checker verifies by enumerating
$S_\Omega$ for the windows of the rows on which $\varpi$ is not zero modulo $\kappa$.
\end{proposition}

\begin{proof}
A unit pivot solves its row for one variable, and forgetting that variable is a bijection between the
solutions of the system and those of the system with the variable eliminated, so unit pivots preserve
solvability. When every remaining coefficient is divisible by $p$, a remaining row reads $p\,{\chi'}^{\mathsf
T}x \equiv \gamma' \pmod{p^\tau}$. If $p \nmid \gamma'$ it has no solution. Otherwise it holds exactly
when ${\chi'}^{\mathsf T}x \equiv \gamma'/p \pmod{p^{\tau-1}}$, and a solution modulo $p^{\tau-1}$ lifts
by any choice of representatives. Every system modulo $1$ is solvable, so induction on $\tau$ and on the
number of rows gives completeness, with one Gaussian elimination per level. Every solution of $\Xi$
solves every global system, which gives soundness. For the certificate, let $\Theta =
\Pi_1\Sigma\Pi_2$ be the Smith normal form over the integers. The system is solvable exactly when every
congruence $\Sigma_{ii}\,(\Pi_2x)_i \equiv \vartheta'_i \pmod\kappa$ is, with $\vartheta' = \Pi_1^{-1}\vartheta$,
that is, when $\gcd(\kappa,\Sigma_{ii})$ divides $\vartheta'_i$, where $\gcd(\kappa,0)=\kappa$. For an
$i$ where it does not, let $\varpi^{\mathsf T}$ be $\kappa/\gcd(\kappa,\Sigma_{ii})$ times row $i$ of
$\Pi_1^{-1}$. Then $\varpi^{\mathsf T}\Theta \equiv 0$ and $\varpi^{\mathsf T}\vartheta \not\equiv 0
\pmod\kappa$. Conversely such a $\varpi$ excludes every solution, so the checker needs only the
relations it combines and the two congruences.
\end{proof}

\begin{proposition}[the summation argument]
\label{prop:count}
Let $\mathcal H$ be an $h$-uniform hypergraph on a point set $\mathcal P$, with one variable
$x_{\mathfrak e}$ per edge $\mathfrak e$, and let $\mathrm{Count}(\mathcal H)$ be the system of the
relations $\sum_{\mathfrak e\ni\mathfrak p}x_{\mathfrak e} = 1$ for the points $\mathfrak p \in \mathcal P$. For every $\kappa \ge 2$
the sum of these relations reads $h\sum_{\mathfrak e}x_{\mathfrak e} \equiv |\mathcal P| \pmod\kappa$,
which has no solution exactly when $\gcd(h,\kappa) \nmid |\mathcal P|$. If every point lies in the same
number of edges and a prime $p$ does not divide that number, then $\mathrm{Count}(\mathcal H)$ is
solvable over GF($p$). For $h=4$ and $|\mathcal P| \equiv 2 \pmod 4$ the contradiction therefore
exists modulo $\kappa$ exactly when $4 \mid \kappa$, so the ring $\mathbb{Z}/4\mathbb{Z}$ provides it
and no prime field does.
\end{proposition}

\begin{proof}
Each edge lies in $h$ relations, which gives the sum, and $h\,\eta \equiv |\mathcal P|$ is solvable in
$\eta$ exactly when $\gcd(h,\kappa)$ divides $|\mathcal P|$. For the second claim, setting every
$x_{\mathfrak e}$ to the inverse of the common number of edges at a point satisfies every relation.
For $h = 4$ and $|\mathcal P| \equiv 2 \pmod 4$, of the values $\gcd(4,\kappa) \in \{1,2,4\}$ only $4$
fails to divide $|\mathcal P|$.
\end{proof}

\paragraph*{Experiments.} The counting families are the triple partition and the 4-set partition on
random hypergraphs in which every point lies in six edges, with $3 \nmid |\mathcal P|$ and
$|\mathcal P| \equiv 2 \pmod 4$ respectively, with each exactly-one
constraint encoded pairwise, sequentially, with commanders or in binary. The reading runs in
Python over windows of the clauses, over GF(2), GF(3) and a large prime field and over the rings
$\mathbb{Z}/p^\tau\mathbb{Z}$, also after bounded variable elimination \cite{een2005}, which removes the
auxiliary variables of the sequential, commander and binary encodings. The baselines on the
clauses are CaDiCaL, Kissat, CryptoMiniSat and Lingeling, pgpbs in its automatic mode, in integer
mode and told the modulus, and Sat4j with cutting planes, RoundingSat, Exact and SCIP reading the
clauses as linear constraints. Xorcle, which runs conflict-driven search over disjunctions of
parity equations \cite{beame2026,xorcle}, ran on the same ladders. The baselines on the native
cardinality constraints are Sat4j, RoundingSat, Exact, SCIP and HiGHS. Every call had a limit of
\Rg{limit}~s on the awake clock of the author's Mac, with \Rg{seeds} seeds per size, and a size
counts as decided when every seed decided it. Table~\ref{tab:rings} gives the result.

\begin{table}[t]
\centering
\caption{Largest size decided within \Rg{limit}~s on every seed, for the counting families in
four encodings. The top of each ladder is the largest size run. The reading column gives the
fastest reading at that size, after variable elimination on the sequential encodings. Best CNF is
the best baseline on the clauses other than Xorcle, and best native the best solver given the
cardinality constraints.}
\label{tab:rings}
\footnotesize\setlength{\tabcolsep}{3pt}
\begin{tabular}{@{}lrrrrr@{}}
\toprule
Family, encoding & top & reading (time) & best CNF & Xorcle & best native\\
\midrule
triple partition, pairwise & 320 & 320 (1.7 s) & 80 & 40 & 80\\
triple partition, commander & 320 & 320 (1.3 s) & 80 & 40 & 80\\
triple partition, binary & 320 & 320 (1.5 s) & 80 & 40 & 80\\
triple partition, sequential & 320 & 320 (1.2 s) & 320 & 40 & 80\\
\midrule
4-set partition, pairwise & 642 & 642 (25.9 s) & 162 & 82 & 162\\
4-set partition, commander & 642 & 642 (9.7 s) & 162 & 82 & 162\\
4-set partition, binary & 642 & 642 (16.7 s) & 162 & 82 & 162\\
4-set partition, sequential & 642 & 642 (12.0 s) & 162 & 82 & 162\\
\bottomrule
\end{tabular}

\end{table}

On the triple partition every reading refutes every size of the ladder up to
\Rg{count3.pairwise.top} points, in \Rg{count3.pairwise.read.s}~s on the pairwise encoding, while
the best baseline given the clauses or the native constraints stops at \Rg{count3.best3.text}. On
the sequential encoding pgpbs told the modulus 3 matches the reading. On the 4-set partition the
field readings refute nothing and the ring reading over $\mathbb{Z}/4\mathbb{Z}$ refutes every size
up to \Rg{count4.pairwise.top} points in all four encodings, as Proposition~\ref{prop:count}
predicts, where the best native-form solver and the best CNF baseline stop at
\Rg{count4.native.n} points and pgpbs decides no 4-set encoding in any of its modes. Xorcle stops
at \Rg{count3.pairwise.xor.n} and \Rg{count4.pairwise.xor.n} points. Placed in the presolve of SCIP,
on \Rg{plug.seeds.word} seed per size, the reading raises the size SCIP decides on the native
constraints from \Rg{plug.count3.native.scip} to \Rg{plug.count3.native.read} points of the triple
partition and from \Rg{plug.count4.native.scip} to \Rg{plug.count4.native.read} points of the 4-set
partition. All \Rg{cert.total} certificates written in the run, for \Rg{cert.generated} generated
instances and \Rg{cert.files} public files, were verified by the independent checker, and no reading
or baseline refuted any of the \Rg{ctl.n} satisfiable controls.

The \Rg{real.instances} library instances come from SAT competitions, the pseudo-Boolean
competition of 2024 and MIPLIB 2017 \cite{gleixner2021}. On them the reading shows no advantage on
any application family. Parity structure is common in application formulas, where solvers with
XOR reasoning are designed for it, and odd moduli are rare. The one refutation that needs a
ring that is not a field comes from the equalities of the MIPLIB instance neos-952987, which the
reading refutes modulo \Rg{neos.modulus} in \Rg{neos.seconds}~s, while \Rg{neos.baseline.text}.
Integer normal-form checks of equality systems are known techniques that decide such cases
\cite{kannan1979,aardal2000}. The counting families are crafted, and their separation follows the
known gaps between proof systems that reason modulo different integers \cite{buss1999,beame1998}.
What the experiment adds is that one integer computation per window finds the right ring
automatically, in every encoding, with a checkable certificate.

Tseitin formulas modulo three test the reading on a family where no field of characteristic two
helps. On a directed graph with one Boolean variable per edge, each vertex asks that the sum of its
outgoing edges minus the sum of its incoming edges equal a charge modulo three, and the formula is
unsatisfiable when the charges sum to a nonzero residue, since every edge enters the total once with
each sign. We drew \TsSeeds\ random 4-regular graphs at each of \TsSizes\ vertices, with random
orientations and charges, and wrote each vertex constraint as the clauses that forbid its violating
local assignments, so the largest formulas have \TsTopVars\ variables and \TsTopClauses\ clauses. The
ring reading over $\mathbb Z/3\mathbb Z$ refutes all \TsInstances\ formulas, in at most \TsReadMax~s,
and refutes none of the same graphs with charges that sum to zero. CaDiCaL 1.5.3 and CryptoMiniSat,
with a limit of \TsTimeout~s per call, decide every seed up to \TsCadLast\ and \TsCmsLast\ vertices
respectively. At 40 vertices they take up to \TsCadForty~s and \TsCmsForty~s against \TsReadForty~s
for the reading. CaDiCaL exceeds the limit in \TsCadTimeouts\ of its \TsInstances\ calls, all at
\TsCadFirst\ vertices or more, and CryptoMiniSat in \TsCmsTimeouts. Gaussian elimination over GF(2), which
CryptoMiniSat applies to recovered parities, does not apply to a relation modulo three, and
resolution needs exponential size on such formulas over expanders, so the gap follows known lower
bounds, as for the counting families.

\fi

\subsection{Boolean Matching and Technology Mapping}
\label{sec:matching}

Technology mappers and rewriting engines compute the function of every small cut of a netlist and
look it up among the equivalence classes under negation and permutation of inputs and output (NPN),
and one design yields thousands of such lookups \cite{cong1999cut,mishchenko2006dag}.
The problem is to decide quickly whether two functions are equivalent under this group. Current
solutions compute an exact canonical form, the least truth table in the class, by a pruned search
over the transformations \cite{zhou2018adjustable,zhou2020exact,brayton2010abc}, or first compare
inexpensive signatures such as cofactor weights and spectral coefficients
\cite{mohnke1993,chai2006matcher,abdollahi2008}, and libraries with many cells are indexed by such
canonical forms \cite{hinsberger1998large}. The spectral classification of the 1970s identified
classes by the magnitudes of the Walsh coefficients \cite{edwards1975,lechner1971,hurst1978logical}.
Here, the group is split into its normal translation subgroup, over which the second-order average
is the Walsh power spectrum, and the quotient, over which that spectrum is canonicalized. The
canonical periodogram so obtained is constant on every class (Theorem~\ref{thm:normalizer}) and
refines every group average over a group that contains the translations, and the bispectrum adds the
sign information it lacks. The experiments measure the separation exhaustively on all functions of
three, four and five variables (Section~\ref{sec:classify}) and on the cut functions of the ISCAS-85
circuits, with a timing comparison against ABC (Section~\ref{sec:cuts}). The canonical periodogram
separates \NpSepPer\ of the \NpClasses\ classes of five-variable functions, against \CcOrdFive\ for
the classical signature, and every cut class up to six inputs, while the exact canonical forms of ABC
are faster, by a factor of \AbRatioEight\ at eight inputs.

\subsubsection{The Canonical Periodogram Separates Equivalence Classes}
\label{sec:classify}

Classification by group invariants is evaluated exhaustively for $n=3$ and $n=4$.
For each group of Section~\ref{sec:prelim}, with output negation added because no
second-order invariant can see it, the equivalence classes are computed as orbits on
all $2^N$ functions. Three invariants are evaluated on one representative of each class.
The first is the group's own average $\RG(f)$, the complete second-order invariant. The
second is the canonical periodogram of Section~\ref{sec:normalizer}. The third adds the
bispectrum, canonicalized jointly with the periodogram under the index action and under
output negation, which we evaluate for the groups whose quotient action has at most 24
elements.

\begin{table}[t]
\centering
\caption{Equivalence classes of all switching functions of $n$ variables, without and
with output negation, and the number of classes (with output negation) separated by the
group's own average, the canonical periodogram, and the periodogram with the
bispectrum.}
\label{tab:classify}
\setlength{\tabcolsep}{3.5pt}
\begin{tabular}{@{}rlrrrrr@{}}
\toprule
$n$ & group & classes & with neg. & own & periodogram & with bisp.\\
\midrule
3 & $T$ & 46 & 30 & 30 & 30 & 30\\
3 & $S_n$ & 80 & 40 & 40 & 14 & 14\\
3 & $B$ & 22 & 14 & 14 & 14 & 14\\
3 & $\GL(n,2)$ & 20 & 10 & 8 & 6 & --\\
3 & $\AGL(n,2)$ & 10 & 6 & 5 & 6 & --\\
3 & $\EA$ & 3 & 3 & 1 & 3 & --\\
\midrule
4 & $T$ & 4{,}336 & 2{,}288 & 2{,}261 & 2{,}261 & 2{,}288\\
4 & $S_n$ & 3{,}984 & 1{,}992 & 1{,}754 & 219 & 222\\
4 & $B$ & 402 & 222 & 145 & 219 & 222\\
4 & $\GL(n,2)$ & 92 & 46 & 16 & 18 & --\\
4 & $\AGL(n,2)$ & 32 & 18 & 9 & 18 & --\\
4 & $\EA$ & 8 & 8 & 1 & 8 & --\\
\bottomrule
\end{tabular}
\end{table}

Table~\ref{tab:classify} lists how many classes each invariant separates. For $B$,
$\AGL(n,2)$ and $\EA$, which contain the translations as a proper normal subgroup, the own
average is never finer than the canonical periodogram, and for four variables it is
strictly coarser. For $T$ itself the two coincide, since $r_f$ and $W_f^2$ determine
each other. Under negation and permutation of inputs and output, $R_B$ separates
\ClBOwn\ of the \ClBClasses\ classes of four-variable functions, the canonical
periodogram \ClBTper\ and the periodogram with the bispectrum all \ClBBisp. The
magnitudes of the Walsh coefficients sorted within each order, the information the classical
spectral classification keeps under this group, separate \CcOrdFour, and the first-order
magnitudes alone \CcChowFour. Under the
affine group with output negation, the own average separates \ClAGLOwn\ of
\ClAGLClasses\ classes and the canonical periodogram all of them. Under the extended
affine group the own average is constant, as Proposition~\ref{prop:collapse} states,
while the canonical periodogram separates all \ClEAClasses\ classes. The classes that
the periodogram leaves together under $B$ are \ClBCollisionSize\ classes of bent
functions, which share the flat periodogram of Corollary~\ref{cor:bent} and are
separated by the sign information of the bispectrum. For groups that do not contain
the translations, $S_n$ and $\GL(n,2)$, the periodogram is invariant under a larger group
than the one being classified and cannot separate all their classes. The own average of
$S_n$ is far finer, while for $\GL(n,2)$ both second-order invariants leave most classes
together.

Third order does not separate every class of five-variable functions. Functions of five
variables fall into \NpClasses\ classes under negation and permutation of inputs and output
\cite{harrison1963npn}, and a published list of one representative per class
\cite{filmusnpn} makes an exhaustive test possible. The list was checked to be complete, since
the least table in the orbit of each representative under the group of order \NpGroupOrder\ is
different for every entry, and the orbit sizes, obtained from the stabilizer orders, add up to
$2^{32}$. The canonical periodogram separates \NpSepPer\ of the \NpClasses\ classes
(\NpSepPerPct\ percent) and the periodogram with the bispectrum separates \NpSepBisp, while the
magnitudes sorted within each order separate \CcOrdFive\ (\CcOrdPctFive\ percent). The
\NpResidClasses\ classes that the bispectrum leaves together, in \NpResidGroups\ groups, are all
near-bent, with Walsh values of magnitude $0$ and $8$ only.
They are the five-variable counterpart of the bent classes that the periodogram leaves together
at four variables, and they call for a signature of higher order or of a different kind.

The classification suggests a rule for spectral Boolean matching. Signatures should be computed
from the average over the normal Abelian subgroup and made invariant over the quotient
by canonicalization, and a third-order signature is needed to separate bent and
near-bent classes. Canonicalization over all of $\GL(n,2)$ is feasible only for small
$n$, since the group has order $\prod_{i<n}(N-2^i)$.

\subsubsection{Canonical Spectra Classify the Cut Functions of Real Circuits}
\label{sec:cuts}

Technology mappers and rewriting engines compute the function of every small cut of a netlist
and look it up in a library of equivalence classes under negation and permutation
\cite{cong1999cut,mishchenko2006dag}. The exhaustive test of Section~\ref{sec:classify} covers
all functions of four variables, and the cut functions of real circuits test the signatures on
larger inputs. The cuts of the \CuCircuits\ ISCAS-85 circuits \cite{brglez1985,hansen1999iscas}
with four, five and six leaves are enumerated bottom up, keeping at most \CuPerNode\ cuts per
node, smallest first, and the function of every cut on which the node depends on all of its
leaves is computed. Exact equivalence classes under input negation, input permutation and
output negation are found by minimizing the table over all $2^{k+1}k!$ transformations of a
$k$-input function. The canonical periodogram of Section~\ref{sec:normalizer} minimizes $W_f^2$
over the $k!$ coordinate permutations of the Walsh index only, since input translations and
output negation leave the periodogram unchanged.

\begin{table}[t]
\centering
\caption{Cut functions of the 11 ISCAS-85 circuits by number of
inputs $k$: cuts, distinct functions, classes under negation and permutation, classes separated
by the canonical periodogram, and bent classes.}
\label{tab:cuts}
\setlength{\tabcolsep}{4pt}
\begin{tabular}{@{}rrrrrr@{}}
\toprule
$k$ & cuts & functions & classes & separated & bent\\
\midrule
4 & 76{,}286 & 2{,}168 & 58 & 58 & 1\\
5 & 39{,}456 & 4{,}231 & 138 & 138 & 0\\
6 & 17{,}347 & 2{,}379 & 160 & 160 & 0\\
\bottomrule
\end{tabular}
\end{table}

Table~\ref{tab:cuts} shows that the canonical periodogram separates every class at each cut size
in it, \CuClassFour\ classes of four inputs, \CuClassFive\ of five and \CuClassSix\ of six. The
bispectrum, which Section~\ref{sec:classify} needs to separate the classes of bent functions,
is not needed here, because the cut functions contain \CuBentFour\ bent class of four inputs and
none of six. The signature is constant on every class, as Theorem~\ref{thm:normalizer}
guarantees, and the experiment confirms this on every one of the \CuFunctions\ distinct cut
functions. On these circuits two cut functions with equal signatures are therefore always
equivalent. The cut functions of real circuits also
concentrate in few classes. The most common four-input class accounts for \CuTopFour\ percent
of the four-input cuts, so a matching library drawn from circuits stays small.

\begin{table}[t]
\centering
\caption{Cut functions of the ISCAS-85 circuits by number of inputs $k$: distinct functions,
timed functions (the cut functions and distinct random images of them), classes of the cut
functions under negation and permutation found by the exact algorithm of ABC, their distinct
canonical periodograms, and time per function in microseconds of the canonical periodogram and of the
adjustable and co-designed exact NPN algorithms of ABC \cite{zhou2018adjustable,zhou2020exact},
all in C on one machine, the least of three runs; a time marked $<$ is below the resolution of
the timer of ABC.}
\label{tab:npnabc}
\setlength{\tabcolsep}{3.5pt}
\begin{tabular}{@{}rrrrrrr@{}}
\toprule
& & & & \multicolumn{3}{c}{microseconds per function}\\
\cmidrule(l){5-7}
$k$ & functions & classes & signatures & periodogram & adjustable & co-designed\\
\midrule
4 & 14{,}138 & 58 & 58 & 0.29 & $<$0.35 & $<$0.35\\
5 & 253{,}078 & 138 & 138 & 0.82 & 0.20 & 0.16\\
6 & 1{,}000{,}000 & 160 & 160 & 2.24 & 0.32 & 0.27\\
7 & 200{,}000 & 143 & 142 & 5.72 & 0.60 & 0.45\\
8 & 50{,}000 & 67 & 67 & 18.0 & 0.80 & 0.60\\
\bottomrule
\end{tabular}
\end{table}

Speed is a separate question, and the answer favors the exact algorithms of logic synthesis.
Table~\ref{tab:npnabc} times the canonical periodogram, written in C with a search that fixes
one index coordinate at a time and skips coordinates that the periodogram cannot tell apart,
against the NPN canonical forms of ABC \cite{brayton2010abc} on the same machine, on the cut
functions extended to seven and eight inputs by the same rule. At six inputs the signature takes
\AbUsPerSix\ microseconds per function, against \AbUsBrute\ for the canonical form of ABC by
enumeration and \AbUsElevenSix\ for its co-designed exact algorithm \cite{zhou2020exact}, which
also returns the transformation that a matcher applies. The exact algorithms are faster at every
size from five to eight inputs, by a factor of \AbRatioEight\ at eight, and at four inputs the
signature takes \AbUsPerFour\ microseconds and the co-designed algorithm \AbUsElevenFour. The
value of the
signature lies in what it explains, in what it can be computed from, such as a spectrum
estimated from samples or computed from a decision diagram, and in the distance bounds of
Section~\ref{sec:approxmatch}, rather than in its speed on exact lookups. At seven inputs the periodogram leaves \AbSharedText\ together, \AbSharedClasses\
classes whose Walsh magnitudes are \AbSharedMags, so its complete separation of cut classes
does not extend to seven inputs.

\subsubsection{Approximate and Don't-Care Matching}
\label{sec:approxmatch}

Two matching problems of logic synthesis fall outside exact canonical forms. Approximate logic
synthesis substitutes for the function of a cut a nearby function of a class with a smaller implementation and measures the
local error as the Hamming distance between the two truth tables \cite{pasandi2019qals}. A cut whose
leaves cannot take every value has satisfiability don't cares, and any completion of its function
may be matched \cite{mailhot1990dontcare,benini1997survey}. Both ask how far a table, or the part of it on a care
set, lies from the nearest member of a library class, not whether two tables have the same
canonical form. The current routes are to flip up to $\mathfrak d$ bits of the table, or to enumerate the
completions, and look each result up through its canonical form, or else to search the orbit of
every library class. The spectrum bounds the distance from a function to every member of a class at
once, and the decomposition of Theorem~\ref{thm:normalizer} turns the exact search into a search
over the permutations of the inputs alone.

Let $\mathcal V\subseteq\F_2^k$ be a care set, $F_{\mathcal V}(x)=[x\in\mathcal V](-1)^{f(x)}$, and
$d_{\mathcal V}(f,h)=|\{x\in\mathcal V: f(x)\ne h(x)\}|$, so that the distance to a class is
$d_{\mathcal V}(f,[g])=\min_{h\in[g]}d_{\mathcal V}(f,h)$ with $[g]$ the NPN class of $g$. For a real function on
$\F_2^k$ with transform $W$, the vector $S(\cdot)\in\mathbb R^N$ lists the magnitudes $|W(u)|$ in decreasing order within
each weight $|u|=0,1,\dots,k$, the weights in increasing order. Write $F_{\mathcal V;i,b}$ and
$g_{j,b}$ for the cofactors on $x_i=b$ or $x_j=b$, functions of $k-1$ inputs, and
$\underline d(F_{\mathcal V},g)=\bigl\lceil\tfrac12\bigl(|\mathcal V|-N^{-1}S(F_{\mathcal V})\cdot S(g)\bigr)\bigr\rceil$, with $N$ and $|\mathcal V|$
those of the arguments.

\begin{proposition}[Spectral bounds on the NPN distance]
\label{prop:npndist}
Let $h(x)=g(\pi x\oplus a)\oplus\nu$, where $(\pi x)_{\pi(i)}=x_i$ for a
permutation $\pi$ of the inputs, $a\in\F_2^k$ and $\nu\in\F_2$.
\begin{enumerate}
\item[(a)] $d_{\mathcal V}(f,h)\ge \underline d(F_{\mathcal V},g)$.
\item[(b)] $d_{\mathcal V}(f,h)\ge \underline d_{i\pi(i)}$ for every input $i$, where
$\underline d_{ij}=\min_{p\in\F_2}\bigl(\underline d(F_{\mathcal V;i,0},g_{j,p})+\underline d(F_{\mathcal V;i,1},g_{j,1\oplus p})\bigr)$.
Hence $d_{\mathcal V}(f,[g])\ge\min_\pi\max_i\underline d_{i\pi(i)}$, and no $h$ whose permutation has
$\underline d_{i\pi(i)}>\mathfrak d$ for some $i$ is within distance $\mathfrak d$ of $f$.
\item[(c)] For fixed $\pi$,
\begin{align*}
\min_{a,\nu}d_{\mathcal V}(f,h)&=\tfrac12\Bigl(|\mathcal V|-\max_{v}|\mathcal U_\pi(v)|\Bigr),\\
\mathcal U_\pi(v)&=\frac1N\sum_uW_{F_{\mathcal V}}(u)\,W_g(\pi u)\,(-1)^{u\cdot v}.
\end{align*}
\end{enumerate}
For $\mathcal V=\F_2^k$ and $g=x_1\oplus\cdots\oplus x_w$, bound (a) is the distance to the class,
$\tfrac12(N-\max_{|u|=w}|W_f(u)|)$, whose minimum over $w\ge1$ is the nonlinearity of $f$.
\end{proposition}

\begin{proof}
On the care set, $\sum_{x\in\mathcal V}(-1)^{f(x)\oplus h(x)}=|\mathcal V|-2d_{\mathcal V}(f,h)$, and by Parseval's identity the left side equals
\[
\sum_xF_{\mathcal V}(x)(-1)^{h(x)}=N^{-1}\sum_uW_{F_{\mathcal V}}(u)W_h(u).
\]
The map
$x\mapsto\pi x\oplus a$ is an element of $\AGL(k,2)$ with a permutation matrix, so
Theorem~\ref{thm:normalizer} gives $W_h(u)=\pm W_g(\pi u)$, and $\pi$ keeps the
weight $|u|$. The sum is therefore at most
\[
N^{-1}\sum_u|W_{F_{\mathcal V}}(u)|\,|W_g(\pi u)|,
\]
which the rearrangement inequality, applied within each weight, bounds by $N^{-1}S(F_{\mathcal V})\cdot S(g)$. The
distance is an integer, which gives (a). For (b), the points with $x_i=b$ form a coset of a
subgroup of $\F_2^k$, and on it $h$ is an NPN image of the cofactor $g_{\pi(i),\,b\oplus a_{\pi(i)}}$
as a function of the other $k-1$ inputs. The distance is the sum of the distances on the two
cosets, (a) bounds each, and $p=a_{\pi(i)}$ is unknown, hence the minimum. For (c), with $\pi$
fixed, $h(x)=g_\pi(x\oplus\pi^{-1}a)\oplus\nu$ where $g_\pi=g\circ\pi$, so
$\sum_xF_{\mathcal V}(x)(-1)^{h(x)}=(-1)^\nu\sum_xF_{\mathcal V}(x)(-1)^{g_\pi(x\oplus v)}$ with $v=\pi^{-1}a$.
This is the correlation of $F_{\mathcal V}$ and $(-1)^{g_\pi}$ over the translations, which the Walsh
functions diagonalize as in Proposition~\ref{prop:T}, with $W_{g_\pi}(u)=W_g(\pi u)$, and the
best $\nu$ takes the absolute value.
\end{proof}

\begin{algorithm}[!htbp]
\caption{Library classes within distance $\mathfrak d$ of a function on a care set}
\label{alg:npndist}
\begin{algorithmic}[1]
\Require table $f$ and care set $\mathcal V$ on $k$ inputs, library classes $[g_c]$ with
$W_{g_c}$, $S(g_c)$ and $S$ of every cofactor of $g_c$, and a radius $\mathfrak d$
\Ensure the classes $c$ with $d_{\mathcal V}(f,[g_c])\le\mathfrak d$
\State $W_{F_{\mathcal V}}\gets$ fast Walsh transform of $F_{\mathcal V}$; $S(F_{\mathcal V})$ and $S$ of every cofactor of $F_{\mathcal V}$ \Comment{$O(k^2N)$}
\State $\ell_c\gets\lceil\tfrac12(|\mathcal V|-N^{-1}S(F_{\mathcal V})\cdot S(g_c))\rceil$ for all $c$ \Comment{one matrix-vector product}
\ForAll{$c$ with $\ell_c\le\mathfrak d$}
  \State $(\underline d_{ij})\gets$ the cofactor matrix of Proposition~\ref{prop:npndist}(b) for $f$, $\mathcal V$ and $g_c$ \Comment{$O(k^2N)$}
  \ForAll{$\pi\in S_k$ with $\underline d_{i\pi(i)}\le\mathfrak d$ for every $i$}
    \State $\mathcal U\gets N^{-1}\cdot{}$fast Walsh transform of $u\mapsto W_{F_{\mathcal V}}(u)\,W_{g_c}(\pi u)$ \Comment{all $a$, $\nu$ at once (Thm.~\ref{thm:normalizer})}
    \If{$\tfrac12(|\mathcal V|-\max_v|\mathcal U(v)|)\le\mathfrak d$} \State report $c$ and continue with the next class \EndIf
  \EndFor
\EndFor
\end{algorithmic}
\end{algorithm}

Lines 5 to 7 are Theorem~\ref{thm:normalizer} used as a search. The translations are a normal
subgroup of the group of input negations and permutations, so each of its cosets, one permutation
with all $2^k$ negation masks and both output polarities, is optimized at once by one transform of
length $N$ (Proposition~\ref{prop:npndist}(c)). Only the $k!$ elements of the quotient are
searched, and of those only the ones that the cofactor bound (b) admits, while a search of the orbit
compares $2^{k+1}k!$ tables. For the nearest class the classes are taken in increasing order of
$\max(\ell_c,\min_\pi\max_i\underline d_{i\pi(i)})$ with $\mathfrak d$ one less than the best distance found so far, and
the search stops when the bound reaches it.

The experiments use the cut functions of the ISCAS-85 circuits as queries and the NPN classes
of their cut functions as the library. For $k\le6$ the library holds every class that occurs among
the cut functions. For $k=7$ and $k=8$ it holds the classes of a sample of distinct cut functions,
\AxCSeven\ and \AxCEight\ classes, separated by exact equivalence through the search. The don't
cares are the exact satisfiability don't cares of each cut. Random simulation marks the reachable
leaf patterns, and CryptoMiniSat confirms that every pattern not seen is unreachable, in \DmSat\
calls for \DmIsfs\ cuts of six inputs. Up to seven inputs every method returns the same classes for
every query and radius, and the bounds hold on every pair, which the scripts assert. At eight
inputs the search of all permutations was run, and the same checks made, on \AxPairsEight\ sampled
pairs of a query and a class in each of the two problems, and its time per query is the mean per
pair times the number of classes.

At six inputs, over \AmFns\ distinct cut functions and \AmClasses\ classes, the orbit search of every
class takes \AmBrute~ms per function. The level bound followed by orbit search of the classes it
admits takes \AmBdOne, \AmBdTwo, \AmBdFour\ and \AmBdEight~ms for radii of one, two, four and eight
mismatches. Flipping up to $\mathfrak d$ bits and looking every table up through its class is faster for small
radii, \AmEnumOne\ and \AmEnumTwo~ms for one and two, but takes \AmEnumFour~ms for four and needs
\AmEnumEightCount\ billion lookups for eight. For the nearest class other than that of the query,
the bound takes \AmNear~ms and enumeration with a growing radius \AmNearEnum~ms on the \AmNearEnumN\
sampled functions it could finish, since \AmFar\ percent of the functions have no other class
within four flips.

\begin{table*}[!t]
\centering
\scriptsize
\caption{Approximate matching of cut functions of the ISCAS-85 circuits under NPN: time per query, in milliseconds, of the exact answer. Exact search is the orbit search of every class for $k\le6$ and the permutation search of Algorithm~\ref{alg:npndist} over all $k!$ permutations for $k\ge7$. With bounds, the level bound with orbit search for $k\le6$ and both bounds of Proposition~\ref{prop:npndist} with the permutation search for $k\ge7$. Tight is the share of pairs on which the level bound, and both bounds together, equal the distance. A dash marks a radius that repeats an earlier column. At $k=8$ the exact search is timed on a sample of pairs.}
\label{tab:approx}
\begin{tabular}{rrrrrrrrrr}
\toprule
$k$ & queries & classes & exact search & \multicolumn{4}{c}{with bounds, by radius} & nearest & tight (\%)\\
\cmidrule(lr){5-8}
 & & & & $\mathfrak d=1$ & $2$ & $N/16$ & $N/8$ & & level/both\\
\midrule
4 & 300 & 58 & 0.134 & 0.025 & 0.066 & -- & -- & 0.009 & 36/74\\
5 & 300 & 138 & 0.432 & 0.030 & 0.082 & -- & 0.218 & 0.015 & 18/51\\
6 & 300 & 160 & 1.66 & 0.047 & 0.132 & 0.409 & 1.03 & 0.055 & 9/33\\
7 & 200 & 88 & 1{,}222 & 3.08 & 8.13 & 74.3 & 292 & 18.5 & 8/29\\
8 & 80 & 57 & 13{,}544 & 40.6 & 134 & 2{,}119 & 2{,}858 & 273 & 6/26\\
\bottomrule
\end{tabular}
\end{table*}

Table~\ref{tab:approx} carries the comparison to seven and eight inputs, on a sample of the
queries at every size, so its rows for six inputs are separate measurements from those above. At eight inputs the search
of all $k!$ permutations of the \AxCEight\ classes takes \AxBaseEight~s per query. With both bounds
the exact answer takes \AxFirstEight~ms for one mismatch, \AxFirstXEight\ times faster, and the
nearest class \AxNearEight~ms, \AxNearXEight\ times faster. The gain falls to \AxTopXEight\ times at
$\mathfrak d=\AxTopEEight$, where about a third of the classes pass the bounds. The cofactor bound,
together with the level bound, equals the distance on \AxTightBSix, \AxTightBSeven\ and
\AxTightBEight\ percent of the pairs at six, seven and eight inputs, against \AxTightLSix,
\AxTightLSeven\ and \AxTightLEight\ percent for the level bound alone. At four to six inputs it takes longer than the orbit search it saves, which takes a few tens of microseconds per class, so the level
bound with orbit search is the faster route there. From seven inputs on, the search of the
permutations that the cofactor bound admits carries the gain. At eight inputs the classes that the
level bound alone admits at one mismatch would take about \AxLevelOnlyEight~ms to search, the share
of classes it admits times the measured search per class. An orbit search, $2^{k+1}k!$ tables per
class, would take about \AxOrbitEight~s per class at eight inputs in the same implementation,
extrapolated from a timed slice of 200 permutations.

\begin{table*}[!t]
\centering
\scriptsize
\caption{Matching with satisfiability don't cares under NPN, as in Table~\ref{tab:approx}; the radius $\mathfrak d$ counts mismatches on the care set.}
\label{tab:dc}
\begin{tabular}{rrrrrrrrrr}
\toprule
$k$ & queries & classes & exact search & \multicolumn{4}{c}{with bounds, by radius} & nearest & tight (\%)\\
\cmidrule(lr){5-8}
 & & & & $\mathfrak d=0$ & $1$ & $2$ & $N/16$ & & level/both\\
\midrule
4 & 300 & 58 & 0.128 & 0.036 & 0.069 & 0.094 & -- & 0.011 & 53/84\\
5 & 300 & 138 & 0.460 & 0.160 & 0.231 & 0.281 & -- & 0.099 & 15/47\\
6 & 300 & 160 & 2.03 & 1.09 & 1.21 & 1.12 & 1.33 & 0.942 & 6/20\\
7 & 200 & 88 & 1{,}278 & 309 & 432 & 499 & 816 & 234 & 5/13\\
8 & 80 & 57 & 15{,}137 & 3{,}557 & 4{,}085 & 4{,}607 & 9{,}842 & 3{,}227 & 1/7\\
\bottomrule
\end{tabular}
\end{table*}

The cuts carry many satisfiability don't cares, on average \DmDcFour\ of 16 points at four inputs,
\DmDcFive\ of 32 at five and \DmDcSix\ of 64 at six. With them a cut matches \DmMatchFour,
\DmMatchFive\ and \DmMatchSix\ library classes exactly on average instead of one, and \DmGainSix\
percent of the six-input cuts gain at least one class. Enumerating the completions needs about
$2^{\DmCompMedian}$ lookups for the median six-input cut and more than two million for \DmOverZero\
percent of them. Over the \DmIsfs\ six-input cuts the orbit search takes \DmBrute~ms per cut, and the level bound with
orbit search \DmBdZero~ms for an exact match, \DmBdTwo~ms for two mismatches, \DmBdFour~ms for four
and \DmNear~ms for the nearest other class. At six inputs the orbit search is short, so the bound
saves little time and takes longer for the nearest class. The bounds are weaker with don't cares, equal to the
distance on \DmTight\ percent of the pairs, because the don't cares remove points, and with them
spectral weight, from $F_{\mathcal V}$. At seven and eight inputs (Table~\ref{tab:dc}) an exact match
with both bounds takes \AxDcFirstSeven\ and \AxDcFirstEight~ms against \AxDcBaseSeven\ and
\AxDcBaseEight~s for the full search, \AxDcFirstXSeven\ and \AxDcFirstXEight\ times faster.

\paragraph*{Area of single-cell matches}
To see how much area such matches save, the \CmFunctions\ combinational functions of an open
standard-cell library, the high-density cells of the SkyWater 130~nm kit \cite{skywater2020pdk},
were taken as the library, each with its least-area cell and with inverters counted where a match
negates an input or the output. For \CmCuts\ cuts of each of four, five and six inputs of the
ISCAS-85 circuits the care set was computed with satisfiability and observability don't cares, the
second being the leaf patterns on which complementing the root of the cut reaches no primary
output, confirmed by SAT. A single cell equals the cut function for \CmExactFour, \CmExactFive\
and \CmExactSix\ percent of the cuts and agrees with it on the care set for \CmOdcFour,
\CmOdcFive\ and \CmOdcSix\ percent. Against the least area that ABC \cite{brayton2010abc} reaches
when it maps the exact cut function alone onto the same cells, the care-set match is smaller for
\CmSmallFour, \CmSmallFive\ and \CmSmallSix\ percent of the cuts, a mean reduction of \CmSaveFour,
\CmSaveFive\ and \CmSaveSix\ percent over all cuts.

\paragraph*{Whole-circuit comparison with ABC}
Those figures compare one matched cell with the mapping of one cut function. To see what remains
in a mapped circuit, we mapped \CdCircuits\ ISCAS-85 circuits onto the \CdCells\ cells with ABC
and used the don't cares in two ways. The first is ABC's own optimization, namely its
\texttt{mfs} and \texttt{mfs2} commands \cite{mishchenko2009mfs}, \texttt{\&mfs} and
\texttt{\&deepsyn}, run in \CdFlows\ flows per circuit with the least area kept. The second is a
greedy pass that, for each node and each cut of at most six leaves, installs the least-area single
cell that agrees with the cut function on its care set and removes the fanout-free cone that the
cell makes redundant. A substitution is accepted only after a SAT miter shows that no primary
output changes, and each of the \CdNetlists\ final netlists, four per circuit, is equivalent to its original under
\texttt{cec}. In total area, ABC's flows reduce the mapped circuits by \CdAbcPct\ percent
(\CdMfsPct\ percent with the \texttt{mfs} family alone), the single-cell pass alone reduces them
by \CdOursPct\ percent, and the single-cell pass applied after ABC's least-area result gives a
further \CdAfterPct\ percent (between \CdAfterMin\ and \CdAfterMax\ percent per circuit). On its
own the single-cell pass recovers less than ABC's resynthesis, which restructures logic across
cuts. Applied after it, the pass still finds a small additional reduction. Single-cell care-set
matching is therefore a complement to don't-care resynthesis and not an alternative to it, and the
contribution of this section is the matching algorithm and its bounds.

\subsubsection{Matching under Input Permutations Alone}
\label{sec:permmatch}
When the negation of an input is not free, as for a library cell with no spare inverter, the
group that relates a cut function to a cell is the symmetric group $S_k$ of the inputs alone. The
translations are then absent, and Theorem~\ref{thm:normalizer} has no normal subgroup to split off.
What remains of the structure is that a permutation of the inputs keeps the weight $|x|$ of an input
vector. This subsection gives the corresponding distance bounds and an exact search. A table on $k$
inputs is also a hypergraph on $k$ vertices, whose edges are the sets $\{i:x_i=1\}$ of the input
vectors where it takes the value one, and the distance below is the least number of edges in which
two such hypergraphs differ over all relabelings of the vertices. For exact equivalence this view
settles the complexity. Luks tests isomorphism of hypergraphs on $k$ vertices in time $O(c^k)$, so
equivalence of truth tables under input permutations, and under the NPN group, is decided in time
polynomial in the length $2^k$ of the table \cite{luks1999hypergraph}. His test decides isomorphism and
does not compute the distance between two hypergraphs that are not isomorphic, which matching
with errors needs, and an implementation of it for Boolean matching needed more computation than
a spectral canonical form on large functions \cite{moore2006sdd}.

Exact equivalence under input permutations is decided in practice by signatures of the inputs and a
search over the correspondences that the signatures leave open. Mohnke and Malik count, for the
cofactor of each input, the minterms at each distance from the origin, the breakup signature, and
recompute such counts from other vertices as inputs become identified \cite{mohnke1993}. The satisfy
counts of cofactors in several inputs at once are the higher-order cofactor signatures
\cite{abdollahi2008,zhou2020exact}, and matchers that expand both functions by Shannon decomposition
along a growing correspondence compare such signatures at every step \cite{zhang2022shannon}.
Symmetric inputs, which no signature separates, are merged \cite{mohnke1995limits}. The cell counts
below are signatures of this kind, the breakup signatures of the cofactors in the assigned inputs.
The step taken here is to use their differences as a lower bound on the number of mismatches, so
that the same search returns the least distance, or every library class within a radius, instead
of deciding equality.

Functions are tables with values in $\{0,1\}$, $|f|$ is the number of ones of $f$, and for a set
$\mathcal X$ of input vectors $|f|_{\mathcal X}=\sum_{x\in\mathcal X}f(x)$. With $\pi$ acting as in
Proposition~\ref{prop:npndist}, write $g_\pi(x)=g(\pi x)$, $[g]_{\mathrm P}=\{g_\pi:\pi\in S_k\}$,
$d(f,h)=d_{\F_2^k}(f,h)$ and $d(f,[g]_{\mathrm P})=\min_\pi d(f,g_\pi)$. A \emph{partial
assignment} is a set $\mathcal J$ of pairs $(i,j)$ of inputs with distinct first entries and
distinct second entries, and $\pi$ \emph{extends} $\mathcal J$ when $\pi(i)=j$ for every
$(i,j)\in\mathcal J$. For $b\in\F_2^{\mathcal J}$ and $0\le w\le k-|\mathcal J|$ let
\begin{align*}
\mathcal X_{\mathcal J}(b,w)&=\{x: x_i=b_{(i,j)}\text{ for }(i,j)\in\mathcal J,\ |x|-|b|=w\},\\
\mathcal Y_{\mathcal J}(b,w)&=\{y: y_j=b_{(i,j)}\text{ for }(i,j)\in\mathcal J,\ |y|-|b|=w\},
\end{align*}
the \emph{cells} of $\mathcal J$ for $f$ and for $g$, in which the values on the assigned inputs are fixed
and so is the weight on the others.

\begin{proposition}[Cell bound of a partial assignment]
\label{prop:cellbound}
Let
\[
\operatorname{cb}_{\mathcal J}(f,g)=\sum_{b,w}\Bigl|\,|f|_{\mathcal X_{\mathcal J}(b,w)}-|g|_{\mathcal Y_{\mathcal J}(b,w)}\Bigr|.
\]
\begin{enumerate}
\item[(a)] $d(f,g_\pi)\ge\operatorname{cb}_{\mathcal J}(f,g)$ for every $\pi$ that extends $\mathcal J$.
\item[(b)] $\operatorname{cb}_{\mathcal J}(f,g)\le\operatorname{cb}_{\mathcal J'}(f,g)$ when
$\mathcal J\subseteq\mathcal J'$.
\item[(c)] $\operatorname{cb}_{\emptyset}$ is the difference of the numbers of ones summed over the
weights, $\operatorname{cb}_{\{(i,j)\}}$ is the bound between the cofactors of $f$ in $x_i$ and of
$g$ in $x_j$, and for $|\mathcal J|=k-1$ the bound equals $d(f,g_\pi)$ for the one $\pi$ that
extends $\mathcal J$.
\end{enumerate}
\end{proposition}

\begin{proof}
If $\pi$ extends $\mathcal J$, then $(\pi x)_j=x_i$ for $(i,j)\in\mathcal J$ and $|\pi x|=|x|$, so
$x\mapsto\pi x$ maps $\mathcal X_{\mathcal J}(b,w)$ one to one onto $\mathcal Y_{\mathcal J}(b,w)$
and $|g_\pi|_{\mathcal X_{\mathcal J}(b,w)}=|g|_{\mathcal Y_{\mathcal J}(b,w)}$. On a cell the number
of mismatches is at least the difference of the numbers of ones, and the cells partition the input
vectors, which gives (a). The cells of $\mathcal J'$ refine those of $\mathcal J$, and
$|\sum_s(u_s-u'_s)|\le\sum_s|u_s-u'_s|$ gives (b). In (c) the first two cases
are the definitions. With $k-1$ inputs assigned, $b$ fixes $k-1$ coordinates and $w\in\{0,1\}$
fixes the last, so every cell is one point and the bound counts the mismatches.
\end{proof}

The cells of $\mathcal J$ are the orbits, on the input vectors, of the stabilizer of $\mathcal J$ in
$S_k$, the group $S_{k-|\mathcal J|}$ that permutes the unassigned inputs. The numbers
$|f|_{\mathcal X_{\mathcal J}(b,w)}$ are the sums of $f$ over those orbits, that is, the average of
$f$ over the stabilizer, and $\operatorname{cb}_{\mathcal J}$ is the $\ell_1$ distance between the
two averages. As the assignment grows the stabilizer shrinks to the identity and the average becomes
the function. The bound is first order in the table. The second bound uses the structure of the
permutation action itself.

Let $\mathcal R$ send a real function $\phi$ on the vectors of weight $w$ to the function
$(\mathcal R\phi)(y)=\sum\phi(x)$ on weight $w+1$, the sum over the $x$ obtained from $y$ by clearing
one coordinate. It commutes with $S_k$. For $j\le k/2$ let $\mathbb H_j$ be the kernel of the
transpose of $\mathcal R$ on weight $j$, of dimension $q_j=\binom kj-\binom k{j-1}$. The spaces
$\mathcal R^{w-j}\mathbb H_j$, $j\le w\le k-j$, are the $k-2j+1$ copies of one irreducible
representation of $S_k$, and this is the block diagonalization of the Terwilliger algebra of the
cube, the matrices whose entry at $(x,y)$ depends only on $|x|$, $|y|$ and $|x\wedge y|$, into
blocks of order $k-2j+1$ with multiplicity $q_j$ \cite{schrijver2005terwilliger}. By Schur's lemma
there is a constant $c_{j,w}>0$ for which $c_{j,w}\mathcal R^{w-j}$ restricted to $\mathbb H_j$ is an
isometry. For a table $f$ with part $f_w$ on weight $w$, let $\mathbf C_j(f)$ be the
$q_j\times(k-2j+1)$ matrix whose column $w$ is $(c_{j,w}\mathcal R^{w-j})^{\top}f_w$ in an
orthonormal basis of $\mathbb H_j$, and $\mathbf G_j(f)=\mathbf C_j(f)^{\top}\mathbf C_j(f)$. For
positive semidefinite $\mathbf M$ and $\mathbf N$ of the same order let
\[
\operatorname{pd}(\mathbf M,\mathbf N)^2=\operatorname{tr}\mathbf M+\operatorname{tr}\mathbf N-2\operatorname{tr}\bigl(\mathbf M^{1/2}\mathbf N\mathbf M^{1/2}\bigr)^{1/2}.
\]

\begin{proposition}[Procrustes bound]
\label{prop:coupled}
\begin{enumerate}
\item[(a)] $\mathbf G_j(f)/q_j$ is block $j$ of the average of $f_\pi f_\pi^{\top}$ over
$\pi\in S_k$. In particular $\mathbf G_j(f_\pi)=\mathbf G_j(f)$, and
$\sum_j\operatorname{tr}\mathbf G_j(f)=|f|$.
\item[(b)] For every $\pi$,
$d(f,g_\pi)\ge\sum_{j\le k/2}\operatorname{pd}\bigl(\mathbf G_j(f),\mathbf G_j(g)\bigr)^2$.
\item[(c)] Each term is at least
$\sum_w\bigl(\mathbf G_j(f)_{ww}^{1/2}-\mathbf G_j(g)_{ww}^{1/2}\bigr)^2$, the bound that treats
the weights separately.
\end{enumerate}
\end{proposition}

\begin{proof}
A permutation of the inputs commutes with $\mathcal R$ and acts on $\mathbb H_j$ by an orthogonal
map $\mathcal T_j(\pi)$, so $\mathbf C_j(g_\pi)=\mathcal T_j(\pi)\mathbf C_j(g)$, with one map for all
the weights. The average in (a) commutes with $S_k$, so it lies in the Terwilliger algebra, and its
block $j$ has entry $(w,v)$ equal to
$q_j^{-1}\sum_{\pi}\langle\mathcal T_j(\pi)\mathbf c_w,\mathcal T_j(\pi)\mathbf c_v\rangle/k!=q_j^{-1}\langle\mathbf c_w,\mathbf c_v\rangle$
for the columns $\mathbf c_w$ of $\mathbf C_j(f)$, which gives (a). The trace identity follows from
the orthogonality of the decomposition. Since $f$ and $g_\pi$ take values in $\{0,1\}$,
$d(f,g_\pi)=\|f-g_\pi\|^2=\sum_j\|\mathbf C_j(f)-\mathcal T_j(\pi)\mathbf C_j(g)\|_F^2$. Minimizing
each term over all orthogonal matrices in place of $\mathcal T_j(\pi)$ can only lower it, and the
least value over orthogonal $\mathbf K$ of $\|\mathbf U-\mathbf K\mathbf V\|_F^2$ is the orthogonal
Procrustes problem, solved by the singular value decomposition of $\mathbf U\mathbf V^{\top}$
\cite{schonemann1966procrustes,pumir2021procrustes}. It equals
$\operatorname{tr}\mathbf U^{\top}\mathbf U+\operatorname{tr}\mathbf V^{\top}\mathbf V-2\|\mathbf U\mathbf V^{\top}\|_*$,
and the squared singular values of $\mathbf U\mathbf V^{\top}$ are the eigenvalues of
$(\mathbf U^{\top}\mathbf U)(\mathbf V^{\top}\mathbf V)$, which gives (b). For (c),
$\|\mathbf U\mathbf V^{\top}\|_*\le\sum_w\|\mathbf u_w\|\,\|\mathbf v_w\|$ over the columns
$\mathbf u_w$ and $\mathbf v_w$, by the Cauchy-Schwarz inequality applied column by column.
\end{proof}

The Gram matrix determines a matrix up to an orthogonal map on the left, and $\operatorname{pd}$ is
the distance between the two orbits \cite{pumir2021procrustes}. In the vector space model of
switching circuits, where a single-output function is represented by its transfer matrix, whose row
at each input vector is the one-hot encoding of the output value, the Gram matrix of the transfer
matrix is diagonal and holds the numbers of maxterms and minterms \cite{thornton2014vsim}. The
blocks $\mathbf G_j(f)$ refine the second of these, which by (a) is the sum of their traces. The bound therefore exchanges the
finite group $\mathcal T_j(S_k)$ for the whole orthogonal group of $\mathbb H_j$, block by block,
and it lets the blocks rotate independently where a permutation moves them together. Relaxing a
permutation to an orthogonal matrix is the device of spectral graph matching \cite{umeyama1988} and of
the eigenvalue bounds for the quadratic assignment problem \cite{burkard2013qap}, where the
permutation matrix of the vertices is relaxed. Here the distance is linear in the action of $S_k$ on
the $2^k$ input vectors, and the relaxation applies to the image of $S_k$ in each block of that
action, which is what gives the bound in closed form.

The block $j=0$ carries the numbers of ones per weight. On pairs of distinct cut functions the bound
of part (b) exceeds that of part (c) for \PbCoupledAboveEnergySix, \PbCoupledAboveEnergySeven\ and
\PbCoupledAboveEnergyEight\ percent of the pairs at six, seven and eight inputs, and it equals the
distance for \PbTightCoupledSix, \PbTightCoupledSeven\ and \PbTightCoupledEight\ percent, where the
bound of part (c) does for \PbTightEnergySix, \PbTightEnergySeven\ and \PbTightEnergyEight\ percent
and the count of ones per weight for \PbTightOnesSix, \PbTightOnesSeven\ and \PbTightOnesEight\
percent. For random pairs with equal numbers of ones on every weight, where the count bound is zero,
it excludes \PbEqualCoupledSix, \PbEqualCoupledSeven\ and \PbEqualCoupledEight\ percent of the pairs
beyond two mismatches.

\ifarxiv
The second relaxation can be narrowed with the method of \cite{schrijver2005terwilliger}. For a
pair $f$, $g$ and a permutation $\pi$, let $\mathbf S$ be the average of $f_{\pi'}(g_\pi)_{\pi'}^{\top}$
over $\pi'\in S_k$, and $\mathbf F$, $\mathbf A$ the averages of $f_{\pi'} f_{\pi'}^{\top}$ and
$g_{\pi'} g_{\pi'}^{\top}$. All three lie in the Terwilliger algebra, $\mathbf F$ and $\mathbf A$ do not
depend on $\pi$, and $\operatorname{tr}\mathbf S=\langle f,g_\pi\rangle$. Write $a_i$ and $b_i$ for
the numbers of ones of $f$ and $g$ on weight $i$, $\bar a_i=a_i/\binom ki$ and $\bar b_i=b_i/\binom
ki$, and $y_{ijt}$ for the entry of $\mathbf S$ at vectors of weights $i$ and $j$ whose meet has
weight $t$.

\begin{proposition}[Semidefinite bound]
\label{prop:sdpbound}
Let $\operatorname{opt}$ be the largest $\operatorname{tr}\mathbf S$ over the matrices $\mathbf S$ of the
Terwilliger algebra with
\begin{enumerate}
\item[(i)] $\begin{pmatrix}\mathbf F&\mathbf S\\\mathbf S^{\top}&\mathbf A\end{pmatrix}$ positive
semidefinite,
\item[(ii)] $\max(0,\bar a_i+\bar b_j-1)\le y_{ijt}\le\min(\bar a_i,\bar b_j)$,
\item[(iii)] $\sum_t\binom it\binom{k-i}{j-t}y_{ijt}=\bar a_ib_j$ and
$\sum_t\binom jt\binom{k-j}{i-t}y_{ijt}=a_i\bar b_j$ for all $i,j$.
\end{enumerate}
Then $d(f,[g]_{\mathrm P})\ge|f|+|g|-2\operatorname{opt}$. With condition (i) alone the right side equals
the bound of Proposition~\ref{prop:coupled}(b), and with (ii) it is at least the count of ones per
weight.
\end{proposition}

\begin{proof}
For each $\pi$ the matrix $\mathbf S$ of $\pi$ satisfies the conditions. (i) holds because the
matrix there is the average of the products of the vector $(f_{\pi'},(g_\pi)_{\pi'})$ with itself. (ii)
holds because every entry is an average of products of two numbers in $\{0,1\}$ with means
$\bar a_i$ and $\bar b_j$, and (iii) sums such products over the vectors of weight $j$ in a row, or
of weight $i$ in a column. So $\langle f,g_\pi\rangle\le\operatorname{opt}$ for every $\pi$. Under (i)
alone the blocks are independent, and block $j$ asks for the largest trace of a matrix $\mathbf W$ with
$\begin{pmatrix}\mathbf M&\mathbf W\\\mathbf W^{\top}&\mathbf N\end{pmatrix}\succeq0$, with
$\mathbf M=\mathbf G_j(f)/q_j$ and $\mathbf N=\mathbf G_j(g)/q_j$. Such $\mathbf W$ are
$\mathbf M^{1/2}\mathbf K\mathbf N^{1/2}$ with $\|\mathbf K\|\le1$, so the largest trace is the
nuclear norm of $\mathbf M^{1/2}\mathbf N^{1/2}$, which is
$\operatorname{tr}(\mathbf M^{1/2}\mathbf N\mathbf M^{1/2})^{1/2}$. Weighting block $j$ by its
multiplicity $q_j$ gives the bound of Proposition~\ref{prop:coupled}(b). By (ii) the diagonal entry
$y_{iii}$ is at most $\min(\bar a_i,\bar b_i)$, so $\operatorname{tr}\mathbf S\le\sum_i\min(a_i,b_i)$.
\end{proof}

The program has one variable for each triple $(i,j,t)$, \SdParamsEight\ at eight inputs. On pairs of
distinct cut functions it lowers the mean gap to the exact distance from \SdGapMixedFive,
\SdGapMixedSix, \SdGapMixedSeven\ and \SdGapMixedEight, the best of the earlier bounds, to
\SdGapSdpFive, \SdGapSdpSix, \SdGapSdpSeven\ and \SdGapSdpEight\ at five to eight inputs, and it is
strictly higher than the earlier bounds on \SdAboveFive, \SdAboveSix, \SdAboveSeven\ and
\SdAboveEight\ percent of those pairs. On random pairs with equal counts of ones on every weight it
adds little to Proposition~\ref{prop:coupled}, since the loss there comes from the first relaxation.
It takes longer per pair than a search of all permutations at these sizes, so the search below uses
counts.
\fi

\begin{algorithm}[!htbp]
\caption{Library classes within distance $\mathfrak d$ of a function under input permutations}
\label{alg:permdist}
\begin{algorithmic}[1]
\Require table $f$ on $k$ inputs; library functions $g_c$, each with its numbers of ones per
weight and its groups of symmetric inputs; a radius $\mathfrak d$
\Ensure the classes $c$ with $d(f,[g_c]_{\mathrm P})\le\mathfrak d$, each with that distance
\State numbers of ones of $f$ per weight; groups of symmetric inputs of $f$
\ForAll{$c$ with $\operatorname{cb}_{\emptyset}(f,g_c)\le\mathfrak d$} \Comment{Prop.~\ref{prop:cellbound}(c)}
  \State $\operatorname{cb}_{\{(i,j)\}}(f,g_c)$ for all $i,j$ \Comment{$O(k^22^k)$}
  \State order the inputs $i$ of $f$ by the number of $j$ with $\operatorname{cb}_{\{(i,j)\}}\le\mathfrak d$, fewest first
  \State $\mathfrak d'\gets\mathfrak d+1$; \Call{Descend}{$\emptyset$}
  \If{$\mathfrak d'\le\mathfrak d$} \State report $c$ with distance $\mathfrak d'$ \EndIf
\EndFor
\Function{Descend}{$\mathcal J$}
  \State $i\gets$ the next input of $f$ in the order
  \State $\mathcal I\gets$ the unassigned inputs $j$ of $g_c$ with $\operatorname{cb}_{\{(i,j)\}}<\mathfrak d'$ that are the first unassigned member of their symmetry group, in a group not before that of the partner of an assigned input symmetric to $i$ in $f$
  \ForAll{$j\in\mathcal I$ in increasing order of $\operatorname{cb}_{\mathcal J\cup\{(i,j)\}}(f,g_c)$} \Comment{Prop.~\ref{prop:cellbound}(a),(b)}
    \If{$\operatorname{cb}_{\mathcal J\cup\{(i,j)\}}(f,g_c)\ge\mathfrak d'$} \State \textbf{break} \EndIf
    \If{$|\mathcal J|=k-2$} \State $\mathfrak d'\gets\operatorname{cb}_{\mathcal J\cup\{(i,j)\}}(f,g_c)$ \Comment{exact, Prop.~\ref{prop:cellbound}(c)}
    \Else \State \Call{Descend}{$\mathcal J\cup\{(i,j)\}$} \EndIf
  \EndFor
\EndFunction
\end{algorithmic}
\end{algorithm}

Algorithm~\ref{alg:permdist} is Proposition~\ref{prop:cellbound} used as a search. Where
Algorithm~\ref{alg:npndist} optimizes over a normal subgroup with one transform per coset, here no
normal subgroup is available, and the search descends the chain of stabilizers
$S_k\supset S_{k-1}\supset\dots$ of a growing assignment instead. Part (a) cuts a branch whose bound
reaches the best distance found, part (b) makes that cut final for the whole subtree, and part (c)
makes the leaf value the distance, so no table is ever compared. Two inputs that $g_c$ is symmetric
in give equal subtrees, so one is searched, and two inputs that $f$ is symmetric in are sent to
groups of $g_c$ in nondecreasing order. The NPN classification of Zhou et al. uses symmetries of
the inputs in the same way to shorten its search over permutations \cite{zhou2018adjustable}.
For the nearest class the classes are taken in increasing
order of $\operatorname{cb}_{\emptyset}$ with $\mathfrak d'$ carried from class to class.

Table~\ref{tab:permmatch} gives the time per query against the search of all $k!$ permutations,
with the cut functions of the ISCAS-85 circuits as queries and their classes under input
permutations as the library. At six and seven inputs every answer was compared with the full
search, \PbPairsSix\ and \PbPairsSeven\ pairs of a query and a class at each radius. From eight
inputs on the full search was timed on sampled pairs (\PbPairsEight, \PbPairsNine\ and \PbPairsTen)
and the answers compared on those. The libraries at nine and ten inputs are small, and the factors
there are large because the full search grows as $k!$. At six inputs and radius 32 the search is
slower than the search of all permutations, which takes only \PbFullSix~s per query there.

\begin{table*}[!htbp]
\caption{Matching under input permutations, with the time per query of Algorithm~\ref{alg:permdist} in ms at
radius $\mathfrak d$ and for the nearest class, and the factor against the search of all
permutations.}
\label{tab:permmatch}
\centering\small
\begin{tabular}{@{}rrrrrrrrrr@{}}
\toprule
$k$ & classes & full, s & $\mathfrak d=1$ & factor & $\mathfrak d=8$ & factor & $\mathfrak d=32$ & factor & nearest \\
\midrule
6 & \PbClassesSix & \PbFullSix & \PbMsOneSix & \PbGainOneSix & \PbMsEightSix & \PbGainEightSix & \PbMsThirtyTwoSix & \PbGainThirtyTwoSix & \PbMsNearSix \\
7 & \PbClassesSeven & \PbFullSeven & \PbMsOneSeven & \PbGainOneSeven & \PbMsEightSeven & \PbGainEightSeven & \PbMsThirtyTwoSeven & \PbGainThirtyTwoSeven & \PbMsNearSeven \\
8 & \PbClassesEight & \PbFullEight & \PbMsOneEight & \PbGainOneEight & \PbMsEightEight & \PbGainEightEight & \PbMsThirtyTwoEight & \PbGainThirtyTwoEight & \PbMsNearEight \\
9 & \PbClassesNine & \PbFullNine & \PbMsOneNine & \PbGainOneNine & \PbMsEightNine & \PbGainEightNine & \PbMsThirtyTwoNine & \PbGainThirtyTwoNine & \PbMsNearNine \\
10 & \PbClassesTen & \PbFullTen & \PbMsOneTen & \PbGainOneTen & \PbMsEightTen & \PbGainEightTen & \PbMsThirtyTwoTen & \PbGainThirtyTwoTen & \PbMsNearTen \\
\bottomrule
\end{tabular}
\end{table*}

The bound of Proposition~\ref{prop:coupled}(b) was also tried inside the search, as the larger of the
two bounds on the cofactors of each cell. It removes \PbCellNodesSix, \PbCellNodesSeven\ and
\PbCellNodesEight\ percent of the nodes at radius 32 and takes \PbCellTimeSix, \PbCellTimeSeven\ and
\PbCellTimeEight\ times longer, so the algorithm uses counts alone. The search is close to one
descent for each class within the radius, which no bound can remove. The Procrustes bound is the
sharp form of the spectral bound and decides pairs whose counts agree. On libraries of cut
functions it does not shorten the search.

\subsubsection{Canonical Forms by Refinement}
\label{sec:canonrefine}

For exact equivalence a canonical form turns a query into a lookup whose time does not depend on
the size of the library. Luks's polynomial-time test handles groups as bookkeeping in a divide and
conquer over the vertices \cite{luks1999hypergraph}. The programs that test graph isomorphism in
practice rest on the same idea in the form of refinement and individualization. They color the
vertices by refinement, give one vertex a color of its own when the coloring stops splitting, and
use the automorphisms met at the leaves of the search to skip branches \cite{mckay2014practical}.
Algorithm~\ref{alg:canonrefine} applies this to a table, with the inputs as the vertices and the
input vectors as the edges of the hypergraph of Section~\ref{sec:permmatch}.

Let $v$ be a real function on the input vectors of $k$ inputs and $\mathfrak c$ a labeling of the
inputs. One refinement step gives each input vector $x$ the label of the pair formed by $v(x)$ and
the multiset $\{\mathfrak c(i):x_i=1\}$, then gives each input $i$ the label of the pair formed by
$\mathfrak c(i)$ and the multiset of the labels of the vectors $x$ with $x_i=1$. The step is
repeated until the number of distinct input labels stops growing, and $\operatorname{ref}(v,\mathfrak c)$
denotes the result. Labels are values of one fixed hash function of these pairs, so renaming the
inputs renames the labels and changes nothing else. A leaf rule $L$ turns an ordering of the inputs
into a table. For input permutations $L_{\mathrm P}(\pi)=f_\pi$ with $v=f$. For the NPN group
$v=W_f^2$, the periodogram, and $L_{\mathrm{NPN}}(\pi)$ is the least table among $f(\pi x\oplus a)\oplus\nu$
over $a\in\F_2^k$ and $\nu\in\F_2$.

\begin{algorithm}[!htbp]
\caption{Canonical form by refinement}
\label{alg:canonrefine}
\begin{algorithmic}[1]
\Require a table $f$ on $k$ inputs; the function $v$ and the leaf rule $L$ of the group
\Ensure the canonical form $\operatorname{cf}(f)$
\State $\mathcal A\gets$ the transpositions of two inputs that leave $f$ unchanged
\State $\operatorname{cf}\gets\infty$; $\mathfrak c_0\gets$ one label for all inputs
\State \Call{Search}{$\operatorname{ref}(v,\mathfrak c_0)$, $()$}
\Function{Search}{$\mathfrak c$, $p$} \Comment{$p$: the inputs individualized so far}
  \If{the labels of $\mathfrak c$ are distinct}
    \State $\pi\gets$ the inputs in increasing order of label; $t\gets L(\pi)$
    \If{an earlier leaf $\pi'$ gave $t$} \State add the permutation $\pi'(j)\mapsto\pi(j)$ to $\mathcal A$ \EndIf
    \State $\operatorname{cf}\gets\min(\operatorname{cf},t)$ \Comment{Prop.~\ref{prop:canonrefine}(a),(b)}
    \State \Return
  \EndIf
  \State $C\gets$ the smallest class of inputs with equal labels, the least label on ties
  \ForAll{$u\in C$}
    \If{$u$ is in the orbit of an earlier $u'\in C$ under the elements of $\mathcal A$ that fix $p$}
      \State skip $u$ \Comment{Prop.~\ref{prop:canonrefine}(c)}
    \EndIf
    \State $\mathfrak c'\gets\mathfrak c$ with a new label for $u$; \Call{Search}{$\operatorname{ref}(v,\mathfrak c')$, $(p,u)$}
  \EndFor
\EndFunction
\end{algorithmic}
\end{algorithm}

\begin{proposition}[Canonical forms by refinement]
\label{prop:canonrefine}
\begin{enumerate}
\item[(a)] With $v=f$ and $L_{\mathrm P}$, Algorithm~\ref{alg:canonrefine} returns the same table
for $f$ and $f_\sigma$ for every $\sigma\in S_k$, and that table lies in $[f]_{\mathrm P}$.
\item[(b)] With $v=W_f^2$ and $L_{\mathrm{NPN}}$ it returns the same table for all members of the
NPN class $[f]$, and that table lies in $[f]$.
\item[(c)] Skipping $u$ as in the algorithm does not change the result.
\end{enumerate}
\end{proposition}

\begin{proof}
(a) The labels are computed from $v$ and the incidence of inputs and vectors only, so the search on
$f_\sigma$ is the search on $f$ with every input $i$ renamed $\sigma^{-1}(i)$. Corresponding leaves
order the inputs $\pi$ and $\sigma^{-1}\pi$, and $(f_\sigma)_{\sigma^{-1}\pi}=f_\pi$, so the two
searches meet the same tables and return the same least one, which is $f_\pi$ for some $\pi$.
(b) For $g(x)=f(\sigma x\oplus a)\oplus\nu$, Theorem~\ref{thm:normalizer} gives
$W_g^2(u)=W_f^2(\sigma u)$, since translations and output negation change only signs. The search on
$W_g^2$ is therefore the search on $W_f^2$ with the inputs renamed as in (a). At corresponding
leaves the tables $g(\pi x\oplus a')\oplus\nu'$ are the tables
$f(\sigma\pi x\oplus\sigma a'\oplus a)\oplus\nu\oplus\nu'$, and as $a'$ and $\nu'$ range over
$\F_2^k$ and $\F_2$ so do $\sigma a'\oplus a$ and $\nu\oplus\nu'$, so the leaf rule returns the same
table. (c) Two leaves with the same table differ by a transformation of the group that fixes $f$, so
every $\alpha\in\mathcal A$ is the permutation part of such a transformation. It therefore keeps
$v$ and all labels, and if it fixes $p$ it maps the subtree below $u'$ onto the subtree below
$\alpha(u')=u$ with the same tables at corresponding leaves.
\end{proof}

Table~\ref{tab:canonrefine} gives the results on the cut functions of the ISCAS-85 circuits from four to
ten inputs. Under input permutations the classes agree with the search of all $k!$ permutations up
to eight inputs, and at nine and ten inputs with Algorithm~\ref{alg:permdist} at radius zero on
every function. Under the NPN group they agree with the search of the whole group up to six inputs
and with the number of classes of the exact canonical forms of ABC from four to ten inputs. No image of a
function under the group, \CrInvChecked\ in all, changed its form. The search visits few nodes, at
eight inputs \CrPermNodesEight\ on average under permutations and \CrNpnNodesEight\ under the NPN
group, because refinement on the cut functions leaves few inputs with equal labels. The set
$\mathcal A$ starts with the transpositions of symmetric inputs, the groups that
Algorithm~\ref{alg:permdist} also uses, so the search skips them from the first branch on. A query
by canonical form and lookup takes about as long as Algorithm~\ref{alg:permdist} at radius zero on
these libraries, both in Python, within a factor of three either way.

\begin{table}[!htbp]
\caption{Canonical forms by refinement on the cut functions: the classes and the mean number of
search nodes under input permutations (P) and under the NPN group.}
\label{tab:canonrefine}
\centering\small
\begin{tabular}{@{}rrrrrr@{}}
\toprule
$k$ & \shortstack{func-\\tions} & \shortstack{P\\classes} & \shortstack{P\\nodes} & \shortstack{NPN\\classes} & \shortstack{NPN\\nodes} \\
\midrule
4 & \CrFunctionsFour & \CrPermClassesFour & \CrPermNodesFour & \CrNpnClassesFour & \CrNpnNodesFour \\
5 & \CrFunctionsFive & \CrPermClassesFive & \CrPermNodesFive & \CrNpnClassesFive & \CrNpnNodesFive \\
6 & \CrFunctionsSix & \CrPermClassesSix & \CrPermNodesSix & \CrNpnClassesSix & \CrNpnNodesSix \\
7 & \CrFunctionsSeven & \CrPermClassesSeven & \CrPermNodesSeven & \CrNpnClassesSeven & \CrNpnNodesSeven \\
8 & \CrFunctionsEight & \CrPermClassesEight & \CrPermNodesEight & \CrNpnClassesEight & \CrNpnNodesEight \\
9 & \CrFunctionsNine & \CrPermClassesNine & \CrPermNodesNine & \CrNpnClassesNine & \CrNpnNodesNine \\
10 & \CrFunctionsTen & \CrPermClassesTen & \CrPermNodesTen & \CrNpnClassesTen & \CrNpnNodesTen \\
\bottomrule
\end{tabular}
\end{table}

For a timing against the canonical forms of logic synthesis, Algorithm~\ref{alg:canonrefine} is also
written in C, with the same labels and therefore the same forms as the Python version, and timed
with ABC \cite{brayton2010abc} on the same machine and the same files. Each file holds the cut functions of
$k$ inputs and seeded random NPN images of them, all distinct, up to the sizes in
Table~\ref{tab:canonrefc}, and every time is the least of three runs. ABC's co-designed exact
algorithm~11 \cite{zhou2020exact} is faster than the refinement under the NPN group at every $k$, by a factor that grows with $k$ from
\CrcNpnRatioMin\ to \CrcNpnRatioMax\ where ABC's time is above its resolution, and the two give the
same number of classes on every file. The time of ABC grows little with $k$ on these files, while
that of the refinement grows with the $2^k$ input vectors it hashes at every step. Under input
permutations ABC's semi-canonical form~12 is faster than the refinement by a factor between
\CrcPRatioMin\ and \CrcPRatioMax\ from six inputs on. It gives the exact number of classes on the
cut functions and leaves a few classes split on the files with images, \CrcAbcTwelveClassesSix\
against \CrcPClassesSix\ at six inputs. The refinement is faster than the search of all $k!$
permutations from \CrcKpermFrom\ inputs on, by a factor of \CrcKpermFactorEight\ at eight inputs and
\CrcKpermFactorTen\ at ten. The refinement over the periodogram gives an exact NPN canonical form
directly from the decomposition of Theorem~\ref{thm:normalizer}, refinement over the quotient and
the translations at the leaves, but on these files it does not reach the speed of ABC's specialized
code. In our profile most of its time goes to the refinement steps and to the translations at the
leaves.

\begin{table}[!htbp]
\caption{Time per function in microseconds of the canonical forms by refinement in C against ABC on
the same files: the number of functions in thousands, the NPN form against ABC's exact
algorithm~11, the permutation form against ABC's form~12 (n/a: ABC stops on an assertion below six
inputs), and on the first functions of each file, their number in thousands, the permutation form
against the search of all $k!$ permutations.}
\label{tab:canonrefc}
\centering\small
\setlength{\tabcolsep}{2.5pt}
\begin{tabular}{@{}rrrrrrrrr@{}}
\toprule
$k$ & \shortstack{func-\\tions} & \shortstack{NPN\\refine} & \shortstack{ABC\\11} & \shortstack{P\\refine} & \shortstack{ABC\\12} & \shortstack{first\\func.} & \shortstack{P\\refine} & \shortstack{$k!$\\search} \\
\midrule
4 & \CrcFunctionsFour & \CrcNpnUsFour & \CrcAbcElevenUsFour & \CrcPUsFour & \CrcAbcTwelveUsFour & \CrcSubFour & \CrcPSubUsFour & \CrcKpermUsFour \\
5 & \CrcFunctionsFive & \CrcNpnUsFive & \CrcAbcElevenUsFive & \CrcPUsFive & \CrcAbcTwelveUsFive & \CrcSubFive & \CrcPSubUsFive & \CrcKpermUsFive \\
6 & \CrcFunctionsSix & \CrcNpnUsSix & \CrcAbcElevenUsSix & \CrcPUsSix & \CrcAbcTwelveUsSix & \CrcSubSix & \CrcPSubUsSix & \CrcKpermUsSix \\
7 & \CrcFunctionsSeven & \CrcNpnUsSeven & \CrcAbcElevenUsSeven & \CrcPUsSeven & \CrcAbcTwelveUsSeven & \CrcSubSeven & \CrcPSubUsSeven & \CrcKpermUsSeven \\
8 & \CrcFunctionsEight & \CrcNpnUsEight & \CrcAbcElevenUsEight & \CrcPUsEight & \CrcAbcTwelveUsEight & \CrcSubEight & \CrcPSubUsEight & \CrcKpermUsEight \\
9 & \CrcFunctionsNine & \CrcNpnUsNine & \CrcAbcElevenUsNine & \CrcPUsNine & \CrcAbcTwelveUsNine & \CrcSubNine & \CrcPSubUsNine & \CrcKpermUsNine \\
10 & \CrcFunctionsTen & \CrcNpnUsTen & \CrcAbcElevenUsTen & \CrcPUsTen & \CrcAbcTwelveUsTen & \CrcSubTen & \CrcPSubUsTen & \CrcKpermUsTen \\
\bottomrule
\end{tabular}
\end{table}

\iffull
\subsection{Peeling Group Structure from a Truth Table}
\label{sec:peel}

The iterative peel of group matching removes, one step at a time, the component that a matched
group explains and stops at a residue that no group concentrates, in the manner of the CLEAN
deconvolution of radio astronomy \cite{hogbom1974clean}. On the cube three things need to be
settled before it can run. The first is how components combine, the second is what a white residue
is, and the third is which classical decompositions the peel reproduces. The cut functions of real
circuits then show how often the structure of a function is a mixture of several groups.

Truth tables combine by XOR, the addition of $\F_2^N$. For a group $G$ of maps of $\F_2^n$ write
$\mathrm{Fix}_2(G)\subseteq\F_2^N$ for the truth tables that are constant on every orbit of $G$.

\begin{proposition}[Fixed tables]
\label{prop:fix2}
(i) $\mathrm{Fix}_2(G)$ is a subspace of $\F_2^N$ whose dimension is the number of orbits of $G$ on
$\F_2^n$, with the orbit indicators as a basis. (ii) $\mathrm{Fix}_2(G_1)\cap\mathrm{Fix}_2(G_2)=
\mathrm{Fix}_2(\langle G_1,G_2\rangle)$. (iii) A uniformly random truth table lies in a subspace $V$
of $\F_2^N$ with probability $2^{\dim V-N}$.
\end{proposition}

\begin{proof}
(i) A table is constant on the orbits exactly when it is a sum of orbit indicators, and the indicators
have disjoint supports, so they are independent. (ii) A table constant on the orbits of $G_1$ and of
$G_2$ takes one value along any word in the elements of both, which is an orbit of the generated
group. Conversely an orbit of $G_1$ or of $G_2$ lies inside an orbit of $\langle G_1,G_2\rangle$.
(iii) $V$ holds $2^{\dim V}$ of the $2^N$ tables.
\end{proof}

A table $f$ is a mixture of $G_1$ and $G_2$ when $f\in\mathrm{Fix}_2(G_1)+\mathrm{Fix}_2(G_2)$ and $f$
lies in neither space. The shared space $\mathrm{Fix}_2(\langle G_1,G_2\rangle)$ plays the part that the
intersection of two commutants plays for covariances, peeling a mixture is decoding in the code
$\mathrm{Fix}_2(G_1)+\mathrm{Fix}_2(G_2)$, and the residue is a coset leader. The library used below has
three families for each set $U$ of inputs: $S_U$, the permutations of $U$ (tables symmetric in
$U$), $T_U$, the translations by vectors supported on $U$ (tables that do not depend on $U$), and
$C_U$, the translation by the indicator of $U$ (tables unchanged when all of $U$ is complemented
together). A pair is counted only when $\dim(\mathrm{Fix}_2(G_1)+\mathrm{Fix}_2(G_2))\le N/2$, so that by
Proposition~\ref{prop:fix2}(iii) a random table lies in it with probability at most $2^{-N/2}$.

Three measures of a residue point to the kind of peel to run. For a group $G$ of permutations of
$\F_2^n$ with matrices $P_g$ let $c_g(f)=N^{-1}F^{\mathsf T}P_gF$, and define the projection residual
$\mathrm{res}_G(f)=\|FF^{\mathsf T}-\RG(f)\|_F/N$ and the concentration
$\mathrm{conc}_G(f)=\lambda_{\max}(\RG(f))/N$, where $\lambda_{\max}$ is the largest eigenvalue. Write
$\mathrm{nl}(f)$ for the distance from $f$ to the nearest affine function, $f_0$ and $f_1$ for the
cofactors of $f$ at $x_i=0$ and $x_i=1$, and $\varrho_i(f)=2N^{-1}\sum_{x:\,x_i=0}F(x)F(x\oplus e_i)$
for the correlation of the two cofactors.

\begin{proposition}[Residual, concentration and cofactor correlation]
\label{prop:peelmeasures}
(i) The residual satisfies
\[
\mathrm{res}_G(f)^2=1-\frac{1}{|G|}\sum_{g\in G}c_g(f)^2,
\]
which is zero exactly when every element of $G$ fixes or complements $f$. (ii) The concentration of
the translations is
\[
\mathrm{conc}_T(f)=\frac{\max_aW_f(a)^2}{N^2}=\Bigl(1-\frac{2\,\mathrm{nl}(f)}{N}\Bigr)^2,
\]
which is one for an affine function and at least $N^{-1}$, with equality exactly for a bent function.
(iii) For the translation $\tau_i$ by $e_i$, $c_{\tau_i}(f)=r_f(e_i)/N=\varrho_i(f)$, and the
Boolean derivative $f_0\oplus f_1$ is one on $N(1-\varrho_i(f))/4$ of its $N/2$ points.
\end{proposition}

\begin{proof}
(i) $\RG$ is the orthogonal projection onto the commutant (Section~\ref{sec:commutant}), so
$\|FF^{\mathsf T}-\RG(f)\|_F^2=\|FF^{\mathsf T}\|_F^2-\langle FF^{\mathsf T},\RG(f)\rangle$. The first term
is $N^2$ and the second is $|G|^{-1}\sum_g(F^{\mathsf T}P_gF)^2$. The residual vanishes exactly when
$|c_g(f)|=1$ for all $g$, that is, $P_gF=\pm F$. (ii) By Proposition~\ref{prop:T} the eigenvalues of
$R_T(f)$ are $W_f(a)^2/N$. The coefficient $W_f(a)$ is $N$ minus twice the distance from $f$ to the
linear function $a\cdot x$, and $-W_f(a)$ is the same for its complement, so
$\max_a|W_f(a)|=N-2\,\mathrm{nl}(f)$. Parseval's relation $\sum_aW_f(a)^2=N^2$ gives
$\max_aW_f(a)^2\ge N$, with equality exactly when every $W_f(a)^2=N$, the definition of a bent
function. (iii) The sum defining $r_f(e_i)$ counts each pair $\{x,x\oplus e_i\}$ twice, which gives
the first identity. The pairs on which $f_0$ and $f_1$ differ contribute $-1$ to the sum over the
$N/2$ points with $x_i=0$, so their number $t$ satisfies $N/2-2t=\varrho_i(f)N/2$.
\end{proof}

The projection residual is a commutativity residual taken one group element at a time. It differs
from the sample commutativity residual $\|[\RG(f),FF^{\mathsf T}]\|_F/(\|\RG(f)\|_F\|FF^{\mathsf T}\|_F)$
of \cite[Eq.~(41)]{thornton2026ad}, which commutes the group average with the observation, and the
difference matters for a single table.

\begin{proposition}[Projection and commutativity residuals]
\label{prop:resforms}
(i) For every $g\in G$, $\|[P_g,FF^{\mathsf T}]\|_F^2=2N^2(1-c_g(f)^2)$, so
\[
\mathrm{res}_G(f)^2=\frac{1}{2|G|N^2}\sum_{g\in G}\bigl\|[P_g,FF^{\mathsf T}]\bigr\|_F^2 ,
\]
the mean square over $G$ of the commutator of each element with the observation, divided by $2N^2$.
(ii) $[\RG(f),FF^{\mathsf T}]=0$ exactly when $F$ is an eigenvector of $\RG(f)$. (iii) For the
translations, $[R_T(f),FF^{\mathsf T}]=0$ exactly when $f$ is plateaued, that is, when $W_f(a)^2$
takes a single nonzero value on the indices where it is not zero. It is therefore zero on every
affine and every bent function, while $\mathrm{res}_T(f)$ is zero on an affine function and
$\sqrt{1-1/N}$ on a bent one.
\end{proposition}

\begin{proof}
(i) Write $P=P_g$, which is orthogonal, and $\Pi=FF^{\mathsf T}$. Then
\[
\|P\Pi-\Pi P\|_F^2=\|P\Pi\|_F^2+\|\Pi P\|_F^2-2\operatorname{tr}(\Pi P^{\mathsf T}\Pi P).
\]
The first two terms are $\|\Pi\|_F^2=N^2$, and $\operatorname{tr}(\Pi P^{\mathsf T}\Pi P)=(F^{\mathsf T}P^{\mathsf T}F)
(F^{\mathsf T}PF)=N^2c_g(f)^2$. Averaging over $G$ and comparing with
Proposition~\ref{prop:peelmeasures}(i) gives the formula. (ii) $\RG(f)$ is symmetric. If
$\RG(f)F=\lambda F$ then $\RG(f)\Pi=\lambda\Pi=\Pi\RG(f)$. Conversely, applying both sides of
$\RG(f)\Pi=\Pi\RG(f)$ to $F$ gives $N\RG(f)F=(F^{\mathsf T}\RG(f)F)F$. (iii) By
Proposition~\ref{prop:T}, $F=N^{-1}\sum_aW_f(a)h_a$ and $R_T(f)F=N^{-2}\sum_aW_f(a)^3h_a$, which is a
multiple of $F$ exactly when $W_f(a)^2$ is constant where $W_f(a)\ne0$. An affine function has a
single nonzero coefficient and a bent function has $W_f(a)^2=N$ for every $a$. For the residual,
$c_t(f)=r_f(t)/N$, which is $\pm1$ for every translation when $f$ is affine, and zero for every
$t\ne0$ when $f$ is bent, so Proposition~\ref{prop:peelmeasures}(i) gives $1-1/N$.
\end{proof}

The same computation holds for any finite group acting by unitary matrices on a single complex
observation, with the fourth power of its norm in place of $N^2$ and the squared modulus of the
normalized quadratic form in place of $c_g(f)^2$. The largest term of the mean in (i) is the discrete
commutativity residual of \cite{thornton2026principles} evaluated at $FF^{\mathsf T}$, so
$\mathrm{res}_G(f)$ is at most that residual divided by $\sqrt2$. With one table $FF^{\mathsf T}$ has
rank one, and by (ii) the sample residual of \cite{thornton2026ad} asks whether $F$ is an eigenvector
of its own average rather than whether $G$ fixes or complements $f$. It cannot tell the affine
functions, which every translation fixes or complements, from the bent functions, on which no
translation other than the identity is correlated with $f$. A covariance of rank above one, as in
\cite{thornton2026ad}, does not have this degeneracy, and there the two residuals measure different
things. On truth tables this section uses the projection residual.

The residual and the concentration are the residual of the Reynolds projection and the spectral
concentration of matching a group to a single observation, and the cofactor correlation is the quantity $c_g$ of the
residual for the single group $\langle\tau_i\rangle$. With two observations, group matching compares
the estimates of one candidate group made from each, normalized by their traces, and selects the
group whose estimates agree. Comparing a group's estimate with itself removes the bias of the
concentration toward groups with small orbits but needs a second observation. On the cube the
two cofactors of an input are two observations of a function of the other $n-1$ inputs, at no further
measurement. For a group $G$ of permutations of those inputs' $M=N/2$ points, with $\pm1$ forms
$F_0$ and $F_1$ of the cofactors and $c_g(f_0,f_1)=M^{-1}F_0^{\mathsf T}P_gF_1$, define the
cross-cofactor consistency
\[
D_{\mathrm{CV}}(G;i)=\Bigl\|\frac{\RG(f_0)}{\operatorname{tr}\RG(f_0)}-\frac{\RG(f_1)}{\operatorname{tr}\RG(f_1)}\Bigr\|_F .
\]

\begin{proposition}[Cross-cofactor consistency]
\label{prop:dcvcube}
(i) $D_{\mathrm{CV}}(G;i)^2=|G|^{-1}\sum_{g\in G}\bigl(c_g(f_0)^2+c_g(f_1)^2-2\,c_g(f_0,f_1)^2\bigr)$.
(ii) For the group $T'$ of translations of the other inputs,
$D_{\mathrm{CV}}(T';i)=M^{-2}\bigl\|W_{f_0}^2-W_{f_1}^2\bigr\|_2$, the distance between the normalized
Walsh periodograms of the two cofactors, which is zero exactly when $|W_{f_0}(a)|=|W_{f_1}(a)|$ for
every $a$, in particular when $f_1$ is a translate of $f_0$ or of its complement.
\end{proposition}

\begin{proof}
(i) $\operatorname{tr}\RG(f_b)=\operatorname{tr}(F_bF_b^{\mathsf T})=M$ because $\RG$ keeps the trace. Expanding the square,
$\|\RG(f_0)-\RG(f_1)\|_F^2$ is the sum of $\|\RG(f_b)\|_F^2$ over $b$ less twice
$\langle\RG(f_0),\RG(f_1)\rangle$. Since $\RG$ is the orthogonal projection onto the commutant,
$\langle\RG(f_0),\RG(f_1)\rangle=\langle F_0F_0^{\mathsf T},\RG(f_1)\rangle=|G|^{-1}\sum_g(F_0^{\mathsf T}P_gF_1)^2$,
and $\|\RG(f_b)\|_F^2$ is the same expression with $F_1$ replaced by $F_b$ and $F_0$ by $F_b$.
Dividing by $M^2$ gives (i). (ii) By Proposition~\ref{prop:T} on $n-1$ inputs both matrices are
diagonal in the normalized Walsh basis with entries $W_{f_b}(a)^2/M$, and the Frobenius norm does not
depend on the orthonormal basis, which gives the formula after dividing by the trace $M$. A
translation of $f_0$ by $c$ multiplies $W_{f_0}(a)$ by $(-1)^{a\cdot c}$ and a complement changes its
sign, so neither changes the magnitudes.
\end{proof}

Corollary~\ref{cor:bent} and Proposition~\ref{prop:peelmeasures}(ii) say that the
white residue of the translations is a bent function, at the least concentration. Write
$T_f^{\pm}=\{c:|r_f(c)|=N\}$ for the translations that fix or complement $f$. One translation atom
with a white residue, $r_f(c)=0$ for every $c\notin T_f^{\pm}$, is Carlet's class of partially bent
functions, which are the sums of an affine function on one subspace and a bent function on a
complement \cite{carlet1993partially}. The translations in $T_f$ are the autosymmetry of Luccio and
Pagli \cite{luccio1999autosymmetric}. For odd $n$ no function is bent, and the whitest residues
are concatenations of bent functions on the two halves of the cube.

The smallest groups of the cube give the classical expansions.

\begin{proposition}[Shannon and Davio from one translation]
\label{prop:davio}
Let $\tau_i$ be the translation by $e_i$. (i) $\mathrm{Fix}_2(\langle\tau_i\rangle)$ is the set of
tables that do not depend on $x_i$, and $f=f_0\oplus x_i(f_0\oplus f_1)$, with the cofactors read as
functions of all inputs, writes $f$ as an element of that space plus $x_i$ times another. (ii) Over
the reals the isotypic components of $F$ under $\langle\tau_i\rangle$ are $(F\pm P_{\tau_i}F)/2$, a
part that does not depend on $x_i$ and $(-1)^{x_i}$ times a part that does not depend on $x_i$.
(iii) The restrictions of $f$ to the two cosets $\{x_i=0\}$ and $\{x_i=1\}$ are the Shannon cofactors.
\end{proposition}

\begin{proof}
(i) The orbits of $\langle\tau_i\rangle$ are the pairs $\{x,x\oplus e_i\}$. At $x_i=0$ the right side
is $f_0$, and at $x_i=1$ it is $f_0\oplus f_0\oplus f_1=f_1$. (ii) $(I\pm P_{\tau_i})/2$ are the
projections onto the eigenspaces of $P_{\tau_i}$ for $\pm1$. The vector $(F+P_{\tau_i}F)/2$ is
constant on each pair, and $(F-P_{\tau_i}F)/2$ changes sign within each pair. (iii) is the definition.
\end{proof}

Part (i) is the positive Davio expansion and part (ii) its real form. Shannon expansion at every
level gives a binary decision diagram, Davio expansion gives a functional decision diagram, and a
choice per input gives the ordered Kronecker functional decision diagram \cite{drechsler1994okfdd}.
A peel adds a choice at each node made by the matched group and atoms of larger groups as leaves.
With a bound set $A$ of $m$ inputs and a free set $B$ of the others, the decomposition chart $X_f$ has
the entry $f(a,b)$ in row $a$ and column $b$, and $S_f$ is its $\pm1$ form.

\begin{proposition}[Decomposition charts as peels]
\label{prop:charts}
(i) $f=g(A)\oplus h(B)$ for some $g$ and $h$ exactly when $\rank S_f=1$. (ii) $f=h(g(A),B)$ for some
$g$ and $h$ exactly when every column of $S_f$ lies in $\spn\{\one,G\}$ for one vector $G\in\{\pm1\}^{2^m}$,
that is, when a peel of the constant atom and one further atom leaves no residue.
\end{proposition}

\begin{proof}
(i) If $f=g\oplus h$ then $S_f=GH^{\mathsf T}$ with $G$ and $H$ the $\pm1$ forms of $g$ and $h$.
Conversely, a rank-one matrix with entries $\pm1$ is $uv^{\mathsf T}$ with $u$ one of its columns and
$v$ a $\pm1$ vector, which defines $g$ and $h$. (ii) If $f=h(g(A),B)$, column $b$ of $X_f$ is $0$,
$1$, $g$ or $\neg g$ according to $h(\cdot,b)$, so the columns of $S_f$ are $\pm\one$ or $\pm G$.
Conversely, a $\pm1$ vector $\alpha\one+\beta G$ has $\alpha\pm\beta\in\{\pm1\}$, so it is one of
$\pm\one$ and $\pm G$, every row of $X_f$ is fixed by the value of $g$, and $h$ is read off the two
row patterns.
\end{proof}

Part (ii) is Ashenhurst's simple disjoint decomposition, reprinted in \cite{curtis1962design}. Exact
disjoint-support decomposition and bi-decomposition are served by specialized algorithms
\cite{mishchenko2001bidecomp}, so the measurements below do not time the peel against them. They ask
how often each kind of structure occurs.

\begin{table}[!htbp]
\caption{Group structure of the cut functions: the share with a symmetry from the library, the share that is a mixture of two groups and in neither space, the functions that are a mixture with no single symmetry, the share of partially bent functions, the median concentration $\mathrm{conc}_T(f)$ (white value $N^{-1}$), and the NPN classes whose functions receive different verdicts.}
\label{tab:peelcube}
\centering\small
\setlength{\tabcolsep}{3pt}
\begin{tabular}{@{}rrrrrrrr@{}}
\toprule
$n$ & \shortstack{func-\\tions} & \shortstack{one sym-\\metry (\%)} & \shortstack{mixture\\(\%)} & \shortstack{mixture\\only} & \shortstack{partially\\bent (\%)} & \shortstack{median\\conc.} & \shortstack{NPN classes\\split} \\
\midrule
4 & \PlFunctionsFour & \PlSinglePctFour & \PlMixPctFour & \PlMixOnlyFour & \PlPbPctFour & \PlConcFour & \PlNpnSplitFour/\PlNpnClassesFour \\
5 & \PlFunctionsFive & \PlSinglePctFive & \PlMixPctFive & \PlMixOnlyFive & \PlPbPctFive & \PlConcFive & \PlNpnSplitFive/\PlNpnClassesFive \\
6 & \PlFunctionsSix & \PlSinglePctSix & \PlMixPctSix & \PlMixOnlySix & \PlPbPctSix & \PlConcSix & \PlNpnSplitSix/\PlNpnClassesSix \\
7 & \PlFunctionsSeven & \PlSinglePctSeven & \PlMixPctSeven & \PlMixOnlySeven & \PlPbPctSeven & \PlConcSeven & \PlNpnSplitSeven/\PlNpnClassesSeven \\
\bottomrule
\end{tabular}
\end{table}

Table~\ref{tab:peelcube} gives the measurements on the cut functions of the ISCAS-85 circuits from
four to seven inputs, with \PlGroupsSeven\ groups and \PlPairsSeven\ counted pairs in the library at
seven inputs. Most cut functions have a symmetry from the library, \PlSinglePctFour\% at four inputs
and \PlSinglePctSeven\% at seven. Mixtures of two groups grow with the input count, from
\PlMixPctFour\% to \PlMixPctSeven\% of the functions, and at every size the most common pair is a
symmetric function plus a function that ignores some inputs (\PlMixSTSeven\% of the mixtures at seven
inputs). The other pairs are two symmetric functions and XOR bi-decompositions. Almost every mixed
function also has a symmetry of its own, and a function that is a mixture and nothing else is rare,
\PlMixOnlySeven\ of \PlFunctionsSeven\ at seven inputs. The library is closed under input
permutations, so its verdict (a single symmetry, a mixture, both or neither) is constant on each P
class, which every one of the \PlPClassesSeven\ P classes at seven inputs confirms. Input negation
turns a symmetric pair into a pair that is symmetric up to the negation of one input, so the verdict
is not an NPN invariant: \PlNpnSplitFour\ of the \PlNpnClassesFour\ NPN classes at four inputs and
\PlNpnSplitSeven\ of \PlNpnClassesSeven\ at seven hold functions with different verdicts. The verdict
is therefore a phase-dependent label that refines NPN classes.

The residue of the translations is far from white on these functions. The partially bent functions
are \PlPbPctFour\% of the functions at four inputs, of which \PlBentFour\ are bent, and under two
percent from six inputs on (\PlPbPctEight\% at eight). The median concentration $\mathrm{conc}_T(f)$ is
\PlConcRatioFour\ times the white value $N^{-1}$ at four inputs and \PlConcRatioEight\ times at
eight, and the median of $\max_{c\notin T_f^{\pm}}|r_f(c)|/N$ grows from \PlOffwhiteFour\ to
\PlOffwhiteEight. Bent pulses are rare as well. Every function of two inputs is bent (of odd weight)
or affine, so pulses on two-input subcubes say nothing. A sequence of pulses on a four-input subcube
that are all bent or affine, with at least one bent, occurs in \PlBentPulseFour, \PlBentPulseFive\ and
\PlBentPulseSix\ functions at four, five and six inputs and in none at seven and eight.

\begin{table}[!htbp]
\caption{Pulses on four inputs after expanding $n-4$ inputs of each cut function, with the best set of inputs and expansions per function (Shannon everywhere, positive Davio everywhere, or a choice among Shannon and the two Davio expansions per input): the percentages of constant, affine, totally symmetric, partially symmetric and other pulses, and the number of bent pulses.}
\label{tab:peelpulses}
\centering\small
\setlength{\tabcolsep}{3pt}
\begin{tabular}{@{}rlrrrrrr@{}}
\toprule
$n$ & expansion & const. & affine & sym. & \shortstack{partially\\sym.} & other & bent \\
\midrule
6 & Shannon & \PlPShConstSix & \PlPShAffSix & \PlPShSymSix & \PlPShPartSix & \PlPShOtherSix & \PlPShBentSix \\
 & pos. Davio & \PlPDvConstSix & \PlPDvAffSix & \PlPDvSymSix & \PlPDvPartSix & \PlPDvOtherSix & \PlPDvBentSix \\
 & per input & \PlPKfConstSix & \PlPKfAffSix & \PlPKfSymSix & \PlPKfPartSix & \PlPKfOtherSix & \PlPKfBentSix \\
\midrule
7 & Shannon & \PlPShConstSeven & \PlPShAffSeven & \PlPShSymSeven & \PlPShPartSeven & \PlPShOtherSeven & \PlPShBentSeven \\
 & pos. Davio & \PlPDvConstSeven & \PlPDvAffSeven & \PlPDvSymSeven & \PlPDvPartSeven & \PlPDvOtherSeven & \PlPDvBentSeven \\
 & per input & \PlPKfConstSeven & \PlPKfAffSeven & \PlPKfSymSeven & \PlPKfPartSeven & \PlPKfOtherSeven & \PlPKfBentSeven \\
\midrule
8 & Shannon & \PlPShConstEight & \PlPShAffEight & \PlPShSymEight & \PlPShPartEight & \PlPShOtherEight & \PlPShBentEight \\
 & pos. Davio & \PlPDvConstEight & \PlPDvAffEight & \PlPDvSymEight & \PlPDvPartEight & \PlPDvOtherEight & \PlPDvBentEight \\
 & per input & \PlPKfConstEight & \PlPKfAffEight & \PlPKfSymEight & \PlPKfPartEight & \PlPKfOtherEight & \PlPKfBentEight \\
\bottomrule
\end{tabular}
\end{table}

Table~\ref{tab:peelpulses} follows the iterated peel. For each function, every set of
$n-4$ inputs is expanded with every choice of Shannon, positive Davio or negative Davio expansion per
input, and the choice with the most constant and affine pulses on the remaining four inputs is kept.
The pulses that remain are not bent. At six inputs \PlPKfPartSix\% of them still carry a symmetric
pair or an absent input under the choice per input, so a peel has more to remove there, and Davio expansion leaves more of them totally
symmetric than Shannon expansion does. A choice per input leaves the fewest pulses without structure.

Whether to peel the whole function or its cofactors is a question of description length, the
code-length criterion of model selection. On the whole function the length is the
dimension of the least library space or counted pair space that holds $f$ (or $N$), plus the bits
that name the space (\PlIndexBitsSix\ at six inputs). On the cofactor route it is the sum over the
pulses of the same quantity on four inputs (\PlPulseIndexBits\ bits per pulse) plus the bits that name
the expanded inputs and the expansions (\PlChoiceBitsSix\ at six inputs). Counting coefficients only,
the cofactor route is shorter for \PlDlCofSmallerSix\ of \PlFunctionsSix\ functions at six inputs,
with medians \PlDlCofSix\ against \PlDlWholeSix\ bits. With the bits that name the spaces, the whole
function is shorter for \PlDlWholeSmallerIdxSix\ (\PlDlWholeSmallerIdxPctSix\%), with medians
\PlDlWholeIdxSix\ against \PlDlCofIdxSix\ bits, and for \PlDlWholeSmallerIdxPctSeven\% at seven
inputs. Neither route is shorter for every function, so the rule is to compare the two lengths for
each function and to peel where the shorter one points. The number of peels follows from that
comparison. It is neither minimized nor maximized directly, since few large atoms leave a long
residue and many small atoms each need their own index. A peel is counted only when
Proposition~\ref{prop:fix2}(iii) makes a chance fit unlikely, which is the calibrated stop.

\begin{table}[!htbp]
\caption{Pointers at five inputs. Expansion: Davio against Shannon for each function and input. Route: the whole function against the best expansion of one input. Areas under the ROC curve, the best single-threshold accuracy (\%) and the base rate (\%).}
\label{tab:dcvcube}
\centering\small
\begin{tabular}{@{}llrr@{}}
\toprule
question & pointer & area & accuracy \\
\midrule
expansion & $D_{\mathrm{CV}}(T';i)$, two cofactors & \DcAucDcv & \DcBestDcvPct \\
 & $|\varrho_i|$, one group & \DcAucRho & \DcBestRhoPct \\
 & concentration, one cofactor each & \DcAucPsi & \DcBestPsiPct \\
 & base rate (always Davio) & & \DcBasePct \\
\midrule
route & least $D_{\mathrm{CV}}(G;i)$, two cofactors & \DcAucRoute & \DcRouteBestPct \\
 & $\mathrm{conc}_T(f)$, one observation & \DcAucConc & \DcRouteConcBestPct \\
 & base rate (always whole) & & \DcRouteBasePct \\
\bottomrule
\end{tabular}
\end{table}

Table~\ref{tab:dcvcube} tests three pointers to the expansion at five inputs, over the \DcPairs\
pairs of a function and an input. Davio expansion gives the shorter description in \DcDavio\ pairs
and Shannon expansion in \DcShannon, so \DcBasePct\% of the strict cases favor Davio. Ranked by the
area under the ROC curve, the cross-cofactor consistency $D_{\mathrm{CV}}(T';i)$ of
Proposition~\ref{prop:dcvcube}(ii) separates the two cases (\DcAucDcv) far better than the cofactor
correlation $|\varrho_i|$ (\DcAucRho) or the mean concentration of the two cofactors (\DcAucPsi, read
the other way \DcAucPsiRev). The consistency is zero in \DcZero\ pairs, and Davio expansion is
shorter in all of them, so on this sample a zero is a certificate for Davio expansion. Against the base
rate no single threshold on any of the three does better than always choosing Davio expansion (best
accuracy \DcBestDcvPct\%, \DcBestRhoPct\% and \DcBestPsiPct\%), so the consistency is a ranking for
the cases that a fuller search should examine first, not a rule to use in place of the search.

The same measure points to the route. At five inputs the whole function (the least library space or
pair space plus \DcWholeIndexBits\ bits that name it) is shorter than the best expansion of one input
(two pulses with \DcPulseIndexBits\ index bits each, plus \DcChoiceBits\ bits that name the input and
the expansion) for \DcWholeShorter\ of \DcFunctions\ functions (\DcRouteBasePct\%). The least
$D_{\mathrm{CV}}(G;i)$ over the inputs and the nontrivial library groups on the other four inputs, a
group on which the two cofactors agree, ranks these functions with an area of \DcAucRoute, and its
best threshold is right for \DcRouteBestPct\% of them. The concentration of the translations of $f$,
the single-observation measure, scores \DcAucConc. The least consistency is zero for \DcRouteZero\
functions, and the whole function is shorter for \DcRouteZeroWhole\ of these.

On the decomposition charts with a bound set of $\lfloor n/2\rfloor$ inputs, the peel with the
constant atom and one further atom agrees with Ashenhurst's test on all \PlChartsFive\ charts at five
inputs and all \PlChartsSix\ at six, as Proposition~\ref{prop:charts} requires. On the
\PlNontrivSix\ charts at six inputs that are not trivially decomposable, the one-step spectral peel gives the least-error
approximate XOR bi-decomposition over all $2^{8}$ bound functions on \PlPeelLeastPctSix\% of them, with
a total error \PlPeelAbovePctSix\% above the least. The peel takes the bound function from the
signs of a column of the projection onto the leading left singular space of $S_f$ and then the best
free function. The projection does not depend on the basis that a singular value decomposition
returns. A sign read from one singular vector does, when the leading singular value repeats or an
entry is zero, and on these charts that changed the count between two linear algebra libraries. This is the setting of approximate synthesis by
Boolean matrix factorization \cite{hashemi2018blasys}, where the residue of the peel is the error of
the approximation. Conditional symmetries of cofactors that the whole function lacks
\cite{kravets2000generalized} are the other reason to recurse. The library of three group families,
pairs only, the mixtures up to seven inputs and the uniform code for the index bound what these
measurements show, and the peel was not timed against decomposition or factorization tools.

\subsubsection{AD-CLEAN over the Cube}
\label{sec:adcleancube}

The peels above take one group at a time. The iterative stripping of \cite[Sec.~IV-F]{thornton2026principles},
an analog of the CLEAN algorithm for covariances, removes several components in turn, and it carries
over to the cube with XOR in place of subtraction. Given a table $f$ and a library of spaces, write
$f=a_1\oplus\cdots\oplus a_k\oplus r$ with $a_j\in\mathrm{Fix}_2(G_j)$, and score the model by its
description length, the bits that name the spaces, give the coordinates of the atoms in
$\mathrm{Fix}_2(G_1)+\cdots+\mathrm{Fix}_2(G_k)$ and send the residue $r$. By
Proposition~\ref{prop:fix2}(ii) the coordinates shared by two atoms lie in
$\mathrm{Fix}_2(\langle G_i,G_j\rangle)$ and are counted once. The procedure has four steps.
\begin{enumerate}
\item Set $r\leftarrow f$.
\item For each library space form the element nearest to $r$, which for $\mathrm{Fix}_2(G)$ is the
majority value of $r$ on each orbit, and score the model that adds it. Score pairs of spaces jointly,
since a mixture can lie in $\mathrm{Fix}_2(G_1)+\mathrm{Fix}_2(G_2)$ while neither space alone shortens
the description.
\item If no candidate shortens the description by more than a margin calibrated on uniformly random
tables, stop.
\item Otherwise add the candidate that shortens it most, refit the atoms jointly on the sum of their
spaces, set $r\leftarrow f\oplus a_1\oplus\cdots\oplus a_k$, and return to step 2.
\end{enumerate}
The final $r$ is the algebraic residue. The cofactor route of Proposition~\ref{prop:davio} applies
the same procedure to the tables of a Shannon or Davio expansion at one input, and the route is
chosen by description length as above. The library of the measurements below holds the three
families $S_U$, $T_U$ and $C_U$ and the space $\mathrm{Aff}$ of affine functions. The output is drawn
as a cascade, a stack of truth tables in index order with the table, its atoms and the residue in
one column and the running residues $f\oplus a_1\oplus\cdots\oplus a_j$ in the other, so that each
atom is the XOR of two consecutive running residues.

\begin{figure}[t]
\centering
\includegraphics[width=\textwidth]{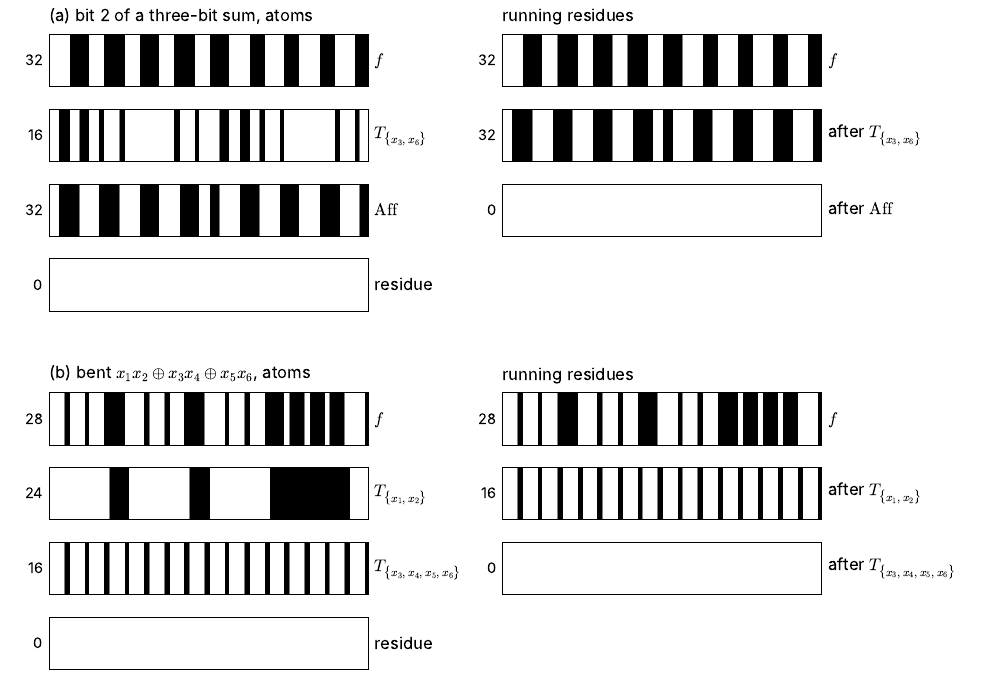}
\caption{Cascades of AD-CLEAN over the cube for (a) bit 2 of the sum of $(x_1,x_2,x_3)$ and
$(x_4,x_5,x_6)$, least significant bits first, and (b) the bent function
$x_1x_2\oplus x_3x_4\oplus x_5x_6$. Each row is a truth table of 64 entries, entry $x$ at position
$\sum_ix_i2^{i-1}$, dark for a one, with its weight at the left. Left: the table, the atoms labeled by
their spaces, and the residue. Right: the running residues after each atom.}
\label{fig:cascade}
\end{figure}

A planted model is recoverable by description length only when it is shorter than the raw table.
The planted scenes were therefore drawn as an atom of one library space or atoms of two spaces whose
sum is not a subspace of either, with \AcNoiseSix\ flipped entries at six inputs, \AcNoiseSeven\ at
seven and \AcNoiseEight\ at eight, and kept only when the planted model is shorter than the raw table
by the stop margin and four bits. With one flipped entry, \AcIdentPairSix\ pairs of spaces meet that
condition at six inputs, \AcIdentPairSeven\ at seven and \AcIdentPairEight\ at eight. The description
found was no longer than the planted one in \AcOptSix\ of \AcScenesSix\ scenes at six inputs
(\AcTrialsSix\ for each kind of scene and noise weight), \AcOptSeven\ of \AcScenesSeven\ at seven and
\AcOptEight\ of \AcScenesEight\ at eight (\AcTrialsSeven\ each). Every shortfall at eight inputs is a
scene of two spaces (\AcOptPairEight\ of \AcPairScenesEight\ reached the planted length, against
\AcOptSingleEight\ of \AcSingleScenesEight\ scenes of one space), and at that size the search over
pairs was limited so that it fits in memory. With noise the description often absorbs a flipped
entry into an atom when that is shorter, so the planted structure itself was found in
\AcStructNoisySix\ of \AcNoisyScenesSix\ noisy scenes at six inputs, and in all \AcPairZeroStructEight\
noise-free scenes of two spaces at eight. No uniformly random table was peeled, \AcNullSix\ of
\AcNullTablesSix\ at six inputs and \AcNullSeven\ of \AcNullTablesSeven\ at seven and at eight.

On seeded samples of \AcCutSix\ cut functions of the ISCAS-85 circuits at each of six, seven and
eight inputs, \AcCutPeeledSix, \AcCutPeeledSeven\ and \AcCutPeeledEight\ functions received at least
one atom and \AcCutTwoSix, \AcCutTwoSeven\ and \AcCutTwoEight\ received two. The median description
was shorter than the raw table by \AcCutSaveSix, \AcCutSaveSeven\ and \AcCutSaveEight\ bits, with a
median residue of \AcCutResSix\%, \AcCutResSeven\% and \AcCutResEight\% of the entries. On
subsamples of \AcCutRouteNSix\ functions the cofactor route was shorter for \AcCutRouteSix,
\AcCutRouteSeven\ and \AcCutRouteEight\ of them. All \AcRefAtoms\ atoms were invariant under their
groups when checked by the reference implementation of group matching, which also agreed with the
projection residuals of \AcRefRes\ residues to within \AcRefMaxDiff.

Figure~\ref{fig:cascade}(a) shows the sum bit. Its two atoms are the XOR of the carry into the bit
with $x_2$, which does not depend on the two most significant inputs $x_3$ and $x_6$, and the affine
function $x_2\oplus x_3\oplus x_6$, with no residue, at \AcSumDl\ bits against \AcSumRaw\ for the raw
table. The split of the affine term $x_2$ between the two atoms is not unique, since the affine
functions of $x_1$, $x_2$, $x_4$ and $x_5$ lie in both spaces. The bent function of
Figure~\ref{fig:cascade}(b) is the whitest residue of the translations
(Proposition~\ref{prop:peelmeasures}(ii)), yet it peels into two atoms with no residue, $x_3x_4\oplus
x_5x_6$, which does not depend on $x_1$ and $x_2$, and $x_1x_2$, which does not depend on the other
four inputs, at \AcBentDl\ bits against \AcBentRaw. A residue that no translation concentrates can
therefore still be a sum of functions of disjoint inputs, which a library with the families $T_U$
detects. The majority of seven inputs gives one symmetric atom at \AcMajDl\ bits against \AcMajRaw,
and the four-to-one multiplexer two atoms that each ignore one pair of data inputs at \AcMuxDl\ bits,
with a Shannon expansion at the first select input at \AcMuxCofDl. The procedure was run only in
Python on functions of at most eight inputs, with the library above, and it was not timed against
decomposition tools.
\fi

\iffull
\subsection{Variable Orders for Multiplexer Trees}
\label{sec:varorder}

The experiment computes, for every function of four inputs and for the distinct cut functions of
four, five and six inputs of the ISCAS-85 circuits (Section~\ref{sec:cuts}), the number of
multiplexers of the reduced tree for all $k!$ orders, the classes of
Proposition~\ref{prop:cosets}, the result of sifting from the natural order, and the greedy orders
of Proposition~\ref{prop:haarenergy}. Table~\ref{tab:varorder} gives the results. The count is
constant on every class wherever the classes were computed, as the proposition requires, and the
classes cut the exact search by a mean factor of \VoReduceAll\ over all functions of four inputs and
of \VoReduceFour\ and \VoReduceFive\ over the cut functions of four and five inputs, which are far
richer in symmetry. The greedy Haar heuristic fails its test. It reaches the optimum for
\VoHaarAll\ percent of the functions of four inputs, against \VoSiftAll\ percent for sifting and
\VoRevAll\ percent for the reverse greedy order, and for \VoHaarSix\ percent of the six-input cuts
against \VoSiftSix\ percent for sifting. Energy at the coarse levels measures how unbalanced the
cofactors are, while the multiplexer count measures how many distinct cofactors there are, and the
two are not aligned. The useful contribution of the group is the exact one, since symmetric inputs can be
fixed relative to each other before any search, as symmetric sifting does heuristically
\cite{panda1994symmetry}, and Proposition~\ref{prop:cosets} extends this to every element of the
stabilizer, negations and products of exchanges included.

\begin{table*}[t]
\centering
\caption{Variable orders for reduced multiplexer trees. Order matters: percentage of functions whose multiplexer count depends on the order. Reduction: mean of $k!$ over the number of order classes under the stabilizer (Proposition~\ref{prop:cosets}). Random: mean ratio of the count of a uniformly random order to the optimum. For sifting and the greedy Haar order (most energy at the root first), the percentage of functions where the order is optimal and the mean ratio to the optimum. Reverse: the greedy order with least energy first. Cut functions are weighted by their numbers of cuts.}
\label{tab:varorder}
\footnotesize
\setlength{\tabcolsep}{2.6pt}
\begin{tabular}{@{}lrrrrrrrrr@{}}
\toprule
& & order & & & \multicolumn{2}{c}{sifting} & \multicolumn{2}{c}{greedy Haar} & reverse\\
\cmidrule(lr){6-7}\cmidrule(lr){8-9}\cmidrule(l){10-10}
functions & distinct & matters (\%) & reduction & random & optimal (\%) & ratio & optimal (\%) & ratio & optimal (\%)\\
\midrule
all functions, $k=4$ & 65{,}536 & 98 & 2.13 & 1.204 & 83 & 1.030 & 19 & 1.200 & 36\\
cut functions, $k=4$ & 2{,}168 & 90 & 5.38 & 1.208 & 99 & 1.002 & 72 & 1.088 & 77\\
cut functions, $k=5$ & 4{,}231 & 93 & 15.07 & 1.302 & 98 & 1.004 & 68 & 1.100 & 65\\
cut functions, $k=6$ & 2{,}379 & 86 & 120.52 & 1.312 & 98 & 1.005 & 77 & 1.066 & 75\\
\bottomrule
\end{tabular}
\end{table*}

\subsection{Blind Group Matching of Circuit Functions}
\label{sec:bgmexp}

Algorithm~\ref{alg:bgm} is exact, and the first test confirms it against enumeration. On all
\BgFour\ functions of four inputs its stabilizer orders in $B$ agree with the enumeration of the
384 elements, with and without output complementation, and on one function of each of the
\BgClasses\ NPN classes its orders in $\AGL(4,2)$ agree with the enumeration of the \BgAffineMaps\
affine maps. The median time per function of four inputs in $B$ is \BgMsFour\ ms. The input
permutations that preserve the periodogram exceed the permutation parts of the stabilizer on
\BgLoose\ percent of the functions of four inputs (a mean of \BgSecondMean\ against \BgProjMean),
while those that preserve the bispectrum equal them on \BgThirdExact\ percent, so the third-order
test removes every false candidate there and the sign system is needed only for the translations.
Table~\ref{tab:bgm} applies the algorithm to the cut functions. Their stabilizers in $B$ are
nontrivial for all but \BgTrivSixB\ percent of the six-input instances, with a median order of
\BgMedSixB, and their stabilizers in $\AGL(6,2)$ have a median order of \BgMedSixAGL. The search
takes a median of \BgMedMsSixAGL\ ms per six-input function in $\AGL(6,2)$ and at most
\BgMaxMsSixAGL\ ms, with at most \BgMaxNodesSixAGL\ search nodes. The colors and the orbit pruning
are what keep it there. The functions that take longest have plateaued spectra with few distinct
values and very large stabilizers, where the colors split least.

Near symmetries are a different matter. The second experiment plants one involution in a random
function of six inputs, an input exchange or a transvection, flips $m$ bits of its table, and ranks
the \mbox{45} candidates of the dictionary by the DC-GEVP of Proposition~\ref{prop:dcgevp}, over
\BgTrials\ trials for each $m$. Without flips the planted candidate lies in the null space every
time. A flip at a point that the involution fixes leaves the symmetry intact, and half of the points
are such points, so after one flip the exact algorithm still finds a nontrivial stabilizer in
\BgExactOne\ percent of the trials, and after two, four and eight flips in \BgExactTwo, \BgExactFour\
and \BgExactEight\ percent. The diagonal ratio $\mathbf X_{kk}/\mathbf Y_{kk}$ ranks the planted
candidate first in \BgDiagOne, \BgDiagTwo, \BgDiagFour\ and \BgDiagEight\ percent of the trials for
one, two, four and eight flips, consistently above the exact rate. The eigenvector of the smallest
eigenvalue does worse, \BgVecOne, \BgVecTwo, \BgVecFour\ and \BgVecEight\ percent, because the
candidates are nearly orthogonal in the Gram matrix $\mathbf Y$ and a combination of several
near-symmetries can have a smaller commutator than the planted one alone. For a single near-symmetry
the diagonal is therefore the better ranking, and the eigenvector matters when the symmetry is a
combination of the dictionary elements. The second-order statistic keeps evidence of a symmetry that
the exact test loses to bit errors, which is the regime of the noisy tables of
Section~\ref{sec:noisy}.

\begin{table*}[t]
\centering
\caption{Blind group matching of the distinct cut functions of the ISCAS-85 circuits with $k$ inputs: stabilizers in $B$ and in $\AGL(k,2)$, weighted by the numbers of cuts, with the search nodes and the time of one function (single thread, reference implementation in Python). Every generator was verified on the truth table.}
\label{tab:bgm}
\small
\setlength{\tabcolsep}{3.5pt}
\begin{tabular}{@{}rlrrrrrrrrr@{}}
\toprule
& & & trivial & median & largest & mean & \multicolumn{2}{c}{nodes} & \multicolumn{2}{c}{time (ms)}\\
\cmidrule(lr){8-9}\cmidrule(l){10-11}
$k$ & group & functions & (\%) & order & order & $\log_2$ order & median & largest & median & largest\\
\midrule
4 & $B$ & 2{,}168 & 0.1 & 4 & 192 & 2.30 & 4 & 18 & 0.20 & 0\\
4 & $\AGL$ & 2{,}168 & 0.0 & 384 & 20{,}160 & 9.11 & 12 & 19 & 0.38 & 1\\
5 & $B$ & 4{,}231 & 0.3 & 8 & 1{,}920 & 3.20 & 7 & 80 & 0.54 & 1\\
5 & $\AGL$ & 4{,}231 & 0.0 & 9{,}216 & 9{,}999{,}360 & 13.97 & 20 & 319 & 0.98 & 10\\
6 & $B$ & 2{,}379 & 0.9 & 36 & 23{,}040 & 5.11 & 12 & 289 & 1.85 & 7\\
6 & $\AGL$ & 2{,}379 & 0.0 & 14{,}450{,}688 & 20{,}158{,}709{,}760 & 22.87 & 36 & 1{,}033 & 3.32 & 115\\
\bottomrule
\end{tabular}
\end{table*}

\fi

\iffull
\subsection{Early Evaluation Reads Cofactors on Cosets of the Late Inputs}
\label{sec:early}

Self-timed circuits use handshakes in place of the clock, and a phased-logic gate fires when every one
of its inputs has arrived. Early evaluation pairs such a master gate with a trigger gate that
watches a subset $V$ of inputs known to arrive early and lets the master fire as soon as the value of
$f$ no longer depends on the late inputs $L$ \cite{thornton2002early}. The trigger fires on an
assignment $u$ to $V$ exactly when the cofactor $f_u$ is constant over the late inputs, and the
coverage of $V$ is the fraction of rows on which it fires. The method of \cite{thornton2002early}
searches the 14 support sets of at most three inputs of a four-input gate, computes the coverage of
each from the cube lists of the ON- and OFF-sets, and weighs it with the arrival times. For the carry
of a full adder with the carry-in late, the trigger on $\{a,b\}$ fires on the generate and kill
assignments and covers half of the rows (Fig.~\ref{fig:early}).

In the language of the framework the rows that agree with $u$ on $V$ form the orbit of the
translations $T_L$ that complement only late inputs, and the trigger tests whether $f$ is constant
on that orbit, that is, whether the $T_L$-average of $F$ at $u$ has magnitude one. The spectrum then
bounds the coverage of every candidate at once.

\begin{figure}[t]
\centering
\begin{tikzpicture}[font=\scriptsize, x=1cm, y=1cm,
  v/.style={circle, draw, inner sep=1pt, minimum size=11pt}]
\foreach \a in {0,1} \foreach \b in {0,1} \foreach \c in {0,1}
  \coordinate (E\a\b\c) at (1.9*\a+0.8*\b, 1.6*\c+0.6*\b);
\foreach \b in {0,1} \foreach \c in {0,1} \draw[gray] (E0\b\c) -- (E1\b\c);
\foreach \a in {0,1} \foreach \c in {0,1} \draw[gray] (E\a0\c) -- (E\a1\c);
\foreach \a/\b/\col in {0/0/green!50!black, 1/1/green!50!black, 0/1/red!70!black, 1/0/red!70!black}
  \draw[line width=1.6pt, \col] (E\a\b0) -- (E\a\b1);
\foreach \a in {0,1} \foreach \b in {0,1} \foreach \c in {0,1} {
  \pgfmathtruncatemacro{\m}{(\a+\b+\c)>=2 ? 1 : 0}
  \node[v, fill=white] at (E\a\b\c) {\m};
}
\node[anchor=north east, inner sep=1pt] at ($(E000)+(-0.12,-0.1)$) {$abc=000$};
\node[anchor=south west, inner sep=1pt] at ($(E111)+(0.12,0.1)$) {$111$};
\node[anchor=west, align=left] at (3.3,1.6) {green: cofactor over $c$\\constant, trigger fires\\($ab=00$ kill, $ab=11$ generate)\\[2pt]red: cofactor balanced,\\master waits for $c$};
\end{tikzpicture}
\caption{Early evaluation of the full adder carry $\mathrm{maj}(a,b,c)$ with the carry-in $c$ late.
Each vertical edge is an orbit of the translation that complements $c$, the rows that agree on the
early inputs $a,b$. The trigger fires on the two orbits where the carry is constant, a coverage of
one half, which equals the bound $\mathcal B_{\{a,b\}}$ of Proposition~\ref{prop:early} because the other two
cofactors are balanced.}
\label{fig:early}
\end{figure}
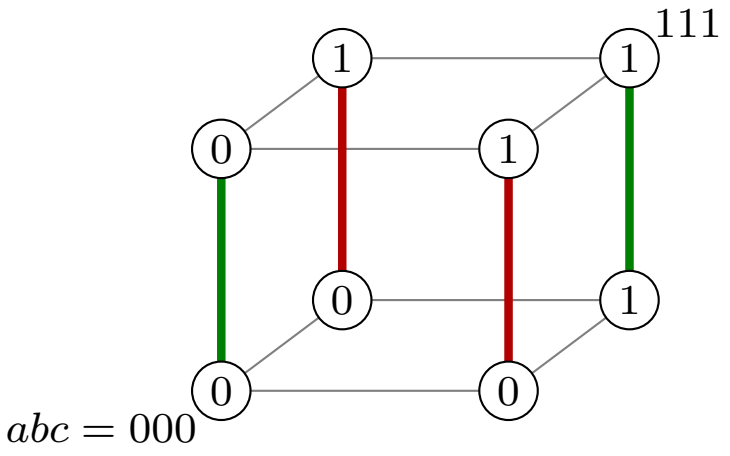

\begin{proposition}[Early-evaluation coverage]
\label{prop:early}
Let $V$ and $L$ partition the inputs. The coverage of $V$ satisfies
\[
\mathrm{cov}(V)\le \mathcal B_V=\frac1{N^2}\sum_{a:\,\supp a\subseteq V}W_f(a)^2,
\]
with equality exactly when every cofactor $f_u$ is constant or balanced, and in particular whenever
$|L|=1$. For an order $o$ that lists the inputs of $V$ first,
$\mathcal B_V=N^{-2}\bigl(W_f(0)^2+\sum_{j<|V|}2^j\mathcal L_j(o)\bigr)$ with the Haar level energies of
Proposition~\ref{prop:haarenergy}. The same holds for the cosets of any subgroup $H$ of $T$, with the
sum taken over the annihilator $H^\perp$.
\end{proposition}

\begin{proof}
Let $\bar F_u$ be the mean of $F$ over the rows that agree with $u$ on $V$. The trigger fires on $u$
exactly when $|\bar F_u|=1$, so $\mathrm{cov}(V)=2^{-|V|}\#\{u:|\bar F_u|=1\}\le2^{-|V|}\sum_u\bar F_u^2$, with
equality exactly when every $\bar F_u^2$ is $0$ or $1$. The sums $2^{|L|}\bar F_u$ form a function of $u$
whose Walsh coefficients are the $W_f(a)$ with $\supp a\subseteq V$, as in the proof of
Proposition~\ref{prop:haarenergy}, so Parseval's identity on $\F_2^{|V|}$ gives
$2^{-|V|}\sum_u\bar F_u^2=\mathcal B_V$. With one late input a cofactor is a function of one variable, which is
constant or balanced. The expression through the Haar levels is Proposition~\ref{prop:haarenergy}
summed over the first $|V|$ levels. For a subgroup $H$ the cosets of $H$ take the place of the cofactors and the
quotient $\F_2^n/H$, whose characters are indexed by $H^\perp$, takes the place of $\F_2^{|V|}$.
\end{proof}

The bound is computed for every candidate set from one Walsh spectrum, with no cube lists, and the
version for subgroups extends early evaluation to triggers that watch XORs of early inputs. For the
full adder carry $\mathcal B_{\{a,b\}}=\EeCarryBound$, equal to its coverage, since the cofactors with respect
to $c$ are constant or balanced. The experiment checks the proposition on every function of three
and four inputs and every set $V$, \EePairsFour\ pairs for four inputs, of which \EeEqualFour\ percent
attain equality, and for the cosets of every subgroup of $T$ on every function of three inputs and
\EeSubFuncs\ seeded functions of four. As a ranking on the \EeFunctions\ distinct four-input cut
functions of the ISCAS-85 circuits, the size of the phased-logic gates of \cite{thornton2002early},
the set with the largest $\mathcal B_V$ has the largest coverage among the sets of its size on \EeHitOne,
\EeHitTwo\ and \EeHitThree\ percent of the \EeInstances\ cut instances for one, two and three early
inputs, losing a mean coverage of \EeLossTwo\ where it does not, and the best coverages average
\EeCovOne, \EeCovTwo\ and \EeCovThree. Arrival times still decide between sizes, as in the weighting
function of \cite{thornton2002early}, but within a size the spectrum ranks the candidates.
A trigger need not observe inputs. A gate that observes $r$ parities of the inputs, which an
XOR stage ahead of the trigger provides, can complete on every coset of the annihilator of
the subspace they span on which $f$ is constant, and the subgroup form of the bound applies to that
subspace. Over the same cut instances the best of the \EeXorSetsOne\ nonzero parities covers more
than the best single input on \EeXorOne\ percent of the instances, and the best of the
\EeXorSetsTwo\ planes of parities more than the best of the \EeUnitSetsTwo\ pairs of inputs on
\EeXorTwo\ percent, with mean gains over all instances of \EeXorGainOne\ and \EeXorGainTwo.

State assignment has the same structure. An assignment of binary codes to the states of a machine
is determined up to the group $B$ acting on the code bits, since permuting or complementing state
variables changes only wiring and inverters, so a search over assignments is a search over the
orbits of $B$, as Boolean matching is a search over NPN classes. Conditions for race-free
assignments of asynchronous machines, which ask that some state variable separate the states of
every pair of transitions in an input column \cite{tracey1966internal,unger1969asynchronous}, are
invariant under $B$, so canonical forms under $B$ reduce that search as well. Burnside's lemma
counts the orbits exactly, as the mean over $B$ of the number of assignments each element fixes. An
element fixes an assignment of $\mathfrak s$ states when it fixes all $\mathfrak s$ codes, and a nonidentity element
fixes an affine subspace of at most $2^{\mathfrak r-1}$ of the $2^{\mathfrak r}$ codes of $\mathfrak r$ state variables, or none.
The action is therefore free when $\mathfrak s>2^{\mathfrak r-1}$, the case of the fewest state variables, and the
orbit count is then the formula $(2^{\mathfrak r}-1)!/\bigl((2^{\mathfrak r}-\mathfrak s)!\,\mathfrak r!\bigr)$ of McCluskey and Unger
\cite{mccluskey1959note}, which gives \AsgFourRows\ assignments for four states and \AsgFiveRows\ for
five. With more state variables, as race-free asynchronous assignments often need, the action is
not free and the formula undercounts. Three states on three variables have \AsgThreeBurn\
essentially different assignments, where the formula gives \AsgThreeMU, and four states have
\AsgFourBurn\ against \AsgFourMU. The two counts agree exactly when $\mathfrak s>2^{\mathfrak r-1}$ in all \AsgCases\ cases
with $2\le\mathfrak r\le\AsgMmax$ and $2\le\mathfrak s\le2^{\mathfrak r}$.
\fi

\iffull
\subsection{Correcting Bit Errors in Stored Truth Tables}
\label{sec:errors}

Truth tables stored in memory lose bits. The configuration memory of a field-programmable gate array
holds each lookup table as a truth table, and radiation upsets flip its bits
\cite{wirthlin2015fpga,kastensmidt2006fpga}. A table measured from a noisy device or rebuilt from
noisy simulation has the same problem. Current solutions add redundancy to the design through periodic
scrubbing from a stored copy of the configuration, error-correcting codes on the memory and triple
modular redundancy of the logic \cite{kastensmidt2006fpga,wirthlin2015fpga}. Here, the stabilizer of
the function supplies the redundancy. Every point of an orbit holds the same bit, or its complement
for a transformation that complements the function, so a majority over the orbit acts as that many
repeated reads (Proposition~\ref{prop:voting}) without a stored copy. Two experiments test this. The
first discovers the symmetries from the noisy table itself on the benchmark outputs
(Section~\ref{sec:noisy}). The second takes them as recorded at mapping time and computes exactly the
protection they give the lookup tables of the ISCAS-85 cut functions (Section~\ref{sec:lut}).
Discovered symmetries lower the balanced error to a median of \NyRatioC\ times the flip rate at
$\epsilon=0.1$ when the minority value is not rarer than the flips, and they give no gain when the mean orbit has
fewer than four points. With recorded symmetries, the probability that a six-input table holds a wrong
bit falls on average to \LuTableSixB\ of its unprotected value. Voting reduces the upsets that reach
the function between scrubs and is not a substitute for scrubbing.

\subsubsection{Orbit Voting Corrects Noisy Truth Tables with Large Orbits}
\label{sec:noisy}

Error correction by orbit voting needs the symmetries of a function, and here they are
discovered from the noisy table itself. The candidate library holds, for every pair of
inputs, the exchange $x_i\leftrightarrow x_j$ and the complemented exchange
$x_i\leftrightarrow x_j\oplus1$, and the translations by every unit vector and every sum of
two unit vectors. A candidate passes a one-sided binomial test when the number of its
2-cycles $\{x,gx\}$ whose observed bits disagree is not significantly larger than the flips
alone produce, at level \NyLevel\ with $\epsilon$ known. Candidates are then visited in
increasing order of their disagreement rate, and each is accepted when joining it to the
elements already accepted leaves every merged orbit as nearly constant as the flips allow,
by a one-sided normal test at the same level on the total minority count of the merged
orbits. The accepted elements generate the group whose orbits vote. The selection needs no
training data and no neural network. The tested vote of Proposition~\ref{prop:testedvote}
runs on the same group at the same level, $\alpha=\NyTestLevel$, and needs no further
pass over the table.

The experiment uses every output of the benchmark circuits with $n\le10$ inputs that some
library element fixes, \NyFunctions\ outputs of \NyCircuits\ circuits with \NyNmin\ to
\NyNmax\ inputs, flip rates from 0.01 to 0.3 and \NyTrials\ noisy tables per output and
rate. A sparse on-set and a scattering of flipped bits look alike, so we measure the
balanced error, the mean of the error rates on the on-set and on the off-set. It equals
$\epsilon$ for the raw table and one half for the constant table equal to the majority
value, so it gives no credit for erasing the rarer value. Table~\ref{tab:noisy} compares the
balanced error after the majority vote and after the tested vote with the flip rate and with
the prediction of Proposition~\ref{prop:voting} for the exact library stabilizer, and
Fig.~\ref{fig:noisy} shows the majority vote output by output.

\begin{table*}[t]
\centering
\caption{Orbit voting on noisy truth tables. Median balanced bit error rate (the mean of the
on-set and off-set error rates) divided by the flip rate $\epsilon$, for the majority vote and
the tested vote over the same discovered symmetries (majority / tested), over the outputs
whose minority value covers at least a fraction $\epsilon$ of the inputs, by the mean orbit
size of the exact library stabilizer, with the ratio predicted for the exact stabilizer in
parentheses. The lower rows give the number of such outputs, the percentage with balanced
error below $\epsilon$, the percentage on which the tested vote is lower and higher, and the
number of outputs whose minority value is rarer.}
\label{tab:noisy}
\ifarxiv\setlength{\tabcolsep}{1.5pt}\else\setlength{\tabcolsep}{5pt}\fi
{\footnotesize
\begin{tabular}{@{}lrrrrr@{}}
\toprule
 & \multicolumn{5}{c}{flip rate $\epsilon$}\\
\cmidrule(l){2-6}
mean orbit & 0.01 & 0.05 & 0.1 & 0.2 & 0.3\\
\midrule
below 4 & 1.50 / 1.38 (1.00) & 1.49 / 1.22 (1.00) & 1.41 / 1.26 (1.00) & 1.26 / 1.20 (1.00) & 1.47 / 1.47 (1.00)\\
4 to 16 & 0.87 / 0.35 (0.06) & 0.57 / 0.32 (0.10) & 0.53 / 0.45 (0.16) & 0.63 / 0.59 (0.34) & 1.02 / 1.01 (0.57)\\
16 to 64 & 0.02 / 0.04 (0.00) & 0.02 / 0.04 (0.00) & 0.06 / 0.04 (0.00) & 0.22 / 0.14 (0.02) & 0.83 / 0.78 (0.10)\\
64 or more & 0.00 / 0.02 (0.00) & 0.00 / 0.01 (0.00) & 0.00 / 0.01 (0.00) & 0.00 / 0.01 (0.00) & 0.05 / 0.05 (0.00)\\
\midrule
all such outputs & 0.32 / 0.22 (0.02) & 0.39 / 0.28 (0.06) & 0.42 / 0.34 (0.15) & 0.53 / 0.45 (0.35) & 0.98 / 0.98 (0.57)\\
outputs & 261 & 195 & 162 & 114 & 84\\
below $\epsilon$ (\%) & 64 / 70 & 66 / 71 & 70 / 73 & 71 / 72 & 52 / 52\\
tested lower / higher (\%) & 32 / 31 & 55 / 26 & 59 / 21 & 49 / 19 & 54 / 13\\
rarer minority & 45 & 111 & 144 & 192 & 222\\
\bottomrule
\end{tabular}}
\end{table*}

\begin{figure}[t]
\centering
\includegraphics[width=0.62\textwidth]{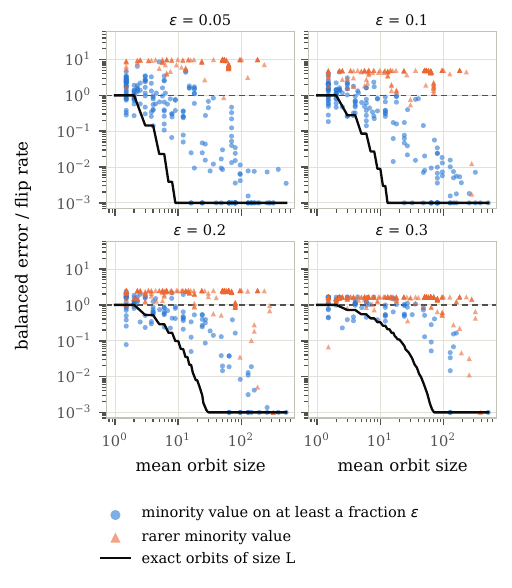}
\caption{Balanced bit error rate after voting over discovered symmetries, divided by the
flip rate, against the mean orbit size of the exact library stabilizer, one point per
output. Values below $10^{-3}$ are drawn at $10^{-3}$. The curves are
$e(L,\epsilon)/\epsilon$ from \eqref{eq:vote} for exact orbits of uniform size $L$, and the
dashed line marks no correction.}
\label{fig:noisy}
\end{figure}

When the minority value covers at least a fraction $\epsilon$ of the inputs, so that it is
not rarer than the flipped bits, voting over the discovered symmetries lowers the balanced
error to a median of \NyRatioC\ times the flip rate at $\epsilon=0.1$, and below the flip
rate on \NyBelowC\ percent of those outputs. The gain follows the orbit size, as the
$(G,L)$ continuum predicts. For the \NyBigCount\ such outputs whose exact stabilizer has
orbits of 16 points or more on average, the median balanced error at $\epsilon=0.1$ is
\NyBigRatioC\ times the flip rate. For mean orbits below 4 voting does not help. Orbits of
one or two points give no correction even for the exact stabilizer, and the near-symmetries
that the selection accepts merge orbits of different values, which raises the balanced
error to \NyRatioSmallC\ times the flip rate. When the minority value is rarer than the
flipped bits, the selection accepts near-symmetries that merge the minority points into
orbits of the majority value, and voting erases them. At $\epsilon=0.1$ the median on-set
error rate of these \NyRestC\ outputs is \NyRestOnErrC. A noisy table alone does not
separate so rare a minority from the flips, and such outputs need their symmetries from
another source, for example the specification of the circuit.

The tested vote keeps nearly all of the correction on large orbits and reduces the damage
that near-symmetries do, most of all on orbits of intermediate size. Over the outputs whose minority value
covers at least a fraction $\epsilon$ of the inputs, it lowers the median balanced error
from \NyRatioA\ to \NyTestA\ times the flip rate at $\epsilon=0.01$, from \NyRatioB\ to
\NyTestB\ at $\epsilon=0.05$ and from \NyRatioC\ to \NyTestC\ at $\epsilon=0.1$, and it
leaves the median unchanged at $\epsilon=0.3$. At $\epsilon=0.1$ its balanced error is lower than that of the
majority vote on \NyTestBetterC\ percent of these outputs and higher on \NyTestWorseC\
percent. The gain comes from the orbits where merges of different values occur. For mean
orbits of 4 to 16 points the median ratio falls from \NyRatioMidA\ to \NyTestMidA\ at
$\epsilon=0.01$ and from \NyRatioMidB\ to \NyTestMidB\ at $\epsilon=0.05$, and for mean
orbits below 4 it falls from \NyRatioSmallC\ to \NyTestSmallC\ at $\epsilon=0.1$ but stays
above one. On
orbits of 16 points or more the median ratio is \NyBigTestC\ against \NyBigRatioC\ for the
majority vote, within the small loss that Proposition~\ref{prop:testedvote} allows on matched
orbits. Orbits of two or three points keep most of the damage, because they cannot separate
a flipped bit from a point of the other value.

\subsubsection{Recorded Symmetries Protect Lookup-Table Contents}
\label{sec:lut}

A $k$-input lookup table of a field-programmable gate array stores the truth table of its
function in configuration memory, and radiation upsets flip its bits
\cite{wirthlin2015fpga,kastensmidt2006fpga}. The symmetries of the function are known when the
design is mapped, so they need not be discovered from noisy data. If the mapper records the
stabilizer of each table, voting over its orbits corrects upset bits from the table's own
contents. The functions here are the cut functions of Section~\ref{sec:cuts}, each weighted by
the number of cuts that realize it. The voting group is the full stabilizer of the function in
the group of input negations and permutations, found by exhaustive search, extended by the
input transformations that complement the function, whose images vote with their bit
complemented. Every quantity is exact. Each point of an orbit carries an independent noisy copy
of the same bit, so Proposition~\ref{prop:voting} gives the error of every bit, and different
orbits fail independently, so the probability that the voted table holds a wrong bit is one
minus the product over orbits of the probability that the orbit is voted correctly.

\begin{figure}[t]
\centering
\includegraphics[width=0.62\textwidth]{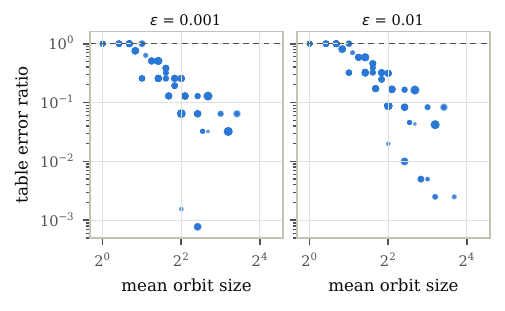}
\caption{Probability that a six-input lookup table holds a wrong bit after orbit voting,
divided by the same probability without voting, against the mean orbit size of the stabilizer
of its function, one point per distinct function, with area growing with the number of cuts
that realize it. The dashed line marks no protection.}
\label{fig:lut}
\end{figure}

The stabilizer is nontrivial for \LuNontrivSix\ percent of the six-input cut instances, because
cut functions of real circuits are rich in symmetric inputs. At an upset probability of $0.01$
per bit a six-input table without voting holds a wrong bit with probability \LuRawSixB. With
voting this probability falls on average to \LuTableSixB\ of its unprotected value, and to at
most one half for \LuHalvedSixB\ percent of the six-input instances. The expected bit error
falls to \LuBitSixB\ of the upset probability. The smaller tables have smaller orbits and
gain less, with average ratios of \LuTableFourB\ for four inputs and \LuTableFiveB\ for five.
Fig.~\ref{fig:lut} shows that the gain follows the mean orbit size, as
Proposition~\ref{prop:observations} predicts. The vote is not a substitute for scrubbing from a stored
copy or modular redundancy. It reduces the number of upsets that reach the function between
scrubs without a copy of the table, and it requires the recorded generators and the logic or
software that applies the vote.

\paragraph*{Comparison with error-correcting codes}
The voting gain must be weighed against the protection that configuration memories already use.
Table~\ref{tab:edac} compares, for the same \EdInstances\ six-input instances, the orbit vote with a
single-error-correcting, double-error-detecting (SECDED) code of length 72 on the 64 bits and with
triple modular redundancy (TMR), and with two ways of spending the same 64 bits on the orbit values,
as repeated copies or as a codeword of an extended BCH code. At upset rates of $10^{-4}$ per bit and
below, the conventional codes win by orders of magnitude, \EdSecdedFour\ for SECDED and \EdTmrFour\
for TMR against \EdOrbitFour\ for the orbit vote, and the vote is better than SECDED on only
\EdBeatsLow\ percent of the instances. The reason is structural. The smallest orbit of the stabilizer
has one point on \EdSmallOne\ percent of the instances and two on \EdSmallTwo\ percent, and a single
upset in an orbit of one or two points cannot be outvoted, so on about \EdSmallLeTwo\ percent of the
tables the error probability after voting is still of first order in $p$, while SECDED and TMR fail
only at second order. At $p=10^{-2}$, far above terrestrial upset rates, the vote is better than
SECDED on \EdBeatsHigh\ percent of the instances, and it needs no extra bits. Spending the 64 bits on
the orbit values instead, which needs logic that maps an input to its orbit, helps little as plain
copies, but as an extended BCH codeword it gives \EdCodeFour\ at $10^{-4}$, within a small factor of
SECDED without its 8 extra bits, and \EdCodeTwo\ at $10^{-2}$, below SECDED and TMR. The tables with
almost no symmetry keep a first-order term even then, so at $10^{-6}$ the code gives \EdCodeSix\
against \EdSecdedSix\ for SECDED. Orbit voting is therefore not a substitute for
SECDED, TMR or scrubbing. Its niche is a table whose stabilizer has no orbit smaller than three points,
or protection between scrubs where no extra bits are available.

Closest in spirit are the in-place methods that reconfigure the contents of lookup tables to mask
faults without changing the function seen at the outputs. They include fault-tolerant Boolean matching in the
resynthesis tool ROSE \cite{hu2008rose}, in-place reconfiguration (IPR), which maximizes identical
configuration bits for complementary inputs \cite{feng2009ipr}, and the maximization of zeros or ones
in the tables with spare XOR gates and carry chains \cite{das2013lut}. These choose among equivalent
implementations at mapping time and rely on logic masking. Orbit voting instead reads the symmetry of
the function and corrects the stored bits at run time.

\paragraph*{Scrubbing with voting between scrubs}
A frame code corrects the configuration only when the scrubber passes, and between two passes a
lookup reads the raw bits. With upsets arriving at a constant rate per bit, the probability that a
bit is wrong grows across the interval from zero to $1-e^{-\mathfrak u}$, where $\mathfrak u$ is the expected number of
upsets per bit in one interval. Averaged over the interval and over the \ScrubInstances\ six-input
instances of Table~\ref{tab:edac}, at $\mathfrak u=10^{-4}$ a raw lookup is wrong with probability
\ScrubRawFour, a lookup that takes the majority of its row's orbit, its own bit breaking ties,
\ScrubVoteFour, and a lookup in three copies voted bit by bit \ScrubTmrFour. The vote keeps a fraction
\ScrubRatio\ of the raw error, which is the share \ScrubSmall\ of rows that lie in orbits of one or two
points, since only those rows keep a first-order term, and \ScrubTenX\ percent of the instances gain a
factor of ten or more. Over the same interval a 72-bit frame collects two or more upsets, which its
SECDED code cannot correct at the next pass, with probability \ScrubFrameFour. Joined with scrubbing,
the vote therefore lowers the error of lookups between passes by the share of rows in larger orbits
at no extra bits, and TMR stays better by orders of magnitude with 128 extra bits. At
$\mathfrak u=10^{-2}$ the raw and voted lookups fail with probabilities \ScrubRawTwo\ and \ScrubVoteTwo, and a frame
collects two upsets in \ScrubFrameTwo\ of the intervals.

\begin{table*}[t]
\centering
\caption{Probability that a six-input lookup table holds a wrong bit for independent upsets of probability $p$ per stored bit, averaged over the 17{,}347 six-input cut instances of the ISCAS-85 circuits. Orbit vote: the stored table voted over the orbits of its stabilizer, no extra bits. Orbit copies and orbit code: the $r$ orbit values stored in the same 64 bits as evenly repeated copies or as an extended BCH codeword of length 64, both needing logic that maps an input to its orbit. SECDED: a $(72,64)$ code with 8 extra bits. TMR: three copies with 128 extra bits. The last column is the percentage of instances on which the orbit vote is better than SECDED.}
\label{tab:edac}
\footnotesize
\setlength{\tabcolsep}{4pt}
\begin{tabular}{@{}lrrrrrrr@{}}
\toprule
$p$ & unprotected & orbit vote & orbit copies & orbit code & SECDED & TMR & vote better (\%)\\
extra bits & 0 & 0 & 0 & 0 & 8 & 128 & \\
\midrule
$10^{-2}$ & $4.7\times10^{-1}$ & $1.4\times10^{-1}$ & $1.1\times10^{-1}$ & $5.1\times10^{-3}$ & $1.6\times10^{-1}$ & $1.9\times10^{-2}$ & 74\\
$10^{-3}$ & $6.2\times10^{-2}$ & $1.6\times10^{-2}$ & $1.3\times10^{-2}$ & $5.8\times10^{-4}$ & $2.4\times10^{-3}$ & $1.9\times10^{-4}$ & 22\\
$10^{-4}$ & $6.4\times10^{-3}$ & $1.6\times10^{-3}$ & $1.3\times10^{-3}$ & $5.9\times10^{-5}$ & $2.5\times10^{-5}$ & $1.9\times10^{-6}$ & 10\\
$10^{-6}$ & $6.4\times10^{-5}$ & $1.6\times10^{-5}$ & $1.3\times10^{-5}$ & $5.9\times10^{-7}$ & $2.6\times10^{-9}$ & $1.9\times10^{-10}$ & 10\\
\bottomrule
\end{tabular}
\end{table*}

\subsubsection{Symmetric Tables Store Fewer Bits}
\label{sec:lutcomp}

Symmetry also compresses a table. If $f$ is fixed by every exchange of inputs within each block of a
partition of its $k$ inputs into blocks of sizes $b_1,\dots,b_q$, it depends only on the weights of
the blocks, and a table with $\prod_i(b_i+1)$ entries addressed by the weights takes the place of the $2^k$
entries of the explicit table. The decompression logic is one population counter per block of two or
more inputs and a multiplexer on the weights. This is transform coding over $\F_2$. Within a block of
$b$ inputs the symmetric functions are spanned by the elementary symmetric functions
$\mathsf e_0,\dots,\mathsf e_b$, the sums of the Reed--Muller monomials over the orbits of $S_b$, and by
Lucas's theorem \cite{lucas1878} $\mathsf e_l(x)=1$ exactly when the binary digits of $l$ are a subset of
those of the weight $w$ of $x$. The value at weight $w$ is therefore
$\bigoplus_{l\subseteq w}\gamma_l$ for the coefficients $\gamma_l$ of $f$ in this basis, a M\"obius
transform over $\F_2$ of the coefficients indexed by the bits of the weight. The coefficients are the
transform coefficients, the population count is the analysis transform, and the radical filtration of
Section~\ref{sec:classical} orders them by degree \cite{berman1967group}. A symmetric function of low
degree has few nonzero coefficients, as in transform coding of signals.

Table~\ref{tab:lutcomp} evaluates the idea in a teaching model of transistor counts, stated in its
caption, and is meant as an illustration rather than a circuit design. A totally symmetric table of
six inputs needs 7 stored bits instead of 64 and about a third of the transistors, while a table with
blocks of two inputs saves little once the counters are included. Over the six-input cut functions,
\LcSymSix\ percent of the instances have a pair of exchangeable inputs, the stored bits fall to a mean
of \LcBitsSix\ of the explicit table, and the transistor count of the model to a mean of
\LcCostSix\ (\LcCostFive\ and \LcCostFour\ for five and four inputs). A programmable lookup table must
hold any function, so the saving applies to fixed-function tables, read-only stores and libraries of
symmetric cells. The reference software checks the identity of Lucas's theorem on every
totally symmetric function of up to ten inputs.

\begin{table*}[t]
\centering
\caption{Transistor counts of a lookup table stored explicitly and stored by the block weights of its input symmetries, in the teaching model of Section~\ref{sec:lutcomp} (six transistors per stored bit, two per 2:1 multiplexer, two per select inverter, 28 per full adder and 18 per half adder). Blocks: sizes of the classes of mutually exchangeable inputs.}
\label{tab:lutcomp}
\small
\begin{tabular}{@{}rlrrrrrrrr@{}}
\toprule
& & \multicolumn{2}{c}{explicit table} & \multicolumn{5}{c}{stored by block weights} & \\
\cmidrule(lr){3-4}\cmidrule(lr){5-9}
$k$ & blocks & bits & transistors & bits & storage & multiplexers & counters & total & ratio\\
\midrule
4 & 4 & 16 & 134 & 5 & 30 & 14 & 64 & 108 & 0.81\\
5 & 5 & 32 & 264 & 6 & 36 & 16 & 74 & 126 & 0.48\\
6 & 6 & 64 & 522 & 7 & 42 & 18 & 102 & 162 & 0.31\\
6 & 3+3 & 64 & 522 & 16 & 96 & 38 & 56 & 190 & 0.36\\
6 & 2+2+2 & 64 & 522 & 27 & 162 & 64 & 54 & 280 & 0.54\\
6 & 4+1+1 & 64 & 522 & 20 & 120 & 48 & 64 & 232 & 0.44\\
6 & 2+1+1+1+1 & 64 & 522 & 48 & 288 & 106 & 18 & 412 & 0.79\\
6 & 1+1+1+1+1+1 & 64 & 522 & 64 & 384 & 138 & 0 & 522 & 1.00\\
\bottomrule
\end{tabular}
\end{table*}

\fi

\iffull
\subsection{Spectra of Wide Functions from Simulation Samples}
\label{sec:sampled}

A function with $20$ to $40$ inputs has a truth table too long to transform, but its Walsh
coefficients of low weight can be estimated from simulation. A symmetry found by a census of the
function makes these estimates more accurate. With $L$ uniform input points $x_1,\dots,x_L$, the
estimate $\widehat W(a)=(N/L)\sum_lF(x_l)(-1)^{a\cdot x_l}$ is unbiased for $W_f(a)$. If $G$
permutes the inputs and fixes $f$, then $W_f$ is constant on the orbit $O$ of $a$ under $G$
(Theorem~\ref{thm:normalizer}), and the mean of $\widehat W$ over $O$ is again unbiased.

\begin{proposition}[Gain of an orbit mean]
\label{prop:sampled}
With $\varsigma=W_f(a)/N$, the variance of $\widehat W(a)$ is $N^2(1-\varsigma^2)/L$, the mean of
$\widehat W$ over $O$ has variance $N^2(1/|O|-\varsigma^2)/L$, and the gain of the orbit mean is
\[
\frac{1-\varsigma^2}{1/|O|-\varsigma^2}\ \ge\ |O| .
\]
\end{proposition}

\begin{proof}
For $a\ne a'$ in $O$, the products $F(x)(-1)^{a\cdot x}$ and $F(x)(-1)^{a'\cdot x}$ multiply to
$(-1)^{(a\oplus a')\cdot x}$, whose mean over uniform $x$ is zero, so one sample has covariance
$-\varsigma^2$ between the two estimates and variance $1-\varsigma^2$ for each. Averaging over $O$ and
over the $L$ samples gives the stated variances.
\end{proof}

The census simulates each output on \SaCensus\ uniform points, keeps the inputs on which the
output depends, and joins two of them when exchanging them leaves the output unchanged on every
point, the classical test for symmetric pairs \cite{pomeranz1994symm,mishchenko2003symm} run on a
sample, so that $G$ is the product of the symmetric groups on the classes, with the exchange of the
two operands added for the multiplier. The functions are the outputs of the ISCAS-85 circuits with
$20$ to $40$ inputs and of the benchmark PLAs with $22$ to $24$ inputs that have a class of two or
more inputs, \SaOutputs\ outputs from \SaSources\ circuits. For every coefficient of weight one or
two whose orbit has two or more elements, \SaR\ repetitions of $L=\SaL$ samples give the realized
gain, and a reference sample of \SaLref\ points gives $\varsigma$. Over \SaCoefs\ coefficients the
median realized gain is \SaGain, the median ratio of realized to predicted gain is \SaRatio, and
the tenth and ninetieth percentiles of that ratio are \SaRatioLo\ and \SaRatioHi. The largest
orbit has \SaMaxOrbit\ elements and the largest gain is \SaMaxGain, and \SaBelow\ coefficients
have a gain below $1.5$, the level our plan set for dropping the method. For \SaBis\ bispectrum
entries $\beta_f(e_i,e_j)$, with $e_i$ the index of the single input $x_i$, above three reference standard deviations, estimated from three
disjoint thirds of each sample and averaged over the orbits of the pairs $(i,j)$, the median
gain is \SaBisGain, and the median ratio of the gain to the orbit size is \SaBisRatio. The gain is
the orbit size, as Proposition~\ref{prop:sampled} predicts, so a symmetry found by a sampled
census lowers the number of simulations a spectral estimate needs by that factor. The census
is itself a sample, and an exchange that changes the output on few points would bias the
averaged estimate, which the reference values bound here.
\fi

\iffull
\subsection{Cryptographic Criteria Are Statements about the Translation Average}
\label{sec:crypto}

A substitution box (S-box) is a vector function $\mathcal F:\F_2^n\to\F_2^{\mathfrak m}$, a small lookup table
in a block cipher, and its component functions are $b\cdot\mathcal F$ for nonzero $b\in\F_2^{\mathfrak m}$. A
message authentication code (MAC) is a keyed function whose output on a message can be computed
only with the key, and HMAC builds one from a hash function by hashing the message with the key
twice, under two fixed paddings \cite{bellare1996keying}. A random bit generator produces a stream
of bits meant to be indistinguishable from independent fair coins, and the statistical tests of
NIST SP 800-22 \cite{nist80022} check streams against that ideal. Each of these is graded by the
translation average of its component functions.

\begin{proposition}[Avalanche, propagation and collisions]
\label{prop:crypto}
(i) A function $f$ satisfies the strict avalanche criterion \cite{webster1985sbox}, that
complementing any one input changes the output for exactly half of the rows, if and only if
$r_f(e_i)=0$ for every unit vector, and the propagation criterion of degree $k$ \cite{preneel1990propagation}
if and only if $r_f(d)=0$ for $1\le\mathrm{wt}(d)\le k$, and $f$ is bent if and only if it satisfies the
propagation criterion of degree $n$. (ii) For $\mathcal F:\F_2^n\to\F_2^{\mathfrak m}$ and a difference $d$, the
number of rows with $\mathcal F(x\oplus d)=\mathcal F(x)$ is
\[
\mathcal N(d)=2^{-\mathfrak m}\sum_{b\in\F_2^{\mathfrak m}}r_{b\cdot\mathcal F}(d),
\]
the average over the group of output masks of the translation averages of the components.
\end{proposition}

\begin{proof}
(i) $f(x)\oplus f(x\oplus d)$ is balanced exactly when $\sum_xF(x)F(x\oplus d)=0$. A function
has $r_f(d)=0$ for every $d\ne0$ exactly when $W_f^2=Hr_f$ is constant, which is bentness.
(ii) The indicator of $\mathcal F(x\oplus d)=\mathcal F(x)$ is $2^{-\mathfrak m}\sum_b(-1)^{b\cdot(\mathcal F(x\oplus
d)\oplus\mathcal F(x))}$, and summing over $x$ gives the autocorrelations.
\end{proof}

The strict avalanche criterion is therefore a condition on the translation average restricted to
the Hamming sphere of radius one, and the propagation criterion on the balls of radius $k$. On the
\CrAllFour\ functions of four inputs, \CrSac\ satisfy the strict avalanche criterion, \CrSacBal\ of
them balanced, and every function that satisfies the propagation criterion of degree two is bent
(\CrPcTwo\ functions). The count $\mathcal N(d)$ is the row $d$, column $0$ entry of the difference
distribution table, which is what a collision search needs, because a difference with large $\mathcal N(d)$ makes a
pair $(x,x\oplus d)$ collide far more often than a random pair. For a toy compression function of
16 to 8 bits, two rounds of the PRESENT S-box \cite{bogdanov2007present} and its bit permutation
followed by the XOR of the two output bytes, the identity agrees with direct counting on every
difference, the best difference makes a pair collide with probability \CrBestRate\ against
$2^{-8}\approx\CrRandomPair$ for a random pair, and \CrAboveFour\ percent of the differences reach four
times the random rate. A real hash function is far too wide to tabulate, so the identity says what
must be estimated, the autocorrelations of the output components at chosen differences, which the
sampled spectra of Section~\ref{sec:sampled} estimate from simulation, rather than how to break any
particular hash. HMAC adds a key to the same structure, and a distinguisher of this kind applies only
when its compression function shows spectral structure at accessible differences, which for
standard hash functions it is designed not to.

\begin{table*}[t]
\centering
\caption{S-boxes graded by the translation average. Nonlinearity $N/2-\max|W|/2$ and linearity $\max|W|$ over all components; differential uniformity, the largest entry of the difference distribution table off $d=0$; absolute indicator, the largest $|r(d)|$ for $d\ne0$; participation ratio $D_2$ of the components as a fraction of $N$ (1 for bent, $1/N$ for affine), smallest and mean; share of (component, input) pairs that meet the strict avalanche criterion; orders of the stabilizers of the components in $\AGL(4,2)$. Random rows give medians over seeded permutations, and the largest stabilizer order over all of them. The last two rows give ranges over the representatives of the 16 affine classes of optimal 4-bit S-boxes, split by the smallest participation ratio of a component.}
\label{tab:sbox}
\scriptsize
\setlength{\tabcolsep}{3pt}
\begin{tabular}{@{}lrrrrrrrl@{}}
\toprule
S-box & NL & $\max|W|$ & diff. unif. & abs. ind. & $D_2/N$ min & $D_2/N$ mean & SAC & $|\Stab_{\AGL}|$\\
\midrule
AES ($n=8$) & 112 & 32 & 4 & 32 & 0.492 & 0.492 & 0.13 & --\\
50 random ($n=8$) & 93 & 70 & 12 & 96 & 0.237 & 0.336 & 0.14 & --\\
\midrule
PRESENT ($n=4$) & 4 & 8 & 4 & 16 & 0.250 & 0.370 & 0.67 & 32 to 384\\
inversion in GF($2^4$) & 4 & 8 & 4 & 8 & 0.400 & 0.400 & 0.60 & 32 to 32\\
1000 random ($n=4$) & 2 & 12 & 6 & 16 & 0.182 & 0.361 & 0.52 & 32 to 10752\\
\midrule
8 of 16 classes & 4 & 8 & 4 & 16 & 0.250 & 0.370--0.390 & 0.58--0.72 & 32 to 384\\
8 of 16 classes & 4 & 8 & 4 & 8 & 0.400 & 0.400 & 0.57--0.63 & 32 to 32\\
\bottomrule
\end{tabular}
\end{table*}

Table~\ref{tab:sbox} grades S-boxes by the quantities of Section~\ref{sec:pr}. An S-box cannot be
bent, since a bent component is never balanced and every component of a permutation is, so good
S-boxes are nearly bent, and the participation ratio of each component, as a fraction of $N$, grades
how near. Every component of the AES S-box \cite{daemen2002rijndael} has the ratio \CrAesPr, against
a median of \CrRandEightPr\ for the mean over components of \CrRandEightCount\ random permutations of
8 bits, and its nonlinearity \CrAesNl\ and differential uniformity \CrAesDu\ \cite{nyberg1993differentially}
are the classical grades of the same spectrum and autocorrelation. The participation ratio weighs
all coefficients, the nonlinearity only the largest, as the R\'enyi orders $2$ and $\infty$ of
Section~\ref{sec:pr}. The PRESENT S-box reaches the largest nonlinearity possible for 4 bits, \CrPresentNl,
with components whose ratio is \CrPresentPrMin\ at the smallest and \CrPresentPrMean\ on average.
The multiset of component ratios is invariant under affine equivalence of S-boxes, so it grades
classes. Over representatives of the \CrOptClasses\ affine classes of optimal 4-bit S-boxes
\cite{leander2007classification}, as listed in \cite{canteaut2019dlct} and checked here to have
nonlinearity 4 and differential uniformity 4, the smallest ratio of a component splits the classes
into \CrOptLow\ with $1/4$ and \CrOptHigh\ whose components all have $2/5$. The absolute indicator,
16 against 8, gives the same split, so on these classes the participation ratio separates nothing
that the autocorrelation does not.

\begin{table}[t]
\centering
\caption{Bit streams cut into $\mathfrak b=2^{16}$ blocks of 8 bits. Serial statistic $\mathfrak b\sum_{a\ne0}\phi(a)^2$, equal to the chi-square of the block counts (255 degrees of freedom, mean 255 for an ideal stream), the largest single term $\mathfrak b\phi(a)^2$, its mask $a$, and the number of coefficients of magnitude one.}
\label{tab:streams}
\footnotesize
\begin{tabular}{@{}lrrrr@{}}
\toprule
stream & serial & largest & mask & $|\phi|=1$\\
\midrule
PCG64, seeded & 242.4 & 8.6 & 166 & 0\\
biased, $\mathbb P(1)=0.6$ & 24558.5 & 2754.6 & 32 & 0\\
LFSR, $x^7+x^6+1$ & 66568.1 & 65536.0 & 193 & 1\\
\bottomrule
\end{tabular}
\end{table}

For a bit stream cut into $\mathfrak b$ blocks of $n$ bits with empirical distribution $\mathsf p$, the serial
statistic $\mathfrak bN\sum_x(\mathsf p(x)-1/N)^2$, a chi-square with $N-1$ degrees of freedom for an ideal stream, is
$\mathfrak b\sum_{a\ne0}\phi(a)^2$ by Parseval's identity, the energy of the periodogram of the block
distribution, with $\phi$ the characteristic function of the generating-function paragraph of
Section~\ref{sec:hashvar}. Table~\ref{tab:streams} shows the difference between the total and the
spectrum. For a seeded PCG64 stream the statistic is \CrPcgSerial\ on 255 degrees of freedom. A
biased stream with $\mathbb P(1)=0.6$ fails, and its coefficients at masks of weight $w$ follow the
piling-up product $0.2^w$ for $w\le3$, \CrBiasW\ measured for $w=1,2,3$, until they reach the noise
floor $\mathfrak b^{-1/2}\approx\CrNoiseFloor$. The maximal-length sequence of $x^7+x^6+1$ fails with one
coefficient of magnitude one at mask \CrLfsrMask, the bits $0$, $6$ and $7$ of each block, which is
the recurrence $s_t=s_{t-1}\oplus s_{t-7}$ itself. The total only says that the stream fails, while the
spectrum names the linear relation, which is what the Berlekamp--Massey algorithm finds for linear
complexity \cite{massey1969shift}.
\fi

\section{Related Work}
\label{sec:related}

Spectral methods for switching functions are classical. The Rademacher--Walsh
spectrum and its invariance properties under negation, permutation and linear
transformation of the inputs were used for classification and threshold-logic
synthesis \cite{edwards1975,lechner1971} and are developed in the books
\cite{karpovsky1976,hurst1985spectral,thornton2001spectral,thornton2014vsim}.
Decision diagrams extend them to functions too large for truth tables
\cite{clarke1993spectral}, and the Fourier analysis of Boolean functions underlies
learning and testing algorithms \cite{kushilevitz1993,odonnell2014}. The counting
of classes under the linear and affine groups goes back to \cite{harrison1964}, the
affine classification of cosets of the first-order Reed--Muller code to
\cite{maiorana1991}, and the cryptographic theory of linear structures, bent
functions and nonlinearity is surveyed in \cite{carlet2021book}, with linear
structures characterized spectrally in \cite{evertse1987,lai1994linear,dubuc2001linear}
and bent functions introduced in \cite{rothaus1976bent}. The commutant of the
hyperoctahedral group is the Bose--Mesner algebra of the Hamming scheme
\cite{delsarte1973,delsarte1998,bannai1984,macwilliams1977}, and finer algebras of the
same kind lead to semidefinite bounds in coding theory \cite{schrijver2005terwilliger}.
Group averaging as a projection onto invariants is standard in invariant theory and in
the use of group representations in statistics \cite{sturmfels2008,diaconis1988}, and
completeness of the triple correlation for Abelian groups is treated in
\cite{yellott1992,kakarala2012bispectrum}. What this paper adds to that background is
the estimation viewpoint of Algebraic Diversity, in which these objects are group
averages of a single observation, together with the collapse along the group lattice,
the normalizer rule, the characteristic-two pairing and the applications.

The spectral classification of the 1970s is the closest precedent for the signatures of
Sections~\ref{sec:normalizer} and~\ref{sec:classify}, and it calls for a direct comparison.
Lechner \cite{lechner1971} and Edwards \cite{edwards1975} defined five operations on the
Rademacher--Walsh spectrum, the permutation and negation of inputs, the negation of the output,
the replacement of an input by its XOR with another input and the addition of an input
to the output, and classified functions by the canonic spectrum that these operations reach. The
books \cite{karpovsky1976,hurst1978logical,hurst1985spectral} made it a design method. The five
operations generate the extended affine group, so the spectral classes are the affine classes of
the cosets of the first-order Reed--Muller code, $8$ for four inputs, $48$ for five
\cite{berlekamp1972rm} and $150{,}357$ for six \cite{maiorana1991}, and their computation is still
being refined \cite{rice2011spectral,soeken2019hybrid}. Three parts of the present treatment are
new against that lineage. The first is the quotient-canonicalization view. The canonic spectrum
is reached by a search over the whole group acting on the signed coefficients, and the magnitudes
that classification reads from it are invariant under every permutation of the $2^n$ indices, a
far larger group than the one being classified. Here the signature is the average over the normal
translation subgroup, which removes the input negations and the output negation in closed form,
followed by a canonicalization over the quotient alone, and Theorem~\ref{thm:normalizer} states
which averages lose information and why. The loss can be measured. Under negation and
permutation of inputs and output, the magnitudes sorted within each spectral order separate
\CcOrdFour\ of the \CcClassesFour\ classes of four-variable functions and \CcOrdFive\ of the
\CcClassesFive\ classes of five-variable functions, against \CcPerFour\ and \CcPerFive\ for the
canonical periodogram. The multiset of all magnitudes takes \CcMagAllFour\ values over the
functions of four variables, one for each spectral class, but only \CcMagFive\ over those of five,
fewer than the $48$ classes. For the extended affine group itself the third-order average is
not available, since adding a linear function to the output moves its index triples off the
zero-sum set and negating the output flips its sign. The products that this group preserves are
those over an even number of Walsh indices with zero sum, of which the periodogram is the first.
Canonicalized over the affine maps of the index, the periodogram separates all \AfClasses\ classes
of five variables, where the magnitudes separate \AfMag. That census is exhaustive, with the
classes computed as the orbits of the affine group on the $2^{26}$ cosets, and the magnitudes
taken together with those of the autocorrelation separate all \AfClasses\ as well. The second is the bispectrum. The magnitudes and the autocorrelation
are second-order quantities and cannot separate classes that share a periodogram, the bent classes
being the first instance. The third-order average separates them, as the completeness of the
triple correlation for Abelian groups suggests \cite{yellott1992,kakarala2012bispectrum}. The
third is the census. Each signature is measured exhaustively, over all \CcClassesFour\ classes of
four variables and all \CcClassesFive\ of five, and what the third-order signature leaves together
is identified as \NpResidClasses\ near-bent classes in \NpResidGroups\ groups.

Linear and affine changes of the input variables have a long history as a tool of logic
design. Spectral methods used them to reduce the size of realizations
\cite{karpovsky1976,hurst1985spectral}, decomposition and synthesis tools chose them from the
spectrum \cite{varma1989}, and the spectral techniques of computer-aided design collect these
uses \cite{thornton2001spectral}. In decision diagrams, linear sifting substitutes for a variable its
XOR with a neighbor when that shrinks the diagram \cite{meinel1997linear}, and linearly
transformed diagrams generalize the variable order to a linear change of coordinates
\cite{gunther1998ltbdd}. The counts of classes under the linear and affine groups that these
methods move within go back to \cite{golomb1959,harrison1964}, and the same freedom has recently
been applied to energy-aware synthesis of adiabatic circuits \cite{thornton2026renesis}. In each
of these lines the transformation is chosen for a size or energy measure. Here the affine group
enters through its translation subgroup, as the group whose average is matched to the Walsh
transform, and through its quotient, which is removed by canonicalization.

Cut enumeration and the matching of cut functions under negation and permutation are the core
of mapping to lookup tables and of rewriting \cite{cong1999cut,mishchenko2006dag}\iffull, and the
protection of lookup-table contents against upsets relies on scrubbing from a stored copy and
on modular redundancy \cite{kastensmidt2006fpga,wirthlin2015fpga}. Orbit voting over recorded
symmetries adds a correction that uses neither\fi.

\iffull Binary decision diagrams \cite{bryant1986bdd} realize Shannon's expansion
\cite{shannon1949synthesis} as a network of multiplexers, which has been mapped to pass-transistor
logic and to timed Shannon circuits for low power \cite{lavagno1995timed,thornton2002timedshannon},
and Haar spectral coefficients have been read from the same diagrams for equivalence checking
\cite{thornton2002haar}. The best variable order is NP-hard to compute \cite{bollig1996np}, and sifting
\cite{rudell1993sifting} with groups of symmetric variables \cite{panda1994symmetry} is the
practical answer. The wreath-product multiresolution analysis of signals \cite{foote2000wreath}
supplies the group behind the Haar transform, which Proposition~\ref{prop:wreath} carries to
switching functions, and Proposition~\ref{prop:cosets} turns the stabilizer into an exact
reduction of the order search. Group discovery from data has a recent line in which a known group is
enforced exactly through group-structured tensor algebras trained on labeled data
\cite{hoyos2026exact}. The blind matching of Section~\ref{sec:bgm} differs in being unsupervised and
in working from a single observation through the commutant of its second-order average, as in the
double-commutator eigenvalue problem \cite{thornton2026gevp}. Reed--Muller codes as powers of the
radical of the group algebra of $\mathbb Z_2^n$ go back to Berman \cite{berman1967group}, and the
counting of symmetric and strongly asymmetric functions to the work on canalyzing functions
\cite{rosenkrantz2019ncf}.\fi
\iffull Spectral criteria for cryptographic Boolean functions, the strict avalanche criterion
\cite{webster1985sbox}, the propagation criterion \cite{preneel1990propagation}, nonlinearity and
differential uniformity \cite{nyberg1993differentially}, and the piling-up lemma
\cite{matsui1993linear}, are classical, and the book of Carlet collects them \cite{carlet2021book}. The
contribution here is to place them as group averages beside the switching-theory results. Early
evaluation in phased logic \cite{thornton2002early} chose triggers by enumerating support sets and
counting coverage from cube lists, and Proposition~\ref{prop:early} supplies a spectral bound that ranks
the candidates and extends them to triggers on XORs of inputs.\fi

Hashing-based counting rests on pairwise independent hash families
\cite{carter1979universal} and on the reduction of approximate counting to
satisfiability \cite{stockmeyer1983,valiant1986unique}. XOR constraints were
introduced for model counting in \cite{gomes2006xor} and made practical in
\cite{chakraborty2013approxmc}. Sparse and structured parity constraints were studied
in \cite{ermon2013wish,ermon2014ldpc,achlioptas2015codes}, which relate the quality of
a hash family to properties of the code it defines. Theorem~\ref{thm:hash} is the dual
statement for a fixed hash and a fixed solution set, with the variance given by the
indicator spectrum of the set on the row space, which is what makes the choice of a
hash matched to the set possible. The analyses of ApproxMC and ApproxMC2 use pairwise independence to bound the
variance of a cell count by its mean \cite{chakraborty2013approxmc,chakraborty2016approxmc2},
which is the property that sparse constraints give up and that
\cite{agrawal2020sparsity} isolates. Adaptive hashing \cite{kuck2019adaptive} also uses
solutions found during counting to shape the constraints, through per-variable marginals, and
states lower bounds. It notes that frozen variables are of no use in a parity constraint, the
parities of weight one among those constant on the solutions. The matched draw of
Proposition~\ref{prop:matched} excludes every parity constant on the pilot, and
Corollary~\ref{cor:pairwise} keeps a two-sided guarantee.

\iffull Parity reasoning in satisfiability solvers recovers XOR constraints from the
clauses and combines conflict-driven search with Gaussian or Gauss--Jordan elimination
\cite{soos2009cms,han2012gauss,soos2019bird,soos2020tinted}. The spectral reading of a
block is a group-theoretic formulation of that recovery that also yields the implied
parities of blocks whose solution sets are not affine.\fi{} Boolean matching and symmetry
detection use signatures and canonical forms under negation and permutation
\cite{mohnke1993,chai2006matcher,abdollahi2008,pomeranz1994symm,mishchenko2003symm}.
The normalizer rule of Section~\ref{sec:normalizer} explains which spectral signatures
lose information and how a third-order signature restores it.

\section{Limitations and Open Problems}
\label{sec:limits}

Approximate and don't-care matching was measured against search in Python on a library of the
classes that occur among the cut functions of one benchmark set, and with satisfiability don't
cares only. It was not timed inside a mapper, against SAT- or BDD-based matching with don't cares,
or with observability don't cares, and the samples at eight inputs are \AxQEight\ queries and
\AxCEight\ classes. The cofactor bound is loose when most points are don't cares, and a sharper
bound that couples the levels, for instance through cofactors of two inputs, is open.

The exact analyses of this paper evaluate truth tables or the local truth tables of
blocks, which limits them to functions and blocks of moderate input count. The
census and the hashing experiment reach 24 inputs. Larger functions call for spectra
computed from decision diagrams \cite{clarke1993spectral} or estimated from samples,
and Proposition~\ref{prop:pairing} gives the variance of the sampled translation
average for that purpose.

The counters of Sections~\ref{sec:counter} and~\ref{sec:core} compute each cell count exactly
from the indicator spectrum, which needs the truth table of the function. A counter for
formulas obtains the cell counts and the pilot from a satisfiability solver, and the clustered
sample models those solutions only coarsely. Corollary~\ref{cor:pairwise} keeps the variance
bound of pairwise independence for any pilot fixed before the hash, and
Proposition~\ref{prop:oraclehull} obtains the implied parities from the formula with at most
$d+2$ solver calls, so neither the guarantee nor the hull depends on uniform solutions. What
the counter for formulas of Section~\ref{sec:satcount} leaves open is the use of matched rows
inside a compiled counter after independent-support reduction, where the hull is
nearly the whole space for formulas whose sampling set is already minimal, its use to estimate
the keys of a locked circuit that remain consistent with oracle observations when the factor
graph is too wide for exact counting \cite{thornton2026versionspace}, and the use of the lower
variance to lower the threshold and with it the number of solver calls, which needs a lower
bound on the concentration of the solution set on the cosets of the pilot span that is not known
in advance. That counter is a harness in Python around CryptoMiniSat with \ScRuns\ calls per rule
and formula, and it covers the formulas that Ganak counts exactly within \ScExactTimeout\
seconds, which leaves out the hardest formulas of the set. The certified hull did not finish within
\ScHullLimit\ seconds on \ScSkipCounter\ of the formulas that Ganak counted, and these are left out
as well, so the comparison covers formulas whose hull is quick to certify. A prefix search that runs past $r$ levels leaves the scope of
Proposition~\ref{prop:matched}. The plug-in rule needs uniformly distributed solutions and, by
the birthday bound, a number of them of the order of $2^{k/2}$.

The fourth moment of a cell count is exact, but on the circuit outputs it lowered the number of
estimates a median counter needs at few levels\iffull{} (Section~\ref{sec:matched})\fi, because the outputs carry
much more additive energy than random sets, and a counter for formulas would also have to estimate
that energy from solutions. The orders above four in Table~\ref{tab:ladder} are placed but not
used here\iffull, and the census behind the sampled spectra of Section~\ref{sec:sampled} is itself
a sample, so an exchange that changes a function on few points goes unnoticed\fi.

\iffull
Spectral elimination decides a formula when the solution set of every block is
affine. For other formulas the implied parities form a sound relaxation that can prove
unsatisfiability or prune a search, but the procedure is not complete. Blocks are
formed by variable sets, so parities implied only by several blocks together are not
detected by the local reading. CryptoMiniSat decides the same families within a few
seconds, so the advantage of the spectral reading lies in its generality rather than in
its speed on these families.
\fi

The complete separation by the periodogram and bispectrum is established
exhaustively for four variables only. The cut functions of the ISCAS-85 circuits extend it to
the \CuClassFive\ and \CuClassSix\ classes of five and six inputs that occur there, with the
cut enumeration limited to \CuPerNode\ cuts per node, which is a statement about those circuits
and not about all functions of five or six inputs, and at seven inputs the periodogram already
leaves \AbSharedClasses\ of their cut classes together. The exact NPN algorithms of ABC are
faster than the signature on these functions. Over all functions of five variables the
bispectrum leaves \NpResidClasses\ near-bent classes together (Section~\ref{sec:classify}). A
signature that separates them without the full canonical form, and an efficient
canonicalization over $\GL(n,2)$, whose order grows as $2^{n^2}$, are open.\iffull{} The symmetry selection for noisy truth tables accepts
near-symmetries, which merge orbits of different values. The tested vote weighs the bias of
each merge against its reduction in variance orbit by orbit, but an orbit of two or three
points cannot separate a flipped bit from a point of the other value, and a rare minority
value is still erased. Weighing merges at the level of the accepted generators, which can
draw on every orbit a generator touches, discovering groups beyond the library by the
double-commutator eigenvalue problem \cite{thornton2026gevp,thornton2026generators}, and
extending the applications to multiple-valued logic through Remark~\ref{rem:prime} are
open. The protection of lookup-table contents assumes independent upsets of equal probability
on every bit, and it does not model upsets that flip several adjacent bits at once or the
memory and logic that store the generators and apply the vote. Against SECDED and triple modular redundancy the
vote loses at realistic upset rates, and the joint design with scrubbing was evaluated with a
model of uniform upsets and reads, not with a device. The greedy Haar order is worse than sifting on every set we
tried, and the reduction of Proposition~\ref{prop:cosets} needs the stabilizer, which takes a search
of its own. The blind matching algorithm is exponential in the worst case and slow on functions with
nearly flat spectra and large stabilizers, and its soft form was tested only with one planted
involution in functions of six inputs. The transistor counts of the compressed tables come from a
teaching model, not from a layout. The Tseitin formulas modulo three were run on random
4-regular graphs only, and obfuscated parities are left to future work. The cryptographic results are
identities and gradings checked on toy and standard S-boxes and a toy compression function, and
say nothing about the security of real hash functions or MACs. The early-evaluation bound ranks
candidate trigger sets but does not model arrival times or the overhead of the trigger gate, and
the state-assignment count was not joined to a synthesis flow. The peel of Section~\ref{sec:peel} was
measured with a library of three group families and pairs of them, mixtures up to seven inputs and a
uniform code for the index, and the cross-cofactor consistency ranks the expansions and the routes
without yet giving a threshold rule that does better than the base rates. AD-CLEAN over the cube
(Section~\ref{sec:adcleancube}) was run in Python only, on functions of at most eight inputs and with
the same library and the affine space, its search over pairs was limited at eight inputs, and it was
not timed against decomposition or factorization tools.\fi

\section{Conclusion}
\label{sec:conclusion}

\iffull
Choosing the group that matches a switching function turns its spectral structure
into algorithms. The translations are matched to the Walsh transform, the larger
affine groups collapse second-order statistics to the weight, and the normal
translation subgroup keeps what the full average discards. A matched stabilizer stands in for
repeated observations, with one evaluation per orbit identifying the function and a vote over
orbits of $s$ points dividing the number of noisy tables by $s$. The same viewpoint gives the
exact variance of XOR hashing and a way to choose hashes from a few solutions that
makes a counter fail less often while keeping the variance bound of pairwise independence, reads implied parities from the spectra of clause blocks,
separates equivalence classes with canonical spectra, including every class of the cut
functions of the ISCAS-85 circuits, and corrects noisy truth tables and lookup-table contents
by voting over orbits, with a tested shrinkage that limits the damage of near-symmetries.
These are the switching-algebra counterparts of the group gain, the reduction in snapshots,
the matched transform and the continuum between group and repetition that algebraic
diversity establishes for signals.
\fi

\appendix

\section{Notation}
\label{app:notation}

Each symbol of the paper with its meaning and where it is defined. Symbols that occur only inside
one proof are defined there and are not listed.

\begin{center}
\small
\begin{longtable}{@{}p{0.26\textwidth}p{0.47\textwidth}p{0.15\textwidth}@{}}
\toprule
symbol & meaning & defined in\\
\midrule
\endhead
$n$, $N=2^n$ & number of inputs, number of rows of the truth table & Sec.~\ref{sec:prelim}\\
$x$, $\F_2^n$ & a row (fully specified input assignment); the Boolean cube & Sec.~\ref{sec:prelim}\\
$a\cdot x$ & parity of the inputs selected by $a$ (an XOR gate) & Sec.~\ref{sec:prelim}\\
$t$, $F$, $S$ & truth vector, sign vector $(-1)^{f}$, ON-set & Sec.~\ref{sec:prelim}\\
$e_i$ & unit vector of input $i$; the translation that complements $x_i$ & Sec.~\ref{sec:prelim}\\
$W_f$, $\Sh$ & Walsh spectrum of $f$, indicator spectrum of $S$ & Sec.~\ref{sec:prelim}\\
$H$ & Sylvester--Hadamard matrix, $W_f=HF$ & Sec.~\ref{sec:prelim}\\
$r_f$ & autocorrelation, the translation average read as a function & Sec.~\ref{sec:prelim}\\
$T$, $S_n$, $B$, $\GL$, $\AGL$, $\EA$, $S_N$, $\Wr_n$ & the groups of Table~\ref{tab:groups} & Table~\ref{tab:groups}\\
$\Stab_G(f)$ & transformations in $G$ that leave $f$ unchanged & Sec.~\ref{sec:prelim}\\
$P_g$ & permutation matrix of a group element on truth vectors & Sec.~\ref{sec:prelim}\\
$\RG(f)$ & group average of $FF^{\mathsf T}$, the Reynolds projection & Sec.~\ref{sec:commutant}\\
$\mathcal C(G)$ & commutant, the matrices that commute with every $P_g$ & Sec.~\ref{sec:commutant}\\
$\varkappa$ & order of a group average & Sec.~\ref{sec:order}\\
$\mathcal Z^\perp$ & annihilator, the parities that are zero on every vector of $\mathcal Z$ & Sec.~\ref{sec:hashvar}\\
$U_S$ & parities constant on $S$; the affine hull is their solution set & Prop.~\ref{prop:hull}\\
$\Phi$, $C_u$, $Z$ & hash matrix (bank of XOR gates), cell, hashed count & Sec.~\ref{sec:hashvar}\\
$\Upsilon_r$, $D_S$, $\mathcal E(S)$ & cell-count probabilities, zero-sum 4-tuples, additive energy & Thm.~\ref{thm:hash4}\\
$\mathrm{cum}_j$, $\mathcal M_c$, $\phi_X$, $\mathsf p$ & cumulants, moment generating function, characteristic function on the group, probability mass function & Sec.~\ref{sec:hashvar}\\
$\mathrm{PR}(\mathfrak w)$, $D_2(f)$ & participation ratio of weights; of the Walsh power spectrum & Sec.~\ref{sec:pr}\\
$H_\alpha$, $q$ & R\'enyi entropy of order $\alpha$; normalized power spectrum & Sec.~\ref{sec:pr}\\
$\mathrm{NL}(f)$ & nonlinearity, distance to the nearest affine function & Sec.~\ref{sec:pr}\\
$o$, $\mathcal D_j(u)$, $\mathcal L_j(o)$ & variable order; Haar coefficient of a multiplexer; Haar level energy & Sec.~\ref{sec:haar}\\
$E_j$ & projection onto the Haar level $j$ & Prop.~\ref{prop:wreath}\\
$\mathcal S_f$ & permutation parts of the stabilizer in $B$ & Prop.~\ref{prop:cosets}\\
$M$, $s(a)$, $T_f$ & spectral map $A^{-\mathsf T}$; sign bit of $W_f(a)$; translations that fix $f$ & Sec.~\ref{sec:bgm}\\
$\mathbf X,\mathbf Y,\mathbf w,\iota$ & matrices, coefficients and eigenvalue of the DC-GEVP & Prop.~\ref{prop:dcgevp}\\
$\mathcal K$, $b_j$, $\mathcal A_j$, $O_j$ & sign system, base, candidate images and orbits of the matcher & Alg.~\ref{alg:bgm}\\
$\mathsf e_k$ & elementary symmetric function of degree $k$ & Sec.~\ref{sec:prelim}\\
$\mathfrak b$ & number of blocks of a bit stream & Sec.~\ref{sec:crypto}\\
$\mathfrak u$ & expected number of upsets per bit between two scrubs & Sec.~\ref{sec:lut}\\
$\mathfrak s$, $\mathfrak r$ & number of states of a machine, number of its state variables & Sec.~\ref{sec:early}\\
$\mathcal V$, $F_{\mathcal V}$ & care set of a cut; $[x\in\mathcal V](-1)^{f(x)}$ & Sec.~\ref{sec:approxmatch}\\
$d_{\mathcal V}(f,h)$ & mismatches of two functions on the care set & Sec.~\ref{sec:approxmatch}\\
$S(\cdot)$ & absolute spectrum sorted within each weight & Sec.~\ref{sec:approxmatch}\\
$\underline d$, $\underline d_{ij}$ & level bound on the distance; cofactor bound for input $i$ sent to $j$ & Prop.~\ref{prop:npndist}\\
$\mathcal U_\pi$ & correlation over the translations for the permutation $\pi$ & Prop.~\ref{prop:npndist}\\
$\mathfrak d$ & radius of a matching query, in mismatches & Alg.~\ref{alg:npndist}\\
$\mathbf D$ & diagonal Walsh form of the translation average & Prop.~\ref{prop:dcgevp}\\
$V$, $L$, $\mathcal B_V$ & early and late inputs; Walsh energy on the early subcube & Prop.~\ref{prop:early}\\
$\mathcal F$, $\mathfrak m$, $\mathcal N(d)$ & vector function (S-box), its output width, collisions at difference $d$ & Prop.~\ref{prop:crypto}\\
\bottomrule
\end{longtable}
\end{center}

\section{Sources of the Benchmark Circuits}
\label{app:sources}

Every circuit used in the census, the hashing experiments and the noisy-table experiments is
public. Table~\ref{tab:sources} gives, for each two-level file, its collection and how the
file relates to its public copy, as the same file byte for byte, a different cover of the same
functions, or the two-level form of a multilevel circuit. The last two relations were checked
by comparing binary decision diagrams output by output, with a negative control in which one
cube is removed. The Berkeley PLA test set is the collection distributed with the Espresso
minimizer \cite{brayton1984espresso}. The repositories are H: \url{https://github.com/ispras/hdl-benchmarks} at commit \texttt{10c3fb51}; E: \url{https://github.com/classabbyamp/espresso-logic} at commit \texttt{85265139}; L: \url{https://github.com/JackHack96/logic-synthesis} at commit \texttt{06845afe}. The ISCAS-85
netlists of the cut experiment come from the first repository, directory
\texttt{iscas85/bench}.

\begin{small}
\begin{longtable}{@{}lrrlll@{}}
\caption{Sources of the two-level benchmark files.}\label{tab:sources}\\\\
\toprule
circuit & $n$ & $m$ & collection & relation & public copy\\\\
\midrule
\endfirsthead
\toprule
circuit & $n$ & $m$ & collection & relation & public copy\\\\
\midrule
\endhead
\texttt{5xp1} & 7 & 10 & LGSynth91 & same file & H:\texttt{lgsynth91/pla/5xp1.pla}\\
\texttt{9sym} & 9 & 1 & LGSynth91 & same file & H:\texttt{lgsynth91/pla/9sym.pla}\\
\texttt{add6} & 12 & 7 & Berkeley test set & same file & L:\texttt{espresso/examples/math/add6}\\
\texttt{addm4} & 9 & 8 & Berkeley test set & same functions & L:\texttt{espresso/examples/math/addm4}\\
\texttt{alu1} & 12 & 8 & Berkeley test set & same file & E:\texttt{examples/alu1}\\
\texttt{alu2} & 10 & 8 & Berkeley test set & same file & E:\texttt{examples/alu2}\\
\texttt{alu2c} & 10 & 6 & LGSynth91 multilevel & two-level form & H:\texttt{lgsynth91/blif/alu2.blif}\\
\texttt{alu3} & 10 & 8 & Berkeley test set & same file & E:\texttt{examples/alu3}\\
\bottomrule
\end{longtable}
\end{small}

\section{Outputs That Depend on Fewer Linear Forms than Inputs}
\label{app:survey}

Table~\ref{tab:hidden} lists the 50 benchmark outputs whose linear support
(Proposition~\ref{thm:support}) has a smaller dimension than the number of inputs they use,
with the codimension of the affine hull of the on-set and the participation ratio.

\begin{longtable}{@{}lrrrrrr@{}}
\caption{Outputs with fewer linear forms than used inputs.}\label{tab:hidden}\\
\toprule
circuit & output & $n$ & used & forms & hull codim. & $D_2$\\
\midrule
\endfirsthead
\toprule
circuit & output & $n$ & used & forms & hull codim. & $D_2$\\
\midrule
\endhead
\texttt{5xp1} & 7 & 7 & 2 & 1 & 1 & 1.00\\
\texttt{9sym} & 0 & 9 & 9 & 8 & 0 & 5.72\\
\texttt{add6} & 0 & 12 & 2 & 1 & 1 & 1.00\\
\texttt{add6} & 1 & 12 & 4 & 3 & 0 & 4.00\\
\texttt{add6} & 2 & 12 & 6 & 5 & 0 & 6.40\\
\texttt{add6} & 3 & 12 & 8 & 7 & 0 & 6.92\\
\texttt{add6} & 4 & 12 & 10 & 9 & 0 & 6.99\\
\texttt{add6} & 5 & 12 & 12 & 11 & 0 & 7.00\\
\bottomrule
\end{longtable}

\section{Hashing Results for Every Output}
\label{app:hash}

Table~\ref{tab:hashall} gives, for each of the 277 outputs of
Section~\ref{sec:hash}, the number of solutions, the number $k$ of hash rows and the
relative standard deviation $\sqrt{\Var Z}/|S|$ of the random mean \eqref{eq:blind}, of
the median and the 99th percentile over random hashes, of the plug-in rule with 32 and
128 solutions, of the hull rule and of the best hash of the pool.

{\footnotesize
\setlength{\tabcolsep}{3pt}
\begin{longtable}{@{}lrrrrrrrrrrr@{}}
\caption{Relative standard deviation of the hashed count per output.}\label{tab:hashall}\\
\toprule
circuit & out & $n$ & $|S|$ & $k$ & mean & median & p99 & $m{=}32$ & $m{=}128$ & hull & best\\
\midrule
\endfirsthead
\toprule
circuit & out & $n$ & $|S|$ & $k$ & mean & median & p99 & $m{=}32$ & $m{=}128$ & hull & best\\
\midrule
\endhead
\texttt{add6} & 0 & 12 & 2048 & 6 & 0.124 & 0.000 & 1.000 & 0.000 & 0.000 & 0.000 & 0.000\\
\texttt{add6} & 1 & 12 & 2048 & 6 & 0.124 & 0.000 & 0.500 & 0.112 & 0.000 & 0.112 & 0.000\\
\texttt{add6} & 2 & 12 & 2048 & 6 & 0.124 & 0.000 & 0.559 & 0.040 & 0.000 & 0.088 & 0.000\\
\texttt{add6} & 3 & 12 & 2048 & 6 & 0.124 & 0.000 & 0.500 & 0.062 & 0.066 & 0.095 & 0.000\\
\texttt{add6} & 4 & 12 & 2048 & 6 & 0.124 & 0.062 & 0.504 & 0.100 & 0.051 & 0.154 & 0.000\\
\texttt{add6} & 5 & 12 & 2048 & 6 & 0.124 & 0.044 & 0.516 & 0.104 & 0.051 & 0.122 & 0.000\\
\texttt{add6} & 6 & 12 & 2016 & 5 & 0.088 & 0.016 & 0.509 & 0.074 & 0.028 & 0.072 & 0.000\\
\texttt{alu1} & 0 & 12 & 3072 & 6 & 0.072 & 0.000 & 0.333 & 0.105 & 0.026 & 0.059 & 0.000\\
\bottomrule
\end{longtable}}

\section{Every Parity Instance}
\label{app:parity}

Table~\ref{tab:parityall} lists every instance of Section~\ref{sec:parity} with the
number of variables and clauses, the verdict of the spectral elimination, the number
of parity equations it collected and the running times in seconds of the three
methods. ``limit'' marks a call stopped at the time limit and ``--'' a call not made.

{\footnotesize
\setlength{\tabcolsep}{3.5pt}
\begin{longtable}{@{}lrrrrlrrrr@{}}
\caption{Parity-structured instances.}\label{tab:parityall}\\
\toprule
family & size & seed & vars & clauses & verdict & eqs & spectral & CaDiCaL & CMS\\
\midrule
\endfirsthead
\toprule
family & size & seed & vars & clauses & verdict & eqs & spectral & CaDiCaL & CMS\\
\midrule
\endhead
tseitin & 20 & 0 & 30 & 80 & unsat & 20 & 0.001 & 0.002 & 0.015\\
tseitin & 20 & 1 & 30 & 80 & unsat & 20 & 0.001 & 0.001 & 0.013\\
tseitin & 20 & 2 & 30 & 80 & unsat & 20 & 0.001 & 0.001 & 0.013\\
tseitin & 40 & 0 & 60 & 160 & unsat & 40 & 0.002 & 0.006 & 0.021\\
tseitin & 40 & 1 & 60 & 160 & unsat & 40 & 0.002 & 0.006 & 0.016\\
tseitin & 40 & 2 & 60 & 160 & unsat & 40 & 0.002 & 0.012 & 0.018\\
tseitin & 60 & 0 & 90 & 240 & unsat & 60 & 0.003 & 0.150 & 0.126\\
tseitin & 60 & 1 & 90 & 240 & unsat & 60 & 0.003 & 0.610 & 0.357\\
\bottomrule
\end{longtable}}

\section{The Periodogram Collision under Negation and Permutation}
\label{app:classify}

Under negation and permutation of the inputs and output, the canonical periodogram of
four-variable functions leaves 4 classes together. Their representatives,
in algebraic normal form with addition over GF(2), are
\begin{itemize}
\item $f=x_{1} + x_{2} + x_{3} + x_{2}x_{3} + x_{4} + x_{1}x_{4}$
\item $f=1 + x_{1} + x_{2} + x_{2}x_{3} + x_{1}x_{4} + x_{3}x_{4}$
\item $f=x_{1} + x_{2}x_{3} + x_{4} + x_{1}x_{4} + x_{2}x_{4} + x_{3}x_{4}$
\item $f=x_{1} + x_{2} + x_{1}x_{2} + x_{3} + x_{1}x_{3} + x_{2}x_{3} + x_{4} + x_{1}x_{4} + x_{2}x_{4} + x_{3}x_{4}$
\end{itemize}
All are quadratic and bent, so their periodograms are flat (Corollary~\ref{cor:bent}),
and the bispectrum separates them.

\section{Orbit Voting for Every Output at Flip Rate 0.1}
\label{app:noisy}

Table~\ref{tab:noisyall} gives, for each output of Section~\ref{sec:noisy}, the number
of library elements that stabilize it, the mean orbit size of the group they generate,
the recall and precision of the accepted set, the bit error rate after the majority vote
and after the tested vote, and the rate predicted from the exact library stabilizer, at the
flip rate $\epsilon=0.1$.

{\footnotesize
\setlength{\tabcolsep}{3.5pt}
\begin{longtable}{@{}lrrrrrrrrr@{}}
\caption{Orbit voting per output.}\label{tab:noisyall}\\
\toprule
circuit & out & $n$ & stab. & orbit & recall & precision & majority & tested & predicted\\
\midrule
\endfirsthead
\toprule
circuit & out & $n$ & stab. & orbit & recall & precision & majority & tested & predicted\\
\midrule
\endhead
\texttt{5xp1} & 0 & 7 & 1 & 1.5 & 0.92 & 0.08 & 0.0872 & 0.0733 & 0.1000\\
\texttt{5xp1} & 3 & 7 & 1 & 2.0 & 0.86 & 0.17 & 0.1094 & 0.0998 & 0.1000\\
\texttt{5xp1} & 4 & 7 & 6 & 6.0 & 0.95 & 0.66 & 0.0609 & 0.0594 & 0.0154\\
\texttt{5xp1} & 5 & 7 & 13 & 12.0 & 0.99 & 0.82 & 0.0663 & 0.0378 & 0.0014\\
\texttt{5xp1} & 6 & 7 & 23 & 24.0 & 0.99 & 1.00 & 0.0000 & 0.0019 & 0.0000\\
\texttt{5xp1} & 7 & 7 & 38 & 64.0 & 1.00 & 1.00 & 0.0000 & 0.0000 & 0.0000\\
\texttt{5xp1} & 8 & 7 & 51 & 64.0 & 1.00 & 1.00 & 0.0000 & 0.0020 & 0.0000\\
\texttt{5xp1} & 9 & 7 & 5 & 5.6 & 0.94 & 0.17 & 0.0584 & 0.0469 & 0.0290\\
\bottomrule
\end{longtable}}

\section{Numerical Checks of the Theory}
\label{app:checks}

Every statement of Section~\ref{sec:theory} is checked numerically. The orbit counts of
Table~\ref{tab:commutant} come from enumerating the groups. The group-enumerated averages
of $T$, $B$, $\GL(n,2)$ and $\AGL(n,2)$ agree with the closed forms of
Propositions~\ref{prop:T} and~\ref{prop:collapse} to within $4.4\times10^{-16}$
for $n\le4$, and the average over translations and modulations returns the identity
exactly. The normalizer covariance \eqref{eq:normalizer}, the effect of adding an affine
function and the covariance of the bispectrum hold exactly on random functions of up to
six variables. Exactly 896 of the $2^{16}$ functions of four variables
have $R_T(f)=I$, and they are the functions with $D_2(f)=N$. The linear support
dimension equals $n-\log_2|\{d:r_f(d)=N\}|$ on random functions of $k$ linear forms
for $n\le10$, the hash identity \eqref{eq:hashvar} and the random mean \eqref{eq:blind}
hold exactly over all row spaces for $n=5,6$ and $k\le3$, and the prime-field orbit
counts of Remark~\ref{rem:prime} for $p=3$, for $\GL$ and $\AGL$ in that order, are 6 and 2 for $n=2$, 6 and 2 for $n=3$.

The Monte Carlo checks follow. Table~\ref{tab:pairmc} compares the variance of an
off-diagonal entry of $R_T$ with \eqref{eq:pairing} for $n=6$ and the cyclic lag average
of the same length.

\begin{table}[htbp]
\centering
\caption{Pairing, with the variance and mean of an off-diagonal entry of $R_T$ for $N=64$, and the
cyclic lag average.}\label{tab:pairmc}
\begin{tabular}{@{}rrrrr@{}}
\toprule
$\mu$ & variance & \eqref{eq:pairing} & mean & $\mu^2$\\
\midrule
0.0 & 0.03110 & 0.03125 & 0.0008 & 0.0000\\
0.4 & 0.03063 & 0.03045 & 0.1609 & 0.1600\\
0.8 & 0.01847 & 0.01845 & 0.6395 & 0.6400\\
\bottomrule
\end{tabular}
\quad
\begin{tabular}{@{}rrr@{}}
\toprule
cyclic lag & variance & predicted\\
\midrule
1 & 0.01555 & 0.01562\\
5 & 0.01556 & 0.01562\\
32 & 0.03055 & 0.03125\\
\bottomrule
\end{tabular}
\end{table}

The collision estimate \eqref{eq:pilot} is unbiased (Table~\ref{tab:pilotmc}, $n=10$,
$m=24$ solutions, on-sets with an implied parity), and orbit voting matches
\eqref{eq:vote}, ties included, on the symmetric function of weight three or four of
eight variables (Table~\ref{tab:votemc}).

\begin{table}[htbp]
\centering
\caption{Collision estimate of the relative hash variance, with the true value, mean estimate and
its standard error.}\label{tab:pilotmc}
\begin{tabular}{@{}lrrr@{}}
\toprule
hash rows (bit masks) & relative variance & mean estimate & s.e.\\
\midrule
{3, 12, 48} & 1.0219 & 1.0241 & 0.0013\\
{5, 10, 100} & 0.0180 & 0.0182 & 0.0010\\
{3, 12, 48} & 1.0069 & 1.0060 & 0.0012\\
{5, 10, 100} & 0.0082 & 0.0081 & 0.0009\\
\bottomrule
\end{tabular}
\end{table}

\begin{table}[htbp]
\centering
\caption{Orbit voting, with the exact error \eqref{eq:vote} and Monte Carlo error per orbit size.}\label{tab:votemc}
\begin{tabular}{@{}rrrr@{}}
\toprule
$\epsilon$ & orbit & exact & Monte Carlo\\
\midrule
0.1 & 1 & 0.10000 & 0.10195\\
0.1 & 8 & 0.00273 & 0.00230\\
0.1 & 28 & 0.00000 & 0.00000\\
0.1 & 56 & 0.00000 & 0.00000\\
0.1 & 70 & 0.00000 & 0.00000\\
0.1 & 56 & 0.00000 & 0.00000\\
0.1 & 28 & 0.00000 & 0.00000\\
0.1 & 8 & 0.00273 & 0.00295\\
0.1 & 1 & 0.10000 & 0.09910\\
0.3 & 1 & 0.30000 & 0.30240\\
0.3 & 8 & 0.12604 & 0.12723\\
0.3 & 28 & 0.01426 & 0.01515\\
0.3 & 56 & 0.00093 & 0.00093\\
0.3 & 70 & 0.00025 & 0.00013\\
0.3 & 56 & 0.00093 & 0.00108\\
0.3 & 28 & 0.01426 & 0.01400\\
0.3 & 8 & 0.12604 & 0.12437\\
0.3 & 1 & 0.30000 & 0.30210\\
\bottomrule
\end{tabular}
\end{table}

The shrinkage weight on an orbit (Proposition~\ref{prop:shrinkvote}) was checked by
enumerating every flip pattern on orbits of 2 to 10 points, every number of
ones up to half the orbit and the flip rates 0.01, 0.05, 0.1, 0.2, 0.3, 170
cases in all. The summed squared error matches \eqref{eq:shrinkrisk} to
$9.7\times10^{-14}$, the mean of $k\,m(1-m)$ matches $(k-1)\nu+B$ to
$7.8\times10^{-14}$, and the minimizer on a grid of step $5\times10^{-4}$
is within $2.5\times10^{-4}$ of \eqref{eq:betastar}. The tested vote on a
constant orbit (Proposition~\ref{prop:testedvote}) was checked by enumerating the number of
flips on orbits of 2 to 64 points at the same flip rates, 315
cases. The largest rejection rate is 0.0099 against the level
$\alpha=0.01$, and the error of the tested vote is within its
bound in every case (Table~\ref{tab:testedcheck}).

\begin{table}[htbp]
\centering
\caption{Tested vote on a constant orbit, with the rejection rate of the purity test, error of the
majority vote, error of the tested vote and the bound $e(k,\epsilon)+\alpha$.}\label{tab:testedcheck}
\begin{tabular}{@{}rrrrrr@{}}
\toprule
$\epsilon$ & orbit & rejection & majority & tested & bound\\
\midrule
0.05 & 3 & 0.0000 & 0.00725 & 0.00725 & 0.01725\\
0.1 & 3 & 0.0000 & 0.02800 & 0.02800 & 0.03800\\
0.2 & 3 & 0.0000 & 0.10400 & 0.10400 & 0.11400\\
0.05 & 4 & 0.0000 & 0.00725 & 0.00725 & 0.01725\\
0.1 & 4 & 0.0000 & 0.02800 & 0.02800 & 0.03800\\
0.2 & 4 & 0.0000 & 0.10400 & 0.10400 & 0.11400\\
0.05 & 8 & 0.0058 & 0.00019 & 0.00222 & 0.01019\\
0.1 & 8 & 0.0046 & 0.00273 & 0.00273 & 0.01273\\
0.2 & 8 & 0.0000 & 0.03334 & 0.03334 & 0.04334\\
0.05 & 16 & 0.0070 & 0.00000 & 0.00181 & 0.01000\\
0.1 & 16 & 0.0033 & 0.00003 & 0.00127 & 0.01003\\
0.2 & 16 & 0.0055 & 0.00424 & 0.00424 & 0.01424\\
0.05 & 32 & 0.0046 & 0.00000 & 0.00089 & 0.01000\\
0.1 & 32 & 0.0033 & 0.00000 & 0.00096 & 0.01000\\
0.2 & 32 & 0.0061 & 0.00009 & 0.00254 & 0.01009\\
0.05 & 64 & 0.0044 & 0.00000 & 0.00065 & 0.01000\\
0.1 & 64 & 0.0099 & 0.00000 & 0.00211 & 0.01000\\
0.2 & 64 & 0.0051 & 0.00000 & 0.00181 & 0.01000\\
\bottomrule
\end{tabular}
\end{table}

\section{Votes and Evaluations of the Observation Count}
\label{app:observations}
The bound and the monotonicity of Lemma~\ref{lem:monotone} were checked in exact rational
arithmetic for votes of up to $64$ copies and in double precision up to $512$ copies, at the
flip rates 0.001, 0.01, 0.02, 0.05, 0.1, 0.2, 0.3, 0.4. The closed-form orbit
numbers used in Fig.~\ref{fig:observations}(b) agree with Burnside's lemma on the enumerated
groups in 35 cases. Table~\ref{tab:copies} lists, for a
per-bit error of $10^{-6}$, the least vote size $L_\delta$, its bound from
Proposition~\ref{prop:observations}, and the number of noisy tables for orbits of $s$ points.

\begin{table}[htbp]
\centering
\caption{Vote size and number of noisy truth tables for a per-bit error of $10^{-6}$.}\label{tab:copies}
\begin{tabular}{@{}rrrrrrr@{}}
\toprule
$\epsilon$ & $L_\delta$ & bound & $s=1$ & $s=4$ & $s=16$ & $s=64$\\
\midrule
0.001 & 5 & 6 & 5 & 2 & 1 & 1\\
0.01 & 7 & 9 & 7 & 2 & 1 & 1\\
0.02 & 9 & 11 & 9 & 3 & 1 & 1\\
0.05 & 15 & 17 & 15 & 4 & 1 & 1\\
0.1 & 23 & 28 & 23 & 6 & 2 & 1\\
0.2 & 51 & 62 & 51 & 13 & 4 & 1\\
0.3 & 131 & 159 & 131 & 33 & 9 & 3\\
0.4 & 553 & 677 & 553 & 139 & 35 & 9\\
\bottomrule
\end{tabular}
\end{table}

\section{Cut Functions per Circuit}
\label{app:cuts}
Table~\ref{tab:cutcirc} lists, for each ISCAS-85 circuit, its inputs, outputs and gates and the
number of cuts with four, five and six leaves on which the node depends on all of its leaves.

\begin{table}[htbp]
\centering
\caption{Cuts of the ISCAS-85 circuits used in Section~\ref{sec:cuts}.}\label{tab:cutcirc}
\begin{tabular}{@{}lrrrrrr@{}}
\toprule
circuit & inputs & outputs & gates & $k=4$ & $k=5$ & $k=6$\\
\midrule
\texttt{c17} & 5 & 2 & 6 & 6 & 0 & 0\\
\texttt{c432} & 36 & 7 & 160 & 527 & 243 & 209\\
\texttt{c499} & 41 & 32 & 202 & 484 & 910 & 1152\\
\texttt{c880} & 60 & 26 & 383 & 1083 & 1582 & 1686\\
\texttt{c1355} & 41 & 32 & 546 & 3283 & 2783 & 1499\\
\texttt{c1908} & 33 & 25 & 880 & 2172 & 385 & 1472\\
\texttt{c2670} & 233 & 140 & 1193 & 4126 & 2035 & 3126\\
\texttt{c3540} & 50 & 22 & 1669 & 8870 & 6313 & 2935\\
\texttt{c5315} & 178 & 123 & 2307 & 12023 & 5889 & 2130\\
\texttt{c6288} & 32 & 32 & 2416 & 22317 & 14445 & 109\\
\texttt{c7552} & 207 & 108 & 3512 & 21395 & 4871 & 3029\\
\bottomrule
\end{tabular}
\end{table}

\section{Lookup-Table Protection by Input Count}
\label{app:lut}
Table~\ref{tab:lutall} gives, for each number of inputs and for the stabilizer with and without
the complementing elements, the share of cut instances whose stabilizer is nontrivial and the
mean ratios of the bit error and of the table error after voting to their values without
voting, weighted by the number of cut instances.

\begin{table}[htbp]
\centering
\small
\caption{Mean bit-error and table-error ratios after orbit voting of lookup-table contents.}\label{tab:lutall}
\setlength{\tabcolsep}{4pt}
\begin{tabular}{@{}rlrrrrrrr@{}}
\toprule
 & & & \multicolumn{2}{c}{$\epsilon=0.001$} & \multicolumn{2}{c}{$\epsilon=0.01$} & \multicolumn{2}{c}{$\epsilon=0.05$}\\
\cmidrule(lr){4-5}\cmidrule(lr){6-7}\cmidrule(l){8-9}
$k$ & group & nontrivial (\%) & bit & table & bit & table & bit & table\\
\midrule
4 & symmetries & 99.9 & 0.654 & 0.654 & 0.662 & 0.664 & 0.696 & 0.710\\
4 & with complementing & 100.0 & 0.589 & 0.589 & 0.597 & 0.600 & 0.633 & 0.650\\
5 & symmetries & 99.7 & 0.493 & 0.495 & 0.501 & 0.515 & 0.535 & 0.611\\
5 & with complementing & 99.7 & 0.470 & 0.472 & 0.478 & 0.493 & 0.511 & 0.588\\
6 & symmetries & 99.1 & 0.258 & 0.261 & 0.265 & 0.297 & 0.295 & 0.466\\
6 & with complementing & 99.1 & 0.255 & 0.258 & 0.261 & 0.293 & 0.290 & 0.459\\
\bottomrule
\end{tabular}
\end{table}

\section{Single-Estimate Failures of the Hashing Counter per Output}
\label{app:counter}
\begin{small}
\setlength{\tabcolsep}{3.5pt}
\begin{longtable}{@{}lrrrrrrrrrr@{}}
\caption{Percent of runs with a single estimate outside the tolerance, per output, for the
random rule (R), the hull rule with matched hashes (H) and its clustered variant (C).}\label{tab:counterall}\\
\toprule
 & & & & & \multicolumn{3}{c}{$c=1.8$} & \multicolumn{3}{c}{$c=1.25$}\\
circuit & out & $n$ & $|S|$ & $k$ & R & H & C & R & H & C\\
\midrule
\endfirsthead
\toprule
circuit & out & $n$ & $|S|$ & $k$ & R & H & C & R & H & C\\
\midrule
\endhead
\texttt{add6} & 0 & 12 & 2048 & 6 & 2 & 0 & 0 & 2 & 0 & 0\\
\texttt{add6} & 1 & 12 & 2048 & 6 & 4 & 0 & 2 & 10 & 2 & 7\\
\texttt{add6} & 2 & 12 & 2048 & 6 & 0 & 1 & 1 & 10 & 8 & 9\\
\texttt{add6} & 3 & 12 & 2048 & 6 & 0 & 2 & 2 & 4 & 7 & 9\\
\texttt{add6} & 4 & 12 & 2048 & 6 & 0 & 3 & 3 & 5 & 8 & 9\\
\texttt{add6} & 5 & 12 & 2048 & 6 & 1 & 1 & 3 & 8 & 8 & 10\\
\texttt{add6} & 6 & 12 & 2016 & 5 & 0 & 3 & 0 & 4 & 6 & 2\\
\texttt{alu1} & 0 & 12 & 3072 & 6 & 0 & 0 & 0 & 1 & 1 & 2\\
\texttt{alu1} & 1 & 12 & 3072 & 6 & 0 & 0 & 0 & 0 & 2 & 1\\
\texttt{alu1} & 2 & 12 & 3072 & 6 & 0 & 0 & 0 & 2 & 2 & 3\\
\texttt{alu1} & 3 & 12 & 3072 & 6 & 0 & 0 & 1 & 0 & 2 & 1\\
\texttt{alu1} & 4 & 12 & 1024 & 5 & 4 & 3 & 4 & 6 & 5 & 6\\
\texttt{alu1} & 5 & 12 & 1024 & 5 & 4 & 1 & 0 & 6 & 3 & 4\\
\texttt{alu1} & 6 & 12 & 1024 & 5 & 4 & 3 & 1 & 8 & 7 & 2\\
\texttt{alu1} & 7 & 12 & 512 & 4 & 1 & 0 & 0 & 1 & 0 & 0\\
\texttt{alu2} & 0 & 10 & 448 & 3 & 0 & 0 & 0 & 5 & 3 & 1\\
\texttt{alu2} & 1 & 10 & 336 & 3 & 4 & 1 & 0 & 9 & 4 & 0\\
\texttt{alu2c} & 0 & 10 & 536 & 4 & 0 & 0 & 0 & 7 & 7 & 11\\
\texttt{alu2c} & 1 & 10 & 534 & 4 & 0 & 0 & 0 & 5 & 5 & 8\\
\texttt{alu2c} & 2 & 10 & 512 & 4 & 3 & 0 & 0 & 3 & 0 & 0\\
\texttt{alu2c} & 3 & 10 & 256 & 3 & 2 & 0 & 0 & 2 & 0 & 0\\
\texttt{alu2c} & 5 & 10 & 256 & 3 & 1 & 0 & 0 & 1 & 0 & 0\\
\texttt{alu3} & 0 & 10 & 640 & 4 & 0 & 0 & 0 & 0 & 0 & 0\\
\texttt{alu3} & 1 & 10 & 704 & 4 & 0 & 0 & 0 & 1 & 2 & 2\\
\texttt{alu3} & 2 & 10 & 464 & 3 & 0 & 0 & 0 & 6 & 2 & 4\\
\texttt{alu3} & 4 & 10 & 682 & 4 & 0 & 0 & 0 & 1 & 1 & 2\\
\texttt{alu3} & 5 & 10 & 468 & 3 & 1 & 0 & 0 & 4 & 4 & 3\\
\texttt{alu3} & 6 & 10 & 680 & 4 & 0 & 1 & 2 & 3 & 5 & 5\\
\texttt{clip} & 0 & 9 & 256 & 3 & 0 & 2 & 0 & 0 & 2 & 0\\
\texttt{clip} & 1 & 9 & 256 & 3 & 1 & 2 & 2 & 7 & 8 & 10\\
\texttt{clip} & 2 & 9 & 256 & 3 & 0 & 0 & 1 & 6 & 7 & 3\\
\texttt{clip} & 3 & 9 & 256 & 3 & 0 & 1 & 0 & 4 & 5 & 8\\
\texttt{clip} & 4 & 9 & 256 & 3 & 2 & 0 & 0 & 4 & 2 & 0\\
\texttt{cm85a} & 0 & 11 & 1264 & 5 & 2 & 0 & 0 & 6 & 1 & 1\\
\texttt{cm85a} & 2 & 11 & 1264 & 5 & 1 & 0 & 0 & 1 & 0 & 1\\
\texttt{sao2} & 2 & 10 & 476 & 3 & 1 & 0 & 0 & 1 & 0 & 0\\
\texttt{sex} & 5 & 9 & 256 & 3 & 4 & 0 & 3 & 11 & 5 & 8\\
\texttt{x2} & 1 & 10 & 768 & 4 & 0 & 0 & 0 & 5 & 6 & 9\\
\texttt{x2} & 5 & 10 & 704 & 4 & 0 & 1 & 1 & 2 & 4 & 2\\
\texttt{x2} & 6 & 10 & 696 & 4 & 0 & 0 & 0 & 4 & 3 & 3\\
\texttt{alu4} & 0 & 14 & 9440 & 8 & 0 & 0 & 1 & 3 & 3 & 1\\
\texttt{alu4} & 1 & 14 & 8192 & 8 & 0 & 0 & 0 & 8 & 10 & 5\\
\texttt{alu4} & 2 & 14 & 9552 & 8 & 0 & 0 & 1 & 4 & 3 & 5\\
\texttt{alu4} & 3 & 14 & 8192 & 8 & 2 & 1 & 1 & 12 & 6 & 10\\
\texttt{alu4} & 4 & 14 & 8192 & 8 & 0 & 0 & 0 & 5 & 6 & 4\\
\texttt{alu4} & 5 & 14 & 8192 & 8 & 0 & 0 & 0 & 7 & 5 & 3\\
\texttt{alu4} & 6 & 14 & 8192 & 8 & 0 & 0 & 0 & 6 & 6 & 6\\
\texttt{alu4} & 7 & 14 & 2304 & 6 & 1 & 1 & 0 & 14 & 9 & 10\\
\texttt{b12} & 0 & 15 & 6144 & 7 & 0 & 1 & 1 & 4 & 5 & 4\\
\texttt{b12} & 1 & 15 & 7424 & 7 & 1 & 1 & 1 & 7 & 7 & 9\\
\texttt{b12} & 2 & 15 & 12800 & 8 & 3 & 3 & 2 & 6 & 5 & 2\\
\texttt{b12} & 3 & 15 & 18432 & 9 & 0 & 0 & 0 & 10 & 8 & 9\\
\texttt{b12} & 4 & 15 & 22528 & 9 & 0 & 0 & 0 & 3 & 6 & 5\\
\texttt{b12} & 5 & 15 & 31744 & 9 & 0 & 0 & 0 & 0 & 0 & 0\\
\texttt{b12} & 6 & 15 & 4864 & 7 & 2 & 0 & 0 & 10 & 3 & 5\\
\texttt{b12} & 7 & 15 & 28672 & 9 & 0 & 0 & 0 & 0 & 0 & 0\\
\texttt{b12} & 8 & 15 & 30464 & 9 & 0 & 0 & 0 & 0 & 0 & 0\\
\texttt{cordic} & 0 & 23 & 7806464 & 17 & 0 & 0 & 0 & 0 & 0 & 0\\
\texttt{cordic} & 1 & 23 & 827904 & 14 & 1 & 2 & 1 & 8 & 8 & 4\\
\texttt{cps} & 0 & 24 & 2032016 & 15 & 0 & 1 & 1 & 5 & 6 & 4\\
\texttt{cps} & 1 & 24 & 4326976 & 17 & 2 & 3 & 0 & 6 & 5 & 1\\
\texttt{cps} & 2 & 24 & 5310032 & 17 & 0 & 0 & 0 & 3 & 3 & 5\\
\texttt{cps} & 3 & 24 & 131072 & 12 & 2 & 0 & 0 & 2 & 0 & 0\\
\texttt{cps} & 4 & 24 & 131072 & 12 & 4 & 0 & 0 & 4 & 0 & 0\\
\texttt{cps} & 5 & 24 & 131072 & 12 & 1 & 0 & 0 & 1 & 0 & 0\\
\texttt{cps} & 6 & 24 & 131072 & 12 & 5 & 0 & 0 & 5 & 0 & 0\\
\texttt{cps} & 23 & 24 & 15728576 & 18 & 0 & 0 & 0 & 0 & 0 & 0\\
\texttt{cps} & 24 & 24 & 3802944 & 16 & 3 & 0 & 1 & 6 & 6 & 5\\
\texttt{cps} & 27 & 24 & 1704288 & 15 & 1 & 0 & 0 & 5 & 5 & 2\\
\texttt{cps} & 28 & 24 & 4325376 & 17 & 1 & 0 & 0 & 4 & 14 & 5\\
\texttt{cps} & 29 & 24 & 655360 & 14 & 4 & 1 & 0 & 6 & 1 & 0\\
\texttt{cps} & 30 & 24 & 524288 & 14 & 6 & 1 & 0 & 7 & 5 & 0\\
\texttt{cps} & 31 & 24 & 526528 & 14 & 4 & 1 & 3 & 5 & 6 & 3\\
\texttt{cps} & 33 & 24 & 786976 & 14 & 1 & 1 & 0 & 7 & 4 & 7\\
\texttt{cps} & 35 & 24 & 524664 & 14 & 4 & 4 & 1 & 7 & 7 & 2\\
\texttt{cps} & 36 & 24 & 262336 & 13 & 3 & 0 & 0 & 3 & 0 & 0\\
\texttt{cps} & 37 & 24 & 3605496 & 16 & 1 & 1 & 0 & 4 & 3 & 3\\
\texttt{cps} & 38 & 24 & 2426072 & 16 & 1 & 3 & 0 & 12 & 16 & 6\\
\texttt{cps} & 39 & 24 & 5375560 & 17 & 0 & 0 & 0 & 2 & 5 & 2\\
\texttt{cps} & 40 & 24 & 5572136 & 17 & 2 & 0 & 0 & 3 & 2 & 2\\
\texttt{cps} & 41 & 24 & 787000 & 14 & 1 & 3 & 0 & 8 & 12 & 3\\
\texttt{cps} & 42 & 24 & 4851712 & 17 & 1 & 0 & 0 & 4 & 5 & 6\\
\texttt{cps} & 43 & 24 & 393280 & 13 & 1 & 0 & 0 & 9 & 2 & 7\\
\texttt{cps} & 44 & 24 & 4851776 & 17 & 3 & 0 & 0 & 6 & 7 & 3\\
\texttt{cps} & 45 & 24 & 525248 & 14 & 3 & 0 & 0 & 3 & 0 & 0\\
\texttt{cps} & 46 & 24 & 131136 & 12 & 5 & 0 & 0 & 5 & 0 & 0\\
\texttt{cps} & 47 & 24 & 524800 & 14 & 3 & 4 & 0 & 9 & 12 & 4\\
\texttt{cps} & 50 & 24 & 2360192 & 16 & 3 & 0 & 0 & 6 & 2 & 5\\
\texttt{cps} & 51 & 24 & 786432 & 14 & 2 & 0 & 0 & 4 & 4 & 1\\
\texttt{cps} & 52 & 24 & 524864 & 14 & 3 & 1 & 2 & 10 & 10 & 5\\
\texttt{cps} & 53 & 24 & 2622336 & 16 & 3 & 0 & 1 & 7 & 3 & 6\\
\texttt{cps} & 54 & 24 & 2097600 & 16 & 0 & 1 & 0 & 2 & 3 & 1\\
\texttt{cps} & 55 & 24 & 262720 & 13 & 3 & 0 & 0 & 3 & 0 & 0\\
\texttt{cps} & 56 & 24 & 704 & 4 & 1 & 0 & 0 & 6 & 10 & 9\\
\texttt{cps} & 57 & 24 & 525184 & 14 & 4 & 0 & 0 & 4 & 0 & 0\\
\texttt{cps} & 59 & 24 & 1311168 & 15 & 3 & 3 & 0 & 12 & 13 & 6\\
\texttt{cps} & 65 & 24 & 1442432 & 15 & 4 & 0 & 0 & 12 & 9 & 2\\
\texttt{cps} & 66 & 24 & 1836352 & 15 & 3 & 0 & 0 & 7 & 7 & 4\\
\texttt{cps} & 67 & 24 & 384 & 3 & 1 & 0 & 0 & 6 & 7 & 1\\
\texttt{cps} & 68 & 24 & 262208 & 13 & 2 & 0 & 0 & 2 & 0 & 0\\
\texttt{cps} & 69 & 24 & 576 & 4 & 0 & 0 & 0 & 1 & 0 & 0\\
\texttt{cps} & 70 & 24 & 524928 & 14 & 5 & 0 & 0 & 5 & 0 & 0\\
\texttt{cps} & 71 & 24 & 256 & 3 & 2 & 0 & 0 & 2 & 0 & 0\\
\texttt{cps} & 72 & 24 & 1180224 & 15 & 3 & 0 & 1 & 9 & 4 & 7\\
\texttt{cps} & 73 & 24 & 1836352 & 15 & 1 & 2 & 0 & 7 & 4 & 1\\
\texttt{cps} & 74 & 24 & 262272 & 13 & 1 & 0 & 0 & 1 & 0 & 0\\
\texttt{cps} & 75 & 24 & 525184 & 14 & 4 & 0 & 0 & 4 & 0 & 0\\
\texttt{cps} & 77 & 24 & 1311232 & 15 & 4 & 1 & 0 & 6 & 3 & 2\\
\texttt{cps} & 78 & 24 & 787008 & 14 & 2 & 0 & 0 & 6 & 6 & 3\\
\texttt{cps} & 79 & 24 & 786752 & 14 & 1 & 0 & 0 & 6 & 7 & 9\\
\texttt{cps} & 80 & 24 & 786816 & 14 & 2 & 0 & 0 & 8 & 5 & 5\\
\texttt{cps} & 81 & 24 & 655680 & 14 & 2 & 1 & 0 & 16 & 5 & 5\\
\texttt{cps} & 82 & 24 & 1442368 & 15 & 1 & 0 & 0 & 5 & 3 & 2\\
\texttt{cps} & 83 & 24 & 786752 & 14 & 1 & 0 & 0 & 7 & 4 & 1\\
\texttt{cps} & 84 & 24 & 1024 & 5 & 5 & 4 & 3 & 10 & 7 & 10\\
\texttt{cps} & 85 & 24 & 262144 & 13 & 2 & 0 & 0 & 2 & 0 & 0\\
\texttt{cps} & 86 & 24 & 786624 & 14 & 1 & 0 & 0 & 6 & 3 & 2\\
\texttt{cps} & 87 & 24 & 524544 & 14 & 1 & 4 & 2 & 4 & 6 & 5\\
\texttt{cps} & 88 & 24 & 393408 & 13 & 2 & 0 & 0 & 9 & 4 & 2\\
\texttt{cps} & 89 & 24 & 16121856 & 18 & 0 & 0 & 0 & 0 & 0 & 0\\
\texttt{cps} & 90 & 24 & 131136 & 12 & 4 & 0 & 0 & 4 & 0 & 0\\
\texttt{cps} & 92 & 24 & 262144 & 13 & 4 & 0 & 0 & 4 & 0 & 0\\
\texttt{cps} & 93 & 24 & 393216 & 13 & 2 & 0 & 0 & 7 & 7 & 1\\
\texttt{cps} & 94 & 24 & 8388608 & 18 & 3 & 0 & 0 & 3 & 0 & 0\\
\texttt{cps} & 95 & 24 & 524288 & 14 & 1 & 0 & 0 & 1 & 0 & 0\\
\texttt{cps} & 96 & 24 & 131072 & 12 & 3 & 0 & 0 & 3 & 0 & 0\\
\texttt{cps} & 97 & 24 & 131072 & 12 & 2 & 0 & 0 & 2 & 0 & 0\\
\texttt{cps} & 98 & 24 & 393216 & 13 & 0 & 0 & 0 & 3 & 5 & 2\\
\texttt{cps} & 99 & 24 & 131072 & 12 & 3 & 0 & 0 & 3 & 0 & 0\\
\texttt{cps} & 100 & 24 & 786432 & 14 & 0 & 0 & 0 & 5 & 4 & 4\\
\texttt{duke2} & 0 & 22 & 364544 & 13 & 2 & 0 & 0 & 7 & 6 & 1\\
\texttt{duke2} & 1 & 22 & 39552 & 10 & 3 & 1 & 0 & 6 & 8 & 3\\
\texttt{duke2} & 2 & 22 & 438272 & 13 & 1 & 1 & 1 & 7 & 7 & 5\\
\texttt{duke2} & 3 & 22 & 524288 & 14 & 1 & 1 & 0 & 5 & 8 & 7\\
\texttt{duke2} & 4 & 22 & 1048576 & 15 & 3 & 0 & 0 & 3 & 0 & 0\\
\texttt{duke2} & 5 & 22 & 5120 & 7 & 2 & 0 & 1 & 6 & 9 & 3\\
\texttt{duke2} & 6 & 22 & 116864 & 11 & 0 & 1 & 0 & 5 & 2 & 1\\
\texttt{duke2} & 7 & 22 & 147456 & 12 & 1 & 0 & 0 & 2 & 2 & 0\\
\texttt{duke2} & 8 & 22 & 1048576 & 15 & 3 & 0 & 0 & 3 & 0 & 0\\
\texttt{duke2} & 9 & 22 & 304896 & 13 & 4 & 3 & 1 & 11 & 11 & 8\\
\texttt{duke2} & 10 & 22 & 524288 & 14 & 3 & 0 & 0 & 10 & 3 & 9\\
\texttt{duke2} & 11 & 22 & 32768 & 10 & 6 & 0 & 0 & 6 & 0 & 0\\
\texttt{duke2} & 12 & 22 & 121984 & 11 & 2 & 1 & 0 & 4 & 6 & 2\\
\texttt{duke2} & 13 & 22 & 393216 & 13 & 1 & 0 & 0 & 5 & 3 & 3\\
\texttt{duke2} & 14 & 22 & 389120 & 13 & 2 & 3 & 0 & 7 & 9 & 6\\
\texttt{duke2} & 15 & 22 & 65536 & 11 & 5 & 0 & 0 & 5 & 0 & 0\\
\texttt{duke2} & 16 & 22 & 119168 & 11 & 3 & 1 & 0 & 5 & 1 & 1\\
\texttt{duke2} & 17 & 22 & 108544 & 11 & 2 & 3 & 0 & 5 & 6 & 1\\
\texttt{duke2} & 18 & 22 & 19456 & 9 & 2 & 3 & 0 & 4 & 4 & 0\\
\texttt{duke2} & 19 & 22 & 65536 & 11 & 1 & 0 & 0 & 1 & 0 & 0\\
\texttt{duke2} & 20 & 22 & 27264 & 9 & 1 & 0 & 0 & 5 & 8 & 3\\
\texttt{duke2} & 21 & 22 & 81920 & 11 & 3 & 1 & 1 & 5 & 4 & 2\\
\texttt{duke2} & 22 & 22 & 131072 & 12 & 4 & 1 & 0 & 12 & 9 & 4\\
\texttt{duke2} & 23 & 22 & 32768 & 10 & 3 & 0 & 0 & 3 & 0 & 0\\
\texttt{duke2} & 24 & 22 & 65536 & 11 & 2 & 3 & 0 & 5 & 11 & 3\\
\texttt{duke2} & 25 & 22 & 65536 & 11 & 1 & 0 & 0 & 1 & 0 & 0\\
\texttt{duke2} & 26 & 22 & 1572864 & 15 & 0 & 0 & 0 & 4 & 4 & 2\\
\texttt{duke2} & 27 & 22 & 163840 & 12 & 0 & 1 & 0 & 4 & 8 & 3\\
\texttt{duke2} & 28 & 22 & 446208 & 13 & 1 & 1 & 0 & 6 & 3 & 3\\
\texttt{misex3} & 0 & 14 & 1536 & 5 & 1 & 0 & 0 & 13 & 7 & 13\\
\texttt{misex3} & 1 & 14 & 1536 & 5 & 2 & 0 & 0 & 12 & 8 & 5\\
\texttt{misex3} & 2 & 14 & 1536 & 5 & 0 & 0 & 0 & 3 & 9 & 10\\
\texttt{misex3} & 3 & 14 & 1536 & 5 & 2 & 0 & 0 & 15 & 2 & 7\\
\texttt{misex3} & 4 & 14 & 1536 & 5 & 1 & 0 & 0 & 6 & 7 & 9\\
\texttt{misex3} & 5 & 14 & 1536 & 5 & 2 & 0 & 2 & 13 & 6 & 9\\
\texttt{misex3} & 6 & 14 & 1536 & 5 & 1 & 0 & 0 & 8 & 11 & 12\\
\texttt{misex3} & 7 & 14 & 1536 & 5 & 0 & 0 & 2 & 8 & 8 & 13\\
\texttt{misex3} & 8 & 14 & 544 & 4 & 1 & 1 & 1 & 10 & 9 & 7\\
\texttt{misex3} & 9 & 14 & 1064 & 5 & 1 & 0 & 2 & 15 & 22 & 15\\
\texttt{misex3} & 13 & 14 & 9132 & 8 & 1 & 0 & 0 & 10 & 10 & 4\\
\texttt{misex3c} & 0 & 14 & 8704 & 8 & 1 & 0 & 0 & 6 & 6 & 5\\
\texttt{misex3c} & 1 & 14 & 9280 & 8 & 0 & 2 & 0 & 6 & 10 & 6\\
\texttt{misex3c} & 2 & 14 & 8512 & 8 & 0 & 1 & 0 & 6 & 8 & 5\\
\texttt{misex3c} & 3 & 14 & 9280 & 8 & 0 & 1 & 0 & 3 & 4 & 6\\
\texttt{misex3c} & 4 & 14 & 8192 & 8 & 1 & 0 & 0 & 3 & 9 & 4\\
\texttt{misex3c} & 5 & 14 & 7680 & 7 & 1 & 0 & 0 & 5 & 3 & 5\\
\texttt{misex3c} & 6 & 14 & 7936 & 7 & 2 & 0 & 0 & 2 & 1 & 1\\
\texttt{misex3c} & 7 & 14 & 7680 & 7 & 0 & 0 & 0 & 0 & 0 & 0\\
\texttt{misex3c} & 8 & 14 & 4128 & 7 & 2 & 0 & 0 & 2 & 0 & 0\\
\texttt{misex3c} & 9 & 14 & 7384 & 7 & 0 & 1 & 0 & 4 & 4 & 2\\
\texttt{misex3c} & 10 & 14 & 6744 & 7 & 0 & 1 & 0 & 2 & 2 & 4\\
\texttt{misex3c} & 11 & 14 & 2460 & 6 & 3 & 1 & 2 & 12 & 6 & 6\\
\texttt{misex3c} & 13 & 14 & 9132 & 8 & 1 & 0 & 1 & 9 & 5 & 4\\
\texttt{pdc} & 0 & 16 & 4696 & 7 & 0 & 0 & 0 & 1 & 1 & 0\\
\texttt{pdc} & 1 & 16 & 4694 & 7 & 0 & 0 & 0 & 4 & 2 & 0\\
\texttt{pdc} & 2 & 16 & 2348 & 6 & 1 & 1 & 0 & 1 & 2 & 1\\
\texttt{pdc} & 3 & 16 & 3139 & 6 & 2 & 4 & 0 & 5 & 7 & 6\\
\texttt{pdc} & 4 & 16 & 792 & 4 & 0 & 0 & 0 & 6 & 4 & 5\\
\texttt{pdc} & 5 & 16 & 1296 & 5 & 2 & 2 & 1 & 9 & 10 & 1\\
\texttt{pdc} & 6 & 16 & 8192 & 8 & 3 & 0 & 0 & 3 & 0 & 0\\
\texttt{pdc} & 7 & 16 & 2348 & 6 & 1 & 1 & 0 & 3 & 1 & 0\\
\texttt{pdc} & 8 & 16 & 3055 & 6 & 0 & 0 & 0 & 5 & 2 & 4\\
\texttt{pdc} & 9 & 16 & 708 & 4 & 2 & 1 & 0 & 12 & 11 & 8\\
\texttt{pdc} & 10 & 16 & 264 & 3 & 5 & 2 & 0 & 5 & 2 & 0\\
\texttt{pdc} & 12 & 16 & 256 & 3 & 1 & 0 & 0 & 1 & 0 & 0\\
\texttt{pdc} & 13 & 16 & 256 & 3 & 1 & 0 & 0 & 1 & 0 & 0\\
\texttt{pdc} & 14 & 16 & 16384 & 9 & 4 & 0 & 0 & 4 & 0 & 0\\
\texttt{pdc} & 15 & 16 & 24576 & 9 & 0 & 0 & 0 & 4 & 7 & 2\\
\texttt{pdc} & 16 & 16 & 872 & 4 & 2 & 0 & 0 & 8 & 1 & 1\\
\texttt{pdc} & 17 & 16 & 872 & 4 & 2 & 1 & 1 & 4 & 4 & 5\\
\texttt{pdc} & 18 & 16 & 872 & 4 & 1 & 0 & 1 & 3 & 3 & 1\\
\texttt{pdc} & 19 & 16 & 872 & 4 & 0 & 1 & 0 & 1 & 6 & 3\\
\texttt{pdc} & 20 & 16 & 704 & 4 & 1 & 1 & 0 & 2 & 4 & 1\\
\texttt{pdc} & 21 & 16 & 384 & 3 & 2 & 0 & 0 & 4 & 3 & 3\\
\texttt{pdc} & 23 & 16 & 789 & 4 & 1 & 0 & 0 & 5 & 4 & 1\\
\texttt{pdc} & 25 & 16 & 512 & 4 & 4 & 0 & 0 & 4 & 0 & 0\\
\texttt{pdc} & 27 & 16 & 256 & 3 & 3 & 0 & 0 & 3 & 0 & 0\\
\texttt{pdc} & 28 & 16 & 5019 & 7 & 2 & 2 & 0 & 4 & 3 & 0\\
\texttt{pdc} & 29 & 16 & 5019 & 7 & 4 & 4 & 1 & 4 & 5 & 2\\
\texttt{pdc} & 30 & 16 & 9270 & 8 & 0 & 2 & 1 & 2 & 6 & 1\\
\texttt{pdc} & 31 & 16 & 288 & 3 & 4 & 1 & 1 & 15 & 9 & 8\\
\texttt{pdc} & 32 & 16 & 1792 & 5 & 1 & 3 & 1 & 2 & 5 & 3\\
\texttt{pdc} & 33 & 16 & 4185 & 7 & 4 & 0 & 0 & 4 & 0 & 0\\
\texttt{pdc} & 34 & 16 & 4697 & 7 & 3 & 5 & 3 & 3 & 5 & 4\\
\texttt{pdc} & 36 & 16 & 5514 & 7 & 1 & 3 & 1 & 9 & 9 & 5\\
\texttt{pdc} & 37 & 16 & 4833 & 7 & 3 & 0 & 0 & 4 & 1 & 0\\
\texttt{pdc} & 38 & 16 & 512 & 4 & 3 & 0 & 0 & 3 & 0 & 0\\
\texttt{pdc} & 39 & 16 & 512 & 4 & 3 & 0 & 0 & 3 & 0 & 0\\
\texttt{spla} & 0 & 16 & 16385 & 9 & 1 & 0 & 0 & 1 & 0 & 0\\
\texttt{spla} & 1 & 16 & 24577 & 9 & 0 & 0 & 0 & 7 & 4 & 4\\
\texttt{spla} & 3 & 16 & 4632 & 7 & 1 & 1 & 0 & 2 & 2 & 0\\
\texttt{spla} & 4 & 16 & 4630 & 7 & 2 & 0 & 0 & 2 & 0 & 0\\
\texttt{spla} & 5 & 16 & 3333 & 6 & 2 & 1 & 0 & 3 & 5 & 1\\
\texttt{spla} & 6 & 16 & 2316 & 6 & 1 & 0 & 0 & 1 & 0 & 0\\
\texttt{spla} & 7 & 16 & 1018 & 4 & 0 & 0 & 0 & 8 & 6 & 5\\
\texttt{spla} & 8 & 16 & 8192 & 8 & 3 & 0 & 0 & 3 & 0 & 0\\
\texttt{spla} & 9 & 16 & 1700 & 5 & 1 & 0 & 0 & 5 & 5 & 6\\
\texttt{spla} & 10 & 16 & 3249 & 6 & 3 & 1 & 1 & 4 & 6 & 4\\
\texttt{spla} & 11 & 16 & 2316 & 6 & 3 & 0 & 0 & 4 & 1 & 0\\
\texttt{spla} & 12 & 16 & 934 & 4 & 0 & 1 & 1 & 8 & 4 & 9\\
\texttt{spla} & 13 & 16 & 384 & 3 & 1 & 0 & 0 & 2 & 4 & 2\\
\texttt{spla} & 14 & 16 & 256 & 3 & 2 & 0 & 0 & 2 & 0 & 0\\
\texttt{spla} & 17 & 16 & 256 & 3 & 2 & 0 & 0 & 2 & 0 & 0\\
\texttt{spla} & 18 & 16 & 256 & 3 & 2 & 0 & 0 & 2 & 0 & 0\\
\texttt{spla} & 23 & 16 & 9270 & 8 & 2 & 0 & 0 & 4 & 1 & 0\\
\texttt{spla} & 24 & 16 & 512 & 4 & 3 & 0 & 0 & 3 & 0 & 0\\
\texttt{spla} & 25 & 16 & 512 & 4 & 5 & 0 & 0 & 5 & 0 & 0\\
\texttt{spla} & 26 & 16 & 4891 & 7 & 3 & 1 & 0 & 4 & 2 & 0\\
\texttt{spla} & 27 & 16 & 4891 & 7 & 2 & 2 & 0 & 3 & 2 & 1\\
\texttt{spla} & 28 & 16 & 4121 & 7 & 5 & 0 & 0 & 5 & 0 & 0\\
\texttt{spla} & 30 & 16 & 4633 & 7 & 3 & 3 & 1 & 4 & 3 & 1\\
\texttt{spla} & 31 & 16 & 1792 & 5 & 1 & 0 & 0 & 2 & 4 & 2\\
\texttt{spla} & 32 & 16 & 5418 & 7 & 2 & 0 & 0 & 6 & 2 & 4\\
\texttt{spla} & 33 & 16 & 4641 & 7 & 3 & 0 & 0 & 3 & 1 & 0\\
\texttt{spla} & 34 & 16 & 512 & 4 & 5 & 0 & 0 & 5 & 0 & 0\\
\texttt{spla} & 35 & 16 & 1180 & 5 & 3 & 0 & 0 & 4 & 3 & 1\\
\texttt{spla} & 36 & 16 & 1180 & 5 & 4 & 0 & 1 & 8 & 2 & 5\\
\texttt{spla} & 37 & 16 & 1132 & 5 & 2 & 0 & 1 & 7 & 1 & 10\\
\texttt{spla} & 38 & 16 & 1132 & 5 & 4 & 1 & 1 & 7 & 5 & 3\\
\texttt{spla} & 42 & 16 & 256 & 3 & 1 & 0 & 0 & 1 & 0 & 0\\
\texttt{spla} & 43 & 16 & 448 & 3 & 1 & 0 & 0 & 6 & 3 & 1\\
\texttt{spla} & 44 & 16 & 704 & 4 & 2 & 0 & 2 & 6 & 3 & 3\\
\texttt{spla} & 45 & 16 & 789 & 4 & 2 & 0 & 0 & 6 & 3 & 1\\
\texttt{t481} & 0 & 16 & 42016 & 10 & 0 & 0 & 0 & 6 & 2 & 2\\
\texttt{table3} & 0 & 14 & 1484 & 5 & 3 & 2 & 1 & 5 & 6 & 1\\
\texttt{table3} & 2 & 14 & 1305 & 5 & 1 & 0 & 0 & 12 & 3 & 11\\
\texttt{table3} & 3 & 14 & 935 & 4 & 0 & 1 & 1 & 1 & 7 & 3\\
\texttt{table3} & 4 & 14 & 1803 & 5 & 1 & 2 & 2 & 5 & 6 & 5\\
\texttt{table3} & 5 & 14 & 895 & 4 & 1 & 3 & 1 & 5 & 7 & 3\\
\texttt{table3} & 10 & 14 & 915 & 4 & 1 & 0 & 0 & 6 & 3 & 1\\
\texttt{table3} & 11 & 14 & 1997 & 5 & 0 & 1 & 0 & 4 & 4 & 3\\
\texttt{table3} & 12 & 14 & 503 & 3 & 0 & 0 & 0 & 9 & 7 & 9\\
\texttt{table3} & 13 & 14 & 906 & 4 & 2 & 1 & 0 & 3 & 2 & 3\\
\texttt{table5} & 1 & 17 & 11765 & 8 & 1 & 0 & 0 & 5 & 7 & 5\\
\texttt{table5} & 2 & 17 & 3645 & 6 & 1 & 0 & 0 & 3 & 11 & 2\\
\texttt{table5} & 3 & 17 & 12920 & 8 & 0 & 1 & 0 & 6 & 8 & 4\\
\texttt{table5} & 4 & 17 & 6334 & 7 & 3 & 0 & 2 & 7 & 2 & 5\\
\texttt{table5} & 5 & 17 & 16513 & 9 & 6 & 0 & 1 & 18 & 12 & 10\\
\texttt{table5} & 6 & 17 & 11054 & 8 & 3 & 2 & 0 & 11 & 8 & 8\\
\texttt{table5} & 7 & 17 & 2306 & 6 & 2 & 2 & 0 & 9 & 11 & 5\\
\texttt{table5} & 8 & 17 & 2306 & 6 & 1 & 0 & 0 & 11 & 7 & 2\\
\texttt{table5} & 9 & 17 & 2362 & 6 & 0 & 1 & 0 & 6 & 7 & 6\\
\texttt{table5} & 10 & 17 & 2644 & 6 & 1 & 3 & 1 & 7 & 17 & 6\\
\texttt{table5} & 11 & 17 & 8996 & 8 & 2 & 1 & 0 & 6 & 7 & 8\\
\texttt{table5} & 12 & 17 & 7670 & 7 & 0 & 2 & 0 & 0 & 7 & 2\\
\texttt{table5} & 13 & 17 & 13867 & 8 & 2 & 4 & 0 & 7 & 9 & 2\\
\texttt{table5} & 14 & 17 & 17025 & 9 & 1 & 1 & 2 & 13 & 10 & 7\\
\bottomrule
\end{longtable}
\end{small}

\section{Misses of One Call of the Prefix Counter per Output}
\label{app:core}
\begin{small}
\setlength{\tabcolsep}{3.5pt}
\begin{longtable}{@{}lrrrrrrrr@{}}
\caption{Percent of the 200 calls per output and rule whose count lies outside
$[|S|/1.8,\,1.8|S|]$, for random rows (R), rows matched to 32 uniform solutions (U), to 32 clustered
solutions (C) and to the exact affine hull (E), where $d$ is the dimension of the affine hull of the
on-set.}\label{tab:coreall}\\
\toprule
circuit & out & $n$ & $|S|$ & $d$ & R & U & C & E\\
\midrule
\endfirsthead
\toprule
circuit & out & $n$ & $|S|$ & $d$ & R & U & C & E\\
\midrule
\endhead
\texttt{add6} & 0 & 12 & 2048 & 11 & 1.0 & 0.0 & 0.0 & 0.0\\
\texttt{add6} & 1 & 12 & 2048 & 12 & 1.0 & 1.5 & 0.0 & 1.5\\
\texttt{add6} & 2 & 12 & 2048 & 12 & 0.0 & 0.0 & 0.0 & 0.0\\
\texttt{add6} & 3 & 12 & 2048 & 12 & 2.0 & 0.5 & 0.5 & 1.0\\
\texttt{add6} & 4 & 12 & 2048 & 12 & 0.0 & 0.5 & 0.0 & 0.5\\
\texttt{add6} & 5 & 12 & 2048 & 12 & 0.0 & 0.5 & 0.5 & 0.0\\
\texttt{add6} & 6 & 12 & 2016 & 12 & 0.0 & 0.0 & 0.5 & 1.5\\
\texttt{alu1} & 0 & 12 & 3072 & 12 & 0.0 & 0.5 & 0.0 & 0.0\\
\texttt{alu1} & 1 & 12 & 3072 & 12 & 0.0 & 0.0 & 0.0 & 0.0\\
\texttt{alu1} & 2 & 12 & 3072 & 12 & 0.0 & 0.0 & 0.0 & 0.0\\
\texttt{alu1} & 3 & 12 & 3072 & 12 & 0.0 & 0.0 & 0.0 & 0.0\\
\texttt{alu1} & 4 & 12 & 1024 & 11 & 2.0 & 0.5 & 1.0 & 2.0\\
\texttt{alu1} & 5 & 12 & 1024 & 11 & 2.5 & 2.5 & 2.5 & 3.0\\
\texttt{alu1} & 6 & 12 & 1024 & 11 & 2.5 & 1.0 & 1.0 & 2.5\\
\texttt{alu1} & 7 & 12 & 512 & 9 & 1.5 & 0.0 & 0.0 & 0.0\\
\texttt{alu2} & 0 & 10 & 448 & 10 & 1.5 & 0.5 & 0.0 & 0.5\\
\texttt{alu2} & 1 & 10 & 336 & 10 & 1.0 & 1.0 & 1.0 & 0.5\\
\texttt{alu2c} & 0 & 10 & 536 & 10 & 0.0 & 0.0 & 0.0 & 0.0\\
\texttt{alu2c} & 1 & 10 & 534 & 10 & 0.0 & 0.0 & 0.0 & 0.0\\
\texttt{alu2c} & 2 & 10 & 512 & 9 & 0.5 & 0.0 & 0.0 & 0.0\\
\texttt{alu2c} & 3 & 10 & 256 & 8 & 1.5 & 0.0 & 0.0 & 0.0\\
\texttt{alu2c} & 5 & 10 & 256 & 8 & 1.0 & 0.0 & 0.0 & 0.0\\
\texttt{alu3} & 0 & 10 & 640 & 10 & 0.0 & 0.0 & 0.0 & 0.0\\
\texttt{alu3} & 1 & 10 & 704 & 10 & 0.0 & 0.0 & 0.0 & 0.0\\
\texttt{alu3} & 2 & 10 & 464 & 10 & 0.0 & 0.0 & 0.0 & 0.0\\
\texttt{alu3} & 4 & 10 & 682 & 10 & 1.5 & 0.5 & 0.0 & 0.5\\
\texttt{alu3} & 5 & 10 & 468 & 10 & 0.0 & 1.0 & 0.5 & 0.0\\
\texttt{alu3} & 6 & 10 & 680 & 10 & 0.5 & 1.5 & 0.0 & 0.5\\
\texttt{clip} & 0 & 9 & 256 & 9 & 1.0 & 0.5 & 0.0 & 0.0\\
\texttt{clip} & 1 & 9 & 256 & 9 & 0.5 & 0.0 & 0.0 & 0.0\\
\texttt{clip} & 2 & 9 & 256 & 9 & 0.0 & 0.0 & 0.0 & 0.0\\
\texttt{clip} & 3 & 9 & 256 & 9 & 0.0 & 0.0 & 0.5 & 0.0\\
\texttt{clip} & 4 & 9 & 256 & 9 & 0.5 & 0.0 & 1.0 & 1.5\\
\texttt{cm85a} & 0 & 11 & 1264 & 11 & 0.0 & 2.0 & 0.5 & 1.5\\
\texttt{cm85a} & 2 & 11 & 1264 & 11 & 1.0 & 0.0 & 0.0 & 0.5\\
\texttt{sao2} & 2 & 10 & 476 & 9 & 1.0 & 0.0 & 0.0 & 0.0\\
\texttt{sex} & 5 & 9 & 256 & 9 & 0.0 & 1.5 & 1.0 & 0.0\\
\texttt{x2} & 1 & 10 & 768 & 10 & 0.0 & 0.0 & 0.0 & 0.0\\
\texttt{x2} & 5 & 10 & 704 & 10 & 0.5 & 1.0 & 1.5 & 0.5\\
\texttt{x2} & 6 & 10 & 696 & 10 & 0.0 & 0.5 & 1.0 & 0.5\\
\texttt{alu4} & 0 & 14 & 9440 & 14 & 0.5 & 0.0 & 0.0 & 1.0\\
\texttt{alu4} & 1 & 14 & 8192 & 14 & 0.0 & 0.0 & 0.0 & 0.0\\
\texttt{alu4} & 2 & 14 & 9552 & 14 & 0.0 & 0.0 & 0.0 & 0.0\\
\texttt{alu4} & 3 & 14 & 8192 & 14 & 0.0 & 0.0 & 0.5 & 0.0\\
\texttt{alu4} & 4 & 14 & 8192 & 14 & 0.0 & 0.0 & 0.0 & 0.0\\
\texttt{alu4} & 5 & 14 & 8192 & 14 & 0.0 & 0.0 & 0.0 & 0.0\\
\texttt{alu4} & 6 & 14 & 8192 & 14 & 0.0 & 0.0 & 0.0 & 0.0\\
\texttt{alu4} & 7 & 14 & 2304 & 14 & 0.0 & 0.5 & 0.5 & 1.0\\
\texttt{b12} & 0 & 15 & 6144 & 14 & 0.0 & 1.5 & 0.5 & 1.5\\
\texttt{b12} & 1 & 15 & 7424 & 14 & 0.5 & 1.0 & 0.5 & 0.5\\
\texttt{b12} & 2 & 15 & 12800 & 15 & 0.5 & 0.5 & 0.0 & 0.5\\
\texttt{b12} & 3 & 15 & 18432 & 15 & 0.0 & 0.0 & 0.0 & 0.0\\
\texttt{b12} & 4 & 15 & 22528 & 15 & 0.5 & 0.0 & 0.0 & 0.0\\
\texttt{b12} & 5 & 15 & 31744 & 15 & 0.0 & 0.0 & 0.0 & 0.0\\
\texttt{b12} & 6 & 15 & 4864 & 13 & 1.0 & 0.0 & 0.5 & 0.0\\
\texttt{b12} & 7 & 15 & 28672 & 15 & 0.0 & 0.0 & 0.0 & 0.0\\
\texttt{b12} & 8 & 15 & 30464 & 15 & 0.0 & 0.0 & 0.0 & 0.0\\
\texttt{cps} & 3 & 24 & 131072 & 17 & 1.0 & 0.0 & 0.0 & 0.0\\
\texttt{cps} & 4 & 24 & 131072 & 17 & 1.0 & 0.0 & 0.0 & 0.0\\
\texttt{cps} & 5 & 24 & 131072 & 17 & 0.5 & 0.0 & 0.0 & 0.0\\
\texttt{cps} & 6 & 24 & 131072 & 17 & 1.0 & 0.0 & 0.0 & 0.0\\
\texttt{cps} & 56 & 24 & 704 & 10 & 1.5 & 0.0 & 0.0 & 0.0\\
\texttt{cps} & 67 & 24 & 384 & 9 & 4.0 & 0.0 & 0.0 & 0.0\\
\texttt{cps} & 69 & 24 & 576 & 10 & 2.5 & 0.5 & 0.0 & 0.0\\
\texttt{cps} & 71 & 24 & 256 & 8 & 1.5 & 0.0 & 0.0 & 0.0\\
\texttt{cps} & 84 & 24 & 1024 & 12 & 2.5 & 1.5 & 0.0 & 2.5\\
\texttt{cps} & 96 & 24 & 131072 & 17 & 1.0 & 0.0 & 0.0 & 0.0\\
\texttt{cps} & 97 & 24 & 131072 & 17 & 2.0 & 0.0 & 0.0 & 0.0\\
\texttt{cps} & 99 & 24 & 131072 & 17 & 1.5 & 0.0 & 0.0 & 0.0\\
\texttt{duke2} & 1 & 22 & 39552 & 20 & 2.0 & 0.5 & 1.5 & 1.0\\
\texttt{duke2} & 5 & 22 & 5120 & 14 & 3.0 & 1.0 & 0.0 & 1.0\\
\texttt{duke2} & 6 & 22 & 116864 & 20 & 1.0 & 1.0 & 1.0 & 1.0\\
\texttt{duke2} & 11 & 22 & 32768 & 15 & 1.5 & 0.0 & 0.0 & 0.0\\
\texttt{duke2} & 12 & 22 & 121984 & 20 & 1.0 & 0.0 & 0.5 & 1.5\\
\texttt{duke2} & 15 & 22 & 65536 & 16 & 2.5 & 0.0 & 0.0 & 0.0\\
\texttt{duke2} & 16 & 22 & 119168 & 19 & 0.5 & 0.5 & 0.0 & 2.0\\
\texttt{duke2} & 17 & 22 & 108544 & 19 & 0.5 & 2.0 & 0.0 & 1.0\\
\texttt{duke2} & 18 & 22 & 19456 & 17 & 3.0 & 2.5 & 0.5 & 1.0\\
\texttt{duke2} & 19 & 22 & 65536 & 16 & 0.5 & 0.0 & 0.0 & 0.0\\
\texttt{duke2} & 20 & 22 & 27264 & 20 & 2.0 & 2.0 & 0.0 & 1.0\\
\texttt{duke2} & 21 & 22 & 81920 & 18 & 1.5 & 1.5 & 0.0 & 2.5\\
\texttt{duke2} & 22 & 22 & 131072 & 19 & 2.5 & 2.0 & 0.0 & 1.5\\
\texttt{duke2} & 23 & 22 & 32768 & 15 & 0.5 & 0.0 & 0.0 & 0.0\\
\texttt{duke2} & 24 & 22 & 65536 & 20 & 0.0 & 0.0 & 0.0 & 1.0\\
\texttt{duke2} & 25 & 22 & 65536 & 16 & 1.0 & 0.0 & 0.0 & 0.0\\
\texttt{misex3} & 0 & 14 & 1536 & 14 & 0.0 & 0.5 & 0.0 & 0.0\\
\texttt{misex3} & 1 & 14 & 1536 & 14 & 1.0 & 0.0 & 0.5 & 1.0\\
\texttt{misex3} & 2 & 14 & 1536 & 14 & 0.0 & 0.5 & 0.5 & 0.5\\
\texttt{misex3} & 3 & 14 & 1536 & 14 & 0.0 & 0.5 & 0.5 & 0.0\\
\texttt{misex3} & 4 & 14 & 1536 & 14 & 0.5 & 0.5 & 0.5 & 1.0\\
\texttt{misex3} & 5 & 14 & 1536 & 14 & 1.5 & 0.0 & 0.5 & 0.5\\
\texttt{misex3} & 6 & 14 & 1536 & 14 & 0.0 & 0.0 & 0.5 & 0.5\\
\texttt{misex3} & 7 & 14 & 1536 & 13 & 0.5 & 0.0 & 0.5 & 0.5\\
\texttt{misex3} & 8 & 14 & 544 & 12 & 0.0 & 0.0 & 0.0 & 0.0\\
\texttt{misex3} & 9 & 14 & 1064 & 14 & 1.5 & 1.0 & 0.0 & 0.5\\
\texttt{misex3} & 13 & 14 & 9132 & 14 & 0.0 & 0.5 & 0.0 & 0.0\\
\texttt{misex3c} & 0 & 14 & 8704 & 14 & 0.0 & 0.0 & 0.0 & 0.0\\
\texttt{misex3c} & 1 & 14 & 9280 & 14 & 0.0 & 0.0 & 0.0 & 0.0\\
\texttt{misex3c} & 2 & 14 & 8512 & 14 & 0.5 & 0.0 & 2.0 & 1.0\\
\texttt{misex3c} & 3 & 14 & 9280 & 14 & 1.0 & 0.0 & 0.0 & 0.0\\
\texttt{misex3c} & 4 & 14 & 8192 & 14 & 0.0 & 1.0 & 0.0 & 0.0\\
\texttt{misex3c} & 5 & 14 & 7680 & 14 & 0.0 & 0.0 & 0.0 & 1.0\\
\texttt{misex3c} & 6 & 14 & 7936 & 14 & 0.5 & 0.0 & 0.0 & 1.0\\
\texttt{misex3c} & 7 & 14 & 7680 & 13 & 0.5 & 0.0 & 0.0 & 0.0\\
\texttt{misex3c} & 8 & 14 & 4128 & 13 & 1.5 & 0.5 & 0.0 & 1.0\\
\texttt{misex3c} & 9 & 14 & 7384 & 14 & 0.0 & 0.0 & 1.0 & 0.0\\
\texttt{misex3c} & 10 & 14 & 6744 & 14 & 1.0 & 1.0 & 1.5 & 0.0\\
\texttt{misex3c} & 11 & 14 & 2460 & 14 & 0.5 & 0.0 & 0.0 & 0.0\\
\texttt{misex3c} & 13 & 14 & 9132 & 14 & 0.0 & 0.0 & 0.0 & 0.0\\
\texttt{pdc} & 0 & 16 & 4696 & 13 & 0.5 & 0.5 & 0.0 & 1.0\\
\texttt{pdc} & 1 & 16 & 4694 & 13 & 0.5 & 0.5 & 0.0 & 1.0\\
\texttt{pdc} & 2 & 16 & 2348 & 12 & 0.5 & 1.5 & 0.0 & 0.5\\
\texttt{pdc} & 3 & 16 & 3139 & 15 & 1.0 & 1.5 & 1.0 & 1.0\\
\texttt{pdc} & 4 & 16 & 792 & 12 & 1.0 & 1.0 & 0.5 & 0.0\\
\texttt{pdc} & 5 & 16 & 1296 & 12 & 3.0 & 0.0 & 0.5 & 1.0\\
\texttt{pdc} & 6 & 16 & 8192 & 13 & 1.0 & 0.0 & 0.0 & 0.0\\
\texttt{pdc} & 7 & 16 & 2348 & 12 & 1.5 & 0.5 & 0.5 & 0.5\\
\texttt{pdc} & 8 & 16 & 3055 & 15 & 3.0 & 2.5 & 0.5 & 3.0\\
\texttt{pdc} & 9 & 16 & 708 & 12 & 3.5 & 0.5 & 0.5 & 0.5\\
\texttt{pdc} & 10 & 16 & 264 & 9 & 1.5 & 1.0 & 0.0 & 1.0\\
\texttt{pdc} & 12 & 16 & 256 & 8 & 1.0 & 0.0 & 0.0 & 0.0\\
\texttt{pdc} & 13 & 16 & 256 & 8 & 0.5 & 0.0 & 0.0 & 0.0\\
\texttt{pdc} & 14 & 16 & 16384 & 14 & 0.0 & 0.0 & 0.0 & 0.0\\
\texttt{pdc} & 15 & 16 & 24576 & 15 & 1.0 & 0.0 & 0.0 & 0.0\\
\texttt{pdc} & 16 & 16 & 872 & 11 & 2.5 & 0.5 & 0.0 & 0.5\\
\texttt{pdc} & 17 & 16 & 872 & 11 & 1.5 & 0.5 & 0.5 & 0.0\\
\texttt{pdc} & 18 & 16 & 872 & 11 & 1.5 & 0.5 & 0.0 & 0.5\\
\texttt{pdc} & 19 & 16 & 872 & 11 & 1.0 & 1.5 & 0.0 & 1.0\\
\texttt{pdc} & 20 & 16 & 704 & 10 & 1.0 & 1.0 & 0.0 & 0.0\\
\texttt{pdc} & 21 & 16 & 384 & 9 & 0.0 & 0.0 & 0.0 & 0.0\\
\texttt{pdc} & 23 & 16 & 789 & 13 & 0.5 & 0.5 & 0.5 & 1.0\\
\texttt{pdc} & 25 & 16 & 512 & 9 & 0.5 & 0.0 & 0.0 & 0.0\\
\texttt{pdc} & 27 & 16 & 256 & 8 & 1.0 & 0.0 & 0.0 & 0.0\\
\texttt{pdc} & 28 & 16 & 5019 & 15 & 2.5 & 2.0 & 0.0 & 2.0\\
\texttt{pdc} & 29 & 16 & 5019 & 15 & 2.5 & 1.0 & 0.0 & 3.0\\
\texttt{pdc} & 30 & 16 & 9270 & 14 & 0.5 & 0.0 & 0.0 & 0.5\\
\texttt{pdc} & 31 & 16 & 288 & 11 & 3.0 & 1.0 & 0.5 & 1.0\\
\texttt{pdc} & 32 & 16 & 1792 & 13 & 0.0 & 0.0 & 0.0 & 1.0\\
\texttt{pdc} & 33 & 16 & 4185 & 14 & 1.0 & 0.0 & 0.0 & 0.5\\
\texttt{pdc} & 34 & 16 & 4697 & 15 & 2.0 & 3.5 & 0.5 & 2.0\\
\texttt{pdc} & 36 & 16 & 5514 & 15 & 2.5 & 2.0 & 0.0 & 1.0\\
\texttt{pdc} & 37 & 16 & 4833 & 14 & 1.0 & 1.0 & 0.0 & 1.5\\
\texttt{pdc} & 38 & 16 & 512 & 9 & 1.5 & 0.0 & 0.0 & 0.0\\
\texttt{pdc} & 39 & 16 & 512 & 9 & 2.5 & 0.0 & 0.0 & 0.0\\
\texttt{spla} & 0 & 16 & 16385 & 15 & 2.0 & 0.0 & 0.0 & 0.5\\
\texttt{spla} & 1 & 16 & 24577 & 15 & 1.0 & 0.0 & 0.0 & 0.0\\
\texttt{spla} & 3 & 16 & 4632 & 13 & 2.0 & 1.0 & 0.5 & 1.5\\
\texttt{spla} & 4 & 16 & 4630 & 13 & 2.0 & 0.5 & 0.0 & 0.5\\
\texttt{spla} & 5 & 16 & 3333 & 15 & 2.5 & 1.5 & 0.5 & 1.0\\
\texttt{spla} & 6 & 16 & 2316 & 12 & 0.5 & 0.5 & 0.0 & 1.0\\
\texttt{spla} & 7 & 16 & 1018 & 12 & 0.5 & 0.0 & 0.0 & 0.5\\
\texttt{spla} & 8 & 16 & 8192 & 13 & 3.0 & 0.0 & 0.0 & 0.0\\
\texttt{spla} & 9 & 16 & 1700 & 12 & 1.5 & 0.0 & 0.5 & 0.5\\
\texttt{spla} & 10 & 16 & 3249 & 15 & 1.0 & 4.0 & 0.0 & 3.0\\
\texttt{spla} & 11 & 16 & 2316 & 12 & 1.5 & 1.0 & 0.0 & 0.5\\
\texttt{spla} & 12 & 16 & 934 & 12 & 1.5 & 0.5 & 0.0 & 0.0\\
\texttt{spla} & 13 & 16 & 384 & 9 & 0.5 & 0.0 & 0.0 & 0.0\\
\texttt{spla} & 14 & 16 & 256 & 8 & 1.5 & 0.0 & 0.0 & 0.0\\
\texttt{spla} & 17 & 16 & 256 & 8 & 1.0 & 0.0 & 0.0 & 0.0\\
\texttt{spla} & 18 & 16 & 256 & 8 & 1.0 & 0.0 & 0.0 & 0.0\\
\texttt{spla} & 23 & 16 & 9270 & 14 & 1.0 & 0.5 & 0.0 & 0.0\\
\texttt{spla} & 24 & 16 & 512 & 9 & 5.0 & 0.0 & 0.0 & 0.0\\
\texttt{spla} & 25 & 16 & 512 & 9 & 1.5 & 0.0 & 0.0 & 0.0\\
\texttt{spla} & 26 & 16 & 4891 & 15 & 1.5 & 1.0 & 0.5 & 1.5\\
\texttt{spla} & 27 & 16 & 4891 & 15 & 4.0 & 2.5 & 0.5 & 3.0\\
\texttt{spla} & 28 & 16 & 4121 & 14 & 1.5 & 0.0 & 0.0 & 1.5\\
\texttt{spla} & 30 & 16 & 4633 & 15 & 1.5 & 0.5 & 0.0 & 2.5\\
\texttt{spla} & 31 & 16 & 1792 & 13 & 1.5 & 0.5 & 0.0 & 0.5\\
\texttt{spla} & 32 & 16 & 5418 & 15 & 2.0 & 0.5 & 0.0 & 2.0\\
\texttt{spla} & 33 & 16 & 4641 & 13 & 2.0 & 0.0 & 0.0 & 0.5\\
\texttt{spla} & 34 & 16 & 512 & 9 & 1.0 & 0.0 & 0.0 & 0.0\\
\texttt{spla} & 35 & 16 & 1180 & 11 & 1.0 & 0.0 & 0.5 & 0.0\\
\texttt{spla} & 36 & 16 & 1180 & 11 & 1.5 & 1.0 & 0.5 & 0.0\\
\texttt{spla} & 37 & 16 & 1132 & 11 & 0.5 & 0.5 & 0.0 & 0.5\\
\texttt{spla} & 38 & 16 & 1132 & 11 & 2.0 & 0.5 & 0.0 & 0.0\\
\texttt{spla} & 42 & 16 & 256 & 8 & 0.5 & 0.0 & 0.0 & 0.0\\
\texttt{spla} & 43 & 16 & 448 & 10 & 1.5 & 0.0 & 0.0 & 0.5\\
\texttt{spla} & 44 & 16 & 704 & 11 & 2.0 & 1.0 & 0.5 & 0.5\\
\texttt{spla} & 45 & 16 & 789 & 13 & 2.0 & 1.0 & 0.0 & 0.5\\
\texttt{t481} & 0 & 16 & 42016 & 16 & 0.0 & 0.0 & 0.0 & 0.0\\
\texttt{table3} & 0 & 14 & 1484 & 14 & 1.0 & 3.5 & 0.5 & 2.5\\
\texttt{table3} & 2 & 14 & 1305 & 14 & 0.5 & 0.5 & 0.5 & 3.5\\
\texttt{table3} & 3 & 14 & 935 & 14 & 1.0 & 0.0 & 0.0 & 1.5\\
\texttt{table3} & 4 & 14 & 1803 & 14 & 1.0 & 3.0 & 1.5 & 3.5\\
\texttt{table3} & 5 & 14 & 895 & 14 & 1.5 & 0.5 & 1.0 & 1.0\\
\texttt{table3} & 10 & 14 & 915 & 14 & 1.0 & 1.5 & 1.0 & 2.0\\
\texttt{table3} & 11 & 14 & 1997 & 14 & 1.5 & 1.5 & 0.0 & 0.5\\
\texttt{table3} & 12 & 14 & 503 & 14 & 0.5 & 0.0 & 1.5 & 0.5\\
\texttt{table3} & 13 & 14 & 906 & 14 & 1.0 & 0.0 & 0.0 & 1.0\\
\texttt{table5} & 1 & 17 & 11765 & 17 & 1.0 & 1.5 & 0.5 & 0.5\\
\texttt{table5} & 2 & 17 & 3645 & 17 & 1.5 & 0.0 & 0.5 & 0.5\\
\texttt{table5} & 3 & 17 & 12920 & 17 & 2.0 & 0.0 & 0.0 & 0.0\\
\texttt{table5} & 4 & 17 & 6334 & 17 & 2.0 & 3.0 & 0.5 & 2.5\\
\texttt{table5} & 5 & 17 & 16513 & 17 & 1.5 & 1.0 & 1.0 & 1.5\\
\texttt{table5} & 6 & 17 & 11054 & 17 & 0.5 & 2.0 & 0.5 & 0.5\\
\texttt{table5} & 7 & 17 & 2306 & 14 & 1.5 & 0.0 & 0.0 & 1.0\\
\texttt{table5} & 8 & 17 & 2306 & 14 & 0.0 & 1.5 & 0.5 & 1.0\\
\texttt{table5} & 9 & 17 & 2362 & 16 & 2.5 & 1.0 & 0.5 & 1.0\\
\texttt{table5} & 10 & 17 & 2644 & 17 & 1.5 & 2.5 & 0.5 & 1.5\\
\texttt{table5} & 11 & 17 & 8996 & 17 & 2.0 & 0.5 & 1.0 & 1.0\\
\texttt{table5} & 12 & 17 & 7670 & 17 & 0.0 & 1.5 & 0.0 & 0.5\\
\texttt{table5} & 13 & 17 & 13867 & 17 & 1.0 & 0.0 & 0.5 & 1.0\\
\texttt{table5} & 14 & 17 & 17025 & 17 & 0.5 & 1.5 & 0.0 & 2.0\\
\bottomrule
\end{longtable}
\end{small}

\section{The Counter on Each Formula}
\label{app:satcount}
\begin{small}
\setlength{\tabcolsep}{3pt}
\begin{longtable}{@{}lrrrrrrrr@{}}
\caption{Each counted formula of the ApproxMC benchmark set, with the size $n$ of the sampling set, dimension
$d$ of the certified affine hull, $\log_2$ of the exact count, solver calls for the hull, and for
random (R) and matched (M) rows the calls of the core outside $[|S|/1.8,\,1.8|S|]$ or over the time
limit of 30 seconds, of 9, and the mean seconds per finished call.}\label{tab:satall}\\
\toprule
formula & $n$ & $d$ & $\log_2|S|$ & hull & R miss & R s & M miss & M s\\
\midrule
\endfirsthead
\toprule
formula & $n$ & $d$ & $\log_2|S|$ & hull & R miss & R s & M miss & M s\\
\midrule
\endhead
\texttt{or-50-5-2} & 100 & 100 & 50.0 & 90 & 9/9 & -- & 9/9 & --\\
\texttt{or-50-10-2-UC-20} & 100 & 51 & 25.4 & 44 & 0/9 & 0.04 & 0/9 & 0.03\\
\texttt{or-60-20-7-UC-30} & 120 & 81 & 37.6 & 73 & 0/9 & 3.26 & 0/9 & 2.64\\
\texttt{or-70-20-4-UC-20} & 140 & 114 & 55.2 & 105 & 9/9 & -- & 9/9 & --\\
\texttt{or-60-20-7-UC-40} & 120 & 49 & 26.7 & 40 & 0/9 & 0.03 & 0/9 & 0.03\\
\texttt{blasted\_case209} & 122 & 120 & 88.0 & 107 & 0/9 & 6.09 & 0/9 & 5.21\\
\texttt{or-70-20-3} & 140 & 140 & 70.0 & 132 & 9/9 & -- & 9/9 & --\\
\texttt{or-100-5-7-UC-30} & 200 & 88 & 44.0 & 82 & 0/9 & 9.85 & 0/9 & 4.63\\
\texttt{or-100-10-10} & 200 & 200 & 100.0 & 189 & 9/9 & -- & 9/9 & --\\
\texttt{11A-1} & 60 & 16 & 10.6 & 3 & 0/9 & 0.12 & 0/9 & 0.11\\
\texttt{blasted\_case64} & 19 & 17 & 15.1 & 8 & 0/9 & 0.02 & 0/9 & 0.01\\
\texttt{s298\_7\_4} & 17 & 17 & 16.0 & 8 & 0/9 & 0.01 & 0/9 & 0.01\\
\texttt{leader\_sync4\_32.pm\_5steps\_7int\_5fr} & 100 & 100 & 100.0 & 90 & 0/9 & 0.45 & 0/9 & 0.44\\
\texttt{blasted\_case140} & 137 & 135 & 103.0 & 127 & 1/9 & 2.69 & 0/9 & 5.70\\
\texttt{blasted\_case21} & 16 & 14 & 13.0 & 7 & 0/9 & 0.01 & 0/9 & 0.01\\
\texttt{blasted\_squaring22} & 61 & 57 & 23.0 & 46 & 0/9 & 0.61 & 0/9 & 0.45\\
\texttt{13A-2} & 70 & 32 & 12.9 & 19 & 0/9 & 0.45 & 0/9 & 0.42\\
\texttt{leader\_sync4\_11.pm\_5steps\_7int\_7fr} & 140 & 140 & 139.5 & 134 & 0/9 & 0.37 & 0/9 & 0.37\\
\texttt{leader\_sync4\_32.pm\_5steps\_7int\_5fr} & 100 & 100 & 100.0 & 90 & 0/9 & 0.45 & 0/9 & 0.45\\
\texttt{blasted\_case\_1\_b14\_1} & 45 & 43 & 42.0 & 35 & 0/9 & 0.01 & 1/9 & 0.01\\
\texttt{s526a\_15\_7} & 24 & 24 & 23.0 & 15 & 0/9 & 0.01 & 0/9 & 0.02\\
\texttt{blasted\_case58} & 19 & 17 & 15.1 & 10 & 0/9 & 0.02 & 0/9 & 0.02\\
\texttt{blasted\_case122} & 27 & 25 & 24.0 & 16 & 0/9 & 0.01 & 0/9 & 0.01\\
\texttt{s344\_7\_4} & 24 & 24 & 23.0 & 15 & 0/9 & 0.01 & 0/9 & 0.01\\
\texttt{blasted\_case\_2\_ptb\_1} & 80 & 78 & 77.0 & 67 & 0/9 & 0.02 & 0/9 & 0.02\\
\texttt{s420\_new1\_15\_7} & 34 & 34 & 33.0 & 27 & 0/9 & 0.02 & 0/9 & 0.01\\
\texttt{sort.sk\_8\_52} & 52 & 51 & 36.3 & 39 & 0/9 & 0.32 & 0/9 & 0.32\\
\texttt{s1423a\_7\_4} & 91 & 91 & 90.6 & 83 & 0/9 & 0.02 & 0/9 & 0.02\\
\texttt{leader\_sync4\_11.pm\_5steps\_7int\_5fr} & 100 & 100 & 97.8 & 86 & 0/9 & 0.29 & 0/9 & 0.29\\
\texttt{17.sk\_3\_45} & 45 & 37 & 37.0 & 30 & 0/9 & 0.05 & 0/9 & 0.05\\
\texttt{blasted\_TR\_b14\_1\_linear} & 204 & 202 & 103.0 & 194 & 9/9 & -- & 9/9 & --\\
\texttt{7.sk\_4\_50} & 50 & 40 & 40.0 & 35 & 0/9 & 0.04 & 0/9 & 0.04\\
\texttt{32.sk\_4\_38} & 38 & 30 & 30.0 & 25 & 1/9 & 0.02 & 0/9 & 0.02\\
\texttt{leader\_sync4\_11.pm\_5steps\_7int\_9fr} & 180 & 180 & 174.7 & 173 & 0/9 & 1.48 & 0/9 & 1.46\\
\texttt{reverse.sk\_11\_258} & 258 & 92 & 92.0 & 87 & 0/9 & 0.44 & 0/9 & 0.46\\
\texttt{s344\_15\_7} & 24 & 24 & 23.0 & 17 & 0/9 & 0.01 & 1/9 & 0.01\\
\texttt{s420\_3\_2} & 34 & 34 & 33.0 & 27 & 0/9 & 0.01 & 0/9 & 0.01\\
\texttt{leader\_sync4\_11.pm\_5steps\_7int\_9fr} & 180 & 180 & 174.5 & 164 & 0/9 & 1.64 & 0/9 & 1.69\\
\texttt{blasted\_squaring25} & 61 & 57 & 24.8 & 46 & 0/9 & 0.60 & 0/9 & 0.46\\
\texttt{09A-1} & 60 & 27 & 13.0 & 10 & 0/9 & 0.13 & 0/9 & 0.13\\
\texttt{enqueueSeqSK.sk\_10\_42} & 42 & 41 & 31.4 & 25 & 1/9 & 0.44 & 0/9 & 0.46\\
\texttt{19.sk\_3\_48} & 48 & 43 & 40.4 & 37 & 0/9 & 0.03 & 0/9 & 0.03\\
\texttt{LoginService.sk\_20\_34} & 34 & 29 & 19.5 & 20 & 0/9 & 0.07 & 1/9 & 0.07\\
\texttt{s510\_15\_7} & 25 & 25 & 23.8 & 18 & 0/9 & 0.01 & 0/9 & 0.01\\
\texttt{blasted\_squaring51} & 42 & 40 & 24.0 & 21 & 0/9 & 0.09 & 0/9 & 0.09\\
\texttt{leader\_sync4\_11.pm\_5steps\_7int\_6fr} & 120 & 120 & 119.0 & 112 & 0/9 & 0.36 & 0/9 & 0.39\\
\texttt{4step} & 165 & 36 & 16.4 & 30 & 0/9 & 0.02 & 0/9 & 0.02\\
\texttt{leader\_sync4\_11.pm\_5steps\_7int\_8fr} & 160 & 160 & 159.7 & 151 & 0/9 & 0.45 & 0/9 & 0.45\\
\texttt{tire-2} & 550 & 93 & 39.4 & 88 & 9/9 & -- & 0/9 & 2.13\\
\texttt{min-1s} & 32 & 32 & 31.0 & 26 & 0/9 & 0.01 & 1/9 & 0.01\\
\texttt{herman3.pm\_20steps\_6int\_1fract\_sta} & 60 & 60 & 60.0 & 53 & 0/9 & 0.12 & 0/9 & 0.12\\
\texttt{tire-1} & 352 & 75 & 29.4 & 70 & 0/9 & 0.84 & 0/9 & 0.12\\
\texttt{13A-3} & 70 & 31 & 11.7 & 10 & 0/9 & 0.66 & 0/9 & 0.67\\
\texttt{blasted\_squaring8} & 72 & 70 & 38.0 & 59 & 0/9 & 0.53 & 0/9 & 0.56\\
\texttt{blasted\_squaring4} & 72 & 70 & 36.0 & 58 & 0/9 & 0.50 & 0/9 & 0.48\\
\texttt{blasted\_squaring21} & 61 & 57 & 23.0 & 47 & 0/9 & 0.69 & 0/9 & 0.44\\
\texttt{herman3.pm\_20steps\_6int\_1fract\_sta} & 60 & 60 & 60.0 & 53 & 0/9 & 0.12 & 0/9 & 0.12\\
\texttt{blasted\_squaring70} & 100 & 98 & 66.0 & 91 & 0/9 & 6.40 & 0/9 & 5.77\\
\texttt{blasted\_squaring16} & 72 & 70 & 40.8 & 51 & 0/9 & 0.92 & 0/9 & 1.00\\
\texttt{07A-2} & 55 & 24 & 10.3 & 7 & 0/9 & 0.11 & 0/9 & 0.12\\
\texttt{min-8s} & 144 & 144 & 140.8 & 137 & 0/9 & 0.04 & 0/9 & 0.04\\
\texttt{min-4s} & 80 & 80 & 77.7 & 73 & 0/9 & 0.02 & 0/9 & 0.02\\
\texttt{08A-1} & 55 & 24 & 11.4 & 6 & 0/9 & 0.10 & 0/9 & 0.10\\
\texttt{10A-1} & 65 & 18 & 11.6 & 4 & 1/9 & 0.04 & 0/9 & 0.04\\
\texttt{12A-1} & 60 & 16 & 10.6 & 3 & 0/9 & 0.13 & 1/9 & 0.12\\
\texttt{5step} & 177 & 34 & 16.3 & 23 & 0/9 & 0.01 & 0/9 & 0.02\\
\texttt{blasted\_TR\_b12\_1\_linear} & 99 & 97 & 59.9 & 89 & 9/9 & -- & 9/9 & --\\
\texttt{tire-4} & 812 & 130 & 46.6 & 125 & 9/9 & -- & 9/9 & --\\
\texttt{min-2s} & 48 & 48 & 46.4 & 39 & 0/9 & 0.01 & 0/9 & 0.01\\
\texttt{tire-3} & 577 & 145 & 37.7 & 140 & 0/9 & 14.14 & 0/9 & 3.53\\
\texttt{min-6s} & 112 & 112 & 109.2 & 105 & 0/9 & 0.03 & 0/9 & 0.03\\
\texttt{min-3s} & 64 & 64 & 62.0 & 57 & 0/9 & 0.02 & 0/9 & 0.02\\
\texttt{blasted\_TR\_device\_1\_linear} & 151 & 145 & 92.0 & 132 & 9/9 & -- & 9/9 & --\\
\texttt{blasted\_TR\_b14\_2\_linear} & 204 & 202 & 103.0 & 195 & 9/9 & -- & 9/9 & --\\
\bottomrule
\end{longtable}
\end{small}


\begin{thebibliography}{100}
\providecommand{\url}[1]{#1}
\csname url@samestyle\endcsname
\providecommand{\newblock}{\relax}
\providecommand{\bibinfo}[2]{#2}
\providecommand{\BIBentrySTDinterwordspacing}{\spaceskip=0pt\relax}
\providecommand{\BIBentryALTinterwordstretchfactor}{4}
\providecommand{\BIBentryALTinterwordspacing}{\spaceskip=\fontdimen2\font plus
\BIBentryALTinterwordstretchfactor\fontdimen3\font minus
  \fontdimen4\font\relax}
\providecommand{\BIBforeignlanguage}[2]{{%
\expandafter\ifx\csname l@#1\endcsname\relax
\typeout{** WARNING: IEEEtran.bst: No hyphenation pattern has been}%
\typeout{** loaded for the language `#1'. Using the pattern for}%
\typeout{** the default language instead.}%
\else
\language=\csname l@#1\endcsname
\fi
#2}}
\providecommand{\BIBdecl}{\relax}
\BIBdecl

\bibitem{valiant1986unique}
L.~G. Valiant and V.~V. Vazirani, ``{NP} is as easy as detecting unique
  solutions,'' \emph{Theoretical Computer Science}, vol.~47, no.~1, pp. 85--93,
  1986.

\bibitem{stockmeyer1983}
L.~Stockmeyer, ``The complexity of approximate counting,'' in \emph{Proceedings
  of the Fifteenth Annual {ACM} Symposium on Theory of Computing ({STOC}
  '83)}.\hskip 1em plus 0.5em minus 0.4em\relax ACM, 1983, pp. 118--126.

\bibitem{gomes2006xor}
C.~P. Gomes, A.~Sabharwal, and B.~Selman, ``Model counting: A new strategy for
  obtaining good bounds,'' in \emph{Proceedings of the 21st National Conference
  on Artificial Intelligence ({AAAI}-06)}.\hskip 1em plus 0.5em minus
  0.4em\relax Boston, MA: AAAI Press, Jul. 2006, pp. 54--61.

\bibitem{chakraborty2013approxmc}
S.~Chakraborty, K.~S. Meel, and M.~Y. Vardi, ``A scalable approximate model
  counter,'' in \emph{Principles and Practice of Constraint Programming ({CP}
  2013)}, ser. Lecture Notes in Computer Science, C.~Schulte, Ed., vol.
  8124.\hskip 1em plus 0.5em minus 0.4em\relax Springer, 2013, pp. 200--216.

\bibitem{thornton2026versionspace}
M.~A. Thornton, ``A version space approach for digital circuit analysis,''
  2026, arXiv:2609.00609.

\bibitem{tseitin1983complexity}
G.~S. Tseitin, ``On the complexity of derivation in propositional calculus,''
  in \emph{Automation of Reasoning 2: Classical Papers on Computational Logic
  1967--1970}, J.~Siekmann and G.~Wrightson, Eds.\hskip 1em plus 0.5em minus
  0.4em\relax Berlin, Heidelberg: Springer, 1983, pp. 466--483, {E}nglish
  translation of the 1968 Russian original.

\bibitem{urquhart1987hard}
A.~Urquhart, ``Hard examples for resolution,'' \emph{Journal of the {ACM}},
  vol.~34, no.~1, pp. 209--219, Jan. 1987.

\bibitem{bensasson2001narrow}
E.~Ben-Sasson and A.~Wigderson, ``Short proofs are narrow---resolution made
  simple,'' \emph{Journal of the {ACM}}, vol.~48, no.~2, pp. 149--169, Mar.
  2001.

\bibitem{mohnke1993}
J.~Mohnke and S.~Malik, ``Permutation and phase independent {B}oolean
  comparison,'' \emph{Integration, the {VLSI} Journal}, vol.~16, no.~2, pp.
  109--129, 1993.

\bibitem{chai2006matcher}
D.~Chai and A.~Kuehlmann, ``Building a better {B}oolean matcher and symmetry
  detector,'' in \emph{Proceedings of the Design, Automation and Test in Europe
  Conference ({DATE} '06)}, vol.~1.\hskip 1em plus 0.5em minus 0.4em\relax
  IEEE, Mar. 2006, pp. 1--6.

\bibitem{abdollahi2008}
A.~Abdollahi and M.~Pedram, ``Symmetry detection and {B}oolean matching
  utilizing a signature-based canonical form of {B}oolean functions,''
  \emph{IEEE Transactions on Computer-Aided Design of Integrated Circuits and
  Systems}, vol.~27, no.~6, pp. 1128--1137, Jun. 2008.

\bibitem{mishchenko2006dag}
A.~Mishchenko, S.~Chatterjee, and R.~K. Brayton, ``{DAG}-aware {AIG} rewriting:
  A fresh look at combinational logic synthesis,'' in \emph{Proc. 43rd Design
  Automation Conf. (DAC)}, 2006, pp. 532--535.

\bibitem{wirthlin2015fpga}
M.~Wirthlin, ``High-reliability {FPGA}-based systems: Space, high-energy
  physics, and beyond,'' \emph{Proc. IEEE}, vol. 103, no.~3, pp. 379--389, Mar.
  2015.

\bibitem{pomeranz1994symm}
I.~Pomeranz and S.~M. Reddy, ``On determining symmetries in inputs of logic
  circuits,'' \emph{IEEE Transactions on Computer-Aided Design of Integrated
  Circuits and Systems}, vol.~13, no.~11, pp. 1428--1434, Nov. 1994.

\bibitem{mishchenko2003symm}
A.~Mishchenko, ``Fast computation of symmetries in {B}oolean functions,''
  \emph{IEEE Transactions on Computer-Aided Design of Integrated Circuits and
  Systems}, vol.~22, no.~11, pp. 1588--1593, Nov. 2003.

\bibitem{stoica2005spectral}
P.~Stoica and R.~Moses, \emph{Spectral Analysis of Signals}.\hskip 1em plus
  0.5em minus 0.4em\relax Upper Saddle River, NJ: Pearson Prentice Hall, 2005.

\bibitem{karpovsky1976}
M.~G. Karpovsky, \emph{Finite Orthogonal Series in the Design of Digital
  Devices: Analysis, Synthesis, and Optimization}.\hskip 1em plus 0.5em minus
  0.4em\relax New York: John Wiley \& Sons, 1976.

\bibitem{hurst1985spectral}
S.~L. Hurst, D.~M. Miller, and J.~C. Muzio, \emph{Spectral Techniques in
  Digital Logic}.\hskip 1em plus 0.5em minus 0.4em\relax London: Academic
  Press, 1985.

\bibitem{thornton2026ad}
M.~A. Thornton, ``Algebraic diversity: Group-theoretic spectral estimation from
  single observations,'' 2026, arXiv:2604.03634v8.

\bibitem{thornton2026principles}
------, ``Algebraic diversity: Principles of a group-theoretic approach to
  signal processing,'' 2026, arXiv:2604.19983v7.

\bibitem{thornton2026unification}
------, ``Unification of signal transform theory,'' 2026, arXiv:2605.11589.

\bibitem{thornton2026continuous}
------, ``Continuous algebraic diversity: Unifying spectral, wavelet, and
  time-frequency analysis via {L}ie group actions,'' 2026, arXiv:2605.00848.

\bibitem{thornton2026gevp}
------, ``Polynomial-time optimal group selection via the double-commutator
  eigenvalue problem,'' 2026, arXiv:2605.00834v3.

\bibitem{thornton2026generators}
------, ``Matched generators for the {K}arhunen-{L}o\`eve transform: A
  double-commutator eigenvalue theory,'' 2026, arXiv:2607.08788.

\bibitem{thornton2026shrinkage}
------, ``Symmetry-aware convex shrinkage for high-dimensional covariance
  estimation,'' 2026, arXiv:2605.17111.

\bibitem{thornton2026quantum}
------, ``Quantum algebraic diversity: Single-copy density matrix estimation
  via group-structured measurements,'' 2026, arXiv:2604.03725.

\bibitem{karnaugh1953map}
M.~Karnaugh, ``The map method for synthesis of combinational logic circuits,''
  \emph{Transactions of the American Institute of Electrical Engineers, Part I:
  Communication and Electronics}, vol.~72, no.~5, pp. 593--599, 1953.

\bibitem{golomb1959}
S.~W. Golomb, ``On the classification of {Boolean} functions,'' \emph{IRE
  Trans. Circuit Theory}, vol.~6, no.~5, pp. 176--186, May 1959.

\bibitem{harrison1964}
M.~A. Harrison, ``On the classification of {B}oolean functions by the general
  linear and affine groups,'' \emph{Journal of the Society for Industrial and
  Applied Mathematics}, vol.~12, no.~2, pp. 285--299, 1964.

\bibitem{edwards1975}
C.~R. Edwards, ``The application of the {R}ademacher-{W}alsh transform to
  {B}oolean function classification and threshold logic synthesis,'' \emph{IEEE
  Transactions on Computers}, vol. C-24, no.~1, pp. 48--62, Jan. 1975.

\bibitem{lechner1971}
R.~J. Lechner, ``Harmonic analysis of switching functions,'' in \emph{Recent
  Developments in Switching Theory}, ser. Electrical Science Series,
  A.~Mukhopadhyay, Ed.\hskip 1em plus 0.5em minus 0.4em\relax New York:
  Academic Press, 1971, pp. 121--228.

\bibitem{maiorana1991}
J.~A. Maiorana, ``A classification of the cosets of the {R}eed-{M}uller code
  {$\mathcal{R}(1,6)$},'' \emph{Mathematics of Computation}, vol.~57, no. 195,
  pp. 403--414, 1991.

\bibitem{thornton2001spectral}
M.~A. Thornton, R.~Drechsler, and D.~M. Miller, \emph{Spectral Techniques in
  {VLSI} {CAD}}.\hskip 1em plus 0.5em minus 0.4em\relax Boston, MA: Kluwer
  Academic Publishers, 2001.

\bibitem{thornton2014vsim}
M.~A. Thornton, \emph{Modeling Digital Switching Circuits with Linear Algebra},
  ser. Synthesis Lectures on Digital Circuits and Systems.\hskip 1em plus 0.5em
  minus 0.4em\relax San Rafael, CA: Morgan \& Claypool, 2014, no.~44,
  {R}eissued by Springer, Cham, doi:10.1007/978-3-031-79867-2.

\bibitem{carlet2021book}
C.~Carlet, \emph{Boolean Functions for Cryptography and Coding Theory}.\hskip
  1em plus 0.5em minus 0.4em\relax Cambridge: Cambridge University Press, 2021.

\bibitem{foote2000wreath}
R.~Foote, G.~Mirchandani, D.~N. Rockmore, D.~Healy, and T.~Olson, ``A wreath
  product group approach to signal and image processing. {I}. {M}ultiresolution
  analysis,'' \emph{IEEE Transactions on Signal Processing}, vol.~48, no.~1,
  pp. 102--132, Jan. 2000.

\bibitem{rosenkrantz2019ncf}
D.~J. Rosenkrantz, M.~V. Marathe, S.~S. Ravi, and R.~E. Stearns, ``Symmetry
  properties of nested canalyzing functions,'' \emph{Discrete Mathematics \&
  Theoretical Computer Science}, vol.~21, no.~4, 2019, article 19.

\bibitem{savicky1994bent}
P.~Savick{\'y}, ``On the bent {Boolean} functions that are symmetric,''
  \emph{European Journal of Combinatorics}, vol.~15, no.~4, pp. 407--410, 1994.

\bibitem{sturmfels2008}
B.~Sturmfels, \emph{Algorithms in Invariant Theory}, 2nd~ed., ser. Texts and
  Monographs in Symbolic Computation.\hskip 1em plus 0.5em minus 0.4em\relax
  Vienna: Springer, 2008.

\bibitem{diaconis1988}
P.~Diaconis, \emph{Group Representations in Probability and Statistics}, ser.
  Lecture Notes--Monograph Series.\hskip 1em plus 0.5em minus 0.4em\relax
  Hayward, CA: Institute of Mathematical Statistics, 1988, vol.~11.

\bibitem{serre1977linear}
J.-P. Serre, \emph{Linear Representations of Finite Groups}, ser. Graduate
  Texts in Mathematics.\hskip 1em plus 0.5em minus 0.4em\relax New York:
  Springer, 1977, vol.~42.

\bibitem{fulton1991rep}
W.~Fulton and J.~Harris, \emph{Representation Theory: A First Course}, ser.
  Graduate Texts in Mathematics.\hskip 1em plus 0.5em minus 0.4em\relax New
  York: Springer, 1991, vol. 129.

\bibitem{burnside1911}
W.~Burnside, \emph{Theory of Groups of Finite Order}, 2nd~ed.\hskip 1em plus
  0.5em minus 0.4em\relax Cambridge: Cambridge University Press, 1911.

\bibitem{neumann1979burnside}
P.~M. Neumann, ``A lemma that is not {Burnside's},'' \emph{Math. Scientist},
  vol.~4, no.~2, pp. 133--141, 1979.

\bibitem{berman1967group}
S.~D. Berman, ``On the theory of group codes,'' \emph{Cybernetics}, vol.~3,
  no.~1, pp. 25--31, 1967, {E}nglish translation of Kibernetika 3(1):31--39,
  1967.

\bibitem{delsarte1973}
P.~Delsarte, \emph{An Algebraic Approach to the Association Schemes of Coding
  Theory}, ser. Philips Research Reports Supplements.\hskip 1em plus 0.5em
  minus 0.4em\relax Philips Research Laboratories, 1973, no.~10, thesis,
  Universit{\'e} Catholique de Louvain, June 1973.

\bibitem{shannon1949synthesis}
C.~E. Shannon, ``The synthesis of two-terminal switching circuits,'' \emph{Bell
  System Technical Journal}, vol.~28, no.~1, pp. 59--98, 1949.

\bibitem{bryant1986bdd}
R.~E. Bryant, ``Graph-based algorithms for {Boolean} function manipulation,''
  \emph{IEEE Transactions on Computers}, vol. C-35, no.~8, pp. 677--691, Aug.
  1986.

\bibitem{lavagno1995timed}
L.~Lavagno, P.~C. McGeer, A.~Saldanha, and A.~L. Sangiovanni-Vincentelli,
  ``Timed {Shannon} circuits: A power-efficient design style and synthesis
  tool,'' in \emph{Proc. 32nd Design Automation Conference (DAC)}, 1995, pp.
  254--260.

\bibitem{thornton2002timedshannon}
M.~A. Thornton, R.~Drechsler, and D.~M. Miller, ``Multi-output timed {Shannon}
  circuits,'' in \emph{Proc. IEEE Computer Society Annual Symposium on VLSI
  (ISVLSI)}, 2002, pp. 47--52.

\bibitem{thornton2002haar}
M.~A. Thornton, R.~Drechsler, and W.~G{\"u}nther, ``Logic circuit equivalence
  checking using {Haar} spectral coefficients and partial {BDDs},'' \emph{VLSI
  Design}, vol. 2002, no.~1, pp. 53--64, 2002.

\bibitem{bollig1996np}
B.~Bollig and I.~Wegener, ``Improving the variable ordering of {OBDDs} is
  {NP}-complete,'' \emph{IEEE Transactions on Computers}, vol.~45, no.~9, pp.
  993--1002, 1996.

\bibitem{rudell1993sifting}
R.~Rudell, ``Dynamic variable ordering for ordered binary decision diagrams,''
  in \emph{Proc. IEEE/ACM International Conference on Computer-Aided Design
  (ICCAD)}, Nov. 1993, pp. 42--47.

\bibitem{panda1994symmetry}
S.~Panda, F.~Somenzi, and B.~F. Plessier, ``Symmetry detection and dynamic
  variable ordering of decision diagrams,'' in \emph{Proc. IEEE/ACM
  International Conference on Computer-Aided Design (ICCAD)}, 1994, pp.
  628--631.

\bibitem{delsarte1998}
P.~Delsarte and V.~I. Levenshtein, ``Association schemes and coding theory,''
  \emph{IEEE Transactions on Information Theory}, vol.~44, no.~6, pp.
  2477--2504, Oct. 1998.

\bibitem{macwilliams1977}
F.~J. MacWilliams and N.~J.~A. Sloane, \emph{The Theory of Error-Correcting
  Codes}, ser. North-Holland Mathematical Library.\hskip 1em plus 0.5em minus
  0.4em\relax Amsterdam: North-Holland, 1977, vol.~16.

\bibitem{rothaus1976bent}
O.~S. Rothaus, ``On ``bent'' functions,'' \emph{Journal of Combinatorial
  Theory, Series A}, vol.~20, no.~3, pp. 300--305, May 1976.

\bibitem{yellott1992}
J.~I. Yellott and G.~J. Iverson, ``Uniqueness properties of higher-order
  autocorrelation functions,'' \emph{Journal of the Optical Society of America
  A}, vol.~9, no.~3, pp. 388--404, Mar. 1992.

\bibitem{kakarala2012bispectrum}
R.~Kakarala, ``The bispectrum as a source of phase-sensitive invariants for
  {F}ourier descriptors: A group-theoretic approach,'' \emph{Journal of
  Mathematical Imaging and Vision}, vol.~44, no.~3, pp. 341--353, 2012.

\bibitem{cameron1999permutation}
P.~J. Cameron, \emph{Permutation Groups}, ser. London Mathematical Society
  Student Texts.\hskip 1em plus 0.5em minus 0.4em\relax Cambridge: Cambridge
  University Press, 1999, vol.~45.

\bibitem{hoyos2026exact}
P.~Hoyos, S.~Ubaru, D.~Huh, V.~Kalantzis, K.~L. Clarkson, M.~Kilmer, H.~Avron,
  and L.~Horesh, ``Exact symmetry as algebra: Group-structured tensors that
  enforce physical selection rules,'' \emph{arXiv preprint arXiv:2605.20440},
  2026.

\bibitem{zhang1995gac}
X.-M. Zhang and Y.~Zheng, ``{GAC} -- the criterion for global avalanche
  characteristics of cryptographic functions,'' \emph{Journal of Universal
  Computer Science}, vol.~1, no.~5, pp. 320--337, 1995.

\bibitem{tao2006additive}
T.~Tao and V.~H. Vu, \emph{Additive Combinatorics}, ser. Cambridge Studies in
  Advanced Mathematics.\hskip 1em plus 0.5em minus 0.4em\relax Cambridge:
  Cambridge University Press, 2006, vol. 105.

\bibitem{gowers2001szemeredi}
W.~T. Gowers, ``A new proof of {S}zemer{\'e}di's theorem,'' \emph{Geometric and
  Functional Analysis}, vol.~11, pp. 465--588, 2001.

\bibitem{webster1985sbox}
A.~F. Webster and S.~E. Tavares, ``On the design of {S}-boxes,'' in
  \emph{Advances in Cryptology, CRYPTO '85}, ser. Lecture Notes in Computer
  Science, vol. 218.\hskip 1em plus 0.5em minus 0.4em\relax Springer, 1986, pp.
  523--534.

\bibitem{blum1990selftesting}
M.~Blum, M.~Luby, and R.~Rubinfeld, ``Self-testing/correcting with applications
  to numerical problems,'' in \emph{Proceedings of the 22nd Annual ACM
  Symposium on Theory of Computing (STOC)}, 1990, pp. 73--83.

\bibitem{bellare1995linearity}
M.~Bellare, D.~Coppersmith, J.~H{\aa}stad, M.~Kiwi, and M.~Sudan, ``Linearity
  testing in characteristic two,'' in \emph{Proceedings of the 36th Annual
  Symposium on Foundations of Computer Science (FOCS)}, 1995, pp. 432--441.

\bibitem{alon2003lowdegree}
N.~Alon, T.~Kaufman, M.~Krivelevich, S.~Litsyn, and D.~Ron, ``Testing
  low-degree polynomials over {GF(2)},'' in \emph{Approximation, Randomization,
  and Combinatorial Optimization (RANDOM-APPROX 2003)}, ser. Lecture Notes in
  Computer Science, vol. 2764.\hskip 1em plus 0.5em minus 0.4em\relax Springer,
  2003, pp. 188--199.

\bibitem{evertse1987}
J.-H. Evertse, ``Linear structures in blockciphers,'' in \emph{Advances in
  Cryptology ({EUROCRYPT} '87)}, ser. Lecture Notes in Computer Science,
  D.~Chaum and W.~L. Price, Eds., vol. 304.\hskip 1em plus 0.5em minus
  0.4em\relax Berlin, Heidelberg: Springer, 1988, pp. 249--266.

\bibitem{lai1994linear}
X.~Lai, ``Additive and linear structures of cryptographic functions,'' in
  \emph{Fast Software Encryption ({FSE} 1994)}, ser. Lecture Notes in Computer
  Science, B.~Preneel, Ed., vol. 1008.\hskip 1em plus 0.5em minus 0.4em\relax
  Springer, 1995, pp. 75--85.

\bibitem{dubuc2001linear}
S.~Dubuc, ``Characterization of linear structures,'' \emph{Designs, Codes and
  Cryptography}, vol.~22, no.~1, pp. 33--45, 2001.

\bibitem{odonnell2014}
R.~O'Donnell, \emph{Analysis of {B}oolean Functions}.\hskip 1em plus 0.5em
  minus 0.4em\relax New York, NY: Cambridge University Press, 2014.

\bibitem{carter1979universal}
J.~L. Carter and M.~N. Wegman, ``Universal classes of hash functions,''
  \emph{Journal of Computer and System Sciences}, vol.~18, no.~2, pp. 143--154,
  1979.

\bibitem{gomes2006nips}
C.~P. Gomes, A.~Sabharwal, and B.~Selman, ``Near-uniform sampling of
  combinatorial spaces using {XOR} constraints,'' in \emph{Advances in Neural
  Information Processing Systems 19 (NIPS 2006)}, 2006, pp. 481--488.

\bibitem{alon1997linear}
N.~Alon, M.~Dietzfelbinger, P.~B. Miltersen, E.~Petrank, and G.~Tardos, ``Is
  linear hashing good?'' in \emph{Proceedings of the 29th Annual ACM Symposium
  on Theory of Computing (STOC)}, 1997, pp. 465--474.

\bibitem{jaber2025linear}
M.~Jaber, V.~M. Kumar, and D.~Zuckerman, ``Linear hashing is optimal,''
  arXiv:2505.14061, 2025.

\bibitem{chakrabarty1998optimal}
K.~Chakrabarty, B.~T. Murray, and J.~P. Hayes, ``Optimal zero-aliasing space
  compaction of test responses,'' \emph{IEEE Trans. Comput.}, vol.~47, no.~11,
  pp. 1171--1187, 1998.

\bibitem{chakrabarty1998parity}
K.~Chakrabarty and J.~P. Hayes, ``Zero-aliasing space compaction of test
  responses using multiple parity signatures,'' \emph{IEEE Trans. Very Large
  Scale Integr. (VLSI) Syst.}, vol.~6, no.~2, pp. 309--313, 1998.

\bibitem{matsui1993linear}
M.~Matsui, ``Linear cryptanalysis method for {DES} cipher,'' in \emph{Advances
  in Cryptology, EUROCRYPT '93}, ser. Lecture Notes in Computer Science, vol.
  765, 1994, pp. 386--397.

\bibitem{chakraborty2016approxmc2}
S.~Chakraborty, K.~S. Meel, and M.~Y. Vardi, ``Algorithmic improvements in
  approximate counting for probabilistic inference: From linear to logarithmic
  {SAT} calls,'' in \emph{Proc. 25th Int. Joint Conf. Artificial Intelligence
  ({IJCAI})}, 2016, pp. 3569--3575.

\bibitem{agrawal2020sparsity}
D.~Agrawal, {Bhavishya}, and K.~S. Meel, ``On the sparsity of {XORs} in
  approximate model counting,'' in \emph{Theory and Applications of
  Satisfiability Testing (SAT)}, ser. Lecture Notes in Computer Science, vol.
  12178, 2020, pp. 250--266.

\bibitem{soos2009cms}
M.~Soos, K.~Nohl, and C.~Castelluccia, ``Extending {SAT} solvers to
  cryptographic problems,'' in \emph{Theory and Applications of Satisfiability
  Testing ({SAT} 2009)}, ser. Lecture Notes in Computer Science, O.~Kullmann,
  Ed., vol. 5584.\hskip 1em plus 0.5em minus 0.4em\relax Springer, 2009, pp.
  244--257.

\bibitem{bell1970vitreous}
R.~J. Bell and P.~Dean, ``Atomic vibrations in vitreous silica,''
  \emph{Discussions of the Faraday Society}, vol.~50, pp. 55--61, 1970.

\bibitem{thouless1974}
D.~J. Thouless, ``Electrons in disordered systems and the theory of
  localization,'' \emph{Physics Reports}, vol.~13, no.~3, pp. 93--142, 1974.

\bibitem{hill1973}
M.~O. Hill, ``Diversity and evenness: A unifying notation and its
  consequences,'' \emph{Ecology}, vol.~54, no.~2, pp. 427--432, Mar. 1973.

\bibitem{kish1965}
L.~Kish, \emph{Survey Sampling}.\hskip 1em plus 0.5em minus 0.4em\relax New
  York: Wiley, 1965.

\bibitem{kong1994sequential}
A.~Kong, J.~S. Liu, and W.~H. Wong, ``Sequential imputations and {Bayesian}
  missing data problems,'' \emph{Journal of the American Statistical
  Association}, vol.~89, no. 425, pp. 278--288, 1994.

\bibitem{roy2007effective}
O.~Roy and M.~Vetterli, ``The effective rank: A measure of effective
  dimensionality,'' in \emph{Proc. 15th European Signal Processing Conference
  (EUSIPCO)}, 2007, pp. 606--610.

\bibitem{renyi1961measures}
A.~R{\'e}nyi, ``On measures of entropy and information,'' in \emph{Proc. Fourth
  Berkeley Symposium on Mathematical Statistics and Probability}, vol.~1, 1961,
  pp. 547--561.

\bibitem{yang1991lgsynth}
S.~Yang, ``Logic synthesis and optimization benchmarks user guide, version
  3.0,'' Microelectronics Center of North Carolina (MCNC), Research Triangle
  Park, NC, Tech. Rep., Jan. 1991.

\bibitem{brayton1984espresso}
R.~K. Brayton, G.~D. Hachtel, C.~T. McMullen, and A.~L.
  Sangiovanni-Vincentelli, \emph{Logic Minimization Algorithms for {VLSI}
  Synthesis}, ser. The Kluwer International Series in Engineering and Computer
  Science.\hskip 1em plus 0.5em minus 0.4em\relax Boston, MA: Kluwer Academic
  Publishers, 1984, vol.~2.

\bibitem{valiant1979enumeration}
L.~G. Valiant, ``The complexity of enumeration and reliability problems,''
  \emph{SIAM J. Comput.}, vol.~8, no.~3, pp. 410--421, 1979.

\bibitem{soos2019bird}
M.~Soos and K.~S. Meel, ``{BIRD}: Engineering an efficient {CNF-XOR} {SAT}
  solver and its applications to approximate model counting,'' in
  \emph{Proceedings of the AAAI Conference on Artificial Intelligence ({AAAI}
  2019)}, vol.~33, no.~01, 2019, pp. 1592--1599.

\bibitem{ermon2014ldpc}
\BIBentryALTinterwordspacing
S.~Ermon, C.~P. Gomes, A.~Sabharwal, and B.~Selman, ``Low-density parity
  constraints for hashing-based discrete integration,'' in \emph{Proceedings of
  the 31st International Conference on Machine Learning ({ICML} 2014)}, ser.
  Proceedings of Machine Learning Research, E.~P. Xing and T.~Jebara, Eds.,
  vol.~32, no.~1.\hskip 1em plus 0.5em minus 0.4em\relax Beijing, China: PMLR,
  2014, pp. 271--279. [Online]. Available:
  \url{https://proceedings.mlr.press/v32/ermon14.html}
\BIBentrySTDinterwordspacing

\bibitem{achlioptas2015codes}
\BIBentryALTinterwordspacing
D.~Achlioptas and P.~Jiang, ``Stochastic integration via error-correcting
  codes,'' in \emph{Proceedings of the Thirty-First Conference on Uncertainty
  in Artificial Intelligence ({UAI} 2015)}, 2015, pp. 22--31. [Online].
  Available: \url{http://auai.org/uai2015/proceedings/papers/248.pdf}
\BIBentrySTDinterwordspacing

\bibitem{soos2024benchmarks}
M.~Soos, ``Benchmarks used for approximate model counting,'' Zenodo dataset,
  version 1, Jan. 2024.

\bibitem{sharma2019ganak}
S.~Sharma, S.~Roy, M.~Soos, and K.~S. Meel, ``{GANAK}: A scalable probabilistic
  exact model counter,'' in \emph{Proc. Int. Joint Conf. Artificial
  Intelligence (IJCAI)}, 2019, pp. 1169--1176.

\bibitem{soos2022arjun}
M.~Soos and K.~S. Meel, ``Arjun: An efficient independent support computation
  technique and its applications to counting and sampling,'' in \emph{Proc.
  IEEE/ACM Int. Conf. Computer-Aided Design (ICCAD)}, 2022, pp. 1--9.

\bibitem{han2012gauss}
C.-S. Han and J.-H.~R. Jiang, ``When {B}oolean satisfiability meets {G}aussian
  elimination in a simplex way,'' in \emph{Computer Aided Verification ({CAV}
  2012)}, ser. Lecture Notes in Computer Science, P.~Madhusudan and S.~A.
  Seshia, Eds., vol. 7358.\hskip 1em plus 0.5em minus 0.4em\relax Berlin,
  Heidelberg: Springer, 2012, pp. 410--426.

\bibitem{soos2020tinted}
M.~Soos, S.~Gocht, and K.~S. Meel, ``Tinted, detached, and lazy {CNF-XOR}
  solving and its applications to counting and sampling,'' in \emph{Computer
  Aided Verification ({CAV} 2020), Part {I}}, ser. Lecture Notes in Computer
  Science, S.~K. Lahiri and C.~Wang, Eds., vol. 12224.\hskip 1em plus 0.5em
  minus 0.4em\relax Cham: Springer, 2020, pp. 463--484.

\bibitem{lv2012gf}
J.~Lv, P.~Kalla, and F.~Enescu, ``Efficient {G}r{\"o}bner basis reductions for
  formal verification of {G}alois field multipliers,'' in \emph{Proc. Design,
  Automation and Test in Europe ({DATE})}, 2012, pp. 899--904.

\bibitem{kaufmann2019sat}
D.~Kaufmann, A.~Biere, and M.~Kauers, ``Verifying large multipliers by
  combining {SAT} and computer algebra,'' in \emph{Proc. 19th Int. Conf. Formal
  Methods in Computer Aided Design ({FMCAD})}.\hskip 1em plus 0.5em minus
  0.4em\relax FMCAD Inc., 2019, pp. 28--36.

\bibitem{biere2024cadical}
A.~Biere, T.~Faller, K.~Fazekas, M.~Fleury, N.~Froleyks, and F.~Pollitt,
  ``{CaDiCaL} 2.0,'' in \emph{Computer Aided Verification ({CAV} 2024), Part
  {I}}, ser. Lecture Notes in Computer Science, A.~Gurfinkel and V.~Ganesh,
  Eds., vol. 14681.\hskip 1em plus 0.5em minus 0.4em\relax Cham: Springer,
  2024, pp. 133--152.

\bibitem{ignatiev2018pysat}
A.~Ignatiev, A.~Morgado, and J.~Marques-Silva, ``{PySAT}: A {P}ython toolkit
  for prototyping with {SAT} oracles,'' in \emph{Theory and Applications of
  Satisfiability Testing ({SAT} 2018)}, ser. Lecture Notes in Computer Science,
  O.~Beyersdorff and C.~M. Wintersteiger, Eds., vol. 10929.\hskip 1em plus
  0.5em minus 0.4em\relax Springer, 2018, pp. 428--437.

\bibitem{buss1999}
S.~R. Buss, D.~Grigoriev, R.~Impagliazzo, and T.~Pitassi, ``Linear gaps between
  degrees for the polynomial calculus modulo distinct primes,'' in \emph{Proc.
  31st Annual {ACM} Symposium on Theory of Computing ({STOC})}, 1999, pp.
  547--556.

\bibitem{beame1998}
P.~Beame and S.~Riis, ``More on the relative strength of counting principles,''
  in \emph{Proof Complexity and Feasible Arithmetics}, ser. DIMACS Series in
  Discrete Mathematics and Theoretical Computer Science, P.~W. Beame and S.~R.
  Buss, Eds.\hskip 1em plus 0.5em minus 0.4em\relax American Mathematical
  Society, 1997, vol.~39, pp. 13--35.

\bibitem{leberre2010}
D.~{Le Berre} and A.~Parrain, ``The {Sat4j} library, release 2.2,''
  \emph{Journal on Satisfiability, Boolean Modeling and Computation}, vol.~7,
  no. 2--3, pp. 59--64, 2010.

\bibitem{elffers2018}
J.~Elffers and J.~Nordstr{\"o}m, ``Divide and conquer: Towards faster
  pseudo-{Boolean} solving,'' in \emph{Proceedings of the Twenty-Seventh
  International Joint Conference on Artificial Intelligence ({IJCAI}
  2018)}.\hskip 1em plus 0.5em minus 0.4em\relax ijcai.org, 2018, pp.
  1291--1299.

\bibitem{devriendt2021}
J.~Devriendt, A.~Gleixner, and J.~Nordstr{\"o}m, ``Learn to relax: Integrating
  0-1 integer linear programming with pseudo-{Boolean} conflict-driven
  search,'' \emph{Constraints}, vol.~26, no.~1, pp. 26--55, 2021.

\bibitem{hojny2025scip10}
\BIBentryALTinterwordspacing
C.~Hojny, M.~Besan{\c{c}}on, K.~Bestuzheva, S.~Borst, J.~Dion{\'i}sio, J.~Ehls,
  L.~Eifler, M.~Ghannam, A.~Gleixner, A.~G{\"o}{\ss}, A.~Hoen, J.~von
  Holly-Ponientzietz, R.~van~der Hulst, D.~Kamp, T.~Koch, K.~Kofler, J.~Lentz,
  M.~L{\"u}bbecke, S.~J. Maher, P.~M. Meinhold, G.~Mexi, T.~Mohr,
  E.~M{\"u}hmer, K.~K. Patel, M.~E. Pfetsch, S.~Pokutta, C.~R. Groba,
  F.~Serrano, Y.~Shinano, M.~Turner, S.~Vigerske, M.~Walter, D.~Weninger, and
  L.~Xu, ``The {SCIP} optimization suite 10.0,'' 2025. [Online]. Available:
  \url{https://arxiv.org/abs/2511.18580}
\BIBentrySTDinterwordspacing

\bibitem{huangfu2018}
Q.~Huangfu and J.~A.~J. Hall, ``Parallelizing the dual revised simplex
  method,'' \emph{Mathematical Programming Computation}, vol.~10, no.~1, pp.
  119--142, 2018.

\bibitem{bryant2022}
R.~E. Bryant, A.~Biere, and M.~J.~H. Heule, ``Clausal proofs for
  pseudo-{Boolean} reasoning,'' in \emph{Tools and Algorithms for the
  Construction and Analysis of Systems, 28th International Conference, {TACAS}
  2022, Proceedings, Part {I}}, ser. Lecture Notes in Computer Science,
  D.~Fisman and G.~Rosu, Eds., vol. 13243.\hskip 1em plus 0.5em minus
  0.4em\relax Springer, 2022, pp. 443--461.

\bibitem{bryant2022pgpbs}
R.~E. Bryant, ``{rebryant/pgpbs-artifact: TACAS22},''
  \url{https://github.com/rebryant/pgpbs-artifact}, 2022, software artifact for
  the TACAS 2022 paper, archived on Zenodo.

\bibitem{een2005}
N.~E{\'e}n and A.~Biere, ``Effective preprocessing in {SAT} through variable
  and clause elimination,'' in \emph{Theory and Applications of Satisfiability
  Testing ({SAT} 2005)}, ser. Lecture Notes in Computer Science, F.~Bacchus and
  T.~Walsh, Eds., vol. 3569.\hskip 1em plus 0.5em minus 0.4em\relax Springer,
  2005, pp. 61--75.

\bibitem{beame2026}
P.~Beame and G.~Sun, ``Extending {CDCL} to disjunctions of parity equations,''
  in \emph{29th International Conference on Theory and Applications of
  Satisfiability Testing ({SAT} 2026)}, ser. Leibniz International Proceedings
  in Informatics (LIPIcs), A.~Ignatiev and S.~Szeider, Eds., vol. 377.\hskip
  1em plus 0.5em minus 0.4em\relax Dagstuhl, Germany: Schloss Dagstuhl --
  Leibniz-Zentrum f{\"u}r Informatik, 2026, pp. 5:1--5:21.

\bibitem{xorcle}
G.~Sun, ``Xorcle,'' \url{https://github.com/glenn-sun/xorcle}, 2026, {T}he
  solver of Beame and Sun's {SAT} 2026 paper. The runs used version 0.1 (commit
  daa0125, March 2026) and the development version 0.2.0-dev (commit af93ea6,
  August 2026). Release 0.2.0 followed in November 2026.

\bibitem{gleixner2021}
A.~M. Gleixner, G.~Hendel, G.~Gamrath, T.~Achterberg, M.~Bastubbe, T.~Berthold,
  P.~Christophel, K.~Jarck, T.~Koch, J.~T. Linderoth, M.~E. L{\"u}bbecke, H.~D.
  Mittelmann, D.~B. {\"O}zyurt, T.~K. Ralphs, D.~Salvagnin, and Y.~Shinano,
  ``{MIPLIB} 2017: Data-driven compilation of the 6th mixed-integer programming
  library,'' \emph{Mathematical Programming Computation}, vol.~13, no.~3, pp.
  443--490, 2021.

\bibitem{kannan1979}
R.~Kannan and A.~Bachem, ``Polynomial algorithms for computing the {Smith} and
  {Hermite} normal forms of an integer matrix,'' \emph{SIAM Journal on
  Computing}, vol.~8, no.~4, pp. 499--507, 1979.

\bibitem{aardal2000}
K.~Aardal, C.~A.~J. Hurkens, and A.~K. Lenstra, ``Solving a system of linear
  {Diophantine} equations with lower and upper bounds on the variables,''
  \emph{Mathematics of Operations Research}, vol.~25, no.~3, pp. 427--442,
  2000.

\bibitem{cong1999cut}
J.~Cong, C.~Wu, and Y.~Ding, ``Cut ranking and pruning: Enabling a general and
  efficient {FPGA} mapping solution,'' in \emph{Proc. ACM/SIGDA Int. Symp.
  Field Programmable Gate Arrays (FPGA)}, 1999, pp. 29--35.

\bibitem{zhou2018adjustable}
X.~Zhou, L.~Wang, P.~Zhao, and A.~Mishchenko, ``Fast adjustable {NPN}
  classification using generalized symmetries,'' in \emph{Proc. 28th Int. Conf.
  Field Programmable Logic and Applications (FPL)}, 2018.

\bibitem{zhou2020exact}
X.~Zhou, L.~Wang, and A.~Mishchenko, ``Fast exact {NPN} classification by
  co-designing canonical form and its computation algorithm,'' \emph{IEEE
  Transactions on Computers}, vol.~69, no.~9, pp. 1293--1307, 2020.

\bibitem{brayton2010abc}
R.~Brayton and A.~Mishchenko, ``{ABC}: An academic industrial-strength
  verification tool,'' in \emph{Computer Aided Verification (CAV)}, ser.
  Lecture Notes in Computer Science, vol. 6174, 2010, pp. 24--40.

\bibitem{hinsberger1998large}
U.~Hinsberger and R.~Kolla, ``{Boolean} matching for large libraries,'' in
  \emph{Proc. 35th Design Automation Conf. (DAC)}, 1998, pp. 206--211.

\bibitem{hurst1978logical}
S.~L. Hurst, \emph{The Logical Processing of Digital Signals}, ser. Computer
  Systems Engineering Series.\hskip 1em plus 0.5em minus 0.4em\relax New York
  and London: Crane, Russak and Edward Arnold, 1978.

\bibitem{harrison1963npn}
M.~A. Harrison, ``The number of equivalence classes of {Boolean} functions
  under groups containing negation,'' \emph{IEEE Trans. Electron. Comput.},
  vol.~12, pp. 559--561, 1963.

\bibitem{filmusnpn}
Y.~Filmus, ``{NPN} equivalence classes of {B}oolean functions,''
  \url{https://yuvalfilmus.cs.technion.ac.il/publications/manuscripts/?id=787},
  list of one representative per class for up to five variables; accessed Nov.
  18, 2026.

\bibitem{brglez1985}
F.~Brglez and H.~Fujiwara, ``A neutral netlist of 10 combinational benchmark
  circuits and a target translator in {Fortran},'' Distributed on tape to
  participants of the Special Session on ATPG and Fault Simulation, IEEE Int.
  Symp. Circuits and Systems (ISCAS), Jun. 1985.

\bibitem{hansen1999iscas}
M.~C. Hansen, H.~Yalcin, and J.~P. Hayes, ``Unveiling the {ISCAS}-85
  benchmarks: A case study in reverse engineering,'' \emph{IEEE Des. Test
  Comput.}, vol.~16, no.~3, pp. 72--80, 1999.

\bibitem{pasandi2019qals}
G.~Pasandi, S.~Nazarian, and M.~Pedram, ``Approximate logic synthesis: A
  reinforcement learning-based technology mapping approach,'' in
  \emph{Proceedings of the 20th International Symposium on Quality Electronic
  Design (ISQED)}, 2019, arXiv:1902.00478.

\bibitem{mailhot1990dontcare}
F.~Mailhot and G.~D. Micheli, ``Technology mapping using {B}oolean matching and
  don't care sets,'' in \emph{Proceedings of the European Design Automation
  Conference (EURO-DAC)}, 1990, pp. 212--216.

\bibitem{benini1997survey}
L.~Benini and G.~D. Micheli, ``A survey of {B}oolean matching techniques for
  library binding,'' \emph{ACM Transactions on Design Automation of Electronic
  Systems}, vol.~2, no.~3, pp. 193--226, 1997.

\bibitem{skywater2020pdk}
{SkyWater Technology Foundry} and {Google}, ``{SkyWater} open source {PDK},
  standard-cell library sky130\_fd\_sc\_hd,''
  \url{https://github.com/google/skywater-pdk}, 2020.

\bibitem{mishchenko2009mfs}
A.~Mishchenko, R.~K. Brayton, J.-H.~R. Jiang, and S.~Jang, ``Scalable
  don't-care-based logic optimization and resynthesis,'' in \emph{Proc.
  ACM/SIGDA Int. Symp. Field Programmable Gate Arrays (FPGA)}, 2009, pp.
  151--160.

\bibitem{luks1999hypergraph}
E.~M. Luks, ``Hypergraph isomorphism and structural equivalence of {Boolean}
  functions,'' in \emph{Proc. 31st Annual ACM Symposium on Theory of Computing
  (STOC)}, 1999, pp. 652--658.

\bibitem{moore2006sdd}
J.~Moore, K.~Fazel, M.~A. Thornton, and D.~M. Miller, ``{Boolean} function
  matching using {Walsh} spectral decision diagrams,'' in \emph{Proc. IEEE
  Dallas/CAS Workshop on Design, Applications, Integration and Software
  (DCAS)}, 2006, pp. 127--130.

\bibitem{zhang2022shannon}
J.~Zhang, W.~Guo, G.~Yang, Y.~Zhu, and X.~Lv, ``A heuristic {Boolean} {NPN}
  equivalent matching verification method based on {Shannon} decomposition,''
  \emph{IEEE Access}, vol.~10, pp. 120\,369--120\,382, 2022.

\bibitem{mohnke1995limits}
J.~Mohnke, P.~Molitor, and S.~Malik, ``Limits of using signatures for
  permutation independent {Boolean} comparison,'' in \emph{Proc. Asia and South
  Pacific Design Automation Conf. (ASP-DAC)}, 1995, pp. 459--464.

\bibitem{schrijver2005terwilliger}
A.~Schrijver, ``New code upper bounds from the {T}erwilliger algebra and
  semidefinite programming,'' \emph{IEEE Transactions on Information Theory},
  vol.~51, no.~8, pp. 2859--2866, Aug. 2005.

\bibitem{schonemann1966procrustes}
P.~H. Sch{\"o}nemann, ``A generalized solution of the orthogonal {Procrustes}
  problem,'' \emph{Psychometrika}, vol.~31, no.~1, pp. 1--10, 1966.

\bibitem{pumir2021procrustes}
T.~Pumir, A.~Singer, and N.~Boumal, ``The generalized orthogonal {Procrustes}
  problem in the high noise regime,'' \emph{Information and Inference: A
  Journal of the IMA}, 2021, arXiv:1907.01145.

\bibitem{umeyama1988}
S.~Umeyama, ``An eigendecomposition approach to weighted graph matching
  problems,'' \emph{IEEE Trans. Pattern Anal. Mach. Intell.}, vol.~10, no.~5,
  pp. 695--703, 1988.

\bibitem{burkard2013qap}
R.~E. Burkard, ``Quadratic assignment problems,'' in \emph{Handbook of
  Combinatorial Optimization}, P.~M. Pardalos, D.-Z. Du, and R.~L. Graham,
  Eds.\hskip 1em plus 0.5em minus 0.4em\relax New York: Springer, 2013, pp.
  2741--2814.

\bibitem{mckay2014practical}
B.~D. McKay and A.~Piperno, ``Practical graph isomorphism, {II},'' \emph{J.
  Symbolic Comput.}, vol.~60, pp. 94--112, 2014, arXiv:1301.1493.

\bibitem{hogbom1974clean}
J.~A. H{\"o}gbom, ``Aperture synthesis with a non-regular distribution of
  interferometer baselines,'' \emph{Astronomy and Astrophysics Supplement
  Series}, vol.~15, p. 417, 1974.

\bibitem{carlet1993partially}
C.~Carlet, ``Partially-bent functions,'' in \emph{Advances in Cryptology,
  CRYPTO '92}, ser. Lecture Notes in Computer Science, vol. 740.\hskip 1em plus
  0.5em minus 0.4em\relax Springer, 1993, pp. 280--291.

\bibitem{luccio1999autosymmetric}
F.~Luccio and L.~Pagli, ``On a new {B}oolean function with applications,''
  \emph{IEEE Transactions on Computers}, vol.~48, no.~3, pp. 296--310, 1999.

\bibitem{drechsler1994okfdd}
R.~Drechsler, A.~Sarabi, M.~Theobald, B.~Becker, and M.~A. Perkowski,
  ``Efficient representation and manipulation of switching functions based on
  ordered {K}ronecker functional decision diagrams,'' in \emph{Proceedings of
  the 31st Design Automation Conference}, 1994, pp. 415--419.

\bibitem{curtis1962design}
H.~A. Curtis, \emph{A New Approach to the Design of Switching Circuits}.\hskip
  1em plus 0.5em minus 0.4em\relax Princeton, NJ: D. Van Nostrand, 1962.

\bibitem{mishchenko2001bidecomp}
A.~Mishchenko, B.~Steinbach, and M.~A. Perkowski, ``An algorithm for
  bi-decomposition of logic functions,'' in \emph{Proceedings of the 38th
  Design Automation Conference}, 2001, pp. 103--108.

\bibitem{hashemi2018blasys}
S.~Hashemi, H.~Tann, and S.~Reda, ``{BLASYS}: Approximate logic synthesis using
  {B}oolean matrix factorization,'' in \emph{Proceedings of the 55th Design
  Automation Conference}, 2018.

\bibitem{kravets2000generalized}
V.~N. Kravets and K.~A. Sakallah, ``Generalized symmetries in {B}oolean
  functions,'' in \emph{Proceedings of the IEEE/ACM International Conference on
  Computer-Aided Design}, 2000, pp. 526--532.

\bibitem{thornton2002early}
M.~A. Thornton, K.~Fazel, R.~B. Reese, and C.~Traver, ``Generalized early
  evaluation in self-timed circuits,'' in \emph{Proc. Design, Automation and
  Test in Europe (DATE)}, 2002, pp. 255--259.

\bibitem{tracey1966internal}
J.~H. Tracey, ``Internal state assignments for asynchronous sequential
  machines,'' \emph{IEEE Transactions on Electronic Computers}, vol. EC-15,
  no.~4, pp. 551--560, 1966.

\bibitem{unger1969asynchronous}
S.~H. Unger, \emph{Asynchronous Sequential Switching Circuits}.\hskip 1em plus
  0.5em minus 0.4em\relax New York: Wiley-Interscience, 1969.

\bibitem{mccluskey1959note}
E.~J. McCluskey and S.~H. Unger, ``A note on the number of internal variable
  assignments for sequential switching circuits,'' \emph{IRE Transactions on
  Electronic Computers}, vol. EC-8, no.~4, pp. 439--440, 1959.

\bibitem{kastensmidt2006fpga}
F.~L. Kastensmidt, L.~Carro, and R.~Reis, \emph{Fault-Tolerance Techniques for
  {SRAM}-Based {FPGAs}}, ser. Frontiers in Electronic Testing.\hskip 1em plus
  0.5em minus 0.4em\relax Dordrecht: Springer, 2006, vol.~32.

\bibitem{hu2008rose}
Y.~Hu, Z.~Feng, L.~He, and R.~Majumdar, ``Robust {FPGA} resynthesis based on
  fault-tolerant {Boolean} matching,'' in \emph{Proc. IEEE/ACM International
  Conference on Computer-Aided Design (ICCAD)}, 2008, pp. 706--713.

\bibitem{feng2009ipr}
Z.~Feng, Y.~Hu, L.~He, and R.~Majumdar, ``{IPR}: In-place reconfiguration for
  {FPGA} fault tolerance,'' in \emph{Proc. IEEE/ACM International Conference on
  Computer-Aided Design (ICCAD)}, 2009, pp. 105--108.

\bibitem{das2013lut}
A.~Das, S.~Venkataraman, and A.~Kumar, ``Improving autonomous soft-error
  tolerance of {FPGA} through {LUT} configuration bit manipulation,'' in
  \emph{Proc. International Conference on Field Programmable Logic and
  Applications (FPL)}, 2013.

\bibitem{lucas1878}
{\'E}.~Lucas, ``Th{\'e}orie des fonctions num{\'e}riques simplement
  p{\'e}riodiques,'' \emph{American Journal of Mathematics}, vol.~1, no.~2, pp.
  184--196, 1878.

\bibitem{bellare1996keying}
M.~Bellare, R.~Canetti, and H.~Krawczyk, ``Keying hash functions for message
  authentication,'' in \emph{Advances in Cryptology, CRYPTO '96}, ser. Lecture
  Notes in Computer Science, N.~Koblitz, Ed., vol. 1109.\hskip 1em plus 0.5em
  minus 0.4em\relax Springer, 1996, pp. 1--15.

\bibitem{nist80022}
L.~E. Bassham, A.~L. Rukhin, J.~Soto, J.~R. Nechvatal, M.~E. Smid, E.~B.
  Barker, S.~D. Leigh, M.~Levenson, M.~Vangel, D.~L. Banks, N.~A. Heckert,
  J.~F. Dray, and S.~Vo, ``A statistical test suite for random and pseudorandom
  number generators for cryptographic applications,'' National Institute of
  Standards and Technology, Gaithersburg, MD, Tech. Rep. SP 800-22 Rev. 1a,
  Apr. 2010.

\bibitem{preneel1990propagation}
B.~Preneel, W.~V. Leekwijck, L.~V. Linden, R.~Govaerts, and J.~Vandewalle,
  ``Propagation characteristics of {Boolean} functions,'' in \emph{Advances in
  Cryptology, EUROCRYPT '90}, ser. Lecture Notes in Computer Science, vol. 473,
  1991, pp. 161--173.

\bibitem{bogdanov2007present}
A.~Bogdanov, L.~R. Knudsen, G.~Leander, C.~Paar, A.~Poschmann, M.~J.~B.
  Robshaw, Y.~Seurin, and C.~Vikkelsoe, ``{PRESENT}: An ultra-lightweight block
  cipher,'' in \emph{Cryptographic Hardware and Embedded Systems, CHES 2007},
  ser. Lecture Notes in Computer Science, vol. 4727, 2007, pp. 450--466.

\bibitem{daemen2002rijndael}
J.~Daemen and V.~Rijmen, \emph{The Design of {Rijndael}: {AES}, the Advanced
  Encryption Standard}, ser. Information Security and Cryptography.\hskip 1em
  plus 0.5em minus 0.4em\relax Berlin, Heidelberg: Springer, 2002.

\bibitem{nyberg1993differentially}
K.~Nyberg, ``Differentially uniform mappings for cryptography,'' in
  \emph{Advances in Cryptology, EUROCRYPT '93}, ser. Lecture Notes in Computer
  Science, vol. 765, 1994, pp. 55--64.

\bibitem{leander2007classification}
G.~Leander and A.~Poschmann, ``On the classification of 4 bit {S}-boxes,'' in
  \emph{Arithmetic of Finite Fields (WAIFI 2007)}, ser. Lecture Notes in
  Computer Science, vol. 4547.\hskip 1em plus 0.5em minus 0.4em\relax Springer,
  2007, pp. 159--176.

\bibitem{canteaut2019dlct}
A.~Canteaut, L.~K{\"o}lsch, C.~Li, C.~Li, K.~Li, L.~Qu, and F.~Wiemer, ``On the
  differential-linear connectivity table of vectorial {B}oolean functions,''
  arXiv:1908.07445, 2019.

\bibitem{massey1969shift}
J.~L. Massey, ``Shift-register synthesis and {BCH} decoding,'' \emph{IEEE
  Transactions on Information Theory}, vol.~15, no.~1, pp. 122--127, 1969.

\bibitem{clarke1993spectral}
E.~M. Clarke, K.~L. McMillan, X.~Zhao, M.~Fujita, and J.~Yang, ``Spectral
  transforms for large {B}oolean functions with applications to technology
  mapping,'' in \emph{Proceedings of the 30th {ACM/IEEE} Design Automation
  Conference ({DAC} '93)}.\hskip 1em plus 0.5em minus 0.4em\relax New York, NY:
  ACM, 1993, pp. 54--60.

\bibitem{kushilevitz1993}
E.~Kushilevitz and Y.~Mansour, ``Learning decision trees using the {F}ourier
  spectrum,'' \emph{SIAM Journal on Computing}, vol.~22, no.~6, pp. 1331--1348,
  1993.

\bibitem{bannai1984}
E.~Bannai and T.~Ito, \emph{Algebraic Combinatorics {I}: Association Schemes},
  ser. Mathematics Lecture Note Series.\hskip 1em plus 0.5em minus 0.4em\relax
  Menlo Park, CA: Benjamin/Cummings, 1984, vol.~58.

\bibitem{berlekamp1972rm}
E.~R. Berlekamp and L.~R. Welch, ``Weight distributions of the cosets of the
  (32,6) {R}eed-{M}uller code,'' \emph{IEEE Transactions on Information
  Theory}, vol.~18, no.~1, pp. 203--207, 1972.

\bibitem{rice2011spectral}
J.~E. Rice, J.~C. Muzio, and N.~Anderson, ``New considerations for spectral
  classification of {B}oolean switching functions,'' \emph{VLSI Design}, vol.
  2011, 2011, {A}rt. no.~356137.

\bibitem{soeken2019hybrid}
M.~Soeken, E.~Testa, and D.~M. Miller, ``A hybrid method for spectral
  translation equivalent {B}oolean functions,'' in \emph{Proc. IEEE Pacific Rim
  Conf. Communications, Computers and Signal Processing (PACRIM)}, 2019, pp.
  1--6.

\bibitem{varma1989}
D.~Varma and E.~A. Trachtenberg, ``Design automation tools for efficient
  implementation of logic functions by decomposition,'' \emph{IEEE Trans.
  Comput.-Aided Design Integr. Circuits Syst.}, vol.~8, no.~8, pp. 901--916,
  Aug. 1989.

\bibitem{meinel1997linear}
C.~Meinel, F.~Somenzi, and T.~Theobald, ``Linear sifting of decision
  diagrams,'' in \emph{Proc. 34th Design Automation Conf. (DAC)}, 1997, pp.
  202--207.

\bibitem{gunther1998ltbdd}
W.~G{\"u}nther and R.~Drechsler, ``{BDD} minimization by linear
  transformations,'' in \emph{Proc. Advanced Computer Systems (ACS)}, Szczecin,
  Poland, 1998, pp. 525--532.

\bibitem{thornton2026renesis}
M.~A. Thornton, ``{RENESIS}: Energy-aware synthesis of adiabatic logic from
  irreversible netlists,'' arXiv:2608.17139, 2026.

\bibitem{ermon2013wish}
\BIBentryALTinterwordspacing
S.~Ermon, C.~P. Gomes, A.~Sabharwal, and B.~Selman, ``Taming the curse of
  dimensionality: Discrete integration by hashing and optimization,'' in
  \emph{Proceedings of the 30th International Conference on Machine Learning
  ({ICML} 2013)}, ser. Proceedings of Machine Learning Research, S.~Dasgupta
  and D.~McAllester, Eds., vol.~28, no.~2.\hskip 1em plus 0.5em minus
  0.4em\relax PMLR, 2013, pp. 334--342. [Online]. Available:
  \url{https://proceedings.mlr.press/v28/ermon13.html}
\BIBentrySTDinterwordspacing

\bibitem{kuck2019adaptive}
J.~Kuck, T.~Dao, S.~Zhao, B.~Bartan, A.~Sabharwal, and S.~Ermon, ``Adaptive
  hashing for model counting,'' in \emph{Proc. 35th Conf. Uncertainty in
  Artificial Intelligence ({UAI} 2019)}, ser. PMLR, vol. 115, 2020, pp.
  271--280.

\end{thebibliography}
\end{document}